\documentclass[runningheads,envcountsect]{llncs}
\usepackage[T1]{fontenc}
\usepackage{xcolor}
\usepackage{hyperref}

\usepackage{booktabs}
\usepackage{color}

\usepackage{xspace}
\usepackage{amsmath}
\usepackage{tikz-cd}
\usepackage{txfonts}
\usepackage{cmll} 

\usepackage{cleveref}
\crefname{theorem}{Thm.}{Thm.}
\crefname{proposition}{Prop.}{Prop.}
\usepackage{hyperref}
\newcommand{\restate}[4]{%
\begin{#1}[#3~\proofnote{proof of~\cref{#2}}]%
\label{#2:proof}%
#4%
\end{#1}%
}
\newcommand{\restatable}[5]{%
\begin{#1}[#4~\proofnote{proof in~\cref{#2:proof}}]%
\label{#2}
#5%
\end{#1}%
\newcommand{#3}{%
\begin{#1}[#4~\proofnote{proof of~\cref{#2}}]%
\label{#2:proof}%
#5%
\end{#1}%
}%
}

\newcommand{\qedhere}{\ensuremath{\hfill\blacksquare}}
\newcommand{\anon}{\cdot}

\newdimen\proofrulebreadth \proofrulebreadth=.05em
\newdimen\proofdotseparation \proofdotseparation=1.25ex
\newdimen\proofrulebaseline \proofrulebaseline=2ex
\newcount\proofdotnumber \proofdotnumber=3
\let\then\relax
\def\hfi{\hskip0pt plus.0001fil}
\mathchardef\squigto="3A3B
\newif\ifinsideprooftree\insideprooftreefalse
\newif\ifonleftofproofrule\onleftofproofrulefalse
\newif\ifproofdots\proofdotsfalse
\newif\ifdoubleproof\doubleprooffalse
\let\wereinproofbit\relax
\newdimen\shortenproofleft
\newdimen\shortenproofright
\newdimen\proofbelowshift
\newbox\proofabove
\newbox\proofbelow
\newbox\proofrulename
\def\shiftproofbelow{\let\next\relax\afterassignment\setshiftproofbelow\dimen0 }
\def\shiftproofbelowneg{\def\next{\multiply\dimen0 by-1 }%
\afterassignment\setshiftproofbelow\dimen0 }
\def\setshiftproofbelow{\next\proofbelowshift=\dimen0 }
\def\setproofrulebreadth{\proofrulebreadth}

\def\prooftree{
\ifnum  \lastpenalty=1
\then   \unpenalty
\else   \onleftofproofrulefalse
\fi
\ifonleftofproofrule
\else   \ifinsideprooftree
        \then   \hskip.5em plus1fil
        \fi
\fi
\bgroup
\setbox\proofbelow=\hbox{}\setbox\proofrulename=\hbox{}%
\let\justifies\proofover\let\leadsto\proofoverdots\let\Justifies\proofoverdbl
\let\using\proofusing\let\[\prooftree
\ifinsideprooftree\let\]\endprooftree\fi
\proofdotsfalse\doubleprooffalse
\let\thickness\setproofrulebreadth
\let\shiftright\shiftproofbelow \let\shift\shiftproofbelow
\let\shiftleft\shiftproofbelowneg
\let\ifwasinsideprooftree\ifinsideprooftree
\insideprooftreetrue
\setbox\proofabove=\hbox\bgroup$\displaystyle 
\let\wereinproofbit\prooftree
\shortenproofleft=0pt \shortenproofright=0pt \proofbelowshift=0pt
\onleftofproofruletrue\penalty1
}

\def\eproofbit{
\ifx    \wereinproofbit\prooftree
\then   \ifcase \lastpenalty
        \then   \shortenproofright=0pt  
        \or     \unpenalty\hfil         
        \or     \unpenalty\unskip       
        \else   \shortenproofright=0pt  
        \fi
\fi
\global\dimen0=\shortenproofleft
\global\dimen1=\shortenproofright
\global\dimen2=\proofrulebreadth
\global\dimen3=\proofbelowshift
\global\dimen4=\proofdotseparation
\global\count255=\proofdotnumber
$\egroup  
\shortenproofleft=\dimen0
\shortenproofright=\dimen1
\proofrulebreadth=\dimen2
\proofbelowshift=\dimen3
\proofdotseparation=\dimen4
\proofdotnumber=\count255
}

\def\proofover{
\eproofbit 
\setbox\proofbelow=\hbox\bgroup 
\let\wereinproofbit\proofover
$\displaystyle
}%
\def\proofoverdbl{
\eproofbit 
\doubleprooftrue
\setbox\proofbelow=\hbox\bgroup 
\let\wereinproofbit\proofoverdbl
$\displaystyle
}%
\def\proofoverdots{
\eproofbit 
\proofdotstrue
\setbox\proofbelow=\hbox\bgroup 
\let\wereinproofbit\proofoverdots
$\displaystyle
}%
\def\proofusing{
\eproofbit 
\setbox\proofrulename=\hbox\bgroup 
\let\wereinproofbit\proofusing
\kern0.3em$
}

\def\endprooftree{
\eproofbit 
  \dimen5 =0pt
\dimen0=\wd\proofabove \advance\dimen0-\shortenproofleft
\advance\dimen0-\shortenproofright
\dimen1=.5\dimen0 \advance\dimen1-.5\wd\proofbelow
\dimen4=\dimen1
\advance\dimen1\proofbelowshift \advance\dimen4-\proofbelowshift
\ifdim  \dimen1<0pt
\then   \advance\shortenproofleft\dimen1
        \advance\dimen0-\dimen1
        \dimen1=0pt
        \ifdim  \shortenproofleft<0pt
        \then   \setbox\proofabove=\hbox{%
                        \kern-\shortenproofleft\unhbox\proofabove}%
                \shortenproofleft=0pt
        \fi
\fi
\ifdim  \dimen4<0pt
\then   \advance\shortenproofright\dimen4
        \advance\dimen0-\dimen4
        \dimen4=0pt
\fi
\ifdim  \shortenproofright<\wd\proofrulename
\then   \shortenproofright=\wd\proofrulename
\fi
\dimen2=\shortenproofleft \advance\dimen2 by\dimen1
\dimen3=\shortenproofright\advance\dimen3 by\dimen4
\ifproofdots
\then
        \dimen6=\shortenproofleft \advance\dimen6 .5\dimen0
        \setbox1=\vbox to\proofdotseparation{\vss\hbox{$\cdot$}\vss}%
        \setbox0=\hbox{%
                \advance\dimen6-.5\wd1
                \kern\dimen6
                $\vcenter to\proofdotnumber\proofdotseparation
                        {\leaders\box1\vfill}$%
                \unhbox\proofrulename}%
\else   \dimen6=\fontdimen22\the\textfont2 
        \dimen7=\dimen6
        \advance\dimen6by.5\proofrulebreadth
        \advance\dimen7by-.5\proofrulebreadth
        \setbox0=\hbox{%
                \kern\shortenproofleft
                \ifdoubleproof
                \then   \hbox to\dimen0{%
                        $\mathsurround0pt\mathord=\mkern-6mu%
                        \cleaders\hbox{$\mkern-2mu=\mkern-2mu$}\hfill
                        \mkern-6mu\mathord=$}%
                \else   \vrule height\dimen6 depth-\dimen7 width\dimen0
                \fi
                \unhbox\proofrulename}%
        \ht0=\dimen6 \dp0=-\dimen7
\fi
\let\doll\relax
\ifwasinsideprooftree
\then   \let\VBOX\vbox
\else   \ifmmode\else$\let\doll=$\fi
        \let\VBOX\vcenter
\fi
\VBOX   {\baselineskip\proofrulebaseline \lineskip.2ex
        \expandafter\lineskiplimit\ifproofdots0ex\else-0.6ex\fi
        \hbox   spread\dimen5   {\hfi\unhbox\proofabove\hfi}%
        \hbox{\box0}%
        \hbox   {\kern\dimen2 \box\proofbelow}}\doll%
\global\dimen2=\dimen2
\global\dimen3=\dimen3
\egroup 
\ifonleftofproofrule
\then   \shortenproofleft=\dimen2
\fi
\shortenproofright=\dimen3
\onleftofproofrulefalse
\ifinsideprooftree
\then   \hskip.5em plus 1fil \penalty2
\fi
}

\newcommand{\smallDerivation}{1}

\newcommand{\indrulename}[1]{\texttt{\textup{#1}}}
\newcommand{\emptyPremise}{\vphantom{@}}
\newcommand{\indrule}[3]{
\ensuremath{
\begin{array}{c}
  \prooftree #2
    \justifies #3
    \thickness=0.05em
    \using \indrulename{#1}
  \endprooftree
\end{array}}}

\newcommand{\derivfrom}[2]{
  \begin{array}{c}
  #1
  \\
  #2
  \end{array}
}
\newcommand{\derivih}[1]{
  \indrule{}{
    \text{\ih}
  }{
    #1
  }
}

\renewcommand{\theenumi}{\arabic{enumi}}
\renewcommand{\theenumii}{\arabic{enumii}}
\renewcommand{\theenumiii}{\arabic{enumiii}}

\makeatletter
\renewcommand\p@enumii{\theenumi.}
\renewcommand\p@enumiii{\theenumi.\theenumii.}
\renewcommand\p@enumiv{\theenumi.\theenumii.\theenumiii.}
\makeatother

\newcommand{\llem}[1]{\label{lemma:#1}}
\newcommand{\rlem}[1]{Lem.~\ref{lemma:#1}}
\newcommand{\ldef}[1]{\label{def:#1}}
\newcommand{\rdef}[1]{Def.~\ref{def:#1}}
\newcommand{\lprop}[1]{\label{prop:#1}}
\newcommand{\rprop}[1]{Prop.~\ref{prop:#1}}
\newcommand{\lthm}[1]{\label{thm:#1}}
\newcommand{\rthm}[1]{Thm.~\ref{thm:#1}}
\newcommand{\lremark}[1]{\label{remark:#1}}
\newcommand{\rremark}[1]{Rem.~\ref{remark:#1}}

\newcommand{\lsec}[1]{\label{section:#1}}
\newcommand{\rsec}[1]{Section~\ref{section:#1}}

\renewcommand{\emptyset}{\varnothing}

\newcommand{\Eg}{{\em E.g.}\xspace}
\newcommand{\eg}{{\em e.g.}\xspace}
\newcommand{\ie}{{\em i.e.}\xspace}
\newcommand{\cf}{{\em cf.}\xspace}
\newcommand{\etal}{{\em et al.}\xspace}
\newcommand{\ih}{IH\xspace}
\newcommand{\ST}{\ |\ }
\newcommand{\HS}{\hspace{.5cm}}

\renewcommand{\emptyset}{\varnothing}
\newcommand{\emptyseq}{\epsilon}
\newcommand{\set}[1]{\{#1\}}
\newcommand{\Nat}{\mathbb{N}}

\newcommand{\proofnote}[1]{\textcolor{red}{{\small \textsc{[#1]}}}}

\newcommand{\lam}[2]{\lambda#1.\,#2}
\newcommand{\sub}[2]{\{#1:=#2\}}
\newcommand{\fv}[1]{\mathsf{fv}(#1)}

\newcommand{\MELL}{\mathsf{MELL}}
\newcommand{\MELLS}{\mathsf{MSCLL}}

\newcommand{\lambdaS}{\lambda^{\osha{}}}

\newcommand{\dom}[1]{\mathsf{dom}(#1)}
\newcommand{\tctx}{\Gamma}
\newcommand{\tctxtwo}{\Delta}
\newcommand{\tctxthree}{\Sigma}

\newcommand{\ctxhole}{\Box}
\newcommand{\gctx}{\mathtt{C}}
\newcommand{\gctxtwo}{\mathtt{D}}
\newcommand{\gctxthree}{\mathtt{E}}
\newcommand{\of}[2]{#1\langle#2\rangle}
\newcommand{\off}[2]{#1\langle\!\langle#2\rangle\!\rangle}

\newcommand{\lvar}{a}
\newcommand{\lvartwo}{b}

\newcommand{\uvar}{u}
\newcommand{\uvartwo}{v}
\newcommand{\uvarthree}{w}

\newcommand{\var}{x}
\newcommand{\vartwo}{y}
\newcommand{\varthree}{z}

\newcommand{\tm}{t}
\newcommand{\tmtwo}{s}
\newcommand{\tmthree}{r}
\newcommand{\tmfour}{p}
\newcommand{\tmg}{g}  
\newcommand{\tmd}{d}

\newcommand{\btyp}{\alpha}
\newcommand{\btyptwo}{\beta}

\newcommand{\negbtyp}[1]{\overline{#1}}
\newcommand{\nbtyp}{\negbtyp{\btyp}}
\newcommand{\nbtyptwo}{\negbtyp{\btyptwo}}

\newcommand{\typ}{A}
\newcommand{\typtwo}{B}
\newcommand{\typthree}{C}

\newcommand{\judg}[1]{\vdash#1}
\newcommand{\judeg}[2]{\vdash^{#1}#2}

\renewcommand{\arg}{(\cdot)}

\newcommand{\tensor}{\otimes}
\newcommand{\limp}{\multimap}
\newcommand{\liff}{\multimapboth}

\newcommand{\lneg}[1]{#1^{\perp}}

\newcommand{\ofc}[1]{{!}{#1}}
\newcommand{\why}[1]{{?}{#1}}

\newcommand{\sha}[1]{\bullet#1}
\newcommand{\ush}[1]{\circ#1}
\newcommand{\wush}[1]{\why{\ush{#1}}}
\newcommand{\osha}[1]{\ofc{\sha{#1}}}

\newcommand{\ruleAx}{\indrulename{ax}}
\newcommand{\ruleCut}{\indrulename{cut}}
\newcommand{\ruleTensor}{\ensuremath{\tensor}}
\newcommand{\ruleParr}{\ensuremath{\parr}}
\newcommand{\ruleW}{\ensuremath{\why{\indrulename{w}}}}
\newcommand{\ruleWs}{\ensuremath{\why{\indrulename{w}^*}}}
\newcommand{\ruleC}{\ensuremath{\why{\indrulename{c}}}}
\newcommand{\ruleCs}{\ensuremath{\why{\indrulename{c}^*}}}
\newcommand{\ruleCn}[1][n]{\ensuremath{\why{\indrulename{c}_{n}}}}

\newcommand{\ruleD}{\ensuremath{\wush{\indrulename{d}}}}
\newcommand{\ruleProm}{\ensuremath{\ofc{\indrulename{p}}}}
\newcommand{\ruleSha}{\ensuremath{\sha{}}}
\newcommand{\ruleUsh}{\ensuremath{\ush{}}}
\newcommand{\ruleMix}{\indrulename{mix}}
\newcommand{\ruleCutd}[1][d]{\indrulename{cut}\ensuremath{^{#1}}}
\newcommand{\ruleMixd}[1][d]{\indrulename{mix}\ensuremath{^{#1}}}
\newcommand{\rulecLvar}{\indrulename{lvar}}
\newcommand{\rulecUvar}{\indrulename{uvar}}
\newcommand{\rulecAbs}{\indrulename{abs}}
\newcommand{\rulecApp}{\indrulename{app}}
\newcommand{\rulecSha}{\indrulename{grant}}
\newcommand{\rulecOpen}{\indrulename{request}}
\newcommand{\rulecProm}{\indrulename{prom}}
\newcommand{\rulecES}{\indrulename{sub}}
\newcommand{\rulelVar}{\indrulename{l-var}}
\newcommand{\rulelAbs}{\indrulename{l-abs}}
\newcommand{\rulelApp}{\indrulename{l-app}}
\newcommand{\rulelES}{\indrulename{l-es}}

\newcommand{\derivjudeg}[3]{#1\mathrel{\rhd}^{#2}#3}
\newcommand{\deriv}{\pi}
\newcommand{\derivtwo}{\rho}
\newcommand{\rep}[2]{({#2})^{#1}}
\newcommand{\sz}[1]{|#1|}
\newcommand{\height}[1]{h(#1)}
\newcommand{\derp}[1]{\left(#1\right)}

\newcommand{\noenv}{\cdot}
\newcommand{\lenv}{\Gamma}
\newcommand{\uenv}{\Delta}

\newcommand{\esub}[2]{[#1/#2]}
\newcommand{\open}[1]{\mathtt{req}(#1)}

\newcommand{\judc}[4]{#1;#2\vdash#3:#4}
\newcommand{\judl}[3]{#1\vdash#2:#3}

\newcommand{\tran}[1]{{#1^{\star}}}

\newcommand{\sctx}{\mathtt{L}}
\newcommand{\sctxtwo}{\mathtt{K}}

\newcommand{\rootto}{\mapsto}

\newcommand{\symSdb}{\mathsf{\sha{db}}}
\newcommand{\symSopen}{\mathsf{\sha{req}}}
\newcommand{\symSls}{\mathsf{\sha{ls}}}
\newcommand{\symSgc}{\mathsf{\sha{gc}}}

\newcommand{\rtoSdb}{\rootto_{\symSdb}}
\newcommand{\rtoSopen}{\rootto_{\symSopen}}
\newcommand{\rtoSls}{\rootto_{\symSls}}
\newcommand{\rtoSgc}{\rootto_{\symSgc}}

\newcommand{\toS}{\to_{\sha{}}}
\newcommand{\toSdb}{\to_{\symSdb}}
\newcommand{\toSopen}{\to_{\symSopen}}
\newcommand{\toSls}{\to_{\symSls}}
\newcommand{\toSgc}{\to_{\symSgc}}

\newcommand{\tosymS}{\leftrightarrow_{\sha{}}}
\newcommand{\tosymcbv}{\leftrightarrow_{\symValue}}

\newcommand{\symTerms}{\mathcal{T}}
\newcommand{\TermsLSC}{\symTerms_{\mathsf{LSC}}}
\newcommand{\TermsS}{\symTerms_{\sha{}}}
\newcommand{\TermsSName}{\traN{\TermsS}}
\newcommand{\TermsSValue}{\traV{\TermsS}}
\newcommand{\TermsSClumsyNeed}{\traCNd{\TermsS}}

\newcommand{\symCtxs}{\mathsf{Ctxs}}
\newcommand{\CtxsS}{\symCtxs_{\sha{}}}
\newcommand{\CtxsSName}{\traN{\CtxsS}}
\newcommand{\CtxsSValue}{\traV{\CtxsS}}
\newcommand{\CtxsSClumsyNeed}{\traCNd{\CtxsS}}

\newcommand{\symSCtxs}{\mathsf{SCtxs}}
\newcommand{\SCtxsS}{\symSCtxs_{\sha{}}}
\newcommand{\SCtxsSName}{\traN{\SCtxsS}}
\newcommand{\SCtxsSValue}{\traV{\SCtxsS}}
\newcommand{\SCtxsSClumsyNeed}{\traCNd{\SCtxsS}}

\newcommand{\symRulenames}{\mathcal{R}}
\newcommand{\RulesS}{\symRulenames_{\sha{}}}
\newcommand{\RulesSName}{\traN{\RulesS}}
\newcommand{\RulesSValue}{\traV{\RulesS}}
\newcommand{\RulesSClumsyNeed}{\traCNd{\RulesS}}

\newcommand{\ars}{\mathcal{X}}
\newcommand{\arstwo}{\mathcal{Y}}
\newcommand{\toars}{\to_\ars}
\newcommand{\toarsinv}{\leftarrow_\ars}
\newcommand{\toarssym}{\leftrightarrow_\ars}
\newcommand{\toarstwo}{\to_\arstwo}

\newcommand{\val}{\mathtt{v}}
\newcommand{\valtwo}{\mathtt{w}}

\newcommand{\symlax}{\texttt{\textup{+}}}
\newcommand{\lval}{\val^\symlax}
\newcommand{\lvaltwo}{\valtwo^\symlax}

\newcommand{\symdb}{\mathsf{db}}
\newcommand{\symls}{\mathsf{ls}}
\newcommand{\symlsv}{\mathsf{lsv}}
\newcommand{\symlsw}{\mathsf{lsw}}
\newcommand{\symgc}{\mathsf{gc}}
\newcommand{\symgclv}{\mathsf{gcv\symlax}}
\newcommand{\symgclvinv}{\symgclv^{-1}}

\newcommand{\rtodb}{\rootto_{\symdb}}
\newcommand{\rtols}{\rootto_{\symls}}
\newcommand{\rtolsv}{\rootto_{\symlsv}}
\newcommand{\rtolsw}{\rootto_{\symlsw}}
\newcommand{\rtogc}{\rootto_{\symgc}}
\newcommand{\rtogclv}{\rootto_{\symgclv}}

\newcommand{\todb}{\to_{\symdb}}
\newcommand{\tols}{\to_{\symls}}
\newcommand{\tolsv}{\to_{\symlsv}}
\newcommand{\tolsw}{\to_{\symlsw}}
\newcommand{\togc}{\to_{\symgc}}
\newcommand{\togclv}{\to_{\symgclv}}
\newcommand{\togclvinv}{\togclv^{-1}}

\newcommand{\tocbn}{\to_{\symName}}
\newcommand{\tocbv}{\to_{\symValue}}
\newcommand{\tocbnd}{\to_{\symNeed}}
\newcommand{\tocbCnd}{\to_{\symClumsyNeed}}
\newcommand{\tocbvUgclvinv}{\rhd_{\symValue}}
\newcommand{\tocbvUgclvinvs}{\tocbvUgclvinv^*}

\newcommand{\symName}{\textcolor{blue}{\mathsf{N}}}
\newcommand{\symValue}{\textcolor{blue}{\mathsf{V}}}
\newcommand{\symClumsyNeed}{\textcolor{blue}{\mathsf{S}}}
\newcommand{\symNeed}{\textcolor{blue}{\mathsf{Nd}}}

\newcommand{\traN}[1]{#1^{\symName}}
\newcommand{\traV}[1]{#1^{\symValue}}
\newcommand{\traVval}[1]{#1^{\textcolor{blue}{(\symValue)}}}
\newcommand{\traValt}[1]{#1^{\textcolor{blue}{\symValue+}}}
\newcommand{\traCNd}[1]{#1^{\symClumsyNeed}}
\newcommand{\traNinv}[1]{#1^{\textcolor{blue}{-\symName}}}
\newcommand{\traVinv}[1]{#1^{\textcolor{blue}{-\symValue}}}
\newcommand{\traCNdinv}[1]{#1^{\textcolor{blue}{-\symClumsyNeed}}}
\newcommand{\traD}[1]{\traCNd{#1}}
\newcommand{\traDinv}[1]{\traCNdinv{#1}}
\newcommand{\traDval}[1]{#1^{\textcolor{blue}{(\symClumsyNeed)}}}

\newcommand{\rtm}{\underline{\tm}}
\newcommand{\rtmtwo}{\underline{\tmtwo}}
\newcommand{\rtmthree}{\underline{\tmthree}}
\newcommand{\rtmfour}{\underline{\tmfour}}
\newcommand{\rsctx}{\underline{\sctx}}
\newcommand{\rgctx}{\underline{\gctx}}

\newcommand{\ann}[2]{{#1}^{#2}}
\newcommand{\lab}{\alpha}
\newcommand{\labtwo}{\beta}
\newcommand{\labthree}{\gamma}
\newcommand{\labfour}{\delta}
\newcommand{\labfive}{\eta}
\newcommand{\llam}[3]{\lambda{#1}^{#3}.\,#2}
\newcommand{\lopen}[2]{\mathtt{req}^{#2}(#1)}
\newcommand{\olopen}[2]{\mathtt{req}^{(#2)}(#1)}
\newcommand{\lesub}[3]{[{#1}^{#3}/#2]}
\newcommand{\symLabeled}{\mathcal{L}}
\newcommand{\symWellLabeled}{\mathcal{WL}}
\newcommand{\symLabeledTerms}{\mathsf{T}^{\symLabeled}}
\newcommand{\symWellLabeledTerms}{\mathsf{T}^{\symWellLabeled}}
\newcommand{\flv}[1]{\mathsf{fv}^{\symLabeled}(#1)}
\newcommand{\unlab}[1]{{#1}^\circ}

\newcommand{\lgctx}{\gctx}
\newcommand{\lgctxtwo}{\gctxtwo}
\newcommand{\lgctxthree}{\gctxthree}
\newcommand{\lsctx}{\sctx}
\newcommand{\lsctxtwo}{\sctxtwo}

\newcommand{\lto}[1]{\to^{#1}}

\newcommand{\toSL}[1]{\toS^{#1}}

\newcommand{\toSdbL}[1]{\toSdb^{#1}}
\newcommand{\toSopenL}[1]{\toSopen^{#1}}
\newcommand{\toSlsL}[1]{\toSls^{#1}}
\newcommand{\toSgcL}[1]{\toSgc^{#1}}

\newcommand{\lrootto}[1]{\mapsto^{#1}}

\newcommand{\rtoSdbL}[1]{\rtoSdb^{#1}}
\newcommand{\rtoSopenL}[1]{\rtoSopen^{#1}}
\newcommand{\rtoSlsL}[1]{\rtoSls^{#1}}
\newcommand{\rtoSgcL}[1]{\rtoSgc^{#1}}

\newcommand{\TermsSL}{\symLabeledTerms_{\sha{}}}
\newcommand{\TermsSWL}{\symWellLabeledTerms_{\sha{}}}

\newcommand{\llambdaS}{\lambda^{\symLabeled}_{\sha{}}}
\newcommand{\symCtxsL}{\mathsf{Ctxs}^{\symLabeled}}
\newcommand{\CtxsSL}{\symCtxsL_{\sha{}}}

\newcommand{\mult}[2]{\#_{#1}(#2)}
\newcommand{\lmult}[1]{\#(#1)}
\newcommand{\multctx}[3]{\#_{#1}^{#2}(#3)}
\newcommand{\lmultctx}[2]{\#^{#1}(#2)}

\newcommand{\wlt}[1]{{#1}\, \mathsf{wl}}

\newcommand{\flatt}{\equiv}
\newcommand{\pflatt}[1]{\stackrel{#1}{\flatt}}

\newcommand{\Objs}{\mathtt{X}}
\newcommand{\Steps}{\mathtt{S}}
\newcommand{\obj}{x}

\newcommand{\redseq}{\rho}
\newcommand{\redseqtwo}{\sigma}
\newcommand{\step}{\mathfrak{r}}
\newcommand{\steptwo}{\mathfrak{s}}
\newcommand{\stepthree}{\mathfrak{t}}
\newcommand{\multistep}{\mathfrak{m}}

\newcommand{\steps}[1]{\mathsf{steps}(#1)}
\newcommand{\labStep}[2]{\mathsf{steps}_{#1}(#2)}
\newcommand{\labels}[1]{\mathsf{lab}(#1)}
\newcommand{\lift}[3]{\mathsf{lift}(#1,#2,#3)}
\newcommand{\resid}[1]{\llbracket {#1}\rrbracket}
\newcommand{\dev}[1]{\twoheadrightarrow^{#1}}
\newcommand{\labord}[1]{\prec_{#1}}
\newcommand{\flatBij}[2]{\phi_{#1,#2}}
\newcommand{\flatBijtwo}[2]{\xi_{#1,#2}}

\newcommand{\lmgctx}{\mathtt{F}}

\newcommand{\mlnode}[1]{\begin{array}{@{}c@{}}#1\end{array}}

\newcommand{\lscUnit}{\mathbf{1}}
\newcommand{\lscunit}{{\ast}}
\newcommand{\lscsym}{{\textsc{\textup{lsc}}}}
\newcommand{\tradlsc}[1]{[\![#1]\!]}
\newcommand{\judlsc}[3]{#1\vdash_{\lscsym} #2:#3}
\newcommand{\tofuse}{\Rrightarrow}
\newcommand{\tofuseone}{\tofuse^1}
\newcommand{\symI}{\textsc{\textup{i}}}

\newcommand{\toSi}{\to_{\sha{\symI}}}
\newcommand{\tolsc}{\to_{\textsc{\textup{lsc}}}}
\newcommand{\tolsci}{\to_{\textsc{\textup{lsc}}\symI}}

\newcommand{\ruleFuseW}{\indrulename{$\tofuse$w}}
\newcommand{\ruleFuseC}{\indrulename{$\tofuse$c}}
\newcommand{\ruleFuseLam}{\indrulename{$\tofuse$abs}}
\newcommand{\ruleFuseAppL}{\indrulename{$\tofuse$appL}}
\newcommand{\ruleFuseAppR}{\indrulename{$\tofuse$appR}}
\newcommand{\ruleFuseEsL}{\indrulename{$\tofuse$esL}}
\newcommand{\ruleFuseEsR}{\indrulename{$\tofuse$esR}}

\newcommand{\aars}{{\mathcal{A}}_{\lambdaS}}  
\newcommand{\aarsObj}{{\mathcal{O}}}
\newcommand{\aarsSteps}{{\mathcal{R}}}
\newcommand{\src}{\mathsf{src}}
\newcommand{\tgt}{\mathsf{tgt}}
\newcommand{\eqclass}[2]{[{#1}]_{#2}}

\newcommand{\mstep}[2]{\langle {#1},\step,{#2}\rangle}
\newcommand{\msteptwo}[2]{\langle {#1},\steptwo,{#2}\rangle}

\newcommand{\axiomName}[1]{\textsf{#1}}

\newcommand{\diagramFuseLSCRel}[6]{
  \begin{array}{ccc}
    #3 & \tofuseone & #4 \vspace{-.5em} \\
    \rotatebox{270}{$#1$} & & \rotatebox{270}{$\tolsci$} \vspace{.25em}\\
    #6 & #2 & #5 \\
  \end{array}
}
\newcommand{\diagramFuseLSC}[4]{\diagramFuseLSCRel{\tolsci^+}{\tofuse}{#1}{#2}{#3}{#4}}
\newcommand{\diagramFuseZeroLSCOne}[4]{\diagramFuseLSCRel{\tolsci}{=}{#1}{#2}{#3}{#4}}
\newcommand{\diagramFuseOneLSCOne}[4]{\diagramFuseLSCRel{\tolsci}{\tofuseone}{#1}{#2}{#3}{#4}}
\newcommand{\diagramFuseManyLSCOne}[4]{\diagramFuseLSCRel{\tolsci}{\tofuse}{#1}{#2}{#3}{#4}}
\newcommand{\diagramFuseOneLSCMany}[4]{\diagramFuseLSCRel{\tolsci^+}{\tofuseone}{#1}{#2}{#3}{#4}}

\newcommand{\TermsBang}{\symTerms_{\symBang}}
\newcommand{\TermsBangInv}{\symTerms_{\sha{}}^{\symBang}}
\newcommand{\der}[1]{\mathtt{der}(#1)}
\newcommand{\symBangGM}{\textcolor{blue}{\mathsf{B^{der}}}}
\newcommand{\TermsBangGM}{\symTerms_{\symBangGM}}
\newcommand{\symBang}{\textcolor{blue}{\mathsf{B}}}
\newcommand{\traB}[1]{#1^{\symBang}}
\newcommand{\traBinv}[1]{#1^{\textcolor{blue}{-\symBang}}}

\newcommand{\CtxsSBang}{\traB{\CtxsS}}
\newcommand{\SCtxsSBang}{\traB{\SCtxsS}}

\newcommand{\rulebVar}{\indrulename{b-var}}
\newcommand{\rulebAbs}{\indrulename{b-abs}}
\newcommand{\rulebApp}{\indrulename{b-app}}

\newcommand{\rulebProm}{\indrulename{b-prom}}
\newcommand{\rulebES}{\indrulename{b-es}}

\newcommand{\symderB}{\mathsf{d!}}
\newcommand{\symdbB}{\mathsf{db}}
\newcommand{\symlsB}{\mathsf{ls!}}
\newcommand{\symgcB}{\mathsf{gc!}}

\newcommand{\rtolsB}{\rootto_{\symlsB}}
\newcommand{\rtogcB}{\rootto_{\symgcB}}
\newcommand{\rtoderB}{\rootto_{\symderB}}

\newcommand{\todbB}{\to_{\symdbB}}
\newcommand{\tolsB}{\to_{\symlsB}}
\newcommand{\togcB}{\to_{\symgcB}}
\newcommand{\toderB}{\to_{\symderB}}
\newcommand{\tonogcB}{\to_{\neg\symgcB}}

\newcommand{\toBangGM}{\to_{\symBangGM}}
\newcommand{\toBang}{\to_{\symBang}}

\newcommand{\derEq}{\ltimes}
\newcommand{\derEqSctx}[1]{\ltimes_{#1}}
\newcommand{\derEqGctx}[1]{\ltimes_{#1}}
\newcommand{\varset}{\mathcal{X}}
\newcommand{\varsettwo}{\mathcal{Y}}

\newcommand{\ruleDerEqVar}{\indrulename{var}}
\newcommand{\ruleDerEqLam}{\indrulename{abs}}
\newcommand{\ruleDerEqApp}{\indrulename{app}}
\newcommand{\ruleDerEqESub}{\indrulename{esub}}
\newcommand{\ruleDerEqOfc}{\indrulename{ofc}}
\newcommand{\ruleDerEqDer}{\indrulename{der}}
\newcommand{\ruleDerEqGc}{\indrulename{gc}}

\newcommand{\toMELLS}[1]{{#1}^{\sha{}}}

\definecolor{darkolivegreen}{rgb}{0.33, 0.42, 0.18}
\definecolor{auburn}{rgb}{0.43, 0.21, 0.1}

\newcommand{\CBN}{{\textcolor{black}{\mathsf{CBN}}}}
\newcommand{\CBV}{{\textcolor{black}{\mathsf{CBV}}}}

\newcommand{\CBNd}{{\textcolor{black}{\mathsf{CBNd}}}}
\newcommand{\CCBNd}{{\textcolor{black}{\mathsf{CBS}}}}
\newcommand{\LL}{\mathsf{LL}}
\newcommand{\Bang}{{\textcolor{black}{\mathsf{Bang}}}}
\newcommand{\LSC}{{\textcolor{black}{\mathsf{LSC}}}}
\newcommand{\CBPV}{{\textcolor{black}{\mathsf{CBPV}}}}

\newcommand{\ops}{\mathcal{A}}
\newcommand{\nfop}{\alpha}
\newcommand{\nfoptwo}{\beta}
\newcommand{\nfS}[1]{\mathcal{N}_{#1}}

\newcommand{\varSymb}{\mathtt{var}}
\newcommand{\lamSymb}{\lambda}
\newcommand{\appSymb}{@}
\newcommand{\openSymb}{\mathtt{req}}
\newcommand{\shaSymb}{\sha{}}
\newcommand{\ofcSymb}{\ofc{}}
\newcommand{\oshaSymb}{\ofc{}\sha{}}

\newcommand{\toss}[1]{\rightsquigarrow_{#1}}
\newcommand{\tossmany}[1]{\rightsquigarrow^*_{#1}}
\newcommand{\tosn}[1]{\rightsquigarrow^{\symName}_{#1}}
\newcommand{\tosv}[1]{\rightsquigarrow^{\symValue}_{#1}}
\newcommand{\tosd}[1]{\rightsquigarrow^{\symClumsyNeed}_{#1}}
\newcommand{\tosnmany}[1]{\mathrel{(\rightsquigarrow^{\symName})^*_{#1}}}

\newcommand{\tosdmany}[1]{\mathrel{(\rightsquigarrow^{\symClumsyNeed})^*_{#1}}}
\newcommand{\tosvUgclvinv}[1]{\blacktriangleright^{\symValue}_{#1}}
\newcommand{\tosvUgclvinvmany}[1]{\mathrel{(\blacktriangleright^{\symValue})^*_{#1}}}
\newcommand{\rulename}{\rho}
\newcommand{\rrulename}{\underline{\rulename}}
\newcommand{\rulenameseq}{\vec{\rulename}}

\newcommand{\symsub}[2]{\mathsf{\varsigma}(#1,#2)}
\newcommand{\symSsub}[2]{\mathsf{\varsigma}(#1,#2)}
\newcommand{\symSid}[1]{\mathsf{\iota}(#1)}
\newcommand{\symid}[1]{\mathsf{\iota}(#1)}

\newcommand{\sequence}[1]{\langle#1\rangle}

\newcommand{\ruleESdb}{\indrulename{E$^\sha{}$-db}}
\newcommand{\ruleESsub}{\indrulename{E$^\sha{}$-$\varsigma$}}
\newcommand{\ruleESofcsub}{\indrulename{E$^\sha{}$-$\ofc{\varsigma}$}}
\newcommand{\ruleESls}{\indrulename{E$^\sha{}$-ls}}
\newcommand{\ruleESgc}{\indrulename{E$^\sha{}$-gc}}
\newcommand{\ruleESid}{\indrulename{E$^\sha{}$-$\iota$}}
\newcommand{\ruleESopen}{\indrulename{E$^\sha{}$-req$\bullet$}}
\newcommand{\ruleESapp}{\indrulename{E$^\sha{}$-app}}
\newcommand{\ruleEScopen}{\indrulename{E$^\sha{}$-req}}
\newcommand{\ruleESsubL}{\indrulename{E$^\sha{}$-esL}}
\newcommand{\ruleESsubR}{\indrulename{E$^\sha{}$-esR}}
\newcommand{\ruleESsubRofc}{\indrulename{E$^\sha{}$-es$\ofc{}$}}

\newcommand{\ruleENdb}{\indrulename{E$^{\symName}$-db}}
\newcommand{\ruleENsub}{\indrulename{E$^{\symName}$-$\varsigma$}}
\newcommand{\ruleENls}{\indrulename{E$^{\symName}$-ls}}
\newcommand{\ruleENgc}{\indrulename{E$^{\symName}$-gc}}
\newcommand{\ruleENapp}{\indrulename{E$^{\symName}$-app}}
\newcommand{\ruleENsubL}{\indrulename{E$^{\symName}$-subL}}

\newcommand{\ruleEVdb}{\indrulename{E$^{\symValue}$-db}}
\newcommand{\ruleEVsub}{\indrulename{E$^{\symValue}$-$\varsigma$}}
\newcommand{\ruleEVlsv}{\indrulename{E$^{\symValue}$-lsv}}
\newcommand{\ruleEVgclv}{\indrulename{E$^{\symValue}$-gcv\symlax}}
\newcommand{\ruleEVapp}{\indrulename{E$^{\symValue}$-app}}
\newcommand{\ruleEVsubR}{\indrulename{E$^{\symValue}$-subR}}
\newcommand{\ruleEVsubL}{\indrulename{E$^{\symValue}$-subL}}

\newcommand{\ruleEDdb}{\indrulename{E$^{\symClumsyNeed}$-db}}
\newcommand{\ruleEDsub}{\indrulename{E$^{\symClumsyNeed}$-$\varsigma$}}
\newcommand{\ruleEDsubES}{\indrulename{E$^{\symClumsyNeed}$-$\varsigma_2$}}
\newcommand{\ruleEDlsw}{\indrulename{E$^{\symClumsyNeed}$-lsw}}
\newcommand{\ruleEDgc}{\indrulename{E$^{\symClumsyNeed}$-gc}}
\newcommand{\ruleEDid}{\indrulename{E$^{\symClumsyNeed}$-$\iota$}}
\newcommand{\ruleEDapp}{\indrulename{E$^{\symClumsyNeed}$-app}}
\newcommand{\ruleEDsubL}{\indrulename{E$^{\symClumsyNeed}$-subL}}
\newcommand{\ruleEDsubR}{\indrulename{E$^{\symClumsyNeed}$-subR}}

\newcommand{\BangCalculus}{Bang-calculus\xspace}

\newcommand{\SharingLinearLogic}{Linear Logic with Restricted Access\xspace}

\newcommand{\bparagraph}[1]{\paragraph{{\bf #1}}}

\newcommand{\eqdef}{:=}

\usepackage{marginnote}

\hypersetup{
  colorlinks = true
}

\usepackage[strict]{changepage}
\newenvironment{acenter}[1]{
\begin{adjustwidth}{#1}{}
\begin{center}
}{
\end{center}
\end{adjustwidth}
}

\definecolor{darkolivegreen}{rgb}{0.33, 0.42, 0.18}
\definecolor{auburn}{rgb}{0.43, 0.21, 0.1}

\newboolean{shortAppendix}
\newboolean{longAppendix}
\setboolean{shortAppendix}{false}
\setboolean{longAppendix}{true}
\makeatletter
\newenvironment{ifShortAppendix}{%
  \ifthenelse{\boolean{shortAppendix}}%
  {\let\ifShortAppendix@i\relax\let\endIfShortAppendix@i\relax}%
  {\def\ifShortAppendix@i{\setbox\z@\vbox\bgroup}%
    \def\endIfShortAppendix@i{\egroup}}%
  \ifShortAppendix@i%
}{%
  \endIfShortAppendix@i%
}
\newenvironment{ifLongAppendix}{%
  \ifthenelse{\boolean{longAppendix}}%
  {\let\ifLongAppendix@i\relax\let\endIfLongAppendix@i\relax}%
  {\def\ifLongAppendix@i{\setbox\z@\vbox\bgroup}%
    \def\endIfLongAppendix@i{\egroup}}%
  \ifLongAppendix@i%
}{%
  \endIfLongAppendix@i%
}
\makeatother

\usepackage{enumitem}
\setlistdepth{9}
\newlist{xenumerate}{enumerate}{9}
\setlist[xenumerate,1]{label*=\arabic*.}
\setlist[xenumerate,2]{label*=\arabic*.}
\setlist[xenumerate,3]{label*=\arabic*.}
\setlist[xenumerate,4]{label*=\arabic*.}
\setlist[xenumerate,5]{label*=\arabic*.}
\setlist[xenumerate,6]{label*=\arabic*.}
\setlist[xenumerate,7]{label*=\arabic*.}
\setlist[xenumerate,8]{label*=\arabic*.}
\setlist[xenumerate,9]{label*=\arabic*.}

\begin{document}

\title{Sharing and Linear Logic with Restricted Access}
%
%
\author{Pablo Barenbaum\inst{1}\orcidID{0009-0003-2494-3345} \and
Eduardo Bonelli\inst{2}\orcidID{0000-0003-1856-2856}}
\authorrunning{P.~Barenbaum and E.~Bonelli}
%
\institute{Universidad Nacional de Quilmes (CONICET) and \\
 Instituto de Ciencias de la Computación, Universidad de Buenos Aires \\
\email{pbarenbaum@dc.uba.ar}\\
\and
Stevens Institute of Technology\\
\email{eabonelli@gmail.com}}
\maketitle              
\begin{abstract}
The two Girard translations provide two different
means of obtaining embeddings of Intuitionistic Logic into Linear Logic,
corresponding to different lambda-calculus calling mechanisms.
The translations, mapping $\typ\to\typtwo$
respectively to $\ofc{\typ}\limp\typtwo$ and $\ofc{(\typ\limp\typtwo)}$,
have been shown to correspond respectively to call-by-name and call-by-value.

In this work, we split the \emph{of-course} modality of linear logic 
into two modalities, written ``$\ofc{}$'' and ``$\sha{}$''.
Intuitively, the modality ``$\ofc{}$'' specifies a subproof
that can be duplicated and erased, but may not necessarily be ``accessed'',
\ie interacted with,
while the combined modality ``$\ofc{\sha{}}$'' specifies a subproof
that can moreover be accessed.
The resulting system, called $\MELLS$,
enjoys cut-elimination and is conservative over $\MELL$.

We study how restricting access to subproofs provides ways to control
\emph{sharing} in evaluation strategies.
For this, we introduce a term-assignment for an intuitionistic fragment of $\MELLS$,
called the $\lambdaS$-calculus, which we show to enjoy
subject reduction, confluence, and strong normalization of the simply typed fragment.
We propose three sound and complete translations that respectively
simulate call-by-name, call-by-value, and a variant of call-by-name
that \emph{shares} the evaluation of its arguments (similarly as in call-by-need).
The translations are extended to simulate the \BangCalculus, as well as weak
reduction strategies.

\end{abstract}



\section{Introduction}
\lsec{introduction}

The propositions-as-types correspondence links \emph{computation} and
\emph{logic},
relating types with propositions, programs with proofs, and program evaluation with
proof normalization.
The prime example is the simply typed $\lambda$-calculus,
which corresponds to intuitionistic propositional logic.
The correspondence has been extended to many other calculi
and logics, including Linear Logic~($\LL$)~\cite{DBLP:journals/tcs/Girard87}.
Linear Logic
proposes a \emph{resource conscious} approach to logic, in that
only formulae prefixed with an \emph{exponential modality} can be
duplicated and erased.
In $\LL$, there are two exponential modalities:
\emph{of-course}~(``$\ofc{}$''), allowing duplication/erasure on the left,
and \emph{why-not}~(``$\why{}$''), allowing duplication/erasure on the right.
These modalities recover the ability to duplicate and erase
formulae in a controlled way, making $\LL$ a suitable language
to model resource-sensitive phenomena such as
concurrency, memory management, and computational complexity.

Girard~\cite{DBLP:journals/tcs/Girard87} discusses two possible
ways of embedding intuitionistic logic into $\LL$,
mapping the intuitionistic implication $\typ\to\typtwo$
respectively to $\ofc{\typ}\limp\typtwo$
and to $\ofc{\typ}\limp\ofc{\typtwo}$,
where ``$\limp$'' stands for the linear implication\footnote{Sometimes
the second embedding is defined as mapping $\typ\to\typtwo$
to $\ofc{(\typ\limp\typtwo)}$. This is just an apparent difference,
which is canceled out by adjusting the translation of sequents accordingly.}.
Maraist \etal~\cite{DBLP:journals/tcs/MaraistOTW99} observed
that these translations can be used to extend the propositions-as-types
correspondence to provide a logical foundation for \emph{evaluation mechanisms}.

The most well-known evaluation mechanism
for the $\lambda$-calculus is perhaps \emph{call-by-name} ($\CBN$),
in which arguments to functions are re-evaluated upon each use.
The theory of $\CBN$ has been thoroughly developed,
\eg in Barendregt's book~\cite{DBLP:books/daglib/0067558}.
On the other hand, in the \emph{call-by-value} ($\CBV$) evaluation
mechanism~\cite{DBLP:journals/tcs/Plotkin75}, arguments to functions
are evaluated once and for all; then their value can be recalled upon each use.
Call-by-value is less deeply studied in the literature than $\CBN$,
but it has been gaining attention since its theory is subtle and corresponds
more closely to the evaluation mechanism behind most programming languages.

\bparagraph{Embeddings Encode Evaluation Mechanisms.}
The first Girard translation $\traN{\arg}$ operates on \emph{formulae}
  by mapping $\typ\to\typtwo$ to $\ofc{\traN{\typ}}\limp\traN{\typtwo}$.
  It operates on \emph{terms} in such a way
  that a $\lambda$-calculus application $\tm\,\tmtwo$
  is mapped to $\traN{\tm}\,\ofc{\traN{\tmtwo}}$
  in a \emph{linear} $\lambda$-calculus\footnote{
    Girard's original translations target Linear Logic presented in sequent
    calculus style. We follow here Maraist~\etal~\cite{DBLP:journals/tcs/MaraistOTW99},
    in which the target language is presented as a linear $\lambda$-calculus,
    in natural deduction style.
  }.
Here ``$\ofc{}$'' is a \emph{term constructor},
which corresponds to an instance of the \emph{$\ofc{}$-introduction} rule,
also called \emph{promotion}.
The key point is that the argument $\tmtwo$ of an application
is prefixed with ``$\ofc{}$''.
This enables the (arbitrary) term $\traN{\tmtwo}$
to be freely copied or discarded in $\traN{\tm}$, as dictated by $\CBN$.
In this sense, the first translation provides a logical foundation for the
$\CBN$ evaluation mechanism.

The second Girard translation $\traV{\arg}$ operates on formulae
  by mapping the intuitionistic implication $\typ\to\typtwo$
  to $\ofc{\traV{\typ}}\limp\ofc{\traV{\typtwo}}$.
  It operates on terms in such a way
  that a $\lambda$-calculus application $\tm\,\tmtwo$
  is mapped to $\der{\traV{\tm}}\,\traV{\tmtwo}$.
  Here, $\der{\cdot}$ stands for an appropriate operation in the target language
  that corresponds to the \emph{$\ofc{}$-elimination} rule,
  also called \emph{dereliction}.
  The key point is that the argument is not prefixed with ``$\ofc{}$'',
  which means that it must be evaluated before being consumed,
  as dictated by $\CBV$.
  This translation prefixes \emph{values}, such as $\lam{\var}{\tmthree}$,
  with a promotion, resulting in
  $\ofc{(\lam{\var}{\traV{\tmthree}})}$. Consequently, only values will end up
  being copied or discarded.


\bparagraph{Linear Logic with Restricted Access.}
In this work, we propose a logical system
that arises from splitting
each exponential modality into two new
modalities.
We refer to this new system as \emph{Linear Logic with Restricted Access},
and formally as $\MELLS$, to reflect that
we study the fragment with \textbf{M}ultiplicative, \textbf{S}haring,
and a\textbf{C}cess connectives.
The \emph{exponential} of-course (``$\ofc{}$'') modality of $\LL$
is split into two modalities in $\MELLS$:
a \emph{sharing} modality ``$\ofc{}$''
and an \emph{access} modality ``$\sha{}$'',
that ``grants'' access to a subproof.
The \emph{sharing of-course} of $\MELLS$
turns out to be weaker than the \emph{exponential of-course} of $\LL$
but, by abuse of notation,
we denote it using the same symbol ``$\ofc{}$''.
Dually, the exponential why-not (``$\why{}$'') modality of $\LL$
is split into a sharing modality ``$\why{}$''
and an access modality ``$\ush{}$'' in $\MELLS$.
The resulting system $\MELLS$ is a conservative extension of
Multiplicative Exponential Linear Logic ($\MELL$), in which the combined
modality ``$\osha{}$'' plays the role of the exponential of-course modality of $\LL$.
The operational intuition is that
the sharing modality ``$\ofc{}$'' in $\MELLS$
specifies an expression that may be duplicated and erased, but it may not necessarily
be used or \emph{accessed}. The combined
modality ``$\ofc{\sha{}}$'' specifies an expression that may, additionally,
be accessed, \ie its contents can be made to interact with the
surrounding computational context.

\bparagraph{Embedding Call-by-Name and Call-by-Value.}
Following, we revisit Girard's translations, but this time targetting
the \emph{$\lambdaS$-calculus}, a linear $\lambda$-calculus based on $\MELLS$.
Our $\CBN$ translation
$\traN{\arg}$
operates on formulae by mapping $\typ\to\typtwo$ to
$\osha{\traN{\typ}}\limp\traN{\typtwo}$.
It operates on
terms by mapping a $\lambda$-calculus application $\tm\,\tmtwo$
to $\traN{\tm}\, \osha{\traN{\tmtwo}}$,
indicating that the argument may be freely copied, discarded, and accessed.
This mimics the original translation of $\CBN$ to $\LL$.
Our $\CBV$ translation $\traV{\arg}$ operates on formulae
by mapping $\typ\to\typtwo$ to
$\osha{\traV{\typ}}\limp\osha{\traV{\typtwo}}$.
It operates on terms by leaving the translation of the argument
of an application intact,
mapping $\tm\,\tmtwo$ to $\open{\der{\traV{\tm}}}\,\traV{\tmtwo}$.
As before, $\der{\cdot}$ is an appropriate operator corresponding to
$\ofc{}$-elimination,
while $\open{\cdot}$ stands for $\sha{}$-elimination.
At the same time, the translation maps
values such as $\lam{\var}{\tm}$ (roughly\footnote{Intuitionistic variables
will be mapped to modal variables.}) to $\osha{\lam{\var}{\traV{\tm}}}$. Thus
an argument $\traV{\tmtwo}$ cannot be copied, discarded, or accessed at all
unless later, after further evaluation, it takes the form $\osha{\val}$,
for some value $\val$, in which case $\val$ can be copied.
This too mimics the original translation from $\CBV$ to $\LL$.
The $\CBN$ and $\CBV$ translations are proved to be sound and complete,
in the sense that two $\lambda$-terms
are interconvertible in the source language if and only if
they are mapped to interconvertible terms in the target language.

\bparagraph{Call-by-Sharing.}
The two translations above suggest a ``missing link'' translation $\traD{\arg}$
that maps $\typ\to\typtwo$ to $\osha{\traD{\typ}}\limp\sha{\traD{\typtwo}}$.
This translation cannot be expressed directly in a linear $\lambda$-calculus,
because the ``$\sha{}$'' modality is used as a stand-alone operator.
A $\lambda$-calculus application $\tm\,\tmtwo$ is now mapped to
$\traD{\tm}\, \ofc{\traD{\tmtwo}}$, meaning that the argument can be copied
and discarded, but \emph{not accessed} yet.
A value such as $\lam{\var}{\tmthree}$ is now (roughly)
mapped to $\sha{\lam{\var}{\traD{\tmthree}}}$.
Thus an argument $\traD{\tmtwo}$ cannot be accessed unless, after further evaluation,
it takes the form $\sha{\val}$, for some value $\val$, in which case $\val$ can be accessed.
The fact that arguments cannot be accessed until they become values
means that the evaluation mechanism can keep \emph{references}
to a single \emph{shared} copy of the argument, until it becomes accessible.
This translation suggests an evaluation mechanism that we dub \emph{call-by-sharing} ($\CCBNd$).

Call-by-sharing bears a strong resemblance to call-by-need ($\CBNd$),
an evaluation mechanism introduced by Wadsworth in 1971~\cite{Wadsworth:Phd:1971}.
Both in $\CBNd$ and in $\CCBNd$, arguments that are not used may be discarded
without being evaluated.
A \emph{reference} to a shared argument may be freely copied,
but the argument itself can only be copied after it has been evaluated to a value.
Nevertheless, there are some subtle differences between $\CBNd$ and $\CCBNd$,
and in particular $\CCBNd$ achieves less sharing than $\CBNd$
(see the discussion in~\rsec{translations}).
Unfortunately, there does not seem to be a way to embed $\CBNd$ into
$\lambdaS$.

\bparagraph{Bang Calculus.}
Another approach towards providing a common framework to explain $\CBV$ and
$\CBN$ is the \BangCalculus~\cite{DBLP:conf/ppdp/EhrhardG16}. It is an untyped
lambda calculus that has explicit constructors in the syntax for
promotion ($\ofc{}$-introduction) and dereliction ($\ofc{}$-elimination).
It was motivated by the fact that Girard's original
$\CBN$ and $\CBV$ translations of the intuitionistic logic into $\LL$ made use
of logical exponentials (promotion and dereliction) that were not reflected in the
syntax. The aim was thus to introduce an intermediate formalism between lambda
calculus and proof nets, a graphical notation for $\LL$
proofs~\cite{DBLP:journals/tcs/Girard87}, that allows explicit use of ``boxes''
to mark values. Soundness and completeness of these translations with respect
to reduction was proved by Guerrieri and
Manzonetto~\cite{DBLP:journals/corr/abs-1904-06845} for slightly different
notion of reduction for the \BangCalculus than that
of~\cite{DBLP:conf/ppdp/EhrhardG16}. The \BangCalculus can in fact be
embedded into our $\lambdaS$-calculus and this embedding is both
sound and complete. 

\bparagraph{Contributions.}
A summary of the contributions are as follows:

\noindent1.
The introduction of a new logic called $\MELLS$ (``\SharingLinearLogic) that enjoys
cut-elimination and is conservative over $\MELL$. It provides a split of
each exponential modality into a \emph{sharing modality}
and an \emph{access modality}.

\noindent2.
A term assignment for $\MELLS$, the \emph{$\lambdaS$-calculus},
which operationally distinguishes between two kinds of expressions.
\emph{Sharable expressions} can be discarded, and \emph{references} to shared
expressions can be duplicated, but they cannot be accessed, and thus they can remain shared.
\emph{Sharable accessible expressions} can moreover be accessed, and they
are copied whenever access to them is requested. This distinction allows to formulate
the $\CCBNd$ evaluation mechanism.

\noindent3.
Translations from $\CBV$ and $\CBN$ to $\lambdaS$ that are sound, complete and
preserve normal forms.

\noindent4.
The presentation of the \emph{call-by-sharing} calculus ($\CCBNd$), and a translation
from $\CCBNd$ to $\lambdaS$ that is sound, complete and preserves normal forms.

\noindent5.
A weak \emph{evaluation mechanism} for $\lambdaS$ that can simulate
weak evaluation strategies in $\CBN$, $\CBV$ and $\CCBNd$, with soundness and completeness
results.

\begin{ifLongAppendix}
\noindent6.
A translation from Ehrhard's \BangCalculus to $\lambdaS$ which is sound,
complete and preserves normal forms.
\end{ifLongAppendix}

\bparagraph{Structure of the paper.}
We review some background notions in~\rsec{preliminaries}.
We present $\MELLS$ in~\rsec{mellsha} and a term
assignment for this logic, the $\lambdaS$-calculus, in~\rsec{lambdaS}.
Definitions of the $\CBV$, $\CBN$, and $\CCBNd$
calculi and their
translations to $\lambdaS$ are presented in~\rsec{translations} together with
results of soundness, completeness, and preservation of normal forms.
\begin{ifLongAppendix}
\rsec{bang} shows how $\lambdaS$ can also embed the \BangCalculus via a sound
and complete translation.
\end{ifLongAppendix}
\rsec{strategies} presents a notion of weak evaluation for $\lambdaS$ which
is shown to simulate weak $\CBV$, $\CBN$, and $\CCBNd$ evaluation.
Finally, we conclude and discuss future work.
Most proofs are omitted from the main body of the paper and can
be found in the appendix.


\section{Preliminary Notions}
\lsec{preliminaries}

In this section we present some background notions and results that we use
throughout the paper.

Recall that an {\em abstract rewriting system} (ARS) is a pair $\ars = (X,\toars)$
where $X$ is a set and ${\toars} \subseteq X^2$ is a binary relation
called {\em reduction}.
We write $\toars^*$ for the reflexive--transitive closure of $\toars$,
$\toars^+$ for the transitive closure,
$\toars^=$ for the reflexive closure,
$\toarssym$ for the symmetric closure,
$\toarsinv$ or $\toars^{-1}$ for the inverse relation,
and $\toars^n$ for the composition of $\toars$ with itself $n$ times.
An ARS is confluent (CR)
if ${\toarsinv^*\,\toars^*} \subseteq {\toars^*\,\toarsinv^*}$
and strongly normalizing (SN)
if there are no infinite reductions $x_1 \toars x_2 \toars \hdots$.

\bparagraph{Abstract Results on Translations}
Given ARSs $\ars = (X,\toars)$ and $\arstwo = (Y,\toarstwo)$,
a {\em translation} $T : \ars \to \arstwo$
is a function $T : X \to Y$,
also written $T$, by abuse of notation.
A translation is {\em sound}
if $x_1 \toars^* x_2$
implies $T(x_1) \toarstwo^* T(x_2)$
for all $x_1,x_2 \in X$,
and {\em complete}
if $T(x_1) \toarstwo^* T(x_2)$
implies $x_1 \toars^* x_2$
for all $x_1,x_2 \in X$.
The following are easy results on a translation $T$:

\restatable{proposition}{prop:sufficient_conditions_translation_sound}{\SufficientConditionsForSoundness}{Conditions for soundness}{
If $x_1 \toars x_2$ implies $T(x_1) \toarstwo^* T(x_2)$
for every $x_1,x_2 \in X$, then $T$ is sound.
}

\restatable{theorem}{thm:sufficient_conditions_translation_complete}{\SufficientConditionsForCompleteness}{Conditions for completeness}{
Let $Y' \subseteq Y$ and let $T^{-1} : Y' \to X$ be a function.
Suppose that
$T^{-1}$ is the left-inverse of $T$,
\ie for all $x \in X$ we have that $T(x) \in Y'$ and $T^{-1}(T(x)) = x$.
Suppose moreover that $T^{-1}$ simulates reduction,
\ie for all $y_1 \in Y'$ and $y_2 \in Y$ such that $y_1 \toarstwo y_2$,
we have that $y_2 \in Y'$ and $T^{-1}(y_1) \toars^* T^{-1}(y_2)$.
Then $T$ is complete.
}

\bparagraph{The Linear Substitution Calculus}
The $\LSC$ is a refinement of the $\lambda$-calculus 
with explicit substitutions,
introduced by Accattoli and Kesner~\cite{DBLP:conf/rta/Accattoli12,DBLP:conf/popl/AccattoliBKL14}
as a variation over a calculus by Milner~\cite{DBLP:journals/entcs/Milner07}.
The set of {\em $\LSC$ terms} ($\TermsLSC$) is defined as follows,
where $\esub{\var}{\tmtwo}$ is called an \emph{explicit substitution} (ES):
\[
  \tm,\tmtwo,\hdots ::=
    \var \mid \lam{\var}{\tm} \mid \tm\,\tmtwo \mid \tm\esub{\var}{\tmtwo}
\]
A {\em context} is a term with a single occurrence of a \emph{hole} ``$\ctxhole$''.
A {\em substitution context} is a list of ESs.
Formally:
\[
  \begin{array}{l@{\HS}rcll}
  \text{General contexts} &
  \gctx & ::= & \ctxhole
    \mid \lam{\var}{\gctx}
    \mid \gctx\,\tm
    \mid \tm\, \gctx
    \mid \gctx\esub{\var}{\tm}
    \mid \tm\esub{\var}{\gctx}
  \\
  \text{Substitution contexts} &
     \sctx & ::= & \ctxhole
             \mid \sctx\esub{\var}{\tm}
  \end{array}
\]
\emph{Free} and \emph{bound} occurrences of variables
are defined as expected, and $\fv{\tm}$ denotes the set of free variables of $\tm$.
Terms are defined up to $\alpha$-renaming of bound variables.

We write $\of{\gctx}{\tm}$ for the
\emph{variable-capturing} substitution of $\ctxhole$ in $\gctx$ by $\tm$.
We write $\off{\gctx}{\tm}$ for the \emph{capture-avoiding}
substitution of $\ctxhole$ in $\gctx$ by $\tm$.
For example,
if $\gctx = \lam{\var}{\ctxhole}$ then $\of{\gctx}{\var} = \lam{\var}{\var}$,
while $\off{\gctx}{\var} = \lam{\vartwo}{\var} \neq \lam{\var}{\var}$.
In the case of substitution contexts,
we usually write $\tm\sctx$ rather than $\of{\sctx}{\tm}$.
The \emph{domain} of $\sctx$, written $\dom{\sctx}$,
is the set of variables bound by $\sctx$.
The $\LSC$ has three rewriting rules, closed by congruence
under arbitrary contexts:
\[
  (\lam{\var}{\tm})\sctx\,\tmtwo    \todb \tm\esub{\var}{\tmtwo}\sctx
  \HS
  \off{\gctx}{\var}\esub{\var}{\tm} \tols \off{\gctx}{\tm}\esub{\var}{\tm}
  \HS
  \tm\esub{\var}{\tmtwo}            \togc \tm \text{\,\,(if $\var\notin\fv{\tm}$)}
\]
\emph{Distance beta} ($\symdb$)
performs a $\beta$-step but creates an ES rather than doing meta-level substitution,
\emph{linear substitution} ($\symls$)
replaces a \emph{single} occurrence of $\var$ to a term $\tm$
when $\var$ is bound to $\tm$
by an ES, and
\emph{garbage collection} ($\symgc$) removes an unreachable ES.
Reduction in $\LSC$ is the union ${\tolsc} \eqdef {\todb \cup \tols \cup \togc}$.

The rewriting rules of $\LSC$ are said to operate ``at a distance'' due to their
peculiar use of contexts. This avoids reduction getting stuck in the
presence of explicit substitutions. For example, if the left-hand side of
$\symdb$ were declared to be $(\lam{\var}{\tm})\,\tmtwo$, then an
expression such as $(\lam{\var}{\tm})\esub{\vartwo}{\tmthree}\,\tmtwo$ would
not be a redex. This notation has a strong connection to proof
nets~\cite{DBLP:conf/ictac/Accattoli18}.

We write $\judlsc{\tctx}{\tm}{\typ}$ is $\tm$ has type $\typ$
under the typing context $\tctx$, with standard simple type assignment rules.
Recall from \cite[Theorem~6.11]{DBLP:journals/corr/abs-2312-13270} that
simply typed terms are SN:
\begin{theorem}
\lthm{simply_typed_lsc_strongly_normalizing}
If $\judlsc{\tctx}{\tm}{\typ}$,
there are no infinite reduction sequences
$\tm \tolsc \tm_1 \tolsc \hdots$.
\end{theorem}


\section{\SharingLinearLogic}
\lsec{mellsha}

We start by defining a one-sided sequent calculus presentation
for a linear logic with \emph{multiplicative} connectives ($\tensor,\parr$,
called \emph{tensor} and \emph{par}),
\emph{\textbf{s}haring} modalities
($\ofc{}$, $\why{}$, called \emph{of-course} and \emph{why-not}),
and \emph{a\textbf{c}cess} modalities
($\sha{}$, $\ush{}$, called \emph{grant} and \emph{demand}),
which we dub $\MELLS$.

\bparagraph{Formulae and Sequent Calculus Presentation.}
We assume given a denumerable set of atomic formulae ($\btyp,\btyptwo,\hdots$),
each of which has a corresponding negative version ($\nbtyp,\nbtyptwo,\hdots$).
The set of formulae is given by the grammar:
\[
  \typ,\typtwo,\hdots ::=
         \btyp
    \mid \nbtyp
    \mid \typ\tensor\typtwo
    \mid \typ\parr\typtwo
    \mid \ofc{\typ}
    \mid \why{\typ}
    \mid \sha{\typ}
    \mid \ush{\typ}
\]
{\em Linear negation} is the involutive operator $\lneg{\arg}$ given by:
\[
  \begin{array}{rcl@{\HS}rcl@{\HS}rcl@{\HS}rcl}
    \lneg{\btyp} & \eqdef & \nbtyp
  &
    \lneg{\nbtyp} & \eqdef & \btyp
  &
    \lneg{(\typ\tensor\typtwo)} & \eqdef & \lneg{\typ}\parr\lneg{\typtwo}
  &
    \lneg{(\typ\parr\typtwo)} & \eqdef & \lneg{\typ}\tensor\lneg{\typtwo}
  \\
    \lneg{(\ofc{\typ})} & \eqdef & \why{\lneg{\typ}}
  &
    \lneg{(\why{\typ})} & \eqdef & \ofc{\lneg{\typ}}
  &
    \lneg{(\sha{\typ})} & \eqdef & \ush{\lneg{\typ}}
  &
    \lneg{(\ush{\typ})} & \eqdef & \sha{\lneg{\typ}}
  \end{array}
\]
Sequents are of the form $\judg{\tctx}$, where $\tctx$ is a finite \emph{multiset} of formulae
(note that we do not include an explicit exchange rule).
If $\tctx = \typ_1,\hdots,\typ_n$ and $\mathfrak{m}$ is one of the modalities,
we write $\mathfrak{m}\tctx$ to stand for $\mathfrak{m}\typ_1,\hdots,\mathfrak{m}\typ_n$,
so for instance
$\ush{\tctx} = \ush{\typ_1},\hdots,\ush{\typ_n}$.
Derivable sequents are given inductively by the following rules:
\[
  \indrule{\ruleAx}{
    \emptyPremise
  }{
    \judg{\typ,\lneg{\typ}}
  }
  \,\,
  \indrule{\ruleCut}{
    \judg{\tctx,\typ}
    \,\,
    \judg{\tctxtwo,\lneg{\typ}}
  }{
    \judg{\tctx,\tctxtwo}
  }
  \,\,
  \indrule{\ruleTensor}{
    \judg{\tctx,\typ}
    \HS
    \judg{\tctxtwo,\typtwo}
  }{
    \judg{\tctx,\tctxtwo,\typ\tensor\typtwo}
  }
  \,\,
  \indrule{\ruleParr}{
    \judg{\tctx,\typ,\typtwo}
  }{
    \judg{\tctx,\typ\parr\typtwo}
  }
  \,\,
  \indrule{\ruleProm}{
    \judg{\why{\tctx},\typ}
  }{
    \judg{\why{\tctx},\ofc{\typ}}
  }
\]
\[
  \indrule{\ruleW}{
    \judg{\tctx}
  }{
    \judg{\tctx,\why{\typ}}
  }
  \,\,
  \indrule{\ruleC}{
    \judg{\tctx,\why{\typ},\why{\typ}}
  }{
    \judg{\tctx,\why{\typ}}
  }
  \,\,
  \indrule{\ruleD}{
    \judg{\tctx,\ush{\typ}}
  }{
    \judg{\tctx,\wush{\typ}}
  }
  \,\,
  \indrule{\ruleSha}{
    \judg{\tctx,\typ}
  }{
    \judg{\tctx,\sha{\typ}}
  }
  \,\,
  \indrule{\ruleUsh}{
    \judg{\tctx,\typ}
  }{
    \judg{\tctx,\ush{\typ}}
  }
\]
Most rules are standard rules
from multiplicative-exponential linear logic~($\MELL$),
including the standard weakening (\ruleW), contraction (\ruleC),
and promotion (\ruleProm) rules.
The atypical rule is dereliction (\ruleD),
which requires the conclusion to be $\tctx,\why{\ush{\typ}}$
instead of the usual $\tctx,\why{\typ}$.
The $\ruleSha$ and $\ruleUsh$ rules
are (trivial) introduction rules for $\sha{}$ and $\ush{}$.

As discussed in the introduction, the intuition is that a proof of $\ofc{\typ}$
lies inside a box which may be duplicated or erased, but
it may not necessarily be possible to \emph{access} the contents of the box,
which means that having a proof does not necessarily enable
one to interact with its contents.
A proof of $\ofc{\sha{\typ}}$ lies inside a box which may
both duplicated, erased, and accessed.
Informally speaking,
the combined modalities $\ofc{\sha{\typ}}$ and $\why{\ush{\typ}}$ in $\MELLS$
play the role of the usual $\ofc{\typ}$ and $\why{\typ}$ modalities in $\MELL$.

\begin{remark}
If we define linear equivalence of formulae $\typ \liff \typtwo$
as usual in linear logic, it is immediate to show that
$\judg{\typ \liff \ush{\typ}}$ holds,
but in general
$\judg{\why{\typ} \liff \why{\ush{\typ}}}$ does not hold.
Hence linear equivalence is not a congruence with respect to
the sharing modalities.
\end{remark}

\bparagraph{Basic Properties.}
Consider the mapping $\toMELLS{\arg}$ from formulae in $\MELL$ to formulae in $\MELLS$ which replaces each occurrence of ``$\ofc{}$'' with ``$\osha{}$'', each occurrence of ``$\why{}$'' with ``$\why{\ush{}}$'', and leaves the other connectives unaltered.
The following theorem is easy to prove by induction on the derivation of the sequents:

\begin{theorem}[Conservativity]
$ \judg{\tctx}$ holds in $\MELL$ if and only if $ \judg{\toMELLS{\tctx}}$ holds in $\MELLS$
\end{theorem}

By the usual techniques, one can show that $\MELLS$ enjoys cut-elimination:

\restatable{theorem}{CutElimination}{\CutElimination}{Cut elimination}{
If $\judg{\tctx}$ is provable in $\MELLS$,
then there is a derivation of $\judg{\tctx}$
without instances of the $\ruleCut$ rule.
}


\section{A Sharing Linear $\lambda$-Calculus}
\lsec{lambdaS}

In this section, we present a \emph{sharing} linear $\lambda$-calculus based
on $\MELLS$, called the $\lambdaS$-calculus.
The relationship between the $\lambdaS$-calculus and $\MELLS$
is akin to that between linear $\lambda$-calculi and $\MELL$.
In particular, typing rules for the $\lambdaS$-calculus are presented
in natural deduction (rather than sequent calculus) style,
and, furthermore, the $\lambdaS$-calculus is intuitionistic
(rather than classical).

\bparagraph{Syntax and Typing System}
We assume given denumerable sets of
{\em linear variables} ($\lvar, \lvartwo, \hdots$)
and
{\em unrestricted variables} ($\uvar, \uvartwo, \hdots$).
The set of {\em types} ($\typ,\typtwo,\hdots$)
and the set $\TermsS$ of {\em $\lambdaS$-terms} ($\tm,\tmtwo,\hdots$),
or just {\em terms}, are given by:
\[
  \typ ::=
    \btyp \mid \typ\limp\typtwo \mid \sha{\typ} \mid \ofc{\typ}
  \HS\HS
  \tm ::=
         \lvar
    \mid \uvar
    \mid \lam{\lvar}{\tm}
    \mid \tm\,\tmtwo
    \mid \sha{\tm}
    \mid \open{\tm}
    \mid \ofc{\tm}
    \mid \tm\esub{\uvar}{\tmtwo}
\]
A term may be a \emph{linear} or an \emph{unrestricted variable},
an \emph{abstraction} $\lam{\lvar}{\tm}$ (binding a linear variable),
an \emph{application} $\tm\,\tmtwo$,
an access \emph{grant} $\sha{\tm}$, an access \emph{request} $\open{\tm}$,
a \emph{promotion} $\ofc{\tm}$ or a \emph{substitution} $\tm\esub{\uvar}{\tmtwo}$
(binding an unrestricted variable).

Free and bound occurrences of variables are defined as expected.
We write $\fv{\tm}$ for the set of free variables of $\tm$.
Terms are defined up to $\alpha$-renaming of bound variables.
By convention, we assume that
$\ofc{\tm\esub{\var}{\tmtwo}}$ stands for $\ofc{(\tm\esub{\var}{\tmtwo})}$.
Similarly,
$\sha{\tm\esub{\var}{\tmtwo}}$, and $\lam{\lvar}{\tm\esub{\var}{\tmtwo}}$
stand, respectively, for
$\sha{(\tm\esub{\var}{\tmtwo})}$,
and $\lam{\lvar}{(\tm\esub{\var}{\tmtwo})}$.

{\em Unrestricted typing environments} ($\uenv,\uenv',\hdots$)
are partial functions mapping unrestricted variables to types,
written $\uvar_1:\typ_1,\hdots,\uvar_n:\typ_n$.
{\em Linear typing environments} ($\lenv,\lenv',\hdots$)
map linear variables to types, written $\lvar_1:\typ_1,\hdots,\lvar_n:\typ_n$.
We assume that typing environments have finite domain.

Typing judgments are of the form $\judc{\uenv}{\lenv}{\tm}{\typ}$. 
Derivable judgments are defined inductively by the following rules.
The types of unrestricted variables in $\uenv$ may be thought of as being
implicitly prefixed by ``$\osha{}$'', as attested by the rule $\rulecES$. 
\[
  \indrule{\rulecLvar}{
  }{
    \judc{\uenv}{\lvar:\typ}{\lvar}{\typ}
  }
  \,\,
  \indrule{\rulecUvar}{
  }{
    \judc{\uenv,\uvar:\typ}{\noenv}{\uvar}{\sha{\typ}}
  }
  \,\,
  \indrule{\rulecApp}{
    \judc{\uenv}{\lenv_1}{\tm}{\typ\limp\typtwo}
    \,\,
    \judc{\uenv}{\lenv_2}{\tmtwo}{\typ}
  }{
    \judc{\uenv}{\lenv_1,\lenv_2}{\tm\,\tmtwo}{\typtwo}
  }
\]
\[
  \indrule{\rulecAbs}{
    \judc{\uenv}{\lenv,\lvar:\typ}{\tm}{\typtwo}
  }{
    \judc{\uenv}{\lenv}{\lam{\lvar}{\tm}}{\typ\limp\typtwo}
  }
  \,\,
  \indrule{\rulecSha}{
    \judc{\uenv}{\lenv}{\tm}{\typ}
  }{
    \judc{\uenv}{\lenv}{\sha{\tm}}{\sha{\typ}}
  }
  \,\,
  \indrule{\rulecOpen}{
    \judc{\uenv}{\lenv}{\tm}{\sha{\typ}}
  }{
    \judc{\uenv}{\lenv}{\open{\tm}}{\typ}
  }
\]
\[
  \indrule{\rulecProm}{
    \judc{\uenv}{\noenv}{\tm}{\typ}
  }{
    \judc{\uenv}{\noenv}{\ofc{\tm}}{\ofc{\typ}}
  }
  \,\,
  \indrule{\rulecES}{
    \judc{\uenv,\uvar:\typ}{\lenv_1}{\tm}{\typtwo}
    \HS
    \judc{\uenv}{\lenv_2}{\tmtwo}{\osha{\typ}}
  }{
    \judc{\uenv}{\lenv_1,\lenv_2}{\tm\esub{\uvar}{\tmtwo}}{\typtwo}
  }
\]

\begin{example}
$\lam{\lvar}{(\osha{\ofc{\uvar}})\esub{\uvar}{\lvar}}$
has type
$\osha{\typ}\limp\osha{\osha{\typ}}$, under the empty contexts.
\end{example}

\bparagraph{Logical Soundness.}
Types of $\lambdaS$ encode \emph{formulae} of $\MELLS$,
while terms encode \emph{proofs}.
Indeed, consider the translation $\tran{\arg}$ on types below,
and the following result:
\[
  \begin{array}{rcl@{\HS}rcl@{\HS}rcl@{\HS}rcl}
    \tran{\btyp}              & \eqdef & \btyp
  &
    \tran{(\typ\limp\typtwo)} & \eqdef & \lneg{\tran{\typ}}\parr\tran{\typtwo}
  &
    \tran{(\sha{\typ})}       & \eqdef & \sha{\tran{\typ}}
  &
    \tran{(\ofc{\typ})}       & \eqdef & \ofc{\tran{\typ}}
  \end{array}
\]

\restatable{proposition}{prop:soundness_of_lambdaS_w_r_t_MELLS}{\SoundessOfLambdaSWRTMELLS}{Logical Soundness of $\lambdaS$}{
If $\judc{\uenv}{\lenv}{\tm}{\typ}$ holds in $\lambdaS$,
then
$\judg{\wush{(\lneg{\tran{\uenv}})},\lneg{\tran{\lenv}},\tran{\typ}}$ holds in $\MELLS$.
}

Note that, by design, $\lambdaS$ is not intended to be \emph{complete}
with respect to $\MELLS$.
For example,
the type $(\ofc{\btyp}\limp\ofc{\btyp}\limp\btyptwo)\limp\ofc{\btyp}\limp\btyptwo$
is not inhabited in $\lambdaS$,
whereas the corresponding formula
$(\ofc{\btyp}\tensor\ofc{\btyp}\tensor\nbtyptwo)\parr\why{\nbtyp}\parr\btyptwo$
is provable in $\MELLS$.

\bparagraph{Reduction Semantics}
Let us write $\CtxsS$ for the the set of
{\em $\lambdaS$-contexts} ($\gctx,\gctx',\hdots$), or just {\em contexts},
which
are $\lambdaS$-terms with a single occurrence of a {\em hole} ``$\ctxhole$'',
and $\SCtxsS$ for the the set of {\em substitution contexts} ($\sctx,\sctx',\hdots$),
which are lists of ESs:
\[
  \begin{array}{rcl@{\HS}rcl}
  \gctx & ::= & \ctxhole
           \mid \lam{\lvar}{\gctx}
           \mid \gctx\,\tm
           \mid \tm\,\gctx
           \mid \sha{\gctx}
           \mid \open{\gctx}
           \mid \ofc{\gctx}
           \mid \gctx\esub{\uvar}{\tm}
           \mid \tm\esub{\uvar}{\gctx}
  &
  \sctx & ::= & \ctxhole \mid \sctx\esub{\uvar}{\tm} \\
  \end{array}
\]
We write $\tm\sub{\lvar}{\tmtwo}$ for the capture-avoiding substitution of
the free occurrences of $\lvar$ in $\tm$ by $\tmtwo$.
The domain of a substitution context ($\dom{\sctx}$),
and the plugging of a term into a context,
both with capture ($\of{\gctx}{\tm}$ and $\tm\sctx$)
and avoiding capture ($\off{\gctx}{\tm}$),
are defined similarly as for the $\LSC$ (see~\rsec{preliminaries}).

There are four \textbf{rewriting rules},
closed by compatibility under arbitrary contexts:
\[
  \begin{array}{r@{\,\,}l@{\,\,}l}
    (\lam{\lvar}{\tm})\sctx\,\tmtwo
    & \toSdb &
    \tm\sub{\lvar}{\tmtwo}\sctx
  \\
    \open{(\sha{\tm})\sctx}
    & \toSopen &
    \tm\sctx
  \\
    \off{\gctx}{\uvar}\esub{\uvar}{(\ofc{(\sha{\tm})\sctx_1})\sctx_2}
    & \toSls &
    \off{\gctx}{(\sha{\tm})\sctx_1}\esub{\uvar}{\ofc{(\sha{\tm})\sctx_1}}\sctx_2
  \\
    \tm\esub{\uvar}{(\ofc{\tmtwo})\sctx}
    & \toSgc &
    \tm\sctx
    \HS\text{(if $\uvar\notin\fv{\tm}$)}
  \end{array}
\]
Reduction in $\lambdaS$ is defined as the union
${\toS} \eqdef {\toSdb \cup \toSopen \cup \toSls \cup \toSgc}$.

The $\symSdb$ rule is a distant $\beta$-rule.
The calculus does not assume that terms are typable; but, in
typable terms, there is exactly one occurrence of $\lvar$ in the body
of $\lam{\lvar}{\tm}$ by linearity.
The $\symSopen$ rule requests access to a term.
The $\symSls$ rule substitutes a single occurrence of $\uvar$
by $(\sha{\tm})\sctx_1$, provided that $\uvar$ is bound to
a term of the form $(\ofc{(\sha{\tm})\sctx_1})\sctx_2$.
Note that the subterm $(\sha{\tm})\sctx_1$ is \emph{copied} by $\symSls$,
while $\sctx_2$ is moved outside so that its bindings remain shared.
The $\symSgc$ rule may erase an unused substitution if
$\uvar$ is bound to a term of the form $(\ofc{\tmtwo})\sctx$.
Allowing to erase \emph{any} unused substitution, \eg with
the usual $\symgc$ rule of $\LSC$, would break subject reduction,
as it could lead to weakening (erasure) of a linear variable.

The $\symSgc$ rule requires that it first be evaluated to a \emph{sharable} term,
of the form $\ofc{\tmtwo'}$, or more in general $(\ofc{\tmtwo'})\sctx$.
A sharable term may not, a priori, be accessed.
For example, in $(\open{\uvar}\,\open{\uvar})\esub{\uvar}{\ofc{\uvartwo}}$
the two occurrences of $\uvar$ cannot be substituted by $\uvartwo$,
because the $\symSls$ rule requires the argument to be a \emph{sharable accessible}
term, of the form $\osha{\tmtwo'}$,
or more in general $(\ofc{(\sha{\tmtwo'})\sctx_1})\sctx_2$.

A first routine result is:

\begin{proposition}[Subject Reduction]
 \lprop{subject_reduction}
   If $\judc{\uenv}{\lenv}{\tm}{\typ}$ and $\tm\toS \tmtwo$, then $\judc{\uenv}{\lenv}{\tmtwo}{\typ}$.
\end{proposition}

\bparagraph{Confluence.}
Confluence is tricky since standard techniques are not immediately applicable.
A variant of Tait--Martin-L\"of's method~\cite[\S~3.2]{DBLP:books/daglib/0067558}
would require to define parallel reduction, which is not immediate
due to the fact that rewrite rules operate at a distance.
The interpretation method, used for confluence of the structural lambda
calculus~\cite{AccattoliK10}, a close relative of $\LSC$,
does not apply since \emph{full composition},
\ie $\tm\esub{\var}{\tmtwo} \to^* \tm\sub{\var}{\tmtwo}$,
does not hold for $\lambdaS$.
An axiomatic rewriting approach based on residuals and orthogonality
fails for $\lambdaS$ too.
A simple example is the following, where we use labels $\lab$ and $\labtwo$ to mark
redexes:
\[
    \tm\esub{\uvartwo}{(\ofc{\sha{\tmtwo}})\sctx}
    \leftarrow^{\lab}_{\symSgc}
    \tm\esub{\ann{\uvar}{\lab}}{(\ofc{\ann{\uvartwo}{\labtwo}})\esub{\uvartwo}{(\ofc{\sha{\tmtwo}})\sctx}}
    \lto{\labtwo}_{\symSls} 
    \tm\esub{\ann{\uvar}{\lab}}{(\ofc{\sha{\tmtwo}})\esub{\uvartwo}{(\ofc{\sha{\tmtwo}})}\sctx}
    \lto{\lab}_{\symSgc}
    \tm\esub{\uvartwo}{\ofc{\sha{\tmtwo}}}\sctx
\]
where, however,
$\tm\esub{\uvartwo}{(\ofc{\sha{\tmtwo}})\sctx}
 \neq
 \tm\esub{\uvartwo}{\ofc{\sha{\tmtwo}}}\sctx$.
Nevertheless, confluence holds for $\lambdaS$ modulo
the congruence generated by
$
  \tm\esub{\uvar}{\tmtwo\esub{\uvartwo}{\tmthree}}
  \flatt
  \tm\esub{\uvar}{\tmtwo}\esub{\uvartwo}{\tmthree}
$, provided that $\uvartwo\notin\fv{\tm}$.
To prove this, we develop a theory of residuals for $\lambdaS$ modulo
$\flatt$. We then resort to the axiomatic rewriting framework due to
Melli\`es~\cite{Mellies:PhD:1996},
verifying that $\lambdaS$ modulo $\flatt$ can be modeled as an
\emph{orthogonal axiomatic rewriting system},
which entails confluence (see~\cite[Theorem~2.4]{Mellies:PhD:1996}):

\begin{proposition}[Confluence \proofnote{Proof in Sec.~\ref{section:appendix:calculus_cr}}]
$\lambdaS$ modulo $\flatt$ is confluent.
\end{proposition}

\bparagraph{Strong Normalization.}
Typable terms in $\lambdaS$ are strongly normalizing.
To prove this, we reduce SN of typed $\lambdaS$
to SN of simply typed $\LSC$.

Let us write $\toSi$ for ${\toS}\!\setminus\!{\toSgc}$
and $\tolsci$ for ${\tolsc}\!\setminus\!{\toSgc}$.
Garbage collection $\toSgc$ can be \emph{postponed}, so
to show there are no infinite reduction sequences 
$\tm \toS \tm_1 \toS \tm_2 \hdots$ we may assume without loss of generality
that the sequence consists of $\toSi$ steps.

We start by defining a translation $\tradlsc{\cdot}$ from $\lambdaS$ to $\LSC$.
Let $\lscUnit$ be any inhabited type in simply typed $\LSC$,
and $\lscunit$ a closed inhabitant of $\lscUnit$ in normal form.
Types and terms are translated as follows (where $\varthree$ is assumed
to be fresh in the ``$\sha{}$'' case):
\[
  \begin{array}{rcl@{\HS}rcl@{\HS}rcl@{\HS}rcl}
  \tradlsc{\btyp}            & \eqdef & \btyp
  &
  \tradlsc{\typ\limp\typtwo} & \eqdef & \tradlsc{\typ} \to \tradlsc{\typtwo}
  &
  \tradlsc{\sha{\typ}}       & \eqdef & \lscUnit \to \tradlsc{\typ}
  &
  \tradlsc{\ofc{\typ}}       & \eqdef & \tradlsc{\typ} \\
  \end{array}
  \vspace{-.25cm}
\]
\[
  \begin{array}{rcl@{\HS}rcl}
    \tradlsc{\lvar}
  & \eqdef &
    \lvar
  &
    \tradlsc{\uvar}
  & \eqdef &
    \uvar
  \\
    \tradlsc{\lam{\lvar}{\tm}}
  & \eqdef &
    \lam{\lvar}{\tradlsc{\tm}}
  &
    \tradlsc{\tm\,\tmtwo}
  & \eqdef &
    \tradlsc{\tm}\,\tradlsc{\tmtwo}
  \end{array}
  \HS
  \begin{array}{rcl@{\HS}rcl}
    \tradlsc{\sha{\tm}}
  & \eqdef &
    \lam{\varthree}{\tradlsc{\tm}}
  &
    \tradlsc{\open{\tm}}
  & \eqdef &
    \tradlsc{\tm}\,\lscunit
  \\
    \tradlsc{\ofc{\tm}} 
  & \eqdef &
    \tradlsc{\tm}
  &
    \tradlsc{\tm\esub{\var}{\tmtwo}} 
  & \eqdef &
    \tradlsc{\tm}\esub{\var}{\tradlsc{\tmtwo}}
  \end{array}
\]
An unrestricted variable $\uvar:\typ$ in the environment
is translated as $\uvar:\lscUnit\to\tradlsc{\typ}$,
while a linear variable $\lvar:\typ$ is translated as $\lvar:\tradlsc{\typ}$.
It is easy to show that the translation preserves typing, in the sense that
$\judc{\uenv}{\lenv}{\tm}{\typ}$
implies $\judlsc{\tradlsc{\uenv},\tradlsc{\lenv}}{\tradlsc{\tm}}{\tradlsc{\typ}}$.

To conclude, it would suffice to show that $\LSC$ simulates $\lambdaS$
reduction;
more precisely that $\tm \toS \tmtwo$
implies $\tradlsc{\tm} \tolsc^+ \tradlsc{\tmtwo}$.
Unfortunately, this is not the case;
for instance, in the following example,
the $\toSls$ rule of $\lambdaS$ \emph{extrudes} the $\esub{\vartwo}{\varthree}$
outside to share the binding,
while the $\tols$ rule of $\LSC$ makes a \emph{copy} of $\esub{\vartwo}{\varthree}$:
\[
  \begin{array}{rll@{\HS}rll}
    \var\esub{\var}{(\osha{\vartwo})\esub{\vartwo}{\varthree}}
  & \toSls &
    (\sha{\vartwo})\esub{\var}{\osha{\vartwo}}\esub{\vartwo}{\varthree}
  &
    \var\esub{\var}{\tm\esub{\vartwo}{\varthree}}
  & \tols &
    (\tm\esub{\vartwo}{\varthree})
      \esub{\var}{\tm\esub{\vartwo}{\varthree}}
  \end{array}
\]
To address this, we define a binary relation ${\tofuse}$ on $\LSC$ terms
called \emph{fusion},
that allows to ``extrude and fuse'' ESs,
given by the reflexive, transitive, and contextual closure of
the following rules, avoiding capture:
\[
  \text{
  $\tm\esub{\var}{\tmtwo}
   \tofuse
   \tm$
  \text{ (if $\var\notin\fv{\tm}$)}
  }
  \HS
  \text{
  $\tm\esub{\var}{\tmtwo}\esub{\vartwo}{\tmtwo}
   \tofuse
   \tm\sub{\var}{\vartwo}\esub{\vartwo}{\tmtwo}$
  }
  \HS
  \text{
  $\of{\gctx}{\tm\esub{\var}{\tmtwo}}
   \tofuse
   \of{\gctx}{\tm}\esub{\var}{\tmtwo}$
  }
\]
The two key technical properties are,
first, that the $\tolsci$ simulates $\toSi$ up to fusion,
more precisely that
$\tm \toSi \tmtwo$ implies $\tradlsc{\tm} \tolsci^+\tofuse \tradlsc{\tmtwo}$.
Second, that fusion can be postponed, \ie that
${\tofuse\toSi} \subseteq {\toSi^+\tofuse}$.
Thus an infinite reduction sequence $\tm \toSi \tm_1 \toSi \tm_2 \hdots$
can be mapped to
$\tradlsc{\tm} \tolsci^+\tofuse \tradlsc{\tm_1} \tolsci^+\tofuse \tradlsc{\tm_2} \hdots$
and by postponing fusion we obtain an infinite reduction in $\LSC$.
If $\tm$ is typable in $\lambdaS$ then $\tradlsc{\tm}$ is typable in $\LSC$,
contradicting \rthm{simply_typed_lsc_strongly_normalizing}. To sum up:
\restatable{theorem}{thm:calculus_sn}{\StrongNormalization}{Termination}{
If $\tm$ is typable in $\lambdaS$ then $\tm$ is $\toS$-SN.
}
To conclude this section, we remark that $\toS$-normal forms can
be characterized inductively. The full characterization can be found
in the appendix (\rsec{appendix:calculus_normal_forms}).


\section{Embedding $\CBN$, $\CBV$ and $\CCBNd$}
\lsec{translations}

In this section, we recall known $\CBN$ and $\CBV$ calculi,
and we introduce a sharing variant of call-by-name
we dub \emph{call-by-sharing} ($\CCBNd$).
Second, we define translations that provide embeddings
into $\lambdaS$, and study their properties.

Our first step is to give precise definitions of each of the calculi we will
work with. These calculi operate on $\LSC$ terms,
and some of them also use the notions of
{\em strict values} ($\val,\valtwo,\hdots$)
and {\em lax values} ($\lval,\lvaltwo,\hdots$), defined as follows:
\[
  \text{Strict values}\,\,\,\,
  \val ::= \lam{\var}{\tm}
  \HS
  \text{Lax values}\,\,\,\,
  \lval ::= \var \mid \lam{\var}{\tm}
\]
We define the following rewriting rules on $\LSC$ terms,
closed by compatibility under arbitrary contexts:
\[
  \begin{array}{rll@{\HS}rll}
    (\lam{\var}{\tm})\sctx\,\tmtwo
    & \todb &
    \tm\esub{\var}{\tmtwo}\sctx
  &
    \off{\gctx}{\var}\esub{\var}{\tm}
    & \tols &
    \off{\gctx}{\tm}\esub{\var}{\tm}
  \\
    \off{\gctx}{\var}\esub{\var}{\val\sctx}
    & \tolsv &
    \off{\gctx}{\val}\esub{\var}{\val}\sctx
  &
    \off{\gctx}{\var}\esub{\var}{\val\sctx}
    & \tolsw &
    \off{\gctx}{\val\sctx}\esub{\var}{\val\sctx}
  \\
    \tm\esub{\var}{\tmtwo}
    & \togc &
    \tm
    \HS\text{(if $\var \notin \fv{\tm}$)}
  &
    \tm\esub{\var}{\lval\sctx}
    & \togclv &
    \tm\sctx
    \HS\text{(if $\var \notin \fv{\tm}$)}
  \end{array}
\]

\begin{definition}[Notions of reduction]
The relations corresponding to $\CBN$ ($\tocbn$),
$\CBV$ ($\tocbv$), and $\CCBNd$ ($\tocbCnd$)
reduction are defined by:
\[
  {\tocbn} \eqdef {\todb \!\cup\! \tols \!\cup\! \togc}
  \HS
  {\tocbv} \eqdef {\todb \!\cup\! \tolsv \!\cup\! \togclv}
  \HS
  {\tocbCnd} \eqdef {\todb \!\cup\! \tolsw \!\cup\! \togc}
\]
\end{definition}

The call-by-name ($\CBN$) calculus $\tocbn$
corresponds to usual reduction in $\LSC$~\cite{DBLP:conf/popl/AccattoliBKL14}.

The call-by-value ($\CBV$) calculus $\tocbv$ is a variant of Accattoli and Paolini's
value-substitution calculus~\cite{DBLP:conf/flops/AccattoliP12} (VSC),
with two differences.
First, the rule $\tolsv$ is \emph{linear}
in that it substitutes one occurrence of a variable $\var$ at a time,
while the corresponding rule of VSC substitutes all occurrences of $\var$ at once.
This difference allows us to present the calculi in a uniform way.
Second, $\tolsv$ allows substituting variables for \emph{strict values} (abstractions),
while VSC allows substituting \emph{lax values} (both abstractions and variables).
This is necessary to be able to define a \emph{complete} embedding
(see also \rremark{w_translation}).

The call-by-sharing ($\CCBNd$) calculus $\tocbCnd$ is a sharing variant of $\CBN$,
in which the argument may be discarded without being evaluated (using $\symgc$).
At the same time, the evaluation of the argument is \emph{shared},
in the sense that $\symlsw$ only allows copying arguments when they have been evaluated
to the form $\val\sctx$.
The $\CCBNd$ calculus bears a strong resemblance to the call-by-need
$\lambda$-calculus of Ariola \etal~\cite{AFMOW95},
which can be obtained by changing $\tolsw$ to $\tolsv$,
\ie call-by-need is ${\tocbnd} \eqdef {\todb \cup \tolsv \cup \togc}$
(see for instance~\cite{DBLP:journals/pacmpl/BalabonskiBBK17}).
Note that $\CCBNd$ achieves \emph{less} sharing that $\CBNd$, because
the $\tolsw$ rule makes two copies of $\sctx$,
whereas $\tolsv$ keeps a single shared copy of $\sctx$.
Unfortunately, it does not seem possible to give a sound embedding of $\CBNd$
into $\lambdaS$.

\begin{remark}
The reduction relations above are \emph{calculi},
\ie orientations of equational theories,
not \emph{evaluation} mechanisms.
We shall turn our attention to evaluation in~\rsec{strategies}.
\end{remark}

Next, we describe the translations
$\traN{\arg}$, $\traV{\arg}$, and $\traCNd{\arg}$.
Each translation
maps a simple type into a $\lambdaS$-type,
an $\LSC$ term into a $\lambdaS$-term,
and an $\LSC$ typing judgment into an $\lambdaS$ typing judgment.

\bparagraph{Embedding Call-by-Name}
The $\CBN$ translation $\traN{\arg}$ is defined on types and terms by:
\[
    \traN{\btyp}
    \eqdef
    \btyp
  \HS
    \traN{(\typ\to\typtwo)}
    \eqdef
    \osha{\traN{\typ}}\limp\traN{\typtwo}
\]
\[
    \traN{\var}
    \eqdef
    \open{\var}
  \HS
    \traN{(\lam{\var}{\tm})}
    \eqdef
    \lam{\lvar}{\traN{\tm}\esub{\var}{\lvar}}
  \HS
    \traN{(\tm\,\tmtwo)}
    \eqdef
    \traN{\tm}\,\osha{\traN{\tmtwo}}
  \HS
    \traN{\tm\esub{\var}{\tmtwo}}
    \eqdef
    \traN{\tm}\esub{\var}{\osha{\traN{\tmtwo}}}
\]
In the abstraction case, $\lvar$ is assumed to be fresh,
\ie $\lvar \notin\fv{\traN{\tm}}$.
The translation is extended
to typing environments:
$\traN{(\var_1:\typ_1,\hdots,\var_n:\typ_n)}
 \eqdef
  \var_1:\traN{\typ_1},\hdots,\var_n:\traN{\typ_n}$,
and judgments:
$\traN{(\judl{\tctx}{\tm}{\typ})}
 \eqdef
 \judc{\traN{\tctx}}{\noenv}{\traN{\tm}}{\traN{\typ}}$.

\restatable{proposition}{prop:typing_the_CBN_translation}{\TypingTheCBNTranslation}{$\CBN$ typing}{
If $\judl{\tctx}{\tm}{\typ}$
then $\judc{\traN{\tctx}}{\noenv}{\traN{\tm}}{\traN{\typ}}$.  
}

\begin{lemma}[$\CBN$ simulation]
\llem{traN_simulation}
If $\tm \tocbn \tmtwo$ then $\traN{\tm} \toS^* \traN{\tmtwo}$.
Furthermore, the reduction uses either at least one, and at most two $\toS$ steps.
\end{lemma}
\begin{proof}
By induction on the derivation of $\tm \tocbn \tmtwo$.
The interesting cases are when there is a $\symdb$, $\symls$, or $\symgc$
step at the root.
  If $(\lam{\var}{\tm})\sctx\,\tmtwo \todb \tm\esub{\var}{\tmtwo}\sctx$,
  then:
  \[
      \traN{((\lam{\var}{\tm})\sctx\,\tmtwo)}
    =
      (\lam{\lvar}{\traN{\tm}\esub{\var}{\lvar}})\traN{\sctx}\,\osha{\traN{\tmtwo}}
    \toSdb
      \traN{\tm}\esub{\var}{\osha{\traN{\tmtwo}}}\traN{\sctx}
    =
      \traN{(\tm\esub{\var}{\tmtwo}\sctx)}
  \]
  If $\off{\gctx}{\var}\esub{\var}{\tm} \tols \off{\gctx}{\tm}\esub{\var}{\tm}$,
  then:
  \[
    \begin{array}{ll}
      \traN{\off{\gctx}{\var}\esub{\var}{\tm}}
    =
      \off{\traN{\gctx}}{\open{\var}}\esub{\var}{\osha{\traN{\tm}}}
    &
    \toSls
      \off{\traN{\gctx}}{\open{\sha{\traN{\tm}}}}\esub{\var}{\osha{\traN{\tm}}}
    \\
    &
    \toSopen
      \off{\traN{\gctx}}{\traN{\tm}}\esub{\var}{\osha{\traN{\tm}}}
    =
      \traN{\off{\gctx}{\tm}\esub{\var}{\tm}}
    \end{array}
  \]
  If $\tm\esub{\var}{\tmtwo} \togc \tm$ with $\var \notin \fv{\tm}$,
  then
  $
    \traN{\tm\esub{\var}{\tmtwo}}
    =
    \traN{\tm}\esub{\var}{\osha{\traN{\tmtwo}}}
    \toSgc
    \traN{\tm}
  $,
  where that $\var\notin\fv{\traN{\tm}}$ because $\fv{\traN{\tm}} = \fv{\tm}$.
  \qedhere
\end{proof}

For completeness, we define an \textbf{inverse $\CBN$ translation}.
We define a subset $\TermsSName \subseteq \TermsS$,
containing the closure by $\toS$-reduction of the image of $\traN{\arg}$:
\[
  \rtm,\rtmtwo,\hdots
     ::= \open{\uvar}
    \mid \open{\sha{\rtm}}
    \mid \lam{\lvar}{\rtm\esub{\uvar}{\lvar}}
    \mid \rtm\,\osha{\rtmtwo}
    \mid \rtm\esub{\uvar}{\osha{\rtmtwo}}
\]
where, in the production $\rtm ::= \lam{\lvar}{\rtm\esub{\uvar}{\lvar}}$
we assume that $\lvar$ is fresh, that is, $\lvar \notin \fv{\rtm}$.
The {\em inverse $\CBN$ translation} is a function
$\traNinv{\arg} : \TermsSName \to \TermsLSC$ defined as follows,
by induction on the derivation of a term with the grammar
above\footnote{Observe that the derivation is unique
since, as can be easily seen, the grammar is unambiguous.}.
\[
  \begin{array}{rcl@{\HS}rcl@{\HS}rcl}
    \traNinv{\open{\var}}
    & \eqdef &
    \var
  &
    \traNinv{\open{\sha{\rtm}}}
    & \eqdef &
    \traNinv{\rtm}
  &
    \traNinv{(\lam{\lvar}{\rtm\esub{\var}{\lvar}})}
    & \eqdef &
    \lam{\var}{\traNinv{\rtm}}
  \\
    \traNinv{(\rtm\,\osha{\rtmtwo})}
    & \eqdef &
    \traNinv{\rtm}\,\traNinv{\rtmtwo}
  &
    \traNinv{\rtm\esub{\var}{\osha{\rtmtwo}}}
    & \eqdef &
    \traNinv{\rtm}\esub{\var}{\traNinv{\rtmtwo}}
  \end{array}
\]
It is easy to check that $\traNinv{\arg}$ is the left-inverse of $\traN{\arg}$,
\ie if $\tm \in \TermsLSC$ then $\traN{\tm} \in \TermsSName$
and $\traNinv{(\traN{\tm})} = \tm$. Moreover:

\restatable{lemma}{lemma:traNinv_simulation}{\traNinvSimulation}{Inverse $\CBN$ simulation}{
Let $\rtm \in \TermsSName$ and $\tmtwo \in \TermsS$
such that $\rtm \toS \tmtwo$.
Then $\tmtwo \in \TermsSName$
and $\traNinv{\rtm} \tocbn^= \traNinv{\tmtwo}$.
}

Using the abstract soundness and completeness results
(\rprop{sufficient_conditions_translation_sound}, \rthm{sufficient_conditions_translation_complete})
together with the lemmas above, we obtain:

\begin{theorem}[Sound and complete $\CBN$ embedding]
Given terms $\tm,\tmtwo \in \TermsLSC$,
$\tm \tocbn^* \tmtwo$
if and only if
$\traN{\tm} \toS^* \traN{\tmtwo}$.
Moreover,
$\tm$ is in $\tocbn$-normal form iff $\traN{\tm}$ is in $\toS$-normal form.
\end{theorem}

\bparagraph{Embedding Call-by-Value}
The $\CBV$ translation $\traV{\arg}$ is defined on types and terms by:
\[
    \traV{\btyp}
    \eqdef
    \btyp
  \HS
    \traV{(\typ\to\typtwo)}
    \eqdef
    \osha{\traV{\typ}}\limp\osha{\traV{\typtwo}}
\]
\[
    \traV{\var}
    \eqdef
    \ofc{\var}
  \HS
    \traV{(\lam{\var}{\tm})}
    \eqdef
    \osha{\lam{\lvar}{\traV{\tm}\esub{\var}{\lvar}}}
  \HS
    \traV{(\tm\,\tmtwo)}
    \eqdef
    \open{\uvar}\esub{\uvar}{\traV{\tm}}\,\traV{\tmtwo}
  \HS
    \traV{\tm\esub{\var}{\tmtwo}}
    \eqdef
    \traV{\tm}\esub{\var}{\traV{\tmtwo}}
\]
where, as for $\CBN$, $\lvar$ is assumed to be fresh in the abstraction case.
The translation is extended
to typing environments:
$\traV{(\var_1:\typ_1,\hdots,\var_n:\typ_n)}
 \eqdef
 \var_1:\traV{\typ_1},\hdots,\var_n:\traV{\typ_n}$,
and judgments:
$\traV{(\judl{\tctx}{\tm}{\typ})}
 \eqdef
 \judc{\traV{\tctx}}{\noenv}{\traV{\tm}}{\osha{\traV{\typ}}}$.

\restatable{proposition}{TypingTheCBVTranslation}{\TypingTheCBVTranslation}{$\CBV$ typing}{
If $\judl{\tctx}{\tm}{\typ}$
then $\judc{\traV{\tctx}}{\noenv}{\traV{\tm}}{\osha{\traV{\typ}}}$.
}

\restatable{lemma}{lemma:traV_simulation}{\traVSimulation}{$\CBV$ simulation}{
If $\tm \tocbv \tmtwo$ then $\traV{\tm} \toS^* \traV{\tmtwo}$.
Furthermore, the reduction uses at least one, and at most four $\toS$ steps.
}

We turn our attention to \textbf{completeness} for $\traV{\arg}$.
A first comment is that the $\CBV$ translation only turns out to be
complete up to garbage collection.
More precisely, soundness with respect to reduction holds, in the sense
that $\tm \tocbv^* \tmtwo$ implies $\traV{\tm} \toS^* \traV{\tmtwo}$,
but completeness only holds in the following weak form:
$\traV{\tm} \toS^* \traV{\tmtwo}$ implies $\tm \tocbvUgclvinvs \tmtwo$,
where $\tocbvUgclvinv \eqdef (\tocbv\cup\togclv^{-1})$.
Resorting to confluence, it is possible recover ``plain'' soundness and completeness
of the translation, \ie with respect to the equational theory and not to reduction;
more precisely
$\tm \tosymcbv^* \tmtwo$ if and only if $\traV{\tm} \tosymS^* \traV{\tmtwo}$.
Besides:

\begin{remark}
\lremark{w_translation}
  The study of completeness motivates
  the fact that in CBV the $\symls$ rule can substitute only {\em strict} values
  (abstractions)
  while the $\symgclv$ rule can erase {\em lax} values
  (abstractions and variables).
  To allow substituting variables,
  the translation of a variable $\var$ should be a term of the form $\osha{\tm}$.
  A preliminary version of this work
  used a CBV translation $\traValt{\arg}$
  similar to $\traV{\arg}$ but with $\traValt{\var} \eqdef \osha{\open{\var}}$.
  However, $\traValt{\arg}$ is not complete.
  In fact,
  it can be checked
  $\var\esub{\var}{\tm}\,\tmtwo \tosymcbv^* \tm\,\tmtwo$
  does not hold in general
  (because $\tm$ may not be convertible to a value),
  while it can be seen that
  $\traValt{(\var\esub{\var}{\tm}\,\tmtwo)} \tosymS^* \traValt{(\tm\,\tmtwo)}$
  always holds.
  On the other hand,
  if the $\symgclv$ rule were not allowed to erase variables,
  this would again lead to incompleteness,
  as
  $\var\esub{\vartwo}{\varthree} \tosymcbv^* \var$
  would not hold,
  but
  $\traV{(\var\esub{\vartwo}{\varthree})} \tosymS^* \traV{\var}$
  would hold, since
  $\traV{(\var\esub{\vartwo}{\varthree})}
   = \ofc{\var}\esub{\vartwo}{\ofc{\varthree}}
   \toSgc \ofc{\var}
   = \traV{\var}$.
\end{remark}

Next, we define an \textbf{inverse $\CBV$ translation}.
First, we define a subset $\TermsSValue \subseteq \TermsS$,
containing the closure by $\toS$-reduction of the image of $\traV{\arg}$,
as well as a subset $\SCtxsSValue \subseteq \SCtxsS$:
\[
  \begin{array}{rrl}
  \rtm,\rtmtwo,\hdots & ::= &
         \ofc{\var}
    \mid \osha{\lam{\lvar}{\rtm\esub{\var}{\lvar}}}
    \mid \open{\uvar}\esub{\uvar}{\rtm}
    \mid \open{\sha{\lam{\lvar}{\rtm\esub{\var}{\lvar}}}}
    \mid \lam{\lvar}{\rtm\esub{\var}{\lvar}}
    \mid \rtm\,\rtmtwo
    \mid \rtm\esub{\uvar}{\rtmtwo}
  \\
  \rsctx & ::= & \ctxhole \mid \rsctx\esub{\uvar}{\rtm}
  \end{array}
\]
where in the occurrences of
$\lam{\lvar}{\rtm\esub{\var}{\lvar}}$
we assume that $\lvar$ is fresh.
The {\em inverse $\CBV$ translation} is a function
$\traVinv{\arg} : \TermsSValue \to \TermsLSC$ defined as follows,
by induction on the derivation of a term with the (unambiguous) grammar above:
\[
  \begin{array}{rcl@{\HS}rcl@{\HS}rcl}
    \traVinv{(\ofc{\var})}
    & \eqdef &
    \var
  &
    \traVinv{(\osha{\lam{\lvar}{\rtm\esub{\var}{\lvar}}})}
    & \eqdef &
    \lam{\var}{\traVinv{\rtm}}
  &
    \traVinv{\rtm\,\rtmtwo}
    & \eqdef &
    \traVinv{\rtm}\,\traVinv{\rtmtwo}
  \\
    \traVinv{(\open{\sha{\lam{\lvar}{\rtm\esub{\var}{\lvar}}}})}
    & \eqdef &
    \lam{\var}{\traVinv{\rtm}}
  &
    \traVinv{(\lam{\lvar}{\rtm\esub{\var}{\lvar}})}
    & \eqdef &
    \lam{\var}{\traVinv{\rtm}}
  \\
    \traVinv{(\open{\uvar}\esub{\uvar}{\rtm})}
    & \eqdef &
    \traVinv{\rtm}
  &
    \traVinv{\rtm\esub{\uvar}{\rtmtwo}}
    & \eqdef &
    \traVinv{\rtm}\esub{\uvar}{\traVinv{\rtmtwo}}
  \end{array}
\]
It is easy to check that $\traVinv{\arg}$ is the left-inverse of $\traV{\arg}$.

\restatable{lemma}{lemma:traVinv_simulation}{\traVinvSimulation}{Inverse $\CBV$ simulation, up to $\symgclv$}{
Let $\rtm \in \TermsSValue$ and $\tmtwo \in \TermsS$
such that $\rtm \toS \tmtwo$.
Then $\tmtwo \in \TermsSValue$
and $\traVinv{\rtm} \tocbvUgclvinvs \traVinv{\tmtwo}$,
where $\tocbvUgclvinv \eqdef (\tocbv\cup\togclv^{-1})$.  
}

\begin{theorem}[Sound and complete $\CBV$ embedding]
  Given terms $\tm,\tmtwo \in \TermsLSC$:
  \begin{enumerate}
    \item 
      $\tm \tocbv^* \tmtwo$
    implies 
    $\traV{\tm} \toS^* \traV{\tmtwo}$
  \item 
    $\traV{\tm} \toS^* \traV{\tmtwo}$
    implies 
    $\tm \tocbvUgclvinvs \tmtwo$
    where $\tocbvUgclvinv \eqdef (\tocbv\cup\togclv^{-1})$.
  \item
    $\tm \tosymcbv^* \tmtwo$
    if and only if
    $\traV{\tm} \tosymS^* \traV{\tmtwo}$
  \item
    $\tm$ is in $\tocbv$-normal form iff $\traV{\tm}$ is in $\toS$-normal form.
  \end{enumerate}
\end{theorem}

\begin{remark}
Arrial~\cite{Arrial:HOR2023} suggests an additional $\CBV$ translation
mapping $\typ\to\typtwo$ to $\ofc{(A\limp \ofc{B})}$.
In our setting, this means mapping $\typ\to\typtwo$ to
$\osha{(\typ\limp\osha{\typtwo})}$. This translation is also sound
and complete but still requires $\tocbvUgclvinv$ to obtain completeness.
\end{remark}

\bparagraph{Embedding Call-by-Sharing}
The $\CCBNd$ translation $\traCNd{\arg}$ is defined on types and terms by:
\[
    \traCNd{\btyp}
    \eqdef
    \btyp
  \HS
    \traCNd{(\typ\to\typtwo)}
    \eqdef
    \osha{\traCNd{\typ}}\limp\sha{\traCNd{\typtwo}}
\]
\[
    \traCNd{\var}
    \eqdef
    \var
  \HS
    \traCNd{(\lam{\var}{\tm})}
    \eqdef
    \sha{\lam{\lvar}{\traCNd{\tm}\esub{\var}{\lvar}}}
  \HS
    \traCNd{(\tm\,\tmtwo)}
    \eqdef
    \open{\traCNd{\tm}}\,\ofc{\traCNd{\tmtwo}}
  \HS
    \traCNd{\tm\esub{\var}{\tmtwo}}
    \eqdef
    \traCNd{\tm}\esub{\var}{\ofc{\traCNd{\tmtwo}}}
\]
where, as before, $\lvar$ is assumed to be fresh in the abstraction case.
The translation is extended to typing environments:
$
    \traCNd{(\var_1:\typ_1,\hdots,\var_n:\typ_n)}
    \eqdef
    \var_1:\traCNd{\typ_1},\hdots,\var_n:\traCNd{\typ_n}
$,
and judgments:
    $\traCNd{(\judl{\tctx}{\tm}{\typ})}
      \eqdef
      \judc{\traCNd{\tctx}}{\noenv}{\traCNd{\tm}}{\sha{\traCNd{\typ}}}$.

\restatable{proposition}{prop:typing_the_CCBNd_translation}{\TypingTheCCBNdTranslation}{$\CCBNd$ typing}{
If $\judl{\tctx}{\tm}{\typ}$
then $\judc{\traCNd{\tctx}}{\noenv}{\traCNd{\tm}}{\sha{\traCNd{\typ}}}$.
}

\begin{lemma}[$\CCBNd$ simulation]
\llem{traCNd_simulation}
If $\tm \tocbCnd \tmtwo$ then $\traCNd{\tm} \toS^* \traCNd{\tmtwo}$.
\end{lemma}
\begin{proof}
By induction on the derivation of $\tm \tocbCnd \tmtwo$.
The interesting cases are when
there is a $\symdb$, $\symlsv$, or $\symgc$ step at the root.
\\
  If $(\lam{\var}{\tm})\sctx\,\tmtwo
       \todb
       \tm\esub{\var}{\tmtwo}\sctx$,
  then:
  \[
    \begin{array}{ll}
      \traCNd{((\lam{\var}{\tm})\sctx\,\tmtwo)}
    =
      \open{
        (\sha{\lam{\lvar}{\traCNd{\tm}\esub{\var}{\lvar}}}) \traCNd{\sctx}}\,\ofc{\traCNd{\tmtwo}}
    &
    \toSopen
        (\lam{\lvar}{\traCNd{\tm}\esub{\var}{\lvar}})\traCNd{\sctx}\,\ofc{\traCNd{\tmtwo}}
    \\
    &
    \toSdb
      \traCNd{\tm}\esub{\var}{\ofc{\traCNd{\tmtwo}}}\traCNd{\sctx}
    =
      \traCNd{(\tm\esub{\var}{\tmtwo}\sctx)}
    \end{array}
  \]
\\
  If $\off{\gctx}{\var}\esub{\var}{\val\sctx}
      \tolsw
      \off{\gctx}{\val\sctx}\esub{\var}{\val\sctx}$,
  note that $\val = \lam{\vartwo}{\tm}$
  so $\traCNd{\val} = \sha{\lam{\lvar}{\traCNd{\tm}\esub{\vartwo}{\lvar}}}$.
  Then:
  \[
    \begin{array}{l}
    \traCNd{(\off{\gctx}{\var}\esub{\var}{\val\sctx})}
    =
      \off{\traCNd{\gctx}}{\var}\esub{\var}{\ofc{(\sha{(\lam{\lvar}{\traCNd{\tm}\esub{\vartwo}{\lvar}})}\traCNd{\sctx})}}
    \\
    \toSls
      \off{\traCNd{\gctx}}{(\sha{(\lam{\lvar}{\traCNd{\tm}\esub{\vartwo}{\lvar}})})\traCNd{\sctx}}\esub{\var}{\ofc{(\sha{(\lam{\lvar}{\traCNd{\tm}\esub{\vartwo}{\lvar}})})\traCNd{\sctx}}}
    =
      \traCNd{(\off{\gctx}{\val\sctx}\esub{\var}{\val\sctx})}
    \end{array}
  \]
\\
  If $\tm\esub{\var}{\tmtwo}
       \togc
       \tm$,
  where $\var \notin \fv{\tm}$,
  then
  $
      \traCNd{(\tm\esub{\var}{\tmtwo})}
    =
      \traCNd{\tm}\esub{\var}{\ofc{\traCNd{\tmtwo}}}
    \toSgc
      \traCNd{\tm}
  $,
  where $\var\notin\fv{\traCNd{\tm}}$
  because $\fv{\traCNd{\tm}} = \fv{\tm}$.
  \qedhere
\end{proof}

For completeness, we define an \textbf{inverse $\CCBNd$ translation}.
First, we define a subset $\TermsSClumsyNeed \subseteq \TermsS$,
containing the closure by $\toS$-reduction of the image
of $\traCNd{\arg}$,
as well as a subset $\SCtxsSClumsyNeed \subseteq \SCtxsS$,
as follows:
\[
  \begin{array}{rcl}
  \rtm,\rtmtwo,\hdots & ::= &
         \var
    \mid \sha{\lam{\lvar}{\rtm\esub{\var}{\lvar}}}
    \mid \open{\rtm}
    \mid \rtm\esub{\uvar}{\ofc{\rtmtwo}}
    \mid \lam{\lvar}{\rtm\esub{\var}{\lvar}}
    \mid \rtm\,\ofc{\rtmtwo}
  \\
  \rsctx & ::= & \ctxhole \mid \rsctx\esub{\uvar}{\ofc{\rtm}}
  \end{array}
\]
where, in the productions
involving a subterm of the form
$\lam{\lvar}{\rtm\esub{\var}{\lvar}}$,
we assume that $\lvar$ is fresh, that is, $\lvar \notin \fv{\rtm}$.

The {\em inverse $\CCBNd$ translation} is a function
$\traCNdinv{\arg} : \TermsSClumsyNeed \to \TermsLSC$ defined as follows,
by induction on the derivation of a term with the (unambiguous) grammar above:
\[
  \begin{array}{rcl@{\HS}rcl@{\HS}rcl}
    \traCNdinv{\var}
    & \eqdef &
    \var
  &
    \traCNdinv{(\sha{\lam{\lvar}{\rtm\esub{\var}{\lvar}}})}
    & \eqdef &
    \lam{\var}{\traCNdinv{\rtm}}
  &
    \traCNdinv{\open{\rtm}}
    & \eqdef &
    \traCNdinv{\rtm}
  \\
    \traCNdinv{(\lam{\lvar}{\rtm\esub{\var}{\lvar}})}
    & \eqdef &
    \lam{\var}{\traCNdinv{\rtm}}
  &
    \traCNdinv{\rtm\esub{\uvar}{\ofc{\rtmtwo}}}
    & \eqdef &
    \traCNdinv{\rtm}\esub{\uvar}{\traCNdinv{\rtmtwo}}
  &
    \traCNdinv{\rtm\,\ofc{\rtmtwo}}
    & \eqdef &
    \traCNdinv{\rtm}\,\traCNdinv{\rtmtwo}
  \end{array}
\]
It is easy to check that $\traCNdinv{\arg}$ is the left-inverse of $\traCNd{\arg}$.

\restatable{lemma}{lemma:traCNdinv_simulation}{\traCNdinvSimulation}{Inverse $\CCBNd$ simulation}{
Let $\rtm \in \TermsSClumsyNeed$ and $\tmtwo \in \TermsS$
such that $\rtm \toS \tmtwo$.
Then $\tmtwo \in \TermsSClumsyNeed$
and $\traCNdinv{\rtm} \tocbnd^= \traCNdinv{\tmtwo}$.
}

\begin{theorem}[Sound and complete $\CCBNd$ embedding]
Let $\tm,\tmtwo \in \TermsLSC$. Then
$\tm \tocbCnd^* \tmtwo$
if and only if
$\traCNd{\tm} \toS^* \traCNd{\tmtwo}$.
Moreover, $\tm$ is in $\tocbCnd$-normal form iff $\traCNd{\tm}$ is in $\toS$-normal form.
\end{theorem}

As previously mentioned, it does not seem possible to embed Wadsworth's
call-by-need (\ie $\CBNd$) in $\lambdaS$.
One could imagine a variant of $\lambdaS$ that includes the following
$\symSls'$ rule rather than $\symSls$:
\[
  \off{\gctx}{\uvar}\esub{\uvar}{(\ofc{(\sha{\tm})\sctx_1})\sctx_2}
  \rootto_{\symSls'}
  \off{\gctx}{\sha{\tm}}\esub{\uvar}{\ofc{(\sha{\tm})}}\sctx_1\sctx_2
\]
The resulting calculus allows embeddings from $\CBN$, $\CBV$, and $\CBNd$.
However, it is not well-behaved, as confluence fails.
Let $\Omega\eqdef (\lam{\lvar}{\lvar\,\lvar})\,(\lam{\lvar}{\lvar\,\lvar})$.
For example,
$\var\esub{\vartwo}{\ofc{\uvar}}\esub{\uvar}{\ofc{(\sha{\uvartwo})\esub{\uvartwo}{\Omega}}}
\to_{\symSgc} \var\esub{\uvar}{\ofc{(\sha{\uvartwo})\esub{\uvartwo}{\Omega}}}
\to_{\symSgc}\var$. But also
$\var\esub{\vartwo}{\ofc{\uvar}}\esub{\uvar}{\ofc{(\sha{\uvartwo})\esub{\uvartwo}{\Omega}}}
\to_{\symSls}
\var\esub{\vartwo}{\ofc{\sha{\uvartwo}}}\esub{\uvar}{\ofc{(\sha{\uvartwo})}}\esub{\uvartwo}{\Omega}
\to_{\symSgc} \var\esub{\uvar}{\ofc{(\sha{\uvartwo})}}\esub{\uvartwo}{\Omega}
\to_{\symSgc} \var\esub{\uvartwo}{\Omega}$.


\section{Simulating Weak Evaluation Strategies}
\lsec{strategies}

In~\rsec{translations}, we have shown that $\CBN$, $\CBV$, and $\CCBNd$
calculi can be embedded in the $\lambdaS$-calculus.
Reduction in these calculi is intended to capture \emph{equivalence},
rather than \emph{evaluation}, of programs.
That is, these calculi are orientations of $\CBN$, $\CBV$, and $\CCBNd$
equational theories rather than evaluation mechanisms.

Reduction in the calculi of~\rsec{translations} is closed by arbitrary contexts.
\Eg in the $\CBV$ \emph{calculus}, a step
$(\lam{\var}{\vartwo})\,\lam{\varthree}{\tm} \tocbv
 (\lam{\var}{\vartwo})\,\lam{\varthree}{\tm'}$
is allowed if $\tm \tocbv \tm'$, while typically call-by-value \emph{evaluation}
would proceed to contract the outermost redex.

In this section, we first define \emph{weak evaluation} relations
$\tosn{}$, $\tosv{}$, and $\tosd{}$ for $\CBN$, $\CBV$, and $\CCBNd$
respectively.
Recall that evaluation is called \emph{weak} if it does not proceed
inside the bodies of $\lambda$-abstractions.
Second, we define an evaluation relation $\toss{}$ for $\lambdaS$,
which is also ``weak'' in that it does not reduce inside
$\lambda$-abstractions, boxes ($\sha{}$), nor promotions ($\ofc{}$).
Finally, we show that evaluation according
to $\tosn{}$, $\tosv{}$, and $\tosd{}$ can be simulated by $\toss{}$
via the translations already introduced in \rsec{translations}.

\bparagraph{Weak $\CBN$ Evaluation}
The one-step weak $\CBN$ evaluation judgment is of the form
$\tm \tosn{\rulename} \tm'$, where $\tm,\tm' \in \TermsLSC$
and the set of \emph{$\CBN$-rulenames} ($\rulename,\rulename',\hdots$)
is given by
$
  \rulename ::= \symdb \mid \symsub{\var}{\tm} \mid \symls \mid \symgc
$.
Weak $\CBN$ evaluation is the union
${\tosn{}} \eqdef {\tosn{\symdb} \cup \tosn{\symls} \cup \tosn{\symgc}}$,
excluding auxiliary $\symsub{\var}{\tm}$ steps.
It is defined by the following inductive rules:
\[
  \indrule{\ruleENdb}{
    \emptyPremise
  }{
    (\lam{\var}{\tm})\sctx\,\tmtwo \tosn{\symdb} \tm\esub{\var}{\tmtwo}\sctx
  }
  \,\,
  \indrule{\ruleENsub}{
    \emptyPremise
  }{
    \var
    \tosn{\symsub{\var}{\tm}}
    \tm
  }
  \,\,
  \indrule{\ruleENls}{
    \tm \tosn{\symsub{\var}{\tmtwo}} \tm'
  }{
    \tm\esub{\var}{\tmtwo}
    \tosn{\symls}
    \tm'\esub{\var}{\tmtwo}
  }
\]
\[
  \indrule{\ruleENgc}{
    \var\notin\fv{\tm}
  }{
    \tm\esub{\var}{\tmtwo}
    \tosn{\symgc}
    \tm
  }
  \,\,
  \indrule{\ruleENapp}{
    \tm \tosn{\rulename} \tm'
  }{
    \tm\,\tmtwo \tosn{\rulename} \tm'\,\tmtwo
  }
  \,\,
  \indrule{\ruleENsubL}{
    \tm \tosn{\rulename} \tm'
    \HS
    \var \notin \fv{\rulename}
  }{
    \tm\esub{\var}{\tmtwo} \tosn{\rulename} \tm'\esub{\var}{\tmtwo}
  }
\]
The $\ruleENdb$, $\ruleENls$, and $\ruleENgc$ rules derive \emph{root reduction}
steps.
The $\ruleENapp$ and $\ruleENsubL$ rules correspond to
congruence closure below weak head evaluation contexts.
The side condition in the $\ruleENsubL$ rule is to avoid
unwanted variable capture.
The somewhat atypical $\ruleENsub$ rule derives steps of the form
$\tm \tosn{\symsub{\var}{\tmtwo}} \tm'$,
which substitute a single free occurrence of $\var$ (in evaluation position)
by $\tmtwo$. This rule works in synchrony with $\ruleENls$ to allow $\symls$
steps:
for example, $\var\,\var\,\vartwo \tosn{\symsub{\var}{\tm}} \tm\,\var\,\vartwo$
and
$(\var\,\var\,\vartwo)\esub{\var}{\tm} \tosn{\symls} (\tm\,\var\,\vartwo)\esub{\var}{\tm}$.
This is inspired the
formulation of strong call-by-need of Balabonski \etal~\cite{DBLP:journals/lmcs/BalabonskiLM23}.

It is straightforward to show that ${\tosn{}} \subseteq {\tocbn}$.
Note also that $\tosn{}$ is non-deterministic, although confluent.
The source of non-determinism is that $\symgc$ steps can be performed in
any order. For example $((\lam{\var}{\var})\,\vartwo)\esub{\varthree}{\tmtwo}$
reduces both with a $\symdb$ and with a $\symgc$ step.

\bparagraph{Weak $\CBV$ Evaluation}
The one-step weak $\CBV$ evaluation judgment is of the form
$\tm \tosv{\rulename} \tm'$,
where $\tm,\tm' \in \TermsLSC$
and the set of \emph{$\CBV$-rulenames} ($\rulename,\rulename',\hdots$)
is given by
$
  \rulename ::= \symdb \mid \symsub{\var}{\val} \mid \symlsv \mid \symgclv
$.
Weak $\CBV$ evaluation is
${\tosv{}} \eqdef {\tosv{\symdb} \cup \tosv{\symlsv} \cup \tosv{\symgclv}}$,
excluding auxiliary $\symsub{\var}{\val}$ steps.
It is defined by the following inductive rules:
\[
  \indrule{\ruleEVdb}{
  }{
    (\lam{\var}{\tm})\sctx\,\tmtwo \tosv{\symdb} \tm\esub{\var}{\tmtwo}\sctx
  }
  \indrule{\ruleEVsub}{
  }{
    \var
    \tosv{\symsub{\var}{\val}}
    \val
  }
\]
\[
  \indrule{\ruleEVgclv}{
    \var\notin\fv{\tm}
  }{
    \tm\esub{\var}{\lval\sctx}
    \tosv{\symgclv}
    \tm\sctx
  }
  \indrule{\ruleEVlsv}{
    \tm \tosv{\symsub{\var}{\val}} \tm'
  }{
    \tm\esub{\var}{\val\sctx}
    \tosv{\symlsv}
    \tm'\esub{\var}{\val}\sctx
  }
  \indrule{\ruleEVapp}{
    \tm \tosv{\rulename} \tm'
  }{
    \tm\,\tmtwo \tosv{\rulename} \tm'\,\tmtwo
  }
\]
\[
  \indrule{\ruleEVsubL}{
    \tm \tosv{\rulename} \tm'
    \HS
    \var \notin \fv{\rulename}
  }{
    \tm\esub{\var}{\tmtwo} \tosv{\rulename} \tm'\esub{\var}{\tmtwo}
  }
  \indrule{\ruleEVsubR}{
    \tmtwo \tosv{\rulename} \tmtwo'
  }{
    \tm\esub{\var}{\tmtwo} \tosv{\rulename} \tm\esub{\var}{\tmtwo'}
  }
\]
Similar remarks as for $\CBN$ apply,
in particular ${\tosv{}} \subseteq {\tocbv}$.
Rules $\ruleEVdb$, $\ruleEVlsv$, and $\ruleEVgclv$
are root reduction rules,
while $\ruleEVapp$, $\ruleEVsubL$, and $\ruleEVsubR$
correspond to congruence closure rules.
The $\ruleEVsub$ rule plays a similar role as the analogue rule in $\CBN$, but only allows
substituting variables for \emph{strict values}.
In this notion of $\CBV$ evaluation, arguments of applications are not evaluated.
The restriction that the argument is a value is
not imposed to contract a $\beta$-like redex,
but rather to perform the substitution.
These ideas can already be found in the $\lambda_{CBV}$-calculus of~\cite{DBLP:conf/tlca/HerbelinZ09}.
Note that in $\CBV$ evaluation ($\tosv{}$)
there is a second source of non-determinism, namely that
$\ruleEVsubL$ and $\ruleEVsubR$ overlap, so the body
and the argument of an ES can be evaluated concurrently.

\bparagraph{Weak $\CCBNd$ Evaluation}
The one-step weak $\CCBNd$ evaluation judgment is of the form
$\tm \tosd{\rulename} \tm'$,
where $\tm,\tm' \in \TermsLSC$
and the set of \emph{$\CCBNd$-rulenames} ($\rulename,\rulename',\hdots$)
is given by
$
  \rulename ::= \symdb
           \mid \symsub{\var}{\val\sctx}
           \mid \symid{\var}
           \mid \symlsw
           \mid \symgc
$.
Weak $\CCBNd$ evaluation is the union
${\tosd{}} \eqdef {\tosd{\symdb} \cup \tosd{\symlsw} \cup \tosd{\symgc}}$,
excluding auxiliary $\symsub{\var}{\val\sctx}$ and $\symid{\var}$ steps.
It is defined by the following inductive rules:
\[
  \indrule{\ruleEDdb}{
  }{
    (\lam{\var}{\tm})\sctx\,\tmtwo \tosd{\symdb} \tm\esub{\var}{\tmtwo}\sctx
  }
  \indrule{\ruleEDsub}{
  }{
    \var
    \tosd{\symsub{\var}{\val\sctx}}
    \val\sctx
  }
  \indrule{\ruleEDsubES}{
  }{
    \tm\esub{\uvar}{\var}
    \tosd{\symsub{\var}{\val\sctx}}
    \tm\esub{\uvar}{\val\sctx}
  }
\]
\[
  \indrule{\ruleEDid}{
    \emptyPremise
  }{
    \var
    \tosd{\symid{\var}}
    \var
  }
  \indrule{\ruleEDlsw}{
    \tm \tosd{\symsub{\var}{\val\sctx}} \tm'
  }{
    \tm\esub{\var}{\val\sctx}
    \tosd{\symlsw}
    \tm'\esub{\var}{\val\sctx}
  }
  \indrule{\ruleEDgc}{
    \var\notin\fv{\tm}
  }{
    \tm\esub{\var}{\tmtwo}
    \tosd{\symgc}
    \tm
  }
\]
\[
  \indrule{\ruleEDsubL}{
    \tm \tosd{\rulename} \tm'
    \HS
    \var \notin \fv{\rulename}
  }{
    \tm\esub{\var}{\tmtwo} \tosd{\rulename} \tm'\esub{\var}{\tmtwo}
  }
  \indrule{\ruleEDapp}{
    \tm \tosd{\rulename} \tm'
  }{
    \tm\,\tmtwo \tosd{\rulename} \tm'\,\tmtwo
  }
  \indrule{\ruleEDsubR}{
    \tm \tosd{\symid{\var}} \tm
    \HS
    \tmtwo \tosd{\rulename} \tmtwo'
  }{
    \tm\esub{\var}{\tmtwo} \tosd{\rulename} \tm\esub{\var}{\tmtwo'}
  }
\]
Similar remarks as for $\CBN$ apply,
in particular ${\tosd{}} \subseteq {\tocbCnd}$.
Rules $\ruleEDdb$, $\ruleEDlsw$, and $\ruleEDgc$
are root reduction rules, while
$\ruleEDapp$, $\ruleEDsubL$, and $\ruleEDsubR$
are congruence closure rules.
The rule $\ruleEDsub$ plays a similar role as the analogue rules in $\CBN$ and $\CBV$,
but only allows substituting variables for terms of the form $\val\sctx$
(known as \emph{answers} in the literature).
The rule $\ruleEDsubES$ is a variant of $\ruleEDsub$ that acts on the argument
of an ES; this rule is not strictly necessary for evaluation, but it is crucial
for the embedding into $\lambdaS$ to be complete.
The $\ruleEDid$ rule is used in synchrony with the congruence rules
to derive steps that are always of the form $\tm \tosd{\symid{\var}} \tm$
indicating that $\var$ occurs in $\tm$ in an evaluation position.
This is used to check whether $\var$ it a \emph{needed} variable.
For example, $\var\,\vartwo \tosd{\symid{\var}} \var\,\vartwo$
and $\varthree\,\varthree \tosd{\symsub{\varthree}{\val}} \val\,\varthree$,
so the fact that $\var$ is needed on the left triggers the evaluation of
the argument:
 $(\var\,\vartwo)\esub{\var}{\varthree\,\varthree}
    \tosd{\symsub{\varthree}{\val}}
    (\var\,\vartwo)\esub{\var}{\val\,\varthree}$.
From this, one obtains the substitution step 
 $(\var\,\vartwo)\esub{\var}{\varthree\,\varthree}\esub{\varthree}{\val}
    \tosd{\symlsw}
    (\var\,\vartwo)\esub{\var}{\val\,\varthree}\esub{\varthree}{\val}$.

\bparagraph{Weak $\lambdaS$-Calculus Evaluation}
The one-step weak $\lambdaS$ evaluation judgment
is of the form $\tm \toss{\rulename} \tm'$,
where $\tm,\tm'\in\TermsS$,
and the set of \emph{rulenames} ($\rulename,\rulename',\hdots$)
is given by
$
  \rulename ::=
         \symSdb
    \mid \symSsub{\uvar}{(\sha{\tm})\sctx}
    \mid \symSid{\uvar}
    \mid \symSls
    \mid \symSgc
    \mid \symSopen
$.
Weak $\lambdaS$ evaluation is the union
${\toss{}} \eqdef {\toss{\symSdb} \cup \toss{\symSls} \cup \toss{\symSgc}}$,
excluding auxiliary $\symsub{\var}{\val\sctx}$ and $\symid{\var}$ steps.
It is defined by the following inductive rules:
\[
  \indrule{\ruleESdb}{
  }{
    (\lam{\lvar}{\tm})\sctx\, \tmtwo \toss{\symSdb} \tm\sub{\lvar}{\tmtwo}\sctx
  }
  \,\,
  \indrule{\ruleESsub}{
  }{
    \uvar
    \toss{\symSsub{\uvar}{(\sha{\tm})\sctx}}
    (\sha{\tm})\sctx
  }
  \,\,
  \indrule{\ruleESofcsub}{
  }{
    \ofc{\uvar}
    \toss{\symSsub{\uvar}{(\sha{\tm})\sctx}}
    \ofc{(\sha{\tm})\sctx}
  }
\]
\[
  \indrule{\ruleESid}{
    \emptyPremise
  }{
    \uvar
    \toss{\symSid{\uvar}}
    \uvar
  }
  \,\,
  \indrule{\ruleESls}{
    \tm \toss{\symSsub{\uvar}{(\sha{\tmtwo})\sctx_1}} \tm'
  }{
    \tm\esub{\uvar}{(\ofc{(\sha{\tmtwo})\sctx_1})\sctx_2}
    \toss{\symSls}
    \tm'\esub{\uvar}{\ofc{(\sha{\tmtwo})\sctx_1}}\sctx_2
  }
  \,\,
  \indrule{\ruleESapp}{
    \tm \toss{\rulename} \tm'
  }{
    \tm\,\tmtwo
    \toss{\rulename}
    \tm'\,\tmtwo
  }
\]
\[
  \indrule{\ruleESgc}{
    \uvar\notin\fv{\tm}
  }{
    \tm\esub{\uvar}{(\ofc{\tmtwo})\sctx}
    \toss{\symSgc}
    \tm\sctx
  }
  \,\,
  \indrule{\ruleESopen}{
    \emptyPremise
  }{
    \open{(\sha{\tm})\sctx}
    \toss{\symSopen}
    \tm\sctx
  }
  \,\,
  \indrule{\ruleEScopen}{
    \tm \toss{\rulename} \tm'
  }{
    \open{\tm}
    \toss{\rulename}
    \open{\tm'}
  }
\]
\[
  \indrule{\ruleESsubL}{
    \tm \toss{\rulename} \tm'
    \quad
    \uvar\notin\fv{\rulename}
  }{
    \tm\esub{\uvar}{\tmtwo}
    \toss{\rulename}
    \tm'\esub{\uvar}{\tmtwo}
  }
  \,\,
  \indrule{\ruleESsubR}{
    \tmtwo \toss{\rulename} \tmtwo'
  }{
    \tm\esub{\uvar}{\tmtwo}
    \toss{\rulename}
    \tm\esub{\uvar}{\tmtwo'}
  }
  \,\,
  \indrule{\ruleESsubRofc}{
    \tm \toss{\symSid{\uvar}} \tm
    \quad
    \tmtwo \toss{\rulename} \tmtwo'
  }{
    \tm\esub{\uvar}{\ofc{\tmtwo}}
    \toss{\rulename}
    \tm\esub{\uvar}{\ofc{\tmtwo'}}
  }
\]
Weak $\lambdaS$ evaluation is a sub-ARS of $\lambdaS$,
in the sense that ${\toss{}} \subseteq {\toS}$.
Rules $\ruleESdb$, $\ruleESls$, $\ruleESgc$, and $\ruleESopen$ are root reduction rules,
while
$\ruleESapp$, $\ruleEScopen$, $\ruleESsubL$, and $\ruleESsubR$
are congruence rules.
Rule $\ruleESsub$ plays a similar role as the analogue rules
in $\CBN$, $\CBV$, and $\CBNd$, but only allows substituting a variable
by a term of the form $(\sha{\tm})\sctx$.
Rule $\ruleESid$ plays a similar role as the analogue rule in $\CBNd$,
used to check whether a variable is in evaluation position.
Note that there are no congruence rules below $\lambda$-abstraction,
box ($\sha{}$), nor promotion ($\ofc{}$).
Evaluation can proceed below promotion in two particular cases.
First, the $\ruleESofcsub$ rule allows to perform substitution immediately
below a promotion; for instance
$(\ofc{\uvar})\esub{\uvar}{\osha{\uvartwo}}
 \toss{\symSls}
 (\osha{\uvartwo})\esub{\uvar}{\osha{\uvartwo}}$.
Second, the $\ruleESsubRofc$ rule allows evaluation below a promotion when
a term is the argument of a ``needed'' substitution; for instance
$(\uvar\uvartwo)\esub{\uvar}{\ofc{((\lam{\lvar}{\lvar})(\sha{\uvarthree}))}}
 \toss{\symSdb}
 (\uvar\uvartwo)\esub{\uvar}{\ofc{\sha{\uvarthree}}}
 \toss{\symSls}
 ((\sha{\uvarthree})\uvartwo)\esub{\uvar}{\ofc{\sha{\uvarthree}}}
$.

The $\lambdaS$-calculus is designed with the goal in mind of providing
a \emph{unifying framework} for call-by-name, call-by-value, and call-by-sharing.
As a matter of fact, weak $\lambdaS$ evaluation
simulates weak $\CBN$, $\CBV$, and $\CCBNd$ evaluation
via the $\traN{\arg}$, $\traV{\arg}$, and $\traCNd{\arg}$ translations
introduced in \rsec{translations}:
\begin{theorem}[Simulation and inverse simulation of evaluation \proofnote{Proofs in~Sec.\ref{section:appendix:strategies}}]
\begin{center}
\begin{tabular}{l@{\,\,}|@{\,\,}l@{\,\,}|@{\,\,}l}
&
  \textup{Soundness}
&
  \textup{Completeness}
\\
\midrule
  $\CBN$
  &
    If $\tm \tosn{} \tmtwo$
    then $\traN{\tm} \tossmany{} \traN{\tmtwo}$.
  &
    If $\traN{\tm} \toss{} \tmtwo$
    then $\tmtwo \in \TermsSName$ and $\tm \mathrel{(\tosn{})^=} \traNinv{\tmtwo}$.
\\
  $\CBV$
  &
    If $\tm \tosv{} \tmtwo$
    then $\traV{\tm} \tossmany{} \traV{\tmtwo}$.
  &
    If $\traV{\tm} \toss{} \tmtwo$
    then $\tmtwo \in \TermsSValue$ and
    $\tm \mathrel{(\tosvUgclvinv{})^=} \traVinv{\tmtwo}$.
\\
  $\CCBNd$
  &
    If $\tm \tosd{} \tmtwo$
    then $\traCNd{\tm} \tossmany{} \traCNd{\tmtwo}$.
  &
    If $\traD{\tm} \toss{} \tmtwo$
    then $\tmtwo \in \TermsSClumsyNeed$ and $\tm \mathrel{(\tosd{})^=} \traDinv{\tmtwo}$.
\end{tabular}
\end{center}
In the $\CBV$ case, completeness only holds for
an extended relation $\tosvUgclvinv{}$, defined as $\tosv{}$
but adding an inverse garbage collection rule that derives
$\tm \tosvUgclvinv{} \tm\esub{\var}{\lval}$ if $\var \notin \fv{\tm}$.
\end{theorem}


\begin{ifLongAppendix}
  \section{Embedding the Bang-Calculus}
  \lsec{bang}
  
This section presents an embedding of the \BangCalculus into $\lambdaS$. Our
presentation of the \BangCalculus uses explicit substitutions and reduction at
a distance. It is based on the formulation of Bucciarelli et
al~\cite{DBLP:conf/flops/BucciarelliKRV20} but where one-shot execution of
explicit substitutions (the $\texttt{s}!$ rule in Sec.~2
of~\cite{DBLP:conf/flops/BucciarelliKRV20}) is replaced by explicit replacement
(rules $\tolsB$ and $\togcB$ below), in unison with the presentation of our
other calculi in this work. A further simplification will be introduced by
dropping the rule for dereliction, which does not change its expressive power,
as explained below.

\bparagraph{Syntax and Reduction Semantics of the \BangCalculus}
\ldef{bang:calculus:syntax}
The set of {\em Bang terms} is denoted by $\TermsBangGM$ and given by:
\[
  \tm,\tmtwo,\hdots ::=
  \var \mid \lam{\var}{\tm} \mid \tm\,\tmtwo \mid \ofc{\tm} \mid
  \der{\tm} \mid \tm\esub{\var}{\tmtwo}
\]
We denote with $\TermsBang$ the set of \emph{simplified} Bang terms; where terms of the form $\der{\tm}$ are disallowed.
{\em Contexts} ($\gctx,\gctxtwo,\hdots$)
and {\em substitution contexts} ($\sctx,\sctxtwo,\hdots$)
are defined as expected,
as well as the notions of free variables of a term ($\fv{\tm}$),
domain of a substitution context ($\dom{\sctx}$),
plugging a term into a context allowing variable-capture
($\of{\gctx}{\tm}$ and $\tm\sctx$)
and avoiding variable-capture ($\off{\gctx}{\tm}$).

There are four rewriting rules, closed by arbitrary contexts:
\[
  \begin{array}{rll@{\HS}rll}
    (\lam{\var}{\tm})\sctx\,\tmtwo
    & \todbB &
    \tm\esub{\var}{\tmtwo}\sctx
  &
    \off{\gctx}{\var}\esub{\var}{(\ofc{\tmtwo})\sctx}
    & \tolsB &
    \off{\gctx}{\tmtwo}\esub{\var}{\ofc{\tmtwo}}\sctx
  \\
    \tm\esub{\var}{(\ofc{\tmtwo})\sctx}
    & \togcB &
    \tm\sctx
    \text{ (if $\var \notin \fv{\tm}$)}
  &
    \der{(\ofc{\tm})\sctx}
    & \toderB &
    \tm\sctx
  \end{array}
\]
Full Bang-reduction is given by the union
$
  {\toBangGM} \eqdef {\todbB \cup \tolsB \cup \togcB \cup \toderB}
$.
Simplified Bang-reduction over simplified Bang terms
${\toBang} \subseteq \TermsBang\times \TermsBang$
is given by ${\toBang} \eqdef {\todbB \cup \tolsB \cup \togcB}$.
Working in the simplified fragment (without the $\der{-}$ constructor)
does not result in any loss of expressivity. Indeed:

\restatable{proposition}{prop:bang_simplified_simulation}{\BangSimplifiedSimulation}{Simplified $\Bang$ simulation}{
$\toBangGM$ and $\toBang$ simulate each other.
}
\begin{proof}
The key to this result is the fact that $\der{(\ofc{\tm})\sctx}$
can be understood as a shorthand for $\var\esub{\var}{(\ofc{\tm})\sctx}$.
The rule $\symderB$ in the \BangCalculus may be simulated by $\symlsB$.
\end{proof}


In the sequel, by \emph{\BangCalculus} we will mean $\toBang$
reduction over simplified Bang terms.

\bparagraph{Typed \BangCalculus}
A typing system for the \BangCalculus is defined as follows. 
The set of {\em types} is given by:
\[
  \typ,\typtwo ::= \btyp \mid \ofc{\typ} \mid \ofc{\typ}\to\typtwo
\]
{\em Typing environments} ($\tctx,\tctxtwo,\hdots$)
are partial functions assigning variables to
types \emph{prefixed with $\ofc{}$},
written $\var_1:\ofc{\typ_1},\hdots,\var_n:\ofc{\typ_n}$.
{\em Typing judgments} are of the form $\judl{\tctx}{\tm}{\typ}$
and defined by:
\[
  \indrule{\rulebVar}{
    \emptyPremise
  }{
    \judl{\tctx,\var:\ofc{\typ}}{\var}{\typ}
  }
  \,\,
  \indrule{\rulebAbs}{
    \judl{\tctx,\var:\ofc{\typ}}{\tm}{\typtwo}
  }{
    \judl{\tctx}{\lam{\var}{\tm}}{\ofc{\typ}\to\typtwo}
  }
  \,\,
  \indrule{\rulebApp}{
    \judl{\tctx}{\tm}{\ofc{\typ}\to\typtwo}
    \HS
    \judl{\tctx}{\tmtwo}{\ofc{\typ}}
  }{
    \judl{\tctx}{\tm\,\tmtwo}{\typtwo}
  }
\]
\[
  \indrule{\rulebProm}{
    \judl{\tctx}{\tm}{\typ}
  }{
    \judl{\tctx}{\ofc{\tm}}{\ofc{\typ}}
  }
  \,\,
  \indrule{\rulebES}{
    \judl{\tctx,\var:\ofc{\typ}}{\tm}{\typtwo}
    \HS
    \judl{\tctx}{\tmtwo}{\ofc{\typ}}
  }{
    \judl{\tctx}{\tm\esub{\var}{\tmtwo}}{\typtwo}
  }
\]
\smallskip

\bparagraph{Embedding the \BangCalculus in $\lambdaS$}
The Bang translation $\traB{\arg}$ is defined on types and terms by:
\[
    \traB{\btyp}
    \eqdef
    \btyp
  \HS
    \traB{(\ofc{\typ})}
    \eqdef
    \osha{\traB{\typ}}
  \HS
    \traB{(\ofc{\typ}\to\typtwo)}
    \eqdef
    \osha{\traB{\typ}}\limp\traB{\typtwo}
\]
\[
  \begin{array}{rcl@{\HS}rcl@{\HS}rcl}
    \traB{\var}
    & \eqdef &
    \open{\var}
  &
    \traB{(\lam{\var}{\tm})}
    & \eqdef &
    \lam{\lvar}{\traB{\tm}\esub{\var}{\lvar}}
  &
    \traB{(\tm\,\tmtwo)}
    & \eqdef &
    \traB{\tm}\,\traB{\tmtwo}
  \\
    \traB{(\ofc{\tm})}
    & \eqdef &
    \osha{\traB{\tm}}
  &
    \traB{\tm\esub{\var}{\tmtwo}}
    & \eqdef &
    \traB{\tm}\esub{\var}{\traB{\tmtwo}}
  \end{array}
\]
The translation is extended to typing environments:
$\traB{(\var_1:\ofc{\typ_1},\hdots,\var_n:\ofc{\typ_n})}
 \eqdef
 \var_1:\traB{\typ_1},\hdots,\var_n:\traB{\typ_n}$,
and judgments:
$\traB{(\judl{\tctx}{\tm}{\typ})}
 \eqdef
 \judc{\traB{\tctx}}{\noenv}{\traB{\tm}}{\traB{\typ}}$.

\restatable{proposition}{prop:typing_the_bang_translation}{\typingTheBangTranslation}{Bang typing}{
If $\judl{\tctx}{\tm}{\typ}$
then $\judc{\traB{\tctx}}{\noenv}{\traB{\tm}}{\traB{\typ}}$.
}

\begin{lemma}[Bang simulation]
\llem{traB_simulation}
If $\tm \toBang \tmtwo$ then $\traB{\tm} \toS^+ \traB{\tmtwo}$.
\end{lemma}
\begin{proof}
By induction on the derivation of $\tm \toBang \tmtwo$.
The interesting cases are when there is a $\symdbB$, $\symlsB$, or
$\symgcB$ step at the root.
  If $(\lam{\var}{\tm})\sctx\,\tmtwo \todbB \tm\esub{\var}{\tmtwo}\sctx$,
  then:
  \[
      \traB{((\lam{\var}{\tm})\sctx\,\tmtwo)}
    =
      (\lam{\lvar}{\traB{\tm}\esub{\var}{\lvar}})\traB{\sctx}\,\traB{\tmtwo}
    \toSdb
      \traB{\tm}\esub{\var}{\traB{\tmtwo}}\traB{\sctx}
    =
      \traB{(\tm\esub{\var}{\tmtwo}\sctx)}
  \]
  If $\off{\gctx}{\var}\esub{\var}{(\ofc{\tmtwo})\sctx} \tolsB \off{\gctx}{\tmtwo}\esub{\var}{\ofc{\tmtwo}}\sctx$,
  then:
  \[
    \begin{array}{ll}
      \traB{\off{\gctx}{\var}\esub{\var}{(\ofc{\tmtwo})\sctx}}
    =
      \off{\traB{\gctx}}{\open{\var}}\esub{\var}{(\osha{\traB{\tmtwo}})\traB{\sctx}}
    &
    \toSls
      \off{\traB{\gctx}}{\open{\sha{\traB{\tmtwo}}}}\esub{\var}{\osha{\traB{\tmtwo}}}\traB{\sctx}
    \\
    &
    \toSopen
      \off{\traB{\gctx}}{\traB{\tmtwo}}\esub{\var}{\osha{\traB{\tmtwo}}}\traB{\sctx}
    =
      \traB{(\off{\gctx}{\tmtwo}\esub{\var}{\tmtwo}\sctx)}
    \end{array}
  \]
\\
  If $\tm\esub{\var}{(\ofc{\tmtwo})\sctx} \togcB \tm\sctx$ with $\var \notin \fv{\tm}$,
  then
  $
    \traB{\tm\esub{\var}{(\ofc{\tmtwo})\sctx}}
    =
    \traB{\tm}\esub{\var}{(\osha{\traB{\tmtwo}})\traB{\sctx}}
    \toSgc
    \traB{\tm}\traB{\sctx}
  $.
  Note that $\var\notin\fv{\traB{\tm}}$
  because $\fv{\traB{\tm}} = \fv{\tm}$.
  \qedhere
\end{proof}

For completeness, we define an \textbf{inverse Bang translation}.
First, we define a subset $\TermsBangInv \subseteq \TermsS$,
containing the closure by $\toS$-reduction of the image of $\traB{\arg}$:
\[
  \rtm,\rtmtwo,\hdots
     ::= \open{\uvar}
    \mid \open{\sha{\rtm}}
    \mid \lam{\lvar}{\rtm\esub{\uvar}{\lvar}}
    \mid \rtm\,\rtmtwo
    \mid \osha{\rtm}
    \mid \rtm\esub{\uvar}{\rtmtwo}
\]
where $\lvar$ is assumed to be fresh in $\lam{\lvar}{\rtm\esub{\uvar}{\lvar}}$.
The {\em inverse Bang translation} is a function
$\traBinv{\anon} : \TermsBangInv \to \TermsLSC$ defined as follows,
by induction on the derivation of a term with the (unambiguous) grammar
above:
\[
  \begin{array}{rcl@{\HS}rcl@{\HS}rcl}
    \traBinv{\open{\var}}
    & \eqdef &
    \var
  &
    \traBinv{\open{\sha{\rtm}}}
    & \eqdef &
    \traBinv{\rtm}
  &
    \traBinv{(\lam{\lvar}{\rtm\esub{\var}{\lvar}})}
    & \eqdef &
    \lam{\var}{\traBinv{\rtm}}
  \\
    \traBinv{(\rtm\,\rtmtwo)}
    & \eqdef &
    \traBinv{\rtm}\,\traBinv{\rtmtwo}
  &
    \traBinv{(\osha{\rtmtwo})}
    & \eqdef &
               \ofc{\traBinv{\rtmtwo}}
  &
    \traBinv{\rtm\esub{\var}{\rtmtwo}}
    & \eqdef &
    \traBinv{\rtm}\esub{\var}{\traBinv{\rtmtwo}}
  \end{array}
\]
It is easy to check that $\traBinv{\arg}$ is the left-inverse of $\traB{\arg}$.

\restatable{lemma}{lemma:traBinv_simulation}{\traBinvSimulation}{Inverse Bang simulation}{
Let $\rtm \in \TermsBang$ and $\tmtwo \in \TermsS$.
If $\rtm \toS \tmtwo$
then $\tmtwo \in \TermsBang$
and $\traBinv{\rtm} \toBang^= \traBinv{\tmtwo}$.
}

\begin{theorem}[Sound and complete Bang embedding]
Let $\tm,\tmtwo \in \TermsBang$. Then
$\tm \toBang^* \tmtwo$
if and only if
$\traB{\tm} \toS^* \traB{\tmtwo}$.
Moreover, $\tm$ is in $\toBang$-normal form iff $\traB{\tm}$ is in $\toS$-normal form.
\end{theorem}

\begin{remark}
Composing our $\CBV$ and $\CBN$ translations with the $\traB{\arg}$ translation
of above and then performing dereliction unfolding (\ie
replacing $\der{\tm}$ with $\uvar\esub{\uvar}{\tm}$) we obtain
 the  call-by-name and call-by-value translations
of~\cite{DBLP:journals/corr/abs-1904-06845} with one minor difference in the
$\CBV$ case:~\cite{DBLP:journals/corr/abs-1904-06845} translates variables to
$\ofc{\var}$, whereas the above composition would translate it  to
$\osha{\open{\var}}$, that is, $\var$ is replaced with its
$\eta_\sha{}$-expansion. Unfortunately, this translation $\traValt{\arg}$
fails to be complete (\cf~\rremark{w_translation}).
\end{remark}


\end{ifLongAppendix}

\section{Related Work and Conclusions}
\lsec{conclusions}

\bparagraph{Related Work.}
The seminal
work~\cite{DBLP:journals/tcs/MaraistOTW99} is the first work to have related
Girard's embeddings of intuitionistic logic into $\LL$ with evaluation
mechanisms\footnote{Although the paper mentions some other authors that had
already hinted at this.}. Call-by-push-value
($\CBPV$)~\cite{DBLP:books/sp/Levy2004,DBLP:journals/lisp/Levy06} is a calculus
that distinguishes \emph{values} from \emph{computations} and allows to subsume
both the $\CBV$ and $\CBN$ evaluation mechanisms.
Ehrhard~\cite{DBLP:conf/esop/Ehrhard16} studied the
connection between
$\CBPV$~\cite{DBLP:books/sp/Levy2004,DBLP:journals/lisp/Levy06} and $\LL$,
producing a calculus which was later modified to become the
\BangCalculus~\cite{DBLP:conf/ppdp/EhrhardG16}. $\CBV$ and $\CBN$ translations
to the \BangCalculus were studied in~\cite{DBLP:conf/ppdp/EhrhardG16}.
Soundness and completeness of these translations with respect to reduction was
proved by Guerrieri and Manzonetto~\cite{DBLP:journals/corr/abs-1904-06845} for
a slightly different notion of reduction for the \BangCalculus than that
of~\cite{DBLP:conf/ppdp/EhrhardG16}. The $\CBV$ translation does not preserve
normal forms; an amended translation that does was studied
in~\cite{DBLP:conf/flops/BucciarelliKRV20,DBLP:journals/iandc/BucciarelliKRV23}.
Intuitionistic truth in terms of classical provability underlies G\"odel's
embedding of intuitionistic logic into (classical) S4.
In~\cite{DBLP:conf/rta/SantoPU19}, the authors consider a program similar to
that of $\CBPV$ but where that target language is a modal lambda calculus.
Promotion and derelection are recast as boxing and unboxing and $\CBV$ and
$\CBN$ are described in terms of a so called \emph{call-by-box} evaluation
mechanism~\cite{DBLP:conf/rta/SantoPU19}. 

\bparagraph{Conclusions.}
This work introduces $\MELLS$, a Sharing Linear Logic.
It arises from splitting each exponential modality~($\ofc{}/\why{}$)
into a sharing modality~($\ofc{}/\why{}$)
and a cloning modality~($\sha{}/\ush{}$).
$\MELLS$ is conservative over $\MELL$ and enjoys cut-elimination.
The usual embeddings of intuitionistic logic into $\LL$ can be restated
in the setting of $\lambdaS$, a Sharing Linear $\lambda$-calculus derived
from $\MELLS$.
The decomposition of the of-course modality allows us
to define an embedding of intuitionistic logic into $\lambdaS$,
corresponding to a \emph{call-by-need} $\lambda$-calculus $\CCBNd$.
The following table summarizes the, sound and complete,
embeddings studied in~\rsec{translations}:
\[
  \begin{array}{c|ccccc}
    &
      \typ\to\typtwo
    & 
      \var
    & 
      \lam{\var}{\tm}
    &
      \tm\,\tmtwo
    &
      \tm\esub{\var}{\tmtwo}
  \\
  \midrule
      \CBN, \traN{\arg}
    & \osha{\traN{\typ}}\to\traN{\typtwo}
    & \open{\var}
    & \lam{\lvar}{\traN{\tm}\esub{\var}{\lvar}}
    & \traN{\tm}\,\osha{\traN{\tmtwo}}
    & \traN{\tm}\esub{\var}{\osha{\traN{\tmtwo}}}
  \\
      \CBV, \traV{\arg}
    & \osha{\traV{\typ}}\to\osha{\traV{\typtwo}}
    & \ofc{\var}
    & \osha{\lam{\lvar}{\traV{\tm}\esub{\var}{\lvar}}}
    & \open{\uvar}\esub{\uvar}{\traV{\tm}}\,\traV{\tmtwo}
    & \traV{\tm}\esub{\var}{\traV{\tmtwo}}
  \\
      \CCBNd, \traCNd{\arg}
    & \osha{\traCNd{\typ}}\to\sha{\traCNd{\typtwo}}
    & \var
    & \sha{\lam{\lvar}{\traCNd{\tm}\esub{\var}{\lvar}}}
    & \open{\traCNd{\tm}}\,\ofc{\traCNd{\tmtwo}}
    & \traCNd{\tm}\esub{\var}{\ofc{\traCNd{\tmtwo}}}
  \end{array}
\]
A weak evaluation mechanism can be defined for $\lambdaS$
that simulates weak evaluation in the original calculi in a sound and complete way.
Moreover, $\MELLS$ also admits a sound and complete embedding of the
\BangCalculus\begin{ifShortAppendix}
(see the companion report~\cite{mells_long})\end{ifShortAppendix}.

  
There are several avenues worth pursuing.
First, developing an appropriate notion of proof nets and semantics for $\MELLS$,
which perhaps would help clarify the somewhat intriguing interaction between
the sharing and access modalities.
Second, studying operational properties of the
$\lambdaS$-calculus such as standardization (as developed for
$\LSC$~\cite{DBLP:conf/popl/AccattoliBKL14}) and solvability.
Additionally, one can consider extending weak evaluation in $\lambdaS$ to \emph{strong} evaluation, to
simulate strong $\CBN$/$\CBV$/$\CCBNd$ evaluation. Also, our use of multiple
exponentials is reminiscent of subexponentials~\cite{DBLP:conf/ppdp/NigamM09},
where instead of one pair of of-course and why-not modalities one introduces a
family of them, each of which cannot be proven equivalent to any other. Further
work is required to determine if there is a rigorous connection with
subexponentials.

It should be noted that our original motivation to study $\MELLS$ was to try
to provide a unified logical account of $\CBN$, $\CBV$, and $\CBNd$.
In~\cite{DBLP:journals/tcs/MaraistOTW99}, an attempt was made at embedding
$\CBNd$ in a linear $\lambda$-calculus, but the target language had to be
changed to become \emph{affine}, allowing weakening of arbitrary propositions.

\subsubsection*{Acknowledgments} 
The first author was partially supported by project grants
PUNQ 2219/22 and PICT-2021-I-INVI-00602.


\bibliographystyle{splncs04}
\bibliography{biblio.bib}

\newpage
\appendix

\section{Appendix: Preliminary Notions}

\subsection{Simple Results about Abstract Translations}

\SufficientConditionsForSoundness

\begin{proof}\label{sufficient_conditions_translation_sound:proof}
By induction
on the number of steps in a reduction sequence $x_1 \toars^* x_2$,
it is immediate to conclude that $T(x_1) \toarstwo^* T(x_2)$.
\end{proof}

\SufficientConditionsForCompleteness

\begin{proof}\label{sufficient_conditions_translation_complete:proof}
First we claim that for all $y_1 \in Y'$ and $y_2 \in Y$
such that $y_1 \toarstwo^* y_2$
we have that $y_2 \in Y'$ and $T^{-1}(y_1) \toars^* T^{-1}(y_2)$.
Indeed, suppose that $y_1 \toarstwo^n y_2$ in $n$ steps.
We proceed by induction on $n$.
The base case ($n = 0$) is immediate, so suppose that
$n > 0$ and that $y_1 \toarstwo y'_1 \toarstwo^{n-1} y_2$.
Since $T^{-1}$ simulates reduction, we have that $y'_1 \in Y'$
and that $T^{-1}(y_1) \toars^* T^{-1}(y'_1)$.
Moreover, by \ih, we have that $y_2 \in Y'$
and $T^{-1}(y'_1) \toars^* T^{-1}(y_2)$.
Composing the reduction sequences we obtain
$T^{-1}(y_1) \toars^* T^{-1}(y_2)$, as required.

Now suppose that $T(x_1) \toarstwo^* T(x_2)$.
Since $T^{-1}$ is the left-inverse of $T$,
we know that $T(x_1) \in Y'$ and $T(x_2) \in Y'$.
By the previous claim,
$T^{-1}(T(x_1)) \toars^* T^{-1}(T(x_2))$.
Finally, since $T^{-1}$ is the left-inverse of $T$,
we have that
$x_1 = T^{-1}(T(x_1)) \toars^* T^{-1}(T(x_2)) = x_2$,
as required.
\end{proof}

\section{Appendix: \SharingLinearLogic}

\begin{definition}[Mix rule]
\[
  \indrule{\ruleMix}{
    \judg{\tctx,\ofc{\typ}}
    \HS
    \judg{\tctxtwo,\rep{n}{\why{\lneg{\typ}}}}
    \HS
    n \geq 0
  }{
    \judg{\tctx,\tctxtwo}
  }
\]
where $\rep{n}{\typ}$ stands for $\typ,\hdots,\typ$ ($n$ times).
\end{definition}

\begin{definition}[Size, degree, height]
\quad
\begin{itemize}
\item
  An instance of a rule in a derivation is said to be {\em cut-like} if it is an
  instance of the $\ruleCut$ rule or of the $\ruleMix$ rule.
\item
  In a cut-like instance of a rule, the {\em eliminated formula}
  is $\typ$ for instances of $\ruleCut$ and $\ofc{\typ}$ for instances of $\ruleMix$.
\item
  The {\em size} of a formula $\typ$ is written $\sz{\typ}$ and defined as follows:
  \[\begin{array}{c}
    \sz{\btyp} = \sz{\btyp} := 1 \\
    \sz{\typ\tensor\typtwo} = \sz{\typ\parr\typtwo} := 1 + \sz{\typ} + \sz{\typtwo} \\
      \sz{\ofc{\typ}} = \sz{\why{\typ}} = \sz{\sha{\typ}} = \sz{\ush{\typ}} := 1 + \sz{\typ}
      \end{array}
  \]
\item
  The {\em degree} of a cut-like instance of a rule is the size of the
  eliminated formula.
\item
  We write $\derivjudeg{\deriv}{d}{\tctx}$
  if $\deriv$ is a derivation of the judgment $\judg{\tctx}$
  such that every cut-like instance of a rule is of degree at most $d$.
  We write $\judeg{d}{\tctx}$ if there is a derivation $\deriv$
  such that $\derivjudeg{\deriv}{d}{\tctx}$.
\item
  The {\em height} of a derivation $\deriv$ is written $\height{\deriv}$
  and defined as usual, regarding the derivation as a finite tree.
\end{itemize}
\end{definition}

\begin{definition}[Principal formula]
Given an instance of an inference rule,
we say that a formula in the conclusion
is a {\em principal formula} according to the following criteria:
\begin{enumerate}
\item
  $\ruleAx$, $\ruleCut$, and $\ruleMix$:
  there are no principal formulae in the conclusion of these rules.
\item
  $\ruleTensor$:
  the instance of $\typ\tensor\typtwo$ is principal.
\item
  $\ruleParr$:
  the instance of $\typ\parr\typtwo$ is principal.
\item
  $\ruleW$ and $\ruleC$:
  the instance of $\why{\typ}$ is principal.
\item
  $\ruleD$:
  the instance of $\wush{\typ}$ is principal.
\item
  $\ruleProm$:
  all the formulae in the conclusion are principal.
\item
  $\ruleSha$:
  the instance of $\sha{\typ}$ is principal.
\item
  $\ruleUsh$:
  the instance of $\ush{\typ}$ is principal.
\end{enumerate}
Note that, following Lincoln~\cite[Section~2.6]{Lincoln1995},
we consider all the formulae in the conclusion of the
$\ruleProm$ rule to be principal.
This is just a convenience in nomenclature to simplify the presentation
of the cut-elimination proof.

Furthermore, we distinguish between {\em commutative}
and {\em principal} cut-like instances of rules:
\begin{enumerate}
\item
  A cut-like instance of a rule is
  {\em left-principal} if the eliminated formula
  is principal in the first premise of the rule.
  A cut-like instance of a rule is
  {\em right-principal} if the negation of the eliminated formula
  is principal in the second premise of the rule.
  A cut-like instance of a rule is {\em principal}
  if it is both left {\em and} right-principal.
\item
  A cut-like instance of a rule is
  {\em left-commutative} if it is not left-principal,
  it is {\em right-commutative} if it is not right-principal,
  and it is {\em commutative} if it is not principal,
  \ie if it is either left {\em or} right-commutative.
\end{enumerate}
\end{definition}

\begin{lemma}[Generalized structural rules]
The following generalized structural rules admissible,
using only instances of structural rules ($\ruleW$ and $\ruleC$):
\[
  \indrule{\ruleWs}{
    \judg{\tctx}
  }{
    \judg{\tctx,\why{\tctxtwo}}
  }
  \indrule{\ruleCs}{
    \judg{\tctx,\why{\tctxtwo},\why{\tctxtwo}}
  }{
    \judg{\tctx,\why{\tctxtwo}}
  }
  \indrule{\ruleCn}{
    \judg{\tctx,\rep{n}{\why{\typ}}}
    \HS
    n \geq 0
  }{
    \judg{\tctx,\why{\typ}}
  }
\]
\end{lemma}
\begin{proof}
Rules $\ruleWs$ and $\ruleCs$ are straightforward by induction on $\tctxtwo$.
Rule $\ruleCn$ is straightforward by induction on $n$.
\end{proof}

\begin{lemma}[Empty mix lemma]
\llem{empty_mixd_lemma}
If $\judeg{d}{\tctx,\ofc{\typ}}$ and $\judeg{d}{\tctxtwo}$
then $\judeg{d}{\tctx,\tctxtwo}$.
\end{lemma}
\begin{proof}
Let $\derivjudeg{\deriv_1}{d}{\tctx,\ofc{\typ}}$
and $\derivjudeg{\deriv_2}{d}{\tctxtwo}$.
We proceed by induction on $\deriv_1$:
\begin{enumerate}
\item $\ruleAx$:
  Then $\tctx$ must be of the form $\why{\lneg{\typ}}$.
  The situation is:
  \[
    \indrule{\ruleAx}{
    }{
      \judeg{d}{\why{\lneg{\typ}},\ofc{\typ}}
    }
  \]
  So it suffices to take:
  \[
    \indrule{\ruleW}{
      \derivfrom{\deriv_2}{\judeg{d}{\tctxtwo}}
    }{
      \judeg{d}{\why{\lneg{\typ}},\tctxtwo}
    }
  \]
\item $\ruleCut$:
  We consider two subcases, depending on whether $\ofc{\typ}$
  occurs in the first premise of the $\ruleCut$ rule,
  or in the second premise of the $\ruleCut$ rule.
  \begin{enumerate}
  \item
    \label{empty_mix_lemma_cut_left}
    If $\ofc{\typ}$ occurs in the first premise of the $\ruleCut$ rule,
    then $\tctx$ is of the form $\tctx = \tctx_1,\tctx_2$
    and there exists a formula $\typtwo$ such that $\sz{\typtwo} \leq d$
    and:
    \[
      \indrule{\ruleCut}{
        \derivfrom{\deriv_{11}}{\judeg{d}{\tctx_1,\typtwo,\ofc{\typ}}}
        \derivfrom{\deriv_{12}}{\judeg{d}{\tctx_2},\lneg{\typtwo}}
      }{
        \judeg{d}{\tctx_1,\tctx_2,\ofc{\typ}}
      }
    \]
    So by \ih on $\deriv_{11}$ we can take:
    \[
      \indrule{\ruleCut}{
        \derivih{\judeg{d}{\tctx_1,\typtwo,\tctxtwo}}
        \derivfrom{\deriv_{12}}{\judeg{d}{\tctx_2},\lneg{\typtwo}}
      }{
        \judeg{d}{\tctx_1,\tctx_2,\tctxtwo}
      }
    \]
  \item
    Symmetric to the previous case (\ref{empty_mix_lemma_cut_left}).
  \end{enumerate}
\item $\ruleMix$:
  We consider two subcases, depending on whether $\ofc{\typ}$
  occurs in the first premise of the $\ruleMix$ rule,
  or in the second premise of the $\ruleMix$ rule.
  \begin{enumerate}
  \item
    \label{empty_mix_lemma_mix_left}
    If $\ofc{\typ}$ occurs in the first premise of the $\ruleMix$ rule,
    then $\tctx$ is of the form $\tctx = \tctx_1,\tctx_2$
    and there exists a formula $\typtwo$ such that $\sz{\ofc{\typtwo}} \leq d$
    and an $n \geq 0$ such that:
    \[
      \indrule{\ruleMix}{
        \derivfrom{\deriv_{11}}{\judeg{d}{\tctx_1,\ofc{\typtwo},\ofc{\typ}}}
        \derivfrom{\deriv_{12}}{\judeg{d}{\tctx_2,\rep{n}{\why{\lneg{\typtwo}}}}}
      }{
        \judeg{d}{\tctx_1,\tctx_2,\ofc{\typ}}
      }
    \]
    So by \ih on $\deriv_{11}$ we can take:
    \[
      \indrule{\ruleMix}{
        \derivih{\judeg{d}{\tctx_1,\ofc{\typtwo},\tctxtwo}}
        \derivfrom{\deriv_{12}}{\judeg{d}{\tctx_2,\rep{n}{\why{\lneg{\typtwo}}}}}
      }{
        \judeg{d}{\tctx_1,\tctx_2,\tctxtwo}
      }
    \]
  \item
    Symmetric to the previous case (\ref{empty_mix_lemma_mix_left}).
  \end{enumerate}
\item $\ruleTensor$:
  We consider two subcases, depending on whether $\ofc{\typ}$
  occurs in the first premise of the $\ruleTensor$ rule,
  or in the second premise of the $\ruleTensor$ rule.
  \begin{enumerate}
  \item
    \label{empty_mix_lemma_tensor_left}
    If $\ofc{\typ}$ occurs in the first premise of the $\ruleTensor$ rule,
    then $\tctx$ is of the form $\tctx = \tctx_1,\tctx_2,\typtwo\tensor\typthree$
    and:
    \[
      \indrule{\ruleTensor}{
        \derivfrom{\deriv_{11}}{\judeg{d}{\tctx_1,\typtwo,\ofc{\typ}}}
        \derivfrom{\deriv_{12}}{\judeg{d}{\tctx_2,\typthree}}
      }{
        \judeg{d}{\tctx_1,\tctx_2,\typtwo\tensor\typthree,\ofc{\typ}}
      }
    \]
    So by \ih on $\deriv_{11}$ we can take:
    \[
      \indrule{\ruleTensor}{
        \derivih{\judeg{d}{\tctx_1,\typtwo,\tctxtwo}}
        \derivfrom{\deriv_{12}}{\judeg{d}{\tctx_2,\typthree}}
      }{
        \judeg{d}{\tctx_1,\tctx_2,\typtwo\tensor\typthree,\tctxtwo}
      }
    \]
  \item
    Symmetric to the previous case (\ref{empty_mix_lemma_tensor_left}).
  \end{enumerate}
\item $\ruleParr$:
  Then $\tctx$ is of the form $\tctx = \tctx',\typtwo\parr\typthree$
  and the situation is:
  \[
    \indrule{\ruleParr}{
      \derivfrom{\deriv'_1}{\judeg{d}{\tctx',\typtwo,\typthree,\ofc{\typ}}}
    }{
      \judeg{d}{\tctx',\typtwo\parr\typthree,\ofc{\typ}}
    }
  \]
  So by \ih on $\deriv'_1$ we can take: 
  \[
    \indrule{\ruleParr}{
      \derivih{\judeg{d}{\tctx',\typtwo,\typthree,\tctxtwo}}
    }{
      \judeg{d}{\tctx',\typtwo\parr\typthree,\tctxtwo}
    }
  \]
\item $\ruleProm$:
  Then $\tctx$ is of the form $\tctx = \why{\tctx'}$ and the situation is:
  \[
    \indrule{\ruleProm}{
      \derivfrom{\deriv'_1}{\judeg{d}{\why{\tctx'},\typ}}
    }{
      \judeg{d}{\why{\tctx'},\ofc{\typ}}
    }
  \]
  So it suffices to take:
  \[
    \indrule{\ruleWs}{
      \judeg{d}{\tctxtwo}
    }{
      \judeg{d}{\why{\tctx'},\tctxtwo}
    }
  \]
\item $\ruleW$:
  Then $\tctx$ is of the form $\tctx = \tctx',\why{\typtwo}$ and the situation is:
  \[
    \indrule{\ruleW}{
      \derivfrom{\deriv'_1}{\judeg{d}{\tctx',\ofc{\typ}}}
    }{
      \judeg{d}{\tctx',\why{\typtwo},\ofc{\typ}}
    }
  \]
  So by \ih on $\deriv'_1$ we can take:
  \[
    \indrule{\ruleW}{
      \derivih{\judeg{d}{\tctx',\tctxtwo}}
    }{
      \judeg{d}{\tctx',\why{\typtwo},\tctxtwo}
    }
  \]
\item $\ruleC$:
  Then $\tctx$ is of the form $\tctx = \tctx',\why{\typtwo}$ and the situation is:
  \[
    \indrule{\ruleC}{
      \derivfrom{\deriv'_1}{\judeg{d}{\tctx',\why{\typtwo},\why{\typtwo},\ofc{\typ}}}
    }{
      \judeg{d}{\tctx',\why{\typtwo},\ofc{\typ}}
    }
  \]
  So by \ih on $\deriv'_1$ we can take:
  \[
    \indrule{\ruleC}{
      \derivfrom{\deriv'_1}{\judeg{d}{\tctx',\why{\typtwo},\why{\typtwo},\tctxtwo}}
    }{
      \judeg{d}{\tctx',\why{\typtwo},\tctxtwo}
    }
  \]
\item $\ruleD$:
  \label{empty_mix_lemma_dereliction}
  Then $\tctx$ is of the form $\tctx = \tctx',\wush{\typtwo}$ and the situation is:
  \[
    \indrule{\ruleD}{
      \derivfrom{\deriv'_1}{\judeg{d}{\tctx',\ush{\typtwo},\ofc{\typ}}}
    }{
      \judeg{d}{\tctx',\wush{\typtwo},\ofc{\typ}}
    }
  \]
  So by \ih on $\deriv'_1$ we can take:
  \[
    \indrule{\ruleD}{
      \derivfrom{\deriv'_1}{\judeg{d}{\tctx',\ush{\typtwo},\tctxtwo}}
    }{
      \judeg{d}{\tctx',\wush{\typtwo},\tctxtwo}
    }
  \]
\item $\ruleSha$:
  Similar to case \ref{empty_mix_lemma_dereliction}.
\item $\ruleUsh$:
  Similar to case \ref{empty_mix_lemma_dereliction}.
\end{enumerate}
\end{proof}

\begin{lemma}[Cut/mix flattening]
\llem{cut_mix_flattening}
The following rules are admissible:
\begin{enumerate}
\item
  {\bf Cut-flattening.}
  \[
    \indrule{\ruleCutd}{
      \judeg{d}{\tctx,\typ}
      \HS
      \judeg{d}{\tctxtwo,\lneg{\typ}}
      \HS
      \sz{\typ} = d + 1
    }{
      \judeg{d}{\tctx,\tctxtwo}
    }
  \]
\item
  {\bf Mix-flattening.}
  \[
    \indrule{\ruleMixd}{
      \judeg{d}{\tctx,\ofc{\typ}}
      \HS
      \judeg{d}{\tctxtwo,\rep{n}{\why{\lneg{\typ}}}}
      \HS
      \sz{\ofc{\typ}} = d + 1
    }{
      \judeg{d}{\tctx,\tctxtwo}
    }
  \]
\end{enumerate}
\end{lemma}
\begin{proof}
In each of the two rules of the statement,
let us write $\deriv_1$ for the derivation of the first premise
and $\deriv_2$ for the derivation of the second premise.
We prove the two items simultaneously,
by induction on
$\height{\deriv_1} + \height{\deriv_2}$.
We proceed by case analysis on the last rule used to construct the
derivations $\deriv_1$ and $\deriv_2$.
As usual, in the case analysis, we distinguish between commutative
and principal cases.
Note that, for cut-flattening,
we only study the left-commutative cases,
since the right-commutative cases are symmetric.
Moreover, for mix-flattening,
in right-commutative cases
we may assume that the $\ruleMixd$ rule we are trying to derive
is left-principal.

\begin{ifShortAppendix}
  We only give the proof of the five most interesting cases.
  The extended version~\cite{mells_long} contains the fully detailed proof.
  \begin{enumerate}
  \item
    {\em Cut-flattening, left-commutative case, $\ruleProm$.}
    We argue that this case is impossible.
    Indeed, $\derivjudeg{\deriv_1}{d}{\tctx,\typ}$
    must be constructed using the $\ruleProm$ rule,
    but all the formulae in the conclusion of $\ruleProm$
    are principal by definition.
  \item
    {\em Mix-flattening, right-commutative case, $\ruleAx$.}
    Then $\derivjudeg{\deriv_2}{d}{\tctxtwo,\rep{n}{\why{\lneg{\typ}}}}$
    is constructed using the $\ruleAx$ rule,
    so the number of formulae in $\tctxtwo,\rep{n}{\why{\lneg{\typ}}}$ is exactly $2$.
    In particular, $n \leq 2$.
    Moreover, $n$ cannot be exactly $2$ because the conclusion
    of the $\ruleAx$ rule cannot be of the form $\why{\lneg{\typ}},\why{\lneg{\typ}}$.
    Therefore $n$ is either $0$ or $1$, and we consider two subcases:
    \begin{enumerate}
    \item
      \label{flattening_mixf_right_ax_zero}
      If $n = 0$, then $\tctxtwo = \typtwo,\lneg{\typtwo}$ and the situation is:
      \[
        \indrule{\ruleMixd}{
          \derivfrom{\deriv_1}{\judeg{d}{\tctx,\ofc{\typ}}}
          \indrule{\ruleAx}{
            \emptyPremise
          }{
            \judeg{d}{\typtwo,\lneg{\typtwo}}
          }
        }{
          \judg{\tctx,\typtwo,\lneg{\typtwo}}
        }
      \]
      Then by \rlem{empty_mixd_lemma} we conclude that
      $\judeg{d}{\tctx,\typtwo,\lneg{\typtwo}}$.
    \item
      If $n = 1$, then $\tctxtwo = \ofc{\typ}$ and the situation is:
      \[
        \indrule{\ruleMixd}{
          \derivfrom{\deriv_1}{\judeg{d}{\tctx,\ofc{\typ}}}
          \indrule{\ruleAx}{
            \emptyPremise
          }{
            \judeg{d}{\ofc{\typ},\why{\lneg{\typ}}}
          }
        }{
          \judg{\tctx,\ofc{\typ}}
        }
      \]
      So we can take $\deriv_1$ itself.
    \end{enumerate}
  \item
    {\em Mix-flattening, right-commutative case, $\ruleCut$.}
    Then $\derivjudeg{\deriv_2}{d}{\tctxtwo,\rep{n}{\why{\lneg{\typ}}}}$
    is constructed using the $\ruleCut$ rule.
    Moreover, we may assume that the instance of $\ruleMixd$
    we are constructing is left-principal,
    since otherwise it would fall into one of the left-commutative cases,
    so $\derivjudeg{\deriv_1}{d}{\tctx,\ofc{\typ}}$
    must necessarily be constructed using the $\ruleProm$ rule
    and $\tctx$ must be of the form $\tctx = \why{\tctx'}$.
    Then the situation is as follows,
    where $\tctxtwo = \tctxtwo_1,\tctxtwo_2$
    and $n = n_1 + n_2$ and $\sz{\typtwo} \leq d$:
    \[
      \indrule{\ruleMixd}{
        \indrule{\ruleProm}{
          \derivfrom{\deriv'_1}{\judeg{d}{\why{\tctx'},\typ}}
        }{
          \judeg{d}{\why{\tctx'},\ofc{\typ}}
        }
        \indrule{\ruleCut}{
          \derivfrom{\deriv_{21}}{\judeg{d}{\tctxtwo_1,\rep{n_1}{\why{\lneg{\typ}}},\typtwo}}
          \derivfrom{\deriv_{22}}{\judeg{d}{\tctxtwo_2,\rep{n_2}{\why{\lneg{\typ}}},\lneg{\typtwo}}}
        }{
          \judeg{d}{\tctxtwo_1,\tctxtwo_2,\rep{n}{\why{\lneg{\typ}}}}
        }
      }{
        \judg{\why{\tctx'},\tctxtwo_1,\tctxtwo_2}
      }
    \]
    Resorting to the \ih, we can take:
    \[
      \indrule{\ruleCs}{
        \indrule{\ruleCut}{
          \derivfrom{\derivtwo_1}{\judeg{d}{\why{\tctx'},\tctxtwo_1,\typtwo}}
          \derivfrom{\derivtwo_2}{\judeg{d}{\why{\tctx'},\tctxtwo_2,\lneg{\typtwo}}}
        }{
          \judeg{d}{\why{\tctx'},\why{\tctx'},\tctxtwo_1,\tctxtwo_2}
        }
      }{
        \judeg{d}{\why{\tctx'},\tctxtwo_1,\tctxtwo_2}
      }
    \]
    where:
    \[
      \begin{array}{rcl}
      \derivtwo_1 & \eqdef &
        \derp{\indrule{\ruleMixd}{
          \indrule{\ruleProm}{
            \derivfrom{\deriv'_1}{\judeg{d}{\why{\tctx'},\typ}}
          }{
            \judeg{d}{\why{\tctx'},\ofc{\typ}}
          }
          \derivfrom{\deriv_{21}}{\judeg{d}{\tctxtwo_1,\typtwo,\rep{n_1}{\why{\lneg{\typ}}}}}
        }{
          \judeg{d}{\why{\tctx'},\tctxtwo_1,\typtwo}
        }}
      \\
      \derivtwo_2 & \eqdef &
        \derp{\indrule{\ruleMixd}{
          \indrule{\ruleProm}{
            \derivfrom{\deriv'_1}{\judeg{d}{\why{\tctx'},\typ}}
          }{
            \judeg{d}{\why{\tctx'},\ofc{\typ}}
          }
          \derivfrom{\deriv_{22}}{\judeg{d}{\tctxtwo_2,\rep{n_2}{\why{\lneg{\typ}}},\lneg{\typtwo}}}
        }{
          \judeg{d}{\why{\tctx'},\tctxtwo_2,\lneg{\typtwo}}
        }}
      \end{array}
    \]
    To apply the \ih, note that sum of the heights of the subderivations decreases.
  \item
    {\em Mix-flattening, right-commutative case, $\ruleMix$.}
    Then $\derivjudeg{\deriv_2}{d}{\tctxtwo,\rep{n}{\why{\lneg{\typ}}}}$ is
    constructed using the $\ruleMix$ rule. Moreover, we may assume that
    the instance of $\ruleMixd$ we are constructing is left-principal,
    since otherwise it would fall into one of the left-commutative cases,
    so $\derivjudeg{\deriv_1}{d}{\tctx,\ofc{\typ}}$ must necessarily be constructed
    using the $\ruleProm$ rule and $\tctx$ must be of the form
    $\tctx = \why{\tctx'}$. Then the situation is as follows, where
    $\tctxtwo = \tctxtwo_1,\tctxtwo_2$ and $n = n_1 + n_2$
    and $\sz{\ofc{\typtwo}} \leq d$ and $m \geq 0$:
    \[
      \indrule{\ruleMixd}{
        \indrule{\ruleProm}{
          \derivfrom{\deriv'_1}{\judeg{d}{\why{\tctx'},\typ}}
        }{
          \judeg{d}{\why{\tctx'},\ofc{\typ}}
        }
        \indrule{\ruleMix}{
          \derivfrom{\deriv_{21}}{\judeg{d}{\tctxtwo_1,\rep{n_1}{\why{\lneg{\typ}}},\ofc{\typtwo}}}
          \derivfrom{\deriv_{22}}{\judeg{d}{\tctxtwo_2,\rep{n_2}{\why{\lneg{\typ}}},\rep{m}{\why{\lneg{\typtwo}}}}}
        }{
          \judeg{d}{\tctxtwo_1,\tctxtwo_2,\rep{n}{\why{\lneg{\typ}}}}
        }
      }{
        \judg{\why{\tctx'},\tctxtwo_1,\tctxtwo_2}
      }
    \]
    Resorting to the \ih, we can take:
    \[
      \indrule{\ruleCs}{
        \indrule{\ruleMix}{
          \derivfrom{\derivtwo_1}{\judeg{d}{\why{\tctx'},\tctxtwo_1,\ofc{\typtwo}}}
          \derivfrom{\derivtwo_2}{\judeg{d}{\why{\tctx'},\tctxtwo_2,\rep{m}{\why{\lneg{\typtwo}}}}}
        }{
          \judeg{d}{\why{\tctx'},\why{\tctx'},\tctxtwo_1,\tctxtwo_2}
        }
      }{
        \judeg{d}{\why{\tctx'},\tctxtwo_1,\tctxtwo_2}
      }
    \]
    where:
    \[
      \begin{array}{rcl}
        \derivtwo_1 & \eqdef &
          \derp{\indrule{\ruleMixd}{
            \indrule{\ruleProm}{
              \derivfrom{\deriv'_1}{\judeg{d}{\why{\tctx'},\typ}}
            }{
              \judeg{d}{\why{\tctx'},\ofc{\typ}}
            }
            \derivfrom{\deriv_{21}}{\judeg{d}{\tctxtwo_1,\ofc{\typtwo},\rep{n_1}{\why{\lneg{\typ}}}}}
          }{
            \judeg{d}{\why{\tctx'},\tctxtwo_1,\ofc{\typtwo}}
          }}
      \\
        \derivtwo_2 & \eqdef &
          \derp{\indrule{\ruleMixd}{
            \indrule{\ruleProm}{
              \derivfrom{\deriv'_1}{\judeg{d}{\why{\tctx'},\typ}}
            }{
              \judeg{d}{\why{\tctx'},\ofc{\typ}}
            }
            \derivfrom{\deriv_{22}}{\judeg{d}{\tctxtwo_2,\rep{n_2}{\why{\lneg{\typ}}},\rep{m}{\why{\lneg{\typtwo}}}}}
          }{
            \judeg{d}{\why{\tctx'},\tctxtwo_2,\rep{m}{\why{\lneg{\typtwo}}}}
          }}
      \end{array}
    \]
    To apply the \ih, note that sum of the heights of the subderivations decreases.
  \item
    {\em Mix-flattening, principal case, $\ruleProm$ / $\ruleD$.}
    Then $\derivjudeg{\deriv_1}{d}{\tctx,\ofc{\typ}}$ is constructed using the
    $\ruleProm$ rule and $\derivjudeg{\deriv_2}{d}{\tctxtwo,\rep{n}{\why{\lneg{\typ}}}}$
    is constructed using the $\ruleD$ rule.
    Since $\judg{\tctx,\ofc{\typ}}$ is the conclusion of an instance of the
    $\ruleProm$ rule, $\tctx$ must be of the form $\tctx = \why{\tctx'}$.
    If $n = 0$, it suffices to resort to \rlem{empty_mixd_lemma} to conclude that
    $\judeg{d}{\why{\tctx'},\tctxtwo}$.
    If $n > 0$, then 
    since $\derivjudeg{\deriv_2}{d}{\tctxtwo,\rep{n}{\why{\lneg{\typ}}}}$
    is the conclusion of an instance of the $\ruleD$ rule with $\why{\lneg{\typ}}$
    as a principal formula, $\lneg{\typ}$ must be of the form
    $\lneg{\typ} = \ush{\typ_1}$, so $\typ = \sha{\lneg{\typ_1}}$.
    Then the situation is:
    \[
      \indrule{\ruleMixd}{
        \indrule{\ruleProm}{
          \derivfrom{\deriv'_1}{\judeg{d}{\why{\tctx'},\sha{\lneg{\typ_1}}}}
        }{
          \judeg{d}{\why{\tctx'},\osha{\lneg{\typ_1}}}
        }
        \indrule{\ruleD}{
          \derivfrom{\deriv'_2}{\judeg{d}{\tctxtwo,\rep{n-1}{\wush{\typ_1}},\ush{\typ_1}}}
        }{
          \judeg{d}{\tctxtwo,\rep{n-1}{\wush{\typ_1}},\wush{\typ_1}}
        }
      }{
        \judg{\why{\tctx'},\tctxtwo}
      }
    \]
    Resorting to the \ih, we can take:
    \[
      \indrule{\ruleCs}{
        \indrule{\ruleCut}{
          \derivfrom{\deriv'_1}{\judeg{d}{\why{\tctx'},\sha{\lneg{\typ_1}}}}
          \indrule{\ruleMixd}{
            \indrule{\ruleProm}{
              \derivfrom{\deriv'_1}{\judeg{d}{\why{\tctx'},\sha{\lneg{\typ_1}}}}
            }{
              \judeg{d}{\why{\tctx'},\osha{\lneg{\typ_1}}}
            }
            \derivfrom{\deriv'_2}{\judeg{d}{\tctxtwo,\rep{n-1}{\wush{\typ_1}},\ush{\typ_1}}}
          }{
            \judeg{d}{\why{\tctx'},\tctxtwo,\ush{\typ_1}}
          }
        }{
          \judeg{d}{\why{\tctx'},\why{\tctx'},\tctxtwo}
        }
      }{
        \judeg{d}{\why{\tctx'},\tctxtwo}
      }
    \]
    To apply the \ih, note that sum of the heights of the subderivations decreases.
    Moreover, note that the last instance of the $\ruleCut$ rule used in the
    derivation is between $\sha{\lneg{\typ_1}}$ and $\ush{\typ_1}$
    and $\sz{\sha{\lneg{\typ_1}}} = \ush{\typ_1} = d$.
  \end{enumerate}
\end{ifShortAppendix}
\begin{ifLongAppendix}
  \begin{enumerate}
  \item
    {\em Cut-flattening, left-commutative case, $\ruleAx$.}
    Then we have that $\derivjudeg{\deriv_1}{d}{\tctx,\typ}$
    is constructed using the $\ruleAx$ rule,
    so $\tctx = \lneg{\typ}$.
    The situation is:
    \[
      \indrule{\ruleCutd}{
        \indrule{\ruleAx}{
          \emptyPremise
        }{
          \judeg{d}{\lneg{\typ},\typ}
        }
        \derivfrom{\deriv_2}{\judeg{d}{\tctxtwo,\lneg{\typ}}}
      }{
        \judg{\lneg{\typ},\tctxtwo}
      }
    \]
    So we can take $\deriv_2$ itself.
  \item
    {\em Cut-flattening, left-commutative case, $\ruleCut$.}
    Then $\derivjudeg{\deriv_1}{d}{\tctx,\typ}$
    is constructed using the $\ruleCut$ rule.
    We consider two subcases, depending on whether the eliminated formula
    $\typ$ occurs in the first premise of the $\ruleCut$ rule
    or in the second premise of the $\ruleCut$ rule:
    \begin{enumerate}
    \item
      \label{flattening_cutd_left_cut_left}
      If $\typ$ occurs in the first premise of the $\ruleCut$ rule
      then $\tctx$ must be of the form $\tctx_1,\tctx_2$,
      and the situation is as follows,
      where $\sz{\typtwo} \leq d$:
      \[
        \indrule{\ruleCutd}{
          \indrule{\ruleCut}{
            \derivfrom{\deriv_{11}}{\judeg{d}{\tctx_1,\typtwo,\typ}}
            \derivfrom{\deriv_{12}}{\judeg{d}{\tctx_2,\lneg{\typtwo}}}
          }{
            \judeg{d}{\tctx_1,\tctx_2,\typ}
          }
          \derivfrom{\deriv_2}{\judeg{d}{\tctxtwo,\lneg{\typ}}}
        }{
          \judg{\tctx_1,\tctx_2,\tctxtwo}
        }
      \]
      Resorting to the \ih, we can take:
      \[
        \indrule{\ruleCut}{
          \indrule{\ruleCutd}{
            \derivfrom{\deriv_{11}}{\judeg{d}{\tctx_1,\typtwo,\typ}}
            \derivfrom{\deriv_2}{\judeg{d}{\tctxtwo,\lneg{\typ}}}
          }{
            \judeg{d}{\tctx_1,\tctxtwo,\typtwo}
          }
          \derivfrom{\deriv_{12}}{\judeg{d}{\tctx_2,\lneg{\typtwo}}}
        }{
          \judeg{d}{\tctx_1,\tctx_2,\tctxtwo}
        }
      \]
      To apply the \ih, note that sum of the heights of the subderivations decreases.
    \item
      If $\typ$ occurs in the second premise of the $\ruleCut$ rule,
      the proof is symmetric to the previous case (\ref{flattening_cutd_left_cut_left}).
    \end{enumerate}
  \item
    {\em Cut-flattening, left-commutative case, $\ruleMix$.}
    Then $\derivjudeg{\deriv_1}{d}{\tctx,\typ}$ is constructed using
    the $\ruleMix$ rule.
    We consider two subcases, depending on whether the eliminated formula
    $\typ$ occurs in the first premise of the $\ruleMix$ rule
    or in the second premise of the $\ruleMix$ rule:
    \begin{enumerate}
    \item
      If $\typ$ occurs in the first premise of the $\ruleMix$ rule,
      then $\tctx$ must be of the form $\tctx_1,\tctx_2$,
      and the situation is as follows,
      where $\sz{\ofc{\typtwo}} \leq d$ and $n \geq 0$:
      \[
        \indrule{\ruleCutd}{
          \indrule{\ruleMix}{
            \derivfrom{\deriv_{11}}{\judeg{d}{\tctx_1,\typ,\ofc{\typtwo}}}
            \derivfrom{\deriv_{12}}{\judeg{d}{\tctx_2,\rep{n}{\why{\lneg{\typtwo}}}}}
          }{
            \judeg{d}{\tctx_1,\tctx_2,\typ}
          }
          \derivfrom{\deriv_2}{\judeg{d}{\tctxtwo,\lneg{\typ}}}
        }{
          \judg{\tctx_1,\tctx_2,\tctxtwo}
        }
      \]
      Resorting to the \ih, we can take:
      \[
        \indrule{\ruleMix}{
          \indrule{\ruleCutd}{
            \derivfrom{\deriv_{11}}{\judeg{d}{\tctx_1,\typ,\ofc{\typtwo}}}
            \derivfrom{\deriv_2}{\judeg{d}{\tctxtwo,\lneg{\typ}}}
          }{
            \judeg{d}{\tctx_1,\ofc{\typtwo},\tctxtwo}
          }
          \derivfrom{\deriv_{12}}{\judeg{d}{\tctx_2,\rep{n}{\why{\lneg{\typtwo}}}}}
        }{
          \judeg{d}{\tctx_1,\tctx_2,\tctxtwo}
        }
      \]
      To apply the \ih, note that sum of the heights of the subderivations decreases.
    \item
      If $\typ$ occurs in the second premise of the \ruleMix rule,
      then $\tctx$ must be of the form $\tctx_1,\tctx_2$ and the
      situation is as follows, where $\sz{\ofc{\typtwo}} \leq d$
      and $n \geq 0$:
      \[
        \indrule{\ruleCutd}{
          \indrule{\ruleMix}{
            \derivfrom{\deriv_{11}}{\judeg{d}{\tctx_1,\ofc{\typtwo}}}
            \derivfrom{\deriv_{12}}{\judeg{d}{\tctx_2,\typ,\rep{n}{\why{\lneg{\typtwo}}}}}
          }{
            \judeg{d}{\tctx_1,\tctx_2,\typ}
          }
          \derivfrom{\deriv_2}{\judeg{d}{\tctxtwo,\lneg{\typ}}}
        }{
          \judg{\tctx_1,\tctx_2,\tctxtwo}
        }
      \]
      Resorting to the \ih, we can take:
      \[
        \indrule{\ruleMix}{
          \derivfrom{\deriv_{11}}{\judeg{d}{\tctx_1,\ofc{\typtwo}}}
          \indrule{\ruleCutd}{
            \derivfrom{\deriv_{12}}{\judeg{d}{\tctx_2,\rep{n}{\why{\lneg{\typtwo}}},\typ}}
            \derivfrom{\deriv_2}{\judeg{d}{\tctxtwo,\lneg{\typ}}}
          }{
            \judeg{d}{\tctx_2,\tctxtwo,\rep{n}{\why{\lneg{\typtwo}}}}
          }
        }{
          \judeg{d}{\tctx_1,\tctx_2,\tctxtwo}
        }
      \]
      To apply the \ih, note that sum of the heights of the subderivations decreases.
    \end{enumerate}
  \item
    {\em Cut-flattening, left-commutative case, $\ruleTensor$.}
    Then $\derivjudeg{\deriv_1}{d}{\tctx,\typ}$
    is constructed using the $\ruleTensor$ rule.
    We consider two subcases, depending on whether the eliminated formula
    $\typ$ occurs in the first premise of the $\ruleTensor$ rule
    or in the second premise of the $\ruleTensor$ rule:
    \begin{enumerate}
    \item
      \label{flattening_cutd_left_tensor_left}
      If $\typ$ occurs in the first premise of the $\ruleTensor$ rule,
      then the situation is:
      \[
        \indrule{\ruleCutd}{
          \indrule{\ruleTensor}{
            \derivfrom{\deriv_{11}}{\judeg{d}{\tctx_1,\typtwo,\typ}}
            \derivfrom{\deriv_{12}}{\judeg{d}{\tctx_2,\typthree}}
          }{
            \judeg{d}{\tctx_1,\tctx_2,\typtwo\tensor\typthree,\typ}
          }
          \derivfrom{\deriv_2}{\judeg{d}{\tctxtwo,\lneg{\typ}}}
        }{
          \judg{\tctx_1,\tctx_2,\typtwo\tensor\typthree,\tctxtwo}
        }
      \]
      Resorting to the \ih, we can take:
      \[
        \indrule{\ruleTensor}{
          \indrule{\ruleCutd}{
            \derivfrom{\deriv_{11}}{\judeg{d}{\tctx_1,\typtwo,\typ}}
            \derivfrom{\deriv_2}{\judeg{d}{\tctxtwo,\lneg{\typ}}}
          }{
            \judeg{d}{\tctx_1,\typtwo,\tctxtwo}
          }
          \derivfrom{\deriv_{12}}{\judeg{d}{\tctx_2,\typthree}}
        }{
          \judeg{d}{\tctx_1,\tctx_2,\typtwo\tensor\typthree,\tctxtwo}
        }
      \]
      To apply the \ih, note that sum of the heights of the subderivations decreases.
    \item
      If $\typ$ occurs on the second premise of the $\ruleTensor$ rule,
      the proof is symmetric to the previous case
      (\ref{flattening_cutd_left_tensor_left}).
    \end{enumerate}
  \item
    {\em Cut-flattening, left-commutative case, $\ruleParr$.}
    Then $\derivjudeg{\deriv_1}{d}{\tctx,\typ}$ is constructed using the
    $\ruleParr$ rule.
    The situation is:
    \[
      \indrule{\ruleCutd}{
        \indrule{\ruleParr}{
          \derivfrom{\deriv'_1}{\judeg{d}{\tctx',\typtwo,\typthree,\typ}}
        }{
          \judeg{d}{\tctx',\typtwo\parr\typthree,\typ}
        }
        \derivfrom{\deriv_2}{\judeg{d}{\tctxtwo,\lneg{\typ}}}
      }{
        \judg{\tctx',\tctxtwo,\typtwo\parr\typthree}
      }
    \]
    Resorting to the \ih, we can take:
    \[
      \indrule{\ruleParr}{
        \indrule{\ruleCutd}{
          \derivfrom{\deriv'_1}{\judeg{d}{\tctx',\typtwo,\typthree,\typ}}
          \derivfrom{\deriv_2}{\judeg{d}{\tctxtwo,\lneg{\typ}}}
        }{
          \judeg{d}{\tctx',\tctxtwo,\typtwo,\typthree}
        }
      }{
        \judeg{d}{\tctx',\tctxtwo,\typtwo\parr\typthree}
      }
    \]
    To apply the \ih, note that sum of the heights of the subderivations decreases.
  \item
    {\em Cut-flattening, left-commutative case, $\ruleProm$.}
    We argue that this case is impossible.
    Indeed, $\derivjudeg{\deriv_1}{d}{\tctx,\typ}$
    must be constructed using the $\ruleProm$ rule,
    but all the formulae in the conclusion of $\ruleProm$
    are principal by definition.
  \item
    {\em Cut-flattening, left-commutative case, $\ruleW$.}
    Then $\derivjudeg{\deriv_1}{d}{\tctx,\typ}$ is constructed using the
    $\ruleW$ rule.
    Note that $\tctx$ must be of the form $\tctx',\why{\typtwo}$.
    The situation is:
    \[
      \indrule{\ruleCutd}{
        \indrule{\ruleW}{
          \derivfrom{\deriv'_1}{\judeg{d}{\tctx',\typ}}
        }{
          \judeg{d}{\tctx',\why{\typtwo},\typ}
        }
        \derivfrom{\deriv_2}{\judeg{d}{\tctxtwo,\lneg{\typ}}}
      }{
        \judg{\tctx',\tctxtwo,\why{\typtwo}}
      }
    \]
    Resorting to the \ih, we can take:
    \[
      \indrule{\ruleW}{
        \indrule{\ruleCutd}{
          \derivfrom{\deriv'_1}{\judeg{d}{\tctx',\typ}}
          \derivfrom{\deriv_2}{\judeg{d}{\tctxtwo,\lneg{\typ}}}
        }{
          \judeg{d}{\tctx',\tctxtwo}
        }
      }{
        \judeg{d}{\tctx',\tctxtwo,\why{\typtwo}}
      }
    \]
    To apply the \ih, note that sum of the heights of the subderivations decreases.
  \item
    {\em Cut-flattening, left-commutative case, $\ruleC$.}
    Then $\derivjudeg{\deriv_1}{d}{\tctx,\typ}$ is constructed using the
    $\ruleC$ rule.
    Note that $\tctx$ must be of the form $\tctx',\why{\typtwo}$.
    The situation is:
    \[
      \indrule{\ruleCutd}{
        \indrule{\ruleC}{
          \derivfrom{\deriv'_1}{\judeg{d}{\tctx',\typ,\why{\typtwo},\why{\typtwo}}}
        }{
          \judeg{d}{\tctx',\typ,\why{\typtwo}}
        }
        \derivfrom{\deriv_2}{\judeg{d}{\tctxtwo,\lneg{\typ}}}
      }{
        \judg{\tctx',\tctxtwo,\why{\typtwo}}
      }
    \]
    Resorting to the \ih, we can take:
    \[
      \indrule{\ruleC}{
        \indrule{\ruleCutd}{
          \derivfrom{\deriv'_1}{\judeg{d}{\tctx',\typ,\why{\typtwo},\why{\typtwo}}}
          \derivfrom{\deriv_2}{\judeg{d}{\tctxtwo,\lneg{\typ}}}
        }{
          \judeg{d}{\tctx',\tctxtwo,\why{\typtwo},\why{\typtwo}}
        }
      }{
        \judeg{d}{\tctx',\tctxtwo,\why{\typtwo}}
      }
    \]
    To apply the \ih, note that sum of the heights of the subderivations decreases.
  \item
    \label{flattening_cutd_left_dereliction}
    {\em Cut-flattening, left-commutative case, $\ruleD$.}
    Then $\derivjudeg{\deriv_1}{d}{\tctx,\typ}$
    is constructed using the $\ruleD$ rule.
    Note that $\tctx$ must be of the form $\tctx',\wush{\typtwo}$.
    The situation is:
    \[
      \indrule{\ruleCutd}{
        \indrule{\ruleD}{
          \derivfrom{\deriv'_1}{\judeg{d}{\tctx',\ush{\typtwo}}}
        }{
          \judeg{d}{\tctx',\wush{\typtwo}}
        }
        \derivfrom{\deriv_2}{\judeg{d}{\tctxtwo}}
      }{
        \judg{\tctx',\wush{\typtwo},\tctxtwo}
      }
    \]
    Resorting to the \ih, we can take:
    \[
      \indrule{\ruleD}{
        \indrule{\ruleCutd}{
          \derivfrom{\deriv'_1}{\judeg{d}{\tctx',\ush{\typtwo}}}
          \derivfrom{\deriv_2}{\judeg{d}{\tctxtwo}}
        }{
          \judeg{d}{\tctx',\ush{\typtwo},\tctxtwo}
        }
      }{
        \judeg{d}{\tctx',\wush{\typtwo},\tctxtwo}
      }
    \]
    To apply the \ih, note that sum of the heights of the subderivations decreases.
  \item
    {\em Cut-flattening, left-commutative case, $\ruleSha$.}
    Similar to case \ref{flattening_cutd_left_dereliction}.
  \item
    {\em Cut-flattening, left-commutative case, $\ruleUsh$.}
    Similar to case \ref{flattening_cutd_left_dereliction}.
  \item
    \label{flattening_cutd_ppal_tensor_parr}
    {\em Cut-flattening, principal case, $\ruleTensor$ / $\ruleParr$.}
    Then $\derivjudeg{\deriv_1}{d}{\tctx,\typ}$
    is constructed using the $\ruleTensor$ rule
    and $\derivjudeg{\deriv_2}{d}{\tctxtwo,\lneg{\typ}}$
    is constructed using the $\ruleParr$ rule.
    Note that $\tctx$ must be of the form $\tctx_1,\tctx_2$
    and $\typ$ must be of the form $\typtwo\tensor\typthree$.
    The situation is:
    \[
      \indrule{\ruleCutd}{
        \indrule{\ruleTensor}{
          \derivfrom{\deriv_{11}}{\judeg{d}{\tctx_1,\typtwo}}
          \derivfrom{\deriv_{12}}{\judeg{d}{\tctx_2,\typthree}}
        }{
          \judeg{d}{\tctx_1,\tctx_2,\typtwo\tensor\typthree}
        }
        \indrule{\ruleParr}{
          \derivfrom{\deriv'_2}{\judeg{d}{\tctxtwo,\lneg{\typtwo},\lneg{\typthree}}}
        }{
          \judeg{d}{\tctxtwo,\lneg{\typtwo}\parr\lneg{\typthree}}
        }
      }{
        \judg{\tctx_1,\tctx_2,\tctxtwo}
      }
    \]
    Then we can take:
    \[
      \indrule{\ruleCut}{
        \derivfrom{\deriv_{11}}{\judeg{d}{\tctx_1,\typtwo}}
        \indrule{\ruleCut}{
          \derivfrom{\deriv_{12}}{\judeg{d}{\tctx_2,\typthree}}
          \derivfrom{\deriv'_2}{\judeg{d}{\tctxtwo,\lneg{\typtwo},\lneg{\typthree}}}
        }{
          \judeg{d}{\tctx_2,\tctxtwo,\lneg{\typtwo}}
        }
      }{
        \judeg{d}{\tctx_1,\tctx_2,\tctxtwo}
      }
    \]
    Note that we do not need to apply the \ih,
    as the sizes of the eliminated formulae strictly decrease.
  \item
    {\em Cut-flattening, principal case, $\ruleProm$ / $\ruleProm$.}
    Then $\derivjudeg{\deriv_1}{d}{\tctx,\typ}$
    and $\derivjudeg{\deriv_2}{d}{\tctxtwo,\lneg{\typ}}$
    are both constructed using the $\ruleProm$ rule.
    Since the conclusion of the $\ruleProm$ rule is of the form
    $\why{\tctxthree},\ofc{\typthree}$,
    we consider two subcases,
    depending on whether $\typ$ is of the form $\ofc{\typ_1}$
    or of the form $\why{\typ_1}$:
    \begin{enumerate}
    \item
      \label{flattening_cutd_ppal_prom_prom_ofc}
      If $\typ$ is of the form $\typ = \ofc{\typ_1}$.
      Then $\tctx$ must be of the form $\why{\tctx'}$.
      Moreover, note that $\lneg{\typ} = \lneg{(\ofc{\typ_1})} = \why{\lneg{\typ_1}}$
      and, since $\lneg{\typ}$ is part of the conclusion of $\deriv_2$,
      which is also constructed using the $\ruleProm$ rule,
      we know that $\lneg{\typ}$ must be of the form $\why{\lneg{\typ_1}}$.
      Moreover, since $\judg{\tctxtwo,\why{\lneg{\typ_1}}}$ is the conclusion of an
      instance of the $\ruleProm$ rule, $\tctxtwo$ must be of the form
      $\tctxtwo = \why{\tctxtwo'},\ofc{\typtwo}$.
      In summary, the situation is:
      \[
        \indrule{\ruleCutd}{
          \indrule{\ruleProm}{
            \derivfrom{\deriv'_1}{\judeg{d}{\why{\tctx'},\typ_1}}
          }{
            \judeg{d}{\why{\tctx'},\ofc{\typ_1}}
          }
          \indrule{\ruleProm}{
            \derivfrom{\deriv'_2}{\judeg{d}{\why{\tctxtwo'},\typtwo,\why{\lneg{\typ_1}}}}
          }{
            \judeg{d}{\why{\tctxtwo'},\ofc{\typtwo},\why{\lneg{\typ_1}}}
          }
        }{
          \judg{\why{\tctx'},\why{\tctxtwo'},\ofc{\typtwo}}
        }
      \]
      Resorting to the \ih, we can take:
      \[
        \indrule{\ruleProm}{
          \indrule{\ruleCutd}{
            \indrule{\ruleProm}{
              \derivfrom{\deriv'_1}{\judeg{d}{\why{\tctx'},\typ_1}}
            }{
              \judeg{d}{\why{\tctx'},\sha{\typ_1}}
            }
            \derivfrom{\deriv'_2}{\judeg{d}{\why{\tctxtwo'},\typtwo,\why{\lneg{\typ_1}}}}
          }{
            \judeg{d}{\why{\tctx'},\why{\tctxtwo'},\typtwo}
          }
        }{
          \judeg{d}{\why{\tctx'},\why{\tctxtwo'},\ofc{\typtwo}}
        }
      \]
      To apply the \ih, note that sum of the heights of the subderivations decreases.
    \item
      If $\typ$ is of the form $\typ = \why{\typ_1}$,
      then since $\judg{\tctx,\typ}$ and $\judg{\tctxtwo,\lneg{\typ}}$ 
      are both conclusions of instances of the $\ruleProm$ rule,
      we have that
      $\tctx$ must be of the form $\tctx = \why{\tctx'},\ofc{\typtwo}$,
      and $\tctxtwo$ must be of the form $\tctxtwo = \why{\tctxtwo'}$,
      and the proof is symmetric to the previous case
      (\ref{flattening_cutd_ppal_prom_prom_ofc}).
    \end{enumerate}
  \item
    \label{flattening_cutd_ppal_prom_weakening}
    {\em Cut-flattening, principal case, $\ruleProm$ / $\ruleW$.}
    Then $\derivjudeg{\deriv_1}{d}{\tctx,\typ}$
    is constructed using the $\ruleProm$ rule
    and $\derivjudeg{\deriv_1}{d}{\tctxtwo,\lneg{\typ}}$
    is constructed using the $\ruleW$ rule.
    Note that, since $\lneg{\typ}$ is the principal formula of the $\ruleW$ rule,
    $\lneg{\typ}$ must be of the form $\lneg{\typ} = \why{\typ_1}$.
    Then $\typ = \ofc{\lneg{\typ_1}}$ and,
    since $\judg{\tctx,\typ}$ is the conclusion of an instance of the
    $\ruleProm$ rule, we have that $\tctx$ is of the form $\tctx = \why{\tctx'}$.
    The situation is:
    \[
      \indrule{\ruleCutd}{
        \indrule{\ruleProm}{
          \derivfrom{\deriv'_1}{\judeg{d}{\why{\tctx'},\lneg{\typ_1}}}
        }{
          \judeg{d}{\why{\tctx'},\ofc{\lneg{\typ_1}}}
        }
        \indrule{\ruleW}{
          \derivfrom{\deriv'_2}{\judeg{d}{\tctxtwo}}
        }{
          \judeg{d}{\tctxtwo,\why{\typ_1}}
        }
      }{
        \judg{\why{\tctx'},\tctxtwo}
      }
    \]
    Hence we can take:
    \[
      \indrule{\ruleWs}{ 
        \derivfrom{\deriv'_2}{\judeg{d}{\tctxtwo}}
      }{
        \judeg{d}{\why{\tctx'},\tctxtwo}
      }
    \]
  \item
    \label{flattening_cutd_ppal_prom_contraction}
    {\em Cut-flattening, principal case, $\ruleProm$ / $\ruleC$.}
    Then $\derivjudeg{\deriv_1}{d}{\tctx,\typ}$ is constructed using the $\ruleProm$ rule
    and $\derivjudeg{\deriv_2}{d}{\tctxtwo,\lneg{\typ}}$ is constructed using the $\ruleC$ rule.
    Note that, since $\lneg{\typ}$ is the principal formula of the $\ruleC$ rule,
    $\lneg{\typ}$ must be of the form $\lneg{\typ} = \why{\typ_1}$.
    Then $\typ = \ofc{\lneg{\typ_1}}$ and, since $\judg{\tctx,\typ}$
    is the conclusion of an instance of the $\ruleProm$ rule,
    we have that $\tctx$ is of the form $\tctx = \why{\tctx'}$.
    The situation is:
    \[
      \indrule{\ruleCutd}{
        \indrule{\ruleProm}{
          \derivfrom{\deriv'_1}{\judeg{d}{\why{\tctx'},\lneg{\typ_1}}}
        }{
          \judeg{d}{\why{\tctx'},\ofc{\lneg{\typ_1}}}
        }
        \indrule{\ruleC}{
          \derivfrom{\deriv'_2}{\judeg{d}{\tctxtwo,\why{\typ_1},\why{\typ_1}}}
        }{
          \judeg{d}{\tctxtwo,\why{\typ_1}}
        }
      }{
        \judg{\why{\tctx'},\tctxtwo}
      }
    \]
    Resorting to the \ih, we can take:
    \[
      \indrule{\ruleMixd}{
        \indrule{\ruleProm}{
          \derivfrom{\deriv'_1}{\judeg{d}{\why{\tctx'},\lneg{\typ_1}}}
        }{
          \judeg{d}{\why{\tctx'},\ofc{\lneg{\typ_1}}}
        }
        \derivfrom{\deriv'_2}{\judeg{d}{\tctxtwo,\rep{2}{\why{\typ_1}}}}
      }{
        \judeg{d}{\why{\tctx'},\tctxtwo}
      }
    \]
    To apply the \ih, note that sum of the heights of the subderivations decreases.
  \item
    \label{flattening_cutd_ppal_prom_dereliction}
    {\em Cut-flattening, principal case, $\ruleProm$ / $\ruleD$.}
    Then $\derivjudeg{\deriv_1}{d}{\tctx,\typ}$ is constructed using the $\ruleProm$ rule
    and $\derivjudeg{\deriv_2}{d}{\tctxtwo,\lneg{\typ}}$ is constructed using the $\ruleD$ rule.
    Note that, since $\lneg{\typ}$ is the principal formula of the $\ruleD$ rule,
    $\lneg{\typ}$ must be of the form $\lneg{\typ} = \wush{\typ_1}$.
    Then $\typ = \osha{\lneg{\typ_1}}$ and, since $\judg{\tctx,\typ}$ is the
    conclusion of an instance of the $\ruleProm$ rule, we have that $\tctx$
    is of the form $\tctx = \why{\tctx'}$. The situation is:
    \[
      \indrule{\ruleCutd}{
        \indrule{\ruleProm}{
          \derivfrom{\deriv'_1}{\judeg{d}{\why{\tctx'},\sha{\lneg{\typ_1}}}}
        }{
          \judeg{d}{\why{\tctx'},\osha{\lneg{\typ_1}}}
        }
        \indrule{\ruleD}{
          \derivfrom{\deriv'_2}{\judeg{d}{\tctxtwo,\ush{\typ_1}}}
        }{
          \judeg{d}{\tctxtwo,\wush{\typ_1}}
        }
      }{
        \judg{\why{\tctx'},\tctxtwo}
      }
    \]
    Then we can take:
    \[
      \indrule{\ruleCut}{
        \derivfrom{\deriv'_1}{\judeg{d}{\why{\tctx'},\sha{\lneg{\typ_1}}}}
        \derivfrom{\deriv'_2}{\judeg{d}{\tctxtwo,\ush{\typ_1}}}
      }{
        \judeg{d}{\why{\tctx'},\tctxtwo}
      }
    \]
    Note that we do not need to apply the \ih,
    as the size of the eliminated formula strictly decreases.
  \item
    \label{flattening_cutd_ppal_sha_ush}
    {\em Cut-flattening, principal case, $\ruleSha$ / $\ruleUsh$.}
    Then $\derivjudeg{\deriv_1}{d}{\tctx,\typ}$ is constructed using the $\ruleSha$ rule,
    and $\derivjudeg{\deriv_2}{d}{\tctxtwo,\lneg{\typ}}$ is constructed using the $\ruleUsh$ rule.
    Since $\typ$ is the principal formula of the $\ruleSha$ rule
    we have that $\typ$ must be of the form $\typ = \sha{\typ_1}$
    and $\lneg{\typ} = \ush{\lneg{\typ_1}}$.
    The situation is:
    \[
      \indrule{\ruleCutd}{
        \indrule{\ruleSha}{
          \derivfrom{\deriv'_1}{\judeg{d}{\tctx,\typ_1}}
        }{
          \judeg{d}{\tctx,\sha{\typ_1}}
        }
        \indrule{\ruleUsh}{
          \derivfrom{\deriv'_2}{\judeg{d}{\tctxtwo,\lneg{\typ_1}}}
        }{
          \judeg{d}{\tctxtwo,\ush{\lneg{\typ_1}}}
        }
      }{
        \judg{\tctx,\tctxtwo}
      }
    \]
    Then we can take:
    \[
      \indrule{\ruleCut}{
        \derivfrom{\deriv'_1}{\judeg{d}{\tctx,\typ_1}}
        \derivfrom{\deriv'_2}{\judeg{d}{\tctxtwo,\lneg{\typ_1}}}
      }{
        \judeg{d}{\tctx,\tctxtwo}
      }
    \]
    Note that we do not need to apply the \ih,
    as the size of the eliminated formula strictly decreases.
  \item
    {\em Cut-flattening, remaining principal cases.}
    The remaining principal cases are symmetric to already covered cases:
    the principal $\ruleParr$/$\ruleTensor$ case
    is symmetric to case \ref{flattening_cutd_ppal_tensor_parr};
    the principal $\ruleW$/$\ruleProm$ case
    is symmetric to case \ref{flattening_cutd_ppal_prom_weakening};
    the principal $\ruleC$/$\ruleProm$ case
    is symmetric to case \ref{flattening_cutd_ppal_prom_contraction};
    the principal $\ruleD$/$\ruleProm$ case
    is symmetric to case \ref{flattening_cutd_ppal_prom_dereliction};
    and
    the principal $\ruleUsh$/$\ruleSha$ case
    is symmetric to case \ref{flattening_cutd_ppal_sha_ush}.
  \item
    {\em Mix-flattening, left-commutative case, $\ruleAx$.}
    Then $\derivjudeg{\deriv_1}{d}{\tctx,\ofc{\typ}}$
    is constructed using the $\ruleAx$ rule,
    so $\tctx = \why{\lneg{\typ}}$.
    The situation is:
    \[
      \indrule{\ruleMixd}{
        \indrule{\ruleAx}{
          \emptyPremise
        }{
          \judeg{d}{\why{\lneg{\typ}},\ofc{\typ}}
        }
        \derivfrom{\deriv_2}{\judeg{d}{\tctxtwo,\rep{n}{\why{\lneg{\typ}}}}}
      }{
        \judg{\why{\lneg{\typ}},\tctxtwo}
      }
    \]
    Then we can take:
    \[
      \indrule{\ruleCn}{
        \derivfrom{\deriv_2}{\judeg{d}{\tctxtwo,\rep{n}{\why{\lneg{\typ}}}}}
      }{
        \judeg{d}{\why{\lneg{\typ}},\tctxtwo}
      }
    \]
  \item
    {\em Mix-flattening, right-commutative case, $\ruleAx$.}
    Then $\derivjudeg{\deriv_2}{d}{\tctxtwo,\rep{n}{\why{\lneg{\typ}}}}$
    is constructed using the $\ruleAx$ rule,
    so the number of formulae in $\tctxtwo,\rep{n}{\why{\lneg{\typ}}}$ is exactly $2$.
    In particular, $n \leq 2$.
    Moreover, $n$ cannot be exactly $2$ because the conclusion
    of the $\ruleAx$ rule cannot be of the form $\why{\lneg{\typ}},\why{\lneg{\typ}}$.
    Therefore $n$ is either $0$ or $1$, and we consider two subcases:
    \begin{enumerate}
    \item
      \label{flattening_mixf_right_ax_zero}
      If $n = 0$, then $\tctxtwo = \typtwo,\lneg{\typtwo}$ and the situation is:
      \[
        \indrule{\ruleMixd}{
          \derivfrom{\deriv_1}{\judeg{d}{\tctx,\ofc{\typ}}}
          \indrule{\ruleAx}{
            \emptyPremise
          }{
            \judeg{d}{\typtwo,\lneg{\typtwo}}
          }
        }{
          \judg{\tctx,\typtwo,\lneg{\typtwo}}
        }
      \]
      Then by \rlem{empty_mixd_lemma} we conclude that
      $\judeg{d}{\tctx,\typtwo,\lneg{\typtwo}}$.
    \item
      If $n = 1$, then $\tctxtwo = \ofc{\typ}$ and the situation is:
      \[
        \indrule{\ruleMixd}{
          \derivfrom{\deriv_1}{\judeg{d}{\tctx,\ofc{\typ}}}
          \indrule{\ruleAx}{
            \emptyPremise
          }{
            \judeg{d}{\ofc{\typ},\why{\lneg{\typ}}}
          }
        }{
          \judg{\tctx,\ofc{\typ}}
        }
      \]
      So we can take $\deriv_1$ itself.
    \end{enumerate}
  \item
    {\em Mix-flattening, left-commutative case, $\ruleCut$.}
    Then $\derivjudeg{\deriv_1}{d}{\tctx,\ofc{\typ}}$ is constructed using $\ruleCut$.
    We consider two subcases, depending on whether the eliminated
    formula $\ofc{\typ}$ occurs in the first premise of the $\ruleCut$
    rule or in the second premise of the $\ruleCut$ rule:
    \begin{enumerate}
    \item
      \label{flattening_mixd_left_cut_left}
      If $\ofc{\typ}$ occurs in the first premise of the $\ruleCut$
      rule then $\tctx$ must be of the form $\tctx_1,\tctx_2$ and
      the situation is as follows, where $\sz{\typtwo} \leq d$:
      \[
        \indrule{\ruleMixd}{
          \indrule{\ruleCut}{
            \derivfrom{\deriv_{11}}{\judeg{d}{\tctx_1,\ofc{\typ},\typtwo}}
            \derivfrom{\deriv_{12}}{\judeg{d}{\tctx_2,\lneg{\typtwo}}}
          }{
            \judeg{d}{\tctx_1,\tctx_2,\ofc{\typ}}
          }
          \derivfrom{\deriv_2}{\judeg{d}{\tctxtwo,\rep{n}{\why{\lneg{\typ}}}}}
        }{
          \judg{\tctx_1,\tctx_2,\tctxtwo}
        }
      \]
      Resorting to the \ih, we can take:
      \[
        \indrule{\ruleCut}{
          \indrule{\ruleMixd}{
            \derivfrom{\deriv_{11}}{\judeg{d}{\tctx_1,\typtwo,\ofc{\typ}}}
            \derivfrom{\deriv_2}{\judeg{d}{\tctxtwo,\rep{n}{\why{\lneg{\typ}}}}}
          }{
            \judeg{d}{\tctx_1,\tctxtwo,\typtwo}
          }
          \derivfrom{\deriv_{12}}{\judeg{d}{\tctx_2,\lneg{\typtwo}}}
        }{
          \judeg{d}{\tctx_1,\tctx_2,\tctxtwo}
        }
      \]
      To apply the \ih, note that sum of the heights of the subderivations decreases.
    \item
      Symmetric to the previous case (\ref{flattening_mixd_left_cut_left}).
    \end{enumerate}
  \item
    {\em Mix-flattening, right-commutative case, $\ruleCut$.}
    Then $\derivjudeg{\deriv_2}{d}{\tctxtwo,\rep{n}{\why{\lneg{\typ}}}}$
    is constructed using the $\ruleCut$ rule.
    Moreover, we may assume that the instance of $\ruleMixd$
    we are constructing is left-principal,
    since otherwise it would fall into one of the left-commutative cases,
    so $\derivjudeg{\deriv_1}{d}{\tctx,\ofc{\typ}}$
    must necessarily be constructed using the $\ruleProm$ rule
    and $\tctx$ must be of the form $\tctx = \why{\tctx'}$.
    Then the situation is as follows,
    where $\tctxtwo = \tctxtwo_1,\tctxtwo_2$
    and $n = n_1 + n_2$ and $\sz{\typtwo} \leq d$:
    \[
      \indrule{\ruleMixd}{
        \indrule{\ruleProm}{
          \derivfrom{\deriv'_1}{\judeg{d}{\why{\tctx'},\typ}}
        }{
          \judeg{d}{\why{\tctx'},\ofc{\typ}}
        }
        \indrule{\ruleCut}{
          \derivfrom{\deriv_{21}}{\judeg{d}{\tctxtwo_1,\rep{n_1}{\why{\lneg{\typ}}},\typtwo}}
          \derivfrom{\deriv_{22}}{\judeg{d}{\tctxtwo_2,\rep{n_2}{\why{\lneg{\typ}}},\lneg{\typtwo}}}
        }{
          \judeg{d}{\tctxtwo_1,\tctxtwo_2,\rep{n}{\why{\lneg{\typ}}}}
        }
      }{
        \judg{\why{\tctx'},\tctxtwo_1,\tctxtwo_2}
      }
    \]
    Resorting to the \ih, we can take:
    \[
      \indrule{\ruleCs}{
        \indrule{\ruleCut}{
          \derivfrom{\derivtwo_1}{\judeg{d}{\why{\tctx'},\tctxtwo_1,\typtwo}}
          \derivfrom{\derivtwo_2}{\judeg{d}{\why{\tctx'},\tctxtwo_2,\lneg{\typtwo}}}
        }{
          \judeg{d}{\why{\tctx'},\why{\tctx'},\tctxtwo_1,\tctxtwo_2}
        }
      }{
        \judeg{d}{\why{\tctx'},\tctxtwo_1,\tctxtwo_2}
      }
    \]
    where:
    \[
      \begin{array}{rcl}
      \derivtwo_1 & \eqdef &
        \derp{\indrule{\ruleMixd}{
          \indrule{\ruleProm}{
            \derivfrom{\deriv'_1}{\judeg{d}{\why{\tctx'},\typ}}
          }{
            \judeg{d}{\why{\tctx'},\ofc{\typ}}
          }
          \derivfrom{\deriv_{21}}{\judeg{d}{\tctxtwo_1,\typtwo,\rep{n_1}{\why{\lneg{\typ}}}}}
        }{
          \judeg{d}{\why{\tctx'},\tctxtwo_1,\typtwo}
        }}
      \\
      \derivtwo_2 & \eqdef &
        \derp{\indrule{\ruleMixd}{
          \indrule{\ruleProm}{
            \derivfrom{\deriv'_1}{\judeg{d}{\why{\tctx'},\typ}}
          }{
            \judeg{d}{\why{\tctx'},\ofc{\typ}}
          }
          \derivfrom{\deriv_{22}}{\judeg{d}{\tctxtwo_2,\rep{n_2}{\why{\lneg{\typ}}},\lneg{\typtwo}}}
        }{
          \judeg{d}{\why{\tctx'},\tctxtwo_2,\lneg{\typtwo}}
        }}
      \end{array}
    \]
    To apply the \ih, note that sum of the heights of the subderivations decreases.
  \item
    {\em Mix-flattening, left-commutative case, $\ruleMix$.}
    Then $\derivjudeg{\deriv_1}{d}{\tctx,\ofc{\typ}}$ is constructed using the
    $\ruleMix$ rule. We consider two subcases, depending on whether the
    eliminated formula $\ofc{\typ}$ occurs
    in the first premise of the $\ruleMix$ rule,
    or in the second premise of the $\ruleMix$ rule:
    \begin{enumerate}
    \item
      If $\ofc{\typ}$ occurs in the first premise of the $\ruleMix$ rule
      then $\tctx$ must be of the form $\tctx_1,\tctx_2$ and the situation
      is as follows, where $\sz{\ofc{\typtwo}} \leq d$ and $m \geq 0$:
      \[
        \indrule{\ruleMixd}{
          \indrule{\ruleMix}{
            \derivfrom{\deriv_{11}}{\judeg{d}{\tctx_1,\ofc{\typ},\ofc{\typtwo}}}
            \derivfrom{\deriv_{12}}{\judeg{d}{\tctx_2,\rep{m}{\why{\lneg{\typtwo}}}}}
          }{
            \judeg{d}{\tctx_1,\tctx_2,\ofc{\typ}}
          }
          \derivfrom{\deriv_2}{\judeg{d}{\tctxtwo,\rep{n}{\why{\lnot{\typ}}}}}
        }{
          \judg{\tctx_1,\tctx_2,\tctxtwo}
        }
      \]
      Resorting to the \ih, we can take:
      \[
        \indrule{\ruleMix}{
          \indrule{\ruleMixd}{
            \derivfrom{\deriv_{11}}{\judeg{d}{\tctx_1,\ofc{\typtwo},\ofc{\typ}}}
            \derivfrom{\deriv_2}{\judeg{d}{\tctxtwo,\rep{n}{\why{\lneg{\typ}}}}}
          }{
            \judeg{d}{\tctx_1,\tctxtwo,\ofc{\typtwo}}
          }
          \derivfrom{\deriv_{12}}{\judeg{d}{\tctx_2,\rep{m}{\why{\lneg{\typtwo}}}}}
        }{
          \judeg{d}{\tctx_1,\tctx_2,\tctxtwo}
        }
      \]
      To apply the \ih, note that sum of the heights of the subderivations decreases.
    \item
      If $\ofc{\typ}$ occurs in the second premise of the $\ruleMix$ rule
      then $\tctx$ must be of the form $\tctx_1,\tctx_2$ and
      is as follows, where $\sz{\ofc{\typtwo}} \leq d$ and $m \geq 0$:
      \[
        \indrule{\ruleMixd}{
          \indrule{\ruleMix}{
            \derivfrom{\deriv_{11}}{\judeg{d}{\tctx_1,\ofc{\typtwo}}}
            \derivfrom{\deriv_{12}}{\judeg{d}{\tctx_2,\ofc{\typ},\rep{m}{\why{\lneg{\typtwo}}}}}
          }{
            \judeg{d}{\tctx_1,\tctx_2,\ofc{\typ}}
          }
          \derivfrom{\deriv_2}{\judeg{d}{\tctxtwo,\rep{n}{\why{\lneg{\typ}}}}}
        }{
          \judg{\tctx_1,\tctx_2,\tctxtwo}
        }
      \]
      Resorting to the \ih, we can take:
      \[
        \indrule{\ruleMix}{
          \derivfrom{\deriv_{11}}{\judeg{d}{\tctx_1,\ofc{\typtwo}}}
          \indrule{\ruleMixd}{
            \derivfrom{\deriv_{12}}{\judeg{d}{\tctx_2,\rep{m}{\why{\lneg{\typtwo}}},\ofc{\typ}}}
            \derivfrom{\deriv_2}{\judeg{d}{\tctxtwo,\rep{n}{\why{\lneg{\typ}}}}}
          }{
            \judeg{d}{\tctx_2,\tctxtwo,\rep{m}{\why{\lneg{\typtwo}}}}
          }
        }{
          \judeg{d}{\tctx_1,\tctx_2,\tctxtwo}
        }
      \]
      To apply the \ih, note that sum of the heights of the subderivations decreases.
    \end{enumerate}
  \item
    {\em Mix-flattening, right-commutative case, $\ruleMix$.}
    Then $\derivjudeg{\deriv_2}{d}{\tctxtwo,\rep{n}{\why{\lneg{\typ}}}}$ is
    constructed using the $\ruleMix$ rule. Moreover, we may assume that
    the instance of $\ruleMixd$ we are constructing is left-principal,
    since otherwise it would fall into one of the left-commutative cases,
    so $\derivjudeg{\deriv_1}{d}{\tctx,\ofc{\typ}}$ must necessarily be constructed
    using the $\ruleProm$ rule and $\tctx$ must be of the form
    $\tctx = \why{\tctx'}$. Then the situation is as follows, where
    $\tctxtwo = \tctxtwo_1,\tctxtwo_2$ and $n = n_1 + n_2$
    and $\sz{\ofc{\typtwo}} \leq d$ and $m \geq 0$:
    \[
      \indrule{\ruleMixd}{
        \indrule{\ruleProm}{
          \derivfrom{\deriv'_1}{\judeg{d}{\why{\tctx'},\typ}}
        }{
          \judeg{d}{\why{\tctx'},\ofc{\typ}}
        }
        \indrule{\ruleMix}{
          \derivfrom{\deriv_{21}}{\judeg{d}{\tctxtwo_1,\rep{n_1}{\why{\lneg{\typ}}},\ofc{\typtwo}}}
          \derivfrom{\deriv_{22}}{\judeg{d}{\tctxtwo_2,\rep{n_2}{\why{\lneg{\typ}}},\rep{m}{\why{\lneg{\typtwo}}}}}
        }{
          \judeg{d}{\tctxtwo_1,\tctxtwo_2,\rep{n}{\why{\lneg{\typ}}}}
        }
      }{
        \judg{\why{\tctx'},\tctxtwo_1,\tctxtwo_2}
      }
    \]
    Resorting to the \ih, we can take:
    \[
      \indrule{\ruleCs}{
        \indrule{\ruleMix}{
          \derivfrom{\derivtwo_1}{\judeg{d}{\why{\tctx'},\tctxtwo_1,\ofc{\typtwo}}}
          \derivfrom{\derivtwo_2}{\judeg{d}{\why{\tctx'},\tctxtwo_2,\rep{m}{\why{\lneg{\typtwo}}}}}
        }{
          \judeg{d}{\why{\tctx'},\why{\tctx'},\tctxtwo_1,\tctxtwo_2}
        }
      }{
        \judeg{d}{\why{\tctx'},\tctxtwo_1,\tctxtwo_2}
      }
    \]
    where:
    \[
      \begin{array}{rcl}
        \derivtwo_1 & \eqdef &
          \derp{\indrule{\ruleMixd}{
            \indrule{\ruleProm}{
              \derivfrom{\deriv'_1}{\judeg{d}{\why{\tctx'},\typ}}
            }{
              \judeg{d}{\why{\tctx'},\ofc{\typ}}
            }
            \derivfrom{\deriv_{21}}{\judeg{d}{\tctxtwo_1,\ofc{\typtwo},\rep{n_1}{\why{\lneg{\typ}}}}}
          }{
            \judeg{d}{\why{\tctx'},\tctxtwo_1,\ofc{\typtwo}}
          }}
      \\
        \derivtwo_2 & \eqdef &
          \derp{\indrule{\ruleMixd}{
            \indrule{\ruleProm}{
              \derivfrom{\deriv'_1}{\judeg{d}{\why{\tctx'},\typ}}
            }{
              \judeg{d}{\why{\tctx'},\ofc{\typ}}
            }
            \derivfrom{\deriv_{22}}{\judeg{d}{\tctxtwo_2,\rep{n_2}{\why{\lneg{\typ}}},\rep{m}{\why{\lneg{\typtwo}}}}}
          }{
            \judeg{d}{\why{\tctx'},\tctxtwo_2,\rep{m}{\why{\lneg{\typtwo}}}}
          }}
      \end{array}
    \]
    To apply the \ih, note that sum of the heights of the subderivations decreases.
  \item
    {\em Mix-flattening, left-commutative case, $\ruleTensor$.}
    Then $\derivjudeg{\deriv_1}{d}{\tctx,\ofc{\typ}}$ is constructed
    using the $\ruleTensor$ rule.
    We consider two subcases, depending on whether the eliminated formula
    $\ofc{\typ}$ occurs in the first premise of the $\ruleTensor$ rule or
    in the second premise of the $\ruleTensor$ rule:
    \begin{enumerate}
    \item
      \label{flattening_mixd_left_tensor_left}
      If $\ofc{\typ}$ occurs in the first premise of the $\ruleTensor$ rule,
      then $\tctx$ must be of the form $\tctx_1,\tctx_2,\typtwo\tensor\typthree$
      and the situation is as follows:
      \[
        \indrule{\ruleMixd}{
          \indrule{\ruleTensor}{
            \derivfrom{\deriv_{11}}{\judeg{d}{\tctx_1,\typtwo,\ofc{\typ}}}
            \derivfrom{\deriv_{12}}{\judeg{d}{\tctx_2,\typthree}}
          }{
            \judeg{d}{\tctx_1,\tctx_2,\typtwo\tensor\typthree,\ofc{\typ}}
          }
          \derivfrom{\deriv_2}{\judeg{d}{\tctxtwo,\rep{n}{\why{\lneg{\typ}}}}}
        }{
          \judg{\tctx_1,\tctx_2,\typtwo\tensor\typthree,\tctxtwo}
        }
      \]
      Resorting to the \ih, we can take:
      \[
        \indrule{\ruleTensor}{
          \indrule{\ruleMixd}{
            \derivfrom{\deriv_{11}}{\judeg{d}{\tctx_1,\typtwo,\ofc{\typ}}}
            \derivfrom{\deriv_2}{\judeg{d}{\tctxtwo,\rep{n}{\why{\lneg{\typ}}}}}
          }{
            \judeg{d}{\tctx_1,\typtwo,\tctxtwo}
          }
          \derivfrom{\deriv_{12}}{\judeg{d}{\tctx_2,\typthree}}
        }{
          \judeg{d}{\tctx_1,\tctx_2,\typtwo\tensor\typthree,\tctxtwo}
        }
      \]
      To apply the \ih, note that sum of the heights of the subderivations decreases.
    \item
      Symmetric to the previous case (\ref{flattening_mixd_left_tensor_left}).
    \end{enumerate}
  \item
    {\em Mix-flattening, right-commutative case, $\ruleTensor$.}
    Then $\derivjudeg{\deriv_2}{d}{\tctxtwo,\rep{n}{\why{\lneg{\typ}}}}$
    is constructed using the $\ruleTensor$ rule.
    Moreover, we may assume that the instance of $\ruleMixd$ we are constructing
    is left-principal, since otherwise it would fall into one of the left-commutative
    cases, so $\derivjudeg{\deriv_1}{d}{\tctx,\ofc{\typ}}$ must necessarily be
    constructed using the $\ruleProm$ rule and $\tctx$ must be of the form
    $\tctx = \why{\tctx'}$. Then the situation is as follows, where
    $\tctxtwo = \tctxtwo_1,\tctxtwo_2,\typtwo\tensor\typthree$
    and $n = n_1 + n_2$:
    \[
      \indrule{\ruleMixd}{
        \indrule{\ruleProm}{
          \derivfrom{\deriv'_1}{\judeg{d}{\why{\tctx},\typ}}
        }{
          \judeg{d}{\why{\tctx},\ofc{\typ}}
        }
        \indrule{\ruleTensor}{
          \derivfrom{\deriv_{21}}{\judeg{d}{\tctxtwo_1,\typtwo,\rep{n_1}{\why{\lneg{\typ}}}}}
          \derivfrom{\deriv_{22}}{\judeg{d}{\tctxtwo_2,\typthree,\rep{n_2}{\why{\lneg{\typ}}}}}
        }{
          \judeg{d}{\tctxtwo_1,\tctxtwo_2,\typtwo\tensor\typthree,\rep{n}{\why{\lneg{\typ}}}}
        }
      }{
        \judg{\why{\tctx},\tctxtwo_1,\tctxtwo_2,\typtwo\tensor\typthree}
      }
    \]
    Resorting to the \ih, we can take:
    \[
      \indrule{\ruleCs}{
        \indrule{\ruleTensor}{
          \derivfrom{\derivtwo_1}{\judeg{d}{\why{\tctx},\tctxtwo_1,\typtwo}}
          \derivfrom{\derivtwo_2}{\judeg{d}{\why{\tctx},\tctxtwo_2,\typthree}}
        }{
          \judeg{d}{\why{\tctx},\why{\tctx},\tctxtwo_1,\tctxtwo_2,\typtwo\tensor\typthree}
        }
      }{
        \judeg{d}{\why{\tctx},\tctxtwo_1,\tctxtwo_2,\typtwo\tensor\typthree}
      }
    \]
    where:
    \[
      \begin{array}{rcl}
        \derivtwo_1 & \eqdef &
          \derp{\indrule{\ruleMixd}{
            \indrule{\ruleProm}{
              \derivfrom{\deriv'_1}{\judeg{d}{\why{\tctx},\typ}}
            }{
              \judeg{d}{\why{\tctx},\ofc{\typ}}
            }
            \derivfrom{\deriv_{21}}{\judeg{d}{\tctxtwo_1,\typtwo,\rep{n_1}{\why{\lneg{\typ}}}}}
          }{
            \judeg{d}{\why{\tctx},\tctxtwo_1,\typtwo}
          }}
      \\
        \derivtwo_2 & \eqdef &
          \derp{\indrule{\ruleMixd}{
            \indrule{\ruleProm}{
              \derivfrom{\deriv'_1}{\judeg{d}{\why{\tctx},\typ}}
            }{
              \judeg{d}{\why{\tctx},\ofc{\typ}}
            }
            \derivfrom{\deriv_{22}}{\judeg{d}{\tctxtwo_2,\typthree,\rep{n_2}{\why{\lneg{\typ}}}}}
          }{
            \judeg{d}{\why{\tctx},\tctxtwo_2,\typthree}
          }}
      \end{array}
    \]
    To apply the \ih, note that sum of the heights of the subderivations decreases.
  \item
    {\em Mix-flattening, left-commutative case, $\ruleParr$.}
    Then $\derivjudeg{\deriv_1}{d}{\tctx,\ofc{\typ}}$ is constructed using the $\ruleParr$ rule.
    Then $\tctx$ must be of the form $\tctx = \tctx',\typtwo\parr\typthree$
    and the situation is:
    \[
      \indrule{\ruleMixd}{
        \indrule{\ruleParr}{
          \derivfrom{\deriv'_1}{\judeg{d}{\tctx',\typtwo,\typthree,\ofc{\typ}}}
        }{
          \judeg{d}{\tctx',\typtwo\parr\typthree,\ofc{\typ}}
        }
        \derivfrom{\deriv_2}{\judeg{d}{\tctxtwo,\rep{n}{\why{\lneg{\typ}}}}}
      }{
        \judg{\tctx',\typtwo\parr\typthree,\tctxtwo}
      }
    \]
    Resorting to the \ih, we can take:
    \[
      \indrule{\ruleParr}{
        \indrule{\ruleMixd}{
          \derivfrom{\deriv'_1}{\judeg{d}{\tctx',\typtwo,\typthree,\ofc{\typ}}}
          \derivfrom{\deriv_2}{\judeg{d}{\tctxtwo,\rep{n}{\why{\lneg{\typ}}}}}
        }{
          \judeg{d}{\tctx',\typtwo,\typthree,\tctxtwo}
        }
      }{
        \judeg{d}{\tctx',\typtwo\parr\typthree,\tctxtwo}
      }
    \]
    To apply the \ih, note that sum of the heights of the subderivations decreases.
  \item
    {\em Mix-flattening, right-commutative case, $\ruleParr$.}
    Then $\derivjudeg{\deriv_2}{d}{\tctxtwo,\rep{n}{\why{\lneg{\typ}}}}$
    is constructed using the $\ruleParr$ rule.
    Then $\tctxtwo$ must be of the form $\tctxtwo = \tctxtwo',\typtwo\parr\typthree$
    and the situation is:
    \[
      \indrule{\ruleMixd}{
        \derivfrom{\deriv_1}{\judeg{d}{\tctx,\ofc{\typ}}}
        \indrule{\ruleParr}{
          \derivfrom{\deriv'_2}{\judeg{d}{\tctxtwo',\typtwo,\typthree,\rep{n}{\why{\lneg{\typ}}}}}
        }{
          \judeg{d}{\tctxtwo',\typtwo\parr\typthree,\rep{n}{\why{\lneg{\typ}}}}
        }
      }{
        \judg{\tctx,\tctxtwo',\typtwo\parr\typthree}
      }
    \]
    Resorting to the \ih, we can take:
    \[
      \indrule{\ruleParr}{
        \indrule{\ruleMixd}{
          \derivfrom{\deriv_1}{\judeg{d}{\tctx,\ofc{\typ}}}
          \derivfrom{\deriv'_2}{\judeg{d}{\tctxtwo',\typtwo,\typthree,\rep{n}{\why{\lneg{\typ}}}}}
        }{
          \judeg{d}{\tctx,\tctxtwo',\typtwo,\typthree}
        }
      }{
        \judeg{d}{\tctx,\tctxtwo',\typtwo\parr\typthree}
      }
    \]
    To apply the \ih, note that sum of the heights of the subderivations decreases.
  \item
    \label{flattening_mixd_left_promotion}
    {\em Mix-flattening, left-commutative case, $\ruleProm$.}
    We argue that this case is impossible.
    Indeed, $\derivjudeg{\deriv_1}{\tctx,\ofc{\typ}}$ must be constructed
    using the $\ruleProm$ rule, but all the formulae in the conclusion of
    $\ruleProm$ are principal by definition.
  \item
    {\em Mix-flattening, right-commutative case, $\ruleProm$.}
    Then $\derivjudeg{\deriv_2}{d}{\tctxtwo,\rep{n}{\why{\lneg{\typ}}}}$
    must be constructed using the $\ruleProm$ rule, and it is such
    that $\why{\lneg{\typ}}$ is not a principal formula in the conclusion
    of the $\ruleProm$ rule.
    Recall that all the formulae in the conclusion of the $\ruleProm$ rule
    are principal, so this case is only possible when $n = 0$. 
    Then, since $\judg{\tctxtwo}$ is the conclusion of an instance of the
    $\ruleProm$ rule, we have that $\tctxtwo$ must be of the form
    $\tctxtwo = \why{\tctxtwo'},\ofc{\typtwo}$.
    Then the situation is as follows:
    \[
      \indrule{\ruleMixd}{
        \derivfrom{\deriv_1}{\judeg{d}{\tctx,\ofc{\typ}}}
        \indrule{\ruleProm}{
          \derivfrom{\deriv'_2}{\judeg{d}{\why{\tctxtwo'},\typtwo}}
        }{
          \judeg{d}{\why{\tctxtwo'},\ofc{\typtwo}}
        }
      }{
        \judg{\tctx,\why{\tctxtwo'},\ofc{\typtwo}}
      }
    \]
    Then by \rlem{empty_mixd_lemma} we conclude that
    $\judeg{d}{\tctx,\why{\tctxtwo'},\ofc{\typtwo}}$, as required.
  \item
    \label{flattening_mixd_left_weakening}
    {\em Mix-flattening, left-commutative case, $\ruleW$.}
    Then $\derivjudeg{\deriv_1}{d}{\tctx,\ofc{\typ}}$ must be constructed
    using the $\ruleW$ rule.
    Then $\tctx$ must be of the form $\tctx = \tctx',\why{\typtwo}$
    and the situation is:
    \[
      \indrule{\ruleMixd}{
        \indrule{\ruleW}{
          \derivfrom{\deriv'_1}{\judeg{d}{\tctx',\ofc{\typ}}}
        }{
          \judeg{d}{\tctx',\why{\typtwo},\ofc{\typ}}
        }
        \derivfrom{\deriv_2}{\judeg{d}{\tctxtwo,\rep{n}{\why{\lneg{\typ}}}}}
      }{
        \judg{\tctx',\why{\typtwo},\tctxtwo}
      }
    \]
    Resorting to the \ih, we can take:
    \[
      \indrule{\ruleW}{
        \indrule{\ruleMixd}{
          \derivfrom{\deriv'_1}{\judeg{d}{\tctx',\ofc{\typ}}}
          \derivfrom{\deriv_2}{\judeg{d}{\tctxtwo,\rep{n}{\why{\lneg{\typ}}}}}
        }{
          \judeg{d}{\tctx',\tctxtwo}
        }
      }{
        \judeg{d}{\tctx',\why{\typtwo},\tctxtwo}
      }
    \]
    To apply the \ih, note that sum of the heights of the subderivations decreases.
  \item
    {\em Mix-flattening, right-commutative case, $\ruleW$.}
    Similar to the previous case (\ref{flattening_mixd_left_weakening}).
  \item
    \label{flattening_mixd_left_contraction}
    {\em Mix-flattening, left-commutative case, $\ruleC$.}
    Then $\derivjudeg{\deriv_1}{d}{\tctx,\ofc{\typ}}$ must be constructed
    using the $\ruleC$ rule.
    Then $\tctx$ must be of the form $\tctx = \tctx',\why{\typtwo}$
    and the situation is:
    \[
      \indrule{\ruleMixd}{
        \indrule{\ruleC}{
          \derivfrom{\deriv'_1}{\judeg{d}{\tctx',\why{\typtwo},\why{\typtwo},\ofc{\typ}}}
        }{
          \judeg{d}{\tctx',\why{\typtwo},\ofc{\typ}}
        }
        \derivfrom{\deriv_2}{\judeg{d}{\tctxtwo,\rep{n}{\why{\lneg{\typ}}}}}
      }{
        \judg{\tctx',\why{\typtwo},\tctxtwo}
      }
    \]
    Resorting to the \ih, we can take:
    \[
      \indrule{\ruleC}{
        \indrule{\ruleMixd}{
          \derivfrom{\deriv'_1}{\judeg{d}{\tctx',\why{\typtwo},\why{\typtwo},\ofc{\typ}}}
          \derivfrom{\deriv_2}{\judeg{d}{\tctxtwo,\rep{n}{\why{\lneg{\typ}}}}}
        }{
          \judeg{d}{\tctx',\why{\typtwo},\why{\typtwo},\tctxtwo}
        }
      }{
        \judeg{d}{\tctx',\why{\typtwo},\tctxtwo}
      }
    \]
    To apply the \ih, note that sum of the heights of the subderivations decreases.
  \item
    {\em Mix-flattening, right-commutative case, $\ruleC$.}
    Similar to the previous case (\ref{flattening_mixd_left_contraction}).
  \item
    \label{flattening_mixd_left_dereliction}
    {\em Mix-flattening, left-commutative case, $\ruleD$.}
    Then $\derivjudeg{\deriv_1}{d}{\tctx,\ofc{\typ}}$ must be constructed
    using the $\ruleD$ rule.
    Then $\tctx$ must be of the form $\tctx = \tctx',\wush{\typtwo}$
    and the situation is:
    \[
      \indrule{\ruleMixd}{
        \indrule{\ruleD}{
          \derivfrom{\deriv'_1}{\judeg{d}{\tctx',\ush{\typtwo},\ofc{\typ}}}
        }{
          \judeg{d}{\tctx',\wush{\typtwo},\ofc{\typ}}
        }
        \derivfrom{\deriv_2}{\judeg{d}{\tctxtwo,\rep{n}{\why{\lneg{\typ}}}}}
      }{
        \judg{\tctx',\wush{\typtwo},\tctxtwo}
      }
    \]
    Resorting to the \ih, we can take:
    \[
      \indrule{\ruleD}{
        \indrule{\ruleMixd}{
          \derivfrom{\deriv'_1}{\judeg{d}{\tctx',\ush{\typtwo},\ofc{\typ}}}
          \derivfrom{\deriv_2}{\judeg{d}{\tctxtwo,\rep{n}{\why{\lneg{\typ}}}}}
        }{
          \judeg{d}{\tctx',\ush{\typtwo},\tctxtwo}
        }
      }{
        \judeg{d}{\tctx',\wush{\typtwo},\tctxtwo}
      }
    \]
    To apply the \ih, note that sum of the heights of the subderivations decreases.
  \item
    {\em Mix-flattening, right-commutative case, $\ruleD$.}
    Similar to the previous case (\ref{flattening_mixd_left_dereliction}).
  \item
    {\em Mix-flattening, left-commutative case, $\ruleSha$.}
    Similar to case \ref{flattening_mixd_left_dereliction}.
  \item
    {\em Mix-flattening, right-commutative case, $\ruleSha$.}
    Similar to case \ref{flattening_mixd_left_dereliction}.
  \item
    {\em Mix-flattening, left-commutative case, $\ruleUsh$.}
    Similar to case \ref{flattening_mixd_left_dereliction}.
  \item
    {\em Mix-flattening, right-commutative case, $\ruleUsh$.}
    Similar to case \ref{flattening_mixd_left_dereliction}.
  \item
    {\em Mix-flattening, principal case, $\ruleProm$ / $\ruleProm$.}
    Then $\derivjudeg{\deriv_1}{d}{\tctx,\ofc{\typ}}$ and
    $\derivjudeg{\deriv_2}{d}{\tctxtwo,\rep{n}{\why{\lneg{\typ}}}}$
    are both constructed using the $\ruleProm$ rule.
    Since $\judg{\tctx,\ofc{\typ}}$ is the conclusion of an instance of
    the $\ruleProm$ rule, $\tctx$ must be of the form $\tctx = \why{\tctx'}$.
    Moreover, since $\judg{\tctxtwo,\rep{n}{\why{\lneg{\typ}}}}$ is the conclusion
    of an instance of the $\ruleProm$ rule, with $\why{\lneg{\typ}}$
    as a principal formula, $\tctxtwo$ must be of the form
    $\tctxtwo = \why{\tctxtwo'},\ofc{\typtwo}$.
    If $n = 0$, it suffices to resort to \rlem{empty_mixd_lemma} to conclude that
    $\judeg{d}{\tctx,\why{\tctxtwo'},\ofc{\typtwo}}$.
    If $n > 0$, the situation is:
    \[
      \indrule{\ruleMixd}{
        \indrule{\ruleProm}{
          \derivfrom{\deriv'_1}{\judeg{d}{\why{\tctx'},\typ}}
        }{
          \judeg{d}{\why{\tctx'},\ofc{\typ}}
        }
        \indrule{\ruleProm}{
          \derivfrom{\deriv'_2}{\judeg{d}{\why{\tctxtwo'},\typtwo,\rep{n}{\why{\lneg{\typ}}}}}
        }{
          \judeg{d}{\why{\tctxtwo'},\ofc{\typtwo},\rep{n}{\why{\lneg{\typ}}}}
        }
      }{
        \judg{\why{\tctx'},\why{\tctxtwo'},\ofc{\typtwo}}
      }
    \]
    Resorting to the \ih, we can take:
    \[
      \indrule{\ruleProm}{
        \indrule{\ruleMixd}{
          \indrule{\ruleProm}{
            \derivfrom{\deriv'_1}{\judeg{d}{\why{\tctx'},\typ}}
          }{
            \judeg{d}{\why{\tctx'},\ofc{\typ}}
          }
          \derivfrom{\deriv'_2}{\judeg{d}{\why{\tctxtwo'},\typtwo,\rep{n}{\why{\lneg{\typ}}}}}
        }{
          \judeg{d}{\why{\tctx'},\why{\tctxtwo'},\typtwo}
        }
      }{
        \judeg{d}{\why{\tctx'},\why{\tctxtwo'},\ofc{\typtwo}}
      }
    \]
    To apply the \ih, note that sum of the heights of the subderivations decreases.
  \item
    {\em Mix-flattening, principal case, $\ruleProm$ / $\ruleW$.}
    Then $\derivjudeg{\deriv_1}{d}{\tctx,\ofc{\typ}}$ is constructed using the
    $\ruleProm$ rule and $\derivjudeg{\deriv_2}{d}{\tctxtwo,\rep{n}{\why{\lneg{\typ}}}}$
    is constructed using the $\ruleW$ rule.
    Since $\judg{\tctx,\ofc{\typ}}$ is the conclusion of an instance of the
    $\ruleProm$ rule, $\tctx$ must be of the form $\tctx = \why{\tctx'}$.
    If $n = 0$, it suffices to resort to \rlem{empty_mixd_lemma} to conclude that
    $\judeg{d}{\why{\tctx'},\tctxtwo}$.
    If $n > 0$, the situation is:
    \[
      \indrule{\ruleMixd}{
        \indrule{\ruleProm}{
          \derivfrom{\deriv'_1}{\judeg{d}{\why{\tctx'},\typ}}
        }{
          \judeg{d}{\why{\tctx'},\ofc{\typ}}
        }
        \indrule{\ruleW}{
          \derivfrom{\deriv'_2}{\judeg{d}{\why{\tctx'},\rep{n-1}{\why{\lneg{\typ}}}}}
        }{
          \judeg{d}{\why{\tctx'},\rep{n}{\why{\lneg{\typ}}}}
        }
      }{
        \judg{\why{\tctx'},\tctxtwo}
      }
    \]
    Resorting to the \ih, we can take:
    \[
      \indrule{\ruleMixd}{
        \indrule{\ruleProm}{
          \derivfrom{\deriv'_1}{\judeg{d}{\why{\tctx'},\typ}}
        }{
          \judeg{d}{\why{\tctx'},\ofc{\typ}}
        }
        \derivfrom{\deriv'_2}{\judeg{d}{\why{\tctx'},\rep{n-1}{\why{\lneg{\typ}}}}}
      }{
        \judeg{d}{\why{\tctx'},\tctxtwo}
      }
    \]
    To apply the \ih, note that sum of the heights of the subderivations decreases.
  \item
    {\em Mix-flattening, principal case, $\ruleProm$ / $\ruleC$.}
    Then $\derivjudeg{\deriv_1}{d}{\tctx,\ofc{\typ}}$ is constructed using the
    $\ruleProm$ rule and $\derivjudeg{\deriv_2}{d}{\tctxtwo,\rep{n}{\why{\lneg{\typ}}}}$
    is constructed using the $\ruleC$ rule.
    Since $\judg{\tctx,\ofc{\typ}}$ is the conclusion of an instance of the
    $\ruleProm$ rule, $\tctx$ must be of the form $\tctx = \why{\tctx'}$.
    Then the situation is:
    \[
      \indrule{\ruleMixd}{
        \indrule{\ruleProm}{
          \derivfrom{\deriv'_1}{\judeg{d}{\why{\tctx'},\typ}}
        }{
          \judeg{d}{\why{\tctx'},\ofc{\typ}}
        }
        \indrule{\ruleC}{
          \derivfrom{\deriv'_2}{\judeg{d}{\tctxtwo,\rep{n+1}{\why{\lneg{\typ}}}}}
        }{
          \judeg{d}{\tctxtwo,\rep{n}{\why{\lneg{\typ}}}}
        }
      }{
        \judg{\why{\tctx'},\tctxtwo}
      }
    \]
    Resorting to the \ih, we can take:
    \[
      \indrule{\ruleMixd}{
        \indrule{\ruleProm}{
          \derivfrom{\deriv'_1}{\judeg{d}{\why{\tctx'},\typ}}
        }{
          \judeg{d}{\why{\tctx'},\ofc{\typ}}
        }
        \derivfrom{\deriv'_2}{\judeg{d}{\tctxtwo,\rep{n+1}{\why{\lneg{\typ}}}}}
      }{
        \judeg{d}{\why{\tctx'},\tctxtwo}
      }
    \]
    To apply the \ih, note that sum of the heights of the subderivations decreases.
  \item
    {\em Mix-flattening, principal case, $\ruleProm$ / $\ruleD$.}
    Then $\derivjudeg{\deriv_1}{d}{\tctx,\ofc{\typ}}$ is constructed using the
    $\ruleProm$ rule and $\derivjudeg{\deriv_2}{d}{\tctxtwo,\rep{n}{\why{\lneg{\typ}}}}$
    is constructed using the $\ruleD$ rule.
    Since $\judg{\tctx,\ofc{\typ}}$ is the conclusion of an instance of the
    $\ruleProm$ rule, $\tctx$ must be of the form $\tctx = \why{\tctx'}$.
    If $n = 0$, it suffices to resort to \rlem{empty_mixd_lemma} to conclude that
    $\judeg{d}{\why{\tctx'},\tctxtwo}$.
    If $n > 0$, then 
    since $\derivjudeg{\deriv_2}{d}{\tctxtwo,\rep{n}{\why{\lneg{\typ}}}}$
    is the conclusion of an instance of the $\ruleD$ rule with $\why{\lneg{\typ}}$
    as a principal formula, $\lneg{\typ}$ must be of the form
    $\lneg{\typ} = \ush{\typ_1}$, so $\typ = \sha{\lneg{\typ_1}}$.
    Then the situation is:
    \[
      \indrule{\ruleMixd}{
        \indrule{\ruleProm}{
          \derivfrom{\deriv'_1}{\judeg{d}{\why{\tctx'},\sha{\lneg{\typ_1}}}}
        }{
          \judeg{d}{\why{\tctx'},\osha{\lneg{\typ_1}}}
        }
        \indrule{\ruleD}{
          \derivfrom{\deriv'_2}{\judeg{d}{\tctxtwo,\rep{n-1}{\wush{\typ_1}},\ush{\typ_1}}}
        }{
          \judeg{d}{\tctxtwo,\rep{n-1}{\wush{\typ_1}},\wush{\typ_1}}
        }
      }{
        \judg{\why{\tctx'},\tctxtwo}
      }
    \]
    Resorting to the \ih, we can take:
    \[
      \indrule{\ruleCs}{
        \indrule{\ruleCut}{
          \derivfrom{\deriv'_1}{\judeg{d}{\why{\tctx'},\sha{\lneg{\typ_1}}}}
          \indrule{\ruleMixd}{
            \indrule{\ruleProm}{
              \derivfrom{\deriv'_1}{\judeg{d}{\why{\tctx'},\sha{\lneg{\typ_1}}}}
            }{
              \judeg{d}{\why{\tctx'},\osha{\lneg{\typ_1}}}
            }
            \derivfrom{\deriv'_2}{\judeg{d}{\tctxtwo,\rep{n-1}{\wush{\typ_1}},\ush{\typ_1}}}
          }{
            \judeg{d}{\why{\tctx'},\tctxtwo,\ush{\typ_1}}
          }
        }{
          \judeg{d}{\why{\tctx'},\why{\tctx'},\tctxtwo}
        }
      }{
        \judeg{d}{\why{\tctx'},\tctxtwo}
      }
    \]
    To apply the \ih, note that sum of the heights of the subderivations decreases.
    Moreover, note that the last instance of the $\ruleCut$ rule used in the
    derivation is between $\sha{\lneg{\typ_1}}$ and $\ush{\typ_1}$
    and $\sz{\sha{\lneg{\typ_1}}} = \ush{\typ_1} = d$.
  \item
    {\em Mix-flattening, remaining principal cases.}
    The remaining principal cases are impossible,
    since the principal formula on the left must be of the form $\ofc{\typ}$.
    This includes the cases $\ruleTensor$/$\ruleParr$, $\ruleParr$/$\ruleTensor$,
    $\ruleW$/$\ruleProm$, $\ruleC$/$\ruleProm$, $\ruleD$/$\ruleProm$,
    $\ruleSha$/$\ruleUsh$, and $\ruleUsh$/$\ruleSha$.
  \end{enumerate}
\end{ifLongAppendix}
\end{proof}

\CutElimination

\begin{proof}\label{cut_elimination:proof}
More in general, we show that
if $\judg{\tctx}$ is provable in $\MELLS{+}\ruleMix$,
then there is a derivation of $\judg{\tctx}$
without instances of the $\ruleCut$ and $\ruleMix$ rules.

Let $\deriv$ be a derivation of some depth, say $d$,
\ie $\derivjudeg{\deriv}{d}{\tctx}$.
We proceed by induction on $d$.
If $d = 0$, observe that $\deriv$
has no instances of the $\ruleCut$ and $\ruleMix$ rules.
For $d > 0$, it suffices to note that, in general,
$\judeg{d}{\tctx}$ implies $\judeg{d-1}{\tctx}$.
To see this, proceed by induction on the derivation of $\judeg{d}{\tctx}$.
Most cases are straightforward by resorting to the \ih,
except for the $\ruleCut$ and $\ruleMix$ cases,
in which it suffices to apply the cut/mix flattening lemma
(\rlem{cut_mix_flattening}) when the size of the eliminated
formula is $d$.
\end{proof}


\section{Appendix: A Sharing Linear $\lambda$-Calculus}

\subsection{Logical Soundness}

We extend the translation to environments and judgments:
\[
  \begin{array}{rcl}
    \tran{(\uvar_1:\typ_1,\hdots,\uvar_n:\typ_n)}
  & \eqdef &
    \tran{\typ_1},\hdots,\tran{\typ_n}
  \\
    \tran{(\lvar_1:\typ_1,\hdots,\lvar_n:\typ_n)}
  & \eqdef &
    \tran{\typ_1},\hdots,\tran{\typ_n}
  \\
    \tran{(\judc{\uenv}{\lenv}{\tm}{\typ})}
  & \eqdef &
    \judg{
        \wush{(\lneg{\tran{\uenv}})},
        \lneg{\tran{\lenv}},
        \tran{\typ}
      }
  \end{array}
\]
\SoundessOfLambdaSWRTMELLS

\begin{proof}\label{soundness_of_lambdaS_w_r_t_MELLS:proof}
We prove that if $\judc{\uenv}{\lenv}{\tm}{\typ}$ holds in $\lambdaS$,
then
$\judg{\wush{(\lneg{\tran{\uenv}})},\lneg{\tran{\lenv}},\tran{\typ}}$ holds in $\MELLS$
by induction on the derivation of $\judc{\uenv}{\lenv}{\tm}{\typ}$:
\begin{enumerate}
\item $\rulecLvar$:
  Let $\judc{\uenv}{\lvar:\typ}{\lvar}{\typ}$. Then:
  \[
    \indrule{\ruleWs}{
      \indrule{\ruleAx}{
        \emptyPremise
      }{
        \judg{\lneg{\tran{\typ}},\tran{\typ}}
      }
    }{
      \judg{\wush{\lneg{\tran{\uenv}}},\lneg{\tran{\typ}},\tran{\typ}}
    }
  \]
\item $\rulecUvar$:
  Let $\judc{\uenv,\uvar:\typ}{\noenv}{\uvar}{\sha{\typ}}$. Then:
  \[
    \indrule{\ruleWs}{
      \indrule{\ruleSha}{
        \indrule{\ruleD}{
          \indrule{\ruleAx}{
            \emptyPremise
          }{
            \judg{\lneg{\tran{\typ}},\tran{\typ}}
          }
        }{
          \judg{\wush{\lneg{\tran{\typ}}},\tran{\typ}}
        }
      }{
        \judg{\wush{\lneg{\tran{\typ}}},\sha{\tran{\typ}}}
      }
    }{
      \judg{\wush{\lneg{\tran{\uenv}}},\wush{\lneg{\tran{\typ}}},\sha{\tran{\typ}}}
    }
  \]
\item $\rulecAbs$:
  Let $\judc{\uenv}{\lenv}{\lam{\lvar}{\tm}}{\typ\limp\typtwo}$
  be derived from $\judc{\uenv}{\lenv,\lvar:\typ}{\tm}{\typtwo}$.
  Then:
  \[
    \indrule{\ruleParr}{
      \derivih{\judg{\wush{\lneg{\tran{\uenv}}},\lneg{\tran{\lenv}},\lneg{\tran{\typ}},\tran{\typtwo}}}
    }{
      \judg{\wush{\lneg{\tran{\uenv}}},\lneg{\tran{\lenv}},\lneg{\tran{\typ}}\parr\tran{\typtwo}}
    }
  \]
\item $\rulecApp$:
  Let $\judc{\uenv}{\lenv_1,\lenv_2}{\tm\,\tmtwo}{\typtwo}$
  be derived from $\judc{\uenv}{\lenv_1}{\tm}{\typ\limp\typtwo}$
  and $\judc{\uenv}{\lenv_2}{\tmtwo}{\typtwo}$.
  Then:
  \[
    \indrule{\ruleCs}{
      \indrule{\ruleCut}{
        \derivih{
          \judg{\wush{\lneg{\tran{\uenv}}},
                \lneg{\tran{\lenv_1}},\lneg{\tran{\typ}}\parr\tran{\typtwo}}
        }
        \HS
        \indrule{\ruleTensor}{
          \derivih{
            \judg{\wush{\lneg{\tran{\uenv}}},
                  \lneg{\tran{\lenv_2}},\tran{\typ}}
          }
          \indrule{\ruleAx}{
            \emptyPremise
          }{
            \judg{\tran{\typtwo},\lneg{\tran{\typtwo}}}
          }
        }{
          \judg{\wush{\lneg{\tran{\uenv}}},
                \lneg{\tran{\lenv_2}},\tran{\typtwo},
                \tran{\typ}\tensor\lneg{\tran{\typtwo}}}
        }
      }{
        \judg{\wush{\lneg{\tran{\uenv}}},\wush{\lneg{\tran{\uenv}}},
              \lneg{\tran{\lenv_1}},\lneg{\tran{\lenv_2}},\tran{\typtwo}}
      }
    }{
      \judg{\wush{\lneg{\tran{\uenv}}},
            \lneg{\tran{\lenv_1}},\lneg{\tran{\lenv_2}},\tran{\typtwo}}
    }
  \]
\item $\rulecSha$:
  Let $\judc{\uenv}{\lenv}{\sha{\tm}}{\sha{\typ}}$
  be derived from $\judc{\uenv}{\lenv}{\tm}{\typ}$.
  Then:
  \[
    \indrule{\ruleSha}{
      \judg{\wush{\lneg{\tran{\uenv}}},\lneg{\tran{\lenv}},\tran{\typ}}
    }{
      \judg{\wush{\lneg{\tran{\uenv}}},\lneg{\tran{\lenv}},\sha{\tran{\typ}}}
    }
  \]
\item $\rulecOpen$:
  Let $\judc{\uenv}{\lenv}{\open{\tm}}{\typ}$
  be derived from $\judc{\uenv}{\lenv}{\tm}{\sha{\typ}}$.
  Then:
  \[
    \indrule{\ruleCut}{
      \derivih{
        \judg{\wush{\lneg{\tran{\uenv}}},\lneg{\tran{\lenv}},\sha{\tran{\typ}}}
      }
      \indrule{\ruleUsh}{
        \indrule{\ruleAx}{
          \emptyPremise
        }{
          \judg{\lneg{\tran{\typ}},\tran{\typ}}
        }
      }{ 
        \judg{\ush{\lneg{\tran{\typ}}},\tran{\typ}}
      } 
    }{
      \judg{\wush{\lneg{\tran{\uenv}}},\lneg{\tran{\lenv}},\tran{\typ}}
    }
  \]
\item $\rulecProm$:
  Let $\judc{\uenv}{\noenv}{\ofc{\tm}}{\ofc{\typ}}$
  be derived from $\judc{\uenv}{\noenv}{\tm}{\typ}$.
  Then:
  \[
    \indrule{\ruleProm}{
      \derivih{\judg{\wush{\lneg{\tran{\uenv}}},\tran{\typ}}}
    }{
      \judg{\wush{\lneg{\tran{\uenv}}},\ofc{\tran{\typ}}}
    }
  \]
\item $\rulecES$:
  Let $\judc{\uenv}{\lenv_1,\lenv_2}{\tm\esub{\uvar}{\tmtwo}}{\typtwo}$
  be derived from $\judc{\uenv,\uvar:\typ}{\lenv_1}{\tm}{\typtwo}$
  and $\judc{\uenv}{\lenv_2}{\tmtwo}{\osha{\typ}}$.
  Then:
  \[
    \indrule{\ruleCs}{
      \indrule{\ruleCut}{
        \derivih{
          \judg{\wush{\lneg{\tran{\uenv}}},\wush{\lneg{\tran{\typ}}},
                \lneg{\tran{\lenv_1}},\tran{\typtwo}}
        }
        \derivih{
          \judg{\wush{\lneg{\tran{\uenv}}},\lneg{\tran{\lenv_2}},\osha{\tran{\typ}}}
        }
      }{
        \judg{\wush{\lneg{\tran{\uenv}}},\wush{\lneg{\tran{\uenv}}},
              \lneg{\tran{\lenv_1}},\lneg{\tran{\lenv_2}},\tran{\typtwo}}
      }
    }{
      \judg{\wush{\lneg{\tran{\uenv}}},
            \lneg{\tran{\lenv_1}},\lneg{\tran{\lenv_2}},\tran{\typtwo}}
    }
  \]
\end{enumerate}
\end{proof}


  \subsection{Confluence}
  \lsec{appendix:calculus_cr}
  
We structure this section as follows. First, we recall Melli\`es notion of orthogonal axiomatic rewriting system. We then introduce the labeled $\lambdaS$-calculus, a tool through which we define a notion of residual for $\lambdaS$-calculus. Then we prove two important properties of labeled reduction, namely, finite developments and semantic orthogonality. Finally, we show how to model the $\lambdaS$-calculus as an axiomatic rewrite system and how the above mentioned axioms hold, which entails confluence.

\section{Orthogonal axiomatic rewriting systems}

We recall here the axiomatic rewriting framework due to Melli\`es~\cite{Mellies:PhD:1996}.
We shall use this framework in~\rsec{lambdaS} to prove that the
Sharing Linear $\lambda$-Calculus is confluent, up to a congruence.

An \emph{axiomatic rewriting system} (abbreviated AxRS) is a $5$-uple
$(\Objs,\Steps,\src,\tgt,\cdot\resid{\cdot}\cdot)$
where $\Objs$ is a set whose elements are called \emph{objects},
$\Steps$ is a set whose elements are called \emph{steps},
$\src,\tgt : \Steps \to \Objs$ are functions providing the \emph{source} and the
\emph{target} of each step,
and $\cdot\resid{\cdot}\cdot \subseteq \Steps \times \Steps \times \Steps$
is a ternary \emph{residual relation},
such that $\step\resid{\steptwo}\step'$
may hold only if $\src(\step) = \src(\steptwo)$
and $\src(\step') = \tgt(\steptwo)$.
A \emph{reduction} is either
the empty reduction that starts an ends on an object $\obj \in \Objs$,
written $\emptyseq_{\obj}$,
or a sequence of $n \geq 1$ steps $\step_1\hdots\step_n$
that are \emph{composable}, \ie $\tgt(\step_i) = \src(\step_{i+1})$
for all $i\in1..n-1$.
The $\src$ and $\tgt$ functions are extended to reductions as expected.
The residual function is extended to reductions by declaring
that $\step\resid{\emptyseq_\step}\step$,
and that the binary relation $\resid{\steptwo_1\hdots\steptwo_n}$
is the composition $\resid{\steptwo_1}\circ\hdots\circ\resid{\steptwo_n}$.

A \emph{multistep} is either the empty multistep starting on an object $\obj
\in \Objs$, written $\emptyset_{\obj}$,
or a finite non-empty set of steps $M \subseteq \Steps$
all with the same source.
The notion of \emph{development} is defined as follows:
the empty reduction $\emptyseq_{\obj}$ is a development of the
empty multistep $\emptyset_{\obj}$,
and a reduction $\step_1\hdots\step_n$ is a development of a multistep $M$
if for all $i \in 1..n$ there exists a step $\steptwo \in M$ such that
$\steptwo\resid{\step_1\hdots\step_{i-1}}\step_i$.
A \emph{complete development} of a multistep $\multistep$
is a development after which no residuals of steps in $\multistep$ remain.

An AxRS is \emph{orthogonal} if it verifies the four following properties:
\\
  {\bf 1.} \axiomName{Auto-Erasure.}
  For all $\step \in \Steps$
  there is no $\steptwo \in \Steps$ such that $\step\resid{\step}\steptwo$.
\\
  {\bf 2.} \axiomName{Finite Residuals.}
  The set $\set{\step' \ST \step\resid{\steptwo}\step'}$ is finite
  for all $\step,\steptwo \in \Steps$.
\\
  {\bf 3.} \axiomName{Finite Developments.}
  If $M \subseteq \Steps$ is a finite non-empty set of steps of the same source,
  all developments of $M$ are finite,
  \ie there are no infinite sequences $\step_1\step_2\hdots$ such that
  for all $i \in \Nat$ there exists a step $\steptwo \in M$ such that
  $\steptwo\resid{\step_1\hdots\step_{i-1}}\step_i$.
\\
  {\bf 4.} \axiomName{Semantic Orthogonality.}
  If $\step,\steptwo \in \Steps$ are steps of the same source,
  $\redseq$ is a complete development of $\set{\step' \ST \step\resid{\steptwo}\step'}$
  and
  $\redseqtwo$ is a complete development of $\set{\steptwo' \ST \steptwo\resid{\step}\steptwo'}$,
  then the compositions $\step\redseqtwo$ and $\steptwo\redseq$
  have the same target and
  $\resid{\step\redseqtwo} = \resid{\steptwo\redseq}$
  are equal as binary relations.

\begin{theorem}
\lthm{orthogonal_AxRS_confluence}
Orthogonal AxRSs are confluent.
(See \cite[Theorem~2.4]{Mellies:PhD:1996}).
\end{theorem}

Confluence of orthogonal AxRSs holds in a very strong sense.
In fact, confluence diagrams are closed with complete developments
of the relative projections of the reductions, which entails that
these diagrams are \emph{pushouts} in the categorical sense.

\subsubsection{The Labeled $\lambdaS$-calculus}
\label{sec:labeled_ lambdaS}

\begin{definition}[The $\llambdaS$-calculus]
The set of {\em labeled $\lambdaS$-terms} $\TermsSL$ ---or just {\em labeled terms}---
is given by the following grammar:
\[
  \begin{array}{rrll}
  \\
    \tm,\tmtwo,\hdots
    & ::=  & \lvar
           & \text{linear variable} \\
    & \mid & \uvar
           & \text{unrestricted variable} \\
    & \mid & \ann{\uvar}{\lab}
           & \text{unrestricted variable} \\
    & \mid & \lam{\lvar}{\tm}
           & \text{abstraction} \\
    & \mid & \llam{\lvar}{\tm}{\lab}
           & \text{abstraction} \\
    & \mid & \tm\,\tmtwo
           & \text{application} \\
    & \mid & \sha{\tm}
           & \text{grant} \\
    & \mid & \open{\tm}
           & \text{request} \\
    & \mid & \lopen{\tm}{\lab}
           & \text{request} \\
    & \mid & \ofc{\tm}
           & \text{promotion} \\
    & \mid & \tm\esub{\uvar}{\tmtwo}
           & \text{substitution} \\
    & \mid & \tm\lesub{\uvar}{\tmtwo}{\lab}
           & \text{substitution} \\
  \end{array}
\]
The set of labeled $\lambdaS$-contexts $\CtxsSL$ ---or just {\em labeled contexts}--- is defined as follows:
\[
  \begin{array}{rcl}
  \lgctx & ::= & \ctxhole
           \mid \lam{\lvar}{\lgctx}
           \mid \llam{\lvar}{\lgctx}{\lab}
           \mid \lgctx\,\tm
           \mid \tm\,\lgctx
           \mid \sha{\lgctx}
           \mid \open{\lgctx}
           \mid \lopen{\lgctx}{\lab}
           \mid \ofc{\lgctx}
           \mid \lgctx\esub{\uvar}{\tm}
           \mid \lgctx\lesub{\uvar}{\tm}{\lab}
           \mid \tm\esub{\uvar}{\lgctx}
           \mid \tm\lesub{\uvar}{\lgctx}{\lab}
  \\
  \lsctx & ::= & \ctxhole \mid \lsctx\esub{\uvar}{\tm} \mid \lsctx\lesub{\uvar}{\tm}{\lab}\\
  \end{array}
\]
We write $\ann{\uvar}{(\lab)}$ for a variable $\uvar$ which may or may not be labeled with label $\lab$. We write $\var$ for both linear and (possibly labeled) unrestricted variables. 
Similar with $\lopen{\tm}{(\lab)}$ and $\tm\esub{\ann{\uvar}{(\lab)}}{\tmtwo}$.
Also, $\fv{\tm}$ are the free variables of $\tm$, disregarding labels. For example, $\fv{\uvar\,\lvar}=\fv{\ann{\uvar}{\lab}\,\lvar}=\{\uvar,\lvar\}$. We write $\flv{\tm}$ for the set of free labeled variables. For example, $\flv{\uvar\,\lvar}=\emptyset$ and $\flv{\ann{\uvar}{\lab}\,\lvar}=\{\uvar\}$. Linear substitution over labeled terms $\tm\sub{\lvar}{\tmtwo}$ is defined as expected since linear variables are not decorated with labels. We write $\unlab{\tm}$ to denote the term obtained from $\tm$ by removing all its labels. Thus for example $\unlab{(\llam{\lvar}{(\lvar\, \lopen{\uvar}{\labtwo})}{\lab})}=\lam{\lvar}{(\lvar\, \open{\uvar})}$. A term $\tm$ can be labeled in different ways, leading to different \emph{variants} of $\tm$. We say that $\tm$ is a variant of $\tmtwo$ iff $\unlab{\tm}=\unlab{\tmtwo}$. In particular, $\tm$ is a
variant of itself.
\end{definition}

\begin{definition}[Well-labeled terms]
The set of {\em well-labeled $\lambdaS$-terms} $\TermsSWL$ ---or just {\em well-labeled terms}--- are those labeled terms $\tm\in\TermsSL$ for which the predicate $\wlt{\tm}$ holds, defined as follows:
  \[
  \begin{array}{c}
    \indrule
    {}
    {}
    {\wlt{\lvar}}
    \quad
    \indrule
    {}
    {}
    {\wlt{\ann{\uvar}{(\lab)}}}
    \\
    \\
    \indrule
    {}
    {\wlt{\tm}}
    {\wlt{\lam{\lvar}{\tm}}}
    \quad
    \indrule
    {}
    {\wlt{\tm}
    \quad
    \wlt{\tmtwo}}
    {\wlt{\tm\,\tmtwo}}
    \quad
    \indrule
    {}
    {\wlt{(\lam{\lvar}{\tm})\lsctx}
    \quad
    \wlt{\tmtwo}}
    {\wlt{(\llam{\lvar}{\tm}{\lab})\lsctx\,\tmtwo}}
    \\
    \\
    \indrule
    {}
    {\wlt{\tm}}
    {\wlt{\sha{\tm}}}
    \quad
    \indrule
    {}
    {\wlt{\tm}}
    {\wlt{\open{\tm}}}
    \quad
        \indrule
    {}
    {\wlt{(\sha{\tm})\lsctx}}
    {\wlt{\lopen{(\sha{\tm})\lsctx}{\lab}}}
    \quad
       \indrule
    {}
    {\wlt{\tm}}
    {\wlt{\ofc{\tm}}}
    \\
    \\
        \indrule
    {}
    {\wlt{\tm}
    \quad
    \wlt{(\ofc{(\sha{\tmtwo})\lsctx_1})\lsctx_2}
    \quad
    \uvar\in\flv{\tm}}
    {\wlt{\tm\esub{\uvar}{(\ofc{(\sha{\tmtwo})\lsctx_1})\lsctx_2}}}
    \quad
    \indrule
    {}
    {\wlt{\tm}
    \quad
    \wlt{(\ofc{\tmtwo})\lsctx}
    \quad
    \uvar\notin\fv{\tm}}
    {\wlt{\tm\lesub{\uvar}{(\ofc{\tmtwo})\lsctx}{\lab}}}
    \quad
            \indrule
    {}
    {\wlt{\tm}
    \quad
    \wlt{\tmtwo}
    \quad
    \uvar\notin\flv{\tm}}
    {\wlt{\tm\esub{\uvar}{\tmtwo}}}
  \end{array}
\]


\end{definition}

\begin{definition}[Labeled reduction]
Labeled reduction at the root is defined on labeled terms as follows:
\[
  \begin{array}{rll@{\HS}l}
    (\llam{\lvar}{\tm}{\lab})\lsctx\,\tmtwo
    & \rtoSdbL{\lab} &
    \tm\sub{\lvar}{\tmtwo}\lsctx
    &
    \text{if $\fv{\tmtwo} \cap \dom{\lsctx} = \emptyset$}
  \\
    \lopen{(\sha{\tm})\lsctx}{\lab}
    & \rtoSopenL{\lab} &
    \tm\lsctx
  \\
    \off{\lgctx}{\ann{\uvar}{\lab}}\esub{\uvar}{(\ofc{(\sha{\tm})\lsctx_1})\lsctx_2}
    & \rtoSlsL{\lab} &
    \off{\lgctx}{(\sha{\tm}) \lsctx_1}\esub{\uvar}{\ofc{(\sha{\tm}) \lsctx_1}}\lsctx_2
    &
    \text{if $\uvar \notin \fv{\tm\lsctx_1}$ and $\fv{\lgctx} \cap \dom{\lsctx_1\lsctx_2} = \emptyset$}
  \\
    \tm\esub{\ann{\uvar}{\lab}}{(\ofc{\tmtwo})\lsctx}
    & \rtoSgcL{\lab} &
    \tm\lsctx
    &
    \text{if $\uvar\notin\fv{\tm}$}
  \end{array}
\]
Note that the side conditions of the $\rtoSdbL{\lab}$ and $\rtoSlsL{\lab}$
rules can always be met by $\alpha$-renaming.
We define the \emph{$\lab$-labeled $R$ step relation} $\lto{\lab}_R \eqdef \of{\lgctx}{\lrootto{\lab}_R}$
for each $R \in \set{\symSdb,\symSopen,\symSls,\symSgc}$,
where $\of{\lgctx}{\lrootto{\lab}_R}$ denotes the closure of $\lrootto{\lab}_R$ by
compatibility under arbitrary labeled contexts. An $\lab$-step ($\step,\steptwo,\stepthree,\ldots$) is a tuple of one of the following forms:
\begin{itemize}

\item $\langle  \gctx,   (\llam{\lvar}{\tm}{\lab})\lsctx\,\tmtwo\rangle$ with $\fv{\tmtwo} \cap \dom{\lsctx} = \emptyset$; 

\item $\langle   \lgctx, \lopen{(\sha{\tm})\lsctx}{\lab} \rangle$;

  \item $\langle \lgctx, \lgctxtwo,
    \off{\lgctxtwo}{\ann{\uvar}{\lab}}\esub{\uvar}{(\ofc{(\sha{\tm})\lsctx_1})\lsctx_2}\rangle$ with
   if $\uvar \notin \fv{\tm\lsctx_1}$ and $\fv{\lgctxtwo} \cap \dom{\lsctx_1\lsctx_2} = \emptyset$;

   \item $\langle \lgctx,
    \tm\esub{\ann{\uvar}{\lab}}{(\ofc{\tmtwo})\lsctx}\rangle$ with $\uvar\notin\fv{\tm}$.
\end{itemize}
Its \emph{anchor} is the variable decorated with the label $\lab$, if the step is in $\set{\symSdb,\symSls,\symSgc}$  or the occurrence of `$\mathtt{open}$' that is decorated with the label $\lab$, if the step is a $\symSopen$-step. Its source and target are defined as expected. For example, the source and target of the step  $\langle  \gctx,   (\llam{\lvar}{\tm}{\lab})\lsctx\,\tmtwo\rangle$ is $\of{\gctx}{(\llam{\lvar}{\tm}{\lab})\lsctx\,\tmtwo}$ and $\of{\gctx}{\tm\sub{\lvar}{\tmtwo}\lsctx}$, resp.
We write $\labStep{\lab}{\tm}$ for the set of $\lab$-labeled steps in $\tm$. 

The \emph{$\lab$-step} reduction relation $\toSL{\lab}$ is defined
as the union of the previous relations,
that is,
$\toSL{\lab} \eqdef \toSdbL{\lab} \cup \toSopenL{\lab} \cup \toSlsL{\lab} \cup \toSgcL{\lab}$. 
\end{definition}

\begin{definition}[Lifting of a step]
Let $\tm\in \TermsSL$ be a labeled term and $\step\in\steps{\tm}$ a step in $\tm$. The $\step-\lab$-lift of $\tm$, written $\lift{\tm}{\step}{\lab}$, is the variant of $\tm$ resulting from assigning label $\lab$ to the anchor of $\step$. For example, if $\step=\langle \gctx,(\llam{\lvar}{\tm_1}{(\labtwo)})\sctx\,\tm_2\rangle $, then  $\lift{\tm}{\step}{\lab}=\of{\gctx}{(\llam{\lvar}{\tm_1}{\lab})\sctx\,\tm_2}$. 
\end{definition}

\begin{lemma}
  Let $\tm\in \TermsSWL$ and $\tm \toSL{\lab}\tmtwo$ implies $\tmtwo\in \TermsSWL$.
\end{lemma}

\subsubsection{Finite Developments and Semantic Orthogonality}
\label{sec:labeled_ lambdaS_properties}


\begin{definition}[Variable Multiplicity]
\ldef{variable_multiplicity}
 The multiplicity of a variable $\var$ in a well-labeled term $\tm$, denoted $\mult{\var}{\tm}$, is defined as follows:
  \begin{center}
    $\begin{array}{rcll}
\mult{\lvar}{\lvar} & \eqdef & 1 \\
\mult{\lvar}{\lvartwo} & \eqdef & 0 & \text{if $\lvar\neq\lvartwo$}\\
\mult{\uvar}{\ann{\uvar}{\lab}} & \eqdef & 1 \\
\mult{\uvar}{\uvar} & \eqdef & 0 \\
     \mult{\uvar}{\ann{\uvartwo}{(\lab)}} & \eqdef & 0 & \text{if $\uvar\neq\uvartwo$}\\

\mult{\var}{\lam{\lvar}{\tm}} & \eqdef & \mult{\var}{\tm}\\
\mult{\var}{(\llam{\lvar}{\tm}{\lab})\lsctx\,\tmtwo} & \eqdef & \mult{\var}{(\lam{\lvar}{\tm})\lsctx} + \mult{\var}{\tmtwo} + \mult{\lvar}{\tm}\times \mult{\var}{\tmtwo}\\
\mult{\var}{\tm\,\tmtwo} & \eqdef & \mult{\var}{\tm} + \mult{\var}{\tmtwo}& \text{if $\tm\,\tmtwo$ not a redex}\\
\mult{\var}{\sha{\tm}} & \eqdef & \mult{\var}{\tm}\\
\mult{\var}{\olopen{\tm}{\lab}} & \eqdef & \mult{\var}{\tm}\\
\mult{\var}{\ofc{\tm}} & \eqdef & \mult{\var}{\tm}\\
\mult{\var}{\tm\lesub{\uvar}{\tmtwo}{(\lab)}} & \eqdef &  \mult{\var}{\tm} + \mult{\var}{\tmtwo} + \mult{\uvar}{\tm}\times \mult{\var}{\tmtwo}\\
   \end{array}$
 \end{center}

  \begin{center}
    $\begin{array}{rcll}
\multctx{\var}{\phi}{\ctxhole} & \eqdef & \phi(\var)\\
  \multctx{\var}{\phi}{\lsctx\lesub{\uvar}{\tmtwo}{(\lab)}} & \eqdef &  \multctx{\var}{\phi}{\lsctx} + \mult{\var}{\tmtwo} + \multctx{\uvar}{\phi}{\lsctx}\times \mult{\var}{\tmtwo}\\
   \end{array}$
 \end{center}

\end{definition}

\begin{definition}[Labeled Redex Multiplicity]
\ldef{redex_multiplicity}
  The multiplicity of labeled redexes in a well-labeled term $\tm$, denoted $\lmult{\tm}$, is defined as follows:
  \begin{center}
    $\begin{array}{rcll}
\lmult{\var} & \eqdef & 0 \\

\lmult{\lam{\lvar}{\tm}} & \eqdef & \lmult{\tm}\\
\lmult{(\llam{\lvar}{\tm}{\lab})\lsctx\,\tmtwo} & \eqdef & 1 + \lmult{(\lam{\lvar}{\tm})\lsctx} + \lmult{\tmtwo} + \mult{\lvar}{\tm}\times \lmult{\tmtwo}\\
\lmult{\tm\,\tmtwo} & \eqdef & \lmult{\tm} + \lmult{\tmtwo}& \text{if $\tm\,\tmtwo$ not a redex}\\
\lmult{\sha{\tm}} & \eqdef & \lmult{\tm}\\
\lmult{\open{\tm}} & \eqdef & \lmult{\tm}\\
\lmult{\lopen{(\sha{\tm})\lsctx}{\lab}} & \eqdef & 1 + \lmult{(\sha{\tm})\lsctx}\\
\lmult{\ofc{\tm}} & \eqdef & \lmult{\tm}\\
\lmult{\tm\lesub{\uvar}{\tmtwo}{\lab}} & \eqdef &  1 + \lmult{\tm} +\lmult{\tmtwo}\\
\lmult{\tm\esub{\uvar}{\tmtwo}} & \eqdef & \lmult{\tm} + \lmult{\tmtwo} + \mult{\uvar}{\tm}\times \lmult{\tmtwo} + \mult{\uvar}{\tm}\\
   \end{array}$
 \end{center}

  \begin{center}
    $\begin{array}{rcll}
\lmultctx{\phi}{\ctxhole} & \eqdef & 0\\
       \lmultctx{\phi}{\lsctx\lesub{\uvar}{\tmtwo}{\lab}} & \eqdef &  1 + \lmultctx{\phi}{\lsctx} + \lmult{\tmtwo}\\
       \lmultctx{\phi}{\lsctx\esub{\uvar}{\tmtwo}} & \eqdef &  \lmultctx{\phi}{\lsctx} + \lmult{\tmtwo} + \multctx{\uvar}{\phi}{\lsctx}\times \lmult{\tmtwo} + \multctx{\uvar}{\phi}{\lsctx}\\
       \\ 
   \end{array}$
 \end{center}

\end{definition}

\begin{lemma}
\llem{linear_substitution_and_free_variables}
Let $\tm,\tmtwo\in \TermsSL$. Then $\tm\sub{\lvar}{\tmtwo}=\tm$ if $\lvar\notin\fv{\tm}$.
\end{lemma}

\begin{proof}
By induction on $\tm$.
\end{proof}

\begin{lemma}
\llem{multiplicity_and_free_variables}
$\mult{\var}{\tm}=0$ if $\var\notin\fv{\tm}$.
\end{lemma}

\begin{proof}
By induction on $\tm$.
\end{proof}

Below we use the notation $\mult{\bullet}{\tm}$ for the function that maps a variable $\var$ to $\mult{\var}{\tm}$.
    
\begin{lemma}
\llem{multiplicity_of_substitution_contexts}
$\mult{\var}{\tm\lsctx} = \multctx{\var}{\mult{\bullet}{\tm}}{\lsctx}$
\end{lemma}

\begin{proof}
By induction on $\lsctx$.
\begin{ifLongAppendix}
  \begin{xenumerate}
  \item $  \lsctx = \ctxhole$. Immediate since $\multctx{\var}{\mult{\bullet}{\tm}}{\ctxhole} = \mult{\var}{\tm}$.

  \item $\lsctx = \lsctx_1\lesub{\uvar}{\tmtwo}{(\lab)}$.

    \[\begin{array}{rll}
        & \mult{\var}{\tm \lsctx_1\lesub{\uvar}{\tmtwo}{(\lab)}} \\
        = & \mult{\var}{\tm \lsctx_1} + \mult{\var}{\tmtwo} + \mult{\uvar}{\tm \lsctx_1}\times \mult{\var}{\tmtwo} & (\rdef{variable_multiplicity})\\
        = & \multctx{\var}{\mult{\bullet}{\tm}}{\lsctx_1} + \mult{\var}{\tmtwo} + \multctx{\uvar}{\mult{\bullet}{\tm}}{\lsctx_1}\times \mult{\var}{\tmtwo} & (\ih\times 2) \\
          =  & \multctx{\var}{\mult{\bullet}{\tm}}{\lsctx_1\lesub{\uvar}{\tmtwo}{(\lab)}} & (\rdef{variable_multiplicity})
        \end{array}\]

  \end{xenumerate}
\end{ifLongAppendix}
\end{proof}

\begin{lemma}
\llem{multiplicity_and_non_capturing_replacement}
Suppose $\tm\in \TermsSWL$ and $\var\notin\fv{\tm}\cup\{\uvar\}$. Then
$\mult{\var}{\off{\lgctx}{\ann{\uvar}{\lab}}}  = 
\mult{\var}{\off{\lgctx}{\tm}}$
\end{lemma}
\begin{proof}
  By induction on the size of the labeled context $\lgctx$.
  \begin{ifLongAppendix}
    \begin{xenumerate}
  \item $\lgctx=\ctxhole$. Immediate from \rlem{multiplicity_and_free_variables}.
  \item $\lgctx=\lam{\lvar}{\lgctx_1}$ (the cases $\lgctx=\sha{\lgctx_1}$ and $\lgctx=\ofc{\lgctx_1}$ are similar).
    \[\begin{array}{rll}
        & \mult{\var}{\lam{\lvar}{\off{\lgctx_1}{\ann{\uvar}{\lab}}}} \\
        = & \mult{\var}{\off{\lgctx_1}{\ann{\uvar}{\lab}}} & (\rdef{variable_multiplicity})\\
        = & \mult{\var}{\off{\lgctx_1}{\tm}} & (\ih) \\
        = & \mult{\var}{\lam{\lvar}{\off{\lgctx_1}{\tm}}} & (\rdef{variable_multiplicity})\\
      \end{array}
    \]

  \item $\lgctx=\llam{\lvar}{\lgctx_1}{\lab}$. Not possible since $\off{\lgctx}{\ann{\uvar}{\lab}}$ is well-labeled.
    
\item $\lgctx=\tmtwo\,\lgctx_1$ (the case $\lgctx=\lgctx_1\,\tmtwo$ is similar).
  \begin{xenumerate}
\item $\tmtwo\neq (\llam{\lvar}{\tmthree}{\labtwo})\lsctx$.
        \[\begin{array}{rll}
        & \mult{\var}{\tmtwo\,\off{\lgctx_1}{\ann{\uvar}{\lab}}} \\
        = & \mult{\var}{\tmtwo} + \mult{\var}{\off{\lgctx_1}{\ann{\uvar}{\lab}}} & (\rdef{variable_multiplicity})\\
        = &  \mult{\var}{\tmtwo} + \mult{\var}{\off{\lgctx_1}{\tm}} & (\ih) \\
        = & \mult{\var}{\tmtwo\,\off{\lgctx_1}{\tm}} & (\rdef{variable_multiplicity})\\
      \end{array}
    \]

  \item $\tmtwo=  (\llam{\lvar}{\tmthree}{\labtwo})\lsctx$.

            \[\begin{array}{rll}
        & \mult{\var}{(\llam{\lvar}{\tmthree}{\labtwo})\lsctx\,\off{\lgctx_1}{\ann{\uvar}{\lab}}} \\
        = &  \mult{\var}{ (\lam{\lvar}{\tmthree}) \lsctx}  +  \mult{\var}{\off{\lgctx_1}{\ann{\uvar}{\lab}}}  +  \mult{\lvar}{\tmthree} \times  \mult{\var}{\off{\lgctx_1}{\ann{\uvar}{\lab}}}  & (\rdef{variable_multiplicity})\\
        = &  \mult{\var}{ (\lam{\lvar}{\tmthree}) \lsctx}  +  \mult{\var}{\off{\lgctx_1}{\tm}}  +  \mult{\lvar}{\tmthree} \times  \mult{\var}{\off{\lgctx_1}{\tm}}  &   (\ih\times 2)\\
                        = &  \mult{\var}{(\llam{\lvar}{\tmthree}{\labtwo})\lsctx\,\off{\lgctx_1}{\tm}}  & (\rdef{variable_multiplicity})\\
      \end{array}
    \]
\end{xenumerate}

\item $\lgctx= \lgctx_1\,\tmtwo$. We consider two cases depending on whether $\off{\lgctx_1}{\ann{\uvar}{\lab}}$ is a labeled abstraction or not.
  \begin{xenumerate}

  \item $\off{\lgctx_1}{\ann{\uvar}{\lab}}=(\llam{\lvar}{\tmthree}{\labtwo})\lsctx$. Then one of the following two cases hold:

      \begin{xenumerate}

\item $\lgctx= (\llam{\lvar}{\lgctx_1}{\labtwo})\lsctx\,\tmtwo$

        \[\begin{array}{rll}
        & \mult{\var}{ (\llam{\lvar}{\off{\lgctx_{11}}{\ann{\uvar}{\lab}}}{\labtwo})\lsctx\,\tmtwo} \\
        = &  \mult{\var}{ (\lam{\lvar}{\off{\lgctx_{11}}{\ann{\uvar}{\lab}}})\lsctx}  +  \mult{\var}{\tmtwo}  +  \mult{\lvar}{\off{\lgctx_{11}}{\ann{\uvar}{\lab}}} \times  \mult{\var}{\tmtwo}  & (\rdef{variable_multiplicity})\\
        = &  \mult{\var}{ (\lam{\lvar}{\off{\lgctx_{11}}{\tm}})\lsctx}  +  \mult{\var}{\tmtwo}  +  \mult{\lvar}{\off{\lgctx_{11}}{\tm}} \times  \mult{\var}{\tmtwo}  & (\ih\times 2)\\
        = &\mult{\var}{ (\llam{\lvar}{\off{\lgctx_{11}}{\tm}}{\labtwo})\lsctx\,\tmtwo} & (\rdef{variable_multiplicity})\\
      \end{array}
    \]
    
  \item $\lgctx= (\llam{\lvar}{\tmthree}{\labtwo})\lsctx_1\lesub{\uvartwo}{\lgctx_{11}}{(\labthree)}\lsctx_2\,\tmtwo$

            \[\begin{array}{rll}
        & \mult{\var}{ (\llam{\lvar}{\tmthree}{\labtwo}) \lsctx_1\lesub{\uvartwo}{\off{\lgctx_{11}}{\ann{\uvar}{\lab}}}{(\labthree)}\lsctx_2\,\tmtwo} \\
        = &  \mult{\var}{ (\lam{\lvar}{\tmthree}) \lsctx_1\lesub{\uvartwo}{\off{\lgctx_{11}}{\ann{\uvar}{\lab}}}{(\labthree)}\lsctx_2}  +  \mult{\var}{\tmtwo}  +  \mult{\lvar}{\tmthree} \times  \mult{\var}{\tmtwo}  & (\rdef{variable_multiplicity})\\
        = &  \multctx{\var}{\mult{\bullet}{ (\lam{\lvar}{\tmthree}) \lsctx_1\lesub{\uvartwo}{\off{\lgctx_{11}}{\ann{\uvar}{\lab}}}{(\labthree)}}}{\lsctx_2}  +  \mult{\var}{\tmtwo}  +  \mult{\lvar}{\tmthree} \times  \mult{\var}{\tmtwo}  & (\rlem{multiplicity_of_substitution_contexts})\\
        = &  \multctx{\var}{\mult{\bullet}{ (\lam{\lvar}{\tmthree}) \lsctx_1} + \mult{\bullet}{\off{\lgctx_{11}}{\ann{\uvar}{\lab}}} + \mult{\uvartwo}{ (\lam{\lvar}{\tmthree}) \lsctx_1}\times \mult{\bullet}{\off{\lgctx_{11}}{\ann{\uvar}{\lab}}}}{\lsctx_2}  +  \mult{\var}{\tmtwo}  +  \mult{\lvar}{\tmthree} \times  \mult{\var}{\tmtwo}  &  (\rdef{variable_multiplicity})\\
                = &  \multctx{\var}{\mult{\bullet}{ (\lam{\lvar}{\tmthree}) \lsctx_1} + \mult{\bullet}{\off{\lgctx_{11}}{\tm}} + \mult{\uvartwo}{ (\lam{\lvar}{\tmthree}) \lsctx_1}\times \mult{\bullet}{\off{\lgctx_{11}}{\tm}}}{\lsctx_2}  +  \mult{\var}{\tmtwo}  +  \mult{\lvar}{\tmthree} \times  \mult{\var}{\tmtwo}  &  (\ih\times 2)\\
                        = &  \multctx{\var}{\mult{\bullet}{ (\lam{\lvar}{\tmthree}) \lsctx_1\lesub{\uvartwo}{\off{\lgctx_{11}}{\tm}}{(\labthree)}}}{\lsctx_2}  +  \mult{\var}{\tmtwo}  +  \mult{\lvar}{\tmthree} \times  \mult{\var}{\tmtwo}   & (\rdef{variable_multiplicity})\\
        = &  \mult{\var}{ (\lam{\lvar}{\tmthree}) \lsctx_1\lesub{\uvartwo}{\off{\lgctx_{11}}{\tm}}{(\labthree)}\lsctx_2}  +  \mult{\var}{\tmtwo}  +  \mult{\lvar}{\tmthree} \times  \mult{\var}{\tmtwo}  &  (\rlem{multiplicity_of_substitution_contexts})\\
       =  & \mult{\var}{ (\llam{\lvar}{\tmthree}{\labtwo}) \lsctx_1\lesub{\uvartwo}{\off{\lgctx_{11}}{\tm}}{(\labthree)}\lsctx_2\,\tmtwo}  & (\rdef{variable_multiplicity})\\
      \end{array}
    \]
  \end{xenumerate}

  \item $\off{\lgctx_1}{\ann{\uvar}{\lab}}\neq (\llam{\lvar}{\tmthree}{\labtwo})\lsctx$. Then also $\off{\lgctx_1}{\tm}\neq(\llam{\lvar}{\lgctx_1}{\labtwo})\lsctx$ since $\tm$ well-labeled. Thus we have:

    \[\begin{array}{rll}
        & \mult{\var}{\off{\lgctx_1}{\ann{\uvar}{\lab}}\,\tmtwo\,} \\
        = & \mult{\var}{\off{\lgctx_1}{\ann{\uvar}{\lab}}} + \mult{\var}{\tmtwo} & (\rdef{variable_multiplicity})\\
        = &  \mult{\var}{\off{\lgctx_1}{\tm}} + \mult{\var}{\tmtwo}  & (\ih) \\
        = & \mult{\var}{\off{\lgctx_1}{\tm}\, \tmtwo} & (\rdef{variable_multiplicity})\\
      \end{array}
    \]

 \end{xenumerate}

\item $\lgctx=\lopen{\lgctx_1}{(\labtwo)}$. We need to consider multiple cases since out induction proceeds on well-formed labeled contexts.

  \begin{xenumerate}

  \item $\labtwo$ is not present. 

        \[\begin{array}{rll}
        & \mult{\var}{\open{\off{\lgctx_1}{\ann{\uvar}{\lab}}}} \\
        = & \mult{\var}{\off{\lgctx_1}{\ann{\uvar}{\lab}}} & (\rdef{variable_multiplicity})\\
        = & \mult{\var}{\off{\lgctx_1}{\tm}}  & (\ih)\\
        = & \mult{\var}{\open{\off{\lgctx_1}{\tm}}} & (\rdef{variable_multiplicity})\\
      \end{array}
    \]

  \item $\labtwo$ is present. Note that $\lgctx_1=\ctxhole $ and $\lgctx_1=\ctxhole\lsctx_1$ (with $\lsctx=\lsctx_1\lsctx_2$) is not possible since $\off{\lgctx}{\ann{\uvar}{\lab}}$ is well-labeled. The remaining cases are:
     \begin{xenumerate}

 \item $\lgctx_1=(\sha{\lgctx_2}) \lsctx$.  

          \[\begin{array}{rll}
        & \mult{\var}{\open{(\sha{\off{\lgctx_2}{\ann{\uvar}{\lab}}}) \lsctx}} \\
        = &  \mult{\var}{(\sha{\off{\lgctx_2}{\ann{\uvar}{\lab}}}) \lsctx} & (\rdef{variable_multiplicity})\\
        = &  \multctx{\var}{\mult{\bullet}{\sha{\off{\lgctx_2}{\ann{\uvar}{\lab}}}}}{\lsctx} & (\rlem{multiplicity_of_substitution_contexts})\\
        = &  \multctx{\var}{\mult{\bullet}{\sha{\off{\lgctx_2}{\tm}}}}{\lsctx} & (\ih)\\
        = &  \mult{\var}{(\sha{\off{\lgctx_2}{\tm}})\lsctx} & (\rlem{multiplicity_of_substitution_contexts})\\
        = &  \mult{\var}{\open{(\sha{\off{\lgctx_2}{\tm}})\lsctx}} & (\rdef{variable_multiplicity})\\
      \end{array}
    \]

  \item $\lgctx_1=(\sha{\tmtwo}) \lsctx_1\lesub{\uvartwo}{\lgctx_2}{(\labthree)}\lsctx_2$

              \[\begin{array}{rll}
        & \mult{\var}{\open{(\sha{\tmtwo}) \lsctx_1\lesub{\uvartwo}{\off{\lgctx_2}{\ann{\uvar}{\lab}}}{(\labthree)}\lsctx_2}} \\
        = & \mult{\var}{(\sha{\tmtwo}) \lsctx_1\lesub{\uvartwo}{\off{\lgctx_2}{\ann{\uvar}{\lab}}}{(\labthree)}\lsctx_2} &  (\rdef{variable_multiplicity})\\
       =  & \multctx{\var}{\mult{\bullet}{(\sha{\tmtwo}) \lsctx_1\lesub{\uvartwo}{\off{\lgctx_2}{\ann{\uvar}{\lab}}}{(\labthree)}}}{\lsctx_2} & (\rlem{multiplicity_of_substitution_contexts})\\
       =  & \multctx{\var}{\mult{\bullet}{(\sha{\tmtwo}) \lsctx_1} + \mult{\bullet}{\off{\lgctx_2}{\ann{\uvar}{\lab}}} + \mult{\uvartwo}{(\sha{\tmtwo})\lsctx_1}\times  \mult{\bullet}{\off{\lgctx_2}{\ann{\uvar}{\lab}}}}{\lsctx_2} &  (\rdef{variable_multiplicity})\\

       =  & \multctx{\var}{\mult{\bullet}{(\sha{\tmtwo}) \lsctx_1} + \mult{\bullet}{\off{\lgctx_2}{\tm}} + \mult{\uvartwo}{(\sha{\tmtwo})\lsctx_1}\times  \mult{\bullet}{\off{\lgctx_2}{\tm}}}{\lsctx_2} &  (\ih\times 2)\\
       =  & \multctx{\var}{\mult{\bullet}{(\sha{\tmtwo}) \lsctx_1\lesub{\uvartwo}{\off{\lgctx_2}{\tm}}{(\labthree)}}}{\lsctx_2} & (\rdef{variable_multiplicity})\\
       =  & \mult{\var}{(\sha{\tmtwo}) \lsctx_1\lesub{\uvartwo}{\off{\lgctx_2}{\tm}}{(\labthree)}\lsctx_2} & (\rlem{multiplicity_of_substitution_contexts})\\
       =  & \mult{\var}{\open{(\sha{\tmtwo}) \lsctx_1\lesub{\uvartwo}{\off{\lgctx_2}{\tm}}{(\labthree)}\lsctx_2}} & (\rdef{variable_multiplicity})\\
      \end{array}
    \]
    
 \end{xenumerate}
 \end{xenumerate}
   
\item $\lgctx=\lgctx_1\lesub{\uvarthree}{\tmtwo}{(\labtwo)}$ 
  \[\begin{array}{rll}
              & \mult{\var}{\off{\lgctx_1}{\ann{\uvar}{\lab}}\lesub{\uvarthree}{\tmtwo}{(\labtwo)}} \\
        = & \mult{\var}{\off{\lgctx_1}{\ann{\uvar}{\lab}}} +\mult{\var}{\tmtwo}  + \mult{\uvarthree}{\off{\lgctx_1}{\ann{\uvar}{\lab}}} \times \mult{\var}{\tmtwo} & (\rdef{variable_multiplicity})\\
        = & \mult{\var}{\off{\lgctx_1}{\tm}} +\mult{\var}{\tmtwo}  + \mult{\uvarthree}{\off{\lgctx_1}{\tm}} \times \mult{\var}{\tmtwo} & (\ih\times 2)\\
         =     & \mult{\var}{\off{\lgctx_1}{\tm}\lesub{\uvarthree}{\tmtwo}{(\labtwo)}} & (\rdef{variable_multiplicity})\\
      \end{array}
    \]

  \item $\lgctx =\tmtwo\lesub{\uvarthree}{\lgctx_1}{(\labtwo)}$

      \[\begin{array}{rll}
              & \mult{\var}{\tmtwo\lesub{\uvarthree}{\off{\lgctx_1}{\ann{\uvar}{\lab}}}{(\labtwo)}} \\
        = & \mult{\var}{\tmtwo} + \mult{\var}{\off{\lgctx_1}{\ann{\uvar}{\lab}}}  + \mult{\uvarthree}{\tmtwo}\times \mult{\var}{\off{\lgctx_1}{\ann{\uvar}{\lab}}} & (\rdef{variable_multiplicity})\\
          = & \mult{\var}{\tmtwo} + \mult{\var}{\off{\lgctx_1}{\tm}}  + \mult{\uvarthree}{\tmtwo}\times \mult{\var}{\off{\lgctx_1}{\tm}} & (\ih\times 2) \\
          =     & \mult{\var}{\tmtwo\lesub{\uvarthree}{\off{\lgctx_1}{\tm}}{(\labtwo)}} & (\rdef{variable_multiplicity})\\
      \end{array}
    \]
    
  \end{xenumerate}
  \end{ifLongAppendix}
\end{proof}

\begin{lemma}
  \llem{multiplicity_and_linear_substitution_non_capturing}
  Let $\tm,\tmtwo\in \TermsSWL$. Suppose $\var\notin\fv{\tmtwo}\cup\{\lvar\}$.
  \begin{xenumerate}
\item Then $\mult{\var}{\tm\sub{\lvar}{\tmtwo}}=\mult{\var}{\tm}$.
\item  Suppose also $\dom{\sctx}\cap \fv{\tmtwo}=\emptyset$. Then $\multctx{\var}{\mult{\bullet}{\tm\sub{\lvar}{\tmtwo}}}{\lsctx}= \multctx{\var}{\mult{\bullet}{\tm}}{\lsctx}$.
  \end{xenumerate}
\end{lemma}

\begin{proof}
  The first item is by induction on $\tm$.
\begin{ifLongAppendix}
  \begin{xenumerate}
  \item $\tm=  \lvartwo$. If $\lvartwo\neq\lvar$, then $\lvartwo\sub{\lvar}{\tmtwo}=\lvartwo$ and we conclude immediately.
    If $\lvartwo=\lvar$, then
    \[\begin{array}{rll}
        & \mult{\var}{\lvar\sub{\lvar}{\tmtwo}} \\
        = & \mult{\var}{\tmtwo} \\
        = & 0 & (\var\notin\fv{\tmtwo}, \rlem{linear_substitution_and_free_variables})\\
        = & \mult{\var}{\lvar}  & (\var\neq\lvar)
      \end{array}\] 
  \item $\tm=  \ann{\uvar}{(\lab)}$. Immediate since $\ann{\uvar}{(\lab)}\sub{\lvar}{\tmtwo} =\ann{\uvar}{(\lab)}$
    
      
  \item $\tm= \lam{\lvartwo}{\tmthree}$ (the cases  $\tm=  \tm_1\,\tm_2$  and $\tm=  \sha{\tm_1}$ and $\tm=   \ofc{\tm_1}$ and $\tm=  \lopen{\tm_1}{(\lab)}$ are similar).

        \[\begin{array}{rll}
        & \mult{\var}{(\lam{\lvartwo}{\tmthree})\sub{\lvar}{\tmtwo}} \\
        = & \mult{\var}{\lam{\lvartwo}{\tmthree \sub{\lvar}{\tmtwo}}} \\
        = & \mult{\var}{\tmthree \sub{\lvar}{\tmtwo}} & (\rdef{variable_multiplicity})\\
        = & \mult{\var}{\tmthree} & (\ih) \\
        = & \mult{\var}{\lam{\lvartwo}{\tmthree}} & (\rdef{variable_multiplicity})
          \end{array}\]

      \item $\tm=(\llam{\lvartwo}{\tm_1}{\lab})\lsctx\,\tm_2$
    
        \[\begin{array}{rll}
        & \mult{\var}{((\llam{\lvartwo}{\tm_1}{\lab})\lsctx\,\tm_2)\sub{\lvar}{\tmtwo}} \\
        = & \mult{\var}{(\llam{\lvartwo}{\tm_1 \sub{\lvar}{\tmtwo}}{\lab})\lsctx \sub{\lvar}{\tmtwo}\,\tm_2 \sub{\lvar}{\tmtwo}}  \\
        = & \mult{\var}{(\lam{\lvartwo}{\tm_1 \sub{\lvar}{\tmtwo}})\lsctx \sub{\lvar}{\tmtwo}}  + \mult{\var}{\tm_2 \sub{\lvar}{\tmtwo}}  + \mult{\lvartwo}{\tm_1 \sub{\lvar}{\tmtwo}} \times \mult{\var}{\tm_2 \sub{\lvar}{\tmtwo}} & (\rdef{variable_multiplicity})\\
        = & \mult{\var}{((\lam{\lvartwo}{\tm_1 })\lsctx) \sub{\lvar}{\tmtwo}}  + \mult{\var}{\tm_2 \sub{\lvar}{\tmtwo}}  + \mult{\lvartwo}{\tm_1 \sub{\lvar}{\tmtwo}} \times \mult{\var}{\tm_2 \sub{\lvar}{\tmtwo}} \\
        = & \mult{\var}{(\lam{\lvartwo}{\tm_1 })\lsctx}  + \mult{\var}{\tm_2 }  + \mult{\lvartwo}{\tm_1} \times \mult{\var}{\tm_2} & (\ih\times 4) \\
        = & \mult{\var}{(\llam{\lvartwo}{\tm_1}{\lab})\lsctx\,\tm_2} & (\rdef{variable_multiplicity})\\
          \end{array}\]
        
      \item $\tm=  \tm_1\lesub{\uvar}{\tm_2}{(\lab)}$

               \[\begin{array}{rll}
        & \mult{\var}{\tm_1\lesub{\uvar}{\tm_2}{(\lab)}\sub{\lvar}{\tmtwo}} \\
        = & \mult{\var}{\tm_1 \sub{\lvar}{\tmtwo}\lesub{\uvar}{\tm_2 \sub{\lvar}{\tmtwo}}{(\lab)}} \\
        = & \mult{\var}{\tm_1 \sub{\lvar}{\tmtwo}} + \mult{\var}{\tm_2 \sub{\lvar}{\tmtwo}} +  \mult{\uvar}{\tm_1 \sub{\lvar}{\tmtwo}}\times \mult{\var}{\tm_2 \sub{\lvar}{\tmtwo}} \\
        = & \mult{\var}{\tm_1} + \mult{\var}{\tm_2} +  \mult{\uvar}{\tm_1}\times \mult{\var}{\tm_2} & (\ih\times 4) \\
        = & \mult{\var}{\tm_1\lesub{\uvar}{\tm_2}{(\lab)}} & (\rdef{variable_multiplicity})
         \end{array}\]
    \end{xenumerate}
  
The second item is by induction on $\lsctx$ and uses the first item.
  \begin{xenumerate}
  \item $\lsctx=\ctxhole$.


        \[\begin{array}{rll}
       & \multctx{\var}{\mult{\bullet}{\tm\sub{\lvar}{\tmtwo}}}{\ctxhole}\\
        = & \mult{\var}{\tm\sub{\lvar}{\tmtwo}} & (\rdef{variable_multiplicity}) \\
            = &   \mult{\var}{\tm} & (\text{item 1})\\ 
       = &  \multctx{\var}{\mult{\bullet}{\tm}}{\ctxhole} &   (\rdef{variable_multiplicity})
      \end{array}
    \]
    
  
\item $\lsctx=\lsctx_1\lesub{\uvar}{\tmthree}{(\lab)}$.
       \[\begin{array}{rll}
       & \multctx{\var}{\mult{\bullet}{\tm\sub{\lvar}{\tmtwo}}}{\lsctx_1\lesub{\uvar}{\tmthree}{(\lab)}}\\
           = & \multctx{\var}{\mult{\bullet}{\tm\sub{\lvar}{\tmtwo}}}{\lsctx_1} +  \mult{\var}{\tmthree}  +\multctx{\uvar}{\mult{\bullet}{\tm\sub{\lvar}{\tmtwo}}}{\lsctx_1} \times \mult{\var}{\tmthree} &  (\rdef{variable_multiplicity})\\
           = & \multctx{\var}{\mult{\bullet}{\tm}}{\lsctx_1} +  \mult{\var}{\tmthree}  +\multctx{\uvar}{\mult{\bullet}{\tm}}{\lsctx_1} \times \mult{\var}{\tmthree} & (\ih\times 2)\\
       = &  \multctx{\var}{\mult{\bullet}{\tm}}{\lsctx_1\lesub{\uvar}{\tmthree}{(\lab)}} &   (\rdef{variable_multiplicity})
      \end{array}
    \]
\end{xenumerate}
\end{ifLongAppendix}
\end{proof}

\begin{lemma}
\llem{multiplicity_and_linear_substitution}
Suppose $\var\neq\lvar$. Then $\mult{\var}{\tm}+\mult{\lvar}{\tm}\times \mult{\var}{\tmtwo} = \mult{\var}{t\sub{\lvar}{\tmtwo}}$
\end{lemma}

\begin{proof}
  By induction on $\tm$.
\begin{ifLongAppendix}
  \begin{xenumerate}
  \item $\tm=  \lvartwo$. If $\lvartwo\neq\lvar$, then
    \[\begin{array}{rll}
        & \mult{\var}{\lvartwo}+\mult{\lvar}{\lvartwo}\times \mult{\var}{\tmtwo} \\
        = & \mult{\var}{\lvartwo} \\
        = & \mult{\var}{\lvartwo\sub{\lvar}{\tmtwo}}
      \end{array}\]

    If $\lvartwo=\lvar$, then
    \[\begin{array}{rll}
        & \mult{\var}{\lvartwo}+ \mult{\lvar}{\lvartwo}\times \mult{\var}{\tmtwo} \\
        =& \mult{\var}{\tmtwo} & (\var\neq\lvar)\\
        = & \mult{\var}{\lvartwo\sub{\lvar}{\tmtwo}}
      \end{array}\]

      \item $\tm=  \ann{\uvar}{(\lab)}$
        \[\begin{array}{rll}
            & \mult{\var}{\ann{\uvar}{(\lab)}}+ \mult{\lvar}{\ann{\uvar}{(\lab)}}\times \mult{\var}{\tmtwo} \\
            = &  \mult{\var}{\ann{\uvar}{(\lab)}}\\
            = & \mult{\var}{\ann{\uvar}{(\lab)}\sub{\lvar}{\tmtwo}}
          \end{array}\]
        
      \item $\tm= \lam{\lvartwo}{\tmthree}$.

        \[\begin{array}{rll}
            & \mult{\var}{\lam{\lvartwo}{\tmthree}}+\mult{\lvar}{\lam{\lvartwo}{\tmthree}}\times \mult{\var}{\tmtwo} \\
            =  & \mult{\var}{\tmthree} + \mult{\lvar}{\tmthree}\times \mult{\var}{\tmtwo} & (\rdef{variable_multiplicity})\\
            =  & \mult{\var}{\tmthree\sub{\lvar}{\tmtwo}} & (\ih) \\
            = &  \mult{\var}{\lam{\lvartwo}{(\tmthree \sub{\lvar}{\tmtwo})}} &  (\rdef{variable_multiplicity})\\
            = &  \mult{\var}{(\lam{\lvartwo}{\tmthree})\sub{\lvar}{\tmtwo}} &   
          \end{array}\]
        
      \item $\tm=\tm_1\,\tm_2$.
           \[\begin{array}{rll}
            & \mult{\var}{\tm_1\,\tm_2}+\mult{\lvar}{\tm_1\,\tm_2}\times \mult{\var}{\tmtwo} \\
            =  & \mult{\var}{\tm_1}+\mult{\var}{\tm_2}+(\mult{\lvar}{\tm_1}+\mult{\lvar}{\tm_2})\times \mult{\var}{\tmtwo} & (\rdef{variable_multiplicity})\\
            =  & \mult{\var}{\tm_1}+\mult{\var}{\tm_2}+ \mult{\lvar}{\tm_1}\times \mult{\var}{\tmtwo}+\mult{\lvar}{\tm_2}\times \mult{\var}{\tmtwo} \\
            =  & \mult{\var}{\tm_1\sub{\lvar}{\tmtwo}} + \mult{\var}{\tm_2\sub{\lvar}{\tmtwo}}& (\ih\times 2) \\
            =  & \mult{\var}{\tm_1\sub{\lvar}{\tmtwo}\, \tm_2\sub{\lvar}{\tmtwo}} & (\rdef{variable_multiplicity})\\
            =  & \mult{\var}{(\tm_1\,\tm_2)\sub{\lvar}{\tmtwo}} & \\
             \end{array}\]
           
\item $ \tm= (\llam{\lvartwo}{\tm_1}{\lab})\lsctx\,\tm_2$.

        \[\begin{array}{rll}
            & \mult{\var}{(\llam{\lvartwo}{\tm_1}{\lab})\lsctx\,\tm_2}+\mult{\lvar}{(\llam{\lvartwo}{\tm_1}{\lab})\lsctx\,\tm_2}\times \mult{\var}{\tmtwo} \\
            = & \mult{\var}{(\lam{\lvartwo}{\tm_1})\lsctx} + \mult{\var}{\tm_2} + \mult{\lvartwo}{\tm_1}\times \mult{\var}{\tm_2} +\mult{\lvar}{(\llam{\lvartwo}{\tm_1}{\lab})\lsctx\,\tm_2}\times \mult{\var}{\tmtwo} &   (\rdef{variable_multiplicity}) \\
            = & \mult{\var}{(\lam{\lvartwo}{\tm_1})\lsctx} + \mult{\var}{\tm_2} + \mult{\lvartwo}{\tm_1}\times \mult{\var}{\tm_2} +(\mult{\lvar}{(\lam{\lvartwo}{\tm_1})\lsctx} + \mult{\lvar}{\tm_2} + \mult{\lvartwo}{\tm_1}\times  \mult{\lvar}{\tm_2})\times \mult{\var}{\tmtwo} &   (\rdef{variable_multiplicity}) \\
            = & \mult{\var}{((\lam{\lvartwo}{\tm_1})\lsctx)\sub{\lvar}{\tmtwo}} + \mult{\var}{\tm_2} + \mult{\lvartwo}{\tm_1}\times \mult{\var}{\tm_2} +(\mult{\lvar}{\tm_2} + \mult{\lvartwo}{\tm_1}\times  \mult{\lvar}{\tm_2})\times \mult{\var}{\tmtwo} & (\ih) \\
            = & \mult{\var}{((\lam{\lvartwo}{\tm_1})\lsctx)\sub{\lvar}{\tmtwo}} + \mult{\var}{\tm_2\sub{\lvar}{\tmtwo}} + \mult{\lvartwo}{\tm_1}\times \mult{\var}{\tm_2} +\mult{\lvartwo}{\tm_1}\times  \mult{\lvar}{\tm_2}\times \mult{\var}{\tmtwo} & (\ih) \\
            = &  \mult{\var}{(\lam{\lvartwo}{\tm_1 \sub{\lvar}{\tmtwo}}) \lsctx\sub{\lvar}{\tmtwo}} +  \mult{\var}{\tm_2 \sub{\lvar}{\tmtwo}}  +  \mult{\lvartwo}{\tm_1} \times  \mult{\var}{\tm_2 \sub{\lvar}{\tmtwo}} &   (\rdef{variable_multiplicity}) \\
            = &  \mult{\var}{(\lam{\lvartwo}{\tm_1 \sub{\lvar}{\tmtwo}}) \lsctx\sub{\lvar}{\tmtwo}} +  \mult{\var}{\tm_2 \sub{\lvar}{\tmtwo}}  +  \mult{\lvartwo}{\tm_1 \sub{\lvar}{\tmtwo}} \times  \mult{\var}{\tm_2 \sub{\lvar}{\tmtwo}} &   (\rlem{multiplicity_and_linear_substitution_non_capturing}(1)) \\
            = &  \mult{\var}{(\llam{\lvartwo}{\tm_1 \sub{\lvar}{\tmtwo}}{\lab})\lsctx\sub{\lvar}{\tmtwo}\,\tm_2 \sub{\lvar}{\tmtwo}} &   (\rdef{variable_multiplicity}) \\
            = &  \mult{\var}{((\llam{\lvartwo}{\tm_1}{\lab})\lsctx\,\tm_2)\sub{\lvar}{\tmtwo}} &   (\rdef{variable_multiplicity})
          \end{array}\]
        
      \item $\tm=\sha{\tm_1}$.
        
        \[\begin{array}{rll}
            & \mult{\var}{\sha{\tm_1}}+\mult{\lvar}{\sha{\tm_1}}\times \mult{\var}{\tmtwo} \\
            =  & \mult{\var}{\tm_1} + \mult{\lvar}{\tm_1}\times \mult{\var}{\tmtwo} & (\rdef{variable_multiplicity})\\
            =   & \mult{\var}{\tm_1\sub{\lvar}{\tmtwo}} & (\ih) \\
            =  & \mult{\var}{\sha{\tm_1}\sub{\lvar}{\tmtwo}} & (\rdef{variable_multiplicity})\\
            =  & \mult{\var}{(\sha{\tm_1})\sub{\lvar}{\tmtwo}} & \\
          \end{array}\]
        
        
      \item $\tm=\lopen{\tm_1}{(\lab)}$. 

          \[\begin{array}{rll}
            & \mult{\var}{\lopen{\tm_1}{(\lab)}} +\mult{\lvar}{\lopen{\tm_1}{(\lab)}} \times \mult{\var}{\tmtwo} \\
            =  & \mult{\var}{\tm_1} +\mult{\lvar}{\tm_1}\ \times \mult{\var}{\tmtwo} &   (\rdef{variable_multiplicity})\\
            =  & \mult{\var}{\tm_1\sub{\lvar}{\tmtwo}} & (\ih) \\
            =  & \mult{\var}{\lopen{\tm_1\sub{\lvar}{\tmtwo}}{(\lab)}} &   (\rdef{variable_multiplicity})\\
            =  & \mult{\var}{(\lopen{\tm_1}{(\lab)})\sub{\lvar}{\tmtwo}} & \\
            \end{array}\]
          
\item $\tm=\ofc{\tm_1}$. Same as case $\tm=\sha{\tm_1}$.
        
      \item $\tm=  \tm_1\lesub{\uvar}{\tm_2}{(\lab)}$

                \[\begin{array}{rll}
            & \mult{\var}{\tm_1\lesub{\uvar}{\tm_2}{(\lab)}}+\mult{\lvar}{\tm_1\lesub{\uvar}{\tm_2}{(\lab)}}\times \mult{\var}{\tmtwo} \\
                   = &  \mult{\var}{\tm_1}+\mult{\var}{\tm_2} + \mult{\uvar}{\tm_1}\times \mult{\var}{\tm_2} + \mult{\lvar}{\tm_1\lesub{\uvar}{\tm_2}{(\lab)}}\times \mult{\var}{\tmtwo} & (\rdef{variable_multiplicity})\\
                   = &  \mult{\var}{\tm_1}+\mult{\var}{\tm_2} + \mult{\uvar}{\tm_1}\times \mult{\var}{\tm_2} + (\mult{\lvar}{\tm_1}  + \mult{\lvar}{\tm_2} + \mult{\uvar}{\tm_1}\times \mult{\lvar}{\tm_2}) \times \mult{\var}{\tmtwo} & (\rdef{variable_multiplicity})\\
                   =  &  \mult{\var}{\tm_1\sub{\lvar}{\tmtwo}} + \mult{\var}{\tm_2} + \mult{\uvar}{\tm_1}\times \mult{\var}{\tm_2} +  (\mult{\lvar}{\tm_2} + \mult{\uvar}{\tm_1}\times \mult{\lvar}{\tm_2}) \times \mult{\var}{\tmtwo}  & (\ih)\\
                    =  &  \mult{\var}{\tm_1\sub{\lvar}{\tmtwo}} + \mult{\var}{\tm_2 \sub{\lvar}{\tmtwo}} + \mult{\uvar}{\tm_1}\times \mult{\var}{\tm_2} +  \mult{\uvar}{\tm_1}\times \mult{\lvar}{\tm_2} \times \mult{\var}{\tmtwo}  & (\ih)\\
                    = &  \mult{\var}{\tm_1 \sub{\lvar}{\tmtwo}} + \mult{\var}{\tm_2 \sub{\lvar}{\tmtwo}} + \mult{\uvar}{\tm_1}\times \mult{\var}{\tm_2 \sub{\lvar}{\tmtwo}} & (\ih)\\
                    = &  \mult{\var}{\tm_1 \sub{\lvar}{\tmtwo}} + \mult{\var}{\tm_2 \sub{\lvar}{\tmtwo}} + \mult{\uvar}{\tm_1\sub{\lvar}{\tmtwo}}\times \mult{\var}{\tm_2 \sub{\lvar}{\tmtwo}} & (\rlem{multiplicity_and_linear_substitution_non_capturing}(1))\\
            = &  \mult{\var}{\tm_1 \sub{\lvar}{\tmtwo}\lesub{\uvar}{\tm_2 \sub{\lvar}{\tmtwo}}{(\lab)}} &   (\rdef{variable_multiplicity}) \\
            = &  \mult{\var}{\tm_1\lesub{\uvar}{\tm_2}{(\lab)}\sub{\lvar}{\tmtwo}} &   
          \end{array}\]
      \end{xenumerate}
      \end{ifLongAppendix}
\end{proof}

\begin{lemma}
\llem{multiplicity_and_linear_beta} 
Suppose $\dom{\sctx}\cap \fv{\tmtwo}=\emptyset$. Then  \[\mult{\var}{(\lam{\lvar}{\tm})\lsctx}+\mult{\lvar}{\tm}\times \mult{\var}{\tmtwo}
  =
  \multctx{\var}{\mult{\bullet}{\tm\sub{\lvar}{\tmtwo}}}{\lsctx}\]
\end{lemma}

\begin{proof}
  By induction on $\lsctx$.
  \begin{ifLongAppendix}
Without loss of generality, we assume $\var\neq\lvar$.
  \begin{xenumerate}
  \item $\lsctx=\ctxhole$.


        \[\begin{array}{rll}
       & \mult{\var}{\lam{\lvar}{\tm}}+\mult{\lvar}{\tm}\times \mult{\var}{\tmtwo} \\
        = & \mult{\var}{\tm}+ \mult{\lvar}{\tm}\times \mult{\var}{\tmtwo} & (\rdef{variable_multiplicity}) \\
       = &   \mult{\var}{\tm\sub{\lvar}{\tmtwo}} & (\rlem{multiplicity_and_linear_substitution})\\
       = &  \multctx{\var}{\mult{\bullet}{\tm\sub{\lvar}{\tmtwo}}}{\ctxhole} & (\rdef{variable_multiplicity})
      \end{array}
    \]
    
  
\item $\lsctx=\lsctx_1\lesub{\uvar}{\tmthree}{(\lab)}$.

    \[\begin{array}{rll}
        & \mult{\var}{(\lam{\lvar}{\tm}) \lsctx_1\lesub{\uvar}{\tmthree}{(\lab)}}+ \mult{\lvar}{\tm}\times \mult{\var}{\tmtwo}\\
        = & \mult{\var}{(\lam{\lvar}{\tm}) \lsctx_1} + \mult{\var}{\tmthree} + \mult{\uvar}{(\lam{\lvar}{\tm}) \lsctx_1} \times \mult{\var}{\tmthree}+ \mult{\lvar}{\tm}\times \mult{\var}{\tmtwo} & (\rdef{variable_multiplicity}) \\
        = & \multctx{\var}{\mult{\bullet}{\tm\sub{\lvar}{\tmtwo}}}{\lsctx_1} + \mult{\var}{\tmthree}+ \mult{\uvar}{(\lam{\lvar}{\tm}) \lsctx_1} \times \mult{\var}{\tmthree}& (\ih)\\
       = &  \multctx{\var}{\mult{\bullet}{\tm\sub{\lvar}{\tmtwo}}}{\lsctx_1} + \mult{\var}{\tmthree} + \multctx{\uvar}{\mult{\bullet}{\lam{\lvar}{\tm}}}{\lsctx_1}\times \mult{\var}{\tmthree} & (\rlem{multiplicity_of_substitution_contexts})
\\
       = &  \multctx{\var}{\mult{\bullet}{\tm\sub{\lvar}{\tmtwo}}}{\lsctx_1} + \mult{\var}{\tmthree} +\multctx{\uvar}{\mult{\bullet}{\tm}}{\lsctx_1}\times \mult{\var}{\tmthree} & (\rdef{variable_multiplicity})
\\
       = &  \multctx{\var}{\mult{\bullet}{\tm\sub{\lvar}{\tmtwo}}}{\lsctx_1} + \mult{\var}{\tmthree} + \multctx{\uvar}{\mult{\bullet}{\tm\sub{\lvar}{\tmtwo}}}{\lsctx_1}\times \mult{\var}{\tmthree} & (\rlem{multiplicity_and_linear_substitution_non_capturing}(2))
\\
       = &  \multctx{\var}{\mult{\bullet}{\tm\sub{\lvar}{\tmtwo}}}{\lsctx_1\lesub{\uvar}{\tmthree}{(\lab)}} & (\rdef{variable_multiplicity})
      \end{array}
    \]

  \end{xenumerate}
 \end{ifLongAppendix} 
\end{proof}

\begin{lemma}
\llem{multiplicity_of_term_context_split}
Suppose $\dom{\sctx}\cap \fv{\tm}=\emptyset$. For any $\tmtwo$,
\begin{xenumerate}
  \item $\mult{\var}{\tm\sctx}\leq \mult{\var}{\tm}+\mult{\var}{\tmtwo\sctx}$.
   \item $\lmult{\tm\lsctx} \leq \lmult{\tm} + \lmultctx{\mult{\bullet}{\tmtwo}}{\lsctx}$
\end{xenumerate}

  \end{lemma}

\begin{proof}
  Both items are by induction on $\lsctx$.
  \begin{ifLongAppendix}
  For the first item we have:

  \begin{xenumerate}

  \item $\lsctx =\ctxhole$. Then $\mult{\var}{\tm}\leq \mult{\var}{\tm}+\mult{\var}{\tmtwo}$ is immediate.

    \item $\lsctx = \lsctx_1\lesub{\uvar}{\tmthree}{(\lab)}$
      \[\begin{array}{rll}
          & \mult{\var}{\tm \lsctx_1\lesub{\uvar}{\tmthree}{(\lab)}} \\
          = & \mult{\var}{\tm\lsctx_1} + \mult{\var}{\tmthree} + \mult{\uvar}{\tm \lsctx_1} \times \mult{\var}{\tmthree} & (\rdef{variable_multiplicity})\\
          \leq & \mult{\var}{\tm}+\mult{\var}{\tmtwo\lsctx_1} + \mult{\var}{\tmthree} + \mult{\uvar}{\tm \lsctx_1} \times \mult{\var}{\tmthree} & (\ih)\\
          \leq & \mult{\var}{\tm}+\mult{\var}{\tmtwo\lsctx_1} + \mult{\var}{\tmthree} + (\mult{\uvar}{\tm}+\mult{\uvar}{\tmtwo\lsctx_1})\times \mult{\var}{\tmthree} & (\ih)\\
          = & \mult{\var}{\tm}+\mult{\var}{\tmtwo\lsctx_1} + \mult{\var}{\tmthree} + \mult{\uvar}{\tm}\times \mult{\var}{\tmthree}+\mult{\uvar}{\tmtwo\lsctx_1}\times \mult{\var}{\tmthree} & (\rdef{variable_multiplicity}) \\
          = & \mult{\var}{\tm}+\mult{\var}{\tmtwo\sctx} + \mult{\uvar}{\tm}\times \mult{\var}{\tmthree} &  (\rdef{variable_multiplicity}) \\
          =  & \mult{\var}{\tm}+\mult{\var}{\tmtwo\sctx} & (\uvar\notin\fv{\tm})
        \end{array}\]
      
    \end{xenumerate}

    For the second item we proceed as follows:
      \begin{xenumerate}

  \item $\lsctx =\ctxhole$. Then $\lmult{\tm}= \lmult{\tm}+ 0 = \lmult{\tm}+ \lmultctx{\mult{\bullet}{\tmtwo}}{\ctxhole}$.

    \item $\lsctx = \lsctx_1\esub{\uvar}{\tmthree}$
      \[\begin{array}{rll}
          & \lmult{\tm \lsctx_1\esub{\uvar}{\tmthree}} \\
          = & \lmult{\tm\lsctx_1} + \lmult{\tmthree} + \mult{\uvar}{\tm \lsctx_1} \times \lmult{\tmthree} + \mult{\uvar}{\tm \lsctx_1} & (\rdef{variable_multiplicity})\\
          \leq & \lmult{\tm} + \lmultctx{\mult{\bullet}{\tmtwo}}{\lsctx_1}+  \lmult{\tmthree} + \mult{\uvar}{\tm \lsctx_1} \times \lmult{\tmthree} + \mult{\uvar}{\tm \lsctx_1}& (\ih)\\
          \leq & \lmult{\tm}+\lmultctx{\mult{\bullet}{\tmtwo}}{\lsctx_1}+ \lmult{\tmthree} + (\mult{\uvar}{\tm}+\mult{\uvar}{\tmtwo\lsctx_1})\times \lmult{\tmthree} + \mult{\uvar}{\tm \lsctx_1}& (\text{item 1})\\
                 \leq & \lmult{\tm}+\lmultctx{\mult{\bullet}{\tmtwo}}{\lsctx_1}+ \lmult{\tmthree} + (\mult{\uvar}{\tm}+\multctx{\uvar}{\mult{\bullet}{\tmtwo}}{\lsctx_1})\times \lmult{\tmthree} + \mult{\uvar}{\tm \lsctx_1}& (\rlem{multiplicity_of_substitution_contexts})\\ 
          =  & \lmult{\tm}+\lmultctx{\mult{\bullet}{\tmtwo}}{\lsctx_1}+ \lmult{\tmthree} + \multctx{\uvar}{\mult{\bullet}{\tmtwo}}{\lsctx_1}\times \lmult{\tmthree} + \mult{\uvar}{\tm \lsctx_1}& (\uvar\notin\fv{\tm})\\
          \leq  & \lmult{\tm}+\lmultctx{\mult{\bullet}{\tmtwo}}{\lsctx_1}+ \lmult{\tmthree} + \multctx{\uvar}{\mult{\bullet}{\tmtwo}}{\lsctx_1}\times \lmult{\tmthree} + \mult{\uvar}{\tm}+\mult{\uvar}{\tmtwo\lsctx_1}& (\text{item 1})\\
          \leq  & \lmult{\tm}+\lmultctx{\mult{\bullet}{\tmtwo}}{\lsctx_1}+ \lmult{\tmthree} + \multctx{\uvar}{\mult{\bullet}{\tmtwo}}{\lsctx_1}\times \lmult{\tmthree} + \mult{\uvar}{\tmtwo\lsctx_1} & (\uvar\notin\fv{\tm})\\
          = & \lmult{\tm}+\lmultctx{\mult{\bullet}{\tmtwo}}{\lsctx_1} + \lmult{\tmthree} + \multctx{\uvar}{\mult{\bullet}{\tmtwo}}{\lsctx_1}\times \lmult{\tmthree}+\multctx{\uvar}{\mult{\bullet}{\tmtwo}}{\lsctx_1}& (\rlem{multiplicity_of_substitution_contexts} ) \\
          =  & \lmult{\tm} + \lmultctx{\mult{\bullet}{\tmtwo}}{\lsctx_1\esub{\uvar}{\tmthree}} & (\rdef{variable_multiplicity})
        \end{array}\]

          \item $\lsctx = \lsctx_1\lesub{\uvar}{\tmthree}{\lab}$
      \[\begin{array}{rll}
          & \lmult{\tm \lsctx_1\lesub{\uvar}{\tmthree}{\lab}} \\
          = & 1 + \lmult{\tm\lsctx_1} + \lmult{\tmthree} & (\rdef{variable_multiplicity})\\
          \leq  & \lmult{\tm}+1 + \lmultctx{\mult{\bullet}{\tmtwo}}{\lsctx_1} + \lmult{\tmthree} & (\ih) \\
          =  & \lmult{\tm} + \lmultctx{\mult{\bullet}{\tmtwo}}{\lsctx_1\lesub{\uvar}{\tmthree}{\lab}} & (\rdef{variable_multiplicity})
        \end{array}\]

    \end{xenumerate}
    \end{ifLongAppendix}
\end{proof}

\begin{lemma}
\llem{multiplicity_and_replacement}
  Let $\var\neq\uvar$ and $\uvar\notin\fv{\tm}$. Then
  \[\mult{\var}{\off{\lgctx}{\ann{\uvar}{\lab}}} + \mult{\uvar}{\off{\lgctx}{\ann{\uvar}{\lab}}}\times \mult{\var}{\tm}
  =
  \mult{\var}{\off{\lgctx}{\tm}} + \mult{\uvar}{\off{\lgctx}{\tm}}\times \mult{\var}{\tm}
  \]
\end{lemma}

\begin{proof}
By induction on the size of the labeled context $\lgctx$.
\begin{ifLongAppendix}
\begin{xenumerate}
\item  $\lgctx=\ctxhole$.
  \[\begin{array}{rll}
      & \mult{\var}{\ann{\uvar}{\lab}} + \mult{\uvar}{\ann{\uvar}{\lab}}\times \mult{\var}{\tm} \\
      = & 0 + \mult{\var}{\tm} & (\var\neq\uvar)\\
  = &
      \mult{\var}{\tm} + 0\times \mult{\var}{\tm} & \\
  = &
      \mult{\var}{\tm} + \mult{\uvar}{\tm}\times \mult{\var}{\tm} & (\rlem{multiplicity_and_free_variables})
    \end{array}
  \]

\item $\lgctx=\lam{\lvar}{\lgctx_1}$
  \[\begin{array}{rll}
      & \mult{\var}{\lam{\lvar}{\off{\lgctx_1}{\ann{\uvar}{\lab}}}} + \mult{\uvar}{\lam{\lvar}{\off{\lgctx_1}{\ann{\uvar}{\lab}}}}\times \mult{\var}{\tm} \\
      = & \mult{\var}{\off{\lgctx_1}{\ann{\uvar}{\lab}}} + \mult{\uvar}{\off{\lgctx_1}{\ann{\uvar}{\lab}}}\times \mult{\var}{\tm} & (\rdef{variable_multiplicity})\\
      = &
  \mult{\var}{\off{\lgctx_1}{\tm}} + \mult{\uvar}{\off{\lgctx_1}{\tm}}\times \mult{\var}{\tm} & (\ih)\\
      = &
  \mult{\var}{\lam{\lvar}{\off{\lgctx_1}{\tm}}} + \mult{\uvar}{\lam{\lvar}{\off{\lgctx_1}{\tm}}}\times \mult{\var}{\tm} & (\rdef{variable_multiplicity})
    \end{array}
  \]

  \item $\lgctx=\llam{\lvar}{\lgctx_1}{\labtwo}$. Not possible since $\off{\lgctx}{\ann{\uvar}{\lab}}$ is well-labeled.

  \item $\lgctx=\lgctx_1\,\tmtwo$. We consider two cases depending on whether $\off{\lgctx_1}{\ann{\uvar}{\lab}}$ is a labeled abstraction or not.
  \begin{xenumerate}

  \item $\off{\lgctx_1}{\ann{\uvar}{\lab}}=(\llam{\lvar}{\tmthree}{\labtwo})\lsctx$. Then one of the following two cases hold:

    \begin{xenumerate}

      \item $\lgctx= (\llam{\lvar}{\lgctx_{11}}{\labtwo})\lsctx\,\tmtwo$. We assume without loss of generality, that $\lvar\notin\fv{\tm}$.

    \[\begin{array}{rll}
      & \mult{\var}{(\llam{\lvar}{\off{\lgctx_{11}}{\ann{\uvar}{\lab}}}{\labtwo})\lsctx\,\tmtwo} + \mult{\uvar}{(\llam{\lvar}{\off{\lgctx_{11}}{\ann{\uvar}{\lab}}}{\labtwo})\lsctx\,\tmtwo}\times \mult{\var}{\tm} \\
      = & \mult{\var}{(\lam{\lvar}{\off{\lgctx_{11}}{\ann{\uvar}{\lab}}})\lsctx} + \mult{\var}{\tmtwo} + \mult{\lvar}{\off{\lgctx_{11}}{\ann{\uvar}{\lab}}}\times \mult{\var}{\tmtwo} + \mult{\uvar}{(\llam{\lvar}{\off{\lgctx_{11}}{\ann{\uvar}{\lab}}}{\labtwo})\lsctx\,\tmtwo}\times \mult{\var}{\tm}& (\rdef{variable_multiplicity})\\
      = & \mult{\var}{(\lam{\lvar}{\off{\lgctx_{11}}{\ann{\uvar}{\lab}}})\lsctx} + \mult{\var}{\tmtwo} + \mult{\lvar}{\off{\lgctx_{11}}{\ann{\uvar}{\lab}}}\times \mult{\var}{\tmtwo} + (\mult{\uvar}{(\lam{\lvar}{\off{\lgctx_{11}}{\ann{\uvar}{\lab}}})\lsctx}+\mult{\uvar}{\tmtwo}+\mult{\lvar}{\off{\lgctx_{11}}{\ann{\uvar}{\lab}}}\times \mult{\uvar}{\tmtwo})\times \mult{\var}{\tm}& (\rdef{variable_multiplicity})\\
      = & \mult{\var}{(\lam{\lvar}{\off{\lgctx_{11}}{\tm}})\lsctx} + \mult{\uvar}{(\lam{\lvar}{\off{\lgctx_{11}}{\tm}})\lsctx}\times \mult{\var}{\tm} + \mult{\var}{\tmtwo} + \mult{\lvar}{\off{\lgctx_{11}}{\ann{\uvar}{\lab}}}\times \mult{\var}{\tmtwo} + (\mult{\uvar}{\tmtwo}+\mult{\lvar}{\off{\lgctx_{11}}{\ann{\uvar}{\lab}}}\times \mult{\uvar}{\tmtwo})\times \mult{\var}{\tm}& (\ih)\\
      = & \mult{\var}{(\lam{\lvar}{\off{\lgctx_{11}}{\tm}})\lsctx} + \mult{\uvar}{(\lam{\lvar}{\off{\lgctx_{11}}{\tm}})\lsctx}\times \mult{\var}{\tm} + \mult{\var}{\tmtwo} + \mult{\lvar}{\off{\lgctx_{11}}{\ann{\uvar}{\lab}}}\times \mult{\var}{\tmtwo} + \mult{\uvar}{\tmtwo}\times \mult{\var}{\tm}+\mult{\lvar}{\off{\lgctx_{11}}{\ann{\uvar}{\lab}}}\times \mult{\uvar}{\tmtwo}\times \mult{\var}{\tm}& \\
      = & \mult{\var}{(\lam{\lvar}{\off{\lgctx_{11}}{\tm}})\lsctx} + \mult{\uvar}{(\lam{\lvar}{\off{\lgctx_{11}}{\tm}})\lsctx}\times \mult{\var}{\tm} + \mult{\var}{\tmtwo} + \mult{\lvar}{\off{\lgctx_{11}}{\tm}}\times \mult{\var}{\tmtwo} + \mult{\uvar}{\tmtwo}\times \mult{\var}{\tm}+\mult{\lvar}{\off{\lgctx_{11}}{\tm}}\times \mult{\uvar}{\tmtwo}\times \mult{\var}{\tm}& (\rlem{multiplicity_and_linear_substitution_non_capturing}(1))\\
      = & \mult{\var}{(\llam{\lvar}{\off{\lgctx_{11}}{\tm}}{\labtwo})\lsctx\,\tmtwo}+ \mult{\uvar}{(\lam{\lvar}{\off{\lgctx_{11}}{\tm}})\lsctx}\times \mult{\var}{\tm} + \mult{\uvar}{\tmtwo}\times \mult{\var}{\tm}+\mult{\lvar}{\off{\lgctx_{11}}{\tm}}\times \mult{\uvar}{\tmtwo}\times \mult{\var}{\tm}& (\rdef{variable_multiplicity}))\\
      = &
  \mult{\var}{(\llam{\lvar}{\off{\lgctx_{11}}{\tm}}{\labtwo})\lsctx\,\tmtwo} + \mult{\uvar}{(\llam{\lvar}{\off{\lgctx_{11}}{\tm}}{\labtwo})\lsctx\,\tmtwo}\times \mult{\var}{\tm} & (\rdef{variable_multiplicity})
    \end{array} 
  \]

\item $\lgctx= (\llam{\lvar}{\tmthree}{\labtwo})\lsctx_1\lesub{\uvartwo}{\lgctx_{11}}{(\labthree)}\lsctx_2\,\tmtwo$. Similar to the previous case.

\end{xenumerate}

\item $\off{\lgctx_1}{\ann{\uvar}{\lab}}\neq (\llam{\lvar}{\tmthree}{\labtwo})\lsctx$. 
    \[\begin{array}{rll}
      & \mult{\var}{\off{\lgctx_1}{\ann{\uvar}{\lab}}\,\tmtwo} + \mult{\uvar}{\off{\lgctx_1}{\ann{\uvar}{\lab}}\,\tmtwo}\times \mult{\var}{\tm} \\
      = & \mult{\var}{\off{\lgctx_1}{\ann{\uvar}{\lab}}} +  \mult{\var}{\tmtwo} + (\mult{\uvar}{\off{\lgctx_1}{\ann{\uvar}{\lab}}}+  \mult{\uvar}{\tmtwo})\times \mult{\var}{\tm} & (\rdef{variable_multiplicity})\\
      = & \mult{\var}{\off{\lgctx_1}{\ann{\uvar}{\lab}}} +  \mult{\var}{\tmtwo} + \mult{\uvar}{\off{\lgctx_1}{\ann{\uvar}{\lab}}}\times \mult{\var}{\tm} +  \mult{\uvar}{\tmtwo}\times \mult{\var}{\tm} & \\
      = &
  \mult{\var}{\off{\lgctx_1}{\tm}} + \mult{\uvar}{\off{\lgctx_1}{\tm}}\times \mult{\var}{\tm} + \mult{\var}{\tmtwo} + \mult{\uvar}{\tmtwo}\times \mult{\var}{\tm} & (\ih)\\
      = &
  \mult{\var}{\off{\lgctx_1}{\tm}\,\tmtwo} + \mult{\uvar}{\off{\lgctx_1}{\tm}\,\tmtwo}\times \mult{\var}{\tm} & (\rdef{variable_multiplicity}) 
    \end{array}
  \]
  \end{xenumerate}

\item $\lgctx=\tmtwo\,\lgctx_1$.  Two cases are possible.
  \begin{xenumerate}
  \item $\tmtwo=(\llam{\lvar}{\tmthree}{\labtwo})\lsctx$. Similar to previous cases above.

  \item $\tmtwo\neq (\llam{\lvar}{\tmthree}{\labtwo})\lsctx$. Similar to previous cases above.
  \end{xenumerate}
\item $\lgctx=\sha{\lgctx_1}$. Similar to the case $\lgctx=\lam{\lvar}{\lgctx_1}$.

  
\item $\lgctx=\lopen{\lgctx_1}{(\labtwo)}$. We need to consider multiple cases since our induction proceeds on well-formed labeled contexts.

  \begin{xenumerate}

  \item $\labtwo$ is not present. Similar to the case $\lgctx=\lam{\lvar}{\lgctx_1}$.
  \[\begin{array}{rll}
      & \mult{\var}{\open{\off{\lgctx_1}{\ann{\uvar}{\lab}}}} + \mult{\uvar}{\open{\off{\lgctx_1}{\ann{\uvar}{\lab}}}}\times \mult{\var}{\tm} \\
      = & \mult{\var}{\off{\lgctx_1}{\ann{\uvar}{\lab}}} + \mult{\uvar}{\off{\lgctx_1}{\ann{\uvar}{\lab}}}\times \mult{\var}{\tm} & (\rdef{variable_multiplicity})\\
      = &
  \mult{\var}{\off{\lgctx_1}{\tm}} + \mult{\uvar}{\off{\lgctx_1}{\tm}}\times \mult{\var}{\tm} & (\ih)\\
      = &
  \mult{\var}{\open{\off{\lgctx_1}{\tm}}} + \mult{\uvar}{\open{\off{\lgctx_1}{\tm}}}\times \mult{\var}{\tm} & (\rdef{variable_multiplicity}) 
    \end{array}
  \]
  
  \item $\labtwo$ is present. Note that $\lgctx_1=\ctxhole $ and $\lgctx_1=\ctxhole\lsctx_1$ (with $\lsctx=\lsctx_1\lsctx_2$) is not possible since $\off{\lgctx}{\ann{\uvar}{\lab}}$ is well-labeled. The remaining cases are:
     \begin{xenumerate}

 \item $\lgctx_1=(\sha{\lgctx_2}) \lsctx$.  

     \[\begin{array}{rll}
      & \mult{\var}{\lopen{(\sha{\off{\lgctx_2}{\ann{\uvar}{\lab}}}) \lsctx}{\labtwo}} + \mult{\uvar}{\lopen{(\sha{\off{\lgctx_2}{\ann{\uvar}{\lab}}}) \lsctx}{\labtwo}}\times \mult{\var}{\tm} \\
      = & \mult{\var}{(\sha{\off{\lgctx_1}{\ann{\uvar}{\lab}}})\lsctx} + \mult{\uvar}{(\sha{\off{\lgctx_1}{\ann{\uvar}{\lab}}})\lsctx}\times \mult{\var}{\tm} & (\rdef{variable_multiplicity})\\
      = & \multctx{\var}{\mult{\bullet}{\sha{\off{\lgctx_1}{\ann{\uvar}{\lab}}}}}{\lsctx} + \multctx{\uvar}{\mult{\bullet}{\sha{\off{\lgctx_1}{\ann{\uvar}{\lab}}}}}{\lsctx}\times \mult{\var}{\tm} & (\rlem{multiplicity_of_substitution_contexts} )\\
      = & \multctx{\var}{\mult{\bullet}{\off{\lgctx_1}{\ann{\uvar}{\lab}}}}{\lsctx} + \multctx{\uvar}{\mult{\bullet}{\off{\lgctx_1}{\ann{\uvar}{\lab}}}}{\lsctx}\times \mult{\var}{\tm} & (\rdef{variable_multiplicity})\\
      = & \mult{\var}{(\sha{\off{\lgctx_1}{\tm}})\lsctx} + \mult{\uvar}{(\sha{\off{\lgctx_1}{\tm}})\lsctx}\times \mult{\var}{\tm} & (\ih)\\
      = &
  \mult{\var}{\lopen{(\sha{\off{\lgctx_2}{\tm}}) \lsctx}{\labtwo}} + \mult{\uvar}{\lopen{(\sha{\off{\lgctx_2}{\tm}}) \lsctx}{\labtwo}}\times \mult{\var}{\tm} & (\rdef{variable_multiplicity}) 
    \end{array}
  \]
  
 \item $\lgctx_1=(\sha{\tmtwo}) \lsctx_1\lesub{\uvartwo}{\lgctx_2}{(\labthree)}\lsctx_2$. Similar to the previous case.
    
    \end{xenumerate}
    
    \end{xenumerate}

\item $\lgctx=\ofc{\lgctx_1}$. Similar to the case $\lgctx=\lam{\lvar}{\lgctx_1}$.
  
\item $\lgctx=\lgctx_1\lesub{\uvartwo}{\tmtwo}{(\labtwo)}$

    \[\begin{array}{rll}
     & \mult{\var}{\off{\lgctx_1}{\ann{\uvar}{\lab}}\lesub{\uvartwo}{\tmtwo}{(\labtwo)}} + \mult{\uvar}{\off{\lgctx_1}{\ann{\uvar}{\lab}}\lesub{\uvartwo}{\tmtwo}{(\labtwo)}}\times \mult{\var}{\tm} \\
      = & \mult{\var}{\off{\lgctx_1}{\ann{\uvar}{\lab}}} + \mult{\var}{\tmtwo}+ \mult{\uvartwo}{\off{\lgctx_1}{\ann{\uvar}{\lab}}} \times \mult{\var}{\tmtwo} +  \mult{\uvar}{\off{\lgctx_1}{\ann{\uvar}{\lab}}\lesub{\uvartwo}{\tmtwo}{(\labtwo)}}\times \mult{\var}{\tm} & (\rdef{variable_multiplicity})\\
      = & \mult{\var}{\off{\lgctx_1}{\ann{\uvar}{\lab}}} + \mult{\var}{\tmtwo}+ \mult{\uvartwo}{\off{\lgctx_1}{\ann{\uvar}{\lab}}} \times \mult{\var}{\tmtwo} +  (\mult{\uvar}{\off{\lgctx_1}{\ann{\uvar}{\lab}}} + \mult{\uvar}{\tmtwo}+ \mult{\uvartwo}{\off{\lgctx_1}{\ann{\uvar}{\lab}}} \times \mult{\uvar}{\tmtwo} )\times \mult{\var}{\tm} & (\rdef{variable_multiplicity})\\
      = &
   \mult{\var}{\off{\lgctx_1}{\tm}} + \mult{\uvar}{\off{\lgctx_1}{\tm}}\times \mult{\var}{\tm} + \mult{\var}{\tmtwo}+ \mult{\uvartwo}{\off{\lgctx_1}{\ann{\uvar}{\lab}}} \times \mult{\var}{\tmtwo} +  (\mult{\uvar}{\tmtwo}+ \mult{\uvartwo}{\off{\lgctx_1}{\ann{\uvar}{\lab}}} \times \mult{\uvar}{\tmtwo} )\times \mult{\var}{\tm} & (\ih)\\
      = &
   \mult{\var}{\off{\lgctx_1}{\tm}} + \mult{\uvar}{\off{\lgctx_1}{\tm}}\times \mult{\var}{\tm} + \mult{\var}{\tmtwo}+ \mult{\uvartwo}{\off{\lgctx_1}{\tm}} \times \mult{\var}{\tmtwo} +  (\mult{\uvar}{\tmtwo}+ \mult{\uvartwo}{\off{\lgctx_1}{\ann{\uvar}{\lab}}} \times \mult{\uvar}{\tmtwo} )\times \mult{\var}{\tm} & (\rlem{multiplicity_and_non_capturing_replacement},\star)\\ 
        = &
  \mult{\var}{\off{\lgctx_1}{\tm}\lesub{\uvartwo}{\tmtwo}{(\labtwo)}} +    \mult{\uvar}{\off{\lgctx_1}{\tm}}\times \mult{\var}{\tm} +  (\mult{\uvar}{\tmtwo}+ \mult{\uvartwo}{\off{\lgctx_1}{\ann{\uvar}{\lab}}} \times \mult{\uvar}{\tmtwo} )\times \mult{\var}{\tm} & (\rdef{variable_multiplicity}) \\
        = &
  \mult{\var}{\off{\lgctx_1}{\tm}\lesub{\uvartwo}{\tmtwo}{(\labtwo)}} +    \mult{\uvar}{\off{\lgctx_1}{\tm}}\times \mult{\var}{\tm} +  \mult{\uvar}{\tmtwo}\times \mult{\var}{\tm}+ \mult{\uvartwo}{\off{\lgctx_1}{\ann{\uvar}{\lab}}} \times \mult{\uvar}{\tmtwo} \times \mult{\var}{\tm}& \\
        = &
  \mult{\var}{\off{\lgctx_1}{\tm}\lesub{\uvartwo}{\tmtwo}{(\labtwo)}} +    \mult{\uvar}{\off{\lgctx_1}{\tm}}\times \mult{\var}{\tm} +  \mult{\uvar}{\tmtwo}\times \mult{\var}{\tm}+ \mult{\uvartwo}{\off{\lgctx_1}{\tm}} \times \mult{\uvar}{\tmtwo} \times \mult{\var}{\tm} & (\rlem{multiplicity_and_non_capturing_replacement},\star)\\
      = &
  \mult{\var}{\off{\lgctx_1}{\tm}\lesub{\uvartwo}{\tmtwo}{(\labtwo)}} + \mult{\uvar}{\off{\lgctx_1}{\tm}\lesub{\uvartwo}{\tmtwo}{(\labtwo)}}\times \mult{\var}{\tm} & (\rdef{variable_multiplicity}) 
    \end{array}
  \]

  Note the use in $\star$ of the fact that $\off{\lgctx}{\tm}$ stands for the capture-avoiding replacement of $\ctxhole$ in $\lgctx$ by $\tm$. In particular, $\uvartwo\notin\fv{\tm}$ and $\uvar\neq\uvartwo$.


\item $\lgctx=\tmtwo\lesub{\uvartwo}{\lgctx_1}{(\labtwo)}$

    \[\begin{array}{rll}
     & \mult{\var}{\tmtwo\lesub{\uvartwo}{\off{\lgctx_1}{\ann{\uvar}{\lab}}}{(\labtwo)}} + \mult{\uvar}{\tmtwo\lesub{\uvartwo}{\off{\lgctx_1}{\ann{\uvar}{\lab}}}{(\labtwo)}}\times \mult{\var}{\tm} \\
      = & \mult{\var}{\tmtwo}+ \mult{\var}{\off{\lgctx_1}{\ann{\uvar}{\lab}}} +  \mult{\uvartwo}{\tmtwo}\times \mult{\var}{\off{\lgctx_1}{\ann{\uvar}{\lab}}} + \mult{\uvar}{\tmtwo\lesub{\uvartwo}{\off{\lgctx_1}{\ann{\uvar}{\lab}}}{(\labtwo)}}\times \mult{\var}{\tm}  & (\rdef{variable_multiplicity})\\
      = &\mult{\var}{\tmtwo}+ \mult{\var}{\off{\lgctx_1}{\ann{\uvar}{\lab}}} +  \mult{\uvartwo}{\tmtwo}\times \mult{\var}{\off{\lgctx_1}{\ann{\uvar}{\lab}}} +  (\mult{\uvar}{\tmtwo} + \mult{\uvar}{\off{\lgctx_1}{\ann{\uvar}{\lab}}} + \mult{\uvartwo}{\tmtwo}\times \mult{\uvar}{\off{\lgctx_1}{\ann{\uvar}{\lab}}})\times \mult{\var}{\tm} & (\rdef{variable_multiplicity})\\
        = &\mult{\var}{\tmtwo}+ \mult{\var}{\off{\lgctx_1}{\ann{\uvar}{\lab}}} +  \mult{\uvartwo}{\tmtwo}\times \mult{\var}{\off{\lgctx_1}{\ann{\uvar}{\lab}}} +  \mult{\uvar}{\tmtwo} \times \mult{\var}{\tm} + \mult{\uvar}{\off{\lgctx_1}{\ann{\uvar}{\lab}}} \times \mult{\var}{\tm}+ \mult{\uvartwo}{\tmtwo}\times \mult{\uvar}{\off{\lgctx_1}{\ann{\uvar}{\lab}}}\times \mult{\var}{\tm} & \\
       = &\mult{\var}{\tmtwo}+ \mult{\var}{\off{\lgctx_1}{\ann{\uvar}{\lab}}} +  \mult{\uvartwo}{\tmtwo}\times (\mult{\var}{\off{\lgctx_1}{\tm}} +   \mult{\uvar}{\off{\lgctx_1}{\tm}}\times \mult{\var}{\tm}) +  \mult{\uvar}{\tmtwo} \times \mult{\var}{\tm} + \mult{\uvar}{\off{\lgctx_1}{\ann{\uvar}{\lab}}} \times \mult{\var}{\tm} & \\

        = &\mult{\var}{\tmtwo}+ \mult{\var}{\off{\lgctx_1}{\ann{\uvar}{\lab}}} +  \mult{\uvartwo}{\tmtwo}\times \mult{\var}{\off{\lgctx_1}{\tm}} +  \mult{\uvartwo}{\tmtwo}\times \mult{\uvar}{\off{\lgctx_1}{\tm}}\times \mult{\var}{\tm} +  \mult{\uvar}{\tmtwo} \times \mult{\var}{\tm} + \mult{\uvar}{\off{\lgctx_1}{\ann{\uvar}{\lab}}} \times \mult{\var}{\tm} & \\

        = &\mult{\var}{\tmtwo}+ \mult{\var}{\off{\lgctx_1}{\tm}}  + \mult{\uvar}{\off{\lgctx_1}{\tm}}\times \mult{\var}{\tm} +  \mult{\uvartwo}{\tmtwo}\times \mult{\var}{\off{\lgctx_1}{\tm}} +  \mult{\uvartwo}{\tmtwo}\times \mult{\uvar}{\off{\lgctx_1}{\tm}}\times \mult{\var}{\tm} +  \mult{\uvar}{\tmtwo} \times \mult{\var}{\tm} & \\

               = & \mult{\var}{\tmtwo\lesub{\uvartwo}{\off{\lgctx_1}{\tm}}{(\labtwo)}}  + \mult{\uvar}{\off{\lgctx_1}{\tm}}\times \mult{\var}{\tm} +  \mult{\uvartwo}{\tmtwo}\times  \mult{\uvar}{\off{\lgctx_1}{\tm}}\times \mult{\var}{\tm} +  \mult{\uvar}{\tmtwo} \times \mult{\var}{\tm} & \\
      = &
  \mult{\var}{\tmtwo\lesub{\uvartwo}{\off{\lgctx_1}{\tm}}{(\labtwo)}} + \mult{\uvar}{\tmtwo\lesub{\uvartwo}{\off{\lgctx_1}{\tm}}{(\labtwo)}}\times \mult{\var}{\tm} & (\rdef{variable_multiplicity}) 
    \end{array}
  \]
  
\end{xenumerate}

\end{ifLongAppendix}  
\end{proof}

\begin{lemma}
\llem{context_multiplicity_and_replacement}
Suppose $\dom{\lsctx}\cap (\fv{\lgctx} \cup\{\uvar\})=\emptyset$ and $\var\neq\uvar$ and $\uvar\notin\fv{\tm}$.
\[\mult{\var}{\off{\lgctx}{\ann{\uvar}{\lab}}}+\mult{\var}{\tm\lsctx} + \mult{\uvar}{\off{\lgctx}{\ann{\uvar}{\lab}}}\times \mult{\var}{\tm\lsctx} 
        \geq    \multctx{\var}{\mult{\bullet}{\off{\lgctx}{\tm}\esub{\uvar}{\tm}}}{\lsctx}\]
\end{lemma}

\begin{proof}
By induction on $\lsctx$.
\begin{ifLongAppendix}
  \begin{xenumerate}
\item $\lsctx=  \ctxhole$.

  \[\begin{array}{rll}
     & \mult{\var}{\off{\lgctx}{\ann{\uvar}{\lab}}}+\mult{\var}{\tm} + \mult{\uvar}{\off{\lgctx}{\ann{\uvar}{\lab}}}\times \mult{\var}{\tm} &\ \\
      = &    \mult{\var}{\off{\lgctx}{\tm}}+\mult{\var}{\tm} + \mult{\uvar}{\off{\lgctx}{\tm}}\times \mult{\var}{\tm} & (\rlem{multiplicity_and_replacement})\\
      = &    \mult{\var}{\off{\lgctx}{\tm}\esub{\uvar}{\tm}} & (\rdef{variable_multiplicity})\\
      = &    \multctx{\var}{\mult{\bullet}{\off{\lgctx}{\tm}\esub{\uvar}{\tm}}}{\ctxhole} & (\rdef{variable_multiplicity})
    \end{array}
  \]

\item $\lsctx=  \lsctx_1\lesub{\uvartwo}{\tmtwo}{(\lab)}$

  \[\begin{array}{rll}
     & \mult{\var}{\off{\lgctx}{\ann{\uvar}{\lab}}}+\mult{\var}{\tm \lsctx_1\lesub{\uvartwo}{\tmtwo}{(\lab)}} + \mult{\uvar}{\off{\lgctx}{\ann{\uvar}{\lab}}}\times \mult{\var}{\tm \lsctx_1\lesub{\uvartwo}{\tmtwo}{(\lab)}}  &\ \\
     = & \mult{\var}{\off{\lgctx}{\ann{\uvar}{\lab}}}+\mult{\var}{\tm\lsctx_1} + \mult{\var}{\tmtwo} + \mult{\uvartwo}{\tm\lsctx_1}\times \mult{\var}{\tmtwo} + \mult{\uvar}{\off{\lgctx}{\ann{\uvar}{\lab}}}\times \mult{\var}{\tm \lsctx_1\lesub{\uvartwo}{\tmtwo}{(\lab)}}  & \\
      = & \mult{\var}{\off{\lgctx}{\ann{\uvar}{\lab}}}+\mult{\var}{\tm\lsctx_1} + \mult{\var}{\tmtwo} + \mult{\uvartwo}{\tm\lsctx_1}\times \mult{\var}{\tmtwo} + \mult{\uvar}{\off{\lgctx}{\ann{\uvar}{\lab}}}\times (\mult{\var}{\tm\lsctx_1} + \mult{\var}{\tmtwo} + \mult{\uvartwo}{\tm\lsctx_1}\times \mult{\var}{\tmtwo}) & \\
         \geq  & \multctx{\var}{\mult{\bullet}{\off{\lgctx}{\tm}\esub{\uvar}{\tm}}}{\lsctx_1}  + \mult{\var}{\tmtwo} + \mult{\uvartwo}{\tm\lsctx_1}\times \mult{\var}{\tmtwo} + \mult{\uvar}{\off{\lgctx}{\ann{\uvar}{\lab}}}\times (\mult{\var}{\tmtwo} + \mult{\uvartwo}{\tm\lsctx_1}\times \mult{\var}{\tmtwo}) &  (\ih) \\
         =  & \multctx{\var}{\mult{\bullet}{\off{\lgctx}{\tm}\esub{\uvar}{\tm}}}{\lsctx_1}  + \mult{\var}{\tmtwo} + \mult{\uvartwo}{\tm\lsctx_1}\times \mult{\var}{\tmtwo} + \mult{\uvar}{\off{\lgctx}{\ann{\uvar}{\lab}}}\times \mult{\var}{\tmtwo} + \mult{\uvar}{\off{\lgctx}{\ann{\uvar}{\lab}}}\times \mult{\uvartwo}{\tm\lsctx_1}\times \mult{\var}{\tmtwo} & \\
       >  & \multctx{\var}{\mult{\bullet}{\off{\lgctx}{\tm}\esub{\uvar}{\tm}}}{\lsctx_1}  + \mult{\var}{\tmtwo} + \mult{\uvartwo}{\tm\lsctx_1}\times \mult{\var}{\tmtwo} + \mult{\uvartwo}{\off{\lgctx}{\ann{\uvar}{\lab}}}\times \mult{\var}{\tmtwo} + \mult{\uvar}{\off{\lgctx}{\ann{\uvar}{\lab}}}\times \mult{\uvartwo}{\tm\lsctx_1}\times \mult{\var}{\tmtwo} & (\dom{\lsctx}\cap (\fv{\lgctx} \cup\{\uvar\})=\emptyset)\\
      \geq  &  \multctx{\var}{\mult{\bullet}{\off{\lgctx}{\tm}\esub{\uvar}{\tm}}}{\lsctx_1}  + \mult{\var}{\tmtwo} + \multctx{\uvartwo}{\mult{\bullet}{\off{\lgctx}{\tm}\esub{\uvar}{\tm}}}{\lsctx_1}\times \mult{\var}{\tmtwo} & (\ih)\\
      = &    \multctx{\var}{\mult{\bullet}{\off{\lgctx}{\tm}\esub{\uvar}{\tm}}}{\lsctx_1\lesub{\uvartwo}{\tmtwo}{(\lab)}} & (\rdef{variable_multiplicity})
    \end{array}
  \]

\end{xenumerate}
\end{ifLongAppendix}
\end{proof}

\begin{proposition}
\lprop{reduction_and_variable_multiplicity} 
Let $\tm\in\TermsSWL$. Then $\tm \lto{\lab}_R\tmtwo$ implies $\mult{\var}{\tm}\geq \mult{\var}{\tmtwo}$
\end{proposition}

\begin{proof}
First  we consider each of the four cases for reduction at the root $\tm \lrootto{\lab}_R\tmtwo$:
  \begin{xenumerate}

    \item  Suppose $\tm=   (\llam{\lvar}{\tm_1}{\lab})\lsctx\,\tm_2
    \rtoSdbL{\lab} 
    \tm_1\sub{\lvar}{\tm_2}\lsctx = \tmtwo$ and $\fv{\tm_2} \cap \dom{\lsctx} = \emptyset$.

    \[\begin{array}{rll}
        &  \mult{\var}{ (\llam{\lvar}{\tm_1}{\lab})\lsctx\,\tm_2} \\
        = &  \mult{\var}{(\lam{\lvar}{\tm_1})\lsctx} + \mult{\var}{\tm_2} + \mult{\lvar}{\tm_1}\times \mult{\var}{\tm_2} & (\rdef{variable_multiplicity})\\
        \geq &  \mult{\var}{(\lam{\lvar}{\tm_1})\lsctx} + \mult{\lvar}{\tm_1}\times \mult{\var}{\tm_2} & \\
        =  &    \multctx{\var}{\mult{\bullet}{\tm_1\sub{\lvar}{\tm_2}}}{\lsctx} & (\rlem{multiplicity_and_linear_beta})\\
        =  &   \mult{\var}{\tm_1\sub{\lvar}{\tm_2}\lsctx} & (\rlem{multiplicity_of_substitution_contexts})
        \end{array}\]

   \item Suppose $\tm=\lopen{(\sha{\tm_1})\lsctx}{\lab}
     \rtoSopenL{\lab} 
     \tm_1\lsctx  = \tmtwo$

         \[\begin{array}{rll}
        &  \mult{\var}{\lopen{(\sha{\tm_1})\lsctx}{\lab}} \\
        = &  \mult{\var}{(\sha{\tm_1})\lsctx} & (\rdef{variable_multiplicity})\\
        =  &  \mult{\var}{\tm_1\lsctx} 
        \end{array}\]

     \item Suppose $\tm=\off{\lgctx}{\ann{\uvar}{\lab}}\esub{\uvar}{(\ofc{(\sha{\tm_1})\lsctx_1})\lsctx_2}
    \rtoSlsL{\lab} 
    \off{\lgctx}{(\sha{\tm_1}) \lsctx_1}\esub{\uvar}{\ofc{(\sha{\tm_1}) \lsctx_1}}\lsctx_2=\tmtwo$ and $\uvar \notin \fv{\tm_1}$ and $\fv{\lgctx} \cap \dom{\lsctx_1\lsctx_2} = \emptyset$.

                 \[\begin{array}{rll}
        &  \mult{\var}{\off{\lgctx}{\ann{\uvar}{\lab}}\esub{\uvar}{(\ofc{(\sha{\tm_1})\lsctx_1})\lsctx_2}} \\
        = &  \mult{\var}{\off{\lgctx}{\ann{\uvar}{\lab}}}+ \mult{\var}{(\ofc{(\sha{\tm_1})\lsctx_1})\lsctx_2}  + \mult{\uvar}{\off{\lgctx}{\ann{\uvar}{\lab}}}\times \mult{\var}{(\ofc{(\sha{\tm_1})\lsctx_1})\lsctx_2} & (\rdef{variable_multiplicity})\\
        \geq  &  \multctx{\var}{\mult{\bullet}{\off{\lgctx}{(\sha{\tm_1}) \lsctx_1}\esub{\uvar}{\ofc{(\sha{\tm_1}) \lsctx_1}}}}{\lsctx_2} & (\rlem{context_multiplicity_and_replacement})\\
        =  &  \mult{\var}{\off{\lgctx}{(\sha{\tm_1}) \lsctx_1}\esub{\uvar}{\ofc{(\sha{\tm_1}) \lsctx_1}}\lsctx_2} & (\rlem{multiplicity_of_substitution_contexts})
                   \end{array}
                 \]
                 
\item Suppose $\tm=\tm_1\esub{\ann{\uvar}{\lab}}{(\ofc{\tm_2})\lsctx}
    \rtoSgcL{\lab} 
    \tm_1\lsctx=\tmtwo$ and $\uvar\notin\fv{\tm_1}$.

             \[\begin{array}{rll}
        &  \mult{\var}{\tm_1\esub{\ann{\uvar}{\lab}}{(\ofc{\tm_2})\lsctx}} \\
        = &  \mult{\var}{\tm_1}+\mult{\var}{(\ofc{\tm_2})\lsctx}  + \mult{\uvar}{\tm_1}\times \mult{\var}{(\ofc{\tm_2})\lsctx} & (\rdef{variable_multiplicity})\\
        = &  \mult{\var}{\tm_1}+\mult{\var}{(\ofc{\tm_2})\lsctx} & (\rlem{multiplicity_and_free_variables}) \\
        \geq  &  \mult{\var}{\tm_1\lsctx} & (\rlem{multiplicity_of_term_context_split}(1)) \\
               \end{array}\]
\end{xenumerate}
\begin{ifShortAppendix}
  For the cases where reduction is internal, see the extended
  version~\cite{mells_long}.
\end{ifShortAppendix}
\begin{ifLongAppendix}
Next we consider internal reduction. We proceed by induction on the size of the labeled context $\lgctx$.
\begin{xenumerate}

\item $\lgctx=\lam{\lvar}{\lgctx_1}$. Suppose $\tm  = \lam{\lvar}{\off{\lgctx_1}{\tm_1}} \to_R \lam{\lvar}{\off{\lgctx_1}{\tmtwo_1}} = \tmtwo$ follows from $\tm_1 \to_R \tmtwo_1$.

      \[\begin{array}{rll}
        &  \mult{\var}{ \lam{\lvar}{\off{\lgctx_1}{\tm_1}} } \\
        = &  \mult{\var}{\off{\lgctx_1}{\tm_1} } & (\rdef{variable_multiplicity})\\
        \geq &  \mult{\var}{\off{\lgctx_1}{\tmtwo_1}} & (\ih)\\
        =  &   \mult{\var}{\lam{\lvar}{\off{\lgctx_1}{\tmtwo_1}} } & (\rdef{variable_multiplicity}) 
        \end{array}\]

\item $\lgctx=\llam{\lvar}{\lgctx_1}{\labtwo}$. Not possible since $\tm\in\TermsSWL$. 
\item $\lgctx= (\llam{\lvar}{\lgctx_1}{\labtwo})\lsctx\,\tm_1$. Suppose $\tm  = (\llam{\lvar}{\off{\lgctx_1}{\tm_2}}{\labtwo})\lsctx\,\tm_1 \to_R (\llam{\lvar}{\off{\lgctx_1}{\tmtwo_2}}{\labtwo})\lsctx\,\tm_1 = \tmtwo$ follows from $\tm_2 \to_R \tmtwo_2$.

       \[\begin{array}{rll}
        &  \mult{\var}{ (\llam{\lvar}{\off{\lgctx_1}{\tm_2}}{\labtwo})\lsctx\,\tm_1} \\
        = &  \mult{\var}{(\lam{\lvar}{\off{\lgctx_1}{\tm_2}})\lsctx} + \mult{\var}{\tm_1} + \mult{\lvar}{\off{\lgctx_1}{\tm_2}}\times \mult{\var}{\tm_1}& (\rdef{variable_multiplicity})\\
        = &  \multctx{\var}{\mult{\bullet}{\lam{\lvar}{\off{\lgctx_1}{\tm_2}}}}{\lsctx} + \mult{\var}{\tm_1} + \mult{\lvar}{\off{\lgctx_1}{\tm_2}}\times \mult{\var}{\tm_1}& (\rlem{multiplicity_of_substitution_contexts})\\
        = &  \multctx{\var}{\mult{\bullet}{\off{\lgctx_1}{\tm_2}}}{\lsctx} + \mult{\var}{\tm_1} + \mult{\lvar}{\off{\lgctx_1}{\tm_2}}\times \mult{\var}{\tm_1}& (\rdef{variable_multiplicity})\\
        \geq &  \multctx{\var}{\mult{\bullet}{\off{\lgctx_1}{\tmtwo_2}}}{\lsctx} + \mult{\var}{\tm_1} + \mult{\lvar}{\off{\lgctx_1}{\tmtwo_2}}\times \mult{\var}{\tm_1}& (\ih\times 2) \\
        = &  \multctx{\var}{\mult{\bullet}{\lam{\lvar}{\off{\lgctx_1}{\tmtwo_2}}}}{\lsctx} + \mult{\var}{\tm_1} + \mult{\lvar}{\off{\lgctx_1}{\tmtwo_2}}\times \mult{\var}{\tm_1}&  (\rdef{variable_multiplicity}) \\
        = & \mult{\var}{(\lam{\lvar}{\off{\lgctx_1}{\tmtwo_2}})\lsctx} + \mult{\var}{\tm_1} + \mult{\lvar}{\off{\lgctx_1}{\tmtwo_2}}\times \mult{\var}{\tm_1}& (\rlem{multiplicity_of_substitution_contexts})\\
        =  &   \mult{\var}{(\llam{\lvar}{\off{\lgctx_1}{\tmtwo_2}}{\labtwo})\lsctx\,\tm_1  } & 
         \end{array}\]

     \item $\lgctx= \ctxhole\lsctx_2\,\tm_1$ and $\tm  = \off{\lgctxtwo}{\ann{\uvar}{\lab}}\esub{\uvar}{(\ofc{(\sha{\tm_3})\lsctxtwo_1})\lsctxtwo_2}\lsctx_2\,\tm_1 \lto{\lab}_R \off{\lgctxtwo}{(\sha{\tm_3})\lsctxtwo_1}\esub{\uvar}{(\ofc{(\sha{\tm_3})\lsctxtwo_1})}\lsctxtwo_2\lsctx_2\,\tm_1 = \tmtwo$ and $\off{\lgctxtwo}{\ann{\uvar}{\lab}}=(\llam{\lvar}{\tm_1}{\labtwo}) \lsctx_1$. There are to cases depending on the location of the hole in $\lgctxtwo$.

       \begin{xenumerate}

         \item  $\lgctxtwo=(\llam{\lvar}{\lgctxtwo_1}{\labtwo}) \lsctx_1$.  Let $\lsctx=\lsctx_1\esub{\uvar}{(\ofc{(\sha{\tm_3})\lsctxtwo_1})\lsctxtwo_2}\lsctx_2$. 

       \[\begin{array}{rll}
        &  \mult{\var}{ (\llam{\lvar}{\off{\lgctxtwo_1}{\ann{\uvar}{\lab}}}{\labtwo}) \lsctx_1\esub{\uvar}{(\ofc{(\sha{\tm_3})\lsctxtwo_1})\lsctxtwo_2}\lsctx_2\,\tm_1 } \\
        = &  \mult{\var}{(\lam{\lvar}{\off{\lgctxtwo_1}{\ann{\uvar}{\lab}}})\lsctx} + \mult{\var}{\tm_1} + \mult{\lvar}{\off{\lgctxtwo_1}{\ann{\uvar}{\lab}}}\times \mult{\var}{\tm_1}& (\rdef{variable_multiplicity})\\
           = &  \multctx{\var}{\mult{\bullet}{\lam{\lvar}{\off{\lgctxtwo_1}{\ann{\uvar}{\lab}}}\lsctx_1\esub{\uvar}{(\ofc{(\sha{\tm_3})\lsctxtwo_1})\lsctxtwo_2}}}{\lsctx_2} + \mult{\var}{\tm_1} + \mult{\lvar}{\off{\lgctxtwo_1}{\ann{\uvar}{\lab}}}\times \mult{\var}{\tm_1}& (\rlem{multiplicity_of_substitution_contexts})\\

           = &  \multctx{\var}{\mult{\bullet}{\lam{\lvar}{\off{\lgctxtwo_1}{\ann{\uvar}{\lab}}}\lsctx_1\esub{\uvar}{(\ofc{(\sha{\tm_3})\lsctxtwo_1})\lsctxtwo_2}}}{\lsctx_2} + \mult{\var}{\tm_1} + \mult{\lvar}{\off{\lgctxtwo}{(\sha{\tm_3})\lsctxtwo_1}}\times \mult{\var}{\tm_1}& (\rlem{multiplicity_and_non_capturing_replacement})\\
           \geq &  \multctx{\var}{\mult{\bullet}{\lam{\lvar}{\off{\lgctxtwo_1}{(\sha{\tm_3})\lsctxtwo_1}}\lsctx_1\esub{\uvar}{(\ofc{(\sha{\tm_3})\lsctxtwo_1})}\lsctxtwo_2}}{\lsctx_2} + \mult{\var}{\tm_1} + \mult{\lvar}{\off{\lgctxtwo}{(\sha{\tm_3})\lsctxtwo_1}}\times \mult{\var}{\tm_1}& (\text{root case, item 3})\\
        = & \mult{\var}{(\llam{\lvar}{\off{\lgctxtwo}{(\sha{\tm_3})\lsctxtwo_1}}{\labtwo})\lsctx_1\esub{\uvar}{(\ofc{(\sha{\tm_3})\lsctxtwo_1})}\lsctxtwo_2\lsctx_2} + \mult{\var}{\tm_1} + \mult{\lvar}{\off{\lgctxtwo}{(\sha{\tm_3})\lsctxtwo_1}}\times \mult{\var}{\tm_1}& (\rlem{multiplicity_of_substitution_contexts})\\
        =  &   \mult{\var}{(\llam{\lvar}{\off{\lgctxtwo}{(\sha{\tm_3})\lsctxtwo_1}}{\labtwo})\lsctx_1\esub{\uvar}{(\ofc{(\sha{\tm_3})\lsctxtwo_1})}\lsctxtwo_2\lsctx_2\,\tm_1  } & 
         \end{array}\]       

     \item  $\lgctxtwo=(\llam{\lvar}{\tm_1}{\labtwo}) \lsctx_{11}\lesub{\uvartwo}{\lgctxtwo_1}{(\labthree)}\lsctx_{12}$. Similar to the previous case.

     \end{xenumerate}

          \item $\lgctx= \ctxhole\lsctx_2\,\tm_1$ and $\tm  = (\llam{\lvar}{\tm_2}{\labtwo}) \lsctx_1\lesub{\uvar}{(\ofc{\tm_3})\lsctxtwo}{\lab}\lsctx_2\,\tm_1 \lto{\lab}_R (\llam{\lvar}{\tm_2}{\labtwo}) \lsctx_1\lsctxtwo\lsctx_2\,\tm_1 = \tmtwo$.

       \[\begin{array}{rll}
        &  \mult{\var}{ (\llam{\lvar}{\tm_2}{\labtwo}) \lsctx_1\lesub{\uvar}{(\ofc{\tm_3})\lsctxtwo}{\lab}\lsctx_2\,\tm_1} \\
        = &  \mult{\var}{(\lam{\lvar}{\tm_2}) \lsctx_1\lesub{\uvar}{(\ofc{\tm_3})\lsctxtwo}{\lab}\lsctx_2} + \mult{\var}{\tm_1} + \mult{\lvar}{\tm_2}\times \mult{\var}{\tm_1}& (\rdef{variable_multiplicity})\\
           = &  \multctx{\var}{\mult{\bullet}{(\lam{\lvar}{\tm_2})\lsctx_1\lesub{\uvar}{(\ofc{\tm_3})\lsctxtwo}{\lab}}}{\lsctx_2} + \mult{\var}{\tm_1} + \mult{\lvar}{\tm_2}\times \mult{\var}{\tm_1}& (\rlem{multiplicity_of_substitution_contexts})\\

           \geq &  \multctx{\var}{\mult{\bullet}{(\lam{\lvar}{\tm_2})\lsctx_1\lsctxtwo}}{\lsctx_2} + \mult{\var}{\tm_1} + \mult{\lvar}{\tm_2}\times \mult{\var}{\tm_1}& (\text{root case, item 4})\\
        = & \mult{\var}{(\llam{\lvar}{\tm_2}{\labtwo})\lsctx_1\lsctxtwo\lsctx_2} + \mult{\var}{\tm_1} + \mult{\lvar}{\tm_2}\times \mult{\var}{\tm_1}& (\rlem{multiplicity_of_substitution_contexts})\\
        =  &   \mult{\var}{(\llam{\lvar}{\tm_1}{\labtwo}) \lsctx_1\lsctxtwo\lsctx_2\,\tm_1 } & 
         \end{array}\]

  \item $\lgctx= (\llam{\lvar}{\tm_1}{\labtwo})\lsctx_1\lesub{\uvartwo}{\lgctx_1}{(\labthree)}\lsctx_2\,\tm_2$. Suppose $\tm  =(\llam{\lvar}{\tm_1}{\labtwo})\lsctx_1\lesub{\uvartwo}{\off{\lgctx_1}{\tm_3}}{(\labthree)}\lsctx_2\,\tm_2\to_R (\llam{\lvar}{\tm_1}{\labtwo})\lsctx_1\lesub{\uvartwo}{\off{\lgctx_1}{\tmtwo_3}}{(\labthree)}\lsctx_2\,\tm_2= \tmtwo$ follows from $\tm_3 \to_R \tmtwo_3$.
      \[\begin{array}{rll}
        &  \mult{\var}{ (\llam{\lvar}{\tm_1}{\labtwo})\lsctx_1\lesub{\uvartwo}{\off{\lgctx_1}{\tm_3}}{(\labthree)}\lsctx_2\,\tm_2} \\
          = &  \mult{\var}{(\lam{\lvar}{\tm_1})\lsctx_1\lesub{\uvartwo}{\off{\lgctx_1}{\tm_3}}{(\labthree)}\lsctx_2} + \mult{\var}{\tm_2} + 
            \mult{\lvar}{\tm_1}\times \mult{\var}{\tm_2} & (\rdef{variable_multiplicity})\\
          = &  \multctx{\var}{\mult{\bullet}{(\lam{\lvar}{\tm_1})\lsctx_1\lesub{\uvartwo}{\off{\lgctx_1}{\tm_3}}{(\labthree)}}}{\lsctx_2} + \mult{\var}{\tm_2} + 
            \mult{\lvar}{\tm_1}\times \mult{\var}{\tm_2} & (\rlem{multiplicity_of_substitution_contexts}) \\
          = &  \multctx{\var}{\mult{\bullet}{(\lam{\lvar}{\tm_1})\lsctx_1}+\mult{\bullet}{\off{\lgctx_1}{\tm_3}}+\mult{\uvartwo}{(\lam{\lvar}{\tm_1})\lsctx_1}\times \mult{\bullet}{\off{\lgctx_1}{\tm_3}}}{\lsctx_2} + \mult{\var}{\tm_2} + 
            \mult{\lvar}{\tm_1}\times \mult{\var}{\tm_2} & (\rdef{variable_multiplicity})\\
          \geq &  \multctx{\var}{\mult{\bullet}{(\lam{\lvar}{\tm_1})\lsctx_1}+\mult{\bullet}{\off{\lgctx_1}{\tmtwo_3}}+\mult{\uvartwo}{(\lam{\lvar}{\tm_1})\lsctx_1}\times \mult{\bullet}{\off{\lgctx_1}{\tmtwo_3}}}{\lsctx_2} + \mult{\var}{\tm_2} + 
                 \mult{\lvar}{\tm_1}\times \mult{\var}{\tm_2} & (\ih)\\
                    = &  \multctx{\var}{\mult{\bullet}{(\lam{\lvar}{\tm_1})\lsctx_1\lesub{\uvartwo}{\off{\lgctx_1}{\tmtwo_3}}{(\labthree)}}}{\lsctx_2} + \mult{\var}{\tm_2} + 
                        \mult{\lvar}{\tm_1}\times \mult{\var}{\tm_2} & (\rdef{variable_multiplicity})\\
           = &  \mult{\var}{(\lam{\lvar}{\tm_1})\lsctx_1\lesub{\uvartwo}{\off{\lgctx_1}{\tmtwo_3}}{(\labthree)}\lsctx_2} + \mult{\var}{\tm_2} + 
            \mult{\lvar}{\tm_1}\times \mult{\var}{\tm_2} & (\rlem{multiplicity_of_substitution_contexts}) \\
        =  &    \mult{\var}{ (\llam{\lvar}{\tm_1}{\labtwo})\lsctx_1\lesub{\uvartwo}{\off{\lgctx_1}{\tmtwo_3}}{(\labthree)}\lsctx_2\,\tm_2} & (\rdef{variable_multiplicity})\\
        \end{array}\]
      
\item $\lgctx= (\llam{\lvar}{\tm_1}{\labtwo})\lsctx\,\lgctx_1$. Suppose $\tm  = (\llam{\lvar}{\tm_1}{\labtwo})\lsctx\,\off{\lgctx_1}{\tm_2}\to_R  (\llam{\lvar}{\tm_1}{\labtwo})\lsctx\,\off{\lgctx_1}{\tmtwo_2}= \tmtwo$ follows from $\tm_2 \to_R \tmtwo_2$.

         \[\begin{array}{rll}
        &  \mult{\var}{ (\llam{\lvar}{\tm_1}{\labtwo})\lsctx\,\off{\lgctx_1}{\tm_2}} \\
        = &  \mult{\var}{ (\lam{\lvar}{\tm_1})\lsctx}  + \mult{\var}{ \off{\lgctx_1}{\tm_2}}  + \mult{\lvar}{ \tm_1} \times \mult{\var}{\off{\lgctx_1}{\tm_2}} & (\rdef{variable_multiplicity})\\
        \geq & \mult{\var}{ (\lam{\lvar}{\tm_1})\lsctx}  + \mult{\var}{ \off{\lgctx_1}{\tmtwo_2}}  + \mult{\lvar}{ \tm_1} \times \mult{\var}{\off{\lgctx_1}{\tmtwo_2}}& (\ih\times 2)\\
   = & \mult{\var}{ (\llam{\lvar}{\tm_1}{\labtwo})\lsctx\,\off{\lgctx_1}{\tmtwo_2}} & (\rdef{variable_multiplicity})\\
           \end{array}\]

\item $\lgctx=\lgctx_1\,\tm_1$.  Suppose $\tm  =\off{\lgctx_1}{\tm_2}\,\tm_1\to_R \off{\lgctx_1}{\tmtwo_2}\,\tm_1= \tmtwo$ follows from $\tm_2 \to_R \tmtwo_2$.

      \[\begin{array}{rll}
        &  \mult{\var}{ \off{\lgctx_1}{\tm_2}\,\tm_1} \\
        = &  \mult{\var}{ \off{\lgctx_1}{\tm_2}} + \mult{\var}{\tm_1} & (\rdef{variable_multiplicity})\\
        \geq &  \mult{\var}{ \off{\lgctx_1}{\tmtwo_2}} + \mult{\var}{\tm_1}  & (\ih)\\
        =  &   \mult{\var}{\off{\lgctx_1}{\tmtwo_2}\,\tm_1 } & (\rdef{variable_multiplicity})
        \end{array}\]

\item $\lgctx=\tm_1\,\lgctx_1$. Similar to the case $\lgctx=\lgctx_1\,\tm_1$.

\item $\lgctx=\sha{\lgctx_1}$. Similar to the case $\lgctx=\lam{\lvar}{\lgctx_1}$.

  
\item $\lgctx=\lopen{\lgctx_1}{(\labtwo)}$. We need to consider multiple cases since out induction proceeds on well-formed labeled contexts.

  \begin{xenumerate}

  \item $\labtwo$ is not present. Suppose $\tm  =\open{\off{\lgctx_1}{\tm_1}}\to_R \open{\off{\lgctx_1}{\tmtwo_1}}= \tmtwo$ follows from $\tm_1 \to_R \tmtwo_1$.  Similar to the case $\lgctx=\lam{\lvar}{\lgctx_1}$.
  
  \item $\labtwo$ is present.
     \begin{xenumerate}

 \item $\lgctx_1=(\sha{\lgctx_2}) \lsctx$.  Suppose $\tm  =\lopen{(\sha{\off{\lgctx_2}{\tm_1}}) \lsctx }{\labtwo} \to_R \lopen{(\sha{\off{\lgctx_2}{\tmtwo_1}}) \lsctx}{\labtwo}= \tmtwo$ follows from $\tm_1 \to_R \tmtwo_1$. 

         \[\begin{array}{rll}
        &  \mult{\var}{ (\sha{\off{\lgctx_2}{\tm_1}}) \lsctx } \\
       =  & \multctx{\var}{\mult{\bullet}{\sha{\off{\lgctx_2}{\tm_1}}}}{\lsctx }  & (\rlem{multiplicity_of_substitution_contexts})\\ 
       =  & \multctx{\var}{\mult{\bullet}{\off{\lgctx_2}{\tm_1}}}{\lsctx }  & \\
        \geq & \multctx{\var}{\mult{\bullet}{\off{\lgctx_2}{\tmtwo_1}}}{\lsctx }& (\ih)\\
       =  & \multctx{\var}{\mult{\bullet}{\sha{\off{\lgctx_2}{\tmtwo_1}}}}{\lsctx }  & \\
   = & \mult{\var}{(\sha{\off{\lgctx_2}{\tmtwo_1}}) \lsctx} & (\rlem{multiplicity_of_substitution_contexts})\\ 
           \end{array}\]

          \item $\lgctx_1=\ctxhole\lsctx_2$ and $\tm  = \lopen{(\sha{\off{\lgctxtwo}{\ann{\uvar}{\lab}}})\lsctx_1\esub{\uvar}{(\ofc{(\sha{\tm_3})\lsctxtwo_1})\lsctxtwo_2}\lsctx_2}{\labtwo} \lto{\lab}_R \lopen{(\sha{\off{\lgctxtwo}{(\sha{\tm_3})\lsctxtwo_1}})\lsctx_1\esub{\uvar}{(\ofc{(\sha{\tm_3})\lsctxtwo_1})}\lsctxtwo_2\lsctx_}{\labtwo} = \tmtwo$. Then we reason as in the previous case but using the root case (item 2) instead of the \ih.

   \item $\lgctx_1=\ctxhole\lsctx_2$ and
   $\tm  = \lopen{(\sha{\tm_1}) \lsctx_1\lesub{\uvar}{(\ofc{\tm_3})\lsctxtwo}{\lab}\lsctx_2}{\labtwo} \lto{\lab}_R \lopen{(\sha{\tm_1})\lsctx_1\lsctxtwo\lsctx_2}{\labtwo} = \tmtwo$. Same as previous case.

 \item $\lgctx_1=(\sha{\tm_1}) \lsctx_1\lesub{\uvartwo}{\lgctx_2}{(\labthree)}\lsctx_2$.  Suppose $\tm  =\lopen{(\sha{\tm_1}) \lsctx_1\lesub{\uvartwo}{\off{\lgctx_2}{\tm_2}}{(\labthree)}\lsctx_2}{\labtwo}\to_R \lopen{(\sha{\tm_1}) \lsctx_1\lesub{\uvartwo}{\off{\lgctx_2}{\tmtwo_2}}{(\labthree)}\lsctx_2}{\labtwo}= \tmtwo$ follows from $\tm_2\to_R \tmtwo_2$. Similar to the previous case.
    
    \end{xenumerate}
    
    \end{xenumerate}

\item $\lgctx=\ofc{\lgctx_1}$. Similar to the case $\lgctx=\lam{\lvar}{\lgctx_1}$.
  
\item $\lgctx=\lgctx_1\lesub{\uvartwo}{\tm_1}{(\labtwo)}$. Suppose $\tm  =\off{\lgctx_1}{\tm_2}\lesub{\uvartwo}{\tm_1}{(\labtwo)}\to_R \off{\lgctx_1}{\tmtwo_2}\lesub{\uvartwo}{\tm_1}{(\labtwo)}= \tmtwo$ follows from $\tm_2\to_R \tmtwo_2$. 

        \[\begin{array}{rll}
        &  \mult{\var}{ \off{\lgctx_1}{\tm_2}\lesub{\uvartwo}{\tm_1}{(\labtwo)}} \\
        &  \mult{\var}{ \off{\lgctx_1}{\tm_2}} + \mult{\var}{ \tm_1} + \mult{\uvartwo}{ \off{\lgctx_1}{\tm_2}}\times \mult{\var}{\tm_1}  & (\rdef{variable_multiplicity})\\
        \geq &  \mult{\var}{ \off{\lgctx_1}{\tmtwo_2}} + \mult{\var}{ \tm_1} + \mult{\uvartwo}{ \off{\lgctx_1}{\tmtwo_2}}\times \mult{\var}{\tm_1}  & (\ih\times 2)\\
   = & \mult{\var}{ \off{\lgctx_1}{\tmtwo_2}\lesub{\uvartwo}{\tm_1}{(\labtwo)}} & (\rdef{variable_multiplicity})\\
           \end{array}\]



\item $\lgctx=\tm_1\lesub{\uvartwo}{\lgctx_1}{(\labtwo)}$. Suppose $\tm  =\tm_1\lesub{\uvartwo}{\off{\lgctx_1}{\tm_2}}{(\labtwo)}\to_R \tm_1\lesub{\uvartwo}{\off{\gctx_1}{\tmtwo_2}}{(\labtwo)}= \tmtwo$ follows from $\tm_2\to_R \tmtwo_2$. 
        \[\begin{array}{rll}
        &  \mult{\var}{ \tm_1\lesub{\uvartwo}{\off{\lgctx_1}{\tm_2}}{(\labtwo)}} \\
        & \mult{\var}{ \tm_1}  + \mult{\var}{ \off{\lgctx_1}{\tm_2}}  + \mult{\uvartwo}{ \tm_1} \times \mult{\var}{\off{\lgctx_1}{\tm_2}}  & (\rdef{variable_multiplicity})\\
        \geq &  \mult{\var}{ \tm_1}  + \mult{\var}{ \off{\lgctx_1}{\tmtwo_2}}  + \mult{\uvartwo}{ \tm_1} \times \mult{\var}{\off{\lgctx_1}{\tmtwo_2}}& (\ih\times 2)\\
   = & \mult{\var}{\tm_1\lesub{\uvartwo}{\off{\gctx_1}{\tmtwo_2}}{(\labtwo)}} & (\rdef{variable_multiplicity})\\
           \end{array}\]
\end{xenumerate}
\end{ifLongAppendix}
\end{proof}

\begin{lemma}
\llem{redex_multiplicity_and_linear_substitution}
$\lmult{\tm}+\mult{\lvar}{\tm}\times \lmult{\tmtwo} = \lmult{t\sub{\lvar}{\tmtwo}}$
\end{lemma}

\begin{proof}
  By induction on $\tm$.
\begin{ifLongAppendix}
  \begin{xenumerate}
  \item $\tm=  \lvartwo$. If $\lvartwo\neq\lvar$, then
    \[\begin{array}{rll}
        & \lmult{\lvartwo}+\mult{\lvar}{\lvartwo}\times \lmult{\tmtwo} \\
        = & 0 \\
        = & \lmult{\lvartwo\sub{\lvar}{\tmtwo}}
      \end{array}\]

    If $\lvartwo=\lvar$, then
    \[\begin{array}{rll}
        & \lmult{\lvartwo}+ \mult{\lvar}{\lvartwo}\times \lmult{\tmtwo} \\
        = & \lmult{\tmtwo}\\
        = & \lmult{\lvartwo\sub{\lvar}{\tmtwo}}
      \end{array}\]

      \item $\tm=  \ann{\uvar}{(\lab)}$
        \[\begin{array}{rll}
            & \lmult{\ann{\uvar}{(\lab)}}+ \mult{\lvar}{\ann{\uvar}{(\lab)}}\times \lmult{\tmtwo} \\
            = &  0\\
            = & \lmult{\ann{\uvar}{(\lab)}\sub{\lvar}{\tmtwo}}
          \end{array}\]
        
      \item $\tm= \lam{\lvartwo}{\tmthree}$.

        \[\begin{array}{rll}
            & \lmult{\lam{\lvartwo}{\tmthree}}+\mult{\lvar}{\lam{\lvartwo}{\tmthree}}\times \lmult{\tmtwo} \\
            =  & \lmult{\tmthree} + \mult{\lvar}{\tmthree}\times \lmult{\tmtwo} & (\rdef{redex_multiplicity})\\
            =  & \lmult{\tmthree\sub{\lvar}{\tmtwo}} & (\ih) \\
            = &  \lmult{(\lam{\lvartwo}{\tmthree})\sub{\lvar}{\tmtwo}} &   (\rdef{redex_multiplicity})
          \end{array}\]
        
      \item $\tm=\tm_1\,\tm_2$.
           \[\begin{array}{rll}
            & \lmult{\tm_1\,\tm_2}+\mult{\lvar}{\tm_1\,\tm_2}\times \lmult{\tmtwo} \\
            =  & \lmult{\tm_1}+\lmult{\tm_2}+(\mult{\lvar}{\tm_1}+\mult{\var}{\tm_2})\times \lmult{\tmtwo} & (\rdef{redex_multiplicity})\\
            =  & \lmult{\tm_1}+\lmult{\tm_2}+ \mult{\lvar}{\tm_1}\times \lmult{\tmtwo}+\mult{\var}{\tm_2}\times \lmult{\tmtwo} \\
            =  & \lmult{\tm_1\sub{\lvar}{\tmtwo}} + \lmult{\tm_2\sub{\lvar}{\tmtwo}}& (\ih\times 2) \\
            =  & \lmult{\tm_1\sub{\lvar}{\tmtwo}\, \tm_2\sub{\lvar}{\tmtwo}} & (\rdef{redex_multiplicity})\\
            =  & \lmult{(\tm_1\,\tm_2)\sub{\lvar}{\tmtwo}} & \\
             \end{array}\]
           
\item $ \tm= (\llam{\lvartwo}{\tm_1}{\lab})\lsctx\,\tm_2$.

        \[\begin{array}{rll}
            & \lmult{(\llam{\lvartwo}{\tm_1}{\lab})\lsctx\,\tm_2}+\mult{\lvar}{(\llam{\lvartwo}{\tm_1}{\lab})\lsctx\,\tm_2}\times \lmult{\tmtwo} \\
            = & 1 + \lmult{(\lam{\lvartwo}{\tm_1})\lsctx} + \lmult{\tm_2} + \mult{\lvartwo}{\tm_1}\times \lmult{\tm_2} +\mult{\lvar}{(\llam{\lvartwo}{\tm_1}{\lab})\lsctx\,\tm_2}\times \lmult{\tmtwo} &   (\rdef{redex_multiplicity}) \\
            = & 1 + \lmult{(\lam{\lvartwo}{\tm_1})\lsctx} + \lmult{\tm_2} + \mult{\lvartwo}{\tm_1}\times \lmult{\tm_2} +(\mult{\lvar}{(\lam{\lvartwo}{\tm_1})\lsctx} + \mult{\lvar}{\tm_2} + \mult{\lvartwo}{\tm_1}\times  \mult{\lvar}{\tm_2})\times \lmult{\tmtwo} &   (\rdef{redex_multiplicity}) \\
            = & 1 +\lmult{((\lam{\lvartwo}{\tm_1})\lsctx)\sub{\lvar}{\tmtwo}} + \lmult{\tm_2} + \mult{\lvartwo}{\tm_1}\times \lmult{\tm_2} +(\mult{\lvar}{\tm_2} + \mult{\lvartwo}{\tm_1}\times  \mult{\lvar}{\tm_2})\times \lmult{\tmtwo} & (\ih) \\
            = & 1 + \lmult{((\lam{\lvartwo}{\tm_1})\lsctx)\sub{\lvar}{\tmtwo}} + \lmult{\tm_2\sub{\lvar}{\tmtwo}} + \mult{\lvartwo}{\tm_1}\times \lmult{\tm_2} +\mult{\lvartwo}{\tm_1}\times  \mult{\lvar}{\tm_2}\times \lmult{\tmtwo} & (\ih) \\
            = &  1 + \lmult{(\lam{\lvartwo}{\tm_1 \sub{\lvar}{\tmtwo}}) \lsctx\sub{\lvar}{\tmtwo}} +  \lmult{\tm_2 \sub{\lvar}{\tmtwo}}  +  \mult{\lvartwo}{\tm_1} \times  \lmult{\tm_2 \sub{\lvar}{\tmtwo}} &   (\rdef{redex_multiplicity}) \\
            = &  1 + \lmult{(\lam{\lvartwo}{\tm_1 \sub{\lvar}{\tmtwo}}) \lsctx\sub{\lvar}{\tmtwo}} +  \lmult{\tm_2 \sub{\lvar}{\tmtwo}}  +  \mult{\lvartwo}{\tm_1 \sub{\lvar}{\tmtwo}} \times  \lmult{\tm_2 \sub{\lvar}{\tmtwo}} &   (\rlem{multiplicity_and_linear_substitution_non_capturing}(1)) \\ 
            = &  \lmult{(\llam{\lvartwo}{\tm_1 \sub{\lvar}{\tmtwo}}{\lab})\lsctx\sub{\lvar}{\tmtwo}\,\tm_2 \sub{\lvar}{\tmtwo}} &   (\rdef{variable_multiplicity}) \\
            = &  \lmult{((\llam{\lvartwo}{\tm_1}{\lab})\lsctx\,\tm_2)\sub{\lvar}{\tmtwo}} &   (\rdef{redex_multiplicity})
          \end{array}\]
        
      \item $\tm=\sha{\tm_1}$.
        
        \[\begin{array}{rll}
            & \lmult{\sha{\tm_1}}+\mult{\lvar}{\sha{\tm_1}}\times \lmult{\tmtwo} \\
            =  & \lmult{\tm_1} + \mult{\lvar}{\tm_1}\times \lmult{\tmtwo} &  (\rdef{redex_multiplicity}, \rdef{variable_multiplicity})\\
            = & \lmult{\tm_1\sub{\lvar}{\tmtwo}} & (\ih) \\
            =  & \lmult{\sha{(\tm_1\sub{\lvar}{\tmtwo})}} &  (\rdef{redex_multiplicity})\\
            =  & \lmult{(\sha{\tm_1})\sub{\lvar}{\tmtwo}} & \\
          \end{array}\]
        
      \item $\tm=\open{\tm_1}$. Same as case $\tm=\sha{\tm_1}$.
        
      \item $\tm=\lopen{(\sha{\tm_1})\lsctx}{\lab}$. 

          \[\begin{array}{rll}
            & \lmult{\lopen{(\sha{\tm_1})\lsctx}{\lab}} +\mult{\lvar}{\lopen{(\sha{\tm_1})\lsctx}{\lab}} \times \lmult{\tmtwo} \\
            =  & 1 + \lmult{(\sha{\tm_1})\lsctx} +\mult{\lvar}{(\sha{\tm_1})\lsctx} \times \lmult{\tmtwo} & (\rdef{redex_multiplicity}, \rdef{variable_multiplicity})\\
            =  & 1 + \lmult{((\sha{\tm_1})\lsctx)\sub{\lvar}{\tmtwo}} & (\ih) \\
            =  & \lmult{\lopen{((\sha{\tm_1})\lsctx) \sub{\lvar}{\tmtwo}}{\lab}} & (\rdef{redex_multiplicity})\\
            =  & \lmult{\lopen{(\sha{\tm_1})\lsctx}{\lab}\sub{\lvar}{\tmtwo}} & \\
            \end{array}\]
          
\item $\tm=\ofc{\tm_1}$. Same as case $\tm=\sha{\tm_1}$.
        
      \item $\tm=  \tm_1\esub{\uvar}{\tm_2}$

                \[\begin{array}{rll}
            & \lmult{\tm_1\esub{\uvar}{\tm_2}}+\mult{\lvar}{\tm_1\esub{\uvar}{\tm_2}}\times \lmult{\tmtwo} \\
                   = &  \lmult{\tm_1}+\lmult{\tm_2} + \mult{\uvar}{\tm_1}\times \lmult{\tm_2} + \mult{\uvar}{\tm_1} + \mult{\lvar}{\tm_1\lesub{\uvar}{\tm_2}{(\lab)}}\times \lmult{\tmtwo} & (\rdef{redex_multiplicity})\\
                   = &  \lmult{\tm_1}+\lmult{\tm_2} + \mult{\uvar}{\tm_1}\times \lmult{\tm_2} + \mult{\uvar}{\tm_1}   + (\mult{\lvar}{\tm_1}  + \mult{\lvar}{\tm_2} + \mult{\uvar}{\tm_1}\times \mult{\lvar}{\tm_2}) \times \lmult{\tmtwo} & (\rdef{variable_multiplicity})\\
                   =  &  \lmult{\tm_1\sub{\lvar}{\tmtwo}} +  \lmult{\tm_2} +\mult{\uvar}{\tm_1}\times \lmult{\tm_2} + \mult{\uvar}{\tm_1} + \mult{\uvar}{\tm_1}\times \mult{\var}{\tm_2} +  (\mult{\lvar}{\tm_2} + \mult{\uvar}{\tm_1}\times \mult{\lvar}{\tm_2}) \times \lmult{\tmtwo}  & (\ih)\\
                    =  &  \lmult{\tm_1\sub{\lvar}{\tmtwo}} + \lmult{\tm_2 \sub{\lvar}{\tmtwo}} + \mult{\uvar}{\tm_1}\times \lmult{\tm_2} +  \mult{\uvar}{\tm_1} + \mult{\uvar}{\tm_1}\times \mult{\lvar}{\tm_2} \times \lmult{\tmtwo}  & (\ih)\\
                    = &  \lmult{\tm_1 \sub{\lvar}{\tmtwo}} + \lmult{\tm_2 \sub{\lvar}{\tmtwo}} + \mult{\uvar}{\tm_1}\times \lmult{\tm_2 \sub{\lvar}{\tmtwo}} + \mult{\uvar}{\tm_1}& (\ih)\\
                    = &  \lmult{\tm_1 \sub{\lvar}{\tmtwo}} + \lmult{\tm_2 \sub{\lvar}{\tmtwo}} + \mult{\uvar}{\tm_1\sub{\lvar}{\tmtwo}}\times \lmult{\tm_2 \sub{\lvar}{\tmtwo}} + \mult{\uvar}{\tm_1 \sub{\lvar}{\tmtwo}} & (\rlem{multiplicity_and_linear_substitution_non_capturing}(1)\times 2)\\
            = &  \lmult{\tm_1 \sub{\lvar}{\tmtwo}\esub{\uvar}{\tm_2 \sub{\lvar}{\tmtwo}}} &   (\rdef{variable_multiplicity}) \\
            = &  \lmult{\tm_1\esub{\uvar}{\tm_2}\sub{\lvar}{\tmtwo}} &   
                  \end{array}\]

                      \item $\tm=  \tm_1\lesub{\uvar}{\tm_2}{\lab}$ and $\uvar\notin\fv{\tm_1}$.

                \[\begin{array}{rll}
            & \lmult{\tm_1\lesub{\uvar}{\tm_2}{\lab}}+\mult{\lvar}{\tm_1\lesub{\uvar}{\tm_2}{\lab}}\times \lmult{\tmtwo} \\
                   = &  1 + \lmult{\tm_1}+\lmult{\tm_2} + \mult{\lvar}{\tm_1\lesub{\uvar}{\tm_2}{\lab}}\times \lmult{\tmtwo} & (\rdef{redex_multiplicity}) \\
                   = &  1 + \lmult{\tm_1}+\lmult{\tm_2} + (\mult{\lvar}{\tm_1}+\mult{\lvar}{\tm_2}+\mult{\uvar}{\tm_1}\times\mult{\lvar}{\tm_2})\times \lmult{\tmtwo} &  (\rdef{variable_multiplicity}) \\
                   =  &  1 + \lmult{\tm_1\sub{\lvar}{\tmtwo}} + \lmult{\tm_2} +  (\mult{\lvar}{\tm_2} + \mult{\uvar}{\tm_1}\times \mult{\lvar}{\tm_2}) \times \lmult{\tmtwo}  & (\ih)\\
                    =  &  1 + \lmult{\tm_1\sub{\lvar}{\tmtwo}} + \lmult{\tm_2 \sub{\lvar}{\tmtwo}} +  \mult{\uvar}{\tm_1}\times \mult{\lvar}{\tm_2} \times \lmult{\tmtwo}  & (\ih)\\
                    = &  1 + \lmult{\tm_1 \sub{\lvar}{\tmtwo}} + \lmult{\tm_2 \sub{\lvar}{\tmtwo}} & (\uvar\notin\fv{\tm})\\
            = &  \lmult{\tm_1 \sub{\lvar}{\tmtwo}\lesub{\uvar}{\tm_2 \sub{\lvar}{\tmtwo}}{\lab}} &   (\rdef{redex_multiplicity}) \\
            = &  \lmult{\tm_1\lesub{\uvar}{\tm_2}{(\lab)}\sub{\lvar}{\tmtwo}} &   
                  \end{array}\]

    \end{xenumerate}
    \end{ifLongAppendix}
    \end{proof}

\begin{lemma}
\llem{redex_multiplicity_and_linear_beta}
Suppose $\fv{\tmtwo}\cap\dom{\lsctx}=\emptyset$.
\[\lmult{(\lam{\lvar}{\tm})\lsctx}+\mult{\lvar}{\tm}\times \lmult{\tmtwo}
  =
  \lmult{\tm\sub{\lvar}{\tmtwo}} + \lmultctx{\mult{\bullet}{\tm\sub{\lvar}{\tmtwo}}}{\lsctx}
  \]
\end{lemma}

\begin{proof}
  By induction on $\lsctx$.
  \begin{ifLongAppendix}
    \begin{xenumerate}
  \item $\lsctx=\ctxhole$.

  \[\begin{array}{rll}
        & \lmult{\lam{\lvar}{\tm}}+\mult{\lvar}{\tm}\times \lmult{\tmtwo} \\
        = & \lmult{\tm}+ \mult{\lvar}{\tm}\times \lmult{\tmtwo} & (\rdef{redex_multiplicity}) \\
       = &   \lmult{\tm\sub{\lvar}{\tmtwo}} & (\rlem{redex_multiplicity_and_linear_substitution})\\
      =  & \lmult{\tm\sub{\lvar}{\tmtwo}} + 0 \\
      = &  \lmult{\tm\sub{\lvar}{\tmtwo}}+ \lmultctx{\mult{\bullet}{\tm\sub{\lvar}{\tmtwo}}}{\ctxhole}
      \end{array}
    \]
    

\item $\lsctx=\lsctx_1\lesub{\uvar}{\tmthree}{\lab}$.

        \[\begin{array}{rll}
        & \lmult{(\lam{\lvar}{\tm}) \lsctx_1\lesub{\uvar}{\tmthree}{\lab}}+ \mult{\lvar}{\tm}\times \lmult{\tmtwo}\\
        = & 1 + \lmult{(\lam{\lvar}{\tm}) \lsctx_1} + \lmult{\tmthree} + \mult{\lvar}{\tm}\times \mult{\var}{\tmtwo} & (\rdef{redex_multiplicity}) \\
        = &  1 + \lmult{\tm\sub{\lvar}{\tmtwo}}  + \multctx{\var}{\mult{\bullet}{\tm\sub{\lvar}{\tmtwo}}}{\lsctx_1} + \lmult{\tmthree} & (\ih)\\
       = &  \lmult{\tm\sub{\lvar}{\tmtwo}} +  \lmultctx{\mult{\bullet}{\tm\sub{\lvar}{\tmtwo}}}{\lsctx_1\lesub{\uvar}{\tmthree}{\lab}} & (\rdef{redex_multiplicity})
      \end{array}
    \]

 \item $\lsctx=\lsctx_1\esub{\uvar}{\tm}$

        \[\begin{array}{rll}
        & \lmult{(\lam{\lvar}{\tm}) \lsctx_1\esub{\uvar}{\tmthree}}+ \mult{\lvar}{\tm}\times \lmult{\tmtwo}\\
        = & \lmult{(\lam{\lvar}{\tm}) \lsctx_1} + \lmult{\tmthree} + \mult{\uvar}{(\lam{\lvar}{\tm}) \lsctx_1} \times \lmult{\tmthree}+\mult{\uvar}{(\lam{\lvar}{\tm}) \lsctx_1} + \mult{\lvar}{\tm}\times \lmult{\tmtwo} & (\rdef{redex_multiplicity}) \\
        = &  \lmult{\tm\sub{\lvar}{\tmtwo}}  + \multctx{\var}{\mult{\bullet}{\tm\sub{\lvar}{\tmtwo}}}{\lsctx_1} + \lmult{\tmthree}+ \mult{\uvar}{(\lam{\lvar}{\tm}) \lsctx_1} \times \lmult{\tmthree} + \mult{\uvar}{(\lam{\lvar}{\tm}) \lsctx_1} & (\ih)\\
       = &  \lmult{\tm\sub{\lvar}{\tmtwo}}  + \multctx{\var}{\mult{\bullet}{\tm\sub{\lvar}{\tmtwo}}}{\lsctx_1} + \lmult{\tmthree} + \multctx{\uvar}{\mult{\bullet}{\lam{\lvar}{\tm}}}{\lsctx_1}\times \lmult{\tmthree} + \mult{\uvar}{(\lam{\lvar}{\tm}) \lsctx_1} & (\rlem{multiplicity_of_substitution_contexts})
\\
       = &  \lmult{\tm\sub{\lvar}{\tmtwo}}  + \multctx{\var}{\mult{\bullet}{\tm\sub{\lvar}{\tmtwo}}}{\lsctx_1} + \lmult{\tmthree} +\multctx{\uvar}{\mult{\bullet}{\tm}}{\lsctx_1}\times \lmult{\tmthree} +  \mult{\uvar}{(\lam{\lvar}{\tm}) \lsctx_1} & (\rdef{variable_multiplicity})
\\
       = &   \lmult{\tm\sub{\lvar}{\tmtwo}} +  \lmultctx{\mult{\bullet}{\tm\sub{\lvar}{\tmtwo}}}{\lsctx_1}   + \lmult{\tmthree} + \multctx{\uvar}{\mult{\bullet}{\tm\sub{\lvar}{\tmtwo}}}{\lsctx_1}\times \lmult{\tmthree} +  \multctx{\uvar}{\mult{\bullet}{\tm\sub{\lvar}{\tmtwo}}}{\lsctx_1}& (\rlem{multiplicity_and_linear_substitution_non_capturing}(2))
\\
       = &  \lmult{\tm\sub{\lvar}{\tmtwo}} +  \lmultctx{\mult{\bullet}{\tm\sub{\lvar}{\tmtwo}}}{\lsctx_1\esub{\uvar}{\tmthree}} & (\rdef{redex_multiplicity})
      \end{array}
    \]
  \end{xenumerate}
  \end{ifLongAppendix}
\end{proof}

\begin{lemma}
\llem{redex_multiplicity_of_substitution_contexts}
$\lmult{\tm\lsctx} = \lmult{\tm} + \lmultctx{\mult{\bullet}{\tm}}{\lsctx}$
\end{lemma}

\begin{proof}
By induction on $\lsctx$.
\begin{ifLongAppendix}
  \begin{xenumerate}
  \item $  \lsctx = \ctxhole$. Immediate since $\lmult{\tm} = \lmult{\tm} + 0 = \lmult{\tm} + \lmultctx{\mult{\bullet}{\tm}}{\ctxhole}$.

  \item $\lsctx = \lsctx_1\esub{\uvar}{\tmtwo}$.

    \[\begin{array}{rll}
        & \lmult{\tm \lsctx_1\esub{\uvar}{\tmtwo}} \\
        = & \lmult{\tm \lsctx_1} + \lmult{\tmtwo} + \mult{\uvar}{\tm \lsctx_1}\times \lmult{\tmtwo} + \mult{\uvar}{\tm\lsctx_1} & (\rdef{redex_multiplicity})\\ 
        = & \lmult{\tm} + \lmultctx{\mult{\bullet}{\tm}}{\lsctx_1} + \lmult{\tmtwo} + \mult{\uvar}{\tm \lsctx_1}\times \lmult{\tmtwo} + \mult{\uvar}{\tm\lsctx_1}  & (\ih) \\
        = & \lmult{\tm} + \lmultctx{\mult{\bullet}{\tm}}{\lsctx_1} + \lmult{\tmtwo} + \multctx{\uvar}{\mult{\bullet}{\tm}}{\lsctx_1}\times \lmult{\tmtwo} + \multctx{\uvar}{\mult{\bullet}{\tm}}{\lsctx_1}  & (\rlem{multiplicity_of_substitution_contexts}\times 2) \\
          =  & \lmult{\tm} + \lmultctx{\mult{\bullet}{\tm}}{ \lsctx_1\esub{\uvar}{\tmtwo}}  & (\rdef{redex_multiplicity})
        \end{array}\]

        \item $\lsctx = \lsctx_1\lesub{\uvar}{\tmtwo}{\lab}$.
    \[\begin{array}{rll}
        & \lmult{\tm \lsctx_1\lesub{\uvar}{\tmtwo}{\lab}} \\
        = & 1 + \lmult{\tm \lsctx_1} + \lmult{\tmtwo} & (\rdef{redex_multiplicity}) \\ 
        = & 1 + \lmult{\tm} + \lmultctx{\mult{\bullet}{\tm}}{\lsctx_1} + \lmult{\tmtwo} & (\ih) \\
          =  & \lmult{\tm} + \lmultctx{\mult{\bullet}{\tm}}{ \lsctx_1\lesub{\uvar}{\tmtwo}{\lab}}  & (\rdef{redex_multiplicity})
        \end{array}\]
  \end{xenumerate}
  \end{ifLongAppendix}
  \end{proof}

\begin{lemma}
  \llem{redex_multiplicity_and_replacement}
  Suppose $u\notin\fv{\tm}$. Then,
  \[\lmult{\off{\lgctx}{\ann{\uvar}{\lab}}} + \mult{\uvar}{\off{\lgctx}{\ann{\uvar}{\lab}}}\times \lmult{\tm} + \mult{\uvar}{\off{\lgctx}{\ann{\uvar}{\lab}}} 
        >       \lmult{\off{\lgctx}{\tm}} + \mult{\uvar}{\off{\lgctx}{\tm}} \times \lmult{\tm} + \mult{\uvar}{\off{\lgctx}{\tm}}\]
\end{lemma}

\begin{proof}
By induction on the labeled context $\lgctx$.
\begin{ifLongAppendix}
\begin{xenumerate}
\item  $\lgctx=\ctxhole$.
  \[\begin{array}{rll}
      & \lmult{\ann{\uvar}{\lab}} + \mult{\uvar}{\ann{\uvar}{\lab}}\times \lmult{\tm} +\mult{\uvar}{\ann{\uvar}{\lab}}\\
      = & 0 + \lmult{\tm} + 1\\
  > &
      \lmult{\tm}\\
  = &
      \lmult{\tm} + \mult{\uvar}{\tm}\times \lmult{\tm} + \mult{\uvar}{\tm} & (\rlem{multiplicity_and_free_variables})
    \end{array}
  \]

\item $\lgctx=\lam{\lvar}{\lgctx_1}$
  \[\begin{array}{rll}
      & \lmult{\lam{\lvar}{\off{\lgctx_1}{\ann{\uvar}{\lab}}}} + \mult{\uvar}{\lam{\lvar}{\off{\lgctx_1}{\ann{\uvar}{\lab}}}}\times \lmult{\tm} + \mult{\uvar}{\lam{\lvar}{\off{\lgctx}{\ann{\uvar}{\lab}}}} \\
      = & \lmult{\off{\lgctx_1}{\ann{\uvar}{\lab}}} + \mult{\uvar}{\off{\lgctx_1}{\ann{\uvar}{\lab}}}\times \lmult{\tm} + \mult{\uvar}{\off{\lgctx}{\ann{\uvar}{\lab}}} & (\rdef{variable_multiplicity},\rdef{redex_multiplicity})\\
      > &
  \lmult{\off{\lgctx_1}{\tm}} + \mult{\uvar}{\off{\lgctx_1}{\tm}}\times \lmult{\tm} + \mult{\uvar}{\off{\lgctx}{\tm}} & (\ih)\\
      = &
  \lmult{\lam{\lvar}{\off{\lgctx_1}{\tm}}} + \mult{\uvar}{\lam{\lvar}{\off{\lgctx_1}{\tm}}}\times \lmult{\tm} + \mult{\uvar}{\lam{\lvar}{\off{\lgctx}{\tm}}}  & (\rdef{variable_multiplicity}) 
    \end{array}
  \]

\item $\lgctx= (\llam{\lvar}{\lgctx_1}{\labtwo})\lsctx\,\tmtwo$

  \item $\lgctx= (\llam{\lvar}{\tmthree}{\labtwo})\lsctx_1\lesub{\uvartwo}{\lgctx_1}{(\labthree)}\lsctx_2\,\tmtwo$

\item $\lgctx= (\llam{\lvar}{\tmthree}{\labtwo})\lsctx\,\lgctx_1$
  
\item $\lgctx=\lgctx_1\,\tmtwo$
  \[\begin{array}{rll}
     & \lmult{\off{\lgctx_1}{\ann{\uvar}{\lab}}\,\tmtwo} + \mult{\uvar}{\off{\lgctx_1}{\ann{\uvar}{\lab}}\,\tmtwo}\times \lmult{\tm} + \mult{\uvar}{\off{\lgctx}{\ann{\uvar}{\lab}}\,\tmtwo}\\
      = & \lmult{\off{\lgctx_1}{\ann{\uvar}{\lab}}} +  \lmult{\tmtwo} + (\mult{\uvar}{\off{\lgctx_1}{\ann{\uvar}{\lab}}}+  \mult{\uvar}{\tmtwo})\times \lmult{\tm} +\mult{\uvar}{\off{\lgctx}{\ann{\uvar}{\lab}}} + \mult{\uvar}{\tmtwo} & (\rdef{redex_multiplicity})\\
      = & \lmult{\off{\lgctx_1}{\ann{\uvar}{\lab}}} +  \lmult{\tmtwo} + \mult{\uvar}{\off{\lgctx_1}{\ann{\uvar}{\lab}}}\times \lmult{\tm} +  \mult{\uvar}{\tmtwo}\times \lmult{\tm} + \mult{\uvar}{\off{\lgctx}{\ann{\uvar}{\lab}}} + \mult{\uvar}{\tmtwo}  & \\
      > &
  \lmult{\off{\lgctx_1}{\tm}} + \mult{\uvar}{\off{\lgctx_1}{\tm}}\times \lmult{\tm} + \mult{\uvar}{\off{\lgctx}{\tm}} + \lmult{\tmtwo} + \mult{\uvar}{\tmtwo}\times \lmult{\tm} +  \mult{\uvar}{\tmtwo}& (\ih)\\
      = &
  \lmult{\off{\lgctx_1}{\tm}\,\tmtwo} + \mult{\uvar}{\off{\lgctx_1}{\tm}\,\tmtwo}\times \lmult{\tm} +  \mult{\uvar}{\off{\lgctx}{\tm}\,\tmtwo} & (\rdef{variable_multiplicity},\rdef{redex_multiplicity})
    \end{array}
  \]

\item $\lgctx=\tmtwo\,\lgctx_1$. Similar to the case $\lgctx=\lgctx_1\,\tmtwo$.

\item $\lgctx=\sha{\lgctx_1}$. Similar to the case $\lgctx=\lam{\lvar}{\lgctx_1}$.

  
\item $\lgctx=\lopen{\lgctx_1}{(\labtwo)}$. We need to consider multiple cases since out induction proceeds on well-formed labeled contexts.

  \begin{xenumerate}

  \item $\labtwo$ is not present. Similar to the case $\lgctx=\lam{\lvar}{\lgctx_1}$.
  \[\begin{array}{rll}
     & \lmult{\open{\off{\lgctx_1}{\ann{\uvar}{\lab}}}} + \mult{\uvar}{\open{\off{\lgctx_1}{\ann{\uvar}{\lab}}}}\times \lmult{\tm} + \mult{\uvar}{\open{\off{\lgctx_1}{\ann{\uvar}{\lab}}}} \\
      = & \lmult{\off{\lgctx_1}{\ann{\uvar}{\lab}}} + \mult{\uvar}{\off{\lgctx_1}{\ann{\uvar}{\lab}}}\times \lmult{\tm} +  \mult{\uvar}{\off{\lgctx_1}{\ann{\uvar}{\lab}}} & (\rdef{variable_multiplicity},\rdef{redex_multiplicity})\\
      > &
  \lmult{\off{\lgctx_1}{\tm}} + \mult{\uvar}{\off{\lgctx_1}{\tm}}\times \lmult{\tm} + \mult{\uvar}{\off{\lgctx_1}{\tm}} & (\ih)\\
      = &
  \lmult{\open{\off{\lgctx_1}{\tm}}} + \mult{\uvar}{\open{\off{\lgctx_1}{\tm}}}\times \lmult{\tm} + \mult{\uvar}{\open{\off{\lgctx_1}{\tm}}} & (\rdef{variable_multiplicity},\rdef{redex_multiplicity}) 
    \end{array}
  \]
  
  \item $\labtwo$ is present.
     \begin{xenumerate}

 \item $\lgctx_1=(\sha{\lgctx_2}) \lsctx$.  Similar to the case above.

 \item $\lgctx_1=(\sha{\tmtwo}) \lsctx_1\lesub{\uvartwo}{\lgctx_2}{(\labtwo)}\lsctx_2$. Similar to the case above.
    
    \end{xenumerate}
    
    \end{xenumerate}

\item $\lgctx=\ofc{\lgctx_1}$. Similar to the case $\lgctx=\lam{\lvar}{\lgctx_1}$.
  
\item $\lgctx=\lgctx_1\esub{\uvartwo}{\tmtwo}$

    \[\begin{array}{rll}
      & \lmult{\off{\lgctx_1}{\ann{\uvar}{\lab}}\esub{\uvartwo}{\tmtwo}} + \mult{\uvar}{\off{\lgctx_1}{\ann{\uvar}{\lab}}\esub{\uvartwo}{\tmtwo}}\times \lmult{\tm}  + \mult{\uvar}{\off{\lgctx_1}{\ann{\uvar}{\lab}}\esub{\uvartwo}{\tmtwo}}\\
      = & \lmult{\off{\lgctx_1}{\ann{\uvar}{\lab}}} + \lmult{\tmtwo}+ \mult{\uvartwo}{\off{\lgctx_1}{\ann{\uvar}{\lab}}} \times \lmult{\tmtwo} + \mult{\uvartwo}{\off{\lgctx_1}{\ann{\uvar}{\lab}}}  +  \mult{\uvar}{\off{\lgctx_1}{\ann{\uvar}{\lab}}\esub{\uvartwo}{\tmtwo}}\times \lmult{\tm} +  \mult{\uvar}{\off{\lgctx_1}{\ann{\uvar}{\lab}}\esub{\uvartwo}{\tmtwo}}& (\rdef{redex_multiplicity})\\
      = & \lmult{\off{\lgctx_1}{\ann{\uvar}{\lab}}} + \lmult{\tmtwo}+ \mult{\uvartwo}{\off{\lgctx_1}{\ann{\uvar}{\lab}}} \times \lmult{\tmtwo} + \mult{\uvartwo}{\off{\lgctx_1}{\ann{\uvar}{\lab}}} +  (\mult{\uvar}{\off{\lgctx_1}{\ann{\uvar}{\lab}}} + \mult{\uvar}{\tmtwo}+ \mult{\uvartwo}{\off{\lgctx_1}{\ann{\uvar}{\lab}}} \times \mult{\uvar}{\tmtwo} )\times \lmult{\tm} +  \mult{\uvar}{\off{\lgctx_1}{\ann{\uvar}{\lab}}\esub{\uvartwo}{\tmtwo}} & (\rdef{variable_multiplicity})\\
      = & \lmult{\off{\lgctx_1}{\ann{\uvar}{\lab}}} + \lmult{\tmtwo}+ \mult{\uvartwo}{\off{\lgctx_1}{\ann{\uvar}{\lab}}} \times \lmult{\tmtwo} + \mult{\uvartwo}{\off{\lgctx_1}{\ann{\uvar}{\lab}}} +  (\mult{\uvar}{\off{\lgctx_1}{\ann{\uvar}{\lab}}} + \mult{\uvar}{\tmtwo}+ \mult{\uvartwo}{\off{\lgctx_1}{\ann{\uvar}{\lab}}} \times \mult{\uvar}{\tmtwo} )\times \lmult{\tm} +  \mult{\uvar}{\off{\lgctx_1}{\ann{\uvar}{\lab}}} + \lmult{\tmtwo}+ \mult{\uvartwo}{\off{\lgctx_1}{\ann{\uvar}{\lab}}} \times \mult{\uvar}{\tmtwo} & (\rdef{variable_multiplicity})\\
      > &
   \lmult{\off{\lgctx_1}{\tm}} + \mult{\uvar}{\off{\lgctx_1}{\tm}}\times \lmult{\tm} + \mult{\uvar}{\off{\lgctx_1}{\tm}} + \lmult{\tmtwo}+ \mult{\uvartwo}{\off{\lgctx_1}{\ann{\uvar}{\lab}}}  +  \mult{\uvartwo}{\off{\lgctx_1}{\ann{\uvar}{\lab}}} \times \lmult{\tmtwo} +  (\mult{\uvar}{\tmtwo}+ \mult{\uvartwo}{\off{\lgctx_1}{\ann{\uvar}{\lab}}} \times \mult{\uvar}{\tmtwo} )\times \lmult{\tm} +  \lmult{\tmtwo}+ \mult{\uvartwo}{\off{\lgctx_1}{\ann{\uvar}{\lab}}} \times \mult{\uvar}{\tmtwo} & (\ih)\\
      = &
      \lmult{\off{\lgctx_1}{\tm}} + \mult{\uvar}{\off{\lgctx_1}{\tm}}\times \lmult{\tm} + \mult{\uvar}{\off{\lgctx_1}{\tm}} + \lmult{\tmtwo}+  \mult{\uvartwo}{\off{\lgctx_1}{\tm}}  + \mult{\uvartwo}{\off{\lgctx_1}{\tm}} \times \lmult{\tmtwo} +  (\mult{\uvar}{\tmtwo}+ \mult{\uvartwo}{\off{\lgctx_1}{\ann{\uvar}{\lab}}} \times \mult{\uvar}{\tmtwo} )\times \lmult{\tm} +  \lmult{\tmtwo}+ \mult{\uvartwo}{\off{\lgctx_1}{\ann{\uvar}{\lab}}} \times \mult{\uvar}{\tmtwo}& (\rlem{multiplicity_and_non_capturing_replacement}\times 2)\\
        = &
       \lmult{\off{\lgctx_1}{\tm}\esub{\uvartwo}{\tmtwo}}  + \mult{\uvar}{\off{\lgctx_1}{\tm}}\times \lmult{\tm} + \mult{\uvar}{\off{\lgctx_1}{\tm}} +  (\mult{\uvar}{\tmtwo}+ \mult{\uvartwo}{\off{\lgctx_1}{\tm}} \times \mult{\uvar}{\tmtwo} )\times \lmult{\tm} +  \mult{\uvar}{\tmtwo}+ \mult{\uvartwo}{\off{\lgctx_1}{\ann{\uvar}{\lab}}} \times \mult{\uvar}{\tmtwo}& (\rdef{variable_multiplicity})\\

        = &
       \lmult{\off{\lgctx_1}{\tm}\esub{\uvartwo}{\tmtwo}}  +  \mult{\uvar}{\off{\lgctx_1}{\tm}\esub{\uvartwo}{\tmtwo}}\times \lmult{\tm}  + \mult{\uvar}{\off{\lgctx_1}{\tm}} +   \mult{\uvar}{\tmtwo}+ \mult{\uvartwo}{\off{\lgctx_1}{\ann{\uvar}{\lab}}} \times \mult{\uvar}{\tmtwo}& (\rdef{variable_multiplicity})\\
        = &
       \lmult{\off{\lgctx_1}{\tm}\esub{\uvartwo}{\tmtwo}}  +  \mult{\uvar}{\off{\lgctx_1}{\tm}\esub{\uvartwo}{\tmtwo}}\times \lmult{\tm}  + \mult{\uvar}{\off{\lgctx_1}{\tm}} +   \mult{\uvar}{\tmtwo}+ \mult{\uvartwo}{\off{\lgctx_1}{\tm}} \times \mult{\uvar}{\tmtwo}& (\rlem{multiplicity_and_non_capturing_replacement})\\
      =  &
  \lmult{\off{\lgctx_1}{\tm}\esub{\uvartwo}{\tmtwo}} + \mult{\uvar}{\off{\lgctx_1}{\tm}\esub{\uvartwo}{\tmtwo}}\times \lmult{\tm} +  \mult{\uvar}{\off{\lgctx_1}{\tm}\esub{\uvartwo}{\tmtwo}}& (\uvar\notin\fv{t}) 
    \end{array}
  \]

  \item $\lgctx=\lgctx_1\lesub{\uvartwo}{\tmtwo}{\labtwo}$

    \[\begin{array}{rll}
      & \lmult{\off{\lgctx_1}{\ann{\uvar}{\lab}}\lesub{\uvartwo}{\tmtwo}{\labtwo}} + \mult{\uvar}{\off{\lgctx_1}{\ann{\uvar}{\lab}}\lesub{\uvartwo}{\tmtwo}{\labtwo}}\times \lmult{\tm}  + \mult{\uvar}{\off{\lgctx_1}{\ann{\uvar}{\lab}}\lesub{\uvartwo}{\tmtwo}{\labtwo}}\\
      = & 1 + \lmult{\off{\lgctx_1}{\ann{\uvar}{\lab}}} + \lmult{\tmtwo} +  \mult{\uvar}{\off{\lgctx_1}{\ann{\uvar}{\lab}}\esub{\uvartwo}{\tmtwo}}\times \lmult{\tm} +  \mult{\uvar}{\off{\lgctx_1}{\ann{\uvar}{\lab}}\esub{\uvartwo}{\tmtwo}}& (\rdef{redex_multiplicity})\\
      = & 1 + \lmult{\off{\lgctx_1}{\ann{\uvar}{\lab}}} + \lmult{\tmtwo}+  (\mult{\uvar}{\off{\lgctx_1}{\ann{\uvar}{\lab}}} + \mult{\uvar}{\tmtwo}+ \mult{\uvartwo}{\off{\lgctx_1}{\ann{\uvar}{\lab}}} \times \mult{\uvar}{\tmtwo} )\times \lmult{\tm} +  \mult{\uvar}{\off{\lgctx_1}{\ann{\uvar}{\lab}}\esub{\uvartwo}{\tmtwo}} & (\rdef{variable_multiplicity})\\
      = & 1 + \lmult{\off{\lgctx_1}{\ann{\uvar}{\lab}}} + \lmult{\tmtwo}+  (\mult{\uvar}{\off{\lgctx_1}{\ann{\uvar}{\lab}}} + \mult{\uvar}{\tmtwo}+ \mult{\uvartwo}{\off{\lgctx_1}{\ann{\uvar}{\lab}}} \times \mult{\uvar}{\tmtwo} )\times \lmult{\tm} +  \mult{\uvar}{\off{\lgctx_1}{\ann{\uvar}{\lab}}} + \lmult{\tmtwo}+ \mult{\uvartwo}{\off{\lgctx_1}{\ann{\uvar}{\lab}}} \times \mult{\uvar}{\tmtwo} & (\rdef{variable_multiplicity})\\
      > &
   1 + \lmult{\off{\lgctx_1}{\tm}} + \mult{\uvar}{\off{\lgctx_1}{\tm}}\times \lmult{\tm} + \mult{\uvar}{\off{\lgctx_1}{\tm}} + \lmult{\tmtwo}+ \mult{\uvartwo}{\off{\lgctx_1}{\ann{\uvar}{\lab}}}  +  \mult{\uvartwo}{\off{\lgctx_1}{\ann{\uvar}{\lab}}} \times \lmult{\tmtwo} +  (\mult{\uvar}{\tmtwo}+ \mult{\uvartwo}{\off{\lgctx_1}{\ann{\uvar}{\lab}}} \times \mult{\uvar}{\tmtwo} )\times \lmult{\tm} +  \lmult{\tmtwo}+ \mult{\uvartwo}{\off{\lgctx_1}{\ann{\uvar}{\lab}}} \times \mult{\uvar}{\tmtwo} & (\ih)\\
      = &
      1 + \lmult{\off{\lgctx_1}{\tm}} + \mult{\uvar}{\off{\lgctx_1}{\tm}}\times \lmult{\tm} + \mult{\uvar}{\off{\lgctx_1}{\tm}} + \lmult{\tmtwo}+  \mult{\uvartwo}{\off{\lgctx_1}{\tm}}  + \mult{\uvartwo}{\off{\lgctx_1}{\tm}} \times \lmult{\tmtwo} +  (\mult{\uvar}{\tmtwo}+ \mult{\uvartwo}{\off{\lgctx_1}{\ann{\uvar}{\lab}}} \times \mult{\uvar}{\tmtwo} )\times \lmult{\tm} +  \lmult{\tmtwo}+ \mult{\uvartwo}{\off{\lgctx_1}{\ann{\uvar}{\lab}}} \times \mult{\uvar}{\tmtwo}& (\rlem{multiplicity_and_non_capturing_replacement}\times 2)\\
        = &
      \lmult{\off{\lgctx_1}{\tm}\lesub{\uvartwo}{\tmtwo}{\labtwo}}  + \mult{\uvar}{\off{\lgctx_1}{\tm}}\times \lmult{\tm} + \mult{\uvar}{\off{\lgctx_1}{\tm}} +  (\mult{\uvar}{\tmtwo}+ \mult{\uvartwo}{\off{\lgctx_1}{\tm}} \times \mult{\uvar}{\tmtwo} )\times \lmult{\tm} +  \mult{\uvar}{\tmtwo}+ \mult{\uvartwo}{\off{\lgctx_1}{\ann{\uvar}{\lab}}} \times \mult{\uvar}{\tmtwo}& (\rdef{variable_multiplicity})\\

        = &
       \lmult{\off{\lgctx_1}{\tm}\lesub{\uvartwo}{\tmtwo}{\labtwo}}  +  \mult{\uvar}{\off{\lgctx_1}{\tm}\lesub{\uvartwo}{\tmtwo}{\labtwo}}\times \lmult{\tm}  + \mult{\uvar}{\off{\lgctx_1}{\tm}} +   \mult{\uvar}{\tmtwo}+ \mult{\uvartwo}{\off{\lgctx_1}{\ann{\uvar}{\lab}}} \times \mult{\uvar}{\tmtwo}& (\rdef{variable_multiplicity})\\
        = &
       \lmult{\off{\lgctx_1}{\tm}\lesub{\uvartwo}{\tmtwo}{\labtwo}}  +  \mult{\uvar}{\off{\lgctx_1}{\tm}\lesub{\uvartwo}{\tmtwo}{\labtwo}}\times \lmult{\tm}  + \mult{\uvar}{\off{\lgctx_1}{\tm}} +   \mult{\uvar}{\tmtwo}+ \mult{\uvartwo}{\off{\lgctx_1}{\tm}} \times \mult{\uvar}{\tmtwo}& (\rlem{multiplicity_and_non_capturing_replacement})\\
      =  &
  \lmult{\off{\lgctx_1}{\tm}\lesub{\uvartwo}{\tmtwo}{\labtwo}} + \mult{\uvar}{\off{\lgctx_1}{\tm}\lesub{\uvartwo}{\tmtwo}{\labtwo}}\times \lmult{\tm} +  \mult{\uvar}{\off{\lgctx_1}{\tm}\lesub{\uvartwo}{\tmtwo}{\labtwo}}& (\uvar\notin\fv{t}) 
    \end{array}
  \]



\item $\lgctx=\tmtwo\esub{\uvartwo}{\lgctx_1}$. Similar to the previous case.

  \item $\lgctx=\tmtwo\lesub{\uvartwo}{\lgctx_1}{\labtwo}$. Similar to the previous case.

\end{xenumerate}

  \end{ifLongAppendix}
\end{proof}

\begin{lemma}
\llem{redex_multiplicity_replacement_main}
  Suppose $\dom{\lsctx}\cap(\fv{\lgctx}\cup\{\uvar\})=\emptyset$. Then
  \[
    \lmult{\off{\lgctx}{\ann{\uvar}{\lab}}} + \lmult{\tm\lsctx}  + \mult{\uvar}{\off{\lgctx}{\ann{\uvar}{\lab}}}\times \lmult{\tm\lsctx} + \mult{\uvar}{\off{\lgctx}{\ann{\uvar}{\lab}}}
    >  \lmult{\off{\lgctx}{\tm}\esub{\uvar}{\tm}} + \lmultctx{\mult{\bullet}{\off{\lgctx}{\tm}\esub{\uvar}{\tm}}}{\lsctx}
    \]

\end{lemma}

\begin{proof}
  By induction on $\lsctx$.
  \begin{ifLongAppendix}
    \begin{xenumerate}
  \item $\lsctx = \ctxhole$

    \[\begin{array}{rll}
        & \lmult{\off{\lgctx}{\ann{\uvar}{\lab}}} + \lmult{\tm}  + \mult{\uvar}{\off{\lgctx}{\ann{\uvar}{\lab}}}\times \lmult{\tm} + \mult{\uvar}{\off{\lgctx}{\ann{\uvar}{\lab}}} \\
        >       &  \lmult{\off{\lgctx}{\tm}} + \lmult{\tm} + \mult{\uvar}{\off{\lgctx}{\tm}} \times \lmult{\tm} + \mult{\uvar}{\off{\lgctx}{\tm}} & (\rlem{redex_multiplicity_and_replacement})\\
        =       &  \lmult{\off{\lgctx}{\tm}\esub{\uvar}{\tm}}  & (\rdef{redex_multiplicity})\\
      =  &  \lmult{\off{\lgctx}{\tm}\esub{\uvar}{\tm}} + \lmultctx{\mult{\bullet}{\off{\lgctx}{\tm}\esub{\uvar}{\tm}}}{\ctxhole} & (\rdef{redex_multiplicity})
      \end{array}
     \]

  \item $\lsctx = \lsctx_1\esub{\uvartwo}{\tmtwo}$

        \[\begin{array}{rll}
        & \lmult{\off{\lgctx}{\ann{\uvar}{\lab}}} + \lmult{\tm \lsctx_1\esub{\uvartwo}{\tmtwo}}  + \mult{\uvar}{\off{\lgctx}{\ann{\uvar}{\lab}}}\times \lmult{\tm \lsctx_1\esub{\uvartwo}{\tmtwo}} + \mult{\uvar}{\off{\lgctx}{\ann{\uvar}{\lab}}} \\
            =       &  \lmult{\off{\lgctx}{\ann{\uvar}{\lab}}} + \lmult{\tm \lsctx_1}+\lmult{\tmtwo} + \mult{\uvartwo}{\tm \lsctx_1}\times \lmult{\tmtwo}+ \mult{\uvartwo}{\tm \lsctx_1} + \mult{\uvar}{\off{\lgctx}{\ann{\uvar}{\lab}}}\times (\lmult{\tm \lsctx_1}+\lmult{\tmtwo} + \mult{\uvartwo}{\tm \lsctx_1}\times \lmult{\tmtwo}+ \mult{\uvartwo}{\tm \lsctx_1})+ \mult{\uvar}{\off{\lgctx}{\ann{\uvar}{\lab}}} & \\
            >       &   \lmult{\off{\lgctx}{\tm}\esub{\uvar}{\tm}} +  \lmultctx{\mult{\bullet}{\off{\lgctx}{\tm}\esub{\uvar}{\tm}}}{\lsctx_1} + \lmult{\tmtwo} + \mult{\uvartwo}{\tm \lsctx_1}\times \lmult{\tmtwo}+ \mult{\uvartwo}{\tm \lsctx_1} + \mult{\uvar}{\off{\lgctx}{\ann{\uvar}{\lab}}}\times (\lmult{\tmtwo} + \mult{\uvartwo}{\tm \lsctx_1}\times \lmult{\tmtwo}+ \mult{\uvartwo}{\tm \lsctx_1}) & (\ih) \\
            \geq        &   \lmult{\off{\lgctx}{\tm}\esub{\uvar}{\tm}} +  \lmultctx{\mult{\bullet}{\off{\lgctx}{\tm}\esub{\uvar}{\tm}}}{\lsctx_1} + \lmult{\tmtwo} + \mult{\uvartwo}{\tm \lsctx_1}\times \lmult{\tmtwo}+ \mult{\uvar}{\off{\lgctx}{\ann{\uvar}{\lab}}}\times (\lmult{\tmtwo} + \mult{\uvartwo}{\tm \lsctx_1}\times \lmult{\tmtwo}) +\lmultctx{\mult{\bullet}{\off{\lgctx}{\tm}\esub{\uvar}{\tm}}}{\lsctx_1}& (\rlem{context_multiplicity_and_replacement}) \\

                        \geq        &   \lmult{\off{\lgctx}{\tm}\esub{\uvar}{\tm}} +  \lmultctx{\mult{\bullet}{\off{\lgctx}{\tm}\esub{\uvar}{\tm}}}{\lsctx_1} + \lmult{\tmtwo} + \mult{\uvar}{\off{\lgctx}{\ann{\uvar}{\lab}}}\times \lmult{\tmtwo} + \multctx{\uvartwo}{\mult{\bullet}{\off{\lgctx}{\tm}\esub{\uvar}{\tm}}}{\lsctx_1}\times \lmult{\tmtwo} + \lmultctx{\mult{\bullet}{\off{\lgctx}{\tm}\esub{\uvar}{\tm}}}{\lsctx_1}& (\rlem{context_multiplicity_and_replacement}) \\

        \geq       &  \lmult{\off{\lgctx}{\tm}\esub{\uvar}{\tm}} +  \lmultctx{\mult{\bullet}{\off{\lgctx}{\tm}\esub{\uvar}{\tm}}}{\lsctx_1} + \lmult{\tmtwo} + \multctx{\uvartwo}{\mult{\bullet}{\off{\lgctx}{\tm}\esub{\uvar}{\tm}}}{\lsctx_1}\times \lmult{\tmtwo}+ \multctx{\uvartwo}{\mult{\bullet}{\off{\lgctx}{\tm}\esub{\uvar}{\tm}}}{\lsctx_1}\\
            =  &  \lmult{\off{\lgctx}{\tm}\esub{\uvar}{\tm}} + \lmultctx{\mult{\bullet}{\off{\lgctx}{\tm}\esub{\uvar}{\tm}}}{\lsctx_1\esub{\uvartwo}{\tmtwo}}
      \end{array}
    \]

    We have used the fact that $\mult{\uvartwo}{\off{\lgctx}{\ann{\uvar}{\lab}}}=0$
  \item $\lsctx = \lsctx_1\lesub{\uvartwo}{\tmtwo}{\lab}$

            \[\begin{array}{rll}
        & \lmult{\off{\lgctx}{\ann{\uvar}{\lab}}} + \lmult{\tm \lsctx_1\lesub{\uvartwo}{\tmtwo}{\lab}}  + \mult{\uvar}{\off{\lgctx}{\ann{\uvar}{\lab}}}\times \lmult{\tm \lsctx_1\lesub{\uvartwo}{\tmtwo}{\lab}} + \mult{\uvar}{\off{\lgctx}{\ann{\uvar}{\lab}}} \\
                =       &  \lmult{\off{\lgctx}{\ann{\uvar}{\lab}}} + 1 + \lmult{\tm \lsctx_1}+\lmult{\tmtwo} + \mult{\uvar}{\off{\lgctx}{\ann{\uvar}{\lab}}}\times (1 + \lmult{\tm \lsctx_1}+\lmult{\tmtwo})+ \mult{\uvar}{\off{\lgctx}{\ann{\uvar}{\lab}}} & \\
                            =       &  \lmult{\off{\lgctx}{\ann{\uvar}{\lab}}} + 1 + \lmult{\tm \lsctx_1}+\lmult{\tmtwo} + \mult{\uvar}{\off{\lgctx}{\ann{\uvar}{\lab}}} + \mult{\uvar}{\off{\lgctx}{\ann{\uvar}{\lab}}}\times \lmult{\tm \lsctx_1}+\mult{\uvar}{\off{\lgctx}{\ann{\uvar}{\lab}}}\times \lmult{\tmtwo}+ \mult{\uvar}{\off{\lgctx}{\ann{\uvar}{\lab}}} & \\
                            >       &   \lmult{\off{\lgctx}{\tm}\esub{\uvar}{\tm}} +  \lmultctx{\mult{\bullet}{\off{\lgctx}{\tm}\esub{\uvar}{\tm}}}{\lsctx_1}  + 1 + \lmult{\tmtwo} + \mult{\uvar}{\off{\lgctx}{\ann{\uvar}{\lab}}} + \mult{\uvar}{\off{\lgctx}{\ann{\uvar}{\lab}}}\times \lmult{\tmtwo}+ \mult{\uvar}{\off{\lgctx}{\ann{\uvar}{\lab}}} & (\ih) \\
                            >       &   \lmult{\off{\lgctx}{\tm}\esub{\uvar}{\tm}} +  \lmultctx{\mult{\bullet}{\off{\lgctx}{\tm}\esub{\uvar}{\tm}}}{\lsctx_1}  + 1 + \lmult{\tmtwo} & \\

        =       &  \lmult{\off{\lgctx}{\tm}\esub{\uvar}{\tm}} +  1 + \lmultctx{\mult{\bullet}{\off{\lgctx}{\tm}\esub{\uvar}{\tm}}}{\lsctx_1} + \lmult{\tmtwo} \\
            =  &  \lmult{\off{\lgctx}{\tm}\esub{\uvar}{\tm}} + \lmultctx{\mult{\bullet}{\off{\lgctx}{\tm}\esub{\uvar}{\tm}}}{\lsctx_1\lesub{\uvartwo}{\tmtwo}{\lab}}
      \end{array}
    \]

  \end{xenumerate}
  \end{ifLongAppendix}
  \end{proof}

\begin{proposition}
\lprop{labeled_reduction_is_finite}
$\tm \lto{\lab}_R\tmtwo$ implies $\lmult{\tm}> \lmult{\tmtwo}$
\end{proposition}

\begin{proof}
  First  we consider each of the four cases for reduction at the root $\tm \lrootto{\lab}_R\tmtwo$:
  \begin{xenumerate}

    \item  Suppose $\tm=   (\llam{\lvar}{\tm_1}{\lab})\lsctx\,\tm_2
    \rtoSdbL{\lab}
    \tm_1\sub{\lvar}{\tm_2}\lsctx = \tmtwo$ and $\fv{\tm_2} \cap \dom{\lsctx} = \emptyset$.

    \[\begin{array}{rll}
        &  \lmult{ (\llam{\lvar}{\tm_1}{\lab})\lsctx\,\tm_2} \\
        = &  1 + \lmult{(\lam{\lvar}{\tm_1})\lsctx} + \lmult{\tm_2} + \mult{\lvar}{\tm_1}\times \lmult{\tm_2} & (\rdef{redex_multiplicity})\\
        > &  \lmult{(\lam{\lvar}{\tm_1})\lsctx} + \mult{\lvar}{\tm_1}\times \lmult{\tm_2} & \\
        =  &    \lmult{\tm_1\sub{\lvar}{\tm_2}} + \lmultctx{\mult{\bullet}{\tm_1\sub{\lvar}{\tm_2}}}{\lsctx} & (\rlem{redex_multiplicity_and_linear_beta})\\
        =  &   \lmult{\tm_1\sub{\lvar}{\tm_2}\lsctx} & (\rlem{redex_multiplicity_of_substitution_contexts})
        \end{array}\]

   \item Suppose $\tm=\lopen{(\sha{\tm_1})\lsctx}{\lab}
     \rtoSopenL{\lab} 
     \tm_1\lsctx  = \tmtwo$

         \[\begin{array}{rll}
        &  \lmult{\lopen{(\sha{\tm_1})\lsctx}{\lab}} \\
        = &  1 + \lmult{(\sha{\tm_1})\lsctx} & (\rdef{redex_multiplicity})\\
             =  &  1 + \lmult{\tm_1\lsctx} \\
             > & \lmult{\tm_1\lsctx}
        \end{array}\]

     \item Suppose $\tm=\off{\lgctx}{\ann{\uvar}{\lab}}\esub{\uvar}{(\ofc{(\sha{\tm_1})\lsctx_1})\lsctx_2}
    \rtoSlsL{\lab} 
    \off{\lgctx}{(\sha{\tm_1}) \lsctx_1}\esub{\uvar}{\ofc{(\sha{\tm_1}) \lsctx_1}}\lsctx_2=\tmtwo$ and $\uvar \notin \fv{\tm_1}$ and $\fv{\lgctx} \cap \dom{\lsctx_1\lsctx_2} = \emptyset$.

                 \[\begin{array}{rll}
        &  \lmult{\off{\lgctx}{\ann{\uvar}{\lab}}\esub{\uvar}{(\ofc{(\sha{\tm_1})\lsctx_1})\lsctx_2}} \\
        = &  \lmult{\off{\lgctx}{\ann{\uvar}{\lab}}} + \lmult{(\ofc{(\sha{\tm_1})\lsctx_1})\lsctx_2}  + \mult{\uvar}{\off{\lgctx}{\ann{\uvar}{\lab}}}\times \lmult{(\ofc{(\sha{\tm_1})\lsctx_1})\lsctx_2} + \mult{\uvar}{\off{\lgctx}{\ann{\uvar}{\lab}}} & (\rdef{redex_multiplicity})\\
        >  &  \lmult{\off{\lgctx}{(\sha{\tm_1}) \lsctx_1}\esub{\uvar}{\ofc{(\sha{\tm_1}) \lsctx_1}}} + \lmultctx{\mult{\bullet}{\off{\lgctx}{(\sha{\tm_1}) \lsctx_1}\esub{\uvar}{\ofc{(\sha{\tm_1}) \lsctx_1}}}}{\lsctx_2} & (\rlem{redex_multiplicity_replacement_main})\\
        =  &  \lmult{\off{\lgctx}{(\sha{\tm_1}) \lsctx_1}\esub{\uvar}{\ofc{(\sha{\tm_1}) \lsctx_1}}\lsctx_2} &  (\rlem{redex_multiplicity_of_substitution_contexts})\\
                   \end{array}
                 \]
                 
\item Suppose $\tm=\tm_1\esub{\ann{\uvar}{\lab}}{(\ofc{\tm_2})\lsctx}
    \rtoSgcL{\lab} 
    \tm_1\lsctx=\tmtwo$ and $\uvar\notin\fv{\tm_1}$.

    \[\begin{array}{rll}
        &  \lmult{\tm_1\esub{\ann{\uvar}{\lab}}{(\ofc{\tm_2})\lsctx}} \\
        = &  1+ \lmult{\tm_1} + \lmult{(\ofc{\tm_2})\lsctx}  & (\rdef{redex_multiplicity})\\
        = &  1+ \lmult{\tm_1} + \lmult{\ofc{\tm_2}}  + \lmultctx{\mult{\bullet}{(\ofc{\tm_2})}}{\lsctx}  & (\rlem{redex_multiplicity_of_substitution_contexts}) \\
        > &  \lmult{\tm_1} + \lmultctx{\mult{\bullet}{(\ofc{\tm_2})}}{\lsctx}  & \\
        \geq  &  \lmult{\tm_1\lsctx} & (\rlem{multiplicity_of_term_context_split}(2)) \\
      \end{array}\]
  \end{xenumerate}
  \begin{ifShortAppendix}
  For the cases where reduction is internal, see the extended
  version~\cite{mells_long}.
    \end{ifShortAppendix}
  \begin{ifLongAppendix}
 Next we consider internal reduction.
\begin{xenumerate}

\item $\lgctx=\lam{\lvar}{\lgctx_1}$. Suppose $\tm  = \lam{\lvar}{\off{\lgctx_1}{\tm_1}} \to_R \lam{\lvar}{\off{\lgctx_1}{\tmtwo_1}} = \tmtwo$ follows from $\tm_1 \to_R \tmtwo_1$.

      \[\begin{array}{rll}
        &  \lmult{ \lam{\lvar}{\off{\lgctx_1}{\tm_1}} } \\
        = &  \lmult{\off{\lgctx_1}{\tm_1} } & (\rdef{redex_multiplicity})\\
        > &  \lmult{\off{\lgctx_1}{\tmtwo_1}} & (\ih)\\
        =  &   \lmult{\lam{\lvar}{\off{\lgctx_1}{\tmtwo_1}} } & (\rdef{redex_multiplicity})
        \end{array}\]
       
\item $\lgctx= (\llam{\lvar}{\lgctx_1}{\labtwo})\lsctx\,\tm_1$. Suppose $\tm  = (\llam{\lvar}{\off{\lgctx_1}{\tm_2}}{\labtwo})\lsctx\,\tm_1 \to_R (\llam{\lvar}{\off{\lgctx_1}{\tmtwo_2}}{\labtwo})\lsctx\,\tm_1 = \tmtwo$ follows from $\tm_2 \to_R \tmtwo_2$.

       \[\begin{array}{rll}
        &  \lmult{ (\llam{\lvar}{\off{\lgctx_1}{\tm_2}}{\labtwo})\lsctx\,\tm_1} \\
        = &  1+\lmult{(\lam{\lvar}{\off{\lgctx_1}{\tm_2}})\lsctx} + \lmult{\tm_1} + \mult{\lvar}{\off{\lgctx_1}{\tm_2}}\times \lmult{\tm_1}& (\rdef{redex_multiplicity})\\
        = &  1+\lmult{\lam{\lvar}{\off{\lgctx_1}{\tm_2}}} + \lmultctx{\mult{\bullet}{(\lam{\lvar}{\off{\lgctx_1}{\tm_2}})}}{\lsctx}+ \lmult{\tm_1} + \mult{\lvar}{\off{\lgctx_1}{\tm_2}}\times \lmult{\tm_1}& (\rlem{redex_multiplicity_of_substitution_contexts})\\
        > &  1+\lmult{\lam{\lvar}{\off{\lgctx_1}{\tmtwo_2}}} + \lmultctx{\mult{\bullet}{(\lam{\lvar}{\off{\lgctx_1}{\tmtwo_2}})}}{\lsctx}+ \lmult{\tm_1} + \mult{\lvar}{\off{\lgctx_1}{\tmtwo_2}}\times \lmult{\tm_1}& (\ih\times 2)\\
        = & 1+\lmult{(\lam{\lvar}{\off{\lgctx_1}{\tmtwo_2}})\lsctx} + \lmult{\tm_1} + \mult{\lvar}{\off{\lgctx_1}{\tmtwo_2}}\times \lmult{\tm_1}&  (\rlem{redex_multiplicity_of_substitution_contexts})\\
        =  &    (\llam{\lvar}{\off{\lgctx_1}{\tmtwo_2}}{\labtwo})\lsctx\,\tm_1 & (\rdef{redex_multiplicity})\\
         \end{array}\]

          \item $\lgctx= \ctxhole\lsctx_2\,\tm_1$ and $\tm  = \off{\lgctxtwo}{\ann{\uvar}{\lab}}\esub{\uvar}{(\ofc{(\sha{\tm_3})\lsctxtwo_1})\lsctxtwo_2}\lsctx_2\,\tm_1 \lto{\lab}_R \off{\lgctxtwo}{(\sha{\tm_3})\lsctxtwo_1}\esub{\uvar}{(\ofc{(\sha{\tm_3})\lsctxtwo_1})}\lsctxtwo_2\lsctx_2\,\tm_1 = \tmtwo$ and $\off{\lgctxtwo}{\ann{\uvar}{\lab}}=(\llam{\lvar}{\tm_1}{\labtwo}) \lsctx_1$. There are to cases depending on the location of the hole in $\lgctxtwo$.

       \begin{xenumerate}

         \item  $\lgctxtwo=(\llam{\lvar}{\lgctxtwo_1}{\labtwo}) \lsctx_1$.  Let $\lsctx=\lsctx_1\esub{\uvar}{(\ofc{(\sha{\tm_3})\lsctxtwo_1})\lsctxtwo_2}\lsctx_2$. Similar to the previous case.

     \item  $\lgctxtwo=(\llam{\lvar}{\tm_1}{\labtwo}) \lsctx_{11}\lesub{\uvartwo}{\lgctxtwo_1}{(\labthree)}\lsctx_{12}$. Similar to the previous case.

     \end{xenumerate}

          \item $\lgctx= \ctxhole\lsctx_2\,\tm_1$ and $\tm  = (\llam{\lvar}{\tm_2}{\labtwo}) \lsctx_1\lesub{\uvar}{(\ofc{\tm_3})\lsctxtwo}{\lab}\lsctx_2\,\tm_1 \lto{\lab}_R (\llam{\lvar}{\tm_2}{\labtwo}) \lsctx_1\lsctxtwo\lsctx_2\,\tm_1 = \tmtwo$. Similar to the previous case.

  \item $\lgctx= (\llam{\lvar}{\tm_1}{\labtwo})\lsctx_1\lesub{\uvartwo}{\lgctx_1}{(\labthree)}\lsctx_2\,\tm_2$. Suppose $\tm  =(\llam{\lvar}{\tm_1}{\labtwo})\lsctx_1\lesub{\uvartwo}{\off{\lgctx_1}{\tm_3}}{(\labthree)}\lsctx_2\,\tm_2\to_R (\llam{\lvar}{\tm_1}{\labtwo})\lsctx_1\lesub{\uvartwo}{\off{\lgctx_1}{\tmtwo_3}}{(\labthree)}\lsctx_2\,\tm_2= \tmtwo$ follows from $\tm_3 \to_R \tmtwo_3$.

          \[\begin{array}{rll}
        &  \lmult{ (\llam{\lvar}{\tm_1}{\labtwo})\lsctx_1\lesub{\uvartwo}{\off{\lgctx_1}{\tm_3}}{(\labthree)}\lsctx_2\,\tm_2} \\
          = &  1 + \lmult{(\lam{\lvar}{\tm_1})\lsctx_1\lesub{\uvartwo}{\off{\lgctx_1}{\tm_3}}{(\labthree)}\lsctx_2} + \lmult{\tm_2} + 
            \mult{\lvar}{\tm_1}\times \lmult{\tm_2} & (\rdef{redex_multiplicity})\\
          = &  1 + \lmult{(\lam{\lvar}{\tm_1})\lsctx_1\lesub{\uvartwo}{\off{\lgctx_1}{\tm_3}}{(\labthree)}} + \lmultctx{\mult{\bullet}{(\lam{\lvar}{\tm_1})\lsctx_1\lesub{\uvartwo}{\off{\lgctx_1}{\tm_3}}{(\labthree)}}}{\lsctx_2} + \lmult{\tm_2} + 
            \mult{\lvar}{\tm_1}\times \lmult{\tm_2} & (\rlem{redex_multiplicity_of_substitution_contexts}) \\
          \geq &  1 + \lmult{(\lam{\lvar}{\tm_1})\lsctx_1\lesub{\uvartwo}{\off{\lgctx_1}{\tm_3}}{(\labthree)}} + \lmultctx{\mult{\bullet}{(\lam{\lvar}{\tm_1})\lsctx_1\lesub{\uvartwo}{\off{\lgctx_1}{\tmtwo_3}}{(\labthree)}}}{\lsctx_2} + \lmult{\tm_2} +  
            \mult{\lvar}{\tm_1}\times \lmult{\tm_2} & (\rprop{reduction_and_variable_multiplicity}) \\
          > &  1 + \lmult{(\lam{\lvar}{\tm_1})\lsctx_1\lesub{\uvartwo}{\off{\lgctx_1}{\tmtwo_3}}{(\labthree)}} + \lmultctx{\mult{\bullet}{(\lam{\lvar}{\tm_1})\lsctx_1\lesub{\uvartwo}{\off{\lgctx_1}{\tmtwo_3}}{(\labthree)}}}{\lsctx_2} + \lmult{\tm_2} +  
            \mult{\lvar}{\tm_1}\times \lmult{\tm_2} & (\ih)\\
              = &  1 + \lmult{(\lam{\lvar}{\tm_1})\lsctx_1\lesub{\uvartwo}{\off{\lgctx_1}{\tmtwo_3}}{(\labthree)}} + \lmultctx{\mult{\bullet}{(\lam{\lvar}{\tm_1})\lsctx_1\lesub{\uvartwo}{\off{\lgctx_1}{\tmtwo_3}}{(\labthree)}}}{\lsctx_2} + \lmult{\tm_2} + 
            \mult{\lvar}{\tm_1}\times \lmult{\tm_2} & (\rlem{redex_multiplicity_of_substitution_contexts}) \\
          = &  1 + \lmult{(\lam{\lvar}{\tm_1})\lsctx_1\lesub{\uvartwo}{\off{\lgctx_1}{\tmtwo_3}}{(\labthree)}\lsctx_2} + \lmult{\tm_2} + 
            \mult{\lvar}{\tm_1}\times \lmult{\tm_2} & (\rdef{redex_multiplicity})\\
        =  &    \lmult{ (\llam{\lvar}{\tm_1}{\labtwo})\lsctx_1\lesub{\uvartwo}{\off{\lgctx_1}{\tmtwo_3}}{(\labthree)}\lsctx_2\,\tm_2} & (\rdef{variable_multiplicity})\\
        \end{array}\]
      
\item $\lgctx= (\llam{\lvar}{\tm_1}{\labtwo})\lsctx\,\lgctx_1$. Suppose $\tm  = (\llam{\lvar}{\tm_1}{\labtwo})\lsctx\,\off{\lgctx_1}{\tm_2}\to_R  (\llam{\lvar}{\tm_1}{\labtwo})\lsctx\,\off{\lgctx_1}{\tmtwo_2}= \tmtwo$ follows from $\tm_2 \to_R \tmtwo_2$.

         \[\begin{array}{rll}
        &  \lmult{ (\llam{\lvar}{\tm_1}{\labtwo})\lsctx\,\off{\lgctx_1}{\tm_2}} \\
        = &  1 + \lmult{ (\lam{\lvar}{\tm_1})\lsctx}  + \lmult{ \off{\lgctx_1}{\tm_2}}  + \mult{\lvar}{ \tm_1} \times \lmult{\off{\lgctx_1}{\tm_2}} & (\rdef{redex_multiplicity})\\
        > &  1 + \lmult{ (\lam{\lvar}{\tm_1})\lsctx}  + \lmult{ \off{\lgctx_1}{\tmtwo_2}}  + \mult{\lvar}{ \tm_1} \times \lmult{\off{\lgctx_1}{\tmtwo_2}}& (\ih\times 2)\\
   = & \lmult{ (\llam{\lvar}{\tm_1}{\labtwo})\lsctx\,\off{\lgctx_1}{\tmtwo_2}} & (\rdef{redex_multiplicity})\\
           \end{array}\]
         
\item $\lgctx= (\llam{\lvar}{\tm_1}{\labtwo})\lsctx\,\lgctx_1$. Suppose $\tm  = (\llam{\lvar}{\tm_1}{\labtwo})\lsctx\,\off{\lgctx_1}{\tm_2}\to_R  (\llam{\lvar}{\tm_1}{\labtwo})\lsctx\,\off{\lgctx_1}{\tmtwo_2}= \tmtwo$ follows from $\tm_2 \to_R \tmtwo_2$.
  
\item $\lgctx=\lgctx_1\,\tm_1$.  Suppose $\tm  =\off{\lgctx_1}{\tm_2}\,\tm_1\to_R \off{\lgctx_1}{\tmtwo_2}\,\tm_1= \tmtwo$ follows from $\tm_2 \to_R \tmtwo_2$.
  
\item $\lgctx=\tm_1\,\lgctx_1$. Similar to the case $\lgctx=\lgctx_1\,\tm_1$.

\item $\lgctx=\sha{\lgctx_1}$. Similar to the case $\lgctx=\lam{\lvar}{\lgctx_1}$.

  
\item $\lgctx=\lopen{\lgctx_1}{(\labtwo)}$. We need to consider multiple cases since out induction proceeds on well-formed labeled contexts.

  \begin{xenumerate}

  \item $\labtwo$ is not present. Suppose $\tm  =\open{\off{\lgctx_1}{\tm_1}}\to_R \open{\off{\lgctx_1}{\tmtwo_1}}= \tmtwo$ follows from $\tm_1 \to_R \tmtwo_1$.  Similar to the case $\lgctx=\lam{\lvar}{\lgctx_1}$.
  
  \item $\labtwo$ is present.
     \begin{xenumerate}

     \item $\lgctx_1=(\sha{\lgctx_2}) \lsctx$.  Suppose $\tm  =\lopen{(\sha{\off{\lgctx_2}{\tm_1}}) \lsctx}{\labtwo} \to_R \lopen{(\sha{\off{\lgctx_2}{\tmtwo_1}}) \lsctx}{\labtwo}= \tmtwo$ follows from $\tm_1 \to_R \tmtwo_1$.

       \[\begin{array}{rll}
           & \lmult{\lopen{(\sha{\off{\lgctx_2}{\tm_1}}) \lsctx}{\labtwo}} \\
        = &  1 + \lmult{ (\sha{\off{\lgctx_2}{\tm_1}}) \lsctx } & (\rdef{redex_multiplicity})\\ 
       =  &  1 + \lmult{ \sha{\off{\lgctx_2}{\tm_1}}} + \lmultctx{\mult{\bullet}{ (\sha{\off{\lgctx_2}{\tm_1}})}}{ \lsctx } & (\rlem{redex_multiplicity_of_substitution_contexts})\\ 
       \geq  & 1 + \lmult{ \sha{\off{\lgctx_2}{\tm_1}}} + \lmultctx{\mult{\bullet}{ (\sha{\off{\lgctx_2}{\tmtwo_1}})}}{ \lsctx } & (\rprop{reduction_and_variable_multiplicity})\\ 
             >  & 1 + \lmult{ \sha{\off{\lgctx_2}{\tmtwo_1}}} + \lmultctx{\mult{\bullet}{ (\sha{\off{\lgctx_2}{\tmtwo_1}})}}{ \lsctx } &  (\ih)\\
           = &1 +  \lmult{(\sha{\off{\lgctx_2}{\tmtwo_1}}) \lsctx} & (\rlem{redex_multiplicity_of_substitution_contexts})\\
           = & \lmult{\lopen{(\sha{\off{\lgctx_2}{\tmtwo_1}}) \lsctx}{\labtwo}} & (\rdef{redex_multiplicity})\\
           \end{array}\]

                   \item $\lgctx_1=\ctxhole\lsctx_2$ and $\tm  = \lopen{(\sha{\off{\lgctxtwo}{\ann{\uvar}{\lab}}})\lsctx_1\esub{\uvar}{(\ofc{(\sha{\tm_3})\lsctxtwo_1})\lsctxtwo_2}\lsctx_2}{\labtwo} \lto{\lab}_R \lopen{(\sha{\off{\lgctxtwo}{(\sha{\tm_3})\lsctxtwo_1}})\lsctx_1\esub{\uvar}{(\ofc{(\sha{\tm_3})\lsctxtwo_1})}\lsctxtwo_2\lsctx_2}{\labtwo} = \tmtwo$. Then we reason as in the previous case but using the root case (item 2) instead of the \ih.

   \item $\lgctx_1=\ctxhole\lsctx_2$ and
   $\tm  = \lopen{(\sha{\tm_1}) \lsctx_1\lesub{\uvar}{(\ofc{\tm_3})\lsctxtwo}{\lab}\lsctx_2}{\labtwo} \lto{\lab}_R \lopen{(\sha{\tm_1})\lsctx_1\lsctxtwo\lsctx_2}{\labtwo} = \tmtwo$. Same as previous case.

 \item $\lgctx_1=(\sha{\tm_1}) \lsctx_1\lesub{\uvartwo}{\lgctx_2}{(\labthree)}\lsctx_2$.  Suppose $\tm  = \lopen{(\sha{\tm_1}) \lsctx_1\lesub{\uvartwo}{\off{\lgctx_2}{\tm_2}}{(\labthree)}\lsctx_2}{\labtwo}\to_R (\sha{\tm_1}) \lsctx_1\lesub{\uvartwo}{\off{\lgctx_2}{\tmtwo_2}}{(\labthree)}\lsctx_2= \tmtwo$ follows from $\tm_2\to_R \tmtwo_2$. 
    
    \end{xenumerate}
    
\end{xenumerate}

\item $\lgctx=\ofc{\lgctx_1}$. Similar to the case $\lgctx=\lam{\lvar}{\lgctx_1}$.
  
\item $\lgctx=\lgctx_1\esub{\uvartwo}{\tm_1}$. Suppose $\tm  =\off{\lgctx_1}{\tm_2}\esub{\uvartwo}{\tm_1}\to_R \off{\lgctx_1}{\tmtwo_2}\esub{\uvartwo}{\tm_1}= \tmtwo$ follows from $\tm_2\to_R \tmtwo_2$. 

       \[\begin{array}{rll}
        &  \lmult{ \off{\lgctx_1}{\tm_2}\esub{\uvartwo}{\tm_1}} \\
        = &  \lmult{ \off{\lgctx_1}{\tm_2}} + \lmult{ \tm_1} + \mult{\uvartwo}{ \off{\lgctx_1}{\tm_2}}\times \lmult{\tm_1}  + \mult{\uvartwo}{ \off{\lgctx_1}{\tm_2}}& (\rdef{redex_multiplicity})\\
        \geq &  \lmult{ \off{\lgctx_1}{\tm_2}} + \lmult{ \tm_1} + \mult{\uvartwo}{ \off{\lgctx_1}{\tmtwo_2}}\times \lmult{\tm_1}  + \mult{\uvartwo}{ \off{\lgctx_1}{\tmtwo_2}}& (\rprop{reduction_and_variable_multiplicity})\\ 
        > &  \lmult{ \off{\lgctx_1}{\tmtwo_2}} + \lmult{ \tm_1} + \mult{\uvartwo}{ \off{\lgctx_1}{\tmtwo_2}}\times \lmult{\tm_1}  + \mult{\uvartwo}{ \off{\lgctx_1}{\tmtwo_2}}& (\ih)\\
   = & \lmult{ \off{\lgctx_1}{\tmtwo_2}\esub{\uvartwo}{\tm_1}} & (\rdef{redex_multiplicity})\\
           \end{array}\]

\item $\lgctx=\lgctx_1\lesub{\uvartwo}{\tm_1}{\labtwo}$. Suppose $\tm  =\off{\lgctx_1}{\tm_2}\lesub{\uvartwo}{\tm_1}{\labtwo}\to_R \off{\lgctx_1}{\tmtwo_2}\lesub{\uvartwo}{\tm_1}{\labtwo}= \tmtwo$ follows from $\tm_2\to_R \tmtwo_2$. 

        \[\begin{array}{rll}
        &  \lmult{ \off{\lgctx_1}{\tm_2}\lesub{\uvartwo}{\tm_1}{\labtwo}} \\
       = & 1 + \lmult{ \off{\lgctx_1}{\tm_2}} +\lmult{ \tm_1}    & (\rdef{redex_multiplicity})\\
       > & 1 + \lmult{ \off{\lgctx_1}{\tmtwo_2}}  +\lmult{ \tm_1}   &(\ih)\\
   = & \lmult{\off{\lgctx_1}{\tmtwo_2}\lesub{\uvartwo}{\tm_1}{\labtwo}} & (\rdef{redex_multiplicity})\\
          \end{array}\]
        


\item $\lgctx=\tm_1\esub{\uvartwo}{\lgctx_1}$. Suppose $\tm  =\tm_1\esub{\uvartwo}{\off{\lgctx_1}{\tm_2}}\to_R \tm_1\esub{\uvartwo}{\off{\gctx_1}{\tmtwo_2}}= \tmtwo$ follows from $\tm_2\to_R \tmtwo_2$.

        \[\begin{array}{rll}
        &  \lmult{ \tm_1\esub{\uvartwo}{\off{\lgctx_1}{\tm_2}}} \\
       = & \lmult{ \tm_1}  + \lmult{ \off{\lgctx_1}{\tm_2}}  + \mult{\uvartwo}{ \tm_1} \times \lmult{\off{\lgctx_1}{\tm_2}}  + \mult{\uvartwo}{ \tm_1}& (\rdef{redex_multiplicity})\\
       > & \lmult{ \tm_1}  + \lmult{ \off{\lgctx_1}{\tmtwo_2}}  + \mult{\uvartwo}{ \tm_1} \times \lmult{\off{\lgctx_1}{\tmtwo_2}}  + \mult{\uvartwo}{ \tm_1}&(\ih\times 2)\\
   = & \lmult{\tm_1\esub{\uvartwo}{\off{\gctx_1}{\tmtwo_2}}} & (\rdef{redex_multiplicity})\\
          \end{array}\]

        \item $\lgctx=\tm_1\lesub{\uvartwo}{\lgctx_1}{\labtwo}$. Suppose $\tm  =\tm_1\lesub{\uvartwo}{\off{\lgctx_1}{\tm_2}}{\labtwo}\to_R \tm_1\lesub{\uvartwo}{\off{\gctx_1}{\tmtwo_2}}{\labtwo}= \tmtwo$ follows from $\tm_2\to_R \tmtwo_2$. 

        \[\begin{array}{rll}
        &  \lmult{ \tm_1\lesub{\uvartwo}{\off{\lgctx_1}{\tm_2}}{\labtwo}} \\
       = & 1 + \lmult{ \tm_1}  + \lmult{ \off{\lgctx_1}{\tm_2}}  & (\rdef{redex_multiplicity})\\
       > & 1 + \lmult{ \tm_1}  + \lmult{ \off{\lgctx_1}{\tmtwo_2}}   &(\ih)\\
   = & \lmult{\tm_1\lesub{\uvartwo}{\off{\gctx_1}{\tmtwo_2}}{\labtwo}} & (\rdef{redex_multiplicity})\\
          \end{array}\]
\end{xenumerate}
\end{ifLongAppendix}
\end{proof}

\subsubsection{Semantic Orthogonality}

Semantic orthogonality, namely that if $\tm\lto{\lab}\tmtwo$ and $\tm\lto{\labtwo}\tmthree$, then there exists $\tmfour$ such that $\tmtwo\dev{\labtwo}\tmfour$ and $\tmthree\dev{\lab}\tmfour$,  fails as illustrated below:
   \begin{acenter}{0cm}               
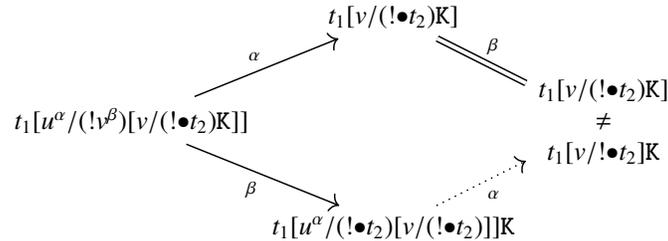
            \begin{tikzcd}[column sep=0em,row sep=1em]
              &                        \tm_{1}\esub{\uvartwo}{(\ofc{\sha{\tm_{2}}})\lsctxtwo} 
  \arrow[rd,equals, "\labtwo"] & \\
              
         \tm_{1}\esub{\ann{\uvar}{\lab}}{(\ofc{\ann{\uvartwo}{\labtwo}})\esub{\uvartwo}{(\ofc{\sha{\tm_{2}}})\lsctxtwo} }
         \arrow[ru, "\lab"]\arrow[rd, "\labtwo"'] & &
         \mlnode{  \tm_{1}\esub{\uvartwo}{(\ofc{\sha{\tm_{2}}})\lsctxtwo} 
           \\
           \neq
           \\
           \tm_{1}\esub{\uvartwo}{\ofc{\sha{\tm_{2}}}}\lsctxtwo }\\
              
              &                         \tm_{1}\esub{\ann{\uvar}{\lab}}{(\ofc{\sha{\tm_{2}}})\esub{\uvartwo}{(\ofc{\sha{\tm_{2}}})} }\lsctxtwo
\arrow[ru,dotted, "\lab"']
            \end{tikzcd}
          \end{acenter}

          This motivates the following notion of \emph{flattening}. $\lambdaS$ rewriting modulo flattening does indeed satisfy semantic orthogonality.
          
\begin{definition}[Flattening]
  \emph{Flattening} is a binary relation $\flatt\subseteq \TermsSL\times \TermsSL$ defined by the following inference rules that prove judgements of the form $\tm\flatt\tmtwo$:
\[\begin{array}{c}
    \indrule{$\flatt$F}
    {\uvartwo\notin\fv{\tm}}
    {\tm\lesub{\uvar}{\tmtwo\lesub{\uvartwo}{\tmthree}{(\labtwo)}}{(\lab)} \flatt
    \tm\lesub{\uvar}{\tmtwo}{(\lab)}\lesub{\uvartwo}{\tmthree}{(\labtwo)}}
    \\
    \\
    \indrule{$\flatt$LV}
    {\phantom{\uvartwo\notin\fv{\tm}}}
    {\lvar\flatt\lvar}
    \quad
    \indrule{$\flatt$UV}
    {\phantom{\uvartwo\notin\fv{\tm}}}
    {\ann{\uvar}{(\lab)}\flatt\ann{\uvar}{(\lab)}}
    \quad
    \indrule{$\flatt$S}
    {\tmtwo\flatt\tm}
    {\tm\flatt\tmtwo}
    \quad
    \indrule{$\flatt$T}
    {\tm\flatt\tmthree
    \quad
    \tmthree\flatt \tmtwo}
    {\tm\flatt\tmtwo}
    \\
    \\
    \indrule{$\flatt$Abs}
    {\tm\flatt\tmtwo}
    {\llam{\lvar}{\tm}{(\lab)}\flatt\llam{\lvar}{\tmtwo}{(\lab)}}
    \quad
    \indrule{$\flatt$App}
    {\tm_1\flatt\tmtwo_1
    \quad
    \tm_2\flatt\tmtwo_2}
    {\tm_1\,\tm_2\flatt\tmtwo_1\,\tmtwo_2}
    \\
    \\
    \indrule{$\flatt$Sh}
    {\tm\flatt\tmtwo}
    {\sha{\tm}\flatt\sha{\tmtwo}}
    \quad
    \indrule{$\flatt$Ofc}
    {\tm\flatt\tmtwo}
    {\ofc{\tm}\flatt\ofc{\tmtwo}}
    \quad
    \indrule{$\flatt$Open}
    {\tm\flatt\tmtwo}
    {\lopen{\tm}{(\lab)}\flatt\lopen{\tmtwo}{(\lab)}}
    \\
    \\
    \indrule{$\flatt$ES}
    {\tm_1\flatt\tmtwo_1
    \quad
    \tm_2\flatt\tmtwo_2}
    {\tm_1\lesub{\uvar}{\tm_2}{(\lab)}\flatt \tmtwo_1\lesub{\uvar}{\tmtwo_2}{(\lab)}}    
  \end{array}
\]
We write $\tm\pflatt{\pi}\tmtwo$ when $\pi$ is a derivation of $\tm\flatt\tmtwo$. We also occasionally write $\tm\flatt\tmtwo$, when there exists $\pi$ such that $\tm\pflatt{\pi}\tmtwo$. Finally, we write $\tm\pflatt{\pi}_1 \tmtwo$ if the rule $ \indrulename{$\flatt$F}$ is used exactly once in $\pi$. 
\end{definition}

\begin{remark}
  \lremark{flatt_one_vs_many}
Suppose  $\tm\pflatt{\pi}\tmtwo$. First note that we may assume, without loss of generality, that if the rule $ \indrulename{$\flatt$F}$ is used exactly once, transitivity is not used at all (since if $ \indrulename{$\flatt$F}$ is not used at all in a derivation of $\tm\flatt\tmtwo$, then $\tm=\tmtwo$). 
Second, it is easy to verify that there exists $n>0$ and $\tmthree_1,\ldots\tmthree_n$ and $\pi_1,\ldots,\pi_{n-1}$ such that $\tmthree_1=\tm$ and $\tmthree_n=\tmtwo$ and $\tmthree_1\pflatt{\pi_1}_1 \tmthree_2, \tmthree_2\pflatt{\pi_2}_1 \tmthree_3,\ldots, \tmthree_{n-1}\pflatt{\pi_{n-1}}_1\tmthree_n$, where $n-1$ is the number of times $\indrulename{$\flatt$F}$ was used in $\pi$. In particular, if $n=1$ then $\tm=\tmtwo$.
\end{remark}

\begin{remark}
\lremark{flatt_is_reflexive}
Note that $\tm\flatt\tm$, for all $\tm\in \TermsSL$ (\ie $\flatt$ is reflexive) as may be verified by straightforward induction  on $\tm$.
\end{remark}

\begin{lemma}
\llem{flatt_preserves_fv}
Suppose $\tm\pflatt{\pi}\tmtwo$. Then $\fv{\tm}=\fv{\tmtwo}$.
\end{lemma}

\begin{proof}
By induction on $\pi$.
\end{proof}

\begin{lemma}
\llem{flatt_preserves_well_labeledness}
Suppose $\tm\pflatt{\pi}\tmtwo$.
 
\begin{xenumerate}
\item If $\tm\in\TermsSWL$, then $\tmtwo\in\TermsSWL$.
   
\item If $\tmtwo\in\TermsSWL$, then $\tm\in\TermsSWL$.
\end{xenumerate}

\end{lemma}

\begin{proof}
  The proof is by induction on $\pi$. The only interesting case is the rule $\indrulename{$\flatt$F}$. 
\[
      \indrule{$\flatt$F}
    {\uvartwo\notin\fv{\tm}}
    {\tm\lesub{\uvar}{\tmtwo\lesub{\uvartwo}{\tmthree}{(\labtwo)}}{(\lab)} \flatt
      \tm\lesub{\uvar}{\tmtwo}{(\lab)}\lesub{\uvartwo}{\tmthree}{(\labtwo)}}
  \]
If the labels are not present, the result is immediate. If the label $\lab$ is present in $\uvar$, then $\tmtwo=(\ofc{\tmtwo_1})\lsctx$ and the result is also immediate since the scope of $\ann{\uvar}{\lab}$ does not change.  Suppose $\uvartwo$ is labeled with $\beta$ on the left-hand side and hence $\uvartwo\notin\fv{\tmtwo}$. Then the condition $\uvartwo\notin\fv{\tm}$ makes sure that $\uvartwo\notin\fv{ \tm\lesub{\uvar}{\tmtwo}{(\lab)}}$ in the right-hand side. Similarly, suppose $\uvartwo$ is labeled with $\beta$ on the right-hand side. Then $\uvartwo\notin\fv{ \tm\lesub{\uvar}{\tmtwo}{(\lab)}}$ and hence, in particular, $\uvartwo\notin\fv{\tmtwo}$.
  
\end{proof}

\begin{lemma}
\llem{flatt_preserved_by_linear_substitution}
Suppose $\tm\pflatt{\pi}\tmtwo$. 
\begin{xenumerate}
\item Then  $\tm\sub{\lvar}{\tmthree}\flatt\tmtwo\sub{\lvar}{\tmthree}$.
\item Then  $\tmthree\sub{\lvar}{\tm}\flatt\tmthree\sub{\lvar}{\tmtwo}$.
  
\end{xenumerate}
\end{lemma}

\begin{proof}
The first item is proved by induction on $\pi$ and resorts to \rremark{flatt_is_reflexive}. The second by induction on $\tmthree$.
\end{proof}

\begin{lemma}
  \llem{flattening_generation_for_contexts}\mbox{}
  
  \begin{xenumerate}
    
  \item  Suppose $\off{\lgctx}{\ann{\uvar}{\lab}}\pflatt{\pi}_1\tm$. Then
  \begin{xenumerate}
    
  \item there exists $\lgctxtwo$ such that $\tm=\off{\lgctxtwo}{\ann{\uvar}{\lab}}$; and
    
  \item $\off{\lgctx}{\tmtwo}\flatt\off{\lgctxtwo}{\tmtwo}$, for all $\tmtwo$.
\end{xenumerate}
  \item Similarly, if  $\tm \pflatt{\pi}_1\off{\lgctx}{\ann{\uvar}{\lab}}$, then
\begin{xenumerate}
    
  \item there exists $\lgctxtwo$ such that $\tm=\off{\lgctxtwo}{\ann{\uvar}{\lab}}$; and
    
  \item $\off{\lgctx}{\tmtwo}\flatt\off{\lgctxtwo}{\tmtwo}$, for all $\tmtwo$.
\end{xenumerate}
    \end{xenumerate}
\end{lemma}

\begin{proof}
  By simultaneous induction on both items.
\begin{ifLongAppendix}
  We consider all possible cases for $\pi$:
  \begin{xenumerate}

  \item $\pi$ ends in $\indrulename{$\flatt$LV}$. Not possible.
    
  \item $\pi$ ends in $\indrulename{$\flatt$F}$. Then $\tm=\tm_1\lesub{\uvartwo}{\tm_2}{(\labtwo)}\lesub{\uvarthree}{\tm_3}{(\labthree)}$ and $\off{\lgctx}{\ann{\uvar}{\lab}}=\tm_1\lesub{\uvartwo}{\tm_2\lesub{\uvarthree}{\tm_3}{(\labthree)}}{(\labtwo)}$. There are three cases.
     \begin{xenumerate}

  \item $\lgctx=\lgctx_1\lesub{\uvartwo}{\tm_2\lesub{\uvarthree}{\tm_3}{(\labthree)}}{(\labtwo)}$. We set $\lgctxtwo\eqdef\lgctx_1\lesub{\uvartwo}{\tm_2}{(\labtwo)}\lesub{\uvarthree}{\tm_3}{(\labthree)}$ and conclude.

      \item $\lgctx=\tm_1\lesub{\uvartwo}{\lgctx_1\lesub{\uvarthree}{\tm_3}{(\labthree)}}{(\labtwo)}$. We set $\lgctxtwo\eqdef\tm_1\lesub{\uvartwo}{\lgctx_1}{(\labtwo)}\lesub{\uvarthree}{\tm_3}{(\labthree)}$ and conclude.
      \item $\lgctx=\tm_1\lesub{\uvartwo}{\tm_2\lesub{\uvarthree}{\lgctx_1}{(\labthree)}}{(\labtwo)}$. We set $\lgctxtwo\eqdef\tm_1\lesub{\uvartwo}{\tm_2}{(\labtwo)}\lesub{\uvarthree}{\lgctx_1}{(\labthree)}$ and conclude.

     \end{xenumerate}

  \item $\pi$ ends in $\indrulename{$\flatt$UV}$. Then $\tm=\ann{\uvar}{\lab}$ and $\lgctx=\ctxhole$. We set $\gctxtwo\eqdef \ctxhole$ and conclude.
    
  \item $\pi$ ends in $\indrulename{$\flatt$S}$. We resort to the \ih w.r.t. item (2).

  \item $\pi$ ends in $\indrulename{$\flatt$T}$. Not possible by \rremark{flatt_one_vs_many}.

  \item $\pi$ ends in $\indrulename{$\flatt$Abs}$ (the cases $\indrulename{$\flatt$Sh}$  
    $\indrulename{$\flatt$Ofc}$, and $\indrulename{$\flatt$Open}$ are similar and omitted). Then $\tm=\llam{\lvar}{\tm_1}{(\labtwo)}$ and $\lgctx =\llam{\lvar}{\lgctx_1}{(\labtwo)}$ and the derivation ends in
  \[
    \indrule{$\flatt$Abs}
    {\tm_1\flatt_1 \off{\lgctx_1}{\ann{\uvar}{\lab}}}
    {\llam{\lvar}{\tm_1}{(\labtwo)}\flatt_1\llam{\lvar}{\off{\lgctx_1}{\ann{\uvar}{\lab}}}{(\labtwo)}}
  \]
  We conclude from the \ih.
  
\item $\pi$ ends in $\indrulename{$\flatt$App}$ (the case $\indrulename{$\flatt$ES}$ is similar and omitted). Then $\tm=\tm_1\,\tm_2$ and there are two cases. 
   \begin{xenumerate}

   \item $\lgctx=\lgctx_1\,\tmtwo_2$. There are two further cases.

     \begin{xenumerate}
       \item The derivation ends in
    \[
    \indrule{$\flatt$App}
    {\off{\lgctx_1}{\ann{\uvar}{\lab}}\flatt_1\tm_1
    \quad
    \tmtwo_2= \tm_2}
  {\off{\lgctx_1}{\ann{\uvar}{\lab}}\,\tmtwo_2\flatt_1 \tm_1\,\tm_2}
\]
     We conclude from the \ih.

\item The derivation ends in
   \[
    \indrule{$\flatt$App}
    {\off{\lgctx_1}{\ann{\uvar}{\lab}}=\tm_1
    \quad
    \tmtwo_2\flatt_1\tm_2}
  {\off{\lgctx_1}{\ann{\uvar}{\lab}}\,\tmtwo_2\flatt_1 \tm_1\,\tm_2}
\]
  We conclude immediately by setting $\lgctxtwo$ to $\lgctx_1\,\tm_2$.
  \end{xenumerate}
\item $\lgctx=\tmtwo_1\,\lgctx_1$. Similar to the previous case.

\end{xenumerate}

  \end{xenumerate}
  
 \end{ifLongAppendix} 
\end{proof}

\begin{lemma}
\llem{flatt_one_step_generation}
Suppose $\tm\pflatt{\pi}_1\tmtwo$. 
 
\begin{xenumerate}
\item Then
\begin{xenumerate}
\item\label{flatt_one_step_generation:ltor_lvar} $\tm=\lvar$ implies $\tmtwo=\lvar$

\item\label{flatt_one_step_generation:ltor_uvar} $\tm=\uvar$ implies $\tmtwo=\uvar$

\item\label{flatt_one_step_generation:ltor_labeled_uvar} $\tm=\ann{\uvar}{\lab}$ implies $\tmtwo=\ann{\uvar}{\lab}$

\item\label{flatt_one_step_generation:ltor_abs} $\tm=\lam{\lvar}{\tm_1}$ implies $\tmtwo=\lam{\lvar}{\tmtwo_1}$ and $\tm_1\flatt_1\tmtwo_1$ 

\item\label{flatt_one_step_generation:ltor_labeled_abs} $\tm=\llam{\lvar}{\tm_1}{\lab}$ implies $\tmtwo=\lam{\lvar}{\tmtwo_1}$ and $\tm_1\flatt_1\tmtwo_1$ 
\item\label{flatt_one_step_generation:ltor_app} $\tm=\tm_1\,\tm_2$ implies $\tmtwo=\tmtwo_1\,\tmtwo_2$ and either
  \begin{xenumerate}
  \item $\tm_1\flatt_1\tmtwo_1$ and  $\tm_2=\tmtwo_2$;  or
  \item $\tm_1=\tmtwo_1$ and $\tm_2\flatt_1\tmtwo_2$.
    \end{xenumerate}

\item\label{flatt_one_step_generation:ltor_sha} $\tm=  \sha{\tm_1}$ implies $\tmtwo=\sha{\tmtwo_1}$ and $\tm_1\flatt_1\tmtwo_1$
  
\item\label{flatt_one_step_generation:ltor_open} $\tm=\open{\tm_1}$ implies $\tmtwo=\open{\tmtwo_1}$ and $\tm_1\flatt_1\tmtwo_1$ 

\item\label{flatt_one_step_generation:ltor_labeled_open} $\tm=\lopen{\tm_1}{\lab}$ implies $\tmtwo=\lopen{\tmtwo_1}{\lab}$ and $\tm_1\flatt_1\tmtwo_1$ 

\item\label{flatt_one_step_generation:ltor_ofc} $\tm=\ofc{\tm_1}$ implies $\tmtwo=\ofc{\tmtwo_1}$ and $\tm_1\flatt_1\tmtwo_1$ 


\item\label{flatt_one_step_generation:ltor_esub} $\tm=\tm_1\lesub{\uvar}{\tm_2}{(\lab)}$ implies either

  \begin{xenumerate}
  \item\label{flatt_one_step_generation:ltor_esub_left} $\tmtwo=\tmtwo_1\lesub{\uvar}{\tmtwo_2}{(\lab)}$ and $\tm_1\flatt_1\tmtwo_1$ and $\tm_2=\tmtwo_2$; or
    \item\label{flatt_one_step_generation:ltor_esub_right} $\tmtwo=\tmtwo_1\lesub{\uvar}{\tmtwo_2}{(\lab)}$ and $\tm_1=\tmtwo_1$ and $\tm_2 \flatt_1 \tmtwo_2$; or
        
      \item\label{flatt_one_step_generation:ltor_esub_root}  $\tm_2=\tm_{21}\lesub{\uvartwo}{\tm_{22}}{(\labtwo)}$ and $\tmtwo= \tm_1\lesub{\uvar}{\tm_{21}}{(\lab)}\lesub{\uvartwo}{\tm_{22}}{(\labtwo)}$ and $\uvartwo\notin\fv{\tm_1}$; or
        
      \item $\tm_1=\tm_{11}\lesub{\uvartwo}{\tm_{12}}{(\labtwo)}$ and $\tmtwo= \tm_{11}\lesub{\uvartwo}{\tm_{21}\lesub{\uvar}{\tm_{2}}{(\lab)}}{(\labtwo)}$ and $\uvar\notin\fv{\tm_{11}}$.
    \end{xenumerate}
\end{xenumerate}

\item Then
\begin{xenumerate}
\item\label{flatt_one_step_generation:rtol_lvar} $\tmtwo=\lvar$ implies $\tm=\lvar$

\item\label{flatt_one_step_generation:rtol_uvar} $\tmtwo=\uvar$ implies $\tm=\uvar$

\item\label{flatt_one_step_generation:rtol_labeled_uvar} $\tmtwo=\ann{\uvar}{\lab}$ implies $\tm=\ann{\uvar}{\lab}$

\item\label{flatt_one_step_generation:rtol_abs} $\tmtwo=\lam{\lvar}{\tmtwo_1}$ implies $\tm=\lam{\lvar}{\tm_1}$ and $\tmtwo_1\flatt_1\tm_1$ 

\item\label{flatt_one_step_generation:rtol_labeled_abs} $\tmtwo=\llam{\lvar}{\tmtwo_1}{\lab}$ implies $\tm=\lam{\lvar}{\tm_1}$ and $\tmtwo_1\flatt_1\tm_1$
  
\item\label{flatt_one_step_generation:rtol_app} $\tmtwo=\tmtwo_1\,\tmtwo_2$ implies $\tm=\tm_1\,\tm_2$ and either  1) $\tmtwo_1\flatt_1\tm_1$ and  $\tmtwo_2=\tm_2$;  or 2) $\tmtwo_1=\tm_1$ and $\tmtwo_2\flatt_1\tm_2$. 

\item\label{flatt_one_step_generation:rtol_sha} $\tmtwo=  \sha{\tmtwo_1}$ implies $\tm=\sha{\tm_1}$ and $\tmtwo_1\flatt_1\tm_1$
  
\item\label{flatt_one_step_generation:rtol_open} $\tmtwo=\open{\tmtwo_1}$ implies $\tm=\open{\tm_1}$ and $\tmtwo_1\flatt_1\tm_1$ 

\item\label{flatt_one_step_generation:rtol_labeled_open} $\tmtwo=\lopen{\tmtwo_1}{\lab}$ implies $\tm=\lopen{\tm_1}{\lab}$ and $\tmtwo_1\flatt_1\tm_1$ 

\item\label{flatt_one_step_generation:rtol_ofc} $\tmtwo=\ofc{\tmtwo_1}$ implies $\tm=\ofc{\tm_1}$ and $\tmtwo_1\flatt_1\tm_1$ 

\item\label{flatt_one_step_generation:rtol_esub} $\tmtwo=\tmtwo_1\lesub{\uvar}{\tmtwo_2}{(\lab)}$ implies either

  \begin{xenumerate}
  \item $\tm=\tm_1\lesub{\uvar}{\tm_2}{(\lab)}$ and $\tmtwo_1\flatt_1\tm_1$ and $\tmtwo_2=\tm_2$; or
    \item $\tm=\tm_1\lesub{\uvar}{\tm_2}{(\lab)}$ and $\tmtwo_1=\tm_1$ and $\tmtwo_2 \flatt_1 \tm_2$; or
        
      \item $\tmtwo_2=\tmtwo_{21}\lesub{\uvartwo}{\tmtwo_{22}}{(\labtwo)}$ and $\tm= \tmtwo_1\lesub{\uvar}{\tmtwo_{21}}{(\lab)}\lesub{\uvartwo}{\tmtwo_{22}}{(\labtwo)}$ and $\uvartwo\notin\fv{\tmtwo_1}$; or
        
      \item $\tmtwo_1=\tmtwo_{11}\lesub{\uvartwo}{\tmtwo_{12}}{(\labtwo)}$ and $\tm= \tmtwo_{11}\lesub{\uvartwo}{\tmtwo_{21}\lesub{\uvar}{\tmtwo_{2}}{(\lab)}}{(\labtwo)}$ and $\uvar\notin\fv{\tmtwo_{11}}$.
    \end{xenumerate}    
\end{xenumerate}

\end{xenumerate}
\end{lemma}

\begin{proof}
Both items are proved by simultaneous induction on $\tm$ and
$\tmtwo$.
\begin{ifLongAppendix}
By \rremark{flatt_one_vs_many} we may assume that $\indrulename{$\flatt$T}$ is not used in $\pi$.
\begin{xenumerate}

  \item Cases for $\tm$.
\begin{xenumerate}
\item $\tm=\lvar$ implies $\tmtwo=\lvar$. Then $\pi$ must end in $\indrulename{$\flatt$LV}$, 
or  $\indrulename{$\flatt$S}$. If it ends in $\indrulename{$\flatt$LV}$, we conclude immediately. If it ends in  
 $\indrulename{$\flatt$S}$, we resort to the \ih on item (2).

\item $\tm=\uvar$ implies $\tmtwo=\uvar$. Similar to the case $\tm=\lvar$.

\item $\tm=\ann{\uvar}{\lab}$ implies $\tmtwo=\ann{\uvar}{\lab}$. Similar to the case $\tm=\lvar$.

\item $\tm=\lam{\lvar}{\tm_1}$. Then $\pi$ must end in $\indrulename{$\flatt$Abs}$, or
 $\indrulename{$\flatt$S}$. If it ends in $\indrulename{$\flatt$Abs}$, we conclude immediately. If it ends in  
 $\indrulename{$\flatt$S}$, we resort to the \ih on item (2). 

\item $\tm=\llam{\lvar}{\tm_1}{\lab}$. Similar to the previous case.

\item $\tm=\tm_1\,\tm_2$.  Then $\pi$ must end in $\indrulename{$\flatt$App}$, or 
 $\indrulename{$\flatt$S}$. If it ends in  
 $\indrulename{$\flatt$S}$, we resort to the \ih on item (2). If it ends in $\indrulename{$\flatt$App}$, two cases are possible. One is when $\pi$ ends in:
 \[
       \indrule{$\flatt$App}
    {\tm_1\flatt_1\tmtwo_1
    \quad
    \tm_2=\tmtwo_2}
    {\tm_1\,\tm_2\flatt_1\tmtwo_1\,\tmtwo_2}
  \]
  Note that by \rremark{flatt_one_vs_many}, $\tm_2\flatt\tmtwo_2$ implies $\tm_2=\tmtwo_2$. The result then holds immediately. 
   The other case is when $\pi$ ends with
   \[
         \indrule{$\flatt$App}
    {\tm_1=\tmtwo_1
    \quad
    \tm_2\flatt_1\tmtwo_2}
  {\tm_1\,\tm_2\flatt_1\tmtwo_1\,\tmtwo_2}
\]
and is treated similarly.

\item $\tm=  \sha{\tm_1}$. Then $\pi$ must end in $\indrulename{$\flatt$Sh}$, or
 $\indrulename{$\flatt$S}$. If it ends in $\indrulename{$\flatt$Sh}$, we conclude immediately. If it ends in  
 $\indrulename{$\flatt$S}$, we resort to the \ih on item (2).

\item $\tm=\open{\tm_1}$. Then $\pi$ must end in $\indrulename{$\flatt$Open}$, or
 $\indrulename{$\flatt$S}$. If it ends in $\indrulename{$\flatt$Open}$, we conclude immediately. If it ends in  
 $\indrulename{$\flatt$S}$, we resort to the \ih on item (2).

\item $\tm=\lopen{\tm_1}{\lab}$. Similar to the previous case.

\item $\tm=\ofc{\tm_1}$. Similar to the case $\tm=  \sha{\tm_1}$. 

\item $\tm=\tm\lesub{\uvar}{\tmtwo}{(\lab)}$.  Then $\pi$ must end in $\indrulename{$\flatt$F}$, 
or $\indrulename{$\flatt$ES}$ or  $\indrulename{$\flatt$S}$. If it ends in  
$\indrulename{$\flatt$S}$, we resort to the \ih on item (2). If $\pi$ must end in $\indrulename{$\flatt$F}$, then~\ref{flatt_one_step_generation:ltor_esub_root} holds.  If $\pi$ must end in $\indrulename{$\flatt$ES}$, then 
either~\ref{flatt_one_step_generation:ltor_esub_left} or~\ref{flatt_one_step_generation:ltor_esub_right}  holds.  
    
\end{xenumerate}

\item The cases for $\tmtwo$ are symmetric to the ones for $\tm$ considered above.










    

\end{xenumerate}

\end{ifLongAppendix}

\end{proof}

\begin{lemma}
\llem{main_lemma_for_flattening_is_strong_bisimulation}
${\flatt_1\toSL{\lab}} \subseteq {\toSL{\lab}\flatt}$
\end{lemma}
\begin{ifShortAppendix}
  Suppose $\tm\flatt_1\tmtwo\toSL{\lab}\tmthree$. We prove that there exists $\tmfour$ such that $\tm\toSL{\lab}\tmfour\flatt\tmthree$.
  We perform induction  on $\tm\in \TermsSWL$.
  Below we only illustrate some interesting cases.
  The extended version~\cite{mells_long} contains the fully detailed proof.
  \begin{enumerate}
  \item Flattening a substitution involved in a $\symSdb$ step:
    \begin{center}
    \begin{tikzcd}[column sep=2em,row sep=2em]
      (\llam{\lvar}{\tm_{1}}{\labtwo}) \lsctx_1\lesub{\uvar}{\tm_3}{(\labthree)}\lesub{\uvartwo}{\tm_4}{(\labfour)}\lsctx_2\,\tm_2 \arrow[r,dash]\arrow[r, dash, shift left=1]\arrow[r, dash, shift right=1, "\!_{1}"' {xshift=14pt,yshift=3pt}] \arrow[d, "\lab"] &    (\llam{\lvar}{\tm_{1}}{\labtwo})\lsctx_1\lesub{\uvar}{\tm_3\lesub{\uvartwo}{\tm_4}{(\labfour)}}{(\labthree)} \lsctx_2\,\tm_2
    \arrow[d, "\lab"]  \\
    \tm_{1}\sub{\lvar}{\tm_2} \lsctx_1\lesub{\uvar}{\tm_3}{(\labthree)}\lesub{\uvartwo}{\tm_4}{(\labfour)}\lsctx_2\arrow[r,dash]\arrow[r, dash, shift left=1]\arrow[r, dash, shift right=1]
    & 
    \tm_{1}\sub{\lvar}{\tm_2}\lsctx_1\lesub{\uvar}{\tm_3\lesub{\uvartwo}{\tm_4}{(\labfour)}}{(\labthree)} \lsctx_2
    \end{tikzcd}
    \end{center}
  \item Flattening a substitution involved in a nested $\symSls$ step:
    \begin{center}
    \begin{tikzcd}[column sep=2em,row sep=2em]
    (\llam{\lvar}{\tm_1}{\labtwo})\lsctx_{11}\lesub{\uvartwo}{\off{\lgctxthree_1}{\ann{\uvar}{\lab}}}{(\labthree)}\lsctx_{12}\esub{\uvar}{(\ofc{(\sha{\tm_{31}})\lsctxtwo_1})\lsctxtwo_2}\lesub{\uvartwo}{\tm_4}{(\labfour)} \lsctx_2 \arrow[r,dash]\arrow[r, dash, shift left=1]\arrow[r, dash, shift right=1, "\!_{1}"' {xshift=24pt,yshift=3pt}] \arrow[d, "\lab"] &             (\llam{\lvar}{\tm_1}{\labtwo}) \lsctx_{11}\lesub{\uvartwo}{\off{\lgctxthree_1}{\ann{\uvar}{\lab}}}{(\labthree)}\lsctx_{12}\esub{\uvar}{(\ofc{(\sha{\tm_{31}})\lsctxtwo_1})\lsctxtwo_2\lesub{\uvartwo}{\tm_4}{(\labfour)}} \lsctx_2   
    \arrow[d, "\lab"]  \\
    (\llam{\lvar}{\tm_1}{\labtwo}) \lsctx_{11}\lesub{\uvartwo}{\off{\lgctxthree_1}{(\sha{\tm_{31}})\lsctxtwo_1}}{(\labthree)}\lsctx_{12} \esub{\uvar}{(\ofc{(\sha{\tm_{31}})\lsctxtwo_1})}\lsctxtwo_2 \lesub{\uvartwo}{\tm_4}{(\labfour)}\lsctx_2\,\tm_2
          \arrow[r,dash]\arrow[r, dash, shift left=1]\arrow[r, dash, shift right=1]
    & 
    (\llam{\lvar}{\tm_1}{\labtwo}) \lsctx_{11}\lesub{\uvartwo}{\off{\lgctxthree_1}{(\sha{\tm_{31}})\lsctxtwo_1}}{(\labthree)}\lsctx_{12} \esub{\uvar}{(\ofc{(\sha{\tm_{31}})\lsctxtwo_1})}\lsctxtwo_2 \lesub{\uvartwo}{\tm_4}{(\labfour)}\lsctx_2\,\tm_2 
    \end{tikzcd}
    \end{center}
  \item Flattening a substitution involved in a nested $\symSgc$ step:
    \begin{center}
    \begin{tikzcd}[column sep=2em,row sep=2em]
    (\llam{\lvar}{\tm_1}{\labtwo})\lsctx_1\lesub{\uvar}{(\ofc{\tmtwo_{111}})\lsctxtwo}{\lab} \lsctx_2\,\tm_2 \arrow[r,dash]\arrow[r, dash, shift left=1]\arrow[r, dash, shift right=1, "\!_{1}"' {xshift=14pt,yshift=3pt}] \arrow[d, "\lab"] &                        (\llam{\lvar}{\tmtwo_{11}}{\labtwo})\lsctx_1\lesub{\uvar}{(\ofc{\tmtwo_{111}})\lsctxtwo }{\lab} \lsctx_2\,\tm_2  
    \arrow[d, "\lab"]  \\
    (\llam{\lvar}{\tm_1}{\labtwo})\lsctx_1\lsctxtwo \lsctx_2\,\tm_2 \arrow[r,dash]\arrow[r, dash, shift left=1]\arrow[r, dash, shift right=1]
    & 
    (\llam{\lvar}{\tmtwo_{11}}{\labtwo})\lsctx_1\lsctxtwo \lsctx_2\,\tm_2  
    \end{tikzcd}
    \end{center}
  \end{enumerate}
\end{ifShortAppendix}
\begin{ifLongAppendix}
  \begin{proof}
  Suppose $\tm\flatt_1\tmtwo\toSL{\lab}\tmthree$. We prove that there exists $\tmfour$ such that $\tm\toSL{\lab}\tmfour\flatt\tmthree$. We perform induction  on $\tm\in \TermsSWL$.

  \begin{xenumerate}
  \item $\tm=\lvar$ (case $\tm=\ann{\uvar}{(\lab)}$ is similar). Then $\tmtwo=\lvar$ by \rlem{flatt_one_step_generation}(\ref{flatt_one_step_generation:ltor_lvar}). The result holds trivially since $\lvar$ is in $\toSL{\lab}$ normal form.
    
  \item $\tm=\lam{\lvar}{\tm_1}$.  Then by \rlem{flatt_one_step_generation}(\ref{flatt_one_step_generation:ltor_abs}), $\tmtwo=\lam{\lvar}{\tmtwo_1}$, for some $\tmtwo_1$, with $\tm_1\flatt_1\tmtwo_1$. Moreover, $\tmtwo\toSL{\lab}\tmthree$ must follow from $\tmtwo_1\toSL{\lab}\tmthree_1$. We conclude from the \ih.
    
  \item $\tm=\tm_1\,\tm_1$. Then by \rlem{flatt_one_step_generation}(\ref{flatt_one_step_generation:ltor_app}), $\tmtwo=\tmtwo_1\,\tmtwo_2$, for some $\tmtwo_1,\tmtwo_2$, with $\tm_1\flatt_1\tmtwo_1$ and $\tm_2=\tmtwo_2$  or $\tm_2\flatt_1\tmtwo_2$ and $\tm_1=\tmtwo_1$. Moreover, since 
    $\tm_1\in \TermsSWL$, there can be no reduction at the root and thus $\tmtwo\toSL{\lab}\tmthree$ must follow from either $\tmtwo_1\toSL{\lab}\tmthree_1$ (in which case $\tmthree=\tmthree_1\tmtwo_2$) or $\tmtwo_2\toSL{\lab}\tmthree_2$ (in which case $\tmthree=\tmtwo_1\tmthree_2$). In all four cases, we conclude from the \ih or immediately. For example, if  $\tm_1\flatt_1\tmtwo_1$  and $\tmtwo_1\toSL{\lab}\tmthree_1$ holds, then the \ih gives us $\tmfour_1$ such that $\tm_1\toSL{\lab}\tmfour_1\flatt\tmthree_1$. From the latter we conclude,  $\tm_1\,\tm_2\toSL{\lab}\tmfour_1\,\tm_2\flatt\tmthree_1\tmtwo_2$.

  \item $\tm= (\llam{\lvar}{\tm_1}{\labtwo})\lsctx\,\tm_2$. Then by \rlem{flatt_one_step_generation}(\ref{flatt_one_step_generation:ltor_app}), $\tmtwo=\tmtwo_1\,\tmtwo_2$, for some $\tmtwo_1,\tmtwo_2$, and two possible cases may arise.

    \begin{xenumerate}

      \item $(\llam{\lvar}{\tm_1}{\labtwo})\lsctx\flatt_1\tmtwo_1$ and $\tm_2=\tmtwo_2$.  We now consider each possible way in which $(\llam{\lvar}{\tm_1}{\labtwo})\lsctx\flatt_1\tmtwo_1$.

    \begin{xenumerate}
    \item $\tmtwo_1= (\llam{\lvar}{\tmtwo_{11}}{\labtwo})\lsctx$ and   $(\llam{\lvar}{\tm_1}{\labtwo})\lsctx\flatt_1\tmtwo_1$ follows from $\tm_1\flatt_1\tmtwo_{11}$.  Suppose $\tmtwo=\of{\lgctx}{\tmg}\toSL{\lab} \of{\lgctx}{\tmd}=\tmthree$, follows from $\tmg \lrootto{\lab}_{\sha{}} \tmd$. We consider each possible case for $\lgctx$

        \begin{xenumerate}
        \item $\lgctx=\ctxhole$. Then $\lab=\labtwo$ and $\tmthree=\tmtwo_{1}\sub{\lvar}{\tm_2}\lsctx$.
             \begin{acenter}{0cm}               
              \begin{tikzcd}[column sep=2em,row sep=2em]
               (\llam{\lvar}{\tm_{1}}{\labtwo})\lsctx\,\tm_2  \arrow[r,dash]\arrow[r, dash, shift left=1]\arrow[r, dash, shift right=1, "\!_{1}"' {xshift=14pt,yshift=3pt}] \arrow[d, "\lab"] &                         (\llam{\lvar}{\tmtwo_{11}}{\labtwo})\lsctx\,\tm_2  
    \arrow[d, "\lab"]  \\
                
           \tm_{1}\sub{\lvar}{\tm_2}\lsctx \arrow[r,dash]\arrow[r, dash, shift left=1]\arrow[r, dash, shift right=1]
           & 
           \tmtwo_{11}\sub{\lvar}{\tm_2}\lsctx
             \\
              \end{tikzcd}
            \end{acenter}
            The equivalence at the bottom follows from \rlem{flatt_preserved_by_linear_substitution}(1).
            
    \item $\lgctx=(\llam{\lvar}{\lgctx_1}{\labtwo})\lsctx\,\tm_2$. We conclude from the \ih on  $\tm_1$ since $\tm_1\flatt_1\tmtwo_{11}=\of{\lgctx_1}{\tmg}\toSL{\lab}\of{\lgctx_1}{\tmd}=\tmthree_1$. 

    \item $\lgctx=\ctxhole\lsctx_2\,\tm_2$.  There are two possibilities for the step  $\tmg \lrootto{\lab}_{\sha{}} \tmd$. It can either be a $ \rtoSlsL{\lab}$-step or a $\rtoSgcL{\lab}$-step.

      \begin{xenumerate}

        \item  The step  $\tmg \lrootto{\lab}_{\sha{}} \tmd$ is a $ \rtoSlsL{\lab}$-step. Then $\lsctx=\lsctx_1 \esub{\uvar}{(\ofc{(\sha{\tm_{21}})\lsctxtwo_1})\lsctxtwo_2}  \lsctx_2$ and the step has the form $\off{\lgctxthree}{\ann{\uvar}{\lab}}\esub{\uvar}{(\ofc{(\sha{\tm_{21}})\lsctxtwo_1})\lsctxtwo_2}  \rtoSlsL{\lab} \off{\lgctxthree}{(\sha{\tm_{21}})\lsctxtwo_1}\esub{\uvar}{\ofc{(\sha{\tm_{21}})\lsctxtwo_1}}\lsctxtwo_2$. Moreover, from $\tm_1\flatt_1\tmtwo_{11}$, also $(\llam{\lvar}{\tm_1}{\labtwo})\lsctx\flatt_1 (\llam{\lvar}{\tmtwo_{11}}{\labtwo})\lsctx = \off{\lgctxthree}{\ann{\uvar}{\lab}}$. Therefore, by
  \rlem{flattening_generation_for_contexts}(1), there exists $\lgctxtwo$ such that $\tm_1=\off{\lgctxtwo}{\ann{\uvar}{\lab}}$.  
      We consider each possible form for $\lgctxthree$.
            \begin{xenumerate}
            \item  $\lgctxthree= (\llam{\lvar}{\lgctxthree_1}{\labtwo})\lsctx_1$ and $\lsctx=\lsctx_1 \esub{\uvar}{(\ofc{(\sha{\tm_{21}})\lsctxtwo_1})\lsctxtwo_2}\lsctx_2$ and $\lvar\notin\fv{(\ofc{(\sha{\tm_{21}})\lsctxtwo_1})}$. From \rlem{flattening_generation_for_contexts}(2), it must be the case that  $\lgctxtwo= (\llam{\lvar}{\lgctxtwo_1}{\labtwo})\lsctx_1$, for some  $\lgctxtwo_1$.

                         \begin{acenter}{-3cm}               
              \begin{tikzcd}[column sep=2em,row sep=2em]
              (\llam{\lvar}{\off{\lgctxtwo_1}{\ann{\uvar}{\lab}}}{\labtwo})\lsctx_1 \esub{\uvar}{(\ofc{(\sha{\tm_{21}})\lsctxtwo_1})\lsctxtwo_2}\lsctx_2\,\tm_2   \arrow[r,dash]\arrow[r, dash, shift left=1]\arrow[r, dash, shift right=1, "\!_{1}"' {xshift=24pt,yshift=3pt}] \arrow[d, "\lab"] &                (\llam{\lvar}{\off{\lgctxthree_1}{\ann{\uvar}{\lab}}}{\labtwo})\lsctx_1 \esub{\uvar}{(\ofc{(\sha{\tm_{21}})\lsctxtwo_1})\lsctxtwo_2}\lsctx_2\,\tm_2        
    \arrow[d, "\lab"]  \\
                
            (\llam{\lvar}{\off{\lgctxtwo_1}{(\sha{\tm_{21}})\lsctxtwo_1}}{\labtwo})\lsctx_1 \esub{\uvar}{(\ofc{(\sha{\tm_{21}})\lsctxtwo_1})}\lsctxtwo_2\lsctx_2\,\tm_2\arrow[r,dash]\arrow[r, dash, shift left=1]\arrow[r, dash, shift right=1]
           & 
            (\llam{\lvar}{\off{\lgctxthree_1}{(\sha{\tm_{21}})\lsctxtwo_1}}{\labtwo})\lsctx_1 \esub{\uvar}{(\ofc{(\sha{\tm_{21}})\lsctxtwo_1})}\lsctxtwo_2\lsctx_2\,\tm_2 
             \\
              \end{tikzcd}
            \end{acenter}
           The bottom equivalence holds from \rlem{flattening_generation_for_contexts}(2).
            
            \item $\lgctxthree= (\llam{\lvar}{\tmtwo_{11}}{\labtwo})\lsctx_{11}\lesub{\uvartwo}{\lgctxthree_1}{(\labthree)}\lsctx_{12}$ and $\lsctx=\lsctx_{11}\lesub{\uvartwo}{\lgctxthree_1}{(\labthree)}\lsctx_{11} \esub{\uvar}{(\ofc{(\sha{\tm_{21}})\lsctxtwo_1})\lsctxtwo_2}\lsctx_2$. From \rlem{flattening_generation_for_contexts}(2), it must be the case that  $\lgctxtwo= (\llam{\lvar}{\tm_{1}}{\labtwo})\lsctx_{11}\lesub{\uvartwo}{\lgctxtwo_1}{(\labthree)}\lsctx_{12}$, for some  $\lgctxtwo_1$.

        \begin{acenter}{-3cm}               
              \begin{tikzcd}[column sep=2em,row sep=2em]
               (\llam{\lvar}{\tm_{1}}{\labtwo})\lsctx_{11}\lesub{\uvartwo}{\off{\lgctxtwo_1}{\ann{\uvar}{\lab}}}{(\labthree)}\lsctx_{12} \esub{\uvar}{(\ofc{(\sha{\tm_{21}})\lsctxtwo_1})\lsctxtwo_2}\lsctx_2\,\tm_2  \arrow[r,dash]\arrow[r, dash, shift left=1]\arrow[r, dash, shift right=1, "\!_{1}"' {xshift=24pt,yshift=3pt}] \arrow[d, "\lab"] &                  (\llam{\lvar}{\tmtwo_{11}}{\labtwo})\lsctx_{11}\lesub{\uvartwo}{\off{\lgctxthree_1}{\ann{\uvar}{\lab}}}{(\labthree)}\lsctx_{12} \esub{\uvar}{(\ofc{(\sha{\tm_{21}})\lsctxtwo_1})\lsctxtwo_2}\lsctx_2\,\tm_2     
    \arrow[d, "\lab"]  \\
                
            (\llam{\lvar}{\tm_{1}}{\labtwo})\lsctx_{11}\lesub{\uvartwo}{\off{\lgctxtwo_1}{(\sha{\tm_{21}})\lsctxtwo_1}}{(\labthree)}\lsctx_{12} \esub{\uvar}{(\ofc{(\sha{\tm_{21}})\lsctxtwo_1})}\lsctxtwo_2\lsctx_2\,\tm_2 \arrow[r,dash]\arrow[r, dash, shift left=1]\arrow[r, dash, shift right=1]
           & 
            (\llam{\lvar}{\tmtwo_{11}}{\labtwo})\lsctx_{11}\lesub{\uvartwo}{\off{\lgctxthree_1}{(\sha{\tm_{21}})\lsctxtwo_1}}{(\labthree)}\lsctx_{12} \esub{\uvar}{(\ofc{(\sha{\tm_{21}})\lsctxtwo_1})}\lsctxtwo_2\lsctx_2\,\tm_2 
             \\
              \end{tikzcd}
            \end{acenter}

                   The bottom equivalence holds from \rlem{flattening_generation_for_contexts}(2).
            
            \end{xenumerate}

          \item The step  $\tmg \lrootto{\lab}_{\sha{}} \tmd$ is a $\rtoSgcL{\lab}$-step. Then  it has the form $(\llam{\lvar}{\tmtwo_{11}}{\labtwo})\lsctx_1\lesub{\uvar}{(\ofc{\tm_{12}})\lsctxtwo}{\lab}  \rtoSgcL{\lab} (\llam{\lvar}{\tmtwo_{11}}{\labtwo})\lsctx_1\lsctxtwo$ and $\uvar\notin\fv{(\llam{\lvar}{\tm_1}{\labtwo})\lsctx_1}$. We make use of \rlem{flatt_preserves_fv} and reason as follows:

              \begin{acenter}{0cm}               
              \begin{tikzcd}[column sep=2em,row sep=2em]
            (\llam{\lvar}{\tm_1}{\labtwo})\lsctx_1\lesub{\uvar}{(\ofc{\tm_{12}})\lsctxtwo}{\lab} \lsctx_2\,\tm_2 \arrow[r,dash]\arrow[r, dash, shift left=1]\arrow[r, dash, shift right=1, "\!_{1}"' {xshift=14pt,yshift=3pt}] \arrow[d, "\lab"] &                        (\llam{\lvar}{\tmtwo_{11}}{\labtwo})\lsctx_1\lesub{\uvar}{(\ofc{\tm_{12}})\lsctxtwo}{\lab} \lsctx_2\,\tm_2  
    \arrow[d, "\lab"]  \\
                
              (\llam{\lvar}{\tm_1}{\labtwo})\lsctx_1\lsctxtwo \lsctx_2\,\tm_2 \arrow[r,dash]\arrow[r, dash, shift left=1]\arrow[r, dash, shift right=1]
           & 
                 (\llam{\lvar}{\tmtwo_{11}}{\labtwo})\lsctx_1\lsctxtwo \lsctx_2\,\tm_2  
             \\
              \end{tikzcd}
            \end{acenter}

            \end{xenumerate}

    \item  $\lgctx=(\llam{\lvar}{\tmtwo_{11}}{\labtwo})\lsctx_1\lesub{\uvar}{\lgctx_1}{(\labthree)}\lsctx_2\,\tm_2$. Then the $\tm\flatt_1\tmtwo$ equivalence and the $\tmtwo\toSL{\lab}\tmthree$ step are disjoint.

                 \begin{acenter}{0cm}               
              \begin{tikzcd}[column sep=2em,row sep=2em]
              (\llam{\lvar}{\tm_1}{\labtwo})\lsctx_1\lesub{\uvar}{\of{\lgctx_1}{\tmg}}{(\labthree)}\lsctx_2\,\tm_2 \arrow[r,dash]\arrow[r, dash, shift left=1]\arrow[r, dash, shift right=1, "\!_{1}"' {xshift=14pt,yshift=3pt}] \arrow[d, "\lab"] &                          (\llam{\lvar}{\tmtwo_{11}}{\labtwo})\lsctx_1\lesub{\uvar}{\of{\lgctx_1}{\tmg}}{(\labthree)}\lsctx_2\,\tm_2   
    \arrow[d, "\lab"]  \\
                
            (\llam{\lvar}{\tm_1}{\labtwo})\lsctx_1\lesub{\uvar}{\of{\lgctx_1}{\tmd}}{(\labthree)}\lsctx_2\,\tm_2  \arrow[r,dash]\arrow[r, dash, shift left=1]\arrow[r, dash, shift right=1]
           & 
               (\llam{\lvar}{\tmtwo_{11}}{\labtwo})\lsctx_1\lesub{\uvar}{\of{\lgctx_1}{\tmd}}{(\labthree)}\lsctx_2\,\tm_2 
             \\
              \end{tikzcd}
            \end{acenter}

    \end{xenumerate}

  \item\label{main_lemma_for_flattening_is_strong_bisimulation_case_F} $\tmtwo_1= (\llam{\lvar}{\tm_{1}}{\labtwo})\lsctx_1\lesub{\uvar}{\tm_3\lesub{\uvartwo}{\tm_4}{(\labfour)}}{(\labthree)} \lsctx_2$ and $\lsctx=\lsctx_1\lesub{\uvar}{\tm_3}{(\labthree)}\lesub{\uvartwo}{\tm_4}{(\labfour)}\lsctx_2$ and $\uvartwo\notin\fv{ (\llam{\lvar}{\tm_{1}}{\labtwo})\lsctx_1}$.  Suppose $\tmtwo=\of{\lgctx}{\tmg}\toSL{\lab} \of{\lgctx}{\tmd}=\tmthree$, follows from $\tmg \lrootto{\lab}_{\sha{}} \tmd$. We consider each possible case for $\lgctx$.

    \begin{xenumerate}

    \item $\lgctx=\ctxhole$. Then $\lab=\labtwo$ and we have
          \begin{acenter}{0cm}               
              \begin{tikzcd}[column sep=2em,row sep=2em]
              (\llam{\lvar}{\tm_{1}}{\labtwo}) \lsctx_1\lesub{\uvar}{\tm_3}{(\labthree)}\lesub{\uvartwo}{\tm_4}{(\labfour)}\lsctx_2\,\tm_2 \arrow[r,dash]\arrow[r, dash, shift left=1]\arrow[r, dash, shift right=1, "\!_{1}"' {xshift=14pt,yshift=3pt}] \arrow[d, "\lab"] &    (\llam{\lvar}{\tm_{1}}{\labtwo})\lsctx_1\lesub{\uvar}{\tm_3\lesub{\uvartwo}{\tm_4}{(\labfour)}}{(\labthree)} \lsctx_2\,\tm_2
    \arrow[d, "\lab"]  \\
                
            \tm_{1}\sub{\lvar}{\tm_2} \lsctx_1\lesub{\uvar}{\tm_3}{(\labthree)}\lesub{\uvartwo}{\tm_4}{(\labfour)}\lsctx_2\arrow[r,dash]\arrow[r, dash, shift left=1]\arrow[r, dash, shift right=1]
           & 
              \tm_{1}\sub{\lvar}{\tm_2}\lsctx_1\lesub{\uvar}{\tm_3\lesub{\uvartwo}{\tm_4}{(\labfour)}}{(\labthree)} \lsctx_2
             \\
              \end{tikzcd}
            \end{acenter}
            
        \item $\lgctx=(\llam{\lvar}{\lgctx_1}{\labtwo})\lsctx_1\lesub{\uvar}{\tm_3\lesub{\uvartwo}{\tm_4}{(\labfour)}}{(\labthree)} \lsctx_2\,\tm_2$. Similar to the case $\lgctx=\ctxhole$. The $\flatt_1$-step and the $\toSL{\lab}$ are disjoint (similar to the case $\lgctx=\ctxhole$). We conclude immediately.
   
            \item $\lgctx=(\llam{\lvar}{\tm_{1}}{\labtwo})\lsctx_{11}\lesub{\uvartwo}{\lgctx_1}{(\labfive)}\lsctx_{12}\lesub{\uvar}{\tm_3\lesub{\uvartwo}{\tm_4}{(\labfour)}}{(\labthree)} \lsctx_2\,\tm_2$. The $\flatt_1$-step and the $\toSL{\lab}$ are disjoint (similar to the case $\lgctx=\ctxhole$). We conclude immediately.

            \item\label{flatt_one_step_generation:ls_or_gc} $\lgctx=\ctxhole \lsctx_2\,\tm_2$. There are two possibilities for the step  $\tmg \lrootto{\lab}_{\sha{}} \tmd$. It can either be a $ \rtoSlsL{\lab}$-step or a $\rtoSgcL{\lab}$-step.

      \begin{xenumerate}

      \item  The step  $\tmg \lrootto{\lab}_{\sha{}} \tmd$ is a $ \rtoSlsL{\lab}$-step. Then the label $\labthree$ is absent, $\tm_3=(\ofc{(\sha{\tm_{31}})\lsctxtwo_1})\lsctxtwo_2$ and thus $\tmtwo=(\llam{\lvar}{\tm_{1}}{\labtwo})\lsctx_1\esub{\uvar}{(\ofc{(\sha{\tm_{31}})\lsctxtwo_1})\lsctxtwo_2\lesub{\uvartwo}{\tm_4}{(\labfour)}} \lsctx_2\,\tm_2$   and the step has the form $\off{\lgctxthree}{\ann{\uvar}{\lab}}\esub{\uvar}{(\ofc{(\sha{\tm_{31}})\lsctxtwo_1})\lsctxtwo_2\lesub{\uvartwo}{\tm_4}{(\labfour)}}  \rtoSlsL{\lab} \off{\lgctxthree}{(\sha{\tm_{11}})\lsctxtwo_1}\esub{\uvar}{\ofc{(\sha{\tm_{31}})\lsctxtwo_1}}\lsctxtwo_2\lesub{\uvartwo}{\tm_4}{(\labfour)}$. 
      We consider each possible form for $\lgctxthree$.
            \begin{xenumerate}
            \item\label{flatt_one_step_generation:F_rtol}  $\lgctxthree= (\llam{\lvar}{\lgctxthree_1}{\labtwo})\lsctx_1$ and $\lsctx=\lsctx_1\esub{\uvar}{(\ofc{(\sha{\tm_{31}})\lsctxtwo_1})\lsctxtwo_2}\lesub{\uvartwo}{\tm_4}{(\labfour)} \lsctx_2 $ and $\lvar\notin\fv{(\ofc{(\sha{\tm_{31}})\lsctxtwo_1})}$. 

                         \begin{acenter}{-2cm}               
              \begin{tikzcd}[column sep=2em,row sep=2em]
               (\llam{\lvar}{\off{\lgctxthree_1}{\ann{\uvar}{\lab}}}{\labtwo})\lsctx_1\esub{\uvar}{(\ofc{(\sha{\tm_{31}})\lsctxtwo_1})\lsctxtwo_2}\lesub{\uvartwo}{\tm_4}{(\labfour)} \lsctx_2 \arrow[r,dash]\arrow[r, dash, shift left=1]\arrow[r, dash, shift right=1, "\!_{1}"' {xshift=24pt,yshift=3pt}] \arrow[d, "\lab"] &             (\llam{\lvar}{\off{\lgctxthree_1}{\ann{\uvar}{\lab}}}{\labtwo})\lsctx_1\esub{\uvar}{(\ofc{(\sha{\tm_{31}})\lsctxtwo_1})\lsctxtwo_2\lesub{\uvartwo}{\tm_4}{(\labfour)}} \lsctx_2   
    \arrow[d, "\lab"]  \\

    (\llam{\lvar}{\off{\lgctxthree_1}{(\sha{\tm_{31}})\lsctxtwo_1}}{\labtwo})\lsctx_1 \esub{\uvar}{(\ofc{(\sha{\tm_{31}})\lsctxtwo_1})}\lsctxtwo_2 \lesub{\uvartwo}{\tm_4}{(\labfour)}\lsctx_2\,\tm_2
          \arrow[r,dash]\arrow[r, dash, shift left=1]\arrow[r, dash, shift right=1]
           & 
              (\llam{\lvar}{\off{\lgctxthree_1}{(\sha{\tm_{31}})\lsctxtwo_1}}{\labtwo})\lsctx_1 \esub{\uvar}{(\ofc{(\sha{\tm_{31}})\lsctxtwo_1})}\lsctxtwo_2 \lesub{\uvartwo}{\tm_4}{(\labfour)}\lsctx_2\,\tm_2 
             \\
              \end{tikzcd}
            \end{acenter}
            
            \item $\lgctxthree= (\llam{\lvar}{\tm_{1}}{\labtwo})\lsctx_{11}\lesub{\uvartwo}{\lgctxthree_1}{(\labthree)}\lsctx_{12}$ and $\lsctx=\lsctx_{11}\lesub{\uvartwo}{\lgctxthree_1}{(\labthree)}\lsctx_{12} \esub{\uvar}{(\ofc{(\sha{\tm_{31}})\lsctxtwo_1})\lsctxtwo_2}\lesub{\uvartwo}{\tm_4}{(\labfour)} \lsctx_2$. 

    \begin{acenter}{-3cm}               
              \begin{tikzcd}[column sep=2em,row sep=2em]
               (\llam{\lvar}{\tm_1}{\labtwo}) \lsctx_{11}\lesub{\uvartwo}{\off{\lgctxthree_1}{\ann{\uvar}{\lab}}}{(\labthree)}\lsctx_{12}\esub{\uvar}{(\ofc{(\sha{\tm_{31}})\lsctxtwo_1})\lsctxtwo_2}\lesub{\uvartwo}{\tm_4}{(\labfour)} \lsctx_2 \arrow[r,dash]\arrow[r, dash, shift left=1]\arrow[r, dash, shift right=1, "\!_{1}"' {xshift=24pt,yshift=3pt}] \arrow[d, "\lab"] &             (\llam{\lvar}{\tm_1}{\labtwo}) \lsctx_{11}\lesub{\uvartwo}{\off{\lgctxthree_1}{\ann{\uvar}{\lab}}}{(\labthree)}\lsctx_{12}\esub{\uvar}{(\ofc{(\sha{\tm_{31}})\lsctxtwo_1})\lsctxtwo_2\lesub{\uvartwo}{\tm_4}{(\labfour)}} \lsctx_2   
    \arrow[d, "\lab"]  \\

    (\llam{\lvar}{\tm_1}{\labtwo}) \lsctx_{11}\lesub{\uvartwo}{\off{\lgctxthree_1}{(\sha{\tm_{31}})\lsctxtwo_1}}{(\labthree)}\lsctx_{12} \esub{\uvar}{(\ofc{(\sha{\tm_{31}})\lsctxtwo_1})}\lsctxtwo_2 \lesub{\uvartwo}{\tm_4}{(\labfour)}\lsctx_2\,\tm_2
          \arrow[r,dash]\arrow[r, dash, shift left=1]\arrow[r, dash, shift right=1]
           & 
              (\llam{\lvar}{\tm_1}{\labtwo}) \lsctx_{11}\lesub{\uvartwo}{\off{\lgctxthree_1}{(\sha{\tm_{31}})\lsctxtwo_1}}{(\labthree)}\lsctx_{12} \esub{\uvar}{(\ofc{(\sha{\tm_{31}})\lsctxtwo_1})}\lsctxtwo_2 \lesub{\uvartwo}{\tm_4}{(\labfour)}\lsctx_2\,\tm_2 
             \\
              \end{tikzcd}
            \end{acenter}
            
            \end{xenumerate}

          \item The step  $\tmg \lrootto{\lab}_{\sha{}} \tmd$ is a $\rtoSgcL{\lab}$-step. Then $\labthree=\lab$ and $\tm_3=(\ofc{\tm_{12}})\lsctxtwo $ and the step has the form $(\llam{\lvar}{\tm_1}{\labtwo})\lsctx_1\lesub{\uvar}{(\ofc{\tm_{12}})\lsctxtwo \lesub{\uvartwo}{\tm_4}{(\labfour)}}{\lab}  \rtoSgcL{\lab} (\llam{\lvar}{\tm_1}{\labtwo})\lsctx_1\lsctxtwo \lesub{\uvartwo}{\tm_4}{(\labfour)}$ and $\uvar\notin\fv{(\llam{\lvar}{\tm_1}{\labtwo})\lsctx_1}$. We make use of \rlem{flatt_preserves_fv} and reason as follows:
              \begin{acenter}{0cm}               
              \begin{tikzcd}[column sep=2em,row sep=2em]
            (\llam{\lvar}{\tm_1}{\labtwo})\lsctx_1\lesub{\uvar}{(\ofc{\tm_{12}})\lsctxtwo}{\lab}\lesub{\uvartwo}{\tm_4}{(\labfour)} \lsctx_2\,\tm_2 \arrow[r,dash]\arrow[r, dash, shift left=1]\arrow[r, dash, shift right=1, "\!_{1}"' {xshift=14pt,yshift=3pt}] \arrow[d, "\lab"] &                        (\llam{\lvar}{\tmtwo_{11}}{\labtwo})\lsctx_1\lesub{\uvar}{(\ofc{\tm_{12}})\lsctxtwo \lesub{\uvartwo}{\tm_4}{(\labfour)}}{\lab} \lsctx_2\,\tm_2  
    \arrow[d, "\lab"]  \\
                
              (\llam{\lvar}{\tm_1}{\labtwo})\lsctx_1\lsctxtwo \lesub{\uvartwo}{\tm_4}{(\labfour)} \lsctx_2\,\tm_2 \arrow[r,equal]
           & 
                 (\llam{\lvar}{\tmtwo_{11}}{\labtwo})\lsctx_1\lsctxtwo \lesub{\uvartwo}{\tm_4}{(\labfour)} \lsctx_2\,\tm_2  
             \\
              \end{tikzcd}
            \end{acenter}

            \end{xenumerate}

            \item\label{flatt_one_step_generation:F_rtol_nested} $\lgctx=(\llam{\lvar}{\tm_{1}}{\labtwo})\lsctx_{1}\lesub{\uvar}{\ctxhole}{(\labthree)} \lsctx_2\,\tm_2$. There are two possibilities for the step  $\tmg \lrootto{\lab}_{\sha{}} \tmd$. It can either be a $ \rtoSlsL{\lab}$-step or a $\rtoSgcL{\lab}$-step.

      \begin{xenumerate}

      \item  The step  $\tmg \lrootto{\lab}_{\sha{}} \tmd$ is a $ \rtoSlsL{\lab}$-step. Then the label $\labfour$ is absent, $\tm_4=(\ofc{(\sha{\tm_{41}})\lsctxtwo_1})\lsctxtwo_2$ and thus $\tmtwo=(\llam{\lvar}{\tm_{1}}{\labtwo})\lsctx_1\lesub{\uvar}{\tm_3\esub{\uvartwo}{(\ofc{(\sha{\tm_{41}})\lsctxtwo_1})\lsctxtwo_2}}{(\labthree)} \lsctx_2\,\tm_2$   and the step has the form $\off{\lgctxthree}{\ann{\uvartwo}{\lab}}\esub{\uvartwo}{(\ofc{(\sha{\tm_{41}})\lsctxtwo_1})\lsctxtwo_2}  \rtoSlsL{\lab} \off{\lgctxthree}{(\sha{\tm_{41}})\lsctxtwo_1}\esub{\uvartwo}{\ofc{(\sha{\tm_{41}})\lsctxtwo_1}}\lsctxtwo_2$. Moreover, by $(\llam{\lvar}{\tm_1}{\labtwo})\lsctx\flatt_1 \off{\lgctxthree}{\ann{\uvar}{\lab}}$ and 
  \rlem{flattening_generation_for_contexts}(1), there exists $\lgctxtwo$ such that $\tmtwo_1=\off{\lgctxtwo}{\ann{\uvar}{\lab}}$.  
      We consider each possible form for $\lgctxthree$.
            \begin{xenumerate}
            \item  $\lgctxthree= (\llam{\lvar}{\lgctxthree_1}{\labtwo})\lsctx_1$. This case is not possible since $\uvartwo\notin\fv{ (\llam{\lvar}{\tm_{1}}{\labtwo})\lsctx_1}$.

            \item $\lgctxthree= (\llam{\lvar}{\tm_{1}}{\labtwo})\lsctx_{11}\lesub{\uvarthree}{\lgctxthree_1}{(\labthree)}\lsctx_{12}$. This case is not possible since $\uvartwo\notin\fv{ (\llam{\lvar}{\tm_{1}}{\labtwo})\lsctx_1}$.

              \item $\lgctxthree= (\llam{\lvar}{\tm_{1}}{\labtwo})\lsctx_{1}\lesub{\uvar}{\lgctxthree_1\esub{\uvartwo}{(\ofc{(\sha{\tm_{41}})\lsctxtwo_1})\lsctxtwo_2}}{(\labthree)} $.
     \begin{acenter}{-3cm}               
              \begin{tikzcd}[column sep=2em,row sep=2em]
            (\llam{\lvar}{\tm_{1}}{\labtwo})\lsctx_{1}\lesub{\uvar}{\off{\lgctxthree_1}{\ann{\uvartwo}{\lab}}}{(\labthree)}\esub{\uvartwo}{(\ofc{(\sha{\tm_{41}})\lsctxtwo_1})\lsctxtwo_2}\lsctx_2\,\tm_2 \arrow[r,dash]\arrow[r, dash, shift left=1]\arrow[r, dash, shift right=1, "\!_{1}"' {xshift=17pt,yshift=3pt}] \arrow[d, "\lab"] &                        (\llam{\lvar}{\tm_{1}}{\labtwo})\lsctx_{1}\lesub{\uvar}{\off{\lgctxthree_1}{\ann{\uvartwo}{\lab}}\esub{\uvartwo}{(\ofc{(\sha{\tm_{41}})\lsctxtwo_1})\lsctxtwo_2}}{(\labthree)}\lsctx_2\,\tm_2  
    \arrow[d, "\lab"]  \\
                
              (\llam{\lvar}{\tm_{1}}{\labtwo})\lsctx_{1}\lesub{\uvar}{\off{\lgctxthree_1}{(\sha{\tm_{41}})\lsctxtwo_1}}{(\labthree)}\esub{\uvartwo}{(\ofc{(\sha{\tm_{41}})\lsctxtwo_1})}\lsctxtwo_2\lsctx_2\,\tm_2 \arrow[r,dash]\arrow[r, dash, shift left=1]\arrow[r, dash, shift right=1]
           & 
                (\llam{\lvar}{\tm_{1}}{\labtwo})\lsctx_{1}\lesub{\uvar}{\off{\lgctxthree_1}{(\sha{\tm_{41}})\lsctxtwo_1}\esub{\uvartwo}{(\ofc{(\sha{\tm_{41}})\lsctxtwo_1})}\lsctxtwo_2}{(\labthree)}\lsctx_2\,\tm_2  
             \\
              \end{tikzcd}
            \end{acenter}

            Note the use of additional instances of $\indrulename{$\flatt$F}$ at the bottom of the diagram. 
            \end{xenumerate}

          \item The step  $\tmg \lrootto{\lab}_{\sha{}} \tmd$ is a $\rtoSgcL{\lab}$-step. Then $\labthree=\lab$ and $\tm_4=(\ofc{\tm_{41}})\lsctxtwo $ and the step has the form $(\llam{\lvar}{\tm_1}{\labtwo})\lsctx_1\lesub{\uvar}{\tm_3\lesub{\uvartwo}{(\ofc{\tm_{41}})\lsctxtwo}{\lab}}{(\labtwo)}  \rtoSgcL{\lab} (\llam{\lvar}{\tm_1}{\labtwo})\lsctx_1 \lesub{\uvar}{\tm_3 \lsctxtwo}{(\labtwo)}$ and $\uvartwo\notin\fv{\tm_3}$.  Moreover, also $\uvartwo\notin\fv{ (\llam{\lvar}{\tm_{1}}{\labtwo})\lsctx_1}$, by hypothesis of this case.

              \begin{acenter}{0cm}               
              \begin{tikzcd}[column sep=2em,row sep=2em]
          (\llam{\lvar}{\tm_1}{\labtwo})\lsctx_1\lesub{\uvar}{\tm_3}{(\labtwo)}\lesub{\uvartwo}{(\ofc{\tm_{41}})\lsctxtwo}{\lab}  \lsctx_2\,\tm_2 \arrow[r,dash]\arrow[r, dash, shift left=1]\arrow[r, dash, shift right=1, "\!_{1}"' {xshift=14pt,yshift=3pt}] \arrow[d, "\lab"] &                       (\llam{\lvar}{\tm_1}{\labtwo})\lsctx_1\lesub{\uvar}{\tm_3\lesub{\uvartwo}{(\ofc{\tm_{41}})\lsctxtwo}{\lab}}{(\labtwo)}  \lsctx_2\,\tm_2  
    \arrow[d, "\lab"]  \\
                
             (\llam{\lvar}{\tm_1}{\labtwo})\lsctx_1\lesub{\uvar}{\tm_3}{(\labtwo)}\lsctxtwo \lsctx_2\,\tm_2 \arrow[r,dash]\arrow[r, dash, shift left=1]\arrow[r, dash, shift right=1]
           & 
                 (\llam{\lvar}{\tm_1}{\labtwo})\lsctx_1\lesub{\uvar}{\tm_3\lsctxtwo}{(\labtwo)}  \lsctx_2\,\tm_2 
             \\
              \end{tikzcd}
            \end{acenter}
   
                      Note the use of additional instances of $\indrulename{$\flatt$F}$ at the bottom of the diagram. 

            \end{xenumerate}

            \item $\lgctx=(\llam{\lvar}{\tm_{1}}{\labtwo})\lsctx_{1}\lesub{\uvar}{\lgctx_1\lesub{\uvartwo}{\tm_4}{(\labfour)}}{(\labthree)} \lsctx_2\,\tm_2$. The $\flatt_1$-step and the $\toSL{\lab}$ are disjoint. We conclude immediately.
                  \item $\lgctx=(\llam{\lvar}{\tm_{1}}{\labtwo})\lsctx_{1}\lesub{\uvar}{\tm_3\lesub{\uvartwo}{\lgctx_1}{(\labfour)}}{(\labthree)} \lsctx_2\,\tm_2$. The $\flatt_1$-step and the $\toSL{\lab}$ are disjoint. We conclude immediately.
            \item $\lgctx=(\llam{\lvar}{\tm_{1}}{\labtwo})\lsctx_{1}\lesub{\uvar}{\tm_3\lesub{\uvartwo}{\tm_4}{(\labfour)}}{(\labthree)} \lsctx_{21}\lesub{\uvartwo}{\lgctx_1}{(\labfive)}\lsctx_{22}\,\tm_2$. The $\flatt_1$-step and the $\toSL{\lab}$ are disjoint. We conclude immediately.

                   \item $\lgctx=(\llam{\lvar}{\tm_{1}}{\labtwo})\lsctx_{1}\lesub{\uvar}{\tm_3\lesub{\uvartwo}{\tm_4}{(\labfour)}}{(\labthree)} \lsctx_2\, \lgctx_1$. The $\flatt_1$-step and the $\toSL{\lab}$ are disjoint. We conclude immediately.
    \end{xenumerate}

  \item\label{main_lemma_for_flattening_is_strong_bisimulation_case_F_symmetric} $\tmtwo_1= (\llam{\lvar}{\tm_{1}}{\labtwo})\lsctx_1\lesub{\uvar}{\tm_2}{(\labthree)}\lesub{\uvartwo}{\tm_3}{(\labfour)}\lsctx_2$ and  $\lsctx= \lsctx_1\lesub{\uvar}{\tm_2\lesub{\uvartwo}{\tm_3}{(\labfour)}}{(\labthree)}\lsctx_2$. Symmetric to case~\ref{main_lemma_for_flattening_is_strong_bisimulation_case_F}.

  \item $\tmtwo_1= (\llam{\lvar}{\tm_{1}}{\labtwo})\lsctx_1\lesub{\uvar}{\tmtwo_{11}}{(\labthree)}\lsctx_2$ and  $\lsctx=\lsctx_1\lesub{\uvar}{\tm_{12}}{(\labthree)}\lsctx_2$ and  $(\llam{\lvar}{\tm_1}{\labtwo})\lsctx\flatt_1\tmtwo_1$ follows from $\tm_{12}\flatt_1\tmtwo_{11}$.  Suppose $\tmtwo=\of{\lgctx}{\tmg}\toSL{\lab} \of{\lgctx}{\tmd}=\tmthree$, follows from $\tmg \lrootto{\lab}_{\sha{}} \tmd$. We consider each possible case for $\lgctx$

    \begin{xenumerate}

    \item $\lgctx=\ctxhole$.  Then $\lab=\labtwo$ and we reason as follows:
    \begin{acenter}{0cm}               
              \begin{tikzcd}[column sep=2em,row sep=2em]
              (\llam{\lvar}{\tm_{1}}{\labtwo}) \lsctx_1\lesub{\uvar}{\tm_{12}}{(\labthree)}\lsctx_2\,\tm_2 \arrow[r,dash]\arrow[r, dash, shift left=1]\arrow[r, dash, shift right=1, "\!_{1}"' {xshift=14pt,yshift=3pt}] \arrow[d, "\lab"] &    (\llam{\lvar}{\tm_{1}}{\labtwo})\lsctx_1\lesub{\uvar}{\tmtwo_{11}}{(\labthree)} \lsctx_2\,\tm_2
    \arrow[d, "\lab"]  \\
                
            \tm_{1}\sub{\lvar}{\tm_2} \lsctx_1\lesub{\uvar}{\tm_{12}}{(\labthree)}\lsctx_2\arrow[r,dash]\arrow[r, dash, shift left=1]\arrow[r, dash, shift right=1]
           & 
              \tm_{1}\sub{\lvar}{\tm_2}\lsctx_1\lesub{\uvar}{\tmtwo_{11}}{(\labthree)} \lsctx_2
             \\
              \end{tikzcd}
            \end{acenter}
            
          \item $\lgctx=(\llam{\lvar}{\lgctx_1}{\labtwo})\lsctx_1\lesub{\uvar}{\tmtwo_{11}}{(\labthree)}\lsctx_2\,\tm_2$. The $\flatt_1$-step and the $\toSL{\lab}$ are disjoint. We conclude immediately.

                \item $\lgctx=(\llam{\lvar}{\tm_1}{\labtwo}) \lsctx_{11}\lesub{\uvartwo}{\lgctx_1}{(\labfour)}\lsctx_{12}\lesub{\uvar}{\tmtwo_{11}}{(\labthree)}\lsctx_2\,\tm_2$. The $\flatt_1$-step and the $\toSL{\lab}$ are disjoint. We conclude immediately.

    \item\label{main_lemma_for_flattening_is_strong_bisimulation_case_ls} $\lgctx=\ctxhole\lsctx_2\,\tm_2$. 
      There are two possibilities for the step  $\tmg \lrootto{\lab}_{\sha{}} \tmd$. It can either be a $ \rtoSlsL{\lab}$-step or a $\rtoSgcL{\lab}$-step.

      \begin{xenumerate}

      \item  The step  $\tmg \lrootto{\lab}_{\sha{}} \tmd$ is a $ \rtoSlsL{\lab}$-step. Then the label $\labthree$ is absent, $\tmtwo_{11}=(\ofc{(\sha{\tmtwo_{111}})\lsctxtwo_1})\lsctxtwo_2$ and thus $\tmtwo=(\llam{\lvar}{\tm_{1}}{\labtwo})\lsctx_1\esub{\uvar}{(\ofc{(\sha{\tmtwo_{111}})\lsctxtwo_1})\lsctxtwo_2} \lsctx_2\,\tm_2$   and the step has the form $\off{\lgctxthree}{\ann{\uvar}{\lab}}\esub{\uvar}{(\ofc{(\sha{\tmtwo_{111}})\lsctxtwo_1})\lsctxtwo_2}  \rtoSlsL{\lab} \off{\lgctxthree}{(\sha{\tmtwo_{111}})\lsctxtwo_1}\esub{\uvar}{\ofc{(\sha{\tmtwo_{111}})\lsctxtwo_1}}\lsctxtwo_2$. We next consider each possible way in which
        \[\tm_{12}\flatt_1(\ofc{(\sha{\tmtwo_{111}})\lsctxtwo_1})\lsctxtwo_2
        \]
        and then, for each of these, each possible form for $\lgctxthree$.

      \begin{xenumerate}
        \item  $\tm_{12}=(\ofc{(\sha{\tm_{121}})\lsctxtwo_1})\lsctxtwo_2$ and  $\tm_{12}\flatt_1(\ofc{(\sha{\tmtwo_{111}})\lsctxtwo_1})\lsctxtwo_2$ follows from $\tm_{121} \flatt_1 \tmtwo_{111}$. Then there are two possible forms for $\lgctxthree$

                    \begin{xenumerate}
            \item  $\lgctxthree= (\llam{\lvar}{\lgctxthree_1}{\labtwo})\lsctx_1$ and $\lsctx=\lsctx_1\esub{\uvar}{(\ofc{(\sha{\tmtwo_{111}})\lsctxtwo_1})\lsctxtwo_2} \lsctx_2 $ and $\lvar\notin\fv{(\ofc{(\sha{\tmtwo_{111}})\lsctxtwo_1})}$.

                         \begin{acenter}{-2cm}               
              \begin{tikzcd}[column sep=2em,row sep=2em]
               (\llam{\lvar}{\off{\lgctxthree_1}{\ann{\uvar}{\lab}}}{\labtwo})\lsctx_1\esub{\uvar}{(\ofc{(\sha{\tmtwo_{111}})\lsctxtwo_1})\lsctxtwo_2} \lsctx_2 \arrow[r,dash]\arrow[r, dash, shift left=1]\arrow[r, dash, shift right=1, "\!_{1}"' {xshift=24pt,yshift=3pt}] \arrow[d, "\lab"] &             (\llam{\lvar}{\off{\lgctxthree_1}{\ann{\uvar}{\lab}}}{\labtwo})\lsctx_1\esub{\uvar}{(\ofc{(\sha{\tmtwo_{111}})\lsctxtwo_1})\lsctxtwo_2} \lsctx_2   
    \arrow[d, "\lab"]  \\

    (\llam{\lvar}{\off{\lgctxthree_1}{(\sha{\tmtwo_{111}})\lsctxtwo_1}}{\labtwo})\lsctx_1 \esub{\uvar}{(\ofc{(\sha{\tmtwo_{111}})\lsctxtwo_1})}\lsctxtwo_2 \lsctx_2\,\tm_2
          \arrow[r,dash]\arrow[r, dash, shift left=1]\arrow[r, dash, shift right=1]
           & 
              (\llam{\lvar}{\off{\lgctxthree_1}{(\sha{\tmtwo_{111}})\lsctxtwo_1}}{\labtwo})\lsctx_1 \esub{\uvar}{(\ofc{(\sha{\tmtwo_{111}})\lsctxtwo_1})}\lsctxtwo_2\lsctx_2\,\tm_2 
             \\
              \end{tikzcd}
            \end{acenter}
            
            \item $\lgctxthree= (\llam{\lvar}{\tm_{1}}{\labtwo})\lsctx_{11}\lesub{\uvartwo}{\lgctxthree_1}{(\labfour)}\lsctx_{12}$ and $\lsctx=\lsctx_{11}\lesub{\uvartwo}{\lgctxthree_1}{(\labfour)}\lsctx_{12} \esub{\uvar}{(\ofc{(\sha{\tmtwo_{111}})\lsctxtwo_1})\lsctxtwo_2} \lsctx_2$. 

    \begin{acenter}{-3cm}               
              \begin{tikzcd}[column sep=2em,row sep=2em]
               (\llam{\lvar}{\tm_1}{\labtwo}) \lsctx_{11}\lesub{\uvartwo}{\off{\lgctxthree_1}{\ann{\uvar}{\lab}}}{(\labfour)}\lsctx_{12}\esub{\uvar}{(\ofc{(\sha{\tmtwo_{111}})\lsctxtwo_1})\lsctxtwo_2} \lsctx_2 \arrow[r,dash]\arrow[r, dash, shift left=1]\arrow[r, dash, shift right=1, "\!_{1}"' {xshift=24pt,yshift=3pt}] \arrow[d, "\lab"] &             (\llam{\lvar}{\tm_1}{\labtwo}) \lsctx_{11}\lesub{\uvartwo}{\off{\lgctxthree_1}{\ann{\uvar}{\lab}}}{(\labfour)}\lsctx_{12}\esub{\uvar}{(\ofc{(\sha{\tmtwo_{111}})\lsctxtwo_1})\lsctxtwo_2} \lsctx_2   
    \arrow[d, "\lab"]  \\

    (\llam{\lvar}{\tm_1}{\labtwo}) \lsctx_{11}\lesub{\uvartwo}{\off{\lgctxthree_1}{(\sha{\tmtwo_{111}})\lsctxtwo_1}}{(\labfour)}\lsctx_{12} \esub{\uvar}{(\ofc{(\sha{\tmtwo_{111}})\lsctxtwo_1})}\lsctxtwo_2 \lsctx_2\,\tm_2
          \arrow[r,dash]\arrow[r, dash, shift left=1]\arrow[r, dash, shift right=1]
           & 
              (\llam{\lvar}{\tm_1}{\labtwo}) \lsctx_{11}\lesub{\uvartwo}{\off{\lgctxthree_1}{(\sha{\tmtwo_{111}})\lsctxtwo_1}}{(\labfour)}\lsctx_{12} \esub{\uvar}{(\ofc{(\sha{\tmtwo_{111}})\lsctxtwo_1})}\lsctxtwo_2 \lsctx_2\,\tm_2 
             \\
              \end{tikzcd}
            \end{acenter}
            
            \end{xenumerate}

        \item  $\tm_{12}=(\ofc{(\sha{\tmtwo_{111}})\lsctxtwo_{11}\lesub{\uvartwo}{\tm_{121}}{(\labfour)}\lsctxtwo_{12}})\lsctxtwo_2$ and  $\tmtwo_{11}=(\ofc{(\sha{\tmtwo_{111}})\lsctxtwo_{11}\lesub{\uvartwo}{\tmtwo_{112}}{(\labfour)}\lsctxtwo_{12}})\lsctxtwo_2$ and $\tm_{121} \flatt_1 \tmtwo_{112}$. The $\flatt_1$-step and the $\toSL{\lab}$ are disjoint. We conclude immediately.

        \item  $\tm_{12}=(\ofc{(\sha{\tmtwo_{111}})\lsctxtwo_{11}\lesub{\uvar}{\tm_{121}}{(\lab)}\lesub{\uvartwo}{\tm_{122}}{(\labfour)}\lsctxtwo_{12}})\lsctxtwo_2$ and  $\tmtwo_{11} =(\ofc{(\sha{\tmtwo_{111}})\lsctxtwo_{11}\lesub{\uvar}{\tm_{121}\lesub{\uvartwo}{\tm_{122}}{(\labfour)}}{(\lab)}\lsctxtwo_{12}})\lsctxtwo_2$. The $\flatt_1$-step and the $\toSL{\lab}$ are disjoint. We conclude immediately.
                \item  $\tm_{12}=(\ofc{(\sha{\tmtwo_{111}})\lsctxtwo_{11}\lesub{\uvar}{\tm_{121}\lesub{\uvartwo}{\tm_{122}}{(\labfour)}}{(\lab)}\lsctxtwo_{12}})\lsctxtwo_2$ and  $\tmtwo_{11} =(\ofc{(\sha{\tmtwo_{21}})\lsctxtwo_{11}\lesub{\uvar}{\tm_{121}}{(\lab)}\lesub{\uvartwo}{\tm_{122}}{(\labfour)}\lsctxtwo_{12}})\lsctxtwo_2$. The $\flatt_1$-step and the $\toSL{\lab}$ are disjoint. We conclude immediately.

               \item  $\tm_{12}=(\ofc{(\sha{\tmtwo_{111}})\lsctxtwo_{1}})\lsctxtwo_{21}\lesub{\uvartwo}{\tm_{121}}{(\labfour)}\lsctxtwo_{12}$ and $\tmtwo_{11}=(\ofc{(\sha{\tmtwo_{111}})\lsctxtwo_{1}})\lsctxtwo_{21}\lesub{\uvartwo}{\tmtwo_{112}}{(\labfour)}\lsctxtwo_{12}$ and $\tm_{121}\flatt_1 \tmtwo_{112}$. The $\flatt_1$-step and the $\toSL{\lab}$ are disjoint. We conclude immediately.

        \item  $\tm_{12}=(\ofc{(\sha{\tmtwo_{111}})\lsctxtwo_{1}}) \lsctxtwo_{21}\lesub{\uvar}{\tm_{121}}{(\lab)}\lesub{\uvartwo}{\tm_{122}}{(\labfour)}\lsctxtwo_{22}$ and  $\tmtwo_{11} =(\ofc{(\sha{\tmtwo_{111}})\lsctxtwo_{1}}) \lsctxtwo_{21}\lesub{\uvar}{\tm_{121}\lesub{\uvartwo}{\tm_{122}}{(\labfour)}}{(\lab)}\lsctxtwo_{22}$. The $\flatt_1$-step and the $\toSL{\lab}$ are disjoint. We conclude immediately.

                \item  $\tm_{12}=(\ofc{(\sha{\tmtwo_{111}})\lsctxtwo_{1}}) \lsctxtwo_{21}\lesub{\uvar}{\tm_{121}\lesub{\uvartwo}{\tm_{122}}{(\labfour)}}{(\lab)}\lsctxtwo_{22}$ and  $\tmtwo_{11} =(\ofc{(\sha{\tmtwo_{21}})\lsctxtwo_{1}}) \lsctxtwo_{21}\lesub{\uvar}{\tm_{121}}{(\lab)}\lesub{\uvartwo}{\tm_{122}}{(\labfour)}\lsctxtwo_{22}$. The $\flatt_1$-step and the $\toSL{\lab}$ are disjoint. We conclude immediately.

                \end{xenumerate}

          \item The step  $\tmg \lrootto{\lab}_{\sha{}} \tmd$ is a $\rtoSgcL{\lab}$-step. Then $\labthree=\lab$ and $\tmtwo_{11}=(\ofc{\tmtwo_{111}})\lsctxtwo $ and the step has the form $(\llam{\lvar}{\tm_1}{\labtwo})\lsctx_1\lesub{\uvar}{(\ofc{\tmtwo_{111}})\lsctxtwo}{\lab}  \rtoSgcL{\lab} (\llam{\lvar}{\tm_1}{\labtwo})\lsctx_1\lsctxtwo $ and $\uvar\notin\fv{(\llam{\lvar}{\tm_1}{\labtwo})\lsctx_1}$. We make use of \rlem{flatt_preserves_fv} and reason as follows:
              \begin{acenter}{0cm}               
              \begin{tikzcd}[column sep=2em,row sep=2em]
            (\llam{\lvar}{\tm_1}{\labtwo})\lsctx_1\lesub{\uvar}{(\ofc{\tmtwo_{111}})\lsctxtwo}{\lab} \lsctx_2\,\tm_2 \arrow[r,dash]\arrow[r, dash, shift left=1]\arrow[r, dash, shift right=1, "\!_{1}"' {xshift=14pt,yshift=3pt}] \arrow[d, "\lab"] &                        (\llam{\lvar}{\tmtwo_{11}}{\labtwo})\lsctx_1\lesub{\uvar}{(\ofc{\tmtwo_{111}})\lsctxtwo }{\lab} \lsctx_2\,\tm_2  
    \arrow[d, "\lab"]  \\
                
              (\llam{\lvar}{\tm_1}{\labtwo})\lsctx_1\lsctxtwo \lsctx_2\,\tm_2 \arrow[r,dash]\arrow[r, dash, shift left=1]\arrow[r, dash, shift right=1]
           & 
                 (\llam{\lvar}{\tmtwo_{11}}{\labtwo})\lsctx_1\lsctxtwo \lsctx_2\,\tm_2  
             \\
              \end{tikzcd}
            \end{acenter}

          \end{xenumerate}
          
      \item $\lgctx=(\llam{\lvar}{\tm_1}{\labtwo})\lsctx_1\lesub{\uvar}{\lgctx_1}{(\labthree)}\lsctx_2\,\tm_2$. We conclude from the \ih.
      \item $\lgctx=(\llam{\lvar}{\tm_1}{\labtwo})\lsctx_1\lesub{\uvar}{\tmtwo_{11}}{(\labthree)}\lsctx_{21}\lesub{\uvartwo}{\lgctx_1}{(\labfour)}\lsctx_{22}\,\tm_2$. The $\flatt_1$-step and the $\toSL{\lab}$ are disjoint. We conclude immediately.
    \end{xenumerate}
    
  \end{xenumerate}

  \item $(\llam{\lvar}{\tm_1}{\labtwo})\lsctx=\tmtwo_1$ and $\tm_2\flatt_1\tmtwo_2$. That is to say, $\tmtwo= (\llam{\lvar}{\tm_1}{\labtwo})\lsctx\,\tmtwo_2$ and   $\tm=(\llam{\lvar}{\tm_1}{\labtwo})\lsctx\,\tm_2\flatt_1\tmtwo$ follows from $\tm_2\flatt_1\tmtwo_{2}$.  Suppose $\tmtwo=\of{\lgctx}{\tmg}\toSL{\lab} \of{\lgctx}{\tmd}=\tmthree$, follows from $\tmg \lrootto{\lab}_{\sha{}} \tmd$. We consider each possible case for $\lgctx$.

        \begin{xenumerate}
        \item $\lgctx=\ctxhole$. Then $\lab=\labtwo$ and $\tmthree=\tmtwo_{1}\sub{\lvar}{\tm_2}\lsctx$.
             \begin{acenter}{0cm}               
              \begin{tikzcd}[column sep=2em,row sep=2em]
               (\llam{\lvar}{\tm_{1}}{\labtwo})\lsctx\,\tm_2  \arrow[r,dash]\arrow[r, dash, shift left=1]\arrow[r, dash, shift right=1, "\!_{1}"' {xshift=14pt,yshift=3pt}] \arrow[d, "\lab"] &                         (\llam{\lvar}{\tm_{1}}{\labtwo})\lsctx\,\tmtwo_2  
    \arrow[d, "\lab"]  \\
                
           \tm_{1}\sub{\lvar}{\tm_2}\lsctx \arrow[r,dash]\arrow[r, dash, shift left=1]\arrow[r, dash, shift right=1]
           & 
           \tm_{1}\sub{\lvar}{\tmtwo_2}\lsctx
              \end{tikzcd}
            \end{acenter}
            The equivalence at the bottom follows from \rlem{flatt_preserved_by_linear_substitution}(2).
            
          \item $\lgctx=(\llam{\lvar}{\lgctx_1}{\labtwo})\lsctx\,\tm_2$. The $\flatt_1$-step and the $\toSL{\lab}$ are disjoint. We conclude immediately.

          \item $\lgctx=\ctxhole\lsctx_2\,\tm_2$.  The $\flatt_1$-step and the $\toSL{\lab}$ are disjoint. We conclude immediately.

    \item $\lgctx=(\llam{\lvar}{\tm_1}{\labtwo})\lsctx_1\lesub{\uvar}{\lgctx_1}{(\labtwo)}\lsctx_2\,\tm_2$.  The $\flatt_1$-step and the $\toSL{\lab}$ are disjoint. We conclude immediately.

            \item $\lgctx=(\llam{\lvar}{\tm_1}{\labtwo})\lsctx\,\lgctx_1$.  We resort to the \ih.

    \end{xenumerate}

      \end{xenumerate}

  \item $\tm=\sha{\tm_1}$. Then by \rlem{flatt_one_step_generation}(\ref{flatt_one_step_generation:ltor_abs}), $\tmtwo=\sha{\tmtwo_1}$, for some $\tmtwo_1$, with $\tm_1\flatt_1\tmtwo_1$. Moreover, $\tmtwo\toSL{\lab}\tmthree$ must follow from $\tmtwo_1\toSL{\lab}\tmthree_1$. We conclude from the \ih.
    
  \item $\tm=\open{\tm_1}$. Same as the previous case.
  \item $\tm=\lopen{(\sha{\tm_1})\lsctx}{\labtwo}$. By~\rlem{flatt_one_step_generation}(\ref{flatt_one_step_generation:ltor_labeled_open}), $\tmtwo=\lopen{\tmtwo_1}{\labtwo}$ and $(\sha{\tm_1})\lsctx\flatt_1\tmtwo_1$. We now consider all possible ways in which  $(\sha{\tm_1})\lsctx\flatt_1\tmtwo_1$.
    \begin{xenumerate}
    \item $\tmtwo_1=(\sha{\tmtwo_{11}})\lsctx$ and $(\sha{\tm_1})\lsctx\flatt_1\tmtwo_1$ follows from $\tm_1\flatt_1\tmtwo_{11}$. Suppose $\tmtwo=\of{\lgctx}{\tmg}\toSL{\lab} \of{\lgctx}{\tmd}=\tmthree$, follows from $\tmg \lrootto{\lab}_{\sha{}} \tmd$. We consider each possible case for $\lgctx$.

      \begin{xenumerate}

      \item $\lgctx=\ctxhole$. Then $\lab=\labtwo$ and we have:
          \begin{acenter}{0cm}               
              \begin{tikzcd}[column sep=2em,row sep=2em]
               \lopen{(\sha{\tm_1})\lsctx}{\labtwo}  \arrow[r,dash]\arrow[r, dash, shift left=1]\arrow[r, dash, shift right=1, "\!_{1}"' {xshift=14pt,yshift=3pt}] \arrow[d, "\lab"] &      \lopen{(\sha{\tmtwo_{11}})\lsctx}{\labtwo}
    \arrow[d, "\lab"]  \\
                
           \tm_{1}\lsctx \arrow[r,dash]\arrow[r, dash, shift left=1]\arrow[r, dash, shift right=1]
           & 
           \tmtwo_{11}\lsctx
              \end{tikzcd}
            \end{acenter}

          \item $\lgctx=(\sha{\lgctx_1})\lsctx$. We use the \ih.
            
          \item $\lgctx=\ctxhole\lsctx_2$ and $\lsctx=\lsctx_1 \esub{\uvar}{(\ofc{(\sha{\tmtwo_{2}})\lsctxtwo_1})\lsctxtwo_2}\lsctx_2$. The $\flatt_1$-step and the $\toSL{\lab}$ are disjoint. We conclude immediately.

  \item $\lgctx= (\sha{\tmtwo_{11}})\lsctx_1\lesub{\uvar}{\lgctx_1}{(\labthree)}\lsctx_2$. The $\flatt_1$-step and the $\toSL{\lab}$ are disjoint. 
         \begin{acenter}{0cm}               
              \begin{tikzcd}[column sep=2em,row sep=2em]
               \lopen{(\sha{\tm_{1}})\lsctx_1\lesub{\uvar}{\of{\lgctx_1}{\tmg}}{(\labthree)}\lsctx_2}{\labtwo} \arrow[r,dash]\arrow[r, dash, shift left=1]\arrow[r, dash, shift right=1, "\!_{1}"' {xshift=14pt,yshift=3pt}] \arrow[d, "\lab"] &      \lopen{(\sha{\tmtwo_{11}})\lsctx_1\lesub{\uvar}{\of{\lgctx_1}{\tmg}}{(\labthree)}\lsctx_2 }{\labtwo}
    \arrow[d, "\lab"]  \\
                
           \lopen{(\sha{\tm_{1}})\lsctx_1\lesub{\uvar}{\of{\lgctx_1}{\tmd}}{(\labthree)}\lsctx_2}{\labtwo} \arrow[r,dash]\arrow[r, dash, shift left=1]\arrow[r, dash, shift right=1]
           & 
            \lopen{(\sha{\tmtwo_{11}})\lsctx_1\lesub{\uvar}{\of{\lgctx_1}{\tmd}}{(\labthree)}\lsctx_2 }{\labtwo}
              \end{tikzcd}
            \end{acenter}

            \end{xenumerate}
            
          \item $\tmtwo_1=(\sha{\tm_{1}})\lsctx_1\lesub{\uvar}{\tmtwo_{11}}{(\labthree)}\lsctx_2$ and $\lsctx=\lsctx_1\lesub{\uvar}{\tm_{2}}{(\labthree)}\lsctx_2 $ and  $(\sha{\tm_1})\lsctx\flatt_1\tmtwo_1$ follows from $\tm_2\flatt_1\tmtwo_{11}$. Suppose $\tmtwo=\of{\lgctx}{\tmg}\toSL{\lab} \of{\lgctx}{\tmd}=\tmthree$, follows from $\tmg \lrootto{\lab}_{\sha{}} \tmd$. We consider each possible case for $\lgctx$. 
            \begin{xenumerate}
            \item $\lgctx=\ctxhole$.   The $\flatt_1$-step and the $\toSL{\lab}$ are disjoint. We conclude immediately.
              \item $\lgctx=(\sha{\lgctx_1})\lsctx_1\lesub{\uvar}{\tmtwo_{11}}{(\labthree)}\lsctx_2$. The $\flatt_1$-step and the $\toSL{\lab}$ are disjoint. We conclude immediately.

              \item $\lgctx=(\sha{\tm_{1}}) \lsctx_{11}\lesub{\uvartwo}{\lgctx_1}{(\labfour)}\lsctx_{12}\lesub{\uvar}{\tmtwo_{11}}{(\labthree)}\lsctx_2$. The $\flatt_1$-step and the $\toSL{\lab}$ are disjoint. We conclude immediately.

              \item $\lgctx=(\sha{\tm_{1}})\lsctx_1\lesub{\uvar}{\lgctx_1}{(\labthree)}\lsctx_2$. We resort to the \ih.

              \item $\lgctx=(\sha{\tm_{1}})\lsctx_1\lesub{\uvar}{\tmtwo_{11}}{(\labthree)}\lsctx_{21}\lesub{\uvartwo}{\lgctx_1}{(\labfour)}\lsctx_{22}$. The $\flatt_1$-step and the $\toSL{\lab}$ are disjoint. We conclude immediately.

            \end{xenumerate}

       \item $\tmtwo_1=(\sha{\tm_{1}})\lsctx_1\lesub{\uvar}{\tm_{21}}{(\labthree)}\lesub{\uvartwo}{\tm_{22}}{(\labfour)}\lsctx_2$ and $\lsctx=\lsctx_1\lesub{\uvar}{\tm_{21}\lesub{\uvartwo}{\tm_{22}}{(\labfour)}}{(\labthree)}\lsctx_2 $ and  $\uvartwo\notin\fv{(\sha{\tm_{1}})\lsctx_1}$. Suppose $\tmtwo=\of{\lgctx}{\tmg}\toSL{\lab} \of{\lgctx}{\tmd}=\tmthree$, follows from $\tmg \lrootto{\lab}_{\sha{}} \tmd$. Similar to case~\ref{main_lemma_for_flattening_is_strong_bisimulation_case_F}.

          \item $\tmtwo_1=(\sha{\tm_{1}})\lsctx_1\lesub{\uvar}{\tm_{21}\lesub{\uvartwo}{\tm_{22}}{(\labfour)}}{(\labthree)}\lsctx_2$ and $\lsctx=\lsctx_1 \lesub{\uvar}{\tm_{21}}{(\labthree)}\lesub{\uvartwo}{\tm_{22}}{(\labfour)}\lsctx_2 $ and  $\uvartwo\notin\fv{(\sha{\tm_{1}})\lsctx_1}$. Suppose $\tmtwo=\of{\lgctx}{\tmg}\toSL{\lab} \of{\lgctx}{\tmd}=\tmthree$, follows from $\tmg \lrootto{\lab}_{\sha{}} \tmd$. Similar to case~\ref{main_lemma_for_flattening_is_strong_bisimulation_case_F_symmetric}.

      \end{xenumerate}

  \item $\tm=\ofc{\tm_1}$. Similar to the case $\tm=\sha{\tm_1}$.
    
  \item $\tm=\tm_1\esub{\uvar}{(\ofc{(\sha{\tm_2})\lsctx_1})\lsctx_2}$ and $\uvar\in\flv{\tm_1}$. By \rlem{flatt_one_step_generation}(\ref{flatt_one_step_generation:ltor_esub}), there are four cases to consider.
    \begin{xenumerate}
    \item\label{flatt_one_step_generation:ltor_esub_left} $\tmtwo=\tmtwo_1\esub{\uvar}{\tmtwo_2}$ and $\tm_1\flatt_1\tmtwo_1$ and $(\ofc{(\sha{\tm_2})\lsctx_1})\lsctx_2=\tmtwo_2$.  Suppose $\tmtwo=\of{\lgctx}{\tmg}\toSL{\lab} \of{\lgctx}{\tmd}=\tmthree$, follows from $\tmg \lrootto{\lab}_{\sha{}} \tmd$. We consider each possible case for $\lgctx$. 

      \begin{xenumerate}
        \item $\lgctx=\ctxhole$.  The step  $\tmg \lrootto{\lab}_{\sha{}} \tmd$ is a $ \rtoSlsL{\lab}$-step. Therefore, the step has the form $\off{\lgctxthree}{\ann{\uvar}{\lab}}\esub{\uvar}{(\ofc{(\sha{\tm_2})\lsctx_1})\lsctx_2} \rtoSlsL{\lab} \off{\lgctxthree}{(\sha{\tm_{2}})\lsctx_1}\esub{\uvar}{(\ofc{(\sha{\tm_2})\lsctx_1})}\lsctx_2$. Moreover, by $\tm_1\flatt_1\tmtwo_1=\off{\lgctxthree}{\ann{\uvar}{\lab}}$ and 
  \rlem{flattening_generation_for_contexts}(1), there exists $\lgctxtwo$ such that $\tm_1=\off{\lgctxtwo}{\ann{\uvar}{\lab}}$.

                         \begin{acenter}{-2cm}               
              \begin{tikzcd}[column sep=2em,row sep=2em]
               \off{\lgctxtwo_1}{\ann{\uvar}{\lab}}\lsctx_1\esub{\uvar}{(\ofc{(\sha{\tm_{2}})\lsctx_1})\lsctx_2} \arrow[r,dash]\arrow[r, dash, shift left=1]\arrow[r, dash, shift right=1, "\!_{1}"' {xshift=24pt,yshift=3pt}] \arrow[d, "\lab"] &            \off{\lgctxthree_1}{\ann{\uvar}{\lab}}\lsctx_1\esub{\uvar}{(\ofc{(\sha{\tm_{2}})\lsctx_1})\lsctx_2}
    \arrow[d, "\lab"]  \\

    \off{\lgctxtwo_1}{(\sha{\tm_{2}})\lsctx_1}\lsctx_1 \esub{\uvar}{(\ofc{(\sha{\tm_{2}})\lsctx_1})}\lsctx_2
          \arrow[r,dash]\arrow[r, dash, shift left=1]\arrow[r, dash, shift right=1]
           & 
              \off{\lgctxthree_1}{(\sha{\tm_{2}})\lsctx_1}\lsctx_1 \esub{\uvar}{(\ofc{(\sha{\tm_{2}})\lsctx_1})}\lsctx_2 
             \\
              \end{tikzcd}
            \end{acenter}

                     The bottom equivalence holds from \rlem{flattening_generation_for_contexts}(2).

          \item $\lgctx=\lgctx_1\esub{\uvar}{\tmtwo_2}$. We use the \ih.
            \item $\lgctx=\tmtwo_1\esub{\uvar}{\lgctx_1}$.  The $\flatt_1$-step and the $\toSL{\lab}$ are disjoint. We conclude immediately.

      \end{xenumerate}
      
    \item\label{flatt_one_step_generation:ltor_esub_right} $\tmtwo=\tmtwo_1\esub{\uvar}{\tmtwo_2}$ and $\tm_1=\tmtwo_1$ and $(\ofc{(\sha{\tm_2})\lsctx_1})\lsctx_2 \flatt_1 \tmtwo_2$. We consider all ways in which  $(\ofc{(\sha{\tm_2})\lsctx_1})\lsctx_2 \flatt_1 \tmtwo_2$. Similar to case~\ref{main_lemma_for_flattening_is_strong_bisimulation_case_F_symmetric}.

        \item\label{flatt_one_step_generation:ltor_esub_root}  $\tmtwo= \tm_1\esub{\uvar}{\tm_{21}}\lesub{\uvartwo}{\tm_{22}}{(\labtwo)}$ and $(\ofc{(\sha{\tm_2})\lsctx_1})\lsctx_2=\tm_{21}\lesub{\uvartwo}{\tm_{22}}{(\labtwo)}$ and  $\uvartwo\notin\fv{\tm_1}$. Then $\tmtwo=\of{\lgctx}{\tmg}\toSL{\lab} \of{\lgctx}{\tmd}=\tmthree$, follows from $\tmg \lrootto{\lab}_{\sha{}} \tmd$. We consider each possible case for $\lgctx$

          \begin{xenumerate}
            \item $\lgctx=\lgctx_1\esub{\uvar}{\tm_{21}}\lesub{\uvartwo}{\tm_{22}}{(\labtwo)}$. The $\flatt_1$-step and the $\toSL{\lab}$ are disjoint. We conclude immediately.
            \item $\lgctx=\ctxhole\lesub{\uvartwo}{\tm_{22}}{(\labtwo)}$. Same as case~\ref{main_lemma_for_flattening_is_strong_bisimulation_case_F_symmetric}.
            \item $\lgctx=\tm_1\esub{\uvar}{\lgctx_1}\lesub{\uvartwo}{\tm_{22}}{(\labtwo)}$. The $\flatt_1$-step and the $\toSL{\lab}$ are disjoint. We conclude immediately.
            \item $\lgctx=\ctxhole$. Same as case~\ref{main_lemma_for_flattening_is_strong_bisimulation_case_F_symmetric}.
             
            \item $\lgctx=\tm_1\esub{\uvar}{\tm_{21}}\lesub{\uvartwo}{\lgctx_1}{(\labtwo)}$. The $\flatt_1$-step and the $\toSL{\lab}$ are disjoint. We conclude immediately.

            \end{xenumerate}
        \item $\tmtwo= \tm_{11}\lesub{\uvartwo}{\tm_{21}\esub{\uvar}{(\ofc{(\sha{\tm_2})\lsctx_1})\lsctx_2}}{(\labtwo)}$ and $\tm_1=\tm_{11}\lesub{\uvartwo}{\tm_{21}}{(\labtwo)}$ and  $\uvar\notin\fv{\tm_{11}}$. Suppose $\tmtwo=\of{\lgctx}{\tmg}\toSL{\lab} \of{\lgctx}{\tmd}=\tmthree$, follows from $\tmg \lrootto{\lab}_{\sha{}} \tmd$. We consider each possible case for $\lgctx$

              \begin{xenumerate}
            \item $\lgctx=\lgctx_1\lesub{\uvartwo}{\tm_{21}\esub{\uvar}{(\ofc{(\sha{\tm_2})\lsctx_1})\lsctx_2}}{(\labtwo)}$. The $\flatt_1$-step and the $\toSL{\lab}$ are disjoint. We conclude immediately.
            \item $\lgctx=\ctxhole$. Same as case~\ref{flatt_one_step_generation:F_rtol}.
            \item $\lgctx=\tm_{11}\lesub{\uvartwo}{\ctxhole}{(\labtwo)}$.  Same as case~\ref{flatt_one_step_generation:F_rtol_nested}.
        \item $\lgctx=\tm_{11}\lesub{\uvartwo}{\lgctx_1\esub{\uvar}{(\ofc{(\sha{\tm_2})\lsctx_1})\lsctx_2}}{(\labtwo)}$. The $\flatt_1$-step and the $\toSL{\lab}$ are disjoint. We conclude immediately.

                \item $\lgctx=\tm_{11}\lesub{\uvartwo}{\tm_{21}\esub{\uvar}{\lgctx_1}}{(\labtwo)}$ and $\of{\lgctx_1}{\tmg}= (\ofc{(\sha{\tm_2})\lsctx_1})\lsctx_2$. The $\flatt_1$-step and the $\toSL{\lab}$ are disjoint. We conclude immediately.

            \end{xenumerate}
            
        \end{xenumerate}
        
      \item $\tm=\tm_1\lesub{\uvar}{(\ofc{\tm_2})\lsctx}{\labtwo}$ and $\uvar\notin\fv{\tm_1}$. By \rlem{flatt_one_step_generation}(\ref{flatt_one_step_generation:ltor_esub}), there are four cases to consider. 

        \begin{xenumerate}
        \item\label{flatt_one_step_generation:ltor_esub_left} $\tmtwo=\tmtwo_1\lesub{\uvar}{\tmtwo_2}{\labtwo}$ and $\tm_1\flatt_1\tmtwo_1$ and $(\ofc{\tm_2})\lsctx=\tmtwo_2$. Suppose $\tmtwo=\of{\lgctx}{\tmg}\toSL{\lab} \of{\lgctx}{\tmd}=\tmthree$, follows from $\tmg \lrootto{\lab}_{\sha{}} \tmd$. We consider each possible case for $\lgctx$. 

      \begin{xenumerate}
      \item $\lgctx=\ctxhole$.  Then $\lab=\labtwo$ and we reason as follows, making use of \rlem{flatt_preserves_fv} and reason as follows:

              \begin{acenter}{0cm}               
              \begin{tikzcd}[column sep=2em,row sep=2em]
            \tm_1\lesub{\uvar}{(\ofc{\tm_{2}})\lsctx}{\lab} \arrow[r,dash]\arrow[r, dash, shift left=1]\arrow[r, dash, shift right=1, "\!_{1}"' {xshift=14pt,yshift=3pt}] \arrow[d, "\lab"] &                        \tmtwo_{1}\lesub{\uvar}{(\ofc{\tm_{2}})\lsctx }{\lab} 
    \arrow[d, "\lab"]  \\
                
              \tm_1\lsctx \arrow[r,dash]\arrow[r, dash, shift left=1]\arrow[r, dash, shift right=1]
           & 
             \tmtwo_{1}\lsctx 
             \\
              \end{tikzcd}
            \end{acenter}

   \item $\lgctx=\lgctx_1\lesub{\uvar}{(\ofc{\tm_2})\lsctx}{\labtwo}$. We resort to the \ih.

      \item $\lgctx=\tmtwo_1\lesub{\uvar}{\lgctx_1}{\labtwo}$. The $\flatt_1$-step and the $\toSL{\lab}$ are disjoint. We conclude immediately.

      \end{xenumerate}

        \item\label{flatt_one_step_generation:ltor_esub_right} $\tmtwo=\tmtwo_1\lesub{\uvar}{\tmtwo_2}{\labtwo}$ and $\tm_1=\tmtwo_1$ and $(\ofc{\tm_2})\lsctx \flatt_1 \tmtwo_2$. We consider all ways in which   $(\ofc{\tm_2})\lsctx \flatt_1 \tmtwo_2$.

      \begin{xenumerate}
        \item  $\tmtwo_2=(\ofc{\tmtwo_{21}})\lsctx$ and  $(\ofc{\tm_2})\lsctx \flatt_1 \tmtwo_2$ follows from $\tm_2 \flatt_1 \tmtwo_{21}$. Suppose $\tmtwo=\of{\lgctx}{\tmg}\toSL{\lab} \of{\lgctx}{\tmd}=\tmthree$, follows from $\tmg \lrootto{\lab}_{\sha{}} \tmd$. We consider each possible case for $\lgctx$. 

      \begin{xenumerate}
      \item $\lgctx=\ctxhole$.  Then $\lab=\labtwo$ and we reason as follows, making use of \rlem{flatt_preserves_fv} and reason as follows:

              \begin{acenter}{0cm}               
              \begin{tikzcd}[column sep=2em,row sep=2em]
            \tm_1\lesub{\uvar}{(\ofc{\tm_{2}})\lsctx}{\labtwo}\arrow[r,dash]\arrow[r, dash, shift left=1]\arrow[r, dash, shift right=1, "\!_{1}"' {xshift=14pt,yshift=3pt}] \arrow[d, "\lab"] &                    \tmtwo_1\lesub{\uvar}{(\ofc{\tmtwo_{21}})\lsctx}{\labtwo}
    \arrow[d, "\lab"]  \\
                
              \tm_1\lsctx \arrow[r,equal]
           & 
             \tmtwo_{1}\lsctx 
             \\
              \end{tikzcd}
            \end{acenter}

          \item $\lgctx=\lgctx_1\lesub{\uvar}{(\ofc{\tmtwo_{21}})\lsctx}{\labtwo}$. The $\flatt_1$-step and the $\toSL{\lab}$ are disjoint. We conclude immediately.

      \item $\lgctx=\tmtwo_1\lesub{\uvar}{(\ofc{\lgctx_1})\lsctx}{\labtwo}$. We resort to the \ih.

            \item\label{flatt_one_step_generation:ltor_esub_right_ls_or_gc} $\lgctx=\tmtwo_1\lesub{\uvar}{\ctxhole\lsctx_2}{\labtwo}$. There are two possibilities for the step  $\tmg \lrootto{\lab}_{\sha{}} \tmd$. It can either be a $ \rtoSlsL{\lab}$-step or a $\rtoSgcL{\lab}$-step.

               \begin{xenumerate}

        \item  The step  $\tmg \lrootto{\lab}_{\sha{}} \tmd$ is a $ \rtoSlsL{\lab}$-step. Then the $ \rtoSlsL{\lab}$-step has the form $\off{\lgctxthree}{\ann{\uvartwo}{\lab}}\esub{\uvartwo}{(\ofc{(\sha{\tm_{3}})\lsctxtwo_1})\lsctxtwo_2}  \rtoSlsL{\lab} \off{\lgctxthree}{(\sha{\tm_{3}})\lsctxtwo_1}\esub{\uvartwo}{\ofc{(\sha{\tm_{3}})\lsctxtwo_1}}\lsctxtwo_2$ and $\off{\lgctxthree}{\ann{\uvartwo}{\lab}} = (\ofc{\tm_{2}})\lsctx_1$.
      We consider each possible form for $\lgctxthree$.
            \begin{xenumerate}
            \item  $\lgctxthree= (\ofc{\lgctxthree_1})\lsctx_1$.  The $\flatt_1$-step and the $\toSL{\lab}$ are disjoint. We conclude immediately.

                         \begin{acenter}{-3cm}               
              \begin{tikzcd}[column sep=2em,row sep=2em]
             \tm_1\lesub{\uvar}{(\ofc{\off{\lgctxthree_1}{\ann{\uvartwo}{\lab}}})\lsctx_1 \esub{\uvartwo}{(\ofc{(\sha{\tm_{21}})\lsctxtwo_1})\lsctxtwo_2}\lsctx_2 }{\labtwo}  \arrow[r,dash]\arrow[r, dash, shift left=1]\arrow[r, dash, shift right=1, "\!_{1}"' {xshift=24pt,yshift=3pt}] \arrow[d, "\lab"] &                \tmtwo_1\lesub{\uvar}{(\ofc{\off{\lgctxthree_1}{\ann{\uvartwo}{\lab}}})\lsctx_1 \esub{\uvartwo}{(\ofc{(\sha{\tm_{21}})\lsctxtwo_1})\lsctxtwo_2}\lsctx_2 }{\labtwo}  
    \arrow[d, "\lab"]  \\
                
                       \tm_1\lesub{\uvar}{(\ofc{\off{\lgctxthree_1}{(\sha{\tm_{21}})\lsctxtwo_1}})\lsctx_1 \esub{\uvartwo}{(\ofc{(\sha{\tm_{21}})\lsctxtwo_1})}\lsctxtwo_2\lsctx_2 }{\labtwo}  \arrow[r,dash]\arrow[r, dash, shift left=1]\arrow[r, dash, shift right=1]
           & 
             \tmtwo_1\lesub{\uvar}{(\ofc{\off{\lgctxthree_1}{(\sha{\tm_{21}})\lsctxtwo_1}})\lsctx_1 \esub{\uvartwo}{(\ofc{(\sha{\tm_{21}})\lsctxtwo_1})}\lsctxtwo_2\lsctx_2 }{\labtwo} 
             \\
              \end{tikzcd}
            \end{acenter}
            
            \item $\lgctxthree= (\ofc{\tm_2})\lsctx_{11}\lesub{\uvartwo}{\lgctxthree_1}{(\labthree)}\lsctx_{12}$ and $\lsctx=\lsctx_{11}\lesub{\uvartwo}{\lgctxthree_1}{(\labthree)}\lsctx_{11} \esub{\uvar}{(\ofc{(\sha{\tm_{11}})\lsctxtwo_1})\lsctxtwo_2}\lsctx_2$. The $\flatt_1$-step and the $\toSL{\lab}$ are disjoint. We conclude immediately.

            \end{xenumerate}

          \item The step  $\tmg \lrootto{\lab}_{\sha{}} \tmd$ is a $\rtoSgcL{\lab}$-step. Then  it has the form $(\ofc{\tm_{12}})\lsctx_1\lesub{\uvartwo}{(\ofc{\tm_{3}})\lsctxtwo}{\lab}  \rtoSgcL{\lab} (\ofc{\tm_{12}})\lsctx_1\lsctxtwo$ and $\uvar\notin\fv{(\ofc{\tm_{12}})\lsctx_1}$. We make use of \rlem{flatt_preserves_fv} and reason as follows:

              \begin{acenter}{0cm}               
              \begin{tikzcd}[column sep=2em,row sep=2em]
            \tm_{1}\lesub{\uvar}{(\ofc{\tm_{12}})\lsctx_1\lesub{\uvartwo}{(\ofc{\tm_{3}})\lsctxtwo}{\lab}\lsctx_2}{\labtwo}  \arrow[r,dash]\arrow[r, dash, shift left=1]\arrow[r, dash, shift right=1, "\!_{1}"' {xshift=14pt,yshift=3pt}] \arrow[d, "\lab"] &                        \tmtwo_{1}\lesub{\uvar}{(\ofc{\tm_{12}})\lsctx_1\lesub{\uvartwo}{(\ofc{\tm_{3}})\lsctxtwo}{\lab}\lsctx_2}{\labtwo}  
    \arrow[d, "\lab"]  \\
                
            \tm_{1}\lesub{\uvar}{(\ofc{\tm_{12}})\lsctx_1\lsctxtwo\lsctx_2}{\labtwo}  \arrow[r,dash]\arrow[r, dash, shift left=1]\arrow[r, dash, shift right=1]
           & 
                  \tmtwo_{1}\lesub{\uvar}{(\ofc{\tm_{12}})\lsctx_1\lsctxtwo\lsctx_2}{\labtwo} 
             \\
              \end{tikzcd}
            \end{acenter}

            \end{xenumerate}

        \item  $\tmtwo_2=(\ofc{\tm_{2}})\lsctx_{11}\lesub{\uvar}{\tmtwo}{(\labthree)}\lesub{\uvartwo}{\tmthree}{(\labfour)}\lsctx_{12}$ and  $(\ofc{\tm_2})\lsctx =(\ofc{\tm_{2}})\lsctx_{11}\lesub{\uvar}{\tmtwo\lesub{\uvartwo}{\tmthree}{(\labfour)}}{(\labthree)}\lsctx_{12}$.  Similar to case~\ref{main_lemma_for_flattening_is_strong_bisimulation_case_F}

                \item  $\tmtwo_2=(\ofc{\tm_{2}})\lsctx_{11}\lesub{\uvar}{\tmtwo\lesub{\uvartwo}{\tmthree}{(\labfour)}}{(\labthree)}\lsctx_{12}$ and  $(\ofc{\tm_2})\lsctx =(\ofc{\tm_{2}})\lsctx_{11}\lesub{\uvar}{\tmtwo}{(\labthree)}\lesub{\uvartwo}{\tmthree}{(\labfour)}\lsctx_{12}$.  Similar to case~\ref{main_lemma_for_flattening_is_strong_bisimulation_case_F_symmetric}

            \end{xenumerate}

        \item  $\tmtwo_2=(\ofc{\tm_{2}})\lsctx_{11}\lesub{\uvartwo}{\tmtwo_{21}}{(\labthree)}\lsctx_{12}$ and $\lsctx=\lsctx_{11}\lesub{\uvartwo}{\tm_{21}}{(\labthree)}\lsctx_{12} $ and  $(\ofc{\tm_2})\lsctx \flatt_1 \tmtwo_2$ follows from $\tm_{21} \flatt_1 \tmtwo_{21}$.  Suppose $\tmtwo=\of{\lgctx}{\tmg}\toSL{\lab} \of{\lgctx}{\tmd}=\tmthree$, follows from $\tmg \lrootto{\lab}_{\sha{}} \tmd$. We consider each possible case for $\lgctx$. 

      \begin{xenumerate}
      \item $\lgctx=\ctxhole$.  Then $\lab=\labtwo$ and we reason as follows, making use of \rlem{flatt_preserves_fv} and reason as follows:

              \begin{acenter}{0cm}               
              \begin{tikzcd}[column sep=2em,row sep=2em]
            \tm_1\lesub{\uvar}{(\ofc{\tm_{2}})\lsctx_{11}\lesub{\uvartwo}{\tmtwo_{21}}{(\labthree)}\lsctx_{12}}{\labtwo}\arrow[r,dash]\arrow[r, dash, shift left=1]\arrow[r, dash, shift right=1, "\!_{1}"' {xshift=14pt,yshift=3pt}] \arrow[d, "\lab"] &                    \tmtwo_1\lesub{\uvar}{(\ofc{\tm_{2}})\lsctx_{11}\lesub{\uvartwo}{\tm_{21}}{(\labthree)}\lsctx_{12}}{\labtwo}
    \arrow[d, "\lab"]  \\
                
              \tm_1 \lsctx_{11}\lesub{\uvartwo}{\tmtwo_{21}}{(\labthree)}\lsctx_{12} \arrow[r,dash]\arrow[r, dash, shift left=1]\arrow[r, dash, shift right=1] 
           & 
             \tmtwo_{1}\lsctx_{11}\lesub{\uvartwo}{\tm_{21}}{(\labthree)}\lsctx_{12}
             \\
              \end{tikzcd}
            \end{acenter}

          \item $\lgctx=\lgctx_1  \lesub{\uvar}{(\ofc{\tm_{2}})\lsctx_{11}\lesub{\uvartwo}{\tm_{21}}{(\labthree)}\lsctx_{12}}{\labtwo}$. The $\flatt_1$-step and the $\toSL{\lab}$ are disjoint. We conclude immediately.
          \item $\lgctx=\tmtwo_1  \lesub{\uvar}{(\ofc{\lgctx_1})\lsctx_{11}\lesub{\uvartwo}{\tm_{21}}{(\labthree)}\lsctx_{12}}{\labtwo}$. The $\flatt_1$-step and the $\toSL{\lab}$ are disjoint. We conclude immediately.

            \item $\lgctx=\tmtwo_1\lesub{\uvar}{\ctxhole\lsctx_{12}}{\labtwo}$. There are two possibilities for the step  $\tmg \lrootto{\lab}_{\sha{}} \tmd$. It can either be a $ \rtoSlsL{\lab}$-step or a $\rtoSgcL{\lab}$-step.  Similar to case~\ref{flatt_one_step_generation:ltor_esub_right_ls_or_gc}

          \item $\lgctx= \tmtwo_1\lesub{\uvar}{(\ofc{\tm_{2}})\lsctx_{111}\lesub{\uvarthree}{\lgctx_1}{(\labfour)}\lsctx_{112}\lesub{\uvartwo}{\tm_{21}}{(\labthree)}\lsctx_{12}}{\labtwo}$. The $\flatt_1$-step and the $\toSL{\lab}$ are disjoint. We conclude immediately.
        
      \item $\lgctx= \tmtwo_1\lesub{\uvar}{(\ofc{\tm_{2}})\lsctx_{11}\lesub{\uvartwo}{\lgctx_1}{(\labthree)}\lsctx_{12}}{\labtwo}$. We resort to the \ih.

      \item $\lgctx= \tmtwo_1\lesub{\uvar}{(\ofc{\tm_{2}})\lsctx_{11}\lesub{\uvartwo}{\tm_{21}}{(\labthree)}\lsctx_{121}\lesub{\uvarthree}{\lgctx_1}{(\labfour)}\lsctx_{122}}{\labtwo}$. The $\flatt_1$-step and the $\toSL{\lab}$ are disjoint. We conclude immediately.
      \end{xenumerate}
        
        \item\label{flatt_one_step_generation:ltor_esub_root}  $\lsctx=\lsctx_{11}\lesub{\uvarthree}{\tm_3\lesub{\uvartwo}{\tm_{22}}{(\labthree)}}{(\labfour)}\lsctx_{12}$ and $\tmtwo= \tm_1\lesub{\uvar}{(\ofc{\tm_2})\lsctx_{11}\lesub{\uvarthree}{\tm_3}{(\labfour)}\lesub{\uvartwo}{\tm_{22}}{(\labthree)}\lsctx_{12}}{\labtwo}$ and $\uvartwo\notin\fv{(\ofc{\tm_2})\lsctx_{11}}$. Similar to case~\ref{main_lemma_for_flattening_is_strong_bisimulation_case_F}
          
        \item $\lsctx=\lsctx_{11}\lesub{\uvarthree}{\tm_3}{(\labfour)}\lesub{\uvartwo}{\tm_{22}}{(\labthree)}\lsctx_{12}$ and  $\tmtwo= \tm_1\lesub{\uvar}{(\ofc{\tm_2})\lsctx_{11}\lesub{\uvarthree}{\tm_3\lesub{\uvartwo}{\tm_{22}}{(\labthree)}}{(\labfour)}\lsctx_{12}}{\labtwo}$ and $\uvartwo\notin\fv{(\ofc{\tm_2})\lsctx_{11}}$. Similar to case~\ref{main_lemma_for_flattening_is_strong_bisimulation_case_F_symmetric}

    \end{xenumerate}

  \item $\tm=\tm_1\esub{\uvar}{\tm_2}$ and $\uvar\notin\flv{\tm_1}$. By \rlem{flatt_one_step_generation}(\ref{flatt_one_step_generation:ltor_esub}), there are four cases to consider.

    \begin{xenumerate}
    \item\label{flatt_one_step_generation:ltor_esub_left} $\tmtwo=\tmtwo_1\esub{\uvar}{\tmtwo_2}$ and $\tm_1\flatt_1\tmtwo_1$ and $\tm_2=\tmtwo_2$. Suppose $\tmtwo=\of{\lgctx}{\tmg}\toSL{\lab} \of{\lgctx}{\tmd}=\tmthree$, follows from $\tmg \lrootto{\lab}_{\sha{}} \tmd$. We consider each possible case for $\lgctx$. 

      \begin{xenumerate}
      \item $\lgctx=\ctxhole$. Not possible.

            \item $\lgctx=\lgctx_1\esub{\uvar}{\tmtwo_2}$. We resort to the \ih.

                  \item $\lgctx=\tmtwo_1\esub{\uvar}{\lgctx_2}$. The $\flatt_1$-step and the $\toSL{\lab}$ are disjoint. We conclude immediately.

      \end{xenumerate}
        
      \item\label{flatt_one_step_generation:ltor_esub_right} $\tmtwo=\tmtwo_1\esub{\uvar}{\tmtwo_2}$ and $\tm_1=\tmtwo_1$ and $\tm_2 \flatt_1 \tmtwo_2$. Suppose $\tmtwo=\of{\lgctx}{\tmg}\toSL{\lab} \of{\lgctx}{\tmd}=\tmthree$, follows from $\tmg \lrootto{\lab}_{\sha{}} \tmd$. We consider each possible case for $\lgctx$. 

      \begin{xenumerate}
      \item $\lgctx=\ctxhole$. Not possible.

            \item $\lgctx=\lgctx_1\esub{\uvar}{\tmtwo_2}$. The $\flatt_1$-step and the $\toSL{\lab}$ are disjoint. We conclude immediately. 

                  \item $\lgctx=\tmtwo_1\esub{\uvar}{\lgctx_2}$.  We resort to the \ih.

      \end{xenumerate}
          
        \item\label{flatt_one_step_generation:ltor_esub_root}  $\tm_2=\tm_{21}\lesub{\uvartwo}{\tm_{22}}{(\labtwo)}$ and $\tmtwo= \tm_1\esub{\uvar}{\tm_{21}}\lesub{\uvartwo}{\tm_{22}}{(\labtwo)}$ and $\uvartwo\notin\fv{\tm_1}$.
          Similar to case~\ref{flatt_one_step_generation:ls_or_gc}.
          
        \item $\tm_1=\tm_{11}\lesub{\uvartwo}{\tm_{12}}{(\labtwo)}$ and $\tmtwo= \tm_{11}\lesub{\uvartwo}{\tm_{21}\esub{\uvar}{\tm_{2}}}{(\labtwo)}$ and $\uvar\notin\fv{\tm_{11}}$. Similar to case~\ref{flatt_one_step_generation:ls_or_gc}.
      \end{xenumerate}
  \end{xenumerate}
  \end{xenumerate}

  \end{proof}
\end{ifLongAppendix}

\begin{proposition}[Flattening is a strong $\toS$-bisimulation]
\lprop{flattening_is_strong_bisimulation}
${\flatt\toSL{\lab}} \subseteq {\toSL{\lab}\flatt}$
\end{proposition}

\begin{proof}
Suppose  $\tm\pflatt{\pi}\tmtwo$. By \rremark{flatt_one_vs_many} there exists $n>0$ and $\tmthree_1,\ldots\tmthree_n$ and $\pi_1,\ldots,\pi_{n-1}$ such that $\tmthree_1=\tm$ and $\tmthree_n=\tmtwo$ and $\tmthree_1\pflatt{\pi_1}_1 \tmthree_2, \tmthree_2\pflatt{\pi_2}_1 \tmthree_3,\ldots, \tmthree_{n-1}\pflatt{\pi_{n-1}}_1\tmthree_n$, where $n-1$ is the number of times $\indrulename{$\flatt$F}$ was used in $\pi$. We therefore prove $(\flatt_1)^{n-1}\toSL{\lab}\subseteq \toSL{\lab}\flatt$, by induction on $n$. If $n=1$, then $\tm=\tmtwo$ and the result hold immediately. If $n=2$, then the result follows from \rlem{main_lemma_for_flattening_is_strong_bisimulation}. Suppose $n>2$. Then we rely on \rlem{main_lemma_for_flattening_is_strong_bisimulation} and the \ih:

         \begin{center}
              \begin{tikzcd}[column sep=4em,row sep=2em]
             \tmthree_1 \arrow[d, "\lab", "\sha{}"'] \arrow[r,dash]\arrow[r, dash, shift left=1, "n-2" {xshift=22pt}]\arrow[r, dash, shift right=1, "\!_{1}"' {xshift=22pt,yshift=3pt}] \arrow[phantom,rd,"\scriptstyle\ih"] &  \tmthree_{n-1}\arrow[r,dash]\arrow[r, dash, shift left=1, "\pi_{n-1}"]\arrow[r, dash, shift right=1, "1"' {xshift=22pt,yshift=3pt}] \arrow[d,  "\lab", "\sha{}"'] \arrow[phantom,rd,"\scriptstyle\rlem{main_lemma_for_flattening_is_strong_bisimulation}"] & \tmthree_n\arrow[d,  "\lab", "\sha{}"']\\
              
             \tmthree_1' \arrow[r,dash]\arrow[r, dash, shift left=1]\arrow[r, dash, shift right=1] & \tmthree_{n-1}' \arrow[r,dash]\arrow[r, dash, shift left=1]\arrow[r, dash, shift right=1] & \tmthree_n'            \\
            \end{tikzcd}
          \end{center}
\end{proof}

\begin{lemma}
  \llem{compatibility_of_labeled_reduction_with_substitution}
  Let $\tm,\tmtwo,\tmthree\in \TermsSWL$ be well-labeled terms. Suppose  $\tm\lto{\lab}\tmtwo$. Then
  \begin{enumerate}
  \item $\tm\sub{\lvar}{\tmthree} \lto{\lab} \tmtwo\sub{\lvar}{\tmthree}$
  \item $\tmthree\sub{\lvar}{\tm} \dev{\lab} \tmthree\sub{\lvar}{\tmtwo}$
    
  \end{enumerate}  
\end{lemma}

\begin{proof}
The first item is by induction on $\tm$ and the second by induction on
$\tmthree$. 
\begin{ifLongAppendix}
  \begin{xenumerate}
\item $\tm=\lvar, \tm=\ann{\uvar}{(\lab)}$. Immediate.
\item $\tm=\lam{\lvartwo}{\tm_1}$. Then the step $\tm\lto{\lab}\tmtwo$ is in $\tm_1$ and we conclude from the \ih.
\item $\tm=\tm_1\,\tm_2$. Then the step $\tm\lto{\lab}\tmtwo$ must be in $\tm_1$ or $\tm_2$, we conclude from the \ih.
     
\item $\tm= (\llam{\lvartwo}{\tm_1}{\labtwo})\lsctx\,\tm_2$.  If the step $\tm\lto{\lab}\tmtwo$ is in $\tm_1$ or $\tm_2$ or $\lsctx$, we conclude from the \ih. Suppose it is at the root. Then  $\labtwo=\lab$ and  $\tm = (\llam{\lvartwo}{\tm_{1}}{\lab})\lsctx\,\tm_{2}
     \rtoSdbL{\lab} 
     \tm_{1}\sub{\lvartwo}{\tm_{2}}\lsctx = \tmtwo$. We conclude from
     \[\begin{array}{rl}
        &  ((\llam{\lvartwo}{\tm_{1}}{\lab})\lsctx\,\tm_{2})\sub{\lvar}{\tmthree}\\
         = & ((\llam{\lvartwo}{\tm_{1}\sub{\lvar}{\tmthree} }{\lab})\lsctx \sub{\lvar}{\tmthree} \,\tm_{2}\sub{\lvar}{\tmthree} )\\
     \rtoSdbL{\lab}&
                     \tm_{1}\sub{\lvar}{\tmthree} \sub{\lvartwo}{\tm_{2}\sub{\lvar}{\tmthree} }\lsctx\sub{\lvar}{\tmthree} \\
         = & (\tm_{1}\sub{\lvartwo}{\tm_{2}}\lsctx) \sub{\lvar}{\tmthree} 
       \end{array}
     \]
     We make use of the substitution lemma stating that $\tm_{1}\sub{\lvar}{\tmthree} \sub{\lvartwo}{\tm_{2}\sub{\lvar}{\tmthree} }=\tm_{1} \sub{\lvartwo}{\tm_{2}}\sub{\lvar}{\tmthree}$, if $\lvartwo\notin\fv{\tmthree}$.
     
\item $\tm=\sha{\tm_1}$. Then the step $\tm\lto{\lab}\tmtwo$ must be in $\tm_1$, we conclude from the \ih.
\item $\tm=\open{\tm_1}$. Then the step $\tm\lto{\lab}\tmtwo$ must be in $\tm_1$, we conclude from the \ih.
\item $\tm=\lopen{(\sha{\tm_1})\lsctx}{\labtwo}$. If the step $\tm\lto{\lab}\tmtwo$ is in $\tm_1$ or $\lsctx$, we conclude from the \ih. Suppose it is at the root. Then $\labtwo=\lab$ and $\tm=  \lopen{(\sha{\tm_1})\lsctx}{\lab}
     \rtoSopenL{\lab} 
    \tm_1lsctx=\tmtwo$. We conclude from
     \[\begin{array}{rl}
        & (\lopen{(\sha{\tm_1})\lsctx}{\lab})\sub{\lvar}{\tmthree}\\
         = &\lopen{(\sha{\tm_1 \sub{\lvar}{\tmthree}})\lsctx \sub{\lvar}{\tmthree}}{\lab}\\
     \rtoSopenL{\lab}&
                    \tm_1\sub{\lvar}{\tmthree}\lsctx\sub{\lvar}{\tmthree} \\
         = & (\tm_1\lsctx) \sub{\lvar}{\tmthree} 
       \end{array}
     \]

\item $\tm=\ofc{\tm_1}$. Then the step $\tm\lto{\lab}\tmtwo$ must be in $\tm_1$, we conclude from the \ih.

\item $\tm=\tm_1\esub{\uvar}{(\ofc{(\sha{\tm_2})\lsctx_1})\lsctx_2}$.  If the step $\tm\lto{\lab}\tmtwo$ is in $\tm_1$ or $\tm_2$ or $\lsctx_1$ or $\lsctx_2$, we conclude from the \ih. Suppose it is at the root. Then $\lab=\labtwo$ and $\tm=  \off{\lgctx}{\ann{\uvar}{\lab}}\esub{\uvar}{(\ofc{(\sha{\tm_2})\lsctx_1})\lsctx_2}
    \rtoSlsL{\lab} 
    \off{\lgctx}{(\sha{\tm_2}) \lsctx_1}\esub{\uvar}{\ofc{(\sha{\tm_2}) \lsctx_1}}\lsctx_2=\tmtwo$, where $\uvar \notin \fv{\tm_2}$ and $\fv{\lgctx} \cap \dom{\lsctx_1\lsctx_2} = \emptyset$. We conclude from
     \[\begin{array}{rl}
        & (\off{\lgctx}{\ann{\uvar}{\lab}}\esub{\uvar}{(\ofc{(\sha{\tm_2})\lsctx_1})\lsctx_2})\sub{\lvar}{\tmthree}\\
         = &\off{\lgctx \sub{\lvar}{\tmthree}}{\ann{\uvar}{\lab}}\esub{\uvar}{(\ofc{(\sha{\tm_2 \sub{\lvar}{\tmthree}})\lsctx_1 \sub{\lvar}{\tmthree}})\lsctx_2 \sub{\lvar}{\tmthree}}\\
     \rtoSlsL{\lab}&
                     \off{\lgctx\sub{\lvar}{\tmthree}}{(\sha{\tm_2\sub{\lvar}{\tmthree}}) \lsctx_1\sub{\lvar}{\tmthree}}\esub{\uvar}{\ofc{(\sha{\tm\sub{\lvar}{\tmthree}}) \lsctx_1\sub{\lvar}{\tmthree}}}\lsctx_2\sub{\lvar}{\tmthree} \\
         = & ( \off{\lgctx}{(\sha{\tm_2}) \lsctx_1}\esub{\uvar}{\ofc{(\sha{\tm_2}) \lsctx_1}}\lsctx_2) \sub{\lvar}{\tmthree} 
       \end{array}
     \]

\item $\tm=\tm_1\lesub{\uvar}{(\ofc{\tm_2})\lsctx}{\labtwo}$. If the step $\tm\lto{\lab}\tmtwo$ is in $\tm_1$ or $\tm_2$ or $\lsctx$, we conclude from the \ih. Suppose it is at the root. Then $\lab=\labtwo$ and $\tm=  \tm_1\esub{\ann{\uvar}{\lab}}{(\ofc{\tm_2})\lsctx}
    \rtoSgcL{\lab} 
    \tm_1\lsctx=\tmtwo$, where
    $\uvar\notin\fv{\tm_1}$. We conclude from
     \[\begin{array}{rl}
        & (\tm_1\esub{\ann{\uvar}{\lab}}{(\ofc{\tm_2})\lsctx})\sub{\lvar}{\tmthree}\\
         = &  \tm_1\sub{\lvar}{\tmthree}\esub{\ann{\uvar}{\lab}}{(\ofc{\tm_2\sub{\lvar}{\tmthree}})\lsctx \sub{\lvar}{\tmthree}}\\
     \rtoSgcL{\lab}&
                     \tm_1\sub{\lvar}{\tmthree}\lsctx\sub{\lvar}{\tmthree} \\
         = & (\tm_1\lsctx) \sub{\lvar}{\tmthree} 
       \end{array}
     \] 
 Note that $\uvar\notin\fv{\tmthree}$.   

 \item $\tm=\tm_1\esub{\uvar}{\tm_2}$ with $\tm_1\in \TermsSWL$, $\tm_2\in \TermsSWL$ and 
   $\uvar\notin\flv{\tm_1}$.  Then the step $\tm\lto{\lab}\tmtwo$ is in $\tm_1$ or $\tm_2$ and we conclude from the \ih.
   
\end{xenumerate}
We now address the second item.

  \begin{xenumerate}
  \item $\tmthree=\lvartwo$. If $\lvar\neq\lvartwo$, then the result is immediate. Otherwise, it follows from the hypothesis  
    $\tm\lto{\lab}\tmtwo$.
    
  \item $\tmthree=\ann{\uvar}{(\lab)}$. Immediate.
    
  \item $\tmthree=\lam{\lvartwo}{\tmthree_1}$. The result follows from the \ih.
    
\item $\tmthree=\tmthree_1\,\tmthree_2$. The result follows from the \ih, applied twice. First to obtain $\tmthree_1\sub{\lvar}{\tm} \dev{\lab} \tmthree_1\sub{\lvar}{\tmtwo}$, then to obtain $\tmthree_2\sub{\lvar}{\tm} \dev{\lab} \tmthree_2\sub{\lvar}{\tmtwo}$. From which we conclude $\tmthree_1\sub{\lvar}{\tm} \,\tmthree_2\sub{\lvar}{\tm} \dev{\lab} \tmthree_1\sub{\lvar}{\tm} \,\tmthree_2\sub{\lvar}{\tmtwo}$.

\item $\tmthree= (\llam{\lvartwo}{\tmthree_1}{\labtwo})\lsctx\,\tmthree_2$. The result follows from the \ih, applied multiple times. First to obtain $\tmthree_1\sub{\lvar}{\tm} \dev{\lab} \tmthree_1\sub{\lvar}{\tmtwo}$, then to obtain $\tmthree_2\sub{\lvar}{\tm} \dev{\lab} \tmthree_2\sub{\lvar}{\tmtwo}$, and finally to obtain $\tmthree_{3i}\sub{\lvar}{\tm} \dev{\lab} \tmthree_{3i}\sub{\lvar}{\tmtwo}$, for each $\tmthree_{3i}$ in $\lsctx=\esub{\uvar_1}{\tmthree_{31}}\ldots\esub{\uvar_n}{\tmthree_{3n}}$. This suffices to conclude that $\tmthree\sub{\lvar}{\tm} \dev{\lab} \tmthree\sub{\lvar}{\tmtwo}$.

\item $\tmthree=\sha{\tmthree_1}$. The result follows from the \ih.
  
\item $\tmthree=\open{\tmthree_1}$. The result follows from the \ih.
  
\item $\tmthree=\lopen{(\sha{\tmthree_1})\lsctx}{\labtwo}$. The result follows from the \ih, applied multiple times.

\item $\tmthree=\ofc{\tmthree_1}$. The result follows from the \ih.
  
\item $\tmthree=\tmthree_1\esub{\uvar}{(\ofc{(\sha{\tmthree_2})\lsctx_1})\lsctx_2}$ and   $\uvar\in\flv{\tmthree_1}$.  The result follows from the \ih, applied multiple times.

\item $\tmthree=\tmthree_1\lesub{\uvar}{(\ofc{\tmthree_2})\lsctx}{(\labtwo)}$ and $\uvar\notin\fv{\tmthree_1}$. The result follows from the \ih, applied multiple times.

 \item $\tmthree=\tmthree_1\esub{\uvar}{\tmthree_2}$ with $\tmthree_1\in \TermsSWL$, $\tmthree_2\in \TermsSWL$ and 
   $\uvar\notin\flv{\tmthree_1}$.  The result follows from the \ih, applied multiple times.
   
\end{xenumerate}
\end{ifLongAppendix}
\end{proof}

\begin{definition}[Multi-hole Labeled contexts]
  \[
    \begin{array}{rcl}
      \lmgctx & ::= & \ctxhole
                      \mid \lam{\lvar}{\lmgctx}
                      \mid \llam{\lvar}{\lmgctx}{\lab}
                      \mid \lmgctx\,\tm
                      \mid \tm\,\lmgctx
                      \mid \lmgctx\,\lmgctx
                      \mid \sha{\lmgctx}
                      \mid \open{\lmgctx}
                      \mid \lopen{\lmgctx}{\lab}
                      \mid \ofc{\lmgctx} \\
              & & 
                  \mid \lmgctx\esub{\uvar}{\tm}
                  \mid \lmgctx\lesub{\uvar}{\tm}{\lab}
                  \mid \lmgctx\lesub{\uvar}{\lmgctx}{\lab}
                  \mid \tm\esub{\uvar}{\lmgctx}
                  \mid \tm\lesub{\uvar}{\lmgctx}{\lab}
                  \mid \lmgctx\lesub{\uvar}{\lmgctx}{\lab}
    \end{array}
  \]
\end{definition}

\begin{lemma}
  \llem{disjoint_ls_and_multihole_contexts}
  Let $\lgctx,\lgctxtwo\in \CtxsSL$ be labeled contexts.  If $\off{\lgctx}{\ann{\uvar}{(\lab)}}=\off{\lgctxtwo}{\ann{\uvartwo}{(\labtwo)}}$, $\lgctx\neq\lgctxtwo$, and $\ann{\uvar}{(\lab)}\neq \ann{\uvartwo}{(\labtwo)}$, then there exists $\lmgctx$ such that $\off{\lgctx}{\ann{\uvar}{(\lab)}}=\off{\lgctxtwo}{\ann{\uvartwo}{(\labtwo)}}=\off{\lmgctx}{\ann{\uvar}{(\lab)},\ann{\uvartwo}{(\labtwo)}}$.
\end{lemma}

\begin{proof}
  By induction on $\lgctx$.
  \begin{xenumerate}
\item $\lgctx=\ctxhole$. Immediate from hypothesis. 
\item $\lgctx=\llam{\lvar}{\lgctx_1}{(\labthree)}$. Then  $\lgctxtwo=\llam{\lvar}{\lgctxtwo_1}{(\labthree)}$ and we conclude from the \ih.
\item $\lgctx=\lgctx_1\,\tm$. Then either  $\lgctxtwo=\lgctxtwo_1\,\tm$ or $\lgctxtwo=\off{\lgctx_1}{\ann{\uvar}{(\lab)}}\,\lgctxtwo_1$ with $\off{\lgctxtwo_1}{\ann{\uvartwo}{(\labtwo)}}=\tm$. In the former we resort to the \ih; in the latter we set $\lmgctx$ to be $\lgctx_1\, \lgctxtwo_1$.
  
\item $\lgctx=\tm\,\lgctx_1$. Similar to previous case.
\item $\lgctx=\sha{\lgctx_1}$. Then  $\lgctxtwo=\sha{\lgctxtwo_1}$ and we conclude from the \ih.
\item $\lgctx=\lopen{\lgctx_1}{(\labthree)}$. Then  $\lgctxtwo=\lopen{\lgctxtwo_1}{(\labthree)}$ and we conclude from the \ih.
\item $\lgctx=\ofc{\lgctx_1}$. Similar to previous case.
  
\item $\lgctx=\lgctx_1\lesub{\uvarthree}{\tm}{(\labthree)}$. Similar to the case for application.
\item $\lgctx=\tm\lesub{\uvarthree}{\lgctx_1}{(\labthree)}$. Similar to the case for application.
 
  \end{xenumerate}
\end{proof}

We next prove Semantic Orthogonality. Note that $\tmfour_1$ has no $\lab$-labels since the only $\lab$-label was reduced in the step $\tm\toSL{\lab}\tmtwo$. Similarly, $\tmfour_2$ has no $\labtwo$-labels since the only $\labtwo$-label was reduced in the step $\tm\toSL{\labtwo}\tmthree$. Moreover, since $\tmfour_1\flatt\tmfour_2$ implies they have the same labels (\cf \rlem{well_named_and_flat_helper}),  $\tmfour_1$ has no $\labtwo$-labels. Thus $\tmtwo \dev{\labtwo}\tmfour_1$ is a complete $\labtwo$-development. Similarly, $\tmthree \dev{\labtwo}\tmfour_2$ is a complete $\lab$-development.

\begin{proposition}[Semantic Orthogonality]
\lprop{perm}
Let $\tm$ be a well-labeled term with a unique occurrence of $\lab$ and a unique occurrence of $\labtwo$. Then,
  
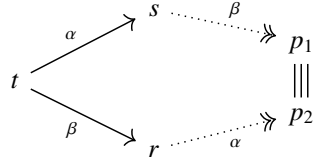
\begin{center}
    \begin{tikzcd}[column sep=4em,row sep=0em]
      & \tmtwo \arrow[rd,dotted, two heads, "\labtwo"] & \\
      & & \tmfour_1 \arrow[dd,dash]\arrow[dd, dash, shift left=1]\arrow[dd, dash, shift right=1]  \\
      \tm \arrow[ruu, "\lab"]\arrow[rdd, "\labtwo"'] & & \\
      & & \tmfour_2 \\
      & \tmthree \arrow[ru,dotted, two heads, "\lab"'] & \\
    \end{tikzcd}
  \end{center}
\end{proposition}

\begin{proof}
  Suppose $\tm=\of{\lgctx}{\tm_1}\lto{\lab}
  \of{\lgctx}{\tmtwo_1}=\tmtwo$, referred to as the $\lab$-step,
  follows from $\tm_1 \lrootto{\lab}  \tmtwo_1$ and
  $\tm=\of{\lgctxtwo}{\tm_2}\lto{\labtwo}
  \of{\lgctxtwo}{\tmthree_1}=\tmthree$, referred to as the
  $\labtwo$-step, follows from $\tm_2 \lrootto{\labtwo}
  \tmthree_1$. We proceed by induction on $\lgctx$ and, for each case,
  consider all possible forms for $\lgctxtwo$.
\begin{ifShortAppendix}
  See the extended
  version~\cite{mells_long} for further details.
  \end{ifShortAppendix}
\begin{ifLongAppendix}          
  \begin{xenumerate}
  \item  $\lgctx=\ctxhole$. Then $\tm=\tm_1$ and $\tmtwo=\tmtwo_1$.
    \begin{xenumerate}
    \item\label{perm:db} The $\tm\lto{\lab}\tmtwo$ step follows from $\tm=
      (\llam{\lvar}{\tm_{11}}{\lab})\lsctx\,\tm_{12}
      \rtoSdbL{\lab} 
      \tm_{11}\sub{\lvar}{\tm_{12}}\lsctx$, with $\fv{\tm_{12}} \cap \dom{\lsctx} = \emptyset$. We next consider all possible forms for $\lgctxtwo$.

      \begin{xenumerate}
      \item  $\lgctxtwo=\ctxhole$. Then $\lab=\labtwo$ and we conclude immediately.

      \item $\lgctxtwo= (\llam{\lvar}{\lgctxtwo_1}{\lab})\lsctx\,\tm_{12}$.  We define $\lgctxtwo_1^\star$ to be $\lgctxtwo_1\sub{\lvar}{\tm_{12}}$ and similarly for $\tm_2^\star$. Note that $\of{\lgctxtwo_1}{\tm_2}^\star=\of{\lgctxtwo_1^\star}{\tm_2^\star}$.

        \begin{center}
          \begin{tikzcd}
            & \of{\lgctxtwo_1^\star}{\tm_2^\star}\lsctx \arrow[rd,dotted, "\labtwo\ (\rlem{compatibility_of_labeled_reduction_with_substitution}(1))"] & \\[-15pt] 
            (\llam{\lvar}{\of{\lgctxtwo_1}{\tm_2}}{\lab})\lsctx\,\tm_{12} \arrow[ru, "\lab"]\arrow[rd, "\labtwo"'] & & \of{\lgctxtwo_1^\star}{\tmthree_1^\star}\lsctx  \\[-15pt] 
            & (\llam{\lvar}{\of{\lgctxtwo_1}{\tmthree_1}}{\lab})\lsctx\,\tm_{12} \arrow[ru,dotted, "\lab"']
          \end{tikzcd}
        \end{center}

      \item\label{perm:db_vs_ls} $\lgctxtwo= \ctxhole\lsctx_2\,\tm_{12}$.
        \begin{xenumerate}
          
        \item The $\tm=\of{\lgctxtwo}{\tm_2}\lto{\labtwo} \of{\lgctxtwo}{\tmthree_1}=\tmthree$ step follows from $\tm_2=\off{\lgctxthree}{\ann{\uvar}{\labtwo}}\esub{\uvar}{(\ofc{(\sha{\tm_{21}})\lsctxtwo_1})\lsctxtwo_2}  \rtoSlsL{\labtwo} \off{\lgctxthree}{(\sha{\tm_{21}})\lsctxtwo_1}\esub{\uvar}{\ofc{(\sha{\tm_{21}})\lsctxtwo_1}}\lsctxtwo_2 = \tmthree_1$. We consider each possible form for $\lgctxthree$.
          \begin{xenumerate}
          \item  $\lgctxthree= (\llam{\lvar}{\lgctxthree_1}{\lab})\lsctx_1$ and $\lsctx=\lsctx_1 \esub{\uvar}{(\ofc{(\sha{\tm_{21}})\lsctxtwo_1})\lsctxtwo_2}\lsctx_2$ and $\lvar\notin\fv{(\ofc{(\sha{\tm_{21}})\lsctxtwo_1})}$.

            \begin{center}
              \begin{tikzcd}[column sep=-5em,row sep=4em]
                &   \off{\lgctxthree_1^\star}{\ann{\uvar}{\labtwo}}\lsctx_1\esub{\uvar}{(\ofc{(\sha{\tm_{21}}) \lsctxtwo_1})\lsctxtwo_2} \lsctx_2  \arrow[rd,dotted, "\labtwo"] & \\
                (\llam{\lvar}{\off{\lgctxthree_1}{\ann{\uvar}{\labtwo}}}{\lab})\lsctx_1\esub{\uvar}{(\ofc{(\sha{\tm_{21}}) \lsctxtwo_1})\lsctxtwo_2} \lsctx_2\,\tm_{12} 
                \arrow[ru, "\lab"]\arrow[rd, "\labtwo"'] & &  \off{\lgctxthree_1^\star}{(\sha{\tm_{21}}) \lsctxtwo_1}\lsctx_1\esub{\uvar}{(\ofc{(\sha{\tm_{21}}) \lsctxtwo_1})}\lsctxtwo_2 \lsctx_2  \\
                &   (\llam{\lvar}{\off{\lgctxthree_1}{(\sha{\tm_{21}}) \lsctxtwo_1}}{\lab})\lsctx_1\esub{\uvar}{\ofc{(\sha{\tm_{21}}) \lsctxtwo_1}}\lsctxtwo_2 \lsctx_2\,\tm_{12}  \arrow[ru,dotted, "\lab"']
              \end{tikzcd}
            \end{center}

          \item $\lgctxthree= (\llam{\lvar}{\tm_{11}}{\lab})\lsctx_{11}\lesub{\uvartwo}{\lgctxthree_1}{(\labthree)}\lsctx_{12}$ and $\lsctx=\lsctx_{11}\lesub{\uvartwo}{\lgctxthree_1}{(\labthree)}\lsctx_{12} \esub{\uvar}{(\ofc{(\sha{\tm_{21}})\lsctxtwo_1})\lsctxtwo_2}\lsctx_2$.

            \begin{center}
              \begin{tikzcd}[column sep=-12em,row sep=4em]
                &    \tm_{11}^\star\lsctx_{11}\lesub{\uvartwo}{\off{\lgctxthree_1}{\ann{\uvar}{\labtwo}}}{(\labthree)}\lsctx_{12}\esub{\uvar}{(\ofc{(\sha{\tm_{21}}) \lsctxtwo_1})\lsctxtwo_2}\lsctx_2  \arrow[rd,dotted, "\labtwo"] & \\
                
                (\llam{\lvar}{\tm_{11}}{\lab})\lsctx_{11}\lesub{\uvartwo}{\off{\lgctxthree_1}{\ann{\uvar}{\labtwo}}}{(\labthree)}\lsctx_{12}\esub{\uvar}{(\ofc{(\sha{\tm_{21}}) \lsctxtwo_1})\lsctxtwo_2} \lsctx_2\,\tm_{12} 
                \arrow[ru, "\lab"]\arrow[rd, "\labtwo"'] & &  \tm_{11}^\star\lsctx_{11}\lesub{\uvartwo}{\off{\lgctxthree_1}{(\sha{\tm_{21}}) \lsctxtwo_1}}{(\labthree)}\lsctx_{12}\esub{\uvar}{\ofc{(\sha{\tm_{21}}) \lsctxtwo_1}}\lsctxtwo_2\lsctx_2    \\
                
                &  (\llam{\lvar}{\tm_{11}}{\lab})\lsctx_{11}\lesub{\uvartwo}{\off{\lgctxthree_1}{(\sha{\tm_{21}}) \lsctxtwo_1}}{(\labthree)}\lsctx_{12}\esub{\uvar}{\ofc{(\sha{\tm_{21}}) \lsctxtwo_1}}\lsctxtwo_2 \lsctx_2\,\tm_{12}  \arrow[ru,dotted, "\lab"']
              \end{tikzcd}
            \end{center}

          \end{xenumerate}

        \item The $\tm=\of{\lgctxtwo}{\tm_2}\lto{\labtwo} \of{\lgctxtwo}{\tmthree_1}=\tmthree$ step follows from $\tm_2=(\llam{\lvar}{\tm_{11}}{\lab})\lsctx_1\lesub{\uvar}{(\ofc{\tm_{21}})\lsctxtwo}{\labtwo} \rtoSgcL{\labtwo}  (\llam{\lvar}{\tm_{11}}{\lab})\lsctx_1\lsctxtwo = \tmthree_1$ and $\lsctx=\lsctx_1\lesub{\uvar}{(\ofc{\tm_{21}})\lsctxtwo}{\labtwo}\lsctx_2$. Note that $\uvar\notin\fv{\tm_{12}}$.

          \begin{center}
            \begin{tikzcd}[column sep=0em,row sep=4em]
              &     \tm_{11}^\star\lsctx_1\lesub{\uvar}{(\ofc{\tm_{21}}) \lsctxtwo}{\beta} \lsctx_2 \arrow[rd,dotted, "\labtwo"] & \\
              
              (\llam{\lvar}{\tm_{11}}{\lab})\lsctx_1\lesub{\uvar}{(\ofc{\tm_{21}}) \lsctxtwo}{\beta} \lsctx_2\,\tm_{12} 
              \arrow[ru, "\lab"]\arrow[rd, "\labtwo"'] & &  \tm_{11}^\star\lsctx_1\lsctxtwo\lsctx_2 \\
              
              &   (\llam{\lvar}{\tm_{11}}{\lab})\lsctx_1\lsctxtwo \lsctx_2\,\tm_{12}  \arrow[ru,dotted, "\lab"']
            \end{tikzcd}
          \end{center}
          
        \end{xenumerate}

      \item $\lgctxtwo= (\llam{\lvar}{\tm_{11}}{\lab})\lsctx_1\lesub{\uvartwo}{\lgctxtwo_1}{(\labthree)}\lsctx_2\,\tm_{12}$

         \begin{center}
            \begin{tikzcd}[column sep=0em,row sep=4em]
              &    \tm_{11}^\star\lsctx_1\lesub{\uvartwo}{\of{\lgctxtwo_1}{\tm_2}}{(\labthree)}\lsctx_2 \arrow[rd,dotted, "\labtwo"] & \\
              
              (\llam{\lvar}{\tm_{11}}{\lab})\lsctx_1\lesub{\uvartwo}{\of{\lgctxtwo_1}{\tm_2}}{(\labthree)}\lsctx_2\,\tm_{12} 
              \arrow[ru, "\lab"]\arrow[rd, "\labtwo"'] & & \tm_{11}^\star\lsctx_1\lesub{\uvartwo}{\of{\lgctxtwo_1}{\tmthree_1}}{(\labthree)}\lsctx_2\\
              
              &   (\llam{\lvar}{\tm_{11}}{\lab})\lsctx_1\lesub{\uvartwo}{\of{\lgctxtwo_1}{\tmthree_1}}{(\labthree)}\lsctx_2\,\tm_{12}  \arrow[ru,dotted, "\lab"']
            \end{tikzcd}
          \end{center}
          
      \item $\lgctxtwo= (\llam{\lvar}{\tm_{11}}{\lab})\lsctx\,\lgctxtwo_1$ and $\of{\lgctxtwo_1}{\tm_2}=\tm_{12}$.

         \begin{center}
            \begin{tikzcd}[column sep=0em,row sep=4em]
              &    \tm_{11}\sub{\lvar}{\of{\lgctxtwo_1}{\tm_2}}\lsctx \arrow[rd,dotted, two heads, "\labtwo\ (\rlem{compatibility_of_labeled_reduction_with_substitution})(2)"] & \\
              
            (\llam{\lvar}{\tm_{11}}{\lab})\lsctx\, \of{\lgctxtwo_1}{\tm_2}
              \arrow[ru, "\lab"]\arrow[rd, "\labtwo"'] & & \tm_{11}\sub{\lvar}{\of{\lgctxtwo_1}{\tmthree_1}}\lsctx\\
              
              &     (\llam{\lvar}{\tm_{11}}{\lab})\lsctx\, \of{\lgctxtwo_1}{\tmthree_1} \arrow[ru,dotted, "\lab"']
            \end{tikzcd}
          \end{center}
          
      \end{xenumerate}

    \item\label{perm:open} The $\tm\lto{\lab}\tmtwo$ step follows from $\tm= \lopen{(\sha{\tm_{11}})\lsctx}{\lab}
      \rtoSopenL{\lab} 
      \tm_{11}\lsctx$

      \begin{xenumerate}
      \item  $\lgctxtwo=\ctxhole$.  Then $\lab=\labtwo$ and we conclude immediately.

      \item $\lgctxtwo=\lopen{(\sha{\lgctxtwo_1}) \lsctx}{\lab}$ and $\of{\lgctxtwo_1}{\tm_2}=\tm_{11}$.

              \begin{center}
            \begin{tikzcd}[column sep=0em,row sep=2em]
              &    \of{\lgctxtwo_1}{\tm_2} \lsctx \arrow[rd,dotted, "\labtwo"] & \\
              
           \lopen{(\sha{\of{\lgctxtwo_1}{\tm_2}}) \lsctx}{\lab}
              \arrow[ru, "\lab"]\arrow[rd, "\labtwo"'] & &  \of{\lgctxtwo_1}{\tmthree_1} \lsctx \\
              
              &    \lopen{(\sha{\of{\lgctxtwo_1}{\tmthree_1}}) \lsctx}{\lab} \arrow[ru,dotted, "\lab"']
            \end{tikzcd}
          \end{center}

      \item $\lgctxtwo=\lopen{\ctxhole\lsctx_2}{\lab}$.  
        \begin{xenumerate}
          
        \item The $\tm=\of{\lgctxtwo}{\tm_2}\lto{\labtwo} \of{\lgctxtwo}{\tmthree_1}=\tmthree$ step follows from $\tm_2=\off{\lgctxthree}{\ann{\uvar}{\labtwo}}\esub{\uvar}{(\ofc{(\sha{\tm_{21}})\lsctxtwo_1})\lsctxtwo_2} \rtoSlsL{\labtwo} \off{\lgctxthree}{(\sha{\tm_{21}})\lsctxtwo_1}\esub{\uvar}{\ofc{(\sha{\tm_{21}})\lsctxtwo_1}}\lsctxtwo_2 = \tmthree_1$. We consider each possible form for $\lgctxthree$.
          \begin{xenumerate}
          \item  $\lgctxthree= (\sha{\lgctxthree_1}) \lsctx_1$ and $\lsctx=\lsctx_1 \esub{\uvar}{(\ofc{(\sha{\tm_{21}})\lsctxtwo_1})\lsctxtwo_2}\lsctx_2$ and $\lvar\notin\fv{(\ofc{(\sha{\tm_{21}})\lsctxtwo_1})}$.

            \begin{center}
              \begin{tikzcd}[column sep=-7em,row sep=4em]
                &      
                \off{\lgctxthree_1}{\ann{\uvar}{\labtwo}}\lsctx_1\esub{\uvar}{(\ofc{(\sha{\tm_{21}}) \lsctxtwo_1})\lsctxtwo_2} \lsctx_2 \arrow[rd,dotted, "\labtwo"] & \\
                
                \lopen{(\sha{\off{\lgctxthree_1}{\ann{\uvar}{\labtwo}}})\lsctx_1\esub{\uvar}{(\ofc{(\sha{\tm_{21}}) \lsctxtwo_1})\lsctxtwo_2} \lsctx_2}{\lab}
                \arrow[ru, "\lab"]\arrow[rd, "\labtwo"'] & &   \off{\lgctxthree_1}{(\sha{\tm_{21}}) \lsctxtwo_1}\lsctx_1\esub{\uvar}{\ofc{(\sha{\tm_{21}}) \lsctxtwo_1}}\lsctxtwo_2 \lsctx_2 \\
                
                &    \lopen{(\sha{\off{\lgctxthree_1}{(\sha{\tm_{21}}) \lsctxtwo_1}})\lsctx_1\esub{\uvar}{\ofc{(\sha{\tm_{21}}) \lsctxtwo_1}}\lsctxtwo_2 \lsctx_2}{\lab} \arrow[ru,dotted, "\lab"']
              \end{tikzcd}
            \end{center}

          \item $\lgctxthree= (\sha{\tm_{11}}) \lsctx_{11}\lesub{\uvartwo}{\lgctxthree_1}{(\labthree)}\lsctx_{12}$ and $\lsctx=\lsctx_{11}\lesub{\uvartwo}{\lgctxthree_1}{(\labthree)}\lsctx_{12} \esub{\uvar}{(\ofc{(\sha{\tm_{21}})\lsctxtwo_1})\lsctxtwo_2}\lsctx_2$.

            \begin{center}
              \begin{tikzcd}[column sep=-12em,row sep=4em]
                &    \tm_{11}^\star\lsctx_{11}\lesub{\uvartwo}{\off{\lgctxthree_1}{\ann{\uvar}{\lab}}}{(\labthree)}\lsctx_{12}\esub{\uvar}{(\ofc{(\sha{\tm_{21}}) \lsctxtwo_1})\lsctxtwo_2}\lsctx_2  \arrow[rd,dotted, "\labtwo"] & \\
                
                \lopen{(\sha{\tm_{11}}) \lsctx_{11}\lesub{\uvartwo}{\lgctxthree_1}{(\labthree)}\lsctx_{12} \esub{\uvar}{(\ofc{(\sha{\tm_{21}}) \lsctxtwo_1})\lsctxtwo_2} \lsctx_2}{\lab} 
                \arrow[ru, "\lab"]\arrow[rd, "\labtwo"'] & &  \tm_{11}^\star\lsctx_{11}\lesub{\uvartwo}{\off{\lgctxthree_1}{(\sha{\tm_{21}}) \lsctxtwo_1}}{(\labthree)}\lsctx_{12}\esub{\uvar}{\ofc{(\sha{\tm_{21}}) \lsctxtwo_1}}\lsctxtwo_2\lsctx_2    \\
                
                &  (\llam{\lvar}{\tm_{11}}{\lab})\lsctx_{11}\lesub{\uvartwo}{\off{\lgctxthree_1}{(\sha{\tm_{21}}) \lsctxtwo_1}}{(\labthree)}\lsctx_{12}\esub{\uvar}{\ofc{(\sha{\tm_{21}}) \lsctxtwo_1}}\lsctxtwo_2 \lsctx_2\,\tm_{12}  \arrow[ru,dotted, "\lab"']
              \end{tikzcd}
            \end{center}

          \end{xenumerate}

        \item The $\tm=\of{\lgctxtwo}{\tm_2}\lto{\labtwo} \of{\lgctxtwo}{\tmthree_1}=\tmthree$ step follows from $\tm_2=(\sha{\tm_{11}})\lsctx_1\lesub{\uvar}{(\ofc{\tm_{21}})\lsctxtwo}{\labtwo} \rtoSgcL{\labtwo} (\sha{\tm_{11}})\lsctx_1\lsctxtwo = \tmthree_1$ and $\lsctx=\lsctx_1\lesub{\uvar}{(\ofc{\tm_{21}})\lsctxtwo}{\labtwo}\lsctx_2$

          \begin{center}
            \begin{tikzcd}[column sep=0em,row sep=4em]
              &       \tm_{11}\lsctx_1\lesub{\uvar}{(\ofc{\tm_{21}})\lsctxtwo}{\labtwo}\lsctx_2 \arrow[rd,dotted, "\labtwo"] & \\
              
              \lopen{(\sha{\tm_{11}})\lsctx_1\lesub{\uvar}{(\ofc{\tm_{21}})\lsctxtwo}{\labtwo}\lsctx_2}{\lab}
              \arrow[ru, "\lab"]\arrow[rd, "\labtwo"'] & &  \tm_{11}\lsctx_1\lsctxtwo\lsctx_2 \\
              
              &    \lopen{(\sha{\tm_{11}})\lsctx_1\lsctxtwo\lsctx_2}{\lab}  \arrow[ru,dotted, "\lab"']
            \end{tikzcd}
          \end{center}
          
        \end{xenumerate}

      \item $\lgctxtwo=\lopen{(\sha{\tm_{11}}) \lsctx_1\lesub{\uvartwo}{\lgctxtwo_1}{(\labthree)}\lsctx_2}{\lab}$. 
        \begin{center}
          \begin{tikzcd}[column sep=0em,row sep=4em]
            &        \tm_{11} \lsctx_1\lesub{\uvartwo}{\of{\lgctxtwo_1}{\tm_2}}{(\labthree)}\lsctx_2\arrow[rd,dotted, "\labtwo"] & \\
            
            \lopen{(\sha{\tm_{11}}) \lsctx_1\lesub{\uvartwo}{\of{\lgctxtwo_1}{\tm_2}}{(\labthree)}\lsctx_2}{\lab}
            \arrow[ru, "\lab"]\arrow[rd, "\labtwo"'] & &  \tm_{11} \lsctx_1\lesub{\uvartwo}{\of{\lgctxtwo_1}{\tmthree_1}}{(\labthree)}\lsctx_2\lsctx_2 \\
            
            &    \lopen{(\sha{\tm_{11}}) \lsctx_1\lesub{\uvartwo}{\of{\lgctxtwo_1}{\tmthree_1}}{(\labthree)}\lsctx_2}{\lab}  \arrow[ru,dotted, "\lab"']
          \end{tikzcd}
        \end{center}

      \end{xenumerate}

      \item\label{perm:ls} The $\tm\lto{\lab}\tmtwo$ step follows from $\tm=
      \off{\lgctx_1}{\ann{\uvar}{\lab}}\esub{\uvar}{(\ofc{(\sha{\tm_{11}})\lsctx_1})\lsctx_2}
      \rtoSlsL{\lab} 
      \off{\lgctx_1}{(\sha{\tm_{11}}) \lsctx_1}\esub{\uvar}{\ofc{(\sha{\tm}) \lsctx_1}}\lsctx_2$ with $\uvar \notin \fv{\tm_{11}}$ and $\fv{\lgctx} \cap \dom{\lsctx_1\lsctx_2} = \emptyset$.

      \begin{xenumerate}
      \item  $\lgctxtwo=\ctxhole$. The step $\tm=\of{\lgctxtwo}{\tm_2}\lto{\labtwo} \of{\lgctxtwo}{\tmthree_1}=\tmthree$ is $\tm_2=\off{\lgctxthree}{\ann{\uvar}{\labtwo}}\esub{\uvar}{(\ofc{(\sha{\tm_{11}})\lsctx_1})\lsctx_2} \rtoSlsL{\labtwo} \off{\lgctxthree}{(\sha{\tm_{11}})\lsctx_1}\esub{\uvar}{\ofc{(\sha{\tm_{11}})\lsctx_1}}\lsctx_2 = \tmthree_1$. We consider each possible form for $\lgctxthree$.

        \begin{xenumerate}
        \item  $\lgctxthree=\lgctx_1$. Then $\lab=\labtwo$ and the result is immediate.
          
        \item  $\lgctxthree\neq \lgctx_1$. Then by \rlem{disjoint_ls_and_multihole_contexts} there exists a multi-hole context $\lmgctx$ such that $\off{\lgctx_1}{\ann{\uvar}{\lab}}=\off{\lgctxthree}{\ann{\uvar}{\labtwo}}=\off{\lmgctx}{\ann{\uvar}{\lab}, \ann{\uvar}{\labtwo}}$.

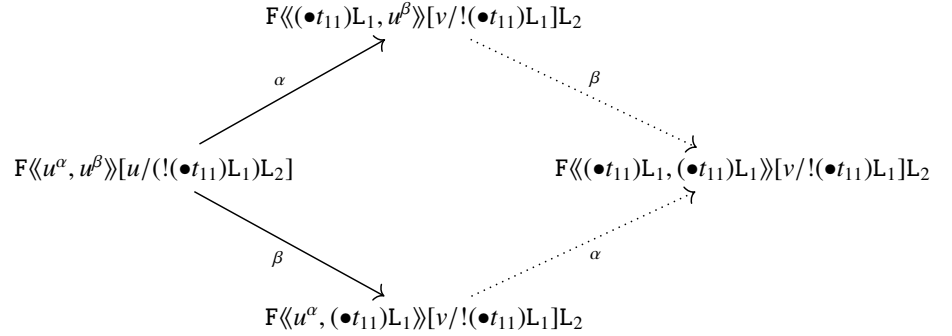
          \begin{center}
            \begin{tikzcd}[column sep=-2em,row sep=4em]
              &        \off{\lmgctx}{(\sha{\tm_{11}})\lsctx_1,\ann{\uvar}{\labtwo}}\esub{\uvartwo}{\ofc{(\sha{\tm_{11}})\lsctx_1}}\lsctx_2 \arrow[rd,dotted, "\labtwo"] & \\
              
              \off{\lmgctx}{\ann{\uvar}{\lab}, \ann{\uvar}{\labtwo}}\esub{\uvar}{(\ofc{(\sha{\tm_{11}})\lsctx_1})\lsctx_2}
              \arrow[ru, "\lab"]\arrow[rd, "\labtwo"'] & &  \off{\lmgctx}{(\sha{\tm_{11}})\lsctx_1 , (\sha{\tm_{11}})\lsctx_1}\esub{\uvartwo}{\ofc{(\sha{\tm_{11}})\lsctx_1}}\lsctx_2 \\
              
              &     \off{\lmgctx}{\ann{\uvar}{\lab} , (\sha{\tm_{11}})\lsctx_1}\esub{\uvartwo}{\ofc{(\sha{\tm_{11}})\lsctx_1}}\lsctx_2 \arrow[ru,dotted, "\lab"']
            \end{tikzcd}
          \end{center}

        \end{xenumerate}

      \item $\lgctxtwo=\lgctxtwo_1 \esub{\uvar}{(\ofc{(\sha{\tm_{11}})\lsctx_1})\lsctx_2}$ and $\of{\lgctxtwo_1}{\tm_2}=\off{\lgctx_1}{\ann{\uvar}{\lab}}$. We consider two further cases.

        \begin{xenumerate}

         \item $\ann{\uvar}{\lab} \in \lgctxtwo_1 $. Consider the context $\lgctxtwo_{11}$ obtained from placing a hole in the unique occurrence of $\ann{\uvar}{\lab}$ in  $\of{\lgctxtwo_1}{\uvartwo}$, for some fresh variable $\uvartwo$.  By   \rlem{disjoint_ls_and_multihole_contexts}
           there exists $\lmgctx$ such that $\off{\lgctx_1}{\ann{\uvar}{\lab}}=\off{\lgctxtwo_{11}}{\uvartwo}=\off{\lmgctx}{\ann{\uvar}{\lab},\uvartwo}$. Moreover, $\of{\lmgctx}{\ann{\uvar}{(\lab)},\tm_2}=\of{\lgctxtwo_1}{\tm_2}$.

           \begin{center}
            \begin{tikzcd}[column sep=-2em,row sep=4em]
              &      \of{\lmgctx}{(\sha{\tm_{11}})\lsctx_1,\tm_2}\esub{\uvar}{(\ofc{(\sha{\tm_{11}})\lsctx_1})}\lsctx_2 \arrow[rd,dotted, "\labtwo"] & \\
              
             \of{\lmgctx}{\ann{\uvar}{\lab},\tm_2}\esub{\uvar}{(\ofc{(\sha{\tm_{11}})\lsctx_1})\lsctx_2}
              \arrow[ru, "\lab"]\arrow[rd, "\labtwo"'] & &  \of{\lmgctx}{(\sha{\tm_{11}})\lsctx_1,\tmthree_1}\esub{\uvar}{(\ofc{(\sha{\tm_{11}})\lsctx_1})}\lsctx_2 \\
              
              &     \of{\lmgctx}{\ann{\uvar}{\lab},\tmthree_1}\esub{\uvar}{(\ofc{(\sha{\tm_{11}})\lsctx_1})\lsctx_2}\arrow[ru,dotted, "\lab"']
            \end{tikzcd}
          \end{center}

        \item\label{perm:ls_vs_all} $\ann{\uvar}{\lab} \in \tm_2$. Then $\lgctx_1=\of{\lgctxtwo_1}{\lgctxtwo_2}$ for some $\lgctxtwo_2$ such that $\off{\lgctxtwo_2}{\ann{\uvar}{\lab}}=\tm_2$.  We consider each possible step for $\tm_2 \lrootto{\labtwo} \tmthree_1$.

          \begin{xenumerate}
          \item The $\tm_2 \lrootto{\labtwo} \tmthree_1$ step follows from $\tm_2=
      (\llam{\lvartwo}{\tm_{21}}{\labtwo})\lsctxtwo\,\tm_{22}
      \rtoSdbL{\labtwo} 
      \tm_{21}\sub{\lvartwo}{\tm_{22}}\lsctxtwo=\tmthree_1$, with $\fv{\tm_{22}} \cap \dom{\lsctx} = \emptyset$. We next consider each possible location of $\ann{\uvar}{\lab}$.

      \begin{xenumerate}
      \item $\ann{\uvar}{\lab}\in \tm_{21}$. Then $\lgctxtwo_2= (\llam{\lvartwo}{\lgctxtwo_{21}}{\labtwo})\lsctxtwo\,\tm_{22}$.

        \begin{acenter}{-3cm}
            \begin{tikzcd}[column sep=-4em,row sep=4em]
              &      \of{\lgctxtwo_1}{(\llam{\lvartwo}{\off{\lgctxtwo_{21}}{(\sha{\tm_{11}})\lsctx_1}}{\labtwo})\lsctxtwo\,\tm_{22}}\esub{\uvar}{(\ofc{(\sha{\tm_{11}})\lsctx_1})}\lsctx_2\arrow[rd,dotted, "\labtwo"] & \\
              
              \of{\lgctxtwo_1}{(\llam{\lvartwo}{\off{\lgctxtwo_{21}}{\ann{\uvar}{\lab}}}{\labtwo})\lsctxtwo\,\tm_{22}}\esub{\uvar}{(\ofc{(\sha{\tm_{11}})\lsctx_1})\lsctx_2}
              \arrow[ru, "\lab"]\arrow[rd, "\labtwo"'] & &   \of{\lgctxtwo_1}{\off{\lgctxtwo_{21}^\star}{(\sha{\tm_{11}})\lsctx_1}\lsctxtwo}\esub{\uvar}{(\ofc{(\sha{\tm_{11}})\lsctx_1})}\lsctx_2\\
              
              &     \of{\lgctxtwo_1}{\off{\lgctxtwo_{21}^\star}{\ann{\uvar}{\lab}}\lsctxtwo}\esub{\uvar}{(\ofc{(\sha{\tm_{11}})\lsctx_1})\lsctx_2}\arrow[ru,dotted, "\lab"']
            \end{tikzcd}
          \end{acenter}

        \item $\ann{\uvar}{\lab}\in \lsctxtwo$. Then $\lgctxtwo_2= (\llam{\lvartwo}{\tm_{21}}{\labtwo})\lsctxtwo_1\lesub{\uvarthree}{\lgctxtwo_{21}}{(\labthree)}\lsctxtwo_2\,\tm_{22}$.

        \begin{acenter}{-3cm}
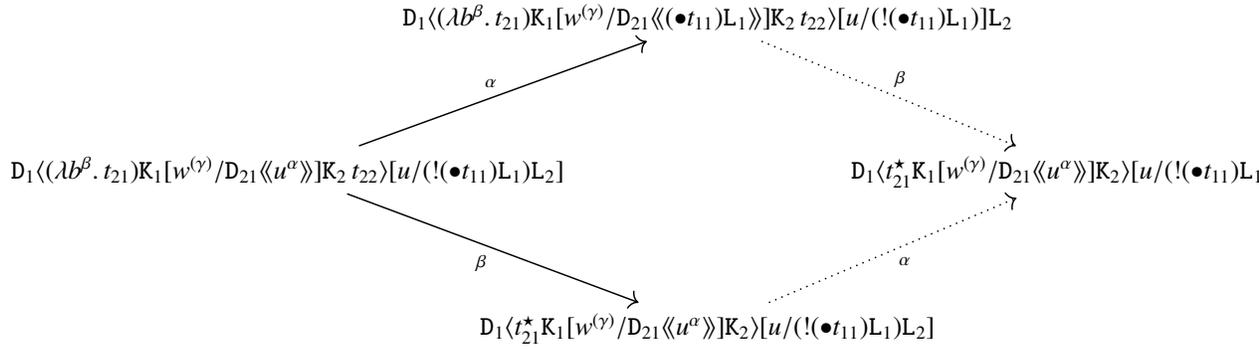
           \begin{tikzcd}[column sep=-7em,row sep=4em]
              &       \of{\lgctxtwo_1}{(\llam{\lvartwo}{\tm_{21}}{\labtwo})\lsctxtwo_1\lesub{\uvarthree}{\off{\lgctxtwo_{21}}{(\sha{\tm_{11}})\lsctx_1}}{(\labthree)}\lsctxtwo_2\,\tm_{22}}\esub{\uvar}{(\ofc{(\sha{\tm_{11}})\lsctx_1})}\lsctx_2\arrow[rd,dotted, "\labtwo"] & \\              
              \of{\lgctxtwo_1}{(\llam{\lvartwo}{\tm_{21}}{\labtwo})\lsctxtwo_1\lesub{\uvarthree}{\off{\lgctxtwo_{21}}{\ann{\uvar}{\lab}}}{(\labthree)}\lsctxtwo_2\,\tm_{22}}\esub{\uvar}{(\ofc{(\sha{\tm_{11}})\lsctx_1})\lsctx_2}
              \arrow[ru, "\lab"]\arrow[rd, "\labtwo"'] & &   \of{\lgctxtwo_1}{\tm_{21}^\star\lsctxtwo_1\lesub{\uvarthree}{\off{\lgctxtwo_{21}}{\ann{\uvar}{\lab}}}{(\labthree)}\lsctxtwo_2}\esub{\uvar}{(\ofc{(\sha{\tm_{11}})\lsctx_1})\lsctx_2}\\              
              &      \of{\lgctxtwo_1}{\tm_{21}^\star\lsctxtwo_1\lesub{\uvarthree}{\off{\lgctxtwo_{21}}{\ann{\uvar}{\lab}}}{(\labthree)}\lsctxtwo_2}\esub{\uvar}{(\ofc{(\sha{\tm_{11}})\lsctx_1})\lsctx_2}\arrow[ru,dotted, "\lab"']
            \end{tikzcd}
       \end{acenter}
        
        \item $\ann{\uvar}{\lab}\in \tm_{22}$. Then $\lgctxtwo_2= (\llam{\lvartwo}{\tm_{21}}{\labtwo})\lsctxtwo\, \lgctxtwo_{21}$.

        \begin{acenter}{-3cm}
            \begin{tikzcd}[column sep=-4em,row sep=4em]
              &     \of{\lgctxtwo_1}{(\llam{\lvartwo}{\tm_{21}}{\labtwo})\lsctxtwo\, \off{\lgctxtwo_{21}}{(\sha{\tm_{11}})\lsctx_1}}\esub{\uvar}{(\ofc{(\sha{\tm_{11}})\lsctx_1})}\lsctx_2\arrow[rd,dotted, "\labtwo"] & \\
              
              \of{\lgctxtwo_1}{(\llam{\lvartwo}{\tm_{21}}{\labtwo})\lsctxtwo\, \off{\lgctxtwo_{21}}{\ann{\uvar}{\lab}}}\esub{\uvar}{(\ofc{(\sha{\tm_{11}})\lsctx_1})\lsctx_2}
              \arrow[ru, "\lab"]\arrow[rd, "\labtwo"'] & &    \of{\lgctxtwo_1}{\tm_{21}\sub{\lvartwo}{\off{\lgctxtwo_{21}}{(\sha{\tm_{11}})\lsctx_1}}\lsctxtwo }\esub{\uvar}{(\ofc{(\sha{\tm_{11}})\lsctx_1})}\lsctx_2\\
              
              &      \of{\lgctxtwo_1}{\tm_{21}\sub{\lvartwo}{\off{\lgctxtwo_{21}}{\ann{\uvar}{\lab}}}\lsctxtwo }\esub{\uvar}{(\ofc{(\sha{\tm_{11}})\lsctx_1})\lsctx_2}\arrow[ru,dotted, two heads, "\lab"']
            \end{tikzcd}
          \end{acenter}
      \end{xenumerate}

      \item The $\tm_2 \lrootto{\labtwo} \tmthree_1$ step follows from $\tm_2= \lopen{(\sha{\tm_{21}})\lsctxtwo}{\labtwo}
      \rtoSopenL{\labtwo} 
      \tm_{21}\lsctxtwo=\tmthree_1$. We next consider each possible location of $\ann{\uvar}{\lab}$.

      \begin{xenumerate}
      \item $\ann{\uvar}{\lab}\in \tm_{21}$. Then $\lgctxtwo_2=\lopen{(\sha{\lgctxtwo_{21}})\lsctxtwo}{\labtwo}$.

        \begin{acenter}{-3cm}
            \begin{tikzcd}[column sep=-4em,row sep=4em]
              &     \of{\lgctxtwo_1}{\lopen{(\sha{\off{\lgctxtwo_{21}}{(\sha{\tm_{11}})\lsctx_1}})\lsctxtwo}{\labtwo}}\esub{\uvar}{(\ofc{(\sha{\tm_{11}})\lsctx_1})}\lsctx_2\arrow[rd,dotted, "\labtwo"] & \\
              
              \of{\lgctxtwo_1}{\lopen{(\sha{\off{\lgctxtwo_{21}}{\ann{\uvar}{\lab}}})\lsctxtwo}{\labtwo}}\esub{\uvar}{(\ofc{(\sha{\tm_{11}})\lsctx_1})\lsctx_2}
              \arrow[ru, "\lab"]\arrow[rd, "\labtwo"'] & &     \of{\lgctxtwo_1}{\off{\lgctxtwo_{21}}{(\sha{\tm_{11}})\lsctx_1}}\lsctxtwo\esub{\uvar}{(\ofc{(\sha{\tm_{11}})\lsctx_1})}\lsctx_2\\
              
              &        \of{\lgctxtwo_1}{\off{\lgctxtwo_{21}}{\ann{\uvar}{\lab}}}\lsctxtwo\esub{\uvar}{(\ofc{(\sha{\tm_{11}})\lsctx_1})\lsctx_2}\arrow[ru,dotted, "\lab"']
            \end{tikzcd}
          \end{acenter}
          
        \item $\ann{\uvar}{\lab}\in \lsctxtwo$. Then $\lgctxtwo_2= \lopen{(\sha{\tm_{21}})\lsctxtwo_1\lesub{\uvarthree}{\lgctxtwo_{21}}{(\labthree)}\lsctxtwo_2}{\labtwo}$.

        \begin{acenter}{-3cm}
            \begin{tikzcd}[column sep=-7em,row sep=4em]
              &       \of{\lgctxtwo_1}{\lopen{(\sha{\tm_{21}})\lsctxtwo_1\lesub{\uvarthree}{\off{\lgctxtwo_{21}}{(\sha{\tm_{11}})\lsctx_1}}{(\labthree)}\lsctxtwo_2}{\labtwo}}\esub{\uvar}{(\ofc{(\sha{\tm_{11}})\lsctx_1})}\lsctx_2\arrow[rd,dotted, "\labtwo"] & \\
              
              \of{\lgctxtwo_1}{\lopen{(\sha{\tm_{21}})\lsctxtwo_1\lesub{\uvarthree}{\off{\lgctxtwo_{21}}{\ann{\uvar}{\lab}}}{(\labthree)}\lsctxtwo_2}{\labtwo}}\esub{\uvar}{(\ofc{(\sha{\tm_{11}})\lsctx_1})\lsctx_2}
              \arrow[ru, "\lab"]\arrow[rd, "\labtwo"'] & &     \of{\lgctxtwo_1}{\tm_{21}\lsctxtwo_1\lesub{\uvarthree}{\off{\lgctxtwo_{21}}{(\sha{\tm_{11}})\lsctx_1}}{(\labthree)}\lsctxtwo_2}\esub{\uvar}{(\ofc{(\sha{\tm_{11}})\lsctx_1})}\lsctx_2\\
              
              &        \of{\lgctxtwo_1}{\tm_{21}\lsctxtwo_1\lesub{\uvarthree}{\off{\lgctxtwo_{21}}{\ann{\uvar}{\lab}}}{(\labthree)}\lsctxtwo_2}\esub{\uvar}{(\ofc{(\sha{\tm_{11}})\lsctx_1})\lsctx_2}\arrow[ru,dotted, "\lab"']
            \end{tikzcd}
          \end{acenter}
          
      \end{xenumerate}

    \item The $\tm_2 \lrootto{\labtwo} \tmthree_1$ step follows from $\tm_2=\off{\lgctxthree}{\ann{\uvartwo}{\labtwo}}\esub{\uvartwo}{(\ofc{(\sha{\tm_{21}})\lsctxtwo_1})\lsctxtwo_2} \rtoSlsL{\labtwo} \off{\lgctxthree}{(\sha{\tm_{21}})\lsctxtwo_1}\esub{\uvartwo}{\ofc{(\sha{\tm_{21}})\lsctxtwo_1}}\lsctxtwo_2 = \tmthree_1$. Recall that $\off{\lgctxtwo_2}{\ann{\uvar}{\lab}}=\tm_2=\off{\lgctxthree}{\ann{\uvartwo}{\labtwo}}\esub{\uvartwo}{(\ofc{(\sha{\tm_{21}})\lsctxtwo_1})\lsctxtwo_2}$. We consider each case for $\ann{\uvar}{\lab}\in\tm_2$.

      \begin{xenumerate}
      \item $\ann{\uvar}{\lab}\in \lgctxthree$. Then there exists $\lgctxthree_1$ such that $\off{\lgctxthree_1}{\ann{\uvar}{\lab}}=\off{\lgctxthree}{\ann{\uvartwo}{\labtwo}}$. Note that  $\lgctxthree_1\neq \lgctxthree$.  By   \rlem{disjoint_ls_and_multihole_contexts}, there exists $\lmgctx$ such that $\off{\lgctxthree_1}{\ann{\uvar}{\lab}}=\off{\lgctxthree}{\ann{\uvartwo}{\labtwo}}=\off{\lmgctx}{\ann{\uvar}{\lab},\ann{\uvartwo}{\labtwo}}$. 
        
        \begin{acenter}{-3cm}
          \begin{tikzcd}[column sep=-7em,row sep=4em]
              &       \of{\lgctxtwo_1}{\off{\lmgctx}{(\sha{\tm_{11}})\lsctx_1,\ann{\uvartwo}{\labtwo}}\esub{\uvartwo}{(\ofc{(\sha{\tm_{21}})\lsctxtwo_1})\lsctxtwo_2}}\esub{\uvar}{(\ofc{(\sha{\tm_{11}})\lsctx_1})}\lsctx_2\arrow[rd,dotted, "\labtwo"] & \\
              
              \of{\lgctxtwo_1}{\off{\lmgctx}{\ann{\uvar}{\lab}, (\sha{\tm_{21}})\lsctxtwo_1}\esub{\uvartwo}{(\ofc{(\sha{\tm_{21}})\lsctxtwo_1})}\lsctxtwo_2}\esub{\uvar}{(\ofc{(\sha{\tm_{11}})\lsctx_1})\lsctx_2}
              \arrow[ru, "\lab"]\arrow[rd, "\labtwo"'] & &     \of{\lgctxtwo_1}{\off{\lmgctx}{(\sha{\tm_{11}})\lsctx_1,\ann{\uvartwo}{\labtwo}}\esub{\uvartwo}{(\ofc{(\sha{\tm_{21}})\lsctxtwo_1})\lsctxtwo_2}}\esub{\uvar}{(\ofc{(\sha{\tm_{11}})\lsctx_1})}\lsctx_2\\
              
              &         \of{\lgctxtwo_1}{\off{\lmgctx}{\ann{\uvar}{\lab}, (\sha{\tm_{21}})\lsctxtwo_1}\esub{\uvartwo}{(\ofc{(\sha{\tm_{21}})\lsctxtwo_1})}\lsctxtwo_2}\esub{\uvar}{(\ofc{(\sha{\tm_{11}})\lsctx_1})\lsctx_2}\arrow[ru,dotted, "\lab"']
            \end{tikzcd}
          \end{acenter}

        \item $\ann{\uvar}{\lab}\in \tm_{21}$. Then  $\off{\lgctxtwo_2}{\ann{\uvar}{\lab}}=\off{\lgctxthree}{\ann{\uvartwo}{\labtwo}}\esub{\uvartwo}{(\ofc{(\sha{\off{\lgctxtwo_{21}}{\ann{\uvar}{\lab}}})\lsctxtwo_1})\lsctxtwo_2}$.

          \begin{acenter}{-3cm}
            \begin{tikzcd}[column sep=-7em,row sep=4em]
              &        \of{\lgctxtwo_1}{\off{\lgctxthree}{\ann{\uvartwo}{\labtwo}}\esub{\uvartwo}{(\ofc{(\sha{\off{\lgctxtwo_{21}}{(\sha{\tm_{11}})\lsctx_1}})\lsctxtwo_1})\lsctxtwo_2}}\esub{\uvar}{(\ofc{(\sha{\tm_{11}})\lsctx_1})}\lsctx_2\arrow[rd,dotted, "\labtwo"] & \\
              
              \of{\lgctxtwo_1}{\off{\lgctxthree}{\ann{\uvartwo}{\labtwo}}\esub{\uvartwo}{(\ofc{(\sha{\off{\lgctxtwo_{21}}{\ann{\uvar}{\lab}}})\lsctxtwo_1})\lsctxtwo_2}}\esub{\uvar}{(\ofc{(\sha{\tm_{11}})\lsctx_1})\lsctx_2}
              \arrow[ru, "\lab"]\arrow[rd, "\labtwo"'] & &      \of{\lgctxtwo_1}{\off{\lgctxthree}{(\sha{\off{\lgctxtwo_{21}}{(\sha{\tm_{11}})\lsctx_1}})\lsctxtwo_1}\esub{\uvartwo}{(\ofc{(\sha{\off{\lgctxtwo_{21}}{(\sha{\tm_{11}})\lsctx_1}})\lsctxtwo_1})\lsctxtwo_2}}\esub{\uvar}{(\ofc{(\sha{\tm_{11}})\lsctx_1})}\lsctx_2\\
              
              &          \of{\lgctxtwo_1}{\off{\lgctxthree}{(\sha{\off{\lgctxtwo_{21}}{\ann{\uvar}{\lab}}})\lsctxtwo_1}\esub{\uvartwo}{(\ofc{(\sha{\off{\lgctxtwo_{21}}{\ann{\uvar}{\lab}}})\lsctxtwo_1})}\lsctxtwo_2}\esub{\uvar}{(\ofc{(\sha{\tm_{11}})\lsctx_1})\lsctx_2}\arrow[ru,dotted, two heads, "\lab"']
            \end{tikzcd}
      \end{acenter}
  
\item $\ann{\uvar}{\lab}\in \lsctxtwo_1$. Then  $\off{\lgctxtwo_2}{\ann{\uvar}{\lab}}=\off{\lgctxthree}{\ann{\uvartwo}{\labtwo}}\esub{\uvartwo}{(\ofc{(\sha{\tm_{21}})\lsctxtwo_{11}\lesub{\uvarthree}{\off{\lgctxtwo_{21}}{\ann{\uvar}{\lab}}}{(\labthree)}\lsctxtwo_{12}}) \lsctxtwo_2}$.

  \begin{acenter}{-3cm}
    \begin{tikzcd}[column sep=-14em,row sep=4em]
              &      \of{\lgctxtwo_1}{\off{\lgctxthree}{\ann{\uvartwo}{\labtwo}}\esub{\uvartwo}{(\ofc{(\sha{\tm_{21}})\lsctxtwo_{11}\lesub{\uvarthree}{\off{\lgctxtwo_{21}}{(\sha{\tm_{11}})\lsctx_1}}{(\labthree)}\lsctxtwo_{12}}) \lsctxtwo_2}}\esub{\uvar}{(\ofc{(\sha{\tm_{11}})\lsctx_1})}\lsctx_2\arrow[rd,dotted, "\labtwo"] & \\
              
              \of{\lgctxtwo_1}{\off{\lgctxthree}{\ann{\uvartwo}{\labtwo}}\esub{\uvartwo}{(\ofc{(\sha{\tm_{21}})\lsctxtwo_{11}\lesub{\uvarthree}{\off{\lgctxtwo_{21}}{\ann{\uvar}{\lab}}}{(\labthree)}\lsctxtwo_{12}}) \lsctxtwo_2}}\esub{\uvar}{(\ofc{(\sha{\tm_{11}})\lsctx_1})\lsctx_2}
              \arrow[ru, "\lab"]\arrow[rd, "\labtwo"'] & &       \of{\lgctxtwo_1}{\off{\lgctxthree}{(\sha{\tm_{21}})\lsctxtwo_{11}\lesub{\uvarthree}{\off{\lgctxtwo_{21}}{\ann{\uvar}{\lab}}}{(\labthree)}\lsctxtwo_{12}}\esub{\uvartwo}{(\ofc{(\sha{\tm_{21}})\lsctxtwo_{11}\lesub{\uvarthree}{\off{\lgctxtwo_{21}}{\ann{\uvar}{\lab}}}{(\labthree)}\lsctxtwo_{12}}) \lsctxtwo_2}}\esub{\uvar}{(\ofc{(\sha{\tm_{11}})\lsctx_1})\lsctx_2}\\
              
              &         \of{\lgctxtwo_1}{\off{\lgctxthree}{(\sha{\tm_{21}})\lsctxtwo_{11}\lesub{\uvarthree}{\off{\lgctxtwo_{21}}{\ann{\uvar}{\lab}}}{(\labthree)}\lsctxtwo_{12}}\esub{\uvartwo}{(\ofc{(\sha{\tm_{21}})\lsctxtwo_{11}\lesub{\uvarthree}{\off{\lgctxtwo_{21}}{\ann{\uvar}{\lab}}}{(\labthree)}\lsctxtwo_{12}}) \lsctxtwo_2}}\esub{\uvar}{(\ofc{(\sha{\tm_{11}})\lsctx_1})\lsctx_2}\arrow[ru,dotted, two heads, "\lab"']
            \end{tikzcd}
      \end{acenter}
    
  \item $\ann{\uvar}{\lab}\in \lsctxtwo_2$.  Then  $\off{\lgctxtwo_2}{\ann{\uvar}{\lab}}=\off{\lgctxthree}{\ann{\uvartwo}{\labtwo}}\esub{\uvartwo}{(\ofc{(\sha{\tm_{21}})\lsctxtwo_1}) \lsctxtwo_{21}\lesub{\uvarthree}{\off{\lgctxtwo_{21}}{\ann{\uvar}{\lab}}}{(\labthree)}\lsctxtwo_{22}}$.

    \begin{acenter}{-3cm}
      \begin{tikzcd}[column sep=-14em,row sep=4em]
              &      \of{\lgctxtwo_1}{\off{\lgctxthree}{\ann{\uvartwo}{\labtwo}}\esub{\uvartwo}{(\ofc{(\sha{\tm_{21}})\lsctxtwo_1}) \lsctxtwo_{21}\lesub{\uvarthree}{\off{\lgctxtwo_{21}}{(\sha{\tm_{11}})\lsctx_1}}{(\labthree)}\lsctxtwo_{22}}}\esub{\uvar}{(\ofc{(\sha{\tm_{11}})\lsctx_1})}\lsctx_2\arrow[rd,dotted, "\labtwo"] & \\
              
              \of{\lgctxtwo_1}{\off{\lgctxthree}{\ann{\uvartwo}{\labtwo}}\esub{\uvartwo}{(\ofc{(\sha{\tm_{21}})\lsctxtwo_1}) \lsctxtwo_{21}\lesub{\uvarthree}{\off{\lgctxtwo_{21}}{\ann{\uvar}{\lab}}}{(\labthree)}\lsctxtwo_{22}}}\esub{\uvar}{(\ofc{(\sha{\tm_{11}})\lsctx_1})\lsctx_2}
              \arrow[ru, "\lab"]\arrow[rd, "\labtwo"'] & &        \of{\lgctxtwo_1}{\off{\lgctxthree}{(\sha{\tm_{21}})\lsctxtwo_1}\esub{\uvartwo}{(\ofc{(\sha{\tm_{21}})\lsctxtwo_1}) }\lsctxtwo_{21}\lesub{\uvarthree}{\off{\lgctxtwo_{21}}{(\sha{\tm_{11}})\lsctx_1}}{(\labthree)}\lsctxtwo_{22}}\esub{\uvar}{(\ofc{(\sha{\tm_{11}})\lsctx_1})}\lsctx_2\\
              
              &        \of{\lgctxtwo_1}{\off{\lgctxthree}{(\sha{\tm_{21}})\lsctxtwo_1}\esub{\uvartwo}{(\ofc{(\sha{\tm_{21}})\lsctxtwo_1}) }\lsctxtwo_{21}\lesub{\uvarthree}{\off{\lgctxtwo_{21}}{\ann{\uvar}{\lab}}}{(\labthree)}\lsctxtwo_{22}}\esub{\uvar}{(\ofc{(\sha{\tm_{11}})\lsctx_1})\lsctx_2}\arrow[ru,dotted, "\lab"']
            \end{tikzcd}
      \end{acenter}
    
              \end{xenumerate}
            \item   The $\tm_2 \lrootto{\labtwo} \tmthree_1$ step follows from  $\tm_2=\tm_{21}\lesub{\uvartwo}{(\ofc{\tm_{22}})\lsctxtwo}{\labtwo} \rtoSgcL{\labtwo} \tm_{21}\lsctxtwo = \tmthree_1$. Then  $\off{\lgctxtwo_2}{\ann{\uvar}{\lab}}=\tm_{21}\lesub{\uvartwo}{(\ofc{\tm_{22}})\lsctxtwo}{\labtwo} \rtoSgcL{\labtwo} \tm_{21}\lsctxtwo$. We consider all possible locations for $\ann{\uvar}{\lab}$.

              \begin{xenumerate}
              \item $\ann{\uvar}{\lab}\in \tm_{21}$. Then $\off{\lgctxtwo_2}{\ann{\uvar}{\lab}}=\off{\lgctxtwo_{21}}{\ann{\uvar}{\lab}}\lesub{\uvartwo}{(\ofc{\tm_{22}})\lsctxtwo}{\labtwo} $, for some $\lgctxtwo_{21}$, and $\off{\lgctxtwo_2}{\ann{\uvar}{\lab}}=\tm_{21}$.

                \begin{acenter}{-3cm}
                  \begin{tikzcd}[column sep=-5em,row sep=4em]
              &        \of{\lgctxtwo_1}{\off{\lgctxtwo_{21}}{\ann{\uvar}{\lab}}\lsctxtwo} \esub{\uvar}{(\ofc{(\sha{\tm_{11}})\lsctx_1})\lsctx_2}\arrow[rd,dotted, "\labtwo"] & \\
              
              \of{\lgctxtwo_1}{\off{\lgctxtwo_{21}}{\ann{\uvar}{\lab}}\lesub{\uvartwo}{(\ofc{\tm_{22}})\lsctxtwo}{\labtwo} }\esub{\uvar}{(\ofc{(\sha{\tm_{11}})\lsctx_1})\lsctx_2}
              \arrow[ru, "\lab"]\arrow[rd, "\labtwo"'] & &          \of{\lgctxtwo_1}{\off{\lgctxtwo_{21}}{(\sha{\tm_{11}})\lsctx_1}\lsctxtwo} \esub{\uvar}{(\ofc{(\sha{\tm_{11}})\lsctx_1})}\lsctx_2\\
              
              &         \of{\lgctxtwo_1}{\off{\lgctxtwo_{21}}{(\sha{\tm_{11}})\lsctx_1}\lesub{\uvartwo}{(\ofc{\tm_{22}})\lsctxtwo}{\labtwo} }\esub{\uvar}{(\ofc{(\sha{\tm_{11}})\lsctx_1})}\lsctx_2\arrow[ru,dotted, "\lab"']
            \end{tikzcd}
      \end{acenter}
    
              \item $\ann{\uvar}{\lab}\in \tm_{22}$. Then $\off{\lgctxtwo_2}{\ann{\uvar}{\lab}}=\tm_{21}\lesub{\uvartwo}{(\ofc{\off{\lgctxtwo_{21}}{\ann{\uvar}{\lab}}})\lsctxtwo}{\labtwo}$, for some $\lgctxtwo_{21}$, and $\off{\lgctxtwo_2}{\ann{\uvar}{\lab}}=\tm_{22}$.

                \begin{acenter}{-3cm}
                  \begin{tikzcd}[column sep=-5em,row sep=4em]
              &       \of{\lgctxtwo_1}{\tm_{21}\lesub{\uvartwo}{(\ofc{\off{\lgctxtwo_{21}}{(\sha{\tm_{11}})\lsctx_1}})\lsctxtwo}{\labtwo}}\esub{\uvar}{(\ofc{(\sha{\tm_{11}})\lsctx_1})}\lsctx_2  \arrow[rd,dotted, "\labtwo"] & \\
              
              \of{\lgctxtwo_1}{\tm_{21}\lesub{\uvartwo}{(\ofc{\off{\lgctxtwo_{21}}{\ann{\uvar}{\lab}}})\lsctxtwo}{\labtwo}}\esub{\uvar}{(\ofc{(\sha{\tm_{11}})\lsctx_1})\lsctx_2}
              \arrow[ru, "\lab"]\arrow[rd, "\labtwo"'] & &          \of{\lgctxtwo_1}{\off{\lgctxtwo_{21}}{(\sha{\tm_{11}})\lsctx_1}\lsctxtwo} \esub{\uvar}{(\ofc{(\sha{\tm_{11}})\lsctx_1})}\lsctx_2\\
              
              &          \of{\lgctxtwo_1}{\tm_{21}\lsctxtwo}\esub{\uvar}{(\ofc{(\sha{\tm_{11}})\lsctx_1})\lsctx_2}\arrow[ru,equal, "\lab"']
            \end{tikzcd}
      \end{acenter}
    
  \item $\ann{\uvar}{\lab}\in \lsctxtwo$. Then $\off{\lgctxtwo_2}{\ann{\uvar}{\lab}}=\tm_{21}\lesub{\uvartwo}{(\ofc{\tm_{22}})\lsctxtwo_1\lesub{\uvarthree}{\off{\lgctxtwo_{21}}{\ann{\uvar}{\lab}}}{(\labthree)}\lsctxtwo_2}{\labtwo} $, for some $\lgctxtwo_{21}$, and $\lsctxtwo=\lsctxtwo_1\lesub{\uvarthree}{\off{\lgctxtwo_{21}}{\ann{\uvar}{\lab}}}{(\labthree)}\lsctxtwo_2$.

    \begin{acenter}{-3cm}
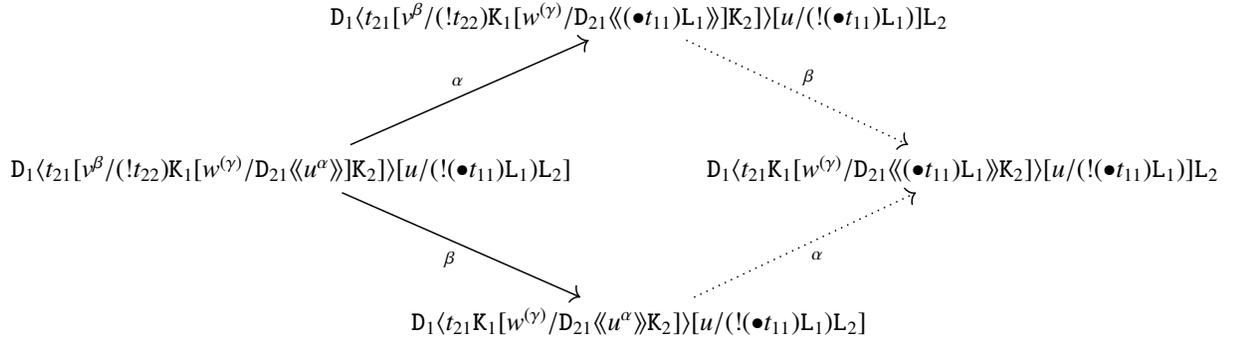
      \begin{tikzcd}[column sep=-10em,row sep=4em]
              &                     \of{\lgctxtwo_1}{\tm_{21}\lesub{\uvartwo}{(\ofc{\tm_{22}})\lsctxtwo_1\lesub{\uvarthree}{\off{\lgctxtwo_{21}}{(\sha{\tm_{11}})\lsctx_1}}{(\labthree)}\lsctxtwo_2}{\labtwo}}\esub{\uvar}{(\ofc{(\sha{\tm_{11}})\lsctx_1})}\lsctx_2
  \arrow[rd,dotted, "\labtwo"] & \\
              
              \of{\lgctxtwo_1}{\tm_{21}\lesub{\uvartwo}{(\ofc{\tm_{22}})\lsctxtwo_1\lesub{\uvarthree}{\off{\lgctxtwo_{21}}{\ann{\uvar}{\lab}}}{(\labthree)}\lsctxtwo_2}{\labtwo}}\esub{\uvar}{(\ofc{(\sha{\tm_{11}})\lsctx_1})\lsctx_2}
              \arrow[ru, "\lab"]\arrow[rd, "\labtwo"'] & &           \of{\lgctxtwo_1}{\tm_{21}\lsctxtwo_1\lesub{\uvarthree}{\off{\lgctxtwo_{21}}{(\sha{\tm_{11}})\lsctx_1}\lsctxtwo_2}{(\labthree)}}\esub{\uvar}{(\ofc{(\sha{\tm_{11}})\lsctx_1})}\lsctx_2\\
              
              &                        \of{\lgctxtwo_1}{\tm_{21}\lsctxtwo_1\lesub{\uvarthree}{\off{\lgctxtwo_{21}}{\ann{\uvar}{\lab}}\lsctxtwo_2}{(\labthree)}}\esub{\uvar}{(\ofc{(\sha{\tm_{11}})\lsctx_1})\lsctx_2}
\arrow[ru,dotted, "\lab"']
            \end{tikzcd}
      \end{acenter}

      \end{xenumerate}

             \end{xenumerate}
          \end{xenumerate}



        \item $\lgctxtwo=\off{\lgctx_1}{\ann{\uvar}{\lab}} \esub{\uvar}{(\ofc{(\sha{\lgctxtwo_1})\lsctx_1})\lsctx_2}$ and $\of{\lgctxtwo_1}{\tm_2}=\tm_{11}$.

          \begin{acenter}{-3cm}
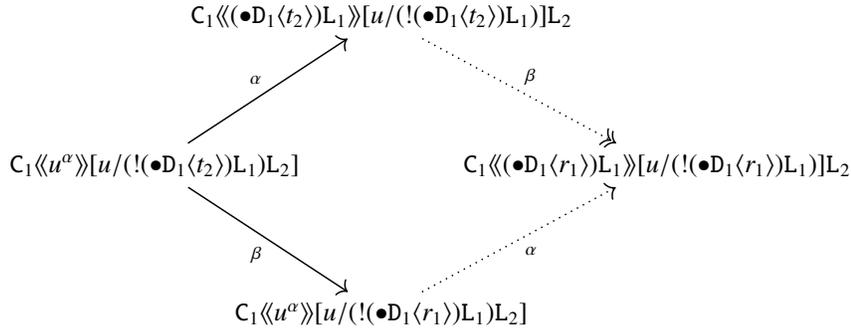
            \begin{tikzcd}[column sep=-5em,row sep=4em]
              &                     \off{\lgctx_1}{(\sha{\of{\lgctxtwo_1}{\tm_2}})\lsctx_1} \esub{\uvar}{(\ofc{(\sha{\of{\lgctxtwo_1}{\tm_2}})\lsctx_1})}\lsctx_2
  \arrow[rd,dotted, two heads, "\labtwo"] & \\
              
             \off{\lgctx_1}{\ann{\uvar}{\lab}} \esub{\uvar}{(\ofc{(\sha{\of{\lgctxtwo_1}{\tm_2}})\lsctx_1})\lsctx_2}
              \arrow[ru, "\lab"]\arrow[rd, "\labtwo"'] & &          \off{\lgctx_1}{(\sha{\of{\lgctxtwo_1}{\tmthree_1}})\lsctx_1} \esub{\uvar}{(\ofc{(\sha{\of{\lgctxtwo_1}{\tmthree_1}})\lsctx_1})}\lsctx_2\\
              
              &                         \off{\lgctx_1}{\ann{\uvar}{\lab}} \esub{\uvar}{(\ofc{(\sha{\of{\lgctxtwo_1}{\tmthree_1}})\lsctx_1})\lsctx_2}
\arrow[ru,dotted, "\lab"']
            \end{tikzcd}
      \end{acenter}
    
        \item $\lgctxtwo=\off{\lgctx_1}{\ann{\uvar}{\lab}} \esub{\uvar}{(\ofc{(\sha{\tm_{11}})\lsctx_{11}\lesub{\uvartwo}{\lgctxtwo_1}{(\labthree)}\lsctx_{12}})\lsctx_2}$ and $\lsctx_1=\lsctx_{11}\lesub{\uvartwo}{\of{\lgctxtwo_1}{\tm_2}}{(\labthree)}\lsctx_{12}$.
            \begin{acenter}{-3cm}               
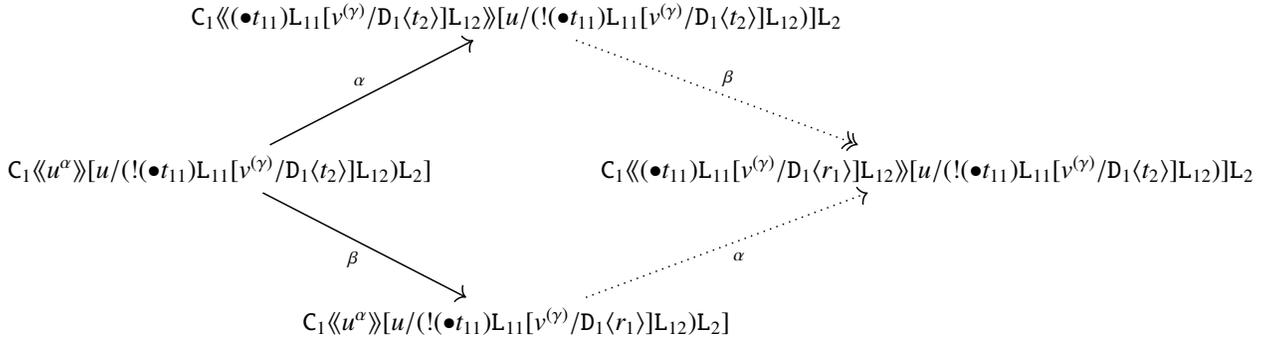
            \begin{tikzcd}[column sep=-10em,row sep=4em]
              &                       \off{\lgctx_1}{(\sha{\tm_{11}})\lsctx_{11}\lesub{\uvartwo}{\of{\lgctxtwo_1}{\tm_2}}{(\labthree)}\lsctx_{12}} \esub{\uvar}{(\ofc{(\sha{\tm_{11}})\lsctx_{11}\lesub{\uvartwo}{\of{\lgctxtwo_1}{\tm_2}}{(\labthree)}\lsctx_{12}})}\lsctx_2
  \arrow[rd,dotted, two heads, "\labtwo"] & \\
              
             \off{\lgctx_1}{\ann{\uvar}{\lab}} \esub{\uvar}{(\ofc{(\sha{\tm_{11}})\lsctx_{11}\lesub{\uvartwo}{\of{\lgctxtwo_1}{\tm_2}}{(\labthree)}\lsctx_{12}})\lsctx_2}
              \arrow[ru, "\lab"]\arrow[rd, "\labtwo"'] & &            \off{\lgctx_1}{(\sha{\tm_{11}})\lsctx_{11}\lesub{\uvartwo}{\of{\lgctxtwo_1}{\tmthree_1}}{(\labthree)}\lsctx_{12}} \esub{\uvar}{(\ofc{(\sha{\tm_{11}})\lsctx_{11}\lesub{\uvartwo}{\of{\lgctxtwo_1}{\tm_2}}{(\labthree)}\lsctx_{12}})}\lsctx_2\\
              
              &                           \off{\lgctx_1}{\ann{\uvar}{\lab}} \esub{\uvar}{(\ofc{(\sha{\tm_{11}})\lsctx_{11}\lesub{\uvartwo}{\of{\lgctxtwo_1}{\tmthree_1}}{(\labthree)}\lsctx_{12}})\lsctx_2}
\arrow[ru,dotted, "\lab"']
            \end{tikzcd}
          \end{acenter}
          
      \item $\lgctxtwo=\off{\lgctx_1}{\ann{\uvar}{\lab}} \esub{\uvar}{\ctxhole\lsctx_{12}\lsctx_2}$. The context $\sctx_1= \lsctx_{11}\esub{\uvartwo}{(\ofc{(\sha{\tm_{21}})\lsctxtwo_1})\lsctxtwo_2} \lsctx_{12}$ and the step $\tm=\of{\lgctxtwo}{\tm_2}\lto{\labtwo} \of{\lgctxtwo}{\tmthree_1}=\tmthree$ is $\tm_2=\off{\lgctxthree}{\ann{\uvartwo}{\labtwo}}\esub{\uvartwo}{(\ofc{(\sha{\tm_{21}})\lsctxtwo_1})\lsctxtwo_2} \rtoSlsL{\labtwo} \off{\lgctxthree}{(\sha{\tm_{21}})\lsctxtwo_1}\esub{\uvartwo}{\ofc{(\sha{\tm_{21}})\lsctxtwo_1}}\lsctxtwo_2 = \tmthree_1$. We consider each possible form for $\lgctxthree$.

        \begin{xenumerate}
        \item $\lgctxthree=\ofc{(\sha{\lgctxthree_1})\lsctx_{11}}$.

           \begin{acenter}{-3cm}               
            \begin{tikzcd}[column sep=-10em,row sep=4em]
              &                       \off{\lgctx_1}{\ann{\uvar}{\lab}} \esub{\uvar}{(\ofc{(\sha{\off{\lgctxthree_1}{\ann{\uvartwo}{\labtwo}})\lsctx_{11}}}\esub{\uvartwo}{(\ofc{(\sha{\tm_{21}})\lsctxtwo_1})\lsctxtwo_2}\lsctx_{12})\lsctx_2}
  \arrow[rd,dotted, two heads, "\labtwo"] & \\
              
            \off{\lgctx_1}{\ann{\uvar}{\lab}} \esub{\uvar}{(\ofc{(\sha{\off{\lgctxthree_1}{\ann{\uvartwo}{\labtwo}})\lsctx_{11}}}\esub{\uvartwo}{(\ofc{(\sha{\tm_{21}})\lsctxtwo_1})\lsctxtwo_2}\lsctx_{12})\lsctx_2}
              \arrow[ru, "\lab"]\arrow[rd, "\labtwo"'] & &            \off{\lgctx_1}{(\sha{\tm_{11}})\lsctx_{11}\lesub{\uvartwo}{\of{\lgctxtwo_1}{\tmthree_1}}{(\labthree)}\lsctx_{12}} \esub{\uvar}{(\ofc{(\sha{\tm_{11}})\lsctx_{11}\lesub{\uvartwo}{\of{\lgctxtwo_1}{\tm_2}}{(\labthree)}\lsctx_{12}})}\lsctx_2\\
              
              &                           \off{\lgctx_1}{\ann{\uvar}{\lab}} \esub{\uvar}{(\ofc{(\sha{\off{\lgctxthree_1}{\ann{\uvartwo}{\labtwo}})\lsctx_{11}}}\esub{\uvartwo}{(\ofc{(\sha{\tm_{21}})\lsctxtwo_1})\lsctxtwo_2}\lsctx_{12})\lsctx_2}
\arrow[ru,dotted, "\lab"']
            \end{tikzcd}
          \end{acenter}
  
        \item $\lgctxthree=\ofc{(\sha{\tm_{11}}) \lsctx_{111}\lesub{\uvarthree}{\lgctxthree_1}{(\labthree)}\lsctx_{112}}$ where $\lsctx_{11}=\lsctx_{111}\lesub{\uvarthree}{\off{\lgctxthree_1}{\ann{\uvartwo}{\labtwo}}}{(\labthree)}\lsctx_{112}$. We write $\lsctx_{11}^\star$ for $\lsctx_{111}\lesub{\uvarthree}{\off{\lgctxthree_1}{(\sha{\tm_{21}})\lsctxtwo_1}}{(\labthree)}\lsctx_{112}$

                 \begin{acenter}{-3cm}               
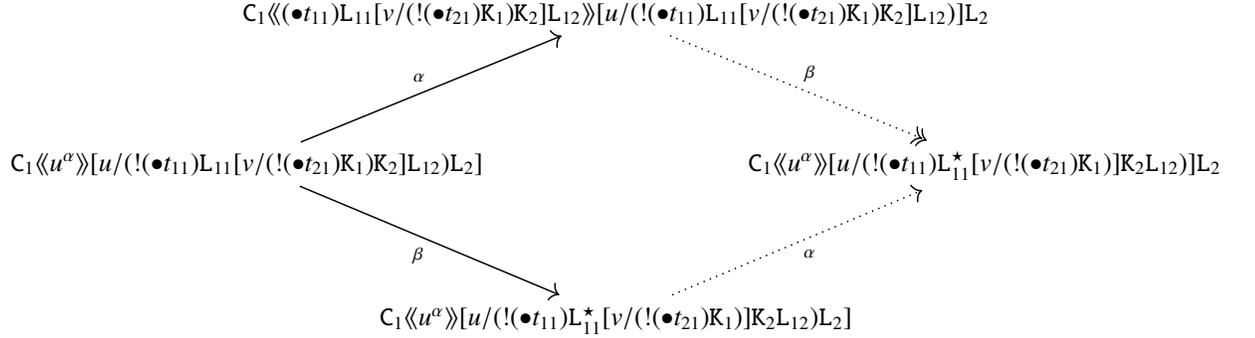
            \begin{tikzcd}[column sep=-10em,row sep=4em]
              &                      \off{\lgctx_1}{(\sha{\tm_{11}}) \lsctx_{11}\esub{\uvartwo}{(\ofc{(\sha{\tm_{21}})\lsctxtwo_1})\lsctxtwo_2}\lsctx_{12}} \esub{\uvar}{(\ofc{(\sha{\tm_{11}}) \lsctx_{11}\esub{\uvartwo}{(\ofc{(\sha{\tm_{21}})\lsctxtwo_1})\lsctxtwo_2}\lsctx_{12}})}\lsctx_2
  \arrow[rd,dotted, two heads, "\labtwo"] & \\
              
            \off{\lgctx_1}{\ann{\uvar}{\lab}} \esub{\uvar}{(\ofc{(\sha{\tm_{11}}) \lsctx_{11}\esub{\uvartwo}{(\ofc{(\sha{\tm_{21}})\lsctxtwo_1})\lsctxtwo_2}\lsctx_{12}})\lsctx_2}
              \arrow[ru, "\lab"]\arrow[rd, "\labtwo"'] & &            \off{\lgctx_1}{\ann{\uvar}{\lab}} \esub{\uvar}{(\ofc{(\sha{\tm_{11}}) \lsctx_{11}^\star\esub{\uvartwo}{(\ofc{(\sha{\tm_{21}})\lsctxtwo_1}) }\lsctxtwo_2\lsctx_{12}})}\lsctx_2\\
              
              &                          \off{\lgctx_1}{\ann{\uvar}{\lab}} \esub{\uvar}{(\ofc{(\sha{\tm_{11}}) \lsctx_{11}^\star\esub{\uvartwo}{(\ofc{(\sha{\tm_{21}})\lsctxtwo_1})}\lsctxtwo_2 \lsctx_{12}})\lsctx_2}
\arrow[ru,dotted, "\lab"']
            \end{tikzcd}
          \end{acenter}

        \end{xenumerate}
        
      \item $\lgctxtwo=\off{\lgctx_1}{\ann{\uvar}{\lab}} \esub{\uvar}{(\ofc{(\sha{\tm_{11}})\lsctx_{1}})\lsctx_{21} \lesub{\uvartwo}{\lgctxtwo_1}{(\labthree)}\lsctx_{22}}$

              \begin{acenter}{-3cm}               
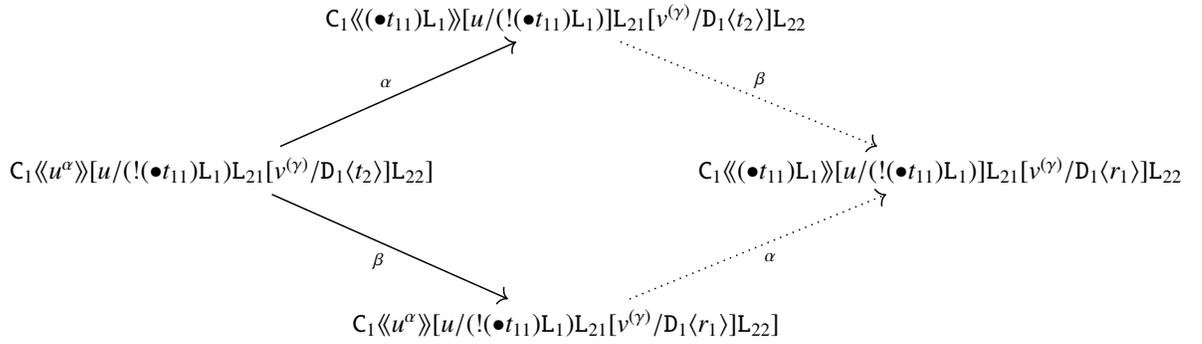
            \begin{tikzcd}[column sep=-5em,row sep=4em]
              &                      \off{\lgctx_1}{(\sha{\tm_{11}})\lsctx_{1}} \esub{\uvar}{(\ofc{(\sha{\tm_{11}})\lsctx_{1}})}\lsctx_{21} \lesub{\uvartwo}{\of{\lgctxtwo_1}{\tm_2}}{(\labthree)}\lsctx_{22}
  \arrow[rd,dotted, "\labtwo"] & \\
              
           \off{\lgctx_1}{\ann{\uvar}{\lab}} \esub{\uvar}{(\ofc{(\sha{\tm_{11}})\lsctx_{1}})\lsctx_{21} \lesub{\uvartwo}{\of{\lgctxtwo_1}{\tm_2}}{(\labthree)}\lsctx_{22}}
              \arrow[ru, "\lab"]\arrow[rd, "\labtwo"'] & &              \off{\lgctx_1}{(\sha{\tm_{11}})\lsctx_{1}} \esub{\uvar}{(\ofc{(\sha{\tm_{11}})\lsctx_{1}})}\lsctx_{21} \lesub{\uvartwo}{\of{\lgctxtwo_1}{\tmthree_1}}{(\labthree)}\lsctx_{22}\\
              
              &                         \off{\lgctx_1}{\ann{\uvar}{\lab}} \esub{\uvar}{(\ofc{(\sha{\tm_{11}})\lsctx_{1}})\lsctx_{21} \lesub{\uvartwo}{\of{\lgctxtwo_1}{\tmthree_1}}{(\labthree)}\lsctx_{22}}
\arrow[ru,dotted, "\lab"']
            \end{tikzcd}
          \end{acenter}

      \item $\lgctxtwo=\off{\lgctx_1}{\ann{\uvar}{\lab}} \esub{\uvar}{\ctxhole\lsctx_{22}}$. The step $\tm=\of{\lgctxtwo}{\tm_2}\lto{\labtwo} \of{\lgctxtwo}{\tmthree_1}=\tmthree$ is $\tm_2=\off{\lgctxthree}{\ann{\uvartwo}{\labtwo}}\esub{\uvartwo}{(\ofc{(\sha{\tm_{21}})\lsctxtwo_1})\lsctxtwo_2} \rtoSlsL{\labtwo} \off{\lgctxthree}{(\sha{\tm_{21}})\lsctxtwo_1}\esub{\uvartwo}{\ofc{(\sha{\tm_{21}})\lsctxtwo_1}}\lsctxtwo_2 = \tmthree_1$. We consider each possible form for $\lgctxthree$.
        \begin{xenumerate}
        \item $\lgctxthree=(\ofc{(\sha{\lgctxthree_1})\lsctx_{1}})\lsctx_{21}$

                  \begin{acenter}{-3cm}               
            \begin{tikzcd}[column sep=-10em,row sep=4em]
              &                     \off{\lgctx_1}{(\sha{\off{\lgctxthree_1}{\ann{\uvartwo}{\labtwo}}})\lsctx_{1}} \esub{\uvar}{(\ofc{(\sha{\off{\lgctxthree_1}{\ann{\uvartwo}{\labtwo}}})\lsctx_{1}})}\lsctx_{21}\esub{\uvartwo}{(\ofc{(\sha{\tm_{21}})\lsctxtwo_1})\lsctxtwo_2}\lsctx_{22}
  \arrow[rd,dotted, two heads, "\labtwo"] & \\
              
           \off{\lgctx_1}{\ann{\uvar}{\lab}} \esub{\uvar}{(\ofc{(\sha{\off{\lgctxthree_1}{\ann{\uvartwo}{\labtwo}}})\lsctx_{1}})\lsctx_{21}\esub{\uvartwo}{(\ofc{(\sha{\tm_{21}})\lsctxtwo_1})\lsctxtwo_2}\lsctx_{22}}
              \arrow[ru, "\lab"]\arrow[rd, "\labtwo"'] & &              \off{\lgctx_1}{(\sha{\off{\lgctxthree_1}{(\sha{\tm_{21}})\lsctxtwo_1}})\lsctx_{1}} \esub{\uvar}{(\ofc{(\sha{\off{\lgctxthree_1}{(\sha{\tm_{21}})\lsctxtwo_1}})\lsctx_{1}})}\lsctx_{21}\esub{\uvartwo}{(\ofc{(\sha{\tm_{21}})\lsctxtwo_1})\lsctxtwo_2}\lsctx_{22}\\
              
              &                        \off{\lgctx_1}{\ann{\uvar}{\lab}} \esub{\uvar}{(\ofc{(\sha{\off{\lgctxthree_1}{(\sha{\tm_{21}})\lsctxtwo_1}})\lsctx_{1}})\lsctx_{21}\esub{\uvartwo}{(\ofc{(\sha{\tm_{21}})\lsctxtwo_1})}\lsctxtwo_2\lsctx_{22}}
\arrow[ru,dotted, "\lab"']
            \end{tikzcd}
          \end{acenter}

        \item $\lgctxthree=(\ofc{(\sha{\tm_{11}}) \lsctx_{11}\lesub{\uvarthree}{\lgctxthree_1}{(\labthree)}\lsctx_{12}})\lsctx_{21}$

                  \begin{acenter}{-3cm}               
            \begin{tikzcd}[column sep=-18em,row sep=4em]
              &                               
           \off{\lgctx_1}{(\sha{\tm_{11}}) \lsctx_{11}\lesub{\uvarthree}{\off{\lgctxthree_1}{\ann{\uvartwo}{\labtwo}}}{(\labthree)}\lsctx_{12}} \esub{\uvar}{(\ofc{(\sha{\tm_{11}}) \lsctx_{11}\lesub{\uvarthree}{\off{\lgctxthree_1}{\ann{\uvartwo}{\labtwo}}}{(\labthree)}\lsctx_{12}})}\lsctx_{21}\esub{\uvartwo}{(\ofc{(\sha{\tm_{21}})\lsctxtwo_1})\lsctxtwo_2}\lsctx_{22}
  \arrow[rd,dotted, two heads, "\labtwo"] & \\
              
           \off{\lgctx_1}{\ann{\uvar}{\lab}} \esub{\uvar}{(\ofc{(\sha{\tm_{11}}) \lsctx_{11}\lesub{\uvarthree}{\off{\lgctxthree_1}{\ann{\uvartwo}{\labtwo}}}{(\labthree)}\lsctx_{12}})\lsctx_{21}\esub{\uvartwo}{(\ofc{(\sha{\tm_{21}})\lsctxtwo_1})\lsctxtwo_2}\lsctx_{22}}
              \arrow[ru, "\lab"]\arrow[rd, "\labtwo"'] & &              \off{\lgctx_1}{(\sha{\tm_{11}}) \lsctx_{11}\lesub{\uvarthree}{\off{\lgctxthree_1}{(\sha{\tm_{21}})\lsctxtwo_1}}{(\labthree)}\lsctx_{12}} \esub{\uvar}{(\ofc{(\sha{\tm_{11}}) \lsctx_{11}\lesub{\uvarthree}{\off{\lgctxthree_1}{(\sha{\tm_{21}})\lsctxtwo_1}}{(\labthree)}\lsctx_{12}})}\lsctx_{21}\esub{\uvartwo}{(\ofc{(\sha{\tm_{21}})\lsctxtwo_1})}\lsctxtwo_2\lsctx_{22}\\
              
              &                                 
           \off{\lgctx_1}{\ann{\uvar}{\lab}} \esub{\uvar}{(\ofc{(\sha{\tm_{11}}) \lsctx_{11}\lesub{\uvarthree}{\off{\lgctxthree_1}{(\sha{\tm_{21}})\lsctxtwo_1}}{(\labthree)}\lsctx_{12}})\lsctx_{21}\esub{\uvartwo}{(\ofc{(\sha{\tm_{21}})\lsctxtwo_1})}\lsctxtwo_2\lsctx_{22}}
\arrow[ru,dotted, "\lab"']
            \end{tikzcd}
          \end{acenter}

        \item $\lgctxthree=(\ofc{(\sha{\tm_{11}}) \lsctx_1}) \lsctx_{211}\lesub{\uvarthree}{\lgctxthree_1}{\labthree}\lsctx_{212}$. The $\lab$ and $\labtwo$-steps are disjoint.

                  \begin{acenter}{-3cm}               
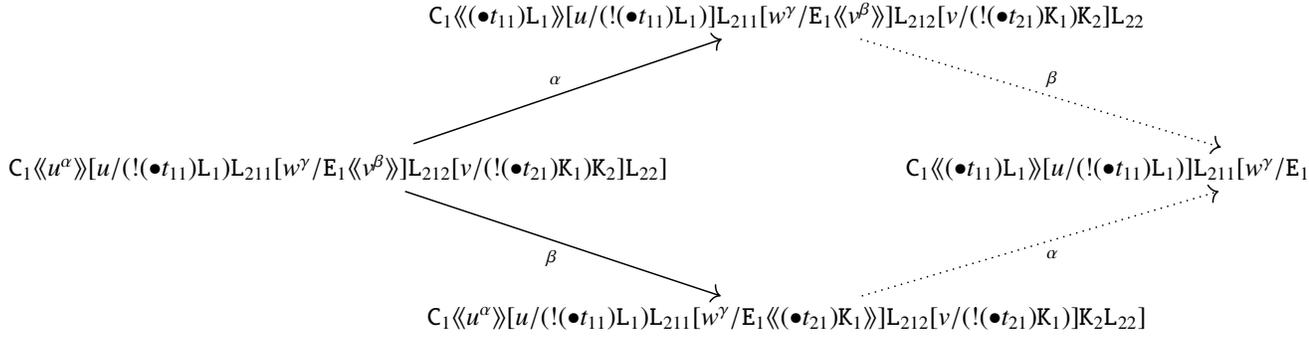
            \begin{tikzcd}[column sep=-10em,row sep=4em]
              &                          
           \off{\lgctx_1}{(\sha{\tm_{11}}) \lsctx_1} \esub{\uvar}{(\ofc{(\sha{\tm_{11}}) \lsctx_1}) }\lsctx_{211}\lesub{\uvarthree}{\off{\lgctxthree_1}{\ann{\uvartwo}{\labtwo}}}{\labthree}\lsctx_{212}\esub{\uvartwo}{(\ofc{(\sha{\tm_{21}})\lsctxtwo_1})\lsctxtwo_2}\lsctx_{22}
  \arrow[rd,dotted, "\labtwo"] & \\
              
           \off{\lgctx_1}{\ann{\uvar}{\lab}} \esub{\uvar}{(\ofc{(\sha{\tm_{11}}) \lsctx_1}) \lsctx_{211}\lesub{\uvarthree}{\off{\lgctxthree_1}{\ann{\uvartwo}{\labtwo}}}{\labthree}\lsctx_{212}\esub{\uvartwo}{(\ofc{(\sha{\tm_{21}})\lsctxtwo_1})\lsctxtwo_2}\lsctx_{22}}
              \arrow[ru, "\lab"]\arrow[rd, "\labtwo"'] & &               \off{\lgctx_1}{(\sha{\tm_{11}}) \lsctx_1} \esub{\uvar}{(\ofc{(\sha{\tm_{11}}) \lsctx_1}) }\lsctx_{211}\lesub{\uvarthree}{\off{\lgctxthree_1}{(\sha{\tm_{21}})\lsctxtwo_1}}{\labthree}\lsctx_{212}\esub{\uvartwo}{(\ofc{(\sha{\tm_{21}})\lsctxtwo_1})}\lsctxtwo_2\lsctx_{22}\\
              
              &                            
           \off{\lgctx_1}{\ann{\uvar}{\lab}} \esub{\uvar}{(\ofc{(\sha{\tm_{11}}) \lsctx_1}) \lsctx_{211}\lesub{\uvarthree}{\off{\lgctxthree_1}{(\sha{\tm_{21}})\lsctxtwo_1}}{\labthree}\lsctx_{212}\esub{\uvartwo}{(\ofc{(\sha{\tm_{21}})\lsctxtwo_1})}\lsctxtwo_2\lsctx_{22}}
\arrow[ru,dotted, "\lab"']
            \end{tikzcd}
          \end{acenter}

        \end{xenumerate}
        
      \end{xenumerate}

    \item The $\tm\lto{\lab}\tmtwo$ step follows from $\tm=\tm_{11}\esub{\ann{\uvar}{\lab}}{(\ofc{\tm_{12}})\lsctx}
      \rtoSgcL{\lab} 
      \tm_{11}\lsctx$, with $\uvar\notin\fv{\tm_{11}}$. We consider each possible form for $\lgctxtwo$:

      \begin{xenumerate}
      \item $\lgctxtwo=\ctxhole$. Then $\lab=\labtwo$ and the result holds immediately.

      \item  $\lgctxtwo=\lgctxtwo_1\esub{\ann{\uvar}{\lab}}{(\ofc{\tm_{12}})\lsctx}$

            \begin{acenter}{-3cm}               
            \begin{tikzcd}[column sep=0em,row sep=4em]
              &                      \of{\lgctxtwo_1}{\tm_2}\lsctx
  \arrow[rd,dotted, "\labtwo"] & \\
              
           \of{\lgctxtwo_1}{\tm_2}\esub{\ann{\uvar}{\lab}}{(\ofc{\tm_{12}})\lsctx}
              \arrow[ru, "\lab"]\arrow[rd, "\labtwo"'] & &              \of{\lgctxtwo_1}{\tmthree_1}\lsctx\\
              
              &                        \of{\lgctxtwo_1}{\tmthree_1}\esub{\ann{\uvar}{\lab}}{(\ofc{\tm_{12}})\lsctx} 
\arrow[ru,dotted, "\lab"']
            \end{tikzcd}
          \end{acenter}
          
      \item  $\lgctxtwo=\tm_{11}\esub{\ann{\uvar}{\lab}}{(\ofc{\lgctxtwo_1})\lsctx}$
\begin{acenter}{-3cm}               
            \begin{tikzcd}[column sep=0em,row sep=4em]
              &                      \tm_{11}\lsctx
  \arrow[rd,equals, "\labtwo"] & \\
              
          \tm_{11}\esub{\ann{\uvar}{\lab}}{(\ofc{\of{\lgctxtwo_1}{\tm_2}})\lsctx}
              \arrow[ru, "\lab"]\arrow[rd, "\labtwo"'] & &              \tm_{11}\lsctx\\
              
              &                       \tm_{11}\esub{\ann{\uvar}{\lab}}{(\ofc{\of{\lgctxtwo_1}{\tmthree_1}})\lsctx}
\arrow[ru,dotted, "\lab"']
            \end{tikzcd}
          \end{acenter}
          
      \item  $\lgctxtwo=\tm_{11}\esub{\ann{\uvar}{\lab}}{(\ofc{\tm_{12}})\lsctx_1\lesub{\uvartwo}{\gctxtwo_1}{(\labthree)}\lsctx_2}$
\begin{acenter}{-3cm}               
            \begin{tikzcd}[column sep=0em,row sep=4em]
              &                        \tm_{11}\lsctx_1\lesub{\uvartwo}{\of{\gctxtwo_1}{\tm_2}}{(\labthree)}\lsctx_2
  \arrow[rd,dotted, "\labtwo"] & \\
              
         \tm_{11}\esub{\ann{\uvar}{\lab}}{(\ofc{\tm_{12}})\lsctx_1\lesub{\uvartwo}{\of{\gctxtwo_1}{\tm_2}}{(\labthree)}\lsctx_2}
              \arrow[ru, "\lab"]\arrow[rd, "\labtwo"'] & &            \tm_{11}\lsctx_1\lesub{\uvartwo}{\of{\gctxtwo_1}{\tmthree_1}}{(\labthree)}\lsctx_2\\
              
              &                         \tm_{11}\esub{\ann{\uvar}{\lab}}{(\ofc{\tm_{12}})\lsctx_1\lesub{\uvartwo}{\of{\gctxtwo_1}{\tmthree_1}}{(\labthree)}\lsctx_2}
\arrow[ru,dotted, "\lab"']
            \end{tikzcd}
          \end{acenter}
        
      \item  $\lgctxtwo=\tm_{11}\esub{\ann{\uvar}{\lab}}{\ctxhole\lsctx_2}$ and $\lsctx= (\ofc{\tm_{12}})\lsctx_1 \esub{\uvartwo}{(\ofc{(\sha{\tm_{21}})\lsctxtwo_1})\lsctxtwo_2} \lsctx_2$. The step $\tm=\of{\lgctxtwo}{\tm_2}\lto{\labtwo} \of{\lgctxtwo}{\tmthree_1}=\tmthree$ is $\tm_2=\off{\lgctxthree}{\ann{\uvartwo}{\labtwo}}\esub{\uvartwo}{(\ofc{(\sha{\tm_{21}})\lsctxtwo_1})\lsctxtwo_2} \rtoSlsL{\labtwo} \off{\lgctxthree}{(\sha{\tm_{21}})\lsctxtwo_1}\esub{\uvartwo}{\ofc{(\sha{\tm_{21}})\lsctxtwo_1}}\lsctxtwo_2 = \tmthree_1$. We consider each possible form for $\lgctxthree$.
        \begin{xenumerate}
        \item $\lgctxthree=(\ofc{\lgctxthree_1})\lsctx_{1}$

          \begin{acenter}{-3cm}               
            \begin{tikzcd}[column sep=-2em,row sep=4em]
              &                        \tm_{11}\lsctx_{1}\esub{\uvartwo}{(\ofc{(\sha{\tm_{21}})\lsctxtwo_1})\lsctxtwo_2} \lsctx_2
  \arrow[rd,equals, "\labtwo"] & \\
              
         \tm_{11}\esub{\ann{\uvar}{\lab}}{(\ofc{\off{\lgctxthree_1}{\ann{\uvartwo}{\labtwo}}})\lsctx_{1}\esub{\uvartwo}{(\ofc{(\sha{\tm_{21}})\lsctxtwo_1})\lsctxtwo_2} \lsctx_2}
         \arrow[ru, "\lab"]\arrow[rd, "\labtwo"'] & &
         \mlnode{ \tm_{11}\lsctx_{1}\esub{\uvartwo}{(\ofc{(\sha{\tm_{21}})\lsctxtwo_1})\lsctxtwo_2} \lsctx_2
           \\
           \flatt
           \\
           \tm_{11}\lsctx_{1}\esub{\uvartwo}{(\ofc{(\sha{\tm_{21}})\lsctxtwo_1})}\lsctxtwo_2 \lsctx_2}\\
              
              &                         \tm_{11}\esub{\ann{\uvar}{\lab}}{(\ofc{\off{\lgctxthree_1}{(\sha{\tm_{21}})\lsctxtwo_1}})\lsctx_{1}\esub{\uvartwo}{(\ofc{(\sha{\tm_{21}})\lsctxtwo_1})}\lsctxtwo_2 \lsctx_2}
\arrow[ru,dotted, "\lab"']
            \end{tikzcd}
          \end{acenter}

        \item $\lgctxthree=(\ofc{\tm_{12} })\lsctx_{11}\lesub{\uvarthree}{\lgctxthree_1}{(\labthree)}\lsctx_{12}$

                   \begin{acenter}{-3cm}               
            \begin{tikzcd}[column sep=-7em,row sep=4em]
              &                       \tm_{11}\lsctx_{11}\lesub{\uvarthree}{\off{\lgctxthree_1}{\ann{\uvartwo}{\labtwo}}}{(\labthree)}\lsctx_{12}\esub{\uvartwo}{(\ofc{(\sha{\tm_{21}})\lsctxtwo_1})\lsctxtwo_2} \lsctx_2
  \arrow[rd,dotted, "\labtwo"] & \\
              
         \tm_{11}\esub{\ann{\uvar}{\lab}}{(\ofc{\tm_{12} })\lsctx_{11}\lesub{\uvarthree}{\off{\lgctxthree_1}{\ann{\uvartwo}{\labtwo}}}{(\labthree)}\lsctx_{12}\esub{\uvartwo}{(\ofc{(\sha{\tm_{21}})\lsctxtwo_1})\lsctxtwo_2} \lsctx_2}
         \arrow[ru, "\lab"]\arrow[rd, "\labtwo"'] & &
         \tm_{11}\lsctx_{11}\lesub{\uvarthree}{\off{\lgctxthree_1}{(\sha{\tm_{21}})\lsctxtwo_1}}{(\labthree)}\lsctx_{12}\esub{\uvartwo}{(\ofc{(\sha{\tm_{21}})\lsctxtwo_1})}\lsctxtwo_2 \lsctx_2\\
              
              &                         \tm_{11}\esub{\ann{\uvar}{\lab}}{(\ofc{\tm_{12} })\lsctx_{11}\lesub{\uvarthree}{\off{\lgctxthree_1}{(\sha{\tm_{21}})\lsctxtwo_1}}{(\labthree)}\lsctx_{12}\esub{\uvartwo}{(\ofc{(\sha{\tm_{21}})\lsctxtwo_1})}\lsctxtwo_2 \lsctx_2}
\arrow[ru,dotted, "\lab"']
            \end{tikzcd}
          \end{acenter}

        \end{xenumerate}

      \item  $\lgctxtwo=\tm_{11}\esub{\ann{\uvar}{\lab}}{\ctxhole\lsctx_2}$ and $\lsctx= \lsctx_1\lesub{\uvartwo}{\gctxtwo_1}{\labtwo}\lsctx_2$. The step $\tm=\of{\lgctxtwo}{\tm_2}\lto{\labtwo} \of{\lgctxtwo}{\tmthree_1}=\tmthree$ is $\tm_2=(\ofc{\tm_{12}})\lsctx_1\lesub{\uvartwo}{(\ofc{\tm_{21}})\lsctxtwo}{\labtwo} \rtoSgcL{\labtwo} (\ofc{\tm_{12}})\lsctx_1\lsctxtwo = \tmthree_1$. 

    \begin{acenter}{-3cm}               
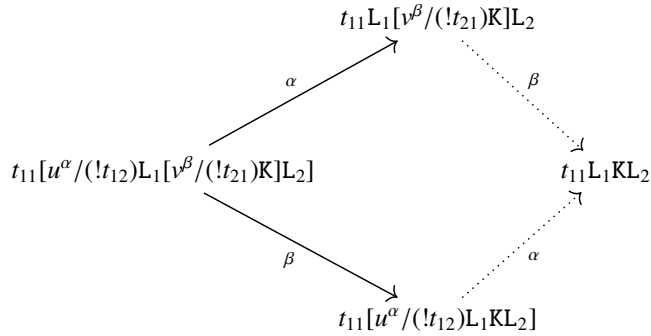
            \begin{tikzcd}[column sep=0em,row sep=4em]
              &                        \tm_{11}\lsctx_1\lesub{\uvartwo}{(\ofc{\tm_{21}})\lsctxtwo}{\labtwo} \lsctx_2
  \arrow[rd,dotted, "\labtwo"] & \\
              
         \tm_{11}\esub{\ann{\uvar}{\lab}}{(\ofc{\tm_{12}})\lsctx_1\lesub{\uvartwo}{(\ofc{\tm_{21}})\lsctxtwo}{\labtwo} \lsctx_2}
         \arrow[ru, "\lab"]\arrow[rd, "\labtwo"'] & &
           \tm_{11}\lsctx_1\lsctxtwo \lsctx_2\\
              
              &                         \tm_{11}\esub{\ann{\uvar}{\lab}}{(\ofc{\tm_{12}})\lsctx_1\lsctxtwo \lsctx_2}
\arrow[ru,dotted, "\lab"']
            \end{tikzcd}
          \end{acenter}

      \end{xenumerate}

    \end{xenumerate}

\item $\lgctx=\lam{\lvar}{\lgctx_1}$. Then both the $\lab$ and $\labtwo$-steps are inside $\of{\lgctx_1}{\tm_1}$ and we conclude from the \ih.

\item $\lgctx= (\llam{\lvar}{\lgctx_1}{\labthree})\lsctx\,\tm_{12}$. We next consider all possible forms for $\lgctxtwo$.

  \begin{xenumerate}
    \item $\lgctxtwo =\ctxhole$. We proceed as in case~\ref{perm:db}.
      \item  $\lgctxtwo=(\llam{\lvar}{\lgctxtwo_1}{\labthree})\lsctx\,\tm_{12}$. Then we conclude from the \ih. 
        
      \item  $\lgctxtwo=(\llam{\lvar}{\of{\lgctx_1}{\tm_1}}{\labthree})\lsctx_1\lesub{\uvar}{\lgctxtwo_1}{(\labthree)}\lsctx_2\,\tm_{12}$. Then we conclude immediately since the $\lab$ and $\labtwo$-steps are disjoint.

      \item  $\lgctxtwo=\ctxhole\, \lsctx_2\,\tm_{12}$ and $\lsctx=\lsctx_1 \esub{\uvartwo}{(\ofc{(\sha{\tm_{21}})\lsctxtwo_1})\lsctxtwo_2}\lsctx_2$. The step $\tm=\of{\lgctxtwo}{\tm_2}\lto{\labtwo} \of{\lgctxtwo}{\tmthree_1}=\tmthree$ is $\tm_2=\off{\lgctxthree}{\ann{\uvartwo}{\labtwo}}\esub{\uvartwo}{(\ofc{(\sha{\tm_{21}})\lsctxtwo_1})\lsctxtwo_2} \rtoSlsL{\labtwo} \off{\lgctxthree}{(\sha{\tm_{21}})\lsctxtwo_1}\esub{\uvartwo}{\ofc{(\sha{\tm_{21}})\lsctxtwo_1}}\lsctxtwo_2 = \tmthree_1$.  We proceed as in case~\ref{perm:ls_vs_all}.

      \item  $\lgctxtwo=\ctxhole\, \lsctx_2\,\tm_{12}$ and $\lsctx=\lsctx_1 \esub{\uvartwo}{(\ofc{\tm_{21}})\lsctxtwo}\lsctx_2$. The step $\tm=\of{\lgctxtwo}{\tm_2}\lto{\labtwo} \of{\lgctxtwo}{\tmthree_1}=\tmthree$ is $\tm_2=(\llam{\lvar}{\off{\lgctx_1}{\tm_1}}{\labthree})\lsctx_1\lesub{\uvar}{(\ofc{\tm_{21}})\lsctxtwo}{\labtwo} \rtoSgcL{\labtwo}  (\llam{\lvar}{\off{\lgctx_1}{\tm_1}}{\labthree})\lsctx_1\lsctxtwo = \tmthree_1$. Then we conclude immediately since the $\lab$ and $\labtwo$-steps are disjoint.
        
           \item  $\lgctxtwo=(\llam{\lvar}{\of{\lgctx_1}{\tm_1}}{\labthree})\lsctx\,\lgctxtwo_1$. Then we conclude immediately since the $\lab$ and $\labtwo$-steps are disjoint.

      \end{xenumerate}

\item $\lgctx= (\llam{\lvar}{\tm_{11}}{\labthree})\lsctx_1\lesub{\uvartwo}{\lgctx_1}{(\labfour)}\lsctx_2\,\tm_{12}$. We next consider all possible forms for $\lgctxtwo$.

  \begin{xenumerate}
    \item $\lgctxtwo =\ctxhole$. We proceed as in case~\ref{perm:db}.
    \item  $\lgctxtwo= (\llam{\lvar}{\lgctxtwo_1}{\labthree})\lsctx_1\lesub{\uvartwo}{\off{\lgctx_1}{\tm_1}}{(\labfour)}\lsctx_2\,\tm_{12}$. Then we conclude immediately since the $\lab$ and $\labtwo$-steps are disjoint.
        
      \item  $\lgctxtwo= (\llam{\lvar}{\tm_{11}}{\labthree}) \lsctx_{11}\lesub{\uvar}{\lgctxtwo_1}{(\labthree)}\lsctx_{12}\lesub{\uvartwo}{\off{\lgctx_1}{\tm_1}}{(\labfour)}\lsctx_2\,\tm_{12} $. Then we conclude immediately since the $\lab$ and $\labtwo$-steps are disjoint.

      \item $\lgctxtwo=(\llam{\lvar}{\tm_{11}}{\labthree})\lsctx_1\lesub{\uvartwo}{\lgctxtwo_1}{(\labfour)}\lsctx_2\,\tm_{12}$. Then we conclude from the \ih.

      \item  $\lgctxtwo= (\llam{\lvar}{\tm_{11}}{\labthree}) \lsctx_1\lesub{\uvartwo}{\off{\lgctx_1}{\tm_1}}{(\labfour)}\lsctx_{21}\lesub{\uvar}{\lgctxtwo_1}{(\labthree)}\lsctx_{22}\,\tm_{12} $. Then we conclude immediately since the $\lab$ and $\labtwo$-steps are disjoint.
        
      \item  $\lgctxtwo=\ctxhole\, \lsctx_{12}\lesub{\uvartwo}{\off{\lgctx_1}{\tm_1}}{(\labfour)}\lsctx_2\,\tm_{12}$ and $\lsctx_1=\lsctx_{11} \esub{\uvartwo}{(\ofc{(\sha{\tm_{21}})\lsctxtwo_1})\lsctxtwo_2}\lsctx_{12}$. The step $\tm=\of{\lgctxtwo}{\tm_2}\lto{\labtwo} \of{\lgctxtwo}{\tmthree_1}=\tmthree$ is $\tm_2=\off{\lgctxthree}{\ann{\uvartwo}{\labtwo}}\esub{\uvartwo}{(\ofc{(\sha{\tm_{21}})\lsctxtwo_1})\lsctxtwo_2} \rtoSlsL{\labtwo} \off{\lgctxthree}{(\sha{\tm_{21}})\lsctxtwo_1}\esub{\uvartwo}{\ofc{(\sha{\tm_{21}})\lsctxtwo_1}}\lsctxtwo_2 = \tmthree_1$. We consider each possible location for $\ann{\uvartwo}{\labtwo}$.
        \begin{xenumerate}
           \item $\ann{\uvartwo}{\labtwo}\in \tm_1$. We proceed as in case~\ref{perm:ls_vs_all}.
        \item $\ann{\uvartwo}{\labtwo}\in \lsctx_{11}$. Then we conclude immediately since the $\lab$ and $\labtwo$-steps are disjoint.
        \end{xenumerate}

      \item  $\lgctxtwo=\ctxhole\, \lsctx_{12}\lesub{\uvartwo}{\off{\lgctx_1}{\tm_1}}{(\labfour)}\lsctx_2\,\tm_{12}$ and $\lsctx_1=\lsctx_{11} \lesub{\uvar}{(\ofc{\tm_{21}})\lsctxtwo}{\labtwo}\lsctx_{12}$. The step $\tm=\of{\lgctxtwo}{\tm_2}\lto{\labtwo} \of{\lgctxtwo}{\tmthree_1}=\tmthree$ is $\tm_2=(\llam{\lvar}{\tm_{11}}{\lab})\lsctx_{11}\lesub{\uvar}{(\ofc{\tm_{21}})\lsctxtwo}{\labtwo} \rtoSgcL{\labtwo}  (\llam{\lvar}{\tm_{11}}{\lab})\lsctx_{11}\lsctxtwo = \tmthree_1$.  Then we conclude immediately since the $\lab$ and $\labtwo$-steps are disjoint.

      \item  $\lgctxtwo=\ctxhole\, \lsctx_{22}\,\tm_{12}$ and $\lsctx_2=\lsctx_{21} \esub{\uvartwo}{(\ofc{(\sha{\tm_{21}})\lsctxtwo_1})\lsctxtwo_2}\lsctx_{22}$. The step $\tm=\of{\lgctxtwo}{\tm_2}\lto{\labtwo} \of{\lgctxtwo}{\tmthree_1}=\tmthree$ is $\tm_2=\off{\lgctxthree}{\ann{\uvartwo}{\labtwo}}\esub{\uvartwo}{(\ofc{(\sha{\tm_{21}})\lsctxtwo_1})\lsctxtwo_2} \rtoSlsL{\labtwo} \off{\lgctxthree}{(\sha{\tm_{21}})\lsctxtwo_1}\esub{\uvartwo}{\ofc{(\sha{\tm_{21}})\lsctxtwo_1}}\lsctxtwo_2 = \tmthree_1$. We consider each possible location for $\ann{\uvartwo}{\labtwo}$.
        \begin{xenumerate}

        \item  $\ann{\uvartwo}{\labtwo}\in\tm_{11}$. Then we conclude immediately since the $\lab$ and $\labtwo$-steps are disjoint.
        \item $\ann{\uvartwo}{\labtwo}\in \lsctx_1$. Then we conclude immediately since the $\lab$ and $\labtwo$-steps are disjoint.

        \item $\ann{\uvartwo}{\labtwo}\in \lgctx_1$. Then we conclude immediately since the $\lab$ and $\labtwo$-steps are disjoint.
              \item $\ann{\uvartwo}{\labtwo}\in \tm_1$.  We proceed as in case~\ref{perm:ls_vs_all}.
        \item $\ann{\uvartwo}{\labtwo}\in \lsctx_{21}$. Then we conclude immediately since the $\lab$ and $\labtwo$-steps are disjoint.
        \end{xenumerate}
        
      \item  $\lgctxtwo = (\llam{\lvar}{\tm_{11}}{\labthree})\lsctx_1\lesub{\uvartwo}{\of{\lgctx_1}{\tm_1}}{(\labfour)}\lsctx_2\,\lgctxtwo_1$. Then we conclude immediately since the $\lab$ and $\labtwo$-steps are disjoint.

      \end{xenumerate}

\item $\lgctx= (\llam{\lvar}{\tm_{11}}{\labthree})\lsctx\,\lgctx_1$. We next consider all possible forms for $\lgctxtwo$.

  \begin{xenumerate}
    \item $\lgctxtwo =\ctxhole$. We proceed as in case~\ref{perm:db}.
      \item  $\lgctxtwo=(\llam{\lvar}{\lgctxtwo_1}{\labthree})\lsctx\, \of{\lgctx_1}{\tm_1}$. Then we conclude immediately since the $\lab$ and $\labtwo$-steps are disjoint.
        
      \item  $\lgctxtwo=(\llam{\lvar}{\tm_{11}}{\labthree})\lsctx_1\lesub{\uvar}{\lgctxtwo_1}{(\labthree)}\, \of{\lgctx_1}{\tm_1}$. Then we conclude immediately since the $\lab$ and $\labtwo$-steps are disjoint.

      \item  $\lgctxtwo=\ctxhole\, \lsctx_2\,\tm_{12}$ and $\lsctx=\lsctx_1 \esub{\uvartwo}{(\ofc{(\sha{\tm_{21}})\lsctxtwo_1})\lsctxtwo_2}\lsctx_2$. The step $\tm=\of{\lgctxtwo}{\tm_2}\lto{\labtwo} \of{\lgctxtwo}{\tmthree_1}=\tmthree$ is $\tm_2=\off{\lgctxthree}{\ann{\uvartwo}{\labtwo}}\esub{\uvartwo}{(\ofc{(\sha{\tm_{21}})\lsctxtwo_1})\lsctxtwo_2} \rtoSlsL{\labtwo} \off{\lgctxthree}{(\sha{\tm_{21}})\lsctxtwo_1}\esub{\uvartwo}{\ofc{(\sha{\tm_{21}})\lsctxtwo_1}}\lsctxtwo_2 = \tmthree_1$. Then we conclude immediately since the $\lab$ and $\labtwo$-steps are disjoint.

              \item   $\lgctxtwo=\ctxhole\, \lsctx_2\,\tm_{12}$ and $\lsctx=\lsctx_1 \lesub{\uvar}{(\ofc{\tm_{21}})\lsctxtwo}{\labtwo}\lsctx_2$.  The step $\tm=\of{\lgctxtwo}{\tm_2}\lto{\labtwo} \of{\lgctxtwo}{\tmthree_1}=\tmthree$ is $\tm_2=(\llam{\lvar}{\tm_{11}}{\lab})\lsctx_{11}\lesub{\uvar}{(\ofc{\tm_{21}})\lsctxtwo}{\labtwo} \rtoSgcL{\labtwo}  (\llam{\lvar}{\tm_{11}}{\lab})\lsctx_{1}\lsctxtwo = \tmthree_1$.  Then we conclude immediately since the $\lab$ and $\labtwo$-steps are disjoint.

\item $\lgctxtwo=(\llam{\lvar}{\tm_{11}}{\labthree})\lsctx\,\lgctxtwo_1$. We conclude from the \ih.
      \end{xenumerate}

\item $\lgctx=\lgctx_1\,\tm_{11}$. We next consider all possible forms for $\lgctxtwo$.

  \begin{xenumerate}
    \item $\lgctxtwo =\ctxhole$. We proceed as in case~\ref{perm:db}.

      \item  $\lgctxtwo=\lgctxtwo_1\,\tm_{11}$. We conclude from the \ih.

       \item  $\lgctxtwo=\of{\lgctx_1}{\tm_1}\, \lgctxtwo_1$. Then we conclude immediately since the $\lab$ and $\labtwo$-steps are disjoint.

      \end{xenumerate}

\item $\lgctx=\tm_{11}\,\lgctx_1$. Same as previous case.

\item $\lgctx=\sha{\lgctx_1}$.  We conclude from the \ih since $\lgctxtwo$ must be of the form $\sha{\lgctxtwo_1}$.

  
\item $\lgctx=\lopen{\lgctx_1}{(\labthree)}$. We next consider all possible forms for $\lgctxtwo$.

  \begin{xenumerate}
    \item $\lgctxtwo =\ctxhole$. We proceed as in case~\ref{perm:open}.

      \item  $\lgctxtwo=\lopen{\lgctxtwo_1}{(\labthree)}$. We conclude from the \ih.

      \end{xenumerate}

\item $\lgctx=\ofc{\lgctx_1}$. Then both the $\lab$ and $\labtwo$-steps are inside $\of{\lgctx_1}{\tm_1}$ and we conclude from the \ih.

\item $\lgctx=\lgctx_1\lesub{\uvartwo}{\tm_{11}}{(\labthree)}$. We next consider all possible forms for $\lgctxtwo$.

  \begin{xenumerate}
    \item $\lgctxtwo =\ctxhole$. We proceed as in case~\ref{perm:ls}.

      \item  $\lgctxtwo=\lgctxtwo_1\lesub{\uvartwo}{\tm_{11}}{(\labthree)}$. We conclude from the \ih.

               \item  $\lgctxtwo=\of{\lgctx_1}{\tm_1}\lesub{\uvartwo}{\lgctxtwo_1}{(\labthree)}$. Then we conclude immediately since the $\lab$ and $\labtwo$-steps are disjoint.

      \end{xenumerate}



\item $\lgctx=\tmtwo\lesub{\uvartwo}{\lgctx_1}{(\labthree)}$. Same as the previous case.
\end{xenumerate}
\end{ifLongAppendix}
\end{proof}

\subsection{The  $\lambdaS$-calculus as an ARS}
\label{sec:lambdaS_as_an_AxRS}

\subsubsection{Residuals after flattening}

\begin{remark}
If $\tm\pflatt{\pi}\tmtwo$, $\step\in\steps{\tm}$, then $\lift{\tm}{\step}{\lab}\pflatt{\pi}\tmtwo'$, for some $\tmtwo'$ variant of $\tmtwo$. 
\end{remark}

\begin{definition}[Residuals after flattening]
\ldef{residuals_after_flattening}
Given $\tm\pflatt{\pi}\tmtwo$, $\step\in\steps{\tm}$ and $\lab\notin\labels{\tm}$, consider $\lift{\tm}{\step}{\lab}\pflatt{\pi}\tmtwo'$, where $\tmtwo'$ is some variant of $\tmtwo$. The set of \emph{residuals of $\step$ after $\tm\pflatt{\pi}\tmtwo$}, is defined as
\[
  \step\resid{\tm\pflatt{\pi}\tmtwo} \eqdef\{\labStep{\lab}{\tmtwo'}\,|\, \lift{\tm}{\step}{\lab}\pflatt{\pi}\tmtwo'\}\]
We write $\step\resid{\tm\pflatt{\pi}\tmtwo}\step'$ if  $\step'\in\step\resid{\tm\pflatt{\pi}\tmtwo}$.
\end{definition}

We must verify that $\step\resid{\tm\pflatt{\pi}\tmtwo}$ in \rdef{residuals_after_flattening}, does not rely on $\pi$. In other words, that $\resid{\tm\pflatt{\pi}\tmtwo}=\resid{\tm\pflatt{\pi'}\tmtwo}$, for any pair of $\tm\pflatt{\pi}\tmtwo$ and $\tm\pflatt{\pi'}\tmtwo$. This requires making sure that 
if there are $\tmtwo'$ and $\tmtwo''$ such that $\lift{\tm}{\step}{\lab}\pflatt{\pi}\tmtwo'$ and $\lift{\tm}{\step}{\lab}\pflatt{\pi'}\tmtwo''$, then $\tmtwo'=\tmtwo''$. By transitivity of $\flatt$, it suffices to check that  $\tm\pflatt{\pi}\tmtwo$ and $\unlab{\tm}=\unlab{\tmtwo}$ implies $\tm=\tmtwo$ (\cf \rlem{well_definedness_of_step_correspondence}). First we introduce some auxiliary notions and results.

\begin{definition}[Well-named labeled term]
 A labeled term $\tm\in\TermsSL$ is \emph{well-named} if 1) all its bound variables are pairwise distinct and 2) all its labels are pairwise distinct.
\end{definition}

\begin{definition}[Label ordering]
Let $\tm\in\TermsSL$ be a well-named term. The \emph{label ordering}, $\labord{\tm}\subseteq\labels{\tm}\times\labels{\tm}$, is the total order on its labels defined as the left-to-right order when reading $\tm$ as a string.
\end{definition}

\begin{lemma}
\llem{well_named_and_flat_helper}
Let $\tm$ be well-named and $\pi\vdash \tm\flatt\tmtwo$. Then:
  \begin{enumerate}
  \item $\tmtwo$ is well-named;
  \item $  \labels{\tm}=\labels{\tmtwo}$; and
  \item $ \labord{\tm}=\labord{\tmtwo}$
\end{enumerate}
\end{lemma}

\begin{proof}
  By induction on $\pi$.
\end{proof}

\begin{definition}[Equally labeled terms]
Let $\tm,\tmtwo\in\TermsSL$ be variants (\ie $\unlab{\tm}=\unlab{\tmtwo}$). We say that $\tm$ and $\tmtwo$ are \emph{equally labeled} if they have labels on exactly the same symbols in $\unlab{\tm}$, although these labels might not be identical.
\end{definition}

For example, $(\llam{\lvar}{\lvar}{\lab})\lesub{\uvar}{\uvartwo}{\labtwo}$ and $(\llam{\lvar}{\lvar}{\labtwo})\lesub{\uvar}{\uvartwo}{\lab}$ are equally labeled.

\begin{lemma}
  \llem{correspondence_correctness_helper}
    Let $\tm,\tmtwo$ be well-named and equally labeled. If
  \begin{enumerate}
  \item $  \labels{\tm}=\labels{\tmtwo}$; and
  \item $ \labord{\tm}=\labord{\tmtwo}$
\end{enumerate}
then $\tm=\tmtwo$
\end{lemma}

\begin{proof}
  By induction on $\tm$.
  \begin{xenumerate}
\item $\tm=\lvar$ or $\tm=\ann{\uvar}{(\lab)}$. The result is immediate.
\item $\tm=\lam{\lvar}{\tm_1}$. From $\unlab{\tm}=\unlab{\tmtwo}$, it must be the case that $ \tmtwo=\lam{\lvar}{\tmtwo_1}$ and $\unlab{\tm_1}=\unlab{\tmtwo_1}$. Moreover, $\tm_1$ and $\tmtwo_1$ must be well-named and equally labeled. We thus resort to the \ih and conclude.

\item $\tm= \llam{\lvar}{\tm_1}{\lab}$. Since $\tm$ and $\tmtwo$ are equally labeled, it must be the case that $ \tmtwo=\llam{\lvar}{\tmtwo_1}{\lab}$ and $\unlab{\tm_1}=\unlab{\tmtwo_1}$. Moreover, it also follows that  $\tm_1$ and $\tmtwo_1$ must be well-named and equally labeled too. We thus resort to the \ih and conclude.
           
\item $\tm=\tm_1\,\tm_2$. From $\unlab{\tm}=\unlab{\tmtwo}$,  it must be the case that $ \tmtwo=\tmtwo_1\,\tmtwo_2$ and $\unlab{\tm_1}=\unlab{\tmtwo_1}$ and $\unlab{\tm_2}=\unlab{\tmtwo_2}$.

  \begin{xenumerate}
  \item $  \labels{\tm}=\labels{\tmtwo}$. First we verify that  $\labels{\tm_1}=\labels{\tmtwo_1}$. Suppose that, on the contrary, there is a $\lab\in \labels{\tm_1}\setminus \labels{\tmtwo_1}$. Then $\lab\in\labels{\tmtwo_2}$. Since $  \labels{\tm}=\labels{\tmtwo}$ and $\tm$ and $\tmtwo$ are equally labeled, there is a $\labtwo\in \labels{\tmtwo_1}\setminus \labels{\tm_1}$. Then $\labtwo\in \tm_2$. Then we have $\lab\labord{\tm}\labtwo$ and $\labtwo\labord{\tmtwo}\lab$, contradicting  $ \labord{\tm}=\labord{\tmtwo}$.

    If we assume $\lab\in  \labels{\tmtwo_1}\setminus \labels{\tm_1}$, rather than $\lab\in \labels{\tm_1}\setminus \labels{\tmtwo_1}$, we reach a contradiction in a similar way. Therefore, $\labels{\tm_1}=\labels{\tmtwo_1}$. A similar argument shows that $\labels{\tm_2}=\labels{\tmtwo_2}$.
    
  \item $ \labord{\tm}=\labord{\tmtwo}$. Note that $ \labord{\tm_1}=\labord{\tmtwo_1}$ and $ \labord{\tm_2}=\labord{\tmtwo_2}$, follow immediately from $ \labord{\tm}=\labord{\tmtwo}$.
  \end{xenumerate}

  We thus conclude from the \ih applied twice.
  
\item $\tm=\sha{\tm_1}$. From $\unlab{\tm}=\unlab{\tmtwo}$,  it must be the case that $ \tmtwo=\sha{\tmtwo_1}$ and $\unlab{\tm_1}=\unlab{\tmtwo_1}$. Moreover, $\labels{\tm_1}=\labels{\tmtwo_1}$ follows from $  \labels{\tm}=\labels{\tmtwo}$. Likewise, $ \labord{\tm_1}=\labord{\tmtwo_1}$  follows from $ \labord{\tm}=\labord{\tmtwo}$. We thus conclude from the \ih.
  
\item $\tm=\open{\tm_1}$. Same as above.
\item $\tm=\lopen{\tm_1}{\lab}$. Same as above.
\item $\tm=\ofc{\tm_1}$. Same as above.
  
\item $\tm=\tm_1\esub{\uvar}{\tm_2}$. This is similar to the case for application. From $\unlab{\tm}=\unlab{\tmtwo}$,  it must be the case that $ \tmtwo=\tmtwo_1\esub{\uvar}{\tmtwo_2}$ and $\unlab{\tm_1}=\unlab{\tmtwo_1}$ and $\unlab{\tm_2}=\unlab{\tmtwo_2}$.

  \begin{xenumerate}
  \item $  \labels{\tm}=\labels{\tmtwo}$. First we verify that  $\labels{\tm_1}=\labels{\tmtwo_1}$. Suppose that, on the contrary, there is a $\lab\in \labels{\tm_1}\setminus \labels{\tmtwo_1}$. Then $\lab\in\labels{\tmtwo_2}$. Since $  \labels{\tm}=\labels{\tmtwo}$ and $\tm$ and $\tmtwo$ are equally labeled, there is a $\labtwo\in \labels{\tmtwo_1}\setminus \labels{\tm_1}$. Then $\labtwo\in \tm_2$. Then we have $\lab\labord{\tm}\labtwo$ and $\labtwo\labord{\tmtwo}\lab$, contradicting  $ \labord{\tm}=\labord{\tmtwo}$.

    If we assume $\lab\in  \labels{\tmtwo_1}\setminus \labels{\tm_1}$, rather than $\lab\in \labels{\tm_1}\setminus \labels{\tmtwo_1}$, we reach a contradiction in a similar way. Therefore, $\labels{\tm_1}=\labels{\tmtwo_1}$. A similar argument shows that $\labels{\tm_2}=\labels{\tmtwo_2}$.
    
  \item $ \labord{\tm}=\labord{\tmtwo}$. Note that $ \labord{\tm_1}=\labord{\tmtwo_1}$ and $ \labord{\tm_2}=\labord{\tmtwo_2}$, follow immediately from $ \labord{\tm}=\labord{\tmtwo}$.
  \end{xenumerate}

  We thus conclude from the \ih applied twice.
  
\item $\tm=\tm_1\lesub{\uvar}{\tm_2}{\lab}$. Same as above.
\end{xenumerate}

\end{proof}

\begin{lemma}[Well-definedness of Step Correspondence]
\llem{well_definedness_of_step_correspondence}
Let $\tm\in\TermsSL$ be well-named having labels exactly at the anchors of all the steps in $\unlab{\tm}$. Let $\tmtwo$ be such that $\tm\pflatt{\pi}\tmtwo$ and $\unlab{\tm}=\unlab{\tmtwo}$. Then $\tm=\tmtwo$.
\end{lemma}

\begin{proof}
By induction on $\pi$ we can prove that $\tm$ and $\tmtwo$ are equally labeled. Indeed, $\flatt$ does not create nor erase steps nor labels. Moreover, from \rlem{well_named_and_flat_helper}, we obtain that $\tmtwo$ is well-named, $\labels{\tm}=\labels{\tmtwo}$ and $\labord{\tm}=\labord{\tmtwo}$.  We then conclude from \rlem{correspondence_correctness_helper}.
\end{proof}

As a consequence of \rlem{well_definedness_of_step_correspondence}, we will henceforth drop the derivation $\pi$ in $\resid{\tm\pflatt{\pi}\tmtwo}$ and write simply $\resid{\tm\flatt\tmtwo}$. The residual relation on flattening $\resid{\tm\flatt\tmtwo}$ is in fact a bijection between steps in $\tm$ and those in $\tmtwo$. We shall name this bijection \emph{step-correspondence}:

\begin{definition}[Step-correspondence]
Let $\tm\flatt\tmtwo$. We define \emph{step-correspondence}, $\flatBij{\tm}{\tmtwo}\subseteq\steps{\tm}\times\steps{\tmtwo}$, as follows: $\flatBij{\tm}{\tmtwo}(\step)=\steptwo$ iff $\step\resid{\tm\flatt\tmtwo}\steptwo$.
\end{definition}

Step-correspondence is well-defined in the sense that it does not depend on the proof of $\tm\flatt\tmtwo$, a consequence of \rlem{well_definedness_of_step_correspondence}.

\begin{lemma}[Well-definedness of step-correspondence ]
  Let $\tm,\tmtwo$ be labeled terms such that $\tm\flatt\tmtwo$. Then $\flatBij{\tm}{\tmtwo}$ is well-defined.
\end{lemma}

\begin{proof}
Assume that all bound variables in $\tm$ are distinct. Let $\tm'$ be the lifting of all the steps in $\tm$. Note that $\tm$ is well-named. Let $\pi$ be any derivation such that $\tm\pflatt{\pi}\tmtwo $. Consider $\tm'\pflatt{\pi}\tmtwo'$. By induction on $\pi$ it is easy to verify that $\tmtwo'$ has labels on the anchors of all steps in $\tmtwo$. Therefore, each step in $\tm'$ has a unique residual in $\tmtwo'$. Moreover, by \rlem{well_definedness_of_step_correspondence}, $\tmtwo'$ is unique for $\tm\flatt\tmtwo$ and $\unlab{\tmtwo'}=\tmtwo$. Thus $\flatBij{\tm}{\tmtwo}$ is well-defined.
\end{proof}

\begin{definition}[Steps modulo flattening and their equivalence]
  Let $\step:\tm_1\to\tmtwo_1$ be a step in $\lambdaS$ and suppose $\tm_1'\pflatt{\pi_1}\tm_1$ and $\tmtwo_1\pflatt{\pi_2}\tmtwo_1'$. We say that $\mstep{\pi_1}{\pi_2}:\eqclass{\tm_1}{\flatt}\toS \eqclass{\tmtwo_1}{\flatt}$ is a \emph{step modulo flattening}:
  \[ \tm_1'\pflatt{\pi_1}\tm_1\toS^{\step} \tmtwo_1\pflatt{\pi_2}\tmtwo_1'
  \]
  Suppose $\msteptwo{\pi_3}{\pi_4}:\eqclass{\tm_2}{\flatt}\toS \eqclass{\tmtwo_2}{\flatt}$. In other words,
  \[\tm_2'\pflatt{\pi_3}\tm_2\toS^{\steptwo} \tmtwo_2\pflatt{\pi_4}\tmtwo_2'
  \]

Assume, moreover, that $\tm_1\flatt\tm_2$ and $\flatBij{\tm_1}{\tm_2}(\step)=\steptwo$. 
  Then we say $\mstep{\pi_1}{\pi_2}$ and $\msteptwo{\pi_3}{\pi_4}$ are \emph{equivalent} and write $\mstep{\pi_1}{\pi_2}\flatt\msteptwo{\pi_3}{\pi_4}$.  Notice that equivalent steps are coinitial and cofinal; the latter follows from \rprop{flattening_is_strong_bisimulation}. 
\end{definition}

\begin{remark}
Equivalence on steps modulo flattening is an equivalence relation.  We write $ \eqclass{\mstep{\pi_1}{\pi_2}}{\flatt}$ for the equivalence class of steps of $\mstep{\pi_1}{\pi_2}$.
\end{remark}

\begin{definition}[$\lambdaS$ as an ARS]
\ldef{lambdaS_as_ARS}
$\lambdaS$ may be modeled as an Axiomatic Rewrite System  $\aars=\langle \aarsObj, \aarsSteps, \src,\tgt,\resid{}\rangle$ where

\begin{itemize}

\item  \textbf{Objects}. Objects are $\flatt$-equivalence classes of terms: $\aarsObj \eqdef \{ \eqclass{\tm}{\flatt}\,|\, \tm\in\TermsS \}$.

\item  \textbf{Steps, Src, Tgt}. Steps are $\lambdaS$-steps modulo flattening
  \[\aarsSteps\eqdef \{\eqclass{\mstep{\pi_1}{\pi_2}}{\flatt}: \eqclass{\tm}{\flatt}\to\eqclass{\tmtwo}{\flatt}\,|\, \step:\tm\to\tmtwo\mbox{ is a step in }\lambdaS\mbox{ and }\tm'\pflatt{\pi_1}\tm \mbox{ and } \tmtwo\pflatt{\pi_2}\tmtwo', \mbox{for some }\tm',\tmtwo'\}
  \]

\item \textbf{Residuals}.  Consider two coinitial steps $\eqclass{\mstep{\pi_1}{\pi_2}}{\flatt}: \eqclass{\tm_1}{\flatt}\to\eqclass{\tmtwo_1}{\flatt}$ and $\eqclass{\msteptwo{\pi_3}{\pi_4}}{\flatt}: \eqclass{\tm_2}{\flatt}\to\eqclass{\tmtwo_2}{\flatt}$ and $\tm_1\flatt\tm_2$:
  \[\begin{array}{l}
      \tm_1'\pflatt{\pi_1}\tm_1\toS^{\step} \tmtwo_1\pflatt{\pi_2}\tmtwo_1'\\
      \tm_2'\pflatt{\pi_3}\tm_2\toS^{\steptwo} \tmtwo_2\pflatt{\pi_4}\tmtwo_2'\\
      
      \end{array}
      \]
      The set of residuals of $\eqclass{\mstep{\pi_1}{\pi_2}}{\flatt}$ after $\eqclass{\msteptwo{\pi_3}{\pi_4}}{\flatt}$ is defined as follows
      \[\mstep{\pi_1}{\pi_2}\resid{\msteptwo{\pi_3}{\pi_4}}\eqdef \flatBij{\tm_1}{\tm_2}(\step)\resid{\steptwo}\]

\end{itemize}
\end{definition}

Note that the notion of residual is well-defined as a consequence of the following result, whose proof is an immediate consequence of the proof of \rprop{flattening_is_strong_bisimulation} (that flattening is a strong $\toS$-bisimulation) whose diagrams are all closed using steps with the same label.

\begin{lemma}
  $\step\resid{\steptwo}\step'$ iff $\flatBij{\tm_1}{\tm_2}{\step}\resid{\flatBij{\tm_1}{\tm_2}{\steptwo}}\flatBijtwo{\tmtwo_1}{\tmtwo_2}{\step'}$

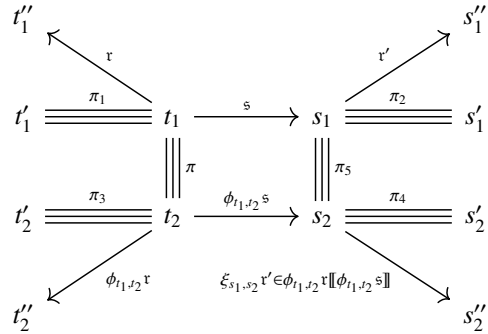
            \begin{center}
              \begin{tikzcd}[column sep=4em,row sep=2em]
               \tm_1'' & & &\tmtwo_1''\\
             \tm_1' \arrow[r,dash]\arrow[r, dash, shift left=1,"\pi_1"]\arrow[r, dash, shift right=1]  &  \tm_1 \arrow[d,dash]\arrow[d, dash, shift left=1,"\pi"]\arrow[d, dash, shift right=1]\arrow[ul, "\step"']\arrow[r, "\steptwo"] & \tmtwo_1 \arrow[ur, "\step'"] \arrow[r,dash]\arrow[r, dash, shift left=1,"\pi_2"]\arrow[r, dash, shift right=1]\arrow[d,dash]\arrow[d, dash, shift left=1,"\pi_5"]\arrow[d, dash, shift right=1]& \tmtwo_1'\\
              
             \tm_2'  \arrow[r,dash]\arrow[r, dash, shift left=1,"\pi_3"]\arrow[r, dash, shift right=1]  &  \tm_2 \arrow[r, "\flatBij{\tm_1}{\tm_2}{\steptwo}"] \arrow[dl, "\flatBij{\tm_1}{\tm_2}{\step}"] & \tmtwo_2  \arrow[r,dash]\arrow[r, dash, shift left=1,"\pi_4"]\arrow[r, dash, shift right=1] \arrow[dr, "\flatBijtwo{\tmtwo_1}{\tmtwo_2}{\step'} \in \flatBij{\tm_1}{\tm_2}{\step}\resid{\flatBij{\tm_1}{\tm_2}{\steptwo}}"']& \tmtwo_2'            \\
             \tm_2'' & & & \tmtwo_2 ''
            \end{tikzcd}
          \end{center}
          
\end{lemma}

\begin{lemma}[Auto-Erasure]
  \llem{auto_erasure}
  $\eqclass{\mstep{\pi_1}{\pi_2}}{\flatt}\resid{\eqclass{\mstep{\pi_1}{\pi_2}}{\flatt}}=\emptyset$.
\end{lemma}
\begin{proof}
  Recall from \rdef{lambdaS_as_ARS} that $\eqclass{\mstep{\pi_1}{\pi_2}}{\flatt}\resid{\eqclass{\mstep{\pi_1}{\pi_2}}{\flatt}}$ is defined as
  \begin{equation}
    \flatBij{\tm_1}{\tm_1}(\step)\resid{\step}
    \label{auto_erasure:eq1}
  \end{equation}
  By taking the trivial derivation of $\tm_1\flatt\tm_1$ that uses reflexivity and resorting to \rlem{well_definedness_of_step_correspondence}, (\ref{auto_erasure:eq1}) is just $\step\resid{\step}$. The latter is immediately seen to be the empty set.

\end{proof}

\begin{lemma}[Finite Residuals]
  \llem{finite_residuals}
For any pair of coinitial steps in $\aars$, $\eqclass{\mstep{\pi_1}{\pi_2}}{\flatt}: \eqclass{\tm_1}{\flatt}\to\eqclass{\tmtwo_1}{\flatt}$ and $\eqclass{\msteptwo{\pi_3}{\pi_4}}{\flatt}: \eqclass{\tm_1}{\flatt}\to\eqclass{\tmtwo_2}{\flatt}$, $\eqclass{\mstep{\pi_1}{\pi_2}}{\flatt}\resid{\eqclass{\msteptwo{\pi_3}{\pi_4}}{\flatt}}$ is finite.
\end{lemma}

\begin{proof}
This is immediate from the fact that $\flatBij{\tm_1}{\tm_2}(\step)\resid{\steptwo}$ is finite.
\end{proof}


We next prove finite developments. Note that since step-correspondence is a bijection (\rlem{well_definedness_of_step_correspondence}), it suffices to prove the following:
\begin{lemma}[Finite Developments]
  \llem{finite_developments}
  Let $\tm\in\TermsSL$. Then $\dev{\lab}$ is strongly-normalizing.
\end{lemma}

\begin{proof}
This is immediate from \rprop{labeled_reduction_is_finite}.
\end{proof}

\begin{lemma}[Semantic Orthogonality]
  \llem{semantic_orthogonality}
 Consider two coinitial steps in $\aars$, say  $\eqclass{\mstep{\pi_1}{\pi_2}}{\flatt}: \eqclass{\tm_1}{\flatt}\to\eqclass{\tmtwo_1}{\flatt}$ and $\eqclass{\msteptwo{\pi_3}{\pi_4}}{\flatt}: \eqclass{\tm_1}{\flatt}\to\eqclass{\tmtwo_2}{\flatt}$. Then
  there exists $\tmthree$ such that
  $\eqclass{\mstep{\pi_1}{\pi_2}}{\flatt}\resid{\eqclass{\msteptwo{\pi_3}{\pi_4}}{\flatt}}: \eqclass{\tmtwo_1}{\flatt}\to\eqclass{\tmthree}{\flatt}$ and 
  $\eqclass{\msteptwo{\pi_3}{\pi_4}}{\flatt}\resid{\eqclass{\mstep{\pi_1}{\pi_2}}{\flatt}}: \eqclass{\tmtwo_2}{\flatt}\to\eqclass{\tmthree}{\flatt}$.
  
\end{lemma}

\begin{proof}
This is an immediate consequence of \rprop{perm}, \rprop{flattening_is_strong_bisimulation} and \rlem{well_definedness_of_step_correspondence}.
\end{proof}


  \subsection{Strong normalization}
  \lsec{appendix:calculus_sn}
  
In this subsection, we show that $\toS$ is strongly normalizing
for the typed fragment of $\lambdaS$ by means of the translation
$\tradlsc{-}$ to the Linear Substitution Calculus (LSC).
Recall that the simply typed LSC \emph{without $\symgc$}
is given by the relation ${\tolsci} \eqdef \todb \cup \tols \cup \togc$
on the set of $\LSC$ terms.

\begin{theorem}[The typed LSC is strongly normalizing]
\lthm{typed_lsc_sn}
If $\judlsc{\tctx}{\tm}{\typ}$,
there are no infinite reduction sequences
$\tm \tolsci \tm_1 \tolsci \tm_2 \tolsci \hdots$.
\end{theorem}
\begin{proof}
See~\cite{DBLP:journals/corr/abs-2312-13270}.
\end{proof}

\subsubsection{Postponement of $\toSgc$}

\begin{lemma}
\llem{source_of_Sgc_step_to_abstraction}
If $\tm \toSgc (\lam{\lvar}{\tmtwo})\sctx$, then one of the following hold:
\begin{enumerate}

\item $\tm=(\lam{\lvar}{\tm_1})\sctx$ and $\tm_1 \toSgc \tmtwo$

\item $\sctx=\sctx_1\esub{\uvar}{\tmthree_2}\sctx_2$ and $\tm=(\lam{\lvar}{\tmtwo}) \sctx_1\esub{\uvar}{\tmthree_1}\sctx_2$ and $\tmthree_1 \toSgc \tmthree_2$

  \item $\sctx=\sctx_1\sctx_2\sctx_3$ and $\tm=(\lam{\lvar}{\tmtwo}) \sctx_1\esub{\uvar}{(\ofc{\tmthree})\sctx_2}\sctx_3$ and $\uvar\notin\fv{(\lam{\lvar}{\tmtwo}) \sctx_1}$.

  \end{enumerate}
  
\end{lemma}

\begin{proof}
  By induction on $\tm$.
  \begin{enumerate}
  \item $\tm=\lvar$ or $\tm=\uvar$. Not possible since there are no $\toSgc$ steps from variables.
    
  \item $\tm=\lam{\lvar}{\tmtwo}$. Then $\sctx=\ctxhole$ and the $\toSgc$ step must be in $\tmtwo$ and case 1 holds.
    
  \item $\tm=\tm_1\,\tm_2$. Not possible since then the $\toSgc$ step would have to be in $\tm_1$ or in $\tm_2$ and $ (\lam{\lvar}{\tmtwo})\sctx$ would have to be an application.
    
  \item $\tm=\sha{\tm_1}$. Not possible since then the $\toSgc$ step would have to be in $\tm_1$ and $ (\lam{\lvar}{\tmtwo})\sctx$ would have to be of the form $\sha{\tmthree}$.
  \item $\tm=\open{\tm_1}$. Not possible since then the $\toSgc$ step would have to be in $\tm_1$ and $ (\lam{\lvar}{\tmtwo})\sctx$ would have to be of the form $\open{\tmthree}$.
  \item $\tm=\ofc{\tm_1}$. Not possible since then the $\toSgc$ step would have to be in $\tm_1$ and $ (\lam{\lvar}{\tmtwo})\sctx$ would have to be of the form $\ofc{\tmthree}$.
  \item $\tm=\tm_1\esub{\uvar}{\tm_2}$. There are three possible cases.
    \begin{enumerate}

      \item The $\toSgc$ step is at the root. Then $\tm_2=(\ofc{\tm_3})\sctx_2$ and $\uvar\notin\fv{\tm_1}$ and $\sctx=\sctx_1\sctx_2$ and $\tm_1=(\lam{\lvar}{\tmtwo})\sctx_1$. Case 3 then holds.

      \item The $\toSgc$ step is in $\tm_1$. Then $\sctx=\sctx_1\esub{\uvar}{\tmthree}$ and  $\tm_1 \toSgc (\lam{\lvar}{\tmtwo})\sctx_1$. We conclude from the \ih.

    \item The $\toSgc$ step is in $\tm_2$. Then $\sctx=\sctx_1\esub{\uvar}{\tmthree}$ and $\tm_2\toSgc \tmthree$. Case 2 holds.

      \end{enumerate}
    \end{enumerate}
\end{proof}

\begin{lemma}
\llem{source_of_Sgc_step_to_sha}
If $\tm \toSgc (\sha{\tmtwo})\sctx$, then one of the following hold:
\begin{enumerate}

\item $\tm=(\sha{\tm_{1}})\sctx$ and $\tm_1 \toSgc \tmtwo$

\item $\sctx=\sctx_1\esub{\uvar}{\tmthree_2}\sctx_2$ and $\tm=(\sha{\tmtwo}) \sctx_1\esub{\uvar}{\tmthree_1}\sctx_2$ and $\tmthree_1 \toSgc \tmthree_2$

  \item $\sctx=\sctx_1\sctx_2\sctx_3$ and $\tm=(\sha{\tmtwo}) \sctx_1\esub{\uvar}{(\ofc{\tmthree})\sctx_2}\sctx_3$ and $\uvar\notin\fv{(\sha{\tmtwo}) \sctx_1}$.

  \end{enumerate}
  
\end{lemma}

\begin{proof}
By induction on $\tm$. Similar to that of~\rlem{source_of_Sgc_step_to_abstraction}.
\end{proof}

\begin{lemma}
\llem{source_of_Sgc_step_to_osha}
If $\tm \toSgc (\ofc{(\sha{\tmtwo})\sctx_1})\sctx_2$, then one of the following hold:
\begin{enumerate}

\item $\tm=(\ofc{(\sha{\tm_1})\sctx_1})\sctx_2$ and $\tm_1 \toSgc \tmtwo$

\item $\sctx_1=\sctx_{11}\esub{\uvar}{\tmthree_2}\sctx_{12}$ and $\tm=(\ofc{(\sha{\tmtwo}) \sctx_{11}\esub{\uvar}{\tmthree_1}\sctx_{12}})\sctx_2 $ and $\tmthree_1 \toSgc \tmthree_2$

\item $\sctx_2=\sctx_{21}\esub{\uvar}{\tmthree_2}\sctx_{22}$ and $\tm=(\ofc{(\sha{\tmtwo}) \sctx_1}) \sctx_{21}\esub{\uvar}{\tmthree_1}\sctx_{22} $ and $\tmthree_1 \toSgc \tmthree_2$

  \item $\sctx_1=\sctx_{11}\sctx_{12}\sctx_{13}$ and $\tm=(\ofc{(\sha{\tmtwo}) \sctx_{11}\esub{\uvar}{(\ofc{\tmthree})\sctx_{12}}\sctx_{13}})\sctx_2$ and $\uvar\notin\fv{(\sha{\tmtwo}) \sctx_{11}}$.

  \item $\sctx_2=\sctx_{21}\sctx_{22}\sctx_{23}$ and $\tm=(\ofc{(\sha{\tmtwo}) \sctx_1}) \sctx_{21}\esub{\uvar}{(\ofc{\tmthree})\sctx_{22}}\sctx_{23}$ and $\uvar\notin\fv{(\ofc{(\sha{\tmtwo}) \sctx_1})\sctx_{21} }$.

  \end{enumerate}
  
\end{lemma}

\begin{proof}
By induction on $\tm$. Similar to that of~\rlem{source_of_Sgc_step_to_abstraction}.
\end{proof}

\begin{lemma}
\llem{postponement_toSgc}
  \quad
  \begin{enumerate}
  \item $\tm \toSgc \tmtwo \toSdb \tmthree$ implies there exists $\tmtwo'$ such that  $\tm\toSdb \tmtwo' \toSgc^*\tmthree $
  \item $\tm \toSgc \tmtwo \toSls \tmthree$ implies there exists $\tmtwo'$ such that $\tm\toSls \tmtwo' \toSgc^+ \tmthree $
  \item $\tm \toSgc \tmtwo \toSopen \tmthree$ implies there exists $\tmtwo'$ such that $\tm\toSopen \tmtwo' \toSgc\tmthree $
\end{enumerate}

\end{lemma}
\begin{ifShortAppendix}
  \begin{proof}
  By induction on $\tm$.
  See the extended version~\cite{mells_long} for the detailed proof.
  \end{proof}
\end{ifShortAppendix}
\begin{ifLongAppendix}
  \begin{proof}
  By induction on $\tm$.
  \begin{xenumerate}
  \item $\tm=\lvar$ or $\tm=\uvar$. This case is not possible since no $\toSgc$-steps are possible from variables.
    
  \item $\tm=\lam{\lvar}{\tm_1}$. Then it must be the case that $\tmtwo=\lam{\lvar}{\tmtwo_1}$ and the step $\tm \toSgc \tmtwo$ follows from $\tm_1 \toSgc \tmtwo_1$, for some term $\tmtwo_1$.  Similarly, it must be the case that $\tmthree=\lam{\lvar}{\tmthree_1}$ and $\tmtwo \to_{\step} \tmthree$ follows from  $\tmtwo_1 \to_{\step} \tmthree_1$, for any $\step\in\{\symSdb,\symSls,\symSopen\}$. We thus obtain the desired result by the \ih.

  \item $\tm=\tm_1\,\tm_2$. Two cases arise depending on whether the $\symSgc$ is internal to $\tm_1$ or to $\tm_2$.

    \begin{xenumerate}

    \item The $\symSgc$ is internal to $\tm_1$. Then $\tmtwo=\tmtwo_1\,\tm_2$ and $\tm_1\,\tm_2 \toSgc \tmtwo_1\,\tm_2$ follows from $\tm_1\toSgc\tmtwo_1$. We next consider each item.

      \begin{xenumerate}

       \item Item 1. There are three possibilities.
        \begin{xenumerate}

        \item The $\symSdb$ step is at the root. Then $\tmtwo_1 = (\lam{\lvar}{\tmtwo_{11}})\sctx$ and $\tmthree=\tmtwo_{11}\sub{\lvar}{\tm_2}\sctx$. By~\rlem{source_of_Sgc_step_to_abstraction}, there are three cases.

          \begin{xenumerate}

  \item $\tm_1=(\lam{\lvar}{\tm_{11}})\sctx$ and $\tm_{11} \toSgc \tmtwo_{11}$.  Then $\tm_1\,\tm_2=(\lam{\lvar}{\tm_{11}})\sctx\,\tm_2 \toSdb \tm_{11}\sub{\lvar}{\tm_2}\sctx\toSgc \tmtwo_{11}\sub{\lvar}{\tm_2}\sctx$.

  \item $\sctx=\sctx_1\esub{\uvar}{\tm_{11}'}\sctx_2$ and $\tm_1=(\lam{\lvar}{\tmtwo_{11}}) \sctx_1\esub{\uvar}{\tm_{11}}\sctx_2$ and $\tm_{11} \toSgc \tm_{11}'$.  Then $\tm_1\,\tm_2=(\lam{\lvar}{\tmtwo_{11}}) \sctx_1\esub{\uvar}{\tm_{11}}\sctx_2\,\tm_2 \toSdb \tmtwo_{11}\sub{\lvar}{\tm_2}\sctx_1\esub{\uvar}{\tm_{11}}\sctx_2\toSgc \tmtwo_{11}\sub{\lvar}{\tm_2}\sctx_1\esub{\uvar}{\tm_{11}'}\sctx_2$.

    \item $\sctx=\sctx_1\sctx_2\sctx_3$ and $\tm_1=(\lam{\lvar}{\tmtwo_{11}}) \sctx_1\esub{\uvar}{(\ofc{\tm_{11}})\sctx_2}\sctx_3$ and $\uvar\notin\fv{(\lam{\lvar}{\tmtwo_{11}}) \sctx_1}$. Then $\tm_1\,\tm_2=(\lam{\lvar}{\tmtwo_{11}}) \sctx_1\esub{\uvar}{(\ofc{\tm_{11}})\sctx_2}\sctx_3\,\tm_2 \toSdb \tmtwo_{11}\sub{\lvar}{\tm_2} \sctx_1\esub{\uvar}{(\ofc{\tm_{11}})\sctx_2}\sctx_3\toSgc \tmtwo_{11}\sub{\lvar}{\tm_2} \sctx_1\sctx_2\sctx_3$. Note that we may assume that $\uvar\notin\fv{\tm_2}$.

    \end{xenumerate}
    
        \item The $\symSdb$ step is internal to $\tm_1$. We use the \ih.

        \item The $\symSdb$ step is internal to $\tm_2$.  Then the $\symSgc$ step and the $\symSdb$ steps are disjoint and we can conclude immediately by swapping them.

          \end{xenumerate}

      \item Item 2. There are two possibilities.

        \begin{xenumerate}

        \item The $\symSls$ step is internal to $\tm_1$.  We use the \ih.

        \item The $\symSls$ step is internal to $\tm_2$.  Then the $\symSgc$ step and the $\symSls$ steps are disjoint and we can conclude immediately by swapping them.

          \end{xenumerate}

        \item Item 3. There are two possibilities

                \begin{xenumerate}

        \item The $\symSopen$ step is internal to $\tm_1$.  We use the \ih.

        \item The $\symSopen$ step is internal to $\tm_2$.  Then the $\symSgc$ step and the $\symSopen$ steps are disjoint and we can conclude immediately by swapping them.

          \end{xenumerate}

        \end{xenumerate}

        \item The $\symSgc$ is internal to $\tm_2$. Similar to the previous case.

      \end{xenumerate}

  \item $\tm=\sha{\tm_1}$. Then it must be the case that $\tmtwo=\sha{\tmtwo_1}$ and the step $\tm \toSgc \tmtwo$ follows from $\tm_1 \toSgc \tmtwo_1$, for some $\tmtwo_1$.  Similarly, it must be the case that $\tmthree=\sha{\tmthree_1}$ and $\tmtwo \to_{\step} \tmthree$ follows from  $\tmtwo_1 \to_{\step} \tmthree_1$, for any $\step\in\{\symSdb,\symSls,\symSopen\}$. We thus obtain the desired result by the \ih.

    \item $\tm=\open{\tm_1}$. Then it must be the case that $\tmtwo=\open{\tmtwo_1}$ and the step $\tm \toSgc \tmtwo$ follows from $\tm_1 \toSgc \tmtwo_1$.  We next consider each item.

      \begin{xenumerate}

       \item Item 1. The $\symSdb$ step must be internal to $\tmtwo_1$.  We use the \ih.

      \item Item 2. The $\symSls$ step mut be internal to $\tmtwo_1$.  We use the \ih.

        \item Item 3. There are two possibilities

                \begin{xenumerate}

        \item The $\symSopen$ step is at the root.  Then $\tmtwo_1=(\sha{\tmtwo_{11}})\sctx$ and   $\tmtwo =\open{(\sha{\tmtwo_{11}})\sctx}
      \toSopen 
      \tmtwo_{11}\sctx=\tmthree$. By~\rlem{source_of_Sgc_step_to_sha}, there are three cases.

          \begin{xenumerate}

  \item $\tm_1=\open{(\sha{\tm_{11}})\sctx}$ and $\tm_{11} \toSgc \tmtwo_{11}$.  Then $\open{\tm_1}=\open{(\sha{\tm_{11}})\sctx} \toSopen \tm_{11}\sctx\toSgc \tmtwo_{11}\sctx$.

  \item $\sctx=\sctx_1\esub{\uvar}{\tm_{11}'}\sctx_2$ and $\tm_1=(\sha{\tmtwo_{11}})\sctx_1\esub{\uvar}{\tm_{11}}\sctx_2$ and $\tm_{11} \toSgc \tm_{11}'$.  Then $\open{\tm_1}=\open{(\sha{\tmtwo_{11}}) \sctx_1\esub{\uvar}{\tm_{11}}\sctx_2} \toSopen \tmtwo_{11}\sctx_1\esub{\uvar}{\tm_{11}}\sctx_2\toSgc \tmtwo_{11}\sctx_1\esub{\uvar}{\tm_{11}'}\sctx_2$.

    \item $\sctx=\sctx_1\sctx_2\sctx_3$ and $\tm_1=(\sha{\tmtwo_{11}}) \sctx_1\esub{\uvar}{(\ofc{\tm_{11}})\sctx_2}\sctx_3$ and $\uvar\notin\fv{(\sha{\tmtwo_{11}}) \sctx_1}$. Then $\open{\tm_1} = \open{(\sha{\tmtwo_{11}}) \sctx_1\esub{\uvar}{(\ofc{\tm_{11}})\sctx_2}\sctx_3} \toSopen \tmtwo_{11} \sctx_1\esub{\uvar}{(\ofc{\tm_{11}})\sctx_2}\sctx_3\toSgc \tmtwo_{11} \sctx_1\sctx_2\sctx_3$. 

    \end{xenumerate}

        \item The $\symSopen$ step is internal to $\tmtwo_1$.  We resort to the \ih.

          \end{xenumerate}

        \end{xenumerate}

      \item $\tm=\ofc{\tm_1}$.  Then it must be the case that $\tmtwo=\ofc{\tmtwo_1}$ and the step $\tm \toSgc \tmtwo$ follows from $\tm_1 \toSgc \tmtwo_1$.  We next consider each item.

      \begin{xenumerate}

       \item Item 1. The $\symSdb$ step must be internal to $\tmtwo_1$.  We use the \ih.

      \item Item 2. The $\symSls$ step must be internal to $\tmtwo_1$. We use the \ih.  

        \item Item 3. The $\symSopen$ step must be internal to $\tmtwo_1$. We use the \ih.
        \end{xenumerate}

      \item $\tm=\tm_1\esub{\uvar}{\tm_2}$. There are three possibilities for the $\symSgc$-step.
              \begin{xenumerate}

              \item The $\symSgc$ step is at the root. Then $\tm_2=(\ofc{\tm_{21}})\sctx$ and $\uvar\notin{\fv{\tm_1}}$ and $\tmtwo = \tm_{1}\sctx$ and $\dom{\sctx}\cap\fv{\tm_1}=\emptyset$.
                                      We next consider each item.

      \begin{xenumerate}

      \item Item 1.
        Three further cases arise.
          \begin{xenumerate}

                 \item The $\symSdb$ step is at the root. This is not possible unless $\sctx=\ctxhole$, in which  case the following case applies.
       \item The $\symSdb$ step is internal to $\tm_{1}$. Then we conclude with $\tm=\tm_1\esub{\uvar}{(\ofc{\tm_{21}})\sctx}\toSdb \tm_1'\esub{\uvar}{(\ofc{\tm_{21}})\sctx} \toSgc \tm_1'\sctx$.
       \item The $\symSdb$ step is internal to $\sctx$.  Then we conclude with $\tm=\tm_1\esub{\uvar}{(\ofc{\tm_{21}})\sctx}\toSdb \tm_1\esub{\uvar}{(\ofc{\tm_{21}})\sctx'} \toSgc \tm_1\sctx'$.
         
  \end{xenumerate}

      \item Item 2.      Three further cases arise.
          \begin{xenumerate}

          \item The $\symSls$ step is at the root of $\tm_{1}\sctx_1 \esub{\uvartwo}{(\ofc{(\sha{\tm_{11}})\sctxtwo_1})\sctxtwo_2}$ and $\sctx=\sctx_1 \esub{\uvartwo}{(\ofc{(\sha{\tm_{11}})\sctxtwo_1})\sctxtwo_2}\,\sctx_2$.  This case is not possible since $\dom{\sctx}\cap\fv{\tm_1}\neq\emptyset$.
          \item The $\symSls$ step is internal to $\tm_{1}$. We use the \ih.
       \item The $\symSls$ step is internal to $\sctx$. The steps are disjoint and we can commute them.      
  \end{xenumerate}

        \item Item 3. The $\symSopen$ step must be in $\tm_{1}$. Then both steps are disjoint and we can commute them. That is $\tm=\tm_1\esub{\uvar}{(\ofc{\tm_{21}})\sctx}\toSopen \tm_1'\esub{\uvar}{(\ofc{\tm_{21}})\sctx} \toSgc \tm_1'\sctx $

        \end{xenumerate}

        \item The $\symSgc$ step is internal to $\tm_1$. In other words, $\tmtwo=\tmtwo_1\esub{\uvar}{\tm_2}$ and the step $\tm \toSgc \tmtwo$ follows from $\tm_1\toSgc\tmtwo_1$. 
          We next consider each item.

      \begin{xenumerate}

      \item Item 1. Two further cases arise.
          \begin{xenumerate}

       \item The $\symSdb$ step is internal to $\tmtwo_1$.  We use the \ih.
       \item The $\symSdb$ step is internal to $\tm_2$.  Then both steps are disjoint and we can commute them.
         
  \end{xenumerate}

  \item Item 2.

        \begin{xenumerate}

        \item  The $\symSls$ step is at the root. Then $\tmtwo_1=\off{\gctx}{\uvar}$ and $\tm_2=(\ofc{(\sha{\tm_{11}})\sctxtwo_1})\sctxtwo_2$ and
          \[\tmtwo=\off{\gctx}{\uvar}\esub{\uvar}{(\ofc{(\sha{\tm_{11}})\sctxtwo_1})\sctxtwo_2} \toSls \off{\gctx}{(\sha{\tm_{11}})\sctxtwo_1}\esub{\uvar}{\ofc{(\sha{\tm_{11}})\sctxtwo_1}}\sctxtwo_2=\tmthree\]
          From  $\tm_1\toSgc\tmtwo_1=\off{\gctx}{\uvar}$, it follows that $\tm_1=\off{\gctxtwo}{\uvar}$, for some context $\gctxtwo$. Then we have $\tm=\tm_1\esub{\uvar}{\tm_2} = \off{\gctxtwo}{\uvar}\esub{\uvar}{(\ofc{(\sha{\tm_{11}})\sctxtwo_1})\sctxtwo_2}\toSls \off{\gctxtwo}{(\sha{\tm_{11}})\sctxtwo_1}\esub{\uvar}{\ofc{(\sha{\tm_{11}})\sctxtwo_1}}\sctxtwo_2 \toSgc  \off{\gctx}{(\sha{\tm_{11}})\sctxtwo_1}\esub{\uvar}{\ofc{(\sha{\tm_{11}})\sctxtwo_1}}\sctxtwo_2$.
          
       \item The $\symSls$ step is internal to $\tmtwo_1$. We use the \ih.
       \item The $\symSls$ step is internal to $\tm_2$. Then both steps are disjoint and we can commute them.
         
  \end{xenumerate}

        \item Item 3. There are two possibilities

                \begin{xenumerate}

        \item The $\symSopen$ step is internal to $\tmtwo_1$. We use the \ih.

        \item The $\symSopen$ step is internal to $\tm_2$.  Then both steps are disjoint and we can commute them.

          \end{xenumerate}

        \end{xenumerate}

        \item The $\symSgc$ step is internal to $\tm_2$.  In other words, $\tmtwo=\tm_1\esub{\uvar}{\tmtwo_2}$ and the step $\tm \toSgc \tmtwo$ follows from $\tm_2\toSgc\tmtwo_2$. 
          We next consider each item.

      \begin{xenumerate}

       \item Item 1. 
       \begin{xenumerate}

       \item The $\symSdb$ step is internal to $\tm_1$.  Then both steps are disjoint and we can commute them.
       \item The $\symSdb$ step is internal to $\tmtwo_2$.   We use the \ih.
         
  \end{xenumerate}

      \item Item 2. There are three possibilities.

                \begin{xenumerate}

        \item The $\symSls$ step is at the root.    Then $\tm_1=\off{\gctx}{\uvar}$ and $\tmtwo_2=(\ofc{(\sha{\tmtwo_{21}})\sctxtwo_1})\sctxtwo_2$ and
          \[\tmtwo=\off{\gctx}{\uvar}\esub{\uvar}{(\ofc{(\sha{\tmtwo_{21}})\sctxtwo_1})\sctxtwo_2} \toSls \off{\gctx}{(\sha{\tmtwo_{21}})\sctxtwo_1}\esub{\uvar}{\ofc{(\sha{\tmtwo_{21}})\sctxtwo_1}}\sctxtwo_2=\tmthree\]
          From  $\tm_2\toSgc (\ofc{(\sha{\tmtwo_{21}})\sctxtwo_1})\sctxtwo_2$ and \rlem{source_of_Sgc_step_to_osha} there are five cases to consider.
          
  \begin{enumerate}

  \item $\tm_2=(\ofc{(\sha{\tm_{21}})\sctxtwo_1})\sctxtwo_2$ and $\tm_{21} \toSgc \tmtwo_{21}$.       Then we have $\tm=\tm_1\esub{\uvar}{\tm_2} = \off{\gctx}{\uvar}\esub{\uvar}{(\ofc{(\sha{\tm_{21}})\sctxtwo_1})\sctxtwo_2}\toSls \off{\gctx}{(\sha{\tm_{21}})\sctxtwo_1}\esub{\uvar}{\ofc{(\sha{\tm_{21}})\sctxtwo_1}}\sctxtwo_2 \toSgc \toSgc  \off{\gctx}{(\sha{\tmtwo_{21}})\sctxtwo_1}\esub{\uvar}{\ofc{(\sha{\tmtwo_{21}})\sctxtwo_1}}\sctxtwo_2$.

  \item $\sctxtwo_1=\sctxtwo_{11}\esub{\uvartwo}{\tmthree_2}\sctxtwo_{12}$ and $\tm_2=(\ofc{(\sha{\tmtwo_{21}}) \sctxtwo_{11}\esub{\uvartwo}{\tmthree_1}\sctxtwo_{12}})\sctxtwo_2 $ and $\tmthree_1 \toSgc \tmthree_2$. Similar to the previous case.

  \item $\sctxtwo_2=\sctxtwo_{21}\esub{\uvartwo}{\tmthree_2}\sctxtwo_{22}$ and $\tm_2=(\ofc{(\sha{\tmtwo_{21}}) \sctxtwo_1}) \sctxtwo_{21}\esub{\uvartwo}{\tmthree_1}\sctxtwo_{22} $ and $\tmthree_1 \toSgc \tmthree_2$.  Similar to the previous case.

    \item $\sctxtwo_1=\sctxtwo_{11}\sctxtwo_{12}\sctxtwo_{13}$ and $\tm_2=(\ofc{(\sha{\tmtwo_{21}}) \sctxtwo_{11}\esub{\uvartwo}{(\ofc{\tmthree})\sctxtwo_{12}}\sctxtwo_{13}})\sctxtwo_2$ and $\uvartwo\notin\fv{(\sha{\tmtwo_{21}}) \sctxtwo_{11}}$.  Similar to the previous case.

    \item $\sctxtwo_2=\sctxtwo_{21}\sctxtwo_{22}\sctxtwo_{23}$ and $\tm_2=(\ofc{(\sha{\tmtwo_{21}}) \sctxtwo_1}) \sctxtwo_{21}\esub{\uvartwo}{(\ofc{\tmthree})\sctxtwo_{22}}\sctxtwo_{23}$ and $\uvartwo\notin\fv{(\ofc{(\sha{\tmtwo_{21}}) \sctxtwo_1})\sctxtwo_{21} }$.  Similar to the previous case.

    \end{enumerate}

        \item The $\symSls$ step is internal to $\tm_1$.   Then both steps are disjoint and we can commute them.

                \item The $\symSls$ step is internal to $\tmtwo_2$.   We resort to the \ih.

          \end{xenumerate}

    \item Item 3. There are two possibilities

                \begin{xenumerate}

        \item The $\symSopen$ step is internal to $\tm_1$.  Then both steps are disjoint and we can commute them.

        \item The $\symSopen$ step is internal to $\tmtwo_2$.   We use the \ih.

          \end{xenumerate}
        \end{xenumerate}
        
          \end{xenumerate}
        \end{xenumerate}
  \end{proof}
\end{ifLongAppendix}

\subsubsection{Fusion}

\begin{definition}[Fusion]
\ldef{fusion}
The binary relation of fusion ${\tofuse} \subseteq \TermsLSC \times \TermsLSC$ 
between terms of the LSC,
called \emph{fusion},
is defined as ${\tofuse} \eqdef (\tofuseone)^*$,
where $\tofuseone$ in turn is the closure by compatibility under arbitrary
contexts of the following rules:
\[
  \begin{array}{l@{\HS}rcll}
    (\ruleFuseW)
  &
    \tm\esub{\var}{\tmtwo}
  & \tofuseone &
    \tm
    & \text{if $\var\notin\fv{\tm}$}
  \\
    (\ruleFuseC)
  &
    \tm\esub{\var}{\tmtwo}\esub{\vartwo}{\tmtwo}
  & \tofuseone &
    \tm\sub{\var}{\vartwo}\esub{\vartwo}{\tmtwo}
  \\
    (\ruleFuseLam)
  &
    \lam{\var}{\tm\esub{\vartwo}{\tmtwo}}
  & \tofuseone &
    (\lam{\var}{\tm})\esub{\vartwo}{\tmtwo}
    & \text{if $\var\notin\fv{\tmtwo}$}
  \\
    (\ruleFuseAppL)
  &
    \tm\esub{\var}{\tmtwo}\,\tmthree
  & \tofuseone &
    (\tm\,\tmthree)\esub{\var}{\tmtwo}
    & \text{if $\var\notin\fv{\tmthree}$}
  \\
    (\ruleFuseAppR)
  &
    \tm\,\tmthree\esub{\var}{\tmtwo}
  & \tofuseone &
    (\tm\,\tmthree)\esub{\var}{\tmtwo}
    & \text{if $\var\notin\fv{\tm}$}
  \\
    (\ruleFuseEsL)
  &
    \tm\esub{\var}{\tmtwo}\esub{\vartwo}{\tmthree}
  & \tofuseone &
    \tm\esub{\vartwo}{\tmthree}\esub{\var}{\tmtwo}
    & \text{if $\var\notin\fv{\tmthree}$ and $\vartwo\notin\fv{\tmtwo}$}
  \\
    (\ruleFuseEsR)
  &
    \tm\esub{\var}{\tmtwo\esub{\vartwo}{\tmthree}}
  & \tofuseone &
    \tm\esub{\var}{\tmtwo}\esub{\vartwo}{\tmthree}
    & \text{if $\vartwo\notin\fv{\tm}$}
  \end{array}
\]
\end{definition}

\begin{remark}
It is easy to check that the relation $\tofuse$ as defined in \rdef{fusion} 
is equivalent to the definition of fusion given in the body of the paper,
which replaces
rules $\ruleFuseLam$, $\ruleFuseAppL$, $\ruleFuseAppR$, $\ruleFuseEsL$, $\ruleFuseEsR$
by a single rule of the form
$\of{\gctx}{\tm\esub{\var}{\tmtwo}} \tofuse \of{\gctx}{\tm}\esub{\var}{\tmtwo}$
provided that $\gctx$ does not bind $\tmtwo$.
\end{remark}

\begin{lemma}[Properties of fusion]
\llem{properties_of_fusion}
\quad
\begin{enumerate}
\item
  $\off{\gctx}{\tm\esub{\var}{\tmtwo}} \tofuse \off{\gctx}{\tm}\esub{\var}{\tmtwo}$
\item
  $\off{\gctx}{\tm\sctx}\esub{\var}{\tm\sctx} \tofuse \off{\gctx}{\tm}\esub{\var}{\tm}\sctx$
\end{enumerate}
\end{lemma}
\begin{proof}
We prove each item separately:
\begin{enumerate}
\item
  By induction on $\gctx$:
  \begin{enumerate}
  \item If $\gctx = \ctxhole$, it is immediate.
  \item If $\gctx = \lam{\var}{\gctx'}$, it follows from the \ih and $\ruleFuseLam$.
  \item If $\gctx = \gctx'\,\tmtwo$, it follows from the \ih and $\ruleFuseAppL$.
  \item If $\gctx = \tmtwo\,\gctx'$, it follows from the \ih and $\ruleFuseAppR$.
  \item If $\gctx = \gctx'\esub{\var}{\tmtwo}$, it follows from the \ih and $\ruleFuseEsL$.
  \item If $\gctx = \tmtwo\esub{\var}{\gctx'}$, it follows from the \ih and $\ruleFuseEsR$.
  \end{enumerate}
\item
  By induction on $\sctx$.
  If $\sctx = \ctxhole$, it is immediate,
  so let $\sctx = \sctx'\esub{\var}{\tmtwo}$.
  Then:
  \[
    \begin{array}{rrll}
      \off{\gctx}{\tm\sctx}\esub{\var}{\tm\sctx}
    & = &
      \off{\gctx}{\tm\sctx'\esub{\var}{\tmtwo}}\esub{\var}{\tm\sctx'\esub{\var}{\tmtwo}}
    \\
    & \tofuse &
      \off{\gctx}{\tm\sctx'}\esub{\var}{\tmtwo}\esub{\var}{\tm\sctx'\esub{\var}{\tmtwo}}
      & \text{by part~1. of this lemma}
    \\
    & \tofuse &
      \off{\gctx}{\tm\sctx'}\esub{\var}{\tmtwo}\esub{\var}{\tm\sctx'}\esub{\var}{\tmtwo}
      & \text{by $\ruleFuseEsR$}
    \\
    & = &
      \off{\gctx}{(\tm\sctx')\sub{\var}{\var'}}\esub{\var'}{\tmtwo}\esub{\var}{\tm\sctx'}\esub{\var}{\tmtwo}
      & \text{by $\alpha$-conversion, where $\var' \notin \fv{\tm\sctx'}$}
    \\
    & \tofuse &
      \off{\gctx}{(\tm\sctx')\sub{\var}{\var'}}\esub{\var}{\tm\sctx'}\esub{\var'}{\tmtwo}\esub{\var}{\tmtwo}
      & \text{by $\ruleFuseEsL$}
    \\
    & \tofuse &
      \off{\gctx}{\tm\sctx'}\esub{\var}{\tm\sctx'}\esub{\var}{\tmtwo}
      & \text{by $\ruleFuseC$}
    \\
    & \tofuse &
      \off{\gctx}{\tm}\esub{\var}{\tm}\sctx'\esub{\var}{\tmtwo}
      & \text{by \ih}
    \\
    & = &
      \off{\gctx}{\tm}\esub{\var}{\tm}\sctx
    \end{array}
  \]
\end{enumerate}
\end{proof}

\begin{lemma}[Reduction before fusion of variables]
\llem{reduction_before_fusion}
\quad
\begin{enumerate}
\item
  Let $\var,\vartwo,\varthree$ be different variables.
  If $\off{\gctx}{\varthree} = \tm\sub{\var}{\vartwo}$
  then there exists a context $\gctx_0$
  such that $\tm = \off{\gctx_0}{\varthree}$
  and $\gctx = \gctx_0\sub{\var}{\vartwo}$.
\item
  Let $\tm\sub{\var}{\vartwo} \tolsci \tmtwo$.
  Then there exists a term $\tmtwo_0$ such that $\tm \tolsci \tmtwo_0$
  and $\tmtwo = \tmtwo_0\sub{\var}{\vartwo}$.
\end{enumerate}
\end{lemma}
\begin{proof}
The first item is by induction on $\gctx$:
  \begin{enumerate}
  \item If $\gctx = \ctxhole$. Then $\off{\gctx}{\varthree} = \varthree = \tm\sub{\var}{\vartwo}$  and $\varthree$ distinct from $\vartwo$ implies that $\tm=\varthree$. We set $\gctx_0\eqdef \ctxhole$ and conclude.
  \item If $\gctx = \lam{\var'}{\gctx'}$. Without loss of generality, we may assume that $\var'$ is distinct from $\var,\vartwo,\varthree$. Then $\off{\gctx}{\varthree}=\off{(\lam{\var'}{\gctx'})}{\varthree}=\lam{\var'}{\off{\gctx'}{\varthree}}=\tm\sub{\var}{\vartwo}$. Thus $\tm=\lam{\var'}{\tm_1}$ and  $\off{\gctx'}{\varthree} =\tm_1\sub{\var}{\vartwo}$. From the \ih there exists a context $\gctx_0$
  such that $\tm_1 = \off{\gctx_0'}{\varthree}$
  and $\gctx' = \gctx_0'\sub{\var}{\vartwo}$. We set $\gctx_0\eqdef \lam{\var'}{\gctx_0'}$ and conclude.
      
  \item If $\gctx = \gctx'\,\tmtwo$ (the case $\gctx = \tmtwo\,\gctx'$ is similar). Then $\off{\gctx}{\varthree}=\off{(\gctx'\,\tmtwo)}{\varthree}=\off{\gctx'}{\varthree}\,\tmtwo=\tm\sub{\var}{\vartwo}$. Thus $\tm=\tm_1\,\tm_2$ and  $\off{\gctx'}{\varthree} =\tm_1\sub{\var}{\vartwo}$ and $\tmtwo=\tm_2\sub{\var}{\vartwo}$. From the \ih there exists a context $\gctx_0$
  such that $\tm_1 = \off{\gctx_0'}{\varthree}$
  and $\gctx' = \gctx_0'\sub{\var}{\vartwo}$. We set $\gctx_0\eqdef \gctx_0'\,\tm_2\sub{\var}{\vartwo}$ and conclude.
  
  \item If $\gctx = \gctx'\esub{\var'}{\tmtwo}$ (the case $\gctx = \tmtwo\esub{\var}{\gctx'}$ is similar). Without loss of generality, we may assume that $\var'$ is distinct from $\var,\vartwo,\varthree$.  Then $\off{\gctx}{\varthree}=\off{(\gctx'\esub{\var'}{\tmtwo})}{\varthree}=\off{\gctx'}{\varthree}\esub{\var'}{\tmtwo}=\tm\sub{\var}{\vartwo}$. Thus $\tm=\tm_1\esub{\var'}{\tm_2}$ and  $\off{\gctx'}{\varthree} =\tm_1\sub{\var}{\vartwo}$ and $\tmtwo=\tm_2\sub{\var}{\vartwo}$. From the \ih there exists a context $\gctx_0$
  such that $\tm_1 = \off{\gctx_0'}{\varthree}$
  and $\gctx' = \gctx_0'\sub{\var}{\vartwo}$. We set $\gctx_0\eqdef \gctx_0'\esub{\var'}{\tm_2\sub{\var}{\vartwo}}$ and conclude.

  \end{enumerate}

  For the second item we proceed as follows. Suppose
  $\tm\sub{\var}{\vartwo} \tolsci \tmtwo$. Then two cases must apply depending on whether the $\tolsci$ step is a $\todb$-step or a $\tols$-step.

  \begin{enumerate}

  \item $\tm\sub{\var}{\vartwo} = \of{\gctx}{(\lam{\var}{\tmtwo})\sctx\,\tmthree}$. Then it must be the case that $\tm = \of{\gctx'}{(\lam{\var}{\tmtwo'})\sctx\,\tmthree'}$ with $\tmtwo=\tmtwo'\sub{\var}{\vartwo}$ and $\tmthree=\tmthree'\sub{\var}{\vartwo}$ and $\gctx=\gctx'\sub{\var}{\vartwo}$. We set $\tmtwo_0\eqdef \of{\gctx'}{(\lam{\var}{\tmtwo'})\sctx\,\tmthree'}$ and conclude.


    \item $\tm\sub{\var}{\vartwo} = \of{\gctx}{\off{\gctxtwo}{\var}\esub{\var}{\tmthree}}$. Then it must be the case that $\tm = \of{\gctx'}{\tm_1\esub{\var}{\tmthree'}}$ with $\off{\gctxtwo}{\var}=\tm_1\sub{\var}{\vartwo}$ and $\tmthree=\tmthree'\sub{\var}{\vartwo}$ and $\gctx=\gctx'\sub{\var}{\vartwo}$. By item 1,   there exists a context $\gctxtwo_0$
  such that $\tm_1 = \off{\gctxtwo_0}{\var}$
  and $\gctxtwo = \gctxtwo_0\sub{\var}{\vartwo}$. We set $\tmtwo_0\eqdef \of{\gctx'}{\off{\gctxtwo_0}{\var}\esub{\var}{\tmthree'}}$ and conclude.

  
\end{enumerate}

\end{proof}

\begin{lemma}[Backwards preservation of abstractions]
\llem{fusion_backwards_abstractions}
If $\tm \tofuseone (\lam{\var}{\tmtwo})\sctx$
then $\tm$ is of the form $\tm = (\lam{\var}{\tm_0})\sctx_0$
where for all $\tmthree$
whose free variables are not bound by $\sctx_0$
we have that $\tm_0\esub{\var}{\tmthree}\sctx_0 \tofuse \tmtwo\esub{\var}{\tmthree}\sctx$.
\end{lemma}
\begin{ifShortAppendix}
  \begin{proof}
  By induction on $\tm$.
  See the extended version~\cite{mells_long} for the detailed proof.
  \end{proof}
\end{ifShortAppendix}
\begin{ifLongAppendix}
  \begin{proof}
  We proceed by induction on $\tm$:
  \begin{enumerate}
  \item
    Variable, $\tm = \var$:
    this case is impossible, as there are no steps $\var \tofuseone (\lam{\var}{\tmtwo})\sctx$.
  \item
    Abstraction, $\tm = \lam{\var}{\tm'}$:
    we consider two subcases, depending on whether the fusion step
    is internal to $\tm'$ or derived from $\ruleFuseLam$ at the root:
    \begin{enumerate}
    \item
      If the fusion step is internal to $\tm'$,
      the situation is that
      we have that $\tm' \tofuseone \tm''$
      and $\tm = \lam{\var}{\tm'} \tofuseone \lam{\var}{\tm''} = (\lam{\var}{\tmtwo})\sctx$
      so $\sctx$ must be empty and $\tm'' = \tmtwo$.
      Taking $\tm_0 := \tm'$ and $\sctx_0 := \ctxhole$
      we have that
      $\tm = (\lam{\var}{\tm_0})\sctx_0$.
      Moreover, given an arbitrary term $\tmthree$
      we have that
      $\tm_0\esub{\var}{\tmthree}\sctx_0
      = \tm'\esub{\var}{\tmthree}
      \tofuseone \tm''\esub{\var}{\tmthree}
      = \tmtwo\esub{\var}{\tmthree}
      = \tmtwo\esub{\var}{\tmthree}\sctx$,
      as required.
    \item
      If the fusion step is derived from $\ruleFuseLam$ at the root,
      then $\tm' = \tm_1\esub{\vartwo}{\tm_2}$
      with $\var \notin \fv{\tm_2}$
      and the step is of the form
      $\tm
       = \lam{\var}{\tm_1\esub{\vartwo}{\tm_2}}
       \tofuseone (\lam{\var}{\tm_1})\esub{\vartwo}{\tm_2}
       = (\lam{\var}{\tmtwo})\sctx$,
      so $\sctx = \ctxhole\esub{\vartwo}{\tm_2}$
      and $\tmtwo = \tm_1$.
      Taking $\tm_0 := \tm' = \tm_1\esub{\vartwo}{\tm_2}$
      and $\sctx_0 = \ctxhole$
      we have that $\tm = (\lam{\var}{\tm_0})\sctx_0$.
      Moreover, given an arbitrary term $\tmthree$
      we have that
      $\tm_0\esub{\var}{\tmthree}\sctx_0
      = \tm_1\esub{\vartwo}{\tm_2}\esub{\var}{\tmthree}
      \tofuseone \tm_1\esub{\var}{\tmthree}\esub{\vartwo}{\tm_2}
      = \tmtwo\esub{\var}{\tmthree}\sctx$
      by $\ruleFuseEsL$.
    \end{enumerate}
  \item
    Application, $\tm = \tm_1\,\tm_2$:
    this case is impossible, as the fusion step
    can be either internal to $\tm_1$, internal to $\tm_2$,
    derived from $\ruleFuseAppL$ at the root,
    or derived from $\ruleFuseAppR$ at the root.
    In any of these cases, the right-hand side is not of the form
    $(\lam{\var}{\tmtwo})\sctx$.
  \item
    Substitution, $\tm = \tm_1\esub{\vartwo}{\tm_2}$:
    we consider six subcases, depending on whether the fusion step is
    internal to $\tm_1$, internal to $\tm_2$,
    or derived from any of the rules $\ruleFuseW$, $\ruleFuseC$,
    $\ruleFuseEsL$, $\ruleFuseEsR$ at the root:
    \begin{enumerate}
    \item
      If the fusion step is internal to $\tm_1$,
      then
      $\tm
       = \tm_1\esub{\vartwo}{\tm_2}
       \tofuseone \tm'_1\esub{\vartwo}{\tm_2}
       = (\lam{\var}{\tmtwo})\sctx$
      with $\tm_1 \tofuseone \tm'_1$.
      Hence $\sctx = \sctx'\esub{\vartwo}{\tm_2}$
      and $\tm'_1 = (\lam{\var}{\tmtwo})\sctx'$.
      Since $\tm_1 \tofuseone \tm'_1 = (\lam{\var}{\tmtwo})\sctx'$
      by \ih we have that
      $\tm_1 = (\lam{\var}{\tm_0})\sctx'_0$.
      Taking $\sctx_0 := \sctx'_0\esub{\vartwo}{\tm_2}$
      we have that
      $\tm
       = \tm_1\esub{\vartwo}{\tm_2}
       = (\lam{\var}{\tm_0})\sctx'_0\esub{\vartwo}{\tm_2}
       = (\lam{\var}{\tm_0})\sctx_0$.
      Moreover, if $\tmthree$ is a term whose free variables are not bound by $\sctx_0$,
      then
      $\tm_0\esub{\var}{\tmthree}\sctx_0
      = \tm_0\esub{\var}{\tmthree}\sctx'_0\esub{\vartwo}{\tm_2}
      \tofuse \tmtwo\esub{\var}{\tmthree}\sctx'\esub{\vartwo}{\tm_2}
      = \tmtwo\esub{\var}{\tmthree}\sctx$
      using the fact that
      $\tm_0\esub{\var}{\tmthree}\sctx'_0 \tofuse \tmtwo\esub{\var}{\tmthree}\sctx'$
      holds by \ih.
    \item
      If the fusion step is internal to $\tm_2$,
      then
      $\tm
      = \tm_1\esub{\vartwo}{\tm_2}
      \tofuseone \tm_1\esub{\vartwo}{\tm'_2}
      = (\lam{\var}{\tmtwo})\sctx$
      with $\tm_2 \tofuseone \tm'_2$.
      Then $\sctx = \sctx'\esub{\vartwo}{\tm'_2}$
      and $\tm_1 = (\lam{\var}{\tmtwo})\sctx'$.
      Taking $\tm_0 := \tmtwo$ and $\sctx_0 := \sctx = \sctx'\esub{\vartwo}{\tm_2}$,
      we have that
      $\tm
      = \tm_1\esub{\vartwo}{\tm_2}
      = (\lam{\var}{\tmtwo})\sctx'\esub{\vartwo}{\tm_2}
      = (\lam{\var}{\tm_0})\sctx_0$.
      Moreover, if $\tmthree$ is a term whose free variables are not bound by $\sctx_0$,
      then
      $\tm_0\esub{\var}{\tmthree}\sctx_0
      = \tmtwo\esub{\var}{\tmthree}\sctx'\esub{\vartwo}{\tm_2}
      \tofuseone \tmtwo\esub{\var}{\tmthree}\sctx'\esub{\vartwo}{\tm'_2}
      = \tmtwo\esub{\var}{\tmthree}\sctx$,
      using the fact that $\tm_2 \tofuseone \tm'_2$.
    \item
      If the fusion step is derived from $\ruleFuseW$ at the root,
      then
      $\tm
      = \tm_1\esub{\vartwo}{\tm_2}
      \tofuseone \tm_1
      = (\lam{\var}{\tmtwo})\sctx$,
      where $\vartwo\notin\fv{\tm_1}$.
      Taking $\tm_0 := \tmtwo$ and $\sctx_0 := \sctx\esub{\vartwo}{\tm_2}$,
      we have that
      $\tm
      = \tm_1\esub{\vartwo}{\tm_2}
      = (\lam{\var}{\tmtwo})\sctx\esub{\vartwo}{\tm_2}
      = (\lam{\var}{\tm_0})\sctx_0$.
      Moreover, if $\tmthree$ is a term whose free variables are not bound by $\sctx$,
      we have that
      $\tm_0\esub{\var}{\tmthree}\sctx_0
      = \tmtwo\esub{\var}{\tmthree}\sctx\esub{\vartwo}{\tm_2}
      \tofuseone \tmtwo\esub{\var}{\tmthree}\sctx$.
      For the last step, note that
      $\vartwo \notin \fv{\tm_1}$ by hypothesis
      and
      $\vartwo \notin \fv{\tmthree}$ because the free variables of $\tmthree$
      are not bound by $\sctx$.
      Hence
      $\vartwo
       \notin \fv{\tm_1} \cup \fv{\tmthree}
       = \fv{(\lam{\var}{\tmtwo})\sctx} \cup \fv{\tmthree}
       = \fv{\tmtwo\esub{\var}{\tmthree}\sctx}$.
    \item
      If the fusion step is derived from $\ruleFuseC$ at the root,
      then
      $\tm
      = \tm_1\esub{\vartwo}{\tm_2}\esub{\varthree}{\tm_2}
      \tofuseone \tm_1\sub{\vartwo}{\varthree}\esub{\varthree}{\tm_2}
      = (\lam{\var}{\tmtwo})\sctx$,
      so $\sctx = \sctx'\esub{\varthree}{\tm_2}$
      and $\tm_1\sub{\vartwo}{\varthree} = (\lam{\var}{\tmtwo})\sctx'$.
      This means that $\tm_1 = (\lam{\var}{\tm_0})\sctx'_0$
      where $\tm_0\sub{\vartwo}{\varthree} = \tmtwo$
      and $\sctx'_0\sub{\vartwo}{\varthree} = \sctx'$.
      Taking $\sctx_0 := \sctx'_0\esub{\vartwo}{\tm_2}\esub{\varthree}{\tm_2}$
      we have that
      $\tm
      = \tm_1\esub{\vartwo}{\tm_2}\esub{\varthree}{\tm_2}
      = (\lam{\var}{\tm_0})\sctx'_0\esub{\vartwo}{\tm_2}\esub{\varthree}{\tm_2}
      = (\lam{\var}{\tm_0})\sctx_0$.
      Moreover, if $\tmthree$ is a term whose free variables are not bound by $\sctx_0$,
      then:
      \[
        \begin{array}{rcll}
          \tm_0\esub{\var}{\tmthree}\sctx_0
        & = &
          \tm_0\esub{\var}{\tmthree}\sctx'_0\esub{\vartwo}{\tm_2}\esub{\varthree}{\tm_2}
        \\
        & \tofuseone &
          (\tm_0\esub{\var}{\tmthree}\sctx'_0)\sub{\vartwo}{\varthree}\esub{\varthree}{\tm_2}
        \\
        & = &
          \tm_0\sub{\vartwo}{\varthree}\esub{\var}{\tmthree\sub{\vartwo}{\varthree}}\sctx'_0\sub{\vartwo}{\varthree}\esub{\varthree}{\tm_2}
        \\
        & = &
          \tm_0\sub{\vartwo}{\varthree}\esub{\var}{\tmthree}\sctx'_0\sub{\vartwo}{\varthree}\esub{\varthree}{\tm_2}
          & \text{as $\vartwo \notin \fv{\tmthree}$}
        \\
        & = &
          \tmtwo\esub{\var}{\tmthree}\sctx'\esub{\varthree}{\tm_2}
        \\
        & = &
          \tmtwo\esub{\var}{\tmthree}\sctx
        \end{array}
      \]
    \item
      If the fusion step is derived from $\ruleFuseEsL$ at the root,
      then $\tm_1 = \tm_{11}\esub{\varthree}{\tm_{12}}$
      where $\vartwo \notin \fv{\tm_{12}}$
      and $\varthree \notin \fv{\tm_2}$,
      and the step is of the form
      $\tm
      = \tm_{11}\esub{\varthree}{\tm_{12}}\esub{\vartwo}{\tm_2}
      \tofuseone \tm_{11}\esub{\vartwo}{\tm_2}\esub{\varthree}{\tm_{12}}
      = (\lam{\var}{\tmtwo})\sctx$.
      Then $\sctx = \sctx'\esub{\vartwo}{\tm_2}\esub{\varthree}{\tm_{12}}$
      and $\tm_{11} = (\lam{\var}{\tmtwo})\sctx'$.
      Taking $\tm_0 := \tmtwo$ and $\sctx_0 := \sctx'\esub{\varthree}{\tm_{12}}\esub{\vartwo}{\tm_2}$
      we have that
      $\tm
      = \tm_{11}\esub{\varthree}{\tm_{12}}\esub{\vartwo}{\tm_2}
      = (\lam{\var}{\tmtwo})\sctx'\esub{\varthree}{\tm_{12}}\esub{\vartwo}{\tm_2}
      = (\lam{\var}{\tm_0})\sctx_0$.
      Moreover, if $\tmthree$ is a term whose free variables are not bound by $\sctx_0$,
      then
      $\tm_0\esub{\var}{\tmthree}\sctx_0
      = \tmtwo\esub{\var}{\tmthree}\sctx'\esub{\varthree}{\tm_{12}}\esub{\vartwo}{\tm_2}
      \tofuseone \tmtwo\esub{\var}{\tmthree}\sctx'\esub{\vartwo}{\tm_2}\esub{\varthree}{\tm_{12}}
      = \tmtwo\esub{\var}{\tmthree}\sctx$.
    \item
      If the fusion step is derived from $\ruleFuseEsR$ at the root,
      then $\tm_2 = \tm_{21}\esub{\varthree}{\tm_{22}}$
      where $\varthree \notin \fv{\tm_1}$,
      and the step is of the form
      $\tm
      = \tm_1\esub{\vartwo}{\tm_{21}\esub{\varthree}{\tm_{22}}}
      \tofuseone \tm_1\esub{\vartwo}{\tm_{21}}\esub{\varthree}{\tm_{22}}
      = (\lam{\var}{\tmtwo})\sctx$.
      Then $\sctx = \sctx'\esub{\vartwo}{\tm_{21}}\esub{\varthree}{\tm_{22}}$
      and $\tm_1 = (\lam{\var}{\tmtwo})\sctx'$.
      Taking $\tm_0 := \tmtwo$ and $\sctx_0 := \sctx'\esub{\vartwo}{\tm_{21}\esub{\varthree}{\tm_{22}}}$
      we have that 
      $\tm
      = \tm_1\esub{\vartwo}{\tm_{21}\esub{\varthree}{\tm_{22}}}
      = (\lam{\var}{\tmtwo})\sctx'\esub{\vartwo}{\tm_{21}\esub{\varthree}{\tm_{22}}}
      = (\lam{\var}{\tm_0})\sctx_0$.
      Moreover, if $\tmthree$ is a term whose free variables are not bound by $\sctx_0$,
      then
      $\tm_0\esub{\var}{\tmthree}\sctx_0
      = \tmtwo\esub{\var}{\tmthree}\sctx'\esub{\vartwo}{\tm_{21}\esub{\varthree}{\tm_{22}}}
      \tofuseone \tmtwo\esub{\var}{\tmthree}\sctx'\esub{\vartwo}{\tm_{21}}\esub{\varthree}{\tm_{22}}
      = \tmtwo\esub{\var}{\tmthree}\sctx$.
    \end{enumerate}
  \end{enumerate}
  \end{proof}
\end{ifLongAppendix}

\begin{lemma}[Backwards preservation of variables]
\llem{fusion_backwards_variables}
If $\tm \tofuseone \off{\gctx}{\var}$
then $\tm$ is of the form $\tm = \off{\gctx_0}{\var,\hdots,\var}$
where $\gctx_0$ is a context with either one or two holes,
and for any term $\tmthree$
whose free variables are not bound by $\gctx_0$
one has that $\off{\gctx_0}{\tmthree,\hdots,\tmthree} \tofuse \off{\gctx}{\tmthree}$.
\end{lemma}
\begin{ifShortAppendix}
  \begin{proof}
  By induction on $\tm$.
  See the extended version~\cite{mells_long} for the detailed proof.
  \end{proof}
\end{ifShortAppendix}
\begin{ifLongAppendix}
  \begin{proof}
  By induction on $\tm$:
  \begin{enumerate}
  \item
    Variable, $\tm = \vartwo$:
    this case is impossible, as there are no steps $\vartwo \tofuse \off{\gctx}{\var}$.
  \item
    Abstraction, $\tm = \lam{\vartwo}{\tm'}$:
    we consider two subcases, depending on whether the fusion step is internal
    to $\tm'$ or derived from $\ruleFuseLam$ at the root:
    \begin{enumerate}
    \item
      If the fusion step is internal to $\tm'$,
      then
      $\tm
      = \lam{\vartwo}{\tm'}
      \tofuseone \lam{\vartwo}{\off{\gctx'}{\var}}
      = \off{\gctx}{\var}$
      where $\gctx = \lam{\vartwo}{\gctx'}$
      and $\tm' \tofuseone \off{\gctx'}{\var}$.
      By \ih, we have that $\tm' = \off{\gctx'_0}{\var,\hdots,\var}$
      and $\off{\gctx'_0}{\tmthree,\hdots,\tmthree} \tofuse \off{\gctx'}{\tmthree}$.
      Taking $\gctx_0 := \lam{\vartwo}{\gctx'_0}$
      we have that
      $\tm
      = \lam{\vartwo}{\tm'}
      = \lam{\vartwo}{\off{\gctx'_0}{\var,\hdots,\var}}
      = \off{\gctx_0}{\var,\hdots,\var}$,
      and
      $\off{\gctx_0}{\tmthree,\hdots,\tmthree}
      = \lam{\vartwo}{\off{\gctx'_0}{\tmthree,\hdots,\tmthree}}
      \tofuse \lam{\vartwo}{\off{\gctx'}{\tmthree}}
      = \off{\gctx}{\tmthree}$.
    \item
      If the fusion step is derived from $\ruleFuseLam$ at the root,
      then $\tm' = \tm_1\esub{\varthree}{\tm_2}$
      with $\vartwo\notin\fv{\tm_2}$,
      and the step is of the form
      $\tm
       = \lam{\vartwo}{\tm_1\esub{\varthree}{\tm_2}}
       \tofuseone (\lam{\vartwo}{\tm_1})\esub{\varthree}{\tm_2}
       = \off{\gctx}{\var}$.
      There are two subcases, depending on whether the hole of $\gctx$
      lies inside $\tm_1$ or inside $\tm_2$.
      We only check the first of these cases, the other one being similar.
      Indeed, suppose that the hole of $\gctx$ lies inside $\tm_1$.
      Then $\gctx = (\lam{\vartwo}{\gctx'})\esub{\varthree}{\tm_2}$
      and $\tm_1 = \off{\gctx'}{\var}$.
      Taking $\gctx_0 := \lam{\vartwo}{\gctx'\esub{\varthree}{\tm_2}}$
      we have that
      $\tm
      = \lam{\vartwo}{\tm_1\esub{\varthree}{\tm_2}}
      = \lam{\vartwo}{\off{\gctx'}{\var}\esub{\varthree}{\tm_2}}
      = \off{\gctx_0}{\var}$
      and
      $\off{\gctx_0}{\tmthree}
      = \lam{\vartwo}{\off{\gctx'}{\tmthree}\esub{\varthree}{\tm_2}}
      \tofuseone (\lam{\vartwo}{\off{\gctx'}{\tmthree}})\esub{\varthree}{\tm_2}
      = \off{\gctx}{\tmthree}$.
    \end{enumerate}
  \item
    Application, $\tm = \tm_1\,\tm_2$:
    we consider four subcases, depending on whether the fusion step is
    internal to $\tm_1$, internal to $\tm_2$,
    derived from $\ruleFuseAppL$ at the root, or
    derived from $\ruleFuseAppR$ at the root:
    \begin{enumerate}
    \item
      \label{fusion_backwards_variables:case_app_internal}
      If the fusion step is internal to $\tm_1$,
      then $\tm = \tm_1\,\tm_2 \tofuseone \tm'_1\,\tm_2 = \off{\gctx}{\var}$
      with $\tm_1 \tofuseone \tm'_1$.
      There are two subcases, depending on whether the hole of $\gctx$
      lies inside $\tm'_1$ or inside $\tm_2$:
      \begin{enumerate}
      \item
        If the hole of $\gctx$ lies inside $\tm'_1$,
        then $\tm'_1 = \off{\gctx'}{\var}$ and $\gctx = \gctx'\,\tm_2$
        and $\tm_1 \tofuseone \off{\gctx'}{\var}$.
        By \ih, we have that $\tm_1 = \off{\gctx'_0}{\var,\hdots,\var}$
        and $\off{\gctx'_0}{\tmthree,\hdots,\tmthree} \tofuse \off{\gctx'}{\tmthree}$.
        Taking $\gctx_0 := \gctx'_0\,\tm_2$
        we have that
        $\tm
        = \tm_1\,\tm_2
        = \off{\gctx'_0}{\var,\hdots,\var}\,\tm_2
        = \off{\gctx_0}{\var,\hdots,\var}$,
        and
        $\off{\gctx_0}{\tmthree,\hdots,\tmthree}
        = \off{\gctx'_0}{\tmthree,\hdots,\tmthree}\,\tm_2
        \tofuse \off{\gctx'}{\tmthree}\,\tm_2
        = \off{\gctx}{\tmthree}$.
      \item
        If the hole of $\gctx$ lies inside $\tm_2$,
        then $\tm_2 = \off{\gctx'}{\var}$ and $\gctx = \tm'_1\,\gctx'$.
        Taking $\gctx_0 := \tm_1\,\gctx'$
        we have that
        $\tm
        = \tm_1\,\tm_2
        = \tm_1\,\off{\gctx'}{\var}
        = \off{\gctx_0}{\var}$
        and
        $\off{\gctx_0}{\tmthree}
        = \tm_1\,\gctx'{\tmthree}
        \tofuseone \tm'_1\,\gctx'{\tmthree}
        = \off{\gctx}{\tmthree}$.
      \end{enumerate}
    \item
      If the fusion step is internal to $\tm_2$,
      the proof is similar to case~\ref{fusion_backwards_variables:case_app_internal}.
    \item
      \label{fusion_backwards_variables:case_app_root}
      If the fusion step is derived from $\ruleFuseAppL$ at the root,
      then $\tm_1 = \tm_{11}\esub{\vartwo}{\tm_{12}}$
      where $\vartwo \notin \fv{\tm_2}$
      and the step is of the form
      $\tm
      = \tm_{11}\esub{\vartwo}{\tm_{12}}\,\tm_2
      \tofuse (\tm_{11}\,\tm_2)\esub{\vartwo}{\tm_{12}}
      = \off{\gctx}{\var}$.
      There are three similar subcases, depending on whether the hole of $\gctx$
      lies inside $\tm_{11}$, inside $\tm_2$, or inside $\tm_{12}$.
      We only check the first of these cases, the other ones being similar.
      Indeed, suppose that the hole of $\gctx$ lies inside $\tm_{11}$.
      Then $\gctx = (\gctx'\,\tm_2)\esub{\vartwo}{\tm_{12}}$
      and $\tm_{11} = \off{\gctx'}{\var}$.
      Taking $\gctx_0 := \gctx'\esub{\vartwo}{\tm_{12}}\,\tm_2$
      we have that
      $\tm
      = \tm_{11}\esub{\vartwo}{\tm_{12}}\,\tm_2
      = \off{\gctx'}{\var}\esub{\vartwo}{\tm_{12}}\,\tm_2
      = \off{\gctx_0}{\var}$
      and that
      $\off{\gctx_0}{\tmthree}
      = \off{\gctx'}{\tmthree}\esub{\vartwo}{\tm_{12}}\,\tm_2
      \tofuseone (\off{\gctx'}{\tmthree}\,\tm_2)\esub{\vartwo}{\tm_{12}}
      = \off{\gctx}{\tmthree}$.
    \item
      If the fusion step is derived from $\ruleFuseAppR$ at the root,
      the proof is similar to case~\ref{fusion_backwards_variables:case_app_root}.
    \end{enumerate}
  \item
    Substitution, $\tm = \tm_1\esub{\vartwo}{\tm_2}$:
    we consider six subcases, depending on whether the fusion step is
    internal to $\tm_1$, internal to $\tm_2$,
    or derived from one of the rules
    $\ruleFuseW$, $\ruleFuseC$, $\ruleFuseEsL$, or $\ruleFuseEsR$ at the root:
    \begin{enumerate}
    \item
      If the fusion step is internal to $\tm_1$, then
      $\tm
      = \tm_1\esub{\vartwo}{\tm_2}
      \tofuseone \tm'_1\esub{\vartwo}{\tm_2}
      = \off{\gctx}{\var}$
      with $\tm_1 \tofuseone \tm'_1$
      and the proof is similar to case~\ref{fusion_backwards_variables:case_app_internal}.
    \item
      If the fusion step is internal to $\tm_2$,
      then
      $\tm
      = \tm_1\esub{\vartwo}{\tm_2}
      \tofuseone \tm_1\esub{\vartwo}{\tm'_2}
      = \off{\gctx}{\var}$
      with $\tm_2 \tofuseone \tm'_2$
      and the proof is similar to case~\ref{fusion_backwards_variables:case_app_internal}.
    \item
      If the fusion step is derived from $\ruleFuseW$ at the root,
      then $\vartwo \notin \fv{\tm_1}$ and
      $\tm
      = \tm_1\esub{\vartwo}{\tm_2}
      \tofuseone \tm_1
      = \off{\gctx}{\var}$.
      Taking $\gctx_0 := \gctx\esub{\vartwo}{\tm_2}$
      we have that
      $\tm
      = \tm_1\esub{\vartwo}{\tm_2}
      = \off{\gctx}{\var}\esub{\vartwo}{\tm_2}
      = \off{\gctx_0}{\var}$
      and
      $\off{\gctx_0}{\tmthree}
      = \off{\gctx}{\tmthree}\esub{\vartwo}{\tm_2}
      \tofuseone \off{\gctx}{\tmthree}$.
      The last step an instance of the $\ruleFuseW$ rule.
      To justify that this step can be applied, note that
      $\vartwo \notin \fv{\tm_1} = \fv{\off{\gctx}{\var}}$
      so in particular $\vartwo \notin \fv{\gctx}$,
      and the free variables of $\tmthree$
      are not bound by $\gctx_0$ by hypothesis,
      so $\vartwo \notin \fv{\tmthree}$.
      Hence $\vartwo \notin \fv{\gctx} \cup \fv{\tmthree} = \fv{\off{\gctx}{\tmthree}}$.
    \item
      If the fusion step is derived from $\ruleFuseC$ at the root,
      then $\tm_1 = \tm'_{1}\esub{\varthree}{\tm_2}$
      and the step is of the form
      $\tm
      = \tm'_{1}\esub{\varthree}{\tm_2}\esub{\vartwo}{\tm_2}
      \tofuseone \tm'_{1}\sub{\varthree}{\vartwo}\esub{\vartwo}{\tm_2}
      = \off{\gctx}{\var}$.
      There are two subcases, depending on whether the hole of $\gctx$
      lies inside $\tm'_{1}\sub{\varthree}{\vartwo}$ or inside $\tm_2$:
      \begin{enumerate}
      \item
        If the hole of $\gctx$ lies inside $\tm'_{1}\sub{\varthree}{\vartwo}$,
        then $\gctx = \gctx'\esub{\vartwo}{\tm_2}$
        and $\off{\gctx'}{\var} = \tm'_{1}\sub{\varthree}{\vartwo}$.
        Since $\var \neq \vartwo$,
        by \rlem{reduction_before_fusion}
        there exists a context $\gctx'_0$ such that
        $\tm'_1 = \off{\gctx'_0}{\var}$
        and $\gctx' = \gctx'_0\sub{\varthree}{\vartwo}$.
        Taking $\gctx_0 := \gctx'_0\esub{\varthree}{\tm_2}\esub{\vartwo}{\tm_2}$
        we have that
        $\tm
        = \tm'_{1}\esub{\varthree}{\tm_2}\esub{\vartwo}{\tm_2}
        = \off{\gctx'_0}{\var}\esub{\varthree}{\tm_2}\esub{\vartwo}{\tm_2}
        = \off{\gctx_0}{\var}$
        and:
        \[
          \begin{array}{rcll}
            \off{\gctx_0}{\tmthree}
          & = &
            \off{\gctx'_0}{\tmthree}\esub{\varthree}{\tm_2}\esub{\vartwo}{\tm_2}
          \\
          & \tofuseone &
            \off{\gctx'_0}{\tmthree}\sub{\varthree}{\vartwo}\esub{\vartwo}{\tm_2}
          \\
          & = &
            \off{\gctx'_0\sub{\varthree}{\vartwo}}{\tmthree\sub{\varthree}{\vartwo}}\esub{\vartwo}{\tm_2}
          \\
          & = &
            \off{\gctx'_0\sub{\varthree}{\vartwo}}{\tmthree}\esub{\vartwo}{\tm_2}
            & \text{as $\varthree \notin \fv{\tmthree}$}
          \\
          & = &
            \off{\gctx'}{\tmthree}\esub{\vartwo}{\tm_2}
          \\
          & = &
            \off{\gctx}{\tmthree}
          \end{array}
        \]
      \item
        If the hole of $\gctx$ lies inside $\tm_2$,
        then $\gctx = \tm'_1\sub{\varthree}{\vartwo}\esub{\vartwo}{\gctx'}$
        and $\off{\gctx'}{\var} = \tm_2$.
        Taking $\gctx_0$ as the two-hole context
        $\gctx_0 := \tm'_1\esub{\varthree}{\gctx'}\esub{\vartwo}{\gctx'}$
        we have that:
        \[
          \begin{array}{rcl}
            \tm
          & = &
            \tm'_1\esub{\varthree}{\tm_2}\esub{\vartwo}{\tm_2}
          \\
          & = &
            \tm'_1\esub{\varthree}{\off{\gctx'}{\var}}\esub{\vartwo}{\off{\gctx'}{\var}}
          \\
          & = &
            \off{(\tm'_1\esub{\varthree}{\gctx'}\esub{\vartwo}{\gctx'})}{\var,\var}
          \\
          & = &
            \off{\gctx_0}{\var,\var}
          \end{array}
        \]
        and that:
        \[
          \begin{array}{rcl}
            \off{\gctx_0}{\tmthree,\tmthree}
          & = &
            \off{(\tm'_1\esub{\varthree}{\gctx'}\esub{\vartwo}{\gctx'})}{\tmthree,\tmthree}
          \\
          & = &
            \tm'_1\esub{\varthree}{\off{\gctx'}{\tmthree}}\esub{\vartwo}{\off{\gctx'}{\tmthree}}
          \\
          & \tofuseone &
            \tm'_1\sub{\varthree}{\vartwo}\esub{\vartwo}{\off{\gctx'}{\tmthree}}
          \\
          & = &
            \off{\gctx}{\tmthree}
          \end{array}
        \]
        Note that this is the only base case in which a context $\gctx_0$ with more
        than one hole is needed.
      \end{enumerate}
    \item
      If the fusion step is derived from $\ruleFuseEsL$ at the root,
      then $\tm_1 = \tm_{11}\esub{\varthree}{\tm_{12}}$
      with $\vartwo \notin \fv{\tm_{12}}$ and $\varthree \notin \fv{\tm_2}$,
      and the step is of the form
      $\tm
      = \tm_{11}\esub{\varthree}{\tm_{12}}\esub{\vartwo}{\tm_2}
      \tofuseone \tm_{11}\esub{\vartwo}{\tm_2}\esub{\varthree}{\tm_{12}}
      = \off{\gctx}{\var}$.
      There are three similar subcases, depending on whether the hole
      of $\gctx$ lies inside $\tm_{11}$, inside $\tm_2$, or inside $\tm_{12}$.
      We only check the first of these cases, the other ones being similar.
      Indeed, suppose that the hole of $\gctx$ lies inside $\tm_{11}$.
      Then $\gctx = \gctx'\esub{\vartwo}{\tm_2}\esub{\varthree}{\tm_{12}}$
      and $\tm_{11} = \off{\gctx'}{\var}$.
      Taking $\gctx_0 := \gctx'\esub{\varthree}{\tm_{12}}\esub{\vartwo}{\tm_2}$
      we have that
      $\tm
      = \tm_{11}\esub{\varthree}{\tm_{12}}\esub{\vartwo}{\tm_2}
      = \off{\gctx'}{\var}\esub{\varthree}{\tm_{12}}\esub{\vartwo}{\tm_2}
      = \off{\gctx_0}{\var}$
      and
      $\off{\gctx_0}{\tmthree}
      = \off{\gctx'}{\tmthree}\esub{\varthree}{\tm_{12}}\esub{\vartwo}{\tm_2}
      \tofuseone \off{\gctx'}{\tmthree}\esub{\vartwo}{\tm_2}\esub{\varthree}{\tm_{12}}
      = \off{\gctx}{\tmthree}$.
    \item
      If the fusion step is derived from $\ruleFuseEsR$ at the root,
      then $\tm_2 = \tm_{21}\esub{\varthree}{\tm_{22}}$
      with $\varthree \notin \fv{\tm_1}$,
      and the step is of the form
      $\tm
      = \tm_1\esub{\vartwo}{\tm_{21}\esub{\varthree}{\tm_{22}}}
      \tofuseone \tm_1\esub{\vartwo}{\tm_{21}}\esub{\varthree}{\tm_{22}}
      = \off{\gctx}{\var}$.
      There are three similar subcases, depending on whether the hole
      of $\gctx$ lies inside $\tm_1$, inside $\tm_{21}$, or inside $\tm_{22}$.
      We only check the first of these cases, the other ones being similar.
      Indeed, suppose that the hole of $\gctx$ lies inside $\tm_1$.
      Then $\gctx = \gctx'\esub{\vartwo}{\tm_{21}}\esub{\varthree}{\tm_{22}}$
      and $\tm_1 = \off{\gctx'}{\var}$.
      Taking $\gctx_0 := \gctx'\esub{\vartwo}{\tm_{21}\esub{\varthree}{\tm_{22}}}$
      we have that
      $\tm
      = \tm_1\esub{\vartwo}{\tm_{21}\esub{\varthree}{\tm_{22}}}
      = \off{\gctx'}{\var}\esub{\vartwo}{\tm_{21}\esub{\varthree}{\tm_{22}}}
      = \off{\gctx_0}{\var}$
      and
      $\off{\gctx_0}{\tmthree}
      = \off{\gctx'}{\tmthree}\esub{\vartwo}{\tm_{21}\esub{\varthree}{\tm_{22}}}
      \tofuseone \off{\gctx'}{\tmthree}\esub{\vartwo}{\tm_{21}}\esub{\varthree}{\tm_{22}}
      = \off{\gctx}{\tmthree}$.
    \end{enumerate}
  \end{enumerate}
  \end{proof}
\end{ifLongAppendix}

\begin{lemma}[Postponement of single fusion step]
\llem{postponement_tofuseone}
${\tofuseone\,\tolsci} \subseteq {\tolsci^+\,\tofuse}$
\end{lemma}
\begin{ifShortAppendix}
  \begin{proof}
  By exhaustive case analysis of the diagrams.
  See the extended version~\cite{mells_long} for the detailed proof.
  \end{proof}
\end{ifShortAppendix}
\begin{ifLongAppendix}
  \begin{proof}
  Let $\tm \tofuseone \tmtwo \tolsci \tmthree$
  and let us check that there exists a term $\tmfour$
  such that $\tm \tolsci^+ \tmfour \tofuse \tmthree$.
  Graphically:
  \[
    \diagramFuseLSC{\tm}{\tmtwo}{\tmthree}{\tmfour}
  \]
  We proceed by induction on $\tm$:
  \begin{enumerate}
  \item
    Variable, $\tm = \var$:
    note that there is no $\tmtwo$ such that $\tm \tofuse \tmtwo$,
    so this case is impossible.
  \item
    Abstraction, $\tm = \lam{\var}{\tm'}$:
    we consider two subcases, depending on whether
    the fusion step $\tm = \lam{\var}{\tm'} \tofuseone \tmtwo$
    is internal to $\tm'$ or derived from $\ruleFuseLam$ at the root:
    \begin{enumerate}
    \item
      If the fusion step is internal,
      then $\tm = \lam{\var}{\tm'} \tofuseone \lam{\var}{\tmtwo'} = \tmtwo$
      with $\tmtwo \tofuseone \tmtwo'$.
      Moreover, the reduction step
      $\tmtwo = \lam{\var}{\tmtwo'} \tolsci \tmthree$
      must be internal, so
      $\tmtwo = \lam{\var}{\tmtwo'} \tolsci \lam{\var}{\tmthree'} = \tmthree$
      with $\tmtwo' \tolsci \tmthree'$. Then:
      \[
         \diagramFuseLSC{\tm'}{\tmtwo'}{\tmthree'}{\tmfour'}
         \HS\text{ by \ih, so }\HS
         \diagramFuseLSC{\lam{\var}{\tm'}}{\lam{\var}{\tmtwo'}}{\lam{\var}{\tmthree'}}{\lam{\var}{\tmfour'}}
      \]
    \item
      If the fusion step is derived from $\ruleFuseLam$ at the root,
      then
      $\tm = \lam{\var}{\tm_1\esub{\vartwo}{\tm_2}}
       \tofuse (\lam{\var}{\tm_1})\esub{\vartwo}{\tm_2} = \tmtwo$.
      There are three further subcases, depending on whether
      the reduction step
       $\tmtwo = (\lam{\var}{\tm_1})\esub{\vartwo}{\tm_2} \tolsci \tmthree$
      is internal to $\tm_1$, internal to $\tm_2$, or a $\tols$ step at the
      root:
      \begin{enumerate}
      \item
        If the reduction step is internal to $\tm_1$,
        then the situation is:
        \[
          \diagramFuseOneLSCOne{
            \lam{\var}{\tm_1\esub{\vartwo}{\tm_2}}
          }{
            (\lam{\var}{\tm_1})\esub{\vartwo}{\tm_2}
          }{
            (\lam{\var}{\tm'_1})\esub{\vartwo}{\tm_2}
          }{
            \lam{\var}{\tm'_1\esub{\vartwo}{\tm_2}}
          }
          \HS\text{where $\tm_1 \tolsci \tm'_1$}
        \]
      \item
        If the reduction step is internal to $\tm_2$,
        it is similar to the previous case.
      \item
        If the reduction step is derived from $\tols$ at the root:
        then $\tm_1 = \off{\gctx}{\vartwo}$ and the situation is:
        \[
          \diagramFuseOneLSCOne{
            \lam{\var}{\off{\gctx}{\vartwo}\esub{\vartwo}{\tm_2}}
          }{
            (\lam{\var}{\off{\gctx}{\vartwo}})\esub{\vartwo}{\tm_2}
          }{
            (\lam{\var}{\off{\gctx}{\tm_2}})\esub{\vartwo}{\tm_2}
          }{
            \lam{\var}{\off{\gctx}{\tm_2}\esub{\vartwo}{\tm_2}}
          }
        \] 
      \end{enumerate}
    \end{enumerate}
  \item
    Application, $\tm = \tm_1\,\tm_2$:
    we consider four subcases, depending on whether the fusion step
    $\tm = \tm_1\,\tm_2 \tofuseone \tmtwo$
    is internal to $\tm_1$, internal to $\tm_2$,
    derived from $\ruleFuseAppL$ at the root,
    or derived from $\ruleFuseAppR$ at the root:
    \begin{enumerate}
    \item
      If the fusion step is internal to $\tm_1$,
      then $\tm = \tm_1\,\tm_2 \tofuseone \tmtwo_1\,\tm_2 = \tmtwo$
      with $\tm_1 \tofuseone \tmtwo_1$.
      There are three subcases, depending on whether the reduction step
      is internal to $\tmtwo_1$, internal to $\tm_2$, or a $\todb$ step at the root:
      \begin{enumerate}
      \item
        \label{postponement_tofuseone:case_app_left_ih}
        If the reduction step is internal to $\tmtwo_1$:
        then $\tmtwo = \tmtwo_1\,\tm_2 \tolsci \tmthree_1\,\tm_2 = \tmthree$
        with $\tmtwo_1 \tolsci \tmthree_1$. Then the diagram can be closed by
        resorting to the \ih:
        \[
           \diagramFuseLSC{\tm_1}{\tmtwo_1}{\tmthree_1}{\tmfour_1}
           \HS\text{ by \ih, so }\HS
           \diagramFuseLSC{\tm_1\,\tm_2}{\tmtwo_1\,\tm_2}{\tmthree_1\,\tm_2}{\tmfour_1\,\tm_2}
        \]
      \item
        \label{postponement_tofuseone:case_app_left_disjoint}
        If the reduction step is internal to $\tm_2$:
        then $\tmtwo = \tmtwo_1\,\tm_2 \tolsci \tmtwo_1\,\tmthree_2 = \tmthree$
        with $\tmtwo_1 \tolsci \tmthree_1$. Then the steps are disjoint and
        the diagram can be immediately closed as follows:
        \[
           \diagramFuseOneLSCOne{\tm_1\,\tm_2}{\tmtwo_1\,\tm_2}{\tmtwo_1\,\tmthree_2}{\tm_1\,\tmthree_2}
        \]
      \item
        If the reduction step is a $\todb$ step at the root:
        then $\tmtwo_1 = (\lam{\var}{\tmtwo'_1})\sctx$.
        Since $\tm_1 \tofuseone \tmtwo_1$,
        by \rlem{fusion_backwards_abstractions}
        we know that $\tm_1$ must be of the form
        $\tm_1 = (\lam{\var}{\tm_0})\sctx_0$
        in such a way that
        $\tm_0\esub{\var}{\tm_2}\sctx_0 \tofuse \tmtwo'_1\esub{\var}{\tm_2}\sctx$.
        Hence the situation is:
        \[
           \diagramFuseLSC{
             (\lam{\var}{\tm_0})\sctx_0\,\tm_2
           }{
             (\lam{\var}{\tmtwo'_1})\sctx\,\tm_2
           }{
             \tmtwo'_1\esub{\var}{\tm_2}\sctx
           }{
             \tm_0\esub{\var}{\tm_2}\sctx_0
           }
        \]
      \end{enumerate}
    \item
      If the fusion step is internal to $\tm_2$,
      then $\tm = \tm_1\,\tm_2 \tofuseone \tm_1\,\tmtwo_2$
      with $\tm_2 \tofuseone \tmtwo_2$.
      There are three subcases, depending on whether the reduction step
      is internal to $\tm_1$, internal to $\tmtwo_2$, or a $\todb$ step at the root:
      \begin{enumerate}
      \item
        If the reduction step is internal to $\tm_1$:
        the steps are disjoint and the diagram can be closed similarly as for
        case~\ref{postponement_tofuseone:case_app_left_disjoint}.
      \item
        If the reduction step is internal to $\tmtwo_1$:
        then the diagram can be closed by resorting to the \ih, similarly as for
        case~\ref{postponement_tofuseone:case_app_left_ih}.
      \item
        If the reduction step is a $\todb$ step at the root:
        then $\tm_1 = (\lam{\var}{\tm'_1})\sctx$. Hence the situation is:
        \[
          \diagramFuseOneLSCOne{
            (\lam{\var}{\tm'_1})\sctx\,\tm_2
          }{
            (\lam{\var}{\tm'_1})\sctx\,\tmtwo_2
          }{
            \tm'_1\esub{\var}{\tmtwo_2}\sctx
          }{
            \tm'_1\esub{\var}{\tm_2}\sctx
          }
        \]
      \end{enumerate}
    \item
      If the fusion step is derived from $\ruleFuseAppL$ at the root,
      then $\tm_1 = \tm_{11}\esub{\var}{\tm_{12}}$
      where $\var \notin \fv{\tm_2}$
      and the step is of the form
      $\tm
      = \tm_{11}\esub{\var}{\tm_{12}}\,\tm_2
      \tofuseone (\tm_{11}\,\tm_2)\esub{\var}{\tm_{12}}
      = \tmtwo$.
      We consider five subcases, depending on whether
      the reduction step
      $(\tm_{11}\,\tm_2)\esub{\var}{\tm_{12}} \tolsci \tmthree$
      is internal to $\tm_{11}$, internal to $\tm_{2}$, internal to $\tm_{12}$,
      a $\tols$ step at the root,
      or a $\todb$ step involving the application $\tm_{11}\,\tm_2$:
      \begin{enumerate}
      \item
        \label{postponement_tofuseone:case_appL_internal}
        If the reduction step is internal to $\tm_{11}$:
        then $\tm_{11} \tolsci \tmthree_{11}$ and the situation is:
        \[
          \diagramFuseOneLSCOne{
            \tm_{11}\esub{\var}{\tm_{12}}\,\tm_2
          }{
            (\tm_{11}\,\tm_2)\esub{\var}{\tm_{12}}
          }{
            (\tmthree_{11}\,\tm_2)\esub{\var}{\tm_{12}}
          }{
            \tmthree_{11}\esub{\var}{\tm_{12}}\,\tm_2
          }
        \]
      \item
        \label{postponement_tofuseone:case_appL_internal_fv}
        If the reduction step is internal to $\tm_{2}$:
        then $\tm_{2} \tolsci \tmthree_{2}$ and the situation is:
        \[
          \diagramFuseOneLSCOne{
            \tm_{11}\esub{\var}{\tm_{12}}\,\tm_2
          }{
            (\tm_{11}\,\tm_2)\esub{\var}{\tm_{12}}
          }{
            (\tm_{11}\,\tmthree_2)\esub{\var}{\tm_{12}}
          }{
            \tm_{11}\esub{\var}{\tm_{12}}\,\tmthree_2
          }
        \]
        for the fusion step at the bottom of the diagram, observe that
        $\var \notin \fv{\tmthree_2}$ because reduction
        does not create free variables.
      \item
        If the reduction step is internal to $\tm_{12}$:
        similar to case~\ref{postponement_tofuseone:case_appL_internal}.
      \item
        \label{postponement_tofuseone:case_appL_ls}
        If the reduction step is a $\tols$ step at the root:
        note that $\var \notin \fv{\tm_2}$
        because the step
        $\tm_{11}\esub{\var}{\tm_{12}}\,\tm_2
         \tofuseone (\tm_{11}\,\tm_2)\esub{\var}{\tm_{12}}$
        is an instance of the $\ruleFuseAppL$ rule.
        Hence in the $\tols$ step
        $(\tm_{11}\,\tm_2)\esub{\var}{\tm_{12}} \tols \tmthree$
        the substituted variable must necessarily lie inside $\tm_{11}$.
        This means that $\tm_{11}$ is of the form $\tm_{11} = \off{\gctx}{\var}$
        and the situation is:
        \[
          \diagramFuseOneLSCOne{
            \off{\gctx}{\var}\esub{\var}{\tm_{12}}\,\tm_2
          }{
            (\off{\gctx}{\var}\,\tm_2)\esub{\var}{\tm_{12}}
          }{
            (\off{\gctx}{\tm_{12}}\,\tm_2)\esub{\var}{\tm_{12}}
          }{
            \off{\gctx}{\tm_{12}}\esub{\var}{\tm_{12}}\,\tm_2
          }
        \]
      \item
        If the reduction step is a $\todb$ step on the left:
        then $\tm_{11} = (\lam{\vartwo}{\tm'_{11}})\sctx$
        and the situation is:
        \[
          \diagramFuseZeroLSCOne{
            (\lam{\vartwo}{\tm'_{11}})\sctx\esub{\var}{\tm_{12}}\,\tm_2
          }{
            ((\lam{\vartwo}{\tm'_{11}})\sctx\,\tm_2)\esub{\var}{\tm_{12}}
          }{
            \tm'_{11}\esub{\vartwo}{\tm_2}\sctx\esub{\var}{\tm_{12}}
          }{
            \tm'_{11}\esub{\vartwo}{\tm_2}\sctx\esub{\var}{\tm_{12}}
          }
        \]
      \end{enumerate}
    \item
      If the fusion step is derived from $\ruleFuseAppR$ at the root,
      then $\tm_2 = \tm_{21}\esub{\var}{\tm_{22}}$
      where $\var \notin \fv{\tm_1}$
      and the step is of the form
      $\tm
      = \tm_1\,\tm_{21}\esub{\var}{\tm_{22}}
      \tofuseone (\tm_1\,\tm_{21})\esub{\var}{\tm_{22}} = \tmtwo$.
      We consider five subcases, depending on whether
      the reduction step
      $(\tm_1\,\tm_{21})\esub{\var}{\tm_{22}} \tolsci \tmthree$
      is internal to $\tm_1$, internal to $\tm_{21}$, internal to $\tm_{22}$,
      a $\tols$ step at the root,
      or a $\todb$ step involving the application $\tm_1\,\tm_{21}$:
      \begin{enumerate}
      \item
        If the reduction step is internal to $\tm_1$:
        similar to case~\ref{postponement_tofuseone:case_appL_internal_fv}.
      \item
        If the reduction step is internal to $\tm_{21}$:
        similar to case~\ref{postponement_tofuseone:case_appL_internal}.
      \item
        If the reduction step is internal to $\tm_{22}$:
        similar to case~\ref{postponement_tofuseone:case_appL_internal}.
      \item
        If the reduction step is a $\tols$ step at the root:
        note that $\var \notin \fv{\tm_1}$ because the step
        $\tm_1\,\tm_{21}\esub{\var}{\tm_{22}}
        \tofuseone (\tm_1\,\tm_{21})\esub{\var}{\tm_{22}}$
        is an instance of the $\ruleFuseAppR$ rule.
        Hence in the $\tols$ step
        $(\tm_1\,\tm_{21})\esub{\var}{\tm_{22}} \tols \tmthree$
        the substituted variable must necessarily lie inside $\tm_{21}$.
        This means that $\tm_{21}$ is of the form $\tm_{21} = \off{\gctx}{\var}$
        and the situation is:
        \[
          \diagramFuseOneLSCOne{
            \tm_1\,\off{\gctx}{\var}\esub{\var}{\tm_{22}}
          }{
            (\tm_1\,\off{\gctx}{\var})\esub{\var}{\tm_{22}}
          }{
            (\tm_1\,\off{\gctx}{\tm_{22}})\esub{\var}{\tm_{22}}
          }{
            \tm_1\,\off{\gctx}{\tm_{22}}\esub{\var}{\tm_{22}}
          }
        \]
      \item
        If the reduction step is a $\todb$ step on the left:
        then $\tm_1 = (\lam{\vartwo}{\tm'_1})\sctx$
        and the situation is:
        \[
          \diagramFuseZeroLSCOne{
            (\lam{\vartwo}{\tm'_1})\sctx\,\tm_{21}\esub{\var}{\tm_{22}}
          }{
            ((\lam{\vartwo}{\tm'_1})\sctx\,\tm_{21})\esub{\var}{\tm_{22}}
          }{
            \tm'_1\esub{\vartwo}{\tm_{21}}\sctx\esub{\var}{\tm_{22}}
          }{
            \tm'_1\esub{\vartwo}{\tm_{21}}\sctx\esub{\var}{\tm_{22}}
          }
        \]
      \end{enumerate}
    \end{enumerate}
  \item
    Substitution, $\tm = \tm_1\esub{\var}{\tm_2}$:
    we consider six subcases, depending on whether the fusion step
    $\tm = \tm_1\esub{\var}{\tm_2} \tofuseone \tmtwo$
    is internal to $\tm_1$, internal to $\tm_2$,
    or derived from one of the rules
    $\ruleFuseW$, $\ruleFuseC$, $\ruleFuseEsL$, or $\ruleFuseEsR$
    at the root:
    \begin{enumerate}
    \item
      If the fusion step is internal to $\tm_1$,
      then
      $\tm = \tm_1\esub{\var}{\tm_2} \tofuseone \tmtwo_1\esub{\var}{\tm_2} = \tmtwo$
      with $\tm_1 \tofuseone \tmtwo_1$.
      There are three subcases, depending on whether the reduction step
      is internal to $\tmtwo_1$, internal to $\tm_2$, or a $\tols$ step at the root:
      \begin{enumerate}
      \item
        If the reduction step is internal to $\tmtwo_1$:
        then the diagram can be closed by resorting to the \ih, similarly as for
        case~\ref{postponement_tofuseone:case_app_left_ih}.
      \item
        If the reduction step is internal to $\tm_2$:
        the steps are disjoint and the diagram can be closed similarly as for
        case~\ref{postponement_tofuseone:case_app_left_disjoint}.
      \item
        If the reduction step is a $\tols$ step at the root:
        then $\tmtwo_1 = \off{\gctx}{\var}$ and $\tm_1 \tofuseone \off{\gctx}{\var}$.
        By \rlem{fusion_backwards_variables}, we have that
        $\tm_1$ is of the form $\off{\gctx_0}{\var,\hdots,\var}$,
        where $\gctx_0$ is a context with either one or two holes
        such that $\off{\gctx_0}{\tm_2,\hdots,\tm_2} \tofuse \off{\gctx}{\tm_2}$.
        Then the situation is:
        \[
          \diagramFuseLSC{
            \off{\gctx_0}{\var,\hdots,\var}\esub{\var}{\tm_2}
          }{
            \off{\gctx}{\var}\esub{\var}{\tm_2}
          }{
            \off{\gctx}{\tm}\esub{\var}{\tm_2}
          }{
            \off{\gctx_0}{\tm_2,\hdots,\tm_2}\esub{\var}{\tm_2}
          }
        \]
      \end{enumerate}
    \item
      If the fusion step is internal to $\tm_2$,
      then $\tm = \tm_1\esub{\var}{\tm_2} \tofuseone \tm_1\esub{\var}{\tmtwo_2}$
      with $\tm_2 \tofuseone \tmtwo_2$. 
      There are three subcases, depending on whether the reduction step is
      internal to $\tm_1$, internal to $\tmtwo_2$, or a $\tols$ step at the root:
      \begin{enumerate}
      \item
        If the reduction step is internal to $\tm_1$:
        the steps are disjoint and the diagram can be closed similarly as for
        case~\ref{postponement_tofuseone:case_app_left_disjoint}.
      \item
        If the reduction step is internal to $\tmtwo_2$:
        then the diagram can be closed by resorting to the \ih, similarly as for
        case~\ref{postponement_tofuseone:case_app_left_ih}.
      \item
        If the reduction step is a $\tols$ step at the root,
        then $\tm_1 = \off{\gctx}{\var}$ and the situation is:
        \[
          \diagramFuseManyLSCOne{
            \off{\gctx}{\var}\esub{\var}{\tm_2}
          }{
            \off{\gctx}{\var}\esub{\var}{\tmtwo_2}
          }{
            \off{\gctx}{\tm_2}\esub{\var}{\tm_2}
          }{
            \off{\gctx}{\tmtwo_2}\esub{\var}{\tmtwo_2}
          }
        \]
      \end{enumerate}
    \item
      If the fusion step is derived from $\ruleFuseW$ at the root,
      then $\var \notin \fv{\tm_1}$ and
      $\tm = \tm_1\esub{\var}{\tm_2} \tofuse \tm_1 = \tmtwo$.
      The situation is:
      \[
        \diagramFuseOneLSCOne{
          \tmtwo\esub{\var}{\tm_2}
        }{
          \tmtwo
        }{
          \tmthree
        }{
          \tmthree\esub{\var}{\tm_2}
        }
      \]
      The fusion step at the bottom is an instance of the $\ruleFuseW$ rule,
      which can be applied because $\var \notin \fv{\tmthree}$ holds,
      given that $\var \notin \fv{\tmtwo}$
      and using the fact that the reduction step $\tmtwo \tolsci \tmthree$
      does not create free variables.
    \item
      If the fusion step is derived from $\ruleFuseC$ at the root,
      then $\tm_1 = \tm'_1\esub{\vartwo}{\tm_2}$
      and
      $\tm
      = \tm'_1\esub{\vartwo}{\tm_2}\esub{\var}{\tm_2}
      \tofuseone \tm'_1\sub{\vartwo}{\var}\esub{\var}{\tm_2}
      = \tmtwo$.
      There are three subcases, depending on whether the step is reduction step
      is internal to $\tm'_1\sub{\vartwo}{\var}$, internal to $\tm_2$,
      or a $\tols$ step at the root:
      \begin{enumerate}
      \item
        If the reduction step is internal to $\tm'_1\sub{\vartwo}{\var}$:
        then $\tm'_1\sub{\vartwo}{\var} \tolsci \tmthree$.
        By \rlem{reduction_before_fusion}
        there exists $\tmthree_0$
        such that $\tm'_1 \tolsci \tmthree_0$
        and $\tmthree = \tmthree_0\sub{\vartwo}{\var}$.
        Hence the situation is:
        \[
          \diagramFuseOneLSCOne{
            \tm'_1\esub{\vartwo}{\tm_2}\esub{\var}{\tm_2}
          }{
            \tm'_1\sub{\vartwo}{\var}\esub{\var}{\tm_2}
          }{
            \tmthree_0\sub{\vartwo}{\var}\esub{\var}{\tm_2}
          }{
            \tmthree_0\esub{\vartwo}{\tm_2}\esub{\var}{\tm_2}
          }
        \]
      \item
        If the reduction step is internal to $\tm_2$:
        then $\tm_2 \tolsci \tmthree_2$ and the situation is:
        \[
          \diagramFuseOneLSCMany{
             \tm'_1\esub{\vartwo}{\tm_2}\esub{\var}{\tm_2}
          }{
             \tm'_1\sub{\vartwo}{\var}\esub{\var}{\tm_2}
          }{
             \tm'_1\sub{\vartwo}{\var}\esub{\var}{\tmthree_2}
          }{
             \tm'_1\esub{\vartwo}{\tmthree_2}\esub{\var}{\tmthree_2}
          }
        \]
      \item
        If the reduction step is a $\tols$ step at the root:
        then $\tm'_1\sub{\vartwo}{\var}\esub{\var}{\tm_2} \tols \tmthree$.
        We consider two further subcases, depending on whether the substituted
        occurrence of $\var$ in $\tm'_1\sub{\vartwo}{\var}$
        corresponds to an occurrence of $\var$ in $\tm'_1$
        or to an occurrence of $\vartwo$ in $\tm'_1$:
        \begin{enumerate}
        \item
          Substitution of an occurrence of $\var$ in $\tm'_1$:
          then $\tm'_1 = \off{\gctx}{\var}$ and the situation is:
          \[
            \diagramFuseOneLSCOne{
              \off{\gctx}{\var}\esub{\vartwo}{\tm_2}\esub{\var}{\tm_2}
            }{
              \off{\gctx\sub{\vartwo}{\var}}{\var}\esub{\var}{\tm_2}
            }{
              \off{\gctx\sub{\vartwo}{\var}}{\tm_2}\esub{\var}{\tm_2}
            }{
              \off{\gctx}{\tm_2}\esub{\vartwo}{\tm_2}\esub{\var}{\tm_2}
            }
          \]
        \item
          Substitution of an occurrence of $\vartwo$ in $\tm'_1$:
          then $\tm'_1 = \off{\gctx}{\vartwo}$ and the situation is:
          \[
            \diagramFuseOneLSCOne{
              \off{\gctx}{\vartwo}\esub{\vartwo}{\tm_2}\esub{\var}{\tm_2}
            }{
              \off{\gctx\sub{\vartwo}{\var}}{\var}\esub{\var}{\tm_2}
            }{
              \off{\gctx\sub{\vartwo}{\var}}{\tm_2}\esub{\var}{\tm_2}
            }{
              \off{\gctx}{\tm_2}\esub{\vartwo}{\tm_2}\esub{\var}{\tm_2}
            }
          \]
        \end{enumerate}
      \end{enumerate}
    \item
      If the fusion step is derived from $\ruleFuseEsL$ at the root,
      then $\tm_1 = \tm_{11}\esub{\vartwo}{\tm_{12}}$
      where $\vartwo \notin \fv{\tm_2}$ and $\var \notin \fv{\tm_{12}}$,
      and the step is of the form
      $\tm
      = \tm_{11}\esub{\vartwo}{\tm_{12}}\esub{\var}{\tm_2}
      \tofuseone \tm_{11}\esub{\var}{\tm_2}\esub{\vartwo}{\tm_{12}}
      = \tmtwo$.
      There are five subcases, depending on whether the reduction step is
      internal to $\tm_{11}$, internal to $\tm_2$, internal to $\tm_{12}$,
      a $\tols$ step contracting $\vartwo$,
      or a $\tols$ step contracting $\var$:
      \begin{enumerate}
      \item
        If the reduction step is internal to $\tm_{11}$:
        then $\tm_{11} \tolsci \tmtwo_{11}$ and the situation is:
        \[
          \diagramFuseOneLSCOne{
            \tm_{11}\esub{\vartwo}{\tm_{12}}\esub{\var}{\tm_2}
          }{
            \tm_{11}\esub{\var}{\tm_2}\esub{\vartwo}{\tm_{12}}
          }{
            \tmtwo_{11}\esub{\var}{\tm_2}\esub{\vartwo}{\tm_{12}}
          }{
            \tmtwo_{11}\esub{\vartwo}{\tm_{12}}\esub{\var}{\tm_2}
          }
        \]
      \item
        If the reduction step is internal to $\tm_2$:
        similar to the previous case.
      \item
        If the reduction step is internal to $\tm_{12}$:
        similar to the previous case.
      \item
        If the reduction step is a $\tols$ step contracting $\vartwo$:
        since $\vartwo \notin \fv{\tm_2}$
        we have that $\tm_{11} = \off{\gctx}{\vartwo}$
        and the situation is:
        \[
          \diagramFuseOneLSCOne{
            \off{\gctx}{\vartwo}\esub{\vartwo}{\tm_{12}}\esub{\var}{\tm_2}
          }{
            \off{\gctx}{\vartwo}\esub{\var}{\tm_2}\esub{\vartwo}{\tm_{12}}
          }{
            \off{\gctx}{\tm_{12}}\esub{\var}{\tm_2}\esub{\vartwo}{\tm_{12}}
          }{
            \off{\gctx}{\tm_{12}}\esub{\vartwo}{\tm_{12}}\esub{\var}{\tm_2}
          }
        \]
      \item
        If the reduction step is a $\tols$ step contracting $\var$:
        then $\tm_{11} = \off{\gctx}{\var}$
        and the situation is:
        \[
          \diagramFuseOneLSCOne{
            \off{\gctx}{\var}\esub{\vartwo}{\tm_{12}}\esub{\var}{\tm_2}
          }{
            \off{\gctx}{\var}\esub{\var}{\tm_2}\esub{\vartwo}{\tm_{12}}
          }{
            \off{\gctx}{\tm_2}\esub{\var}{\tm_2}\esub{\vartwo}{\tm_{12}}
          }{
            \off{\gctx}{\tm_2}\esub{\vartwo}{\tm_{12}}\esub{\var}{\tm_2}
          }
        \]
      \end{enumerate}
    \item
      If the fusion step is derived from $\ruleFuseEsR$ at the root,
      then $\tm_2 = \tm_{21}\esub{\vartwo}{\tm_{22}}$
      where $\vartwo \notin \fv{\tm_1}$
      and the step is of the form
      $\tm = \tm_1\esub{\var}{\tm_{21}\esub{\vartwo}{\tm_{22}}} = \tmtwo$.
      There are five subcases, depending on whether the reduction step is
      internal to $\tm_1$, internal to $\tm_{21}$, internal to $\tm_{22}$,
      a $\tols$ step contracting $\vartwo$,
      or a $\tols$ step contracting $\var$:
      \begin{enumerate}
      \item
        If the reduction step is internal to $\tm_1$:
        then $\tm_1 \tolsci \tmtwo_1$ and the situation is:
        \[
          \diagramFuseOneLSCOne{
            \tm_1\esub{\var}{\tm_{21}\esub{\vartwo}{\tm_{22}}}
          }{
            \tm_1\esub{\var}{\tm_{21}}\esub{\vartwo}{\tm_{22}}
          }{
            \tmtwo_1\esub{\var}{\tm_{21}}\esub{\vartwo}{\tm_{22}}
          }{
            \tmtwo_1\esub{\var}{\tm_{21}\esub{\vartwo}{\tm_{22}}}
          }
        \]
        The step on the bottom is an instance of the $\ruleFuseEsR$ rule,
        which can be applied because $\vartwo \notin \fv{\tmtwo_1}$ holds,
        given that $\vartwo \notin \fv{\tm_1}$
        and using the fact that the reduction step $\tm_1 \tolsci \tmtwo_1$
        does not create free variables.
      \item
        If the reduction step is internal to $\tm_{21}$:
        similar to the previous case.
      \item
        If the reduction step is internal to $\tm_{22}$:
        similar to the previous case.
      \item
        If the reduction step is a $\tols$ step contracting $\vartwo$:
        since $\vartwo \notin \fv{\tm_1}$,
        we have that $\tm_{21} = \off{\gctx}{\vartwo}$, and the situation is:
        \[
          \diagramFuseOneLSCOne{
            \tm_1\esub{\var}{\off{\gctx}{\vartwo}\esub{\vartwo}{\tm_{22}}}
          }{
            \tm_1\esub{\var}{\off{\gctx}{\vartwo}}\esub{\vartwo}{\tm_{22}}
          }{
            \tm_1\esub{\var}{\off{\gctx}{\tm_{22}}}\esub{\vartwo}{\tm_{22}}
          }{
            \tm_1\esub{\var}{\off{\gctx}{\tm_{22}}\esub{\vartwo}{\tm_{22}}}
          }
        \]
      \item
        If the reduction step is a $\tols$ step contracting $\var$:
        then $\tm_1 = \off{\gctx}{\var}$ and the situation is:
        \[
          \diagramFuseManyLSCOne{
            \off{\gctx}{\var}\esub{\var}{\tm_{21}\esub{\vartwo}{\tm_{22}}}
          }{
            \off{\gctx}{\var}\esub{\var}{\tm_{21}}\esub{\vartwo}{\tm_{22}}
          }{
            \off{\gctx}{\tm_{21}}\esub{\var}{\tm_{21}}\esub{\vartwo}{\tm_{22}}
          }{
            \off{\gctx}{\tm_{21}\esub{\vartwo}{\tm_{22}}}\esub{\var}{\tm_{21}\esub{\vartwo}{\tm_{22}}}
          }
        \]
        The step on the bottom is justified by \rlem{properties_of_fusion}.
      \end{enumerate}
    \end{enumerate}
  \end{enumerate}
  \end{proof}
\end{ifLongAppendix}

\begin{lemma}[Postponement of fusion]
\llem{postponement_tofuse}
If $\tm$ is $\tolsci$-SN and $\tm \tofuse\,\tolsci \tmtwo$
then $\tm \tolsci^+\,\tofuse \tmtwo$.
\end{lemma}
\begin{proof}
We prove the statement by first proving two auxiliary claims:
\begin{enumerate}
\item 
  First claim: ${\tofuse\tolsci} \subseteq {\tolsci(\tolsci\cup\tofuseone)^*}$. \\
  Since $\tofuse$ is the reflexive--transitive closure of $\tofuseone$,
  by definition, this is equivalent to showing that
  ${(\tofuseone)^n\tolsci} \subseteq {\tolsci(\tolsci\cup\tofuseone)^*}$
  holds for all $n \in \Nat_0$.
  We proceed by induction on $n$. If $n = 0$, it is immediate to conclude.
  Now assume the property holds for a given $n \in \Nat_0$.
  Then:
  \[
    \begin{array}{rcll}
      {(\tofuseone)^{n+1}\tolsci}
    & = &
      {(\tofuseone)^{n}\tofuseone\tolsci}
    \\
    & \subseteq &
      {(\tofuseone)^{n}\tolsci^+\tofuse}
      & \text{by \rlem{postponement_tofuseone}}
    \\
    & = &
      {(\tofuseone)^{n}\tolsci\tolsci^*\tofuse}
      & \text{since ${\tolsci^+} = {\tolsci\tolsci^*}$}
    \\
    & \subseteq &
      {\tolsci(\tolsci\cup\tofuseone)^*\tolsci^*\tofuse}
      & \text{by \ih}
    \\
    & = &
      {\tolsci(\tolsci\cup\tofuseone)^*\tofuse}
      & \text{since ${\tolsci^*} \subseteq (\tolsci\cup\tofuseone)^*$ and transitivity}
    \\
    & = &
      {\tolsci(\tolsci\cup\tofuseone)^*}
      & \text{since ${\tofuse} = (\tofuseone)^* \subseteq (\tolsci\cup\tofuseone)^*$ and transitivity}
    \end{array}
  \]
\item
  Second claim:
  if $\tm$ is $\tolsci$-SN
  and $\tm \mathrel{(\tolsci\cup\tofuseone)^*} \tmtwo$
  then $\tm \mathrel{\tolsci^*\tofuse} \tmtwo$.
  The relation $\tolsci$ restricted to SN terms is obviously SN,
  so we may proceed by well-founded induction on $\tm$
  with respect to $\tolsci$.

  We know that $\tm \mathrel{(\tolsci\cup\tofuseone)^*} \tmtwo$ holds.
  We consider two cases, depending on whether $\tm \tofuse \tmtwo$ or not:
  \begin{enumerate}
  \item
    If $\tm \tofuse \tmtwo$,
    it is immediate to conclude that $\tm \mathrel{\tolsci^*\tofuse} \tmtwo$.
  \item
    Otherwise, in the reduction $\tm \mathrel{(\tolsci\cup\tofuseone)^*} \tmtwo$
    there must be at least one $\tolsci$ step.
    Considering the first such step,
    this means that $\tm \tofuse\tolsci(\tolsci\cup\tofuseone)^* \tmtwo$.
    By the previous claim we have that
    $\tm \tolsci(\tolsci\cup\tofuseone)^*(\tolsci\cup\tofuseone)^* \tmtwo$
    so, by transitivity,
    $\tm \tolsci(\tolsci\cup\tofuseone)^* \tmtwo$.
    Let $\tm'$ be a term such that
    $\tm \tolsci \tm' (\tolsci\cup\tofuseone)^* \tmtwo$,
    and note that $\tm'$ is $\tolsci$-SN,
    because it is a reduct of $\tm$ which is itself $\tolsci$-SN.
    By \ih we have that $\tm' \tolsci^*\tofuse \tmtwo$,
    so $\tm \tolsci \tm' \tolsci^*\tofuse \tmtwo$.
    This means that $\tm \tolsci^*\tofuse \tmtwo$, as required.
  \end{enumerate}
\item
  Finally, we prove the main statement:
  if $\tm$ is $\tolsci$-SN and $\tm \tofuse\,\tolsci \tmtwo$
  then $\tm \tolsci^+\,\tofuse \tmtwo$. \\
  Suppose that
  $\tm \tofuse\,\tolsci \tmtwo$.
  By the first claim,
  $\tm \mathrel{\tolsci(\tolsci\cup\tofuseone)^*} \tmtwo$.
  Let $\tm'$ be a term such that
  $\tm \tolsci \tm' (\tolsci\cup\tofuseone)^* \tmtwo$,
  and note that $\tm'$ is $\tolsci$-SN, because it is a reduct of
  $\tm$ which is itself $\tolsci$-SN.
  Since $\tm' (\tolsci\cup\tofuseone)^* \tmtwo$,
  by the second claim we have that $\tm' \tolsci^*\tofuse \tmtwo$.
  Hence $\tm \tolsci \tm' \tolsci^*\tofuse \tmtwo$,
  which means in particular that $\tm \tolsci^+\tofuse \tmtwo$,
  as required.
\end{enumerate}
\end{proof}

\begin{lemma}[Fusion preserves SN]
\llem{fusion_preserves_SN}
If $\tm$ is $\tolsci$-SN
then $\tm$ is $(\tolsci\,\tofuse)$-SN.
\end{lemma}
\begin{proof}
We first prove an auxiliary claim:
if $\tm$ is a $\tolsci$-SN term
which is not $(\tolsci\,\tofuse)$-SN,
then it has a reduct $\tm \tolsci \tmtwo$
such that $\tmtwo$ is not $(\tolsci\,\tofuse)$-SN.
Indeed,
suppose that $\tm$ is not $(\tolsci\,\tofuse)$-SN.
Then there is an infinite $(\tolsci\,\tofuse)$-reduction sequence
starting from $\tm$.
Consider in particular a prefix of the sequence
of the form
$\tm \tolsci\,\tofuse \tolsci\,\tofuse \tmthree$.
Let $\tm'$ be a term such that
$\tm \tolsci \tm' \tofuse \tolsci\,\tofuse \tmthree$.
By hypothesis, $\tm$ is $\tolsci$-SN
so its reduct $\tm'$ must also be $\tolsci$-SN,
Which means that we may apply~\rlem{postponement_tofuse}
to postpone the fusion step and obtain that
$\tm \mathrel{\tolsci\,\tolsci^+ \tofuse\,\tofuse} \tmthree$.
Since fusion is transitive, we have that
$\tm \mathrel{\tolsci\,\tolsci^+ \tofuse} \tmthree$.
Since $\tofuse$ is reflexive,
${\tolsci} \subseteq {\tolsci\tofuse}$
and, moreover,
${\tolsci^+} \subseteq {(\tolsci\tofuse)^+}$,
which means that
$\tm \mathrel{\tolsci\,(\tolsci\tofuse)^+\tofuse} \tmthree$.
The last fusion step can be absorbed to the immediately preceding
one by transitivity, so we have that
$\tm \mathrel{\tolsci\,(\tolsci\tofuse)^+} \tmthree$.
Let $\tmtwo$ be a term such that
$\tm \tolsci \tmtwo \mathrel{(\tolsci\tofuse)^+} \tmthree$.
Then $\tm \tolsci \tmtwo$ and
$\tmtwo$ is not $(\tolsci\tofuse)$-SN
because $\tmtwo (\tolsci\tofuse)^+ \tmthree$ where $\tmthree$ in turn
is not $(\tolsci\tofuse)$-SN.
This concludes the proof of the claim. \\
\indent
To prove the statement of the lemma,
let $\tm$ be $\tolsci$-SN
and suppose by contradiction that it is not $(\tolsci\tofuse)$-SN.
By the claim, it has a reduct $\tm \tolsci \tm_1$ which is not
$(\tolsci\tofuse)$-SN.
Iterating this argument, we construct an infinite reduction sequence
$\tm \tolsci \tm_1 \tolsci \tm_2 \hdots$,
meaning that $\tm$ is not $\tolsci$-SN,
contradicting the hypothesis.
\end{proof}

\subsection{Translation of $\lambdaS$ to LSC}

\begin{definition}[From $\lambdaS$ to LSC]
\ldef{from_lambdaS_to_LSC}
We assume that $\lscUnit$ stands for some inhabited type in the LSC,
and $\lscunit$ for a closed inhabitant of $\lscUnit$ in normal form.
Types, terms, and typing contexts, are translated as follows:
\[
  \begin{array}{rcl}
  \tradlsc{\btyp}            & \eqdef & \btyp \\
  \tradlsc{\typ\limp\typtwo} & \eqdef & \tradlsc{\typ} \to \tradlsc{\typtwo} \\
  \tradlsc{\sha{\typ}}       & \eqdef & \lscUnit \to \tradlsc{\typ} \\
  \tradlsc{\ofc{\typ}}       & \eqdef & \tradlsc{\typ} \\
  \end{array}
  \HS
  \begin{array}{rcl}
    \tradlsc{\uvar_1:\typ_1,\hdots,\uvar_n:\typ_n}
  & \eqdef &
    \uvar_1:\lscUnit\to\tradlsc{\typ_1},\hdots,\uvar_n:\lscUnit\to\tradlsc{\typ_n}
  \\
    \tradlsc{\lvar_1:\typ_1,\hdots,\lvar_n:\typ_n}
  & \eqdef &
    \lvar_1:\tradlsc{\typ_1},\hdots,\lvar_n:\tradlsc{\typ_n}
  \end{array}
\]
\[
  \begin{array}{rcll}
    \tradlsc{\lvar}
  & \eqdef &
    \lvar
  \\
    \tradlsc{\uvar}
  & \eqdef &
    \uvar
  \\
    \tradlsc{\lam{\lvar}{\tm}}
  & \eqdef &
    \lam{\lvar}{\tradlsc{\tm}}
  \\
    \tradlsc{\tm\,\tmtwo}
  & \eqdef &
    \tradlsc{\tm}\,\tradlsc{\tmtwo}
  \end{array}
  \begin{array}{rcll}
    \tradlsc{\sha{\tm}}
  & \eqdef &
    \lam{\varthree}{\tradlsc{\tm}}
    & \text{where $\varthree \notin \fv{\tm}$}
  \\
    \tradlsc{\open{\tm}}
  & \eqdef &
    \tradlsc{\tm}\,\lscunit
  \\
    \tradlsc{\ofc{\tm}} 
  & \eqdef &
    \tradlsc{\tm}
  \\
    \tradlsc{\tm\esub{\var}{\tmtwo}} 
  & \eqdef &
    \tradlsc{\tm}\esub{\var}{\tradlsc{\tmtwo}}
  \end{array}
\]
The translation is extended to contexts by declaring that
$\tradlsc{\ctxhole} = \ctxhole$.
In particular,
the translation of a substitution context
$\sctx = \ctxhole\esub{\var_1}{\tm_1}\hdots\esub{\var_n}{\tm_n}$
is a substitution context
$\tradlsc{\sctx} = \ctxhole\esub{\var_1}{\tradlsc{\tm_1}}\hdots\esub{\var_n}{\tradlsc{\tm_n}}$.
\end{definition}

\begin{remark}
$\fv{\tradlsc{\tm}} = \fv{\tm}$
\end{remark}

\begin{lemma}[$\tradlsc{-}$ commutes with substitution]
\llem{tradlsc_substitution}
The following hold:
\begin{enumerate}
\item $\tradlsc{\tm\sctx} = \tradlsc{\tm}\tradlsc{\sctx}$
\item $\tradlsc{\of{\gctx}{\tm}} = \of{\tradlsc{\gctx}}{\tradlsc{\tm}}$
\item $\tradlsc{\tm\sub{\var}{\tmtwo}} = \tradlsc{\tm}\sub{\var}{\tradlsc{\tmtwo}}$
\end{enumerate}
\end{lemma}
\begin{proof}
  The first item is by induction on $\sctx$. If $\sctx=\ctxhole$, then  $\tradlsc{\tm\sctx} = \tradlsc{\tm} = \tradlsc{\tm}\tradlsc{\sctx}$ since $\tradlsc{\ctxhole} = \ctxhole$. If $\sctx=\sctx'\esub{\var}{\tmtwo}$, we reason as follows.
  \[
    \begin{array}{rll}
      & \tradlsc{\tm\sctx} \\
      = & \tradlsc{\tm \sctx'\esub{\var}{\tmtwo}} \\
      =  & \tradlsc{\tm \sctx'}\esub{\var}{\tradlsc{\tmtwo}} & \text{by definition} \\
      = & \tradlsc{\tm} \tradlsc{\sctx'}\esub{\var}{\tradlsc{\tmtwo}} & (\ih)\\
      = & \tradlsc{\tm} \tradlsc{\sctx'\esub{\var}{\tmtwo}} & \text{by definition} \\
      = & \tradlsc{\tm}\tradlsc{\sctx}
    \end{array}
    \]
The second item is by induction on $\gctx$.
\begin{enumerate}
\item $\gctx=\ctxhole$. Then $\tradlsc{\of{\gctx}{\tm}}= \tradlsc{\tm} = \of{\tradlsc{\gctx}}{\tradlsc{\tm}}$, since $\tradlsc{\ctxhole} = \ctxhole$.

\item $\gctx=\lam{\lvar}{\gctx_1}$. Then we reason as follows:
  \[\begin{array}{rll}
      & \tradlsc{\of{\gctx}{\tm}} \\
      = & \tradlsc{\lam{\lvar}{\of{\gctx_1}{\tm}}} \\
      = & \lam{\lvar}{\tradlsc{\of{\gctx_1}{\tm}}} & \text{by definition} \\
      = & \lam{\lvar}{\of{\tradlsc{\gctx_1}}{\tradlsc{\tm}}} & (\ih) \\
      = & \of{\tradlsc{\gctx}}{\tradlsc{\tm}} & \text{by definition} \\
    \end{array}
    \]

  \item  $\gctx=\gctx_1\,\tmtwo$ (the case $\gctx=\tmtwo\, \gctx_1$ is similar). Then we reason as follows:
  \[\begin{array}{rll}
      & \tradlsc{\of{\gctx}{\tm}} \\
      = & \tradlsc{\of{\gctx_1}{\tm}\,\tmtwo} \\
      = & \tradlsc{\of{\gctx_1}{\tm}}\, \tradlsc{\tmtwo} & \text{by definition} \\
      = & \of{\tradlsc{\gctx_1}}{\tradlsc{\tm}}\, \tradlsc{\tmtwo} & (\ih) \\
      = & \of{\tradlsc{\gctx}}{\tradlsc{\tm}} & \text{by definition} \\
    \end{array}
  \]
  
    \item $\gctx=\gctx_1\esub{\var}{\tmtwo}$ (the case $\gctx=\tmtwo\esub{\var}{\gctx_1}$ is similar). Then we reason as follows:
  \[\begin{array}{rll}
      & \tradlsc{\of{\gctx}{\tm}} \\
      = & \tradlsc{\of{\gctx_1}{\tm}\esub{\var}{\tmtwo}} \\
      = & \tradlsc{\of{\gctx_1}{\tm}}\esub{\var}{\tradlsc{\tmtwo}} & \text{by definition} \\
      = & \of{\tradlsc{\gctx_1}}{\tradlsc{\tm}}\esub{\var}{\tradlsc{\tmtwo}} & (\ih) \\
      = & \of{\tradlsc{\gctx}}{\tradlsc{\tm}} & \text{by definition}
    \end{array}
  \]

      \end{enumerate}
The third item is by induction on $\tm$.
\begin{enumerate}
\item $\tm= \vartwo$. We consider two cases.
  \begin{enumerate}

  \item $\vartwo=\var$. Then $\tradlsc{\tm\sub{\var}{\tmtwo}} = \tradlsc{\tmtwo} = \var\sub{\var}{\tradlsc{\tmtwo}} = \tradlsc{\tm}\sub{\var}{\tradlsc{\tmtwo}}$.

    \item $\vartwo\neq \var$. Then $\tradlsc{\tm\sub{\var}{\tmtwo}} = \tradlsc{\vartwo} = \vartwo = \vartwo\sub{\var}{\tradlsc{\tmtwo}} = \tradlsc{\tm}\sub{\var}{\tradlsc{\tmtwo}}$.
    \end{enumerate}
 
  \item $\tm= \lam{\vartwo}{\tmthree}$.  Then
    \[\begin{array}{rll}
        & \tradlsc{\tm\sub{\var}{\tmtwo}}\\
        = & \tradlsc{(\lam{\vartwo}{\tmthree})\sub{\var}{\tmtwo}}\\
        = & \tradlsc{\lam{\vartwo}{\tmthree \sub{\var}{\tmtwo}}}\\
        = & \lam{\vartwo}{\tradlsc{\tmthree \sub{\var}{\tmtwo}}} & \text{by definition} \\
        = & \lam{\vartwo}{\tradlsc{\tmthree}\sub{\var}{\tradlsc{\tmtwo}}} & (\ih) \\
        = & (\lam{\vartwo}{\tradlsc{\tmthree}}) \sub{\var}{\tradlsc{\tmtwo}} \\
        = & \tradlsc{\lam{\vartwo}{\tmthree}} \sub{\var}{\tradlsc{\tmtwo}} & \text{by definition}\\
      \end{array}
      \]

\item $\tm= \tm_1\,\tm_2$ (the case $\tm= \tm_1\esub{\var}{\tm_2}$ is similar).   Then
    \[\begin{array}{rll}
        & \tradlsc{\tm\sub{\var}{\tmtwo}}\\
        = & \tradlsc{(\tm_1\,\tm_2)\sub{\var}{\tmtwo}}\\
        = & \tradlsc{\tm_1\sub{\var}{\tmtwo}\,\tm_2\sub{\var}{\tmtwo}}\\
        = & \tradlsc{\tm_1 \sub{\var}{\tmtwo}}\, \tradlsc{\tm_2 \sub{\var}{\tmtwo}} & \text{by definition} \\
        = & \tradlsc{\tm_1}\sub{\var}{\tradlsc{\tmtwo}}\, \tradlsc{\tm_2}\sub{\var}{\tradlsc{\tmtwo}} & (\ih) \\
        = & (\tradlsc{\tm_1}\,\tradlsc{\tm_2}) \sub{\var}{\tradlsc{\tmtwo}} \\
        = & \tradlsc{\tm_1\,\tm_2} \sub{\var}{\tradlsc{\tmtwo}} & \text{by definition} \\
      \end{array}
      \]
\end{enumerate}
\end{proof}

\begin{lemma}[$\tradlsc{-}$ preserves typing]
\llem{tradlsc_preserves_typing}
If $\judc{\uenv}{\lenv}{\tm}{\typ}$
then $\judlsc{\tradlsc{\uenv},\tradlsc{\lenv}}{\tradlsc{\tm}}{\tradlsc{\typ}}$.
\end{lemma}
\begin{proof}
By induction on the derivation of $\judc{\uenv}{\lenv}{\tm}{\typ}$:
\begin{enumerate}
\item $\rulecLvar$:
  Let $\judc{\uenv}{\lvar:\typ}{\lvar}{\typ}$.
  Then $\judlsc{\tradlsc{\uenv},\lvar:\tradlsc{\typ}}{\lvar}{\tradlsc{\typ}}$
  by $\rulelVar$.
\item $\rulecUvar$:
  Let 
    $\judc{\uenv,\uvar:\typ}{\noenv}{\uvar}{\sha{\typ}}$.
  Then we have that
    $\judlsc{\tradlsc{\uenv},\uvar:\lscUnit\to\tradlsc{\typ}}{\uvar}{\lscUnit\to\tradlsc{\typ}}$,
  so indeed
    $\judlsc{\tradlsc{\uenv,\uvar:\typ}}{\uvar}{\tradlsc{\sha{\typ}}}$
  by $\rulelVar$.
\item $\rulecAbs$:
  Let
    $\judc{\uenv}{\lenv}{\lam{\lvar}{\tm}}{\typ\limp\typtwo}$
  be derived from
    $\judc{\uenv}{\lenv,\lvar:\typ}{\tm}{\typtwo}$.
  By \ih
    $\judlsc{\tradlsc{\uenv},\tradlsc{\lenv},\lvar:\tradlsc{\typ}}{\tradlsc{\tm}}{\tradlsc{\typtwo}}$,
  so
    $\judlsc{\tradlsc{\uenv},\tradlsc{\lenv}}{\lam{\lvar}{\tradlsc{\tm}}}{\tradlsc{\typ}\to\tradlsc{\typtwo}}$
  by $\rulelAbs$,
  which means that
    $\judlsc{\tradlsc{\uenv},\tradlsc{\lenv}}{\tradlsc{\lam{\lvar}{\tm}}}{\tradlsc{\typ\limp\typtwo}}$,
  as required.
\item $\rulecApp$:
  Let
    $\judc{\uenv}{\lenv_1,\lenv_2}{\tm\,\tmtwo}{\typtwo}$
  be derived from
    $\judc{\uenv}{\lenv_1}{\tm}{\typ\limp\typtwo}$
  and
    $\judc{\uenv}{\lenv_2}{\tmtwo}{\typtwo}$.
  By \ih we have that
    $\judlsc{\tradlsc{\uenv},\tradlsc{\lenv_1}}{\tradlsc{\tm}}{\tradlsc{\typ}\to\tradlsc{\typtwo}}$
  and
    $\judlsc{\tradlsc{\uenv},\tradlsc{\lenv_2}}{\tradlsc{\tmtwo}}{\tradlsc{\typtwo}}$.
  By weakening
  we have that
    $\judlsc{\tradlsc{\uenv},\tradlsc{\lenv_1},\tradlsc{\lenv_2}}{\tradlsc{\tm}}{\tradlsc{\typ}\to\tradlsc{\typtwo}}$
  and
    $\judlsc{\tradlsc{\uenv},\tradlsc{\lenv_1},\tradlsc{\lenv_2}}{\tradlsc{\tmtwo}}{\tradlsc{\typtwo}}$.
  Hence, by the $\rulelApp$ rule, we have that
  $\judlsc{\tradlsc{\uenv},\tradlsc{\lenv_1},\tradlsc{\lenv_2}}{\tradlsc{\tm}\,\tradlsc{\tmtwo}}{\tradlsc{\typtwo}}$, as required.
\item $\rulecSha$:
  Let
    $\judc{\uenv}{\lenv}{\sha{\tm}}{\sha{\typ}}$
  be derived from
    $\judc{\uenv}{\lenv}{\tm}{\typ}$.
  By \ih we have that
    $\judlsc{\tradlsc{\uenv},\tradlsc{\lenv}}{\tradlsc{\tm}}{\tradlsc{\typ}}$.
  By weakening
  we have that
    $\judlsc{\tradlsc{\uenv},\tradlsc{\lenv},\varthree:\lscUnit}{\tradlsc{\tm}}{\tradlsc{\typ}}$,
  where $\varthree$ is a fresh variable, not occurring free in $\uenv$, $\lenv$, nor $\tm$.
  Then, by the $\rulelAbs$ rule,
    $\judlsc{\tradlsc{\uenv},\tradlsc{\lenv}}{\lam{\varthree}{\tradlsc{\tm}}}{\lscUnit\to\tradlsc{\typ}}$.
  Hence
    $\judlsc{\tradlsc{\uenv},\tradlsc{\lenv}}{\tradlsc{\sha{\tm}}}{\tradlsc{\sha{\typ}}}$.
\item $\rulecOpen$:
  Let
    $\judc{\uenv}{\lenv}{\open{\tm}}{\typ}$
  be derived from
    $\judc{\uenv}{\lenv}{\tm}{\sha{\typ}}$.
  By \ih,
    $\judlsc{\tradlsc{\uenv},\tradlsc{\lenv}}{\tradlsc{\tm}}{\lscUnit\to\tradlsc{\typ}}$.
  By the $\rulelApp$ rule we have that
    $\judlsc{\tradlsc{\uenv},\tradlsc{\lenv}}{\tradlsc{\tm}\,\lscunit}{\tradlsc{\typ}}$,
  given that $\lscunit$ is assumed to be a closed term of type $\lscUnit$.
  So indeed
    $\judlsc{\tradlsc{\uenv},\tradlsc{\lenv}}{\tradlsc{\open{\tm}}}{\tradlsc{\typ}}$.
\item $\rulecProm$:
  Let
    $\judc{\uenv}{\noenv}{\ofc{\tm}}{\ofc{\typ}}$
  be derived from
    $\judc{\uenv}{\noenv}{\tm}{\typ}$.
  Then by \ih
    $\judlsc{\tradlsc{\uenv}}{\tradlsc{\tm}}{\tradlsc{\typ}}$,
  and we are done because
    $\tradlsc{\ofc{\tm}} = \tradlsc{\tm}$
  and
    $\tradlsc{\ofc{\typ}} = \tradlsc{\typ}$
  by definition.
\item $\rulecES$:
  Let
    $\judc{\uenv}{\lenv_1,\lenv_2}{\tm\esub{\uvar}{\tmtwo}}{\typtwo}$
  be derived from
    $\judc{\uenv,\uvar:\typ}{\lenv_1}{\tm}{\typtwo}$
  and
    $\judc{\uenv}{\lenv_2}{\tmtwo}{\osha{\typ}}$.
  By \ih we have that
    $\judlsc{\tradlsc{\uenv},\uvar:\lscUnit\to\tradlsc{\typ},\tradlsc{\lenv_1}}{\tradlsc{\tm}}{\tradlsc{\typtwo}}$
  and
    $\judlsc{\tradlsc{\uenv},\tradlsc{\lenv_2}}{\tradlsc{\tmtwo}}{\tradlsc{\osha{\typ}}}$,
  where, by definition,
    $\tradlsc{\osha{\typ}} = \lscUnit\to\tradlsc{\typ}$.
  By weakening
  we have that
    $\judlsc{\tradlsc{\uenv},\uvar:\lscUnit\to\tradlsc{\typ},\tradlsc{\lenv_1},\tradlsc{\lenv_2}}{\tradlsc{\tm}}{\tradlsc{\typtwo}}$
  and
    $\judlsc{\tradlsc{\uenv},\tradlsc{\lenv_1},\tradlsc{\lenv_2}}{\tradlsc{\tmtwo}}{\lscUnit\to\tradlsc{\typ}}$.
  Hence by the $\rulelES$ rule
    $\judlsc{\tradlsc{\uenv},\tradlsc{\lenv_1},\tradlsc{\lenv_2}}{\tradlsc{\tm}\esub{\uvar}{\tradlsc{\tmtwo}}}{\tradlsc{\typtwo}}$.
  To conclude, it suffices to observe that
  $\tradlsc{\tm\esub{\uvar}{\tmtwo}} = \tradlsc{\tm}\esub{\uvar}{\tradlsc{\tmtwo}}$.
\end{enumerate}
\end{proof}

\begin{lemma}[Simulation of $\lambdaS$ in LSC, up to fusion]
\llem{simulation_lambdaS_LSC}
If $\tm \toSi \tm'$
then $\tradlsc{\tm} \tolsci^+\tofuse \tradlsc{\tm'}$.
\end{lemma}
\begin{proof}
By induction on the derivation of the step $\tm \toSi \tm'$:
\begin{enumerate}
\item
  Root $\toSdb$ step:
  let
    $(\lam{\lvar}{\tm})\sctx\,\tmtwo
     \rtoSdb
     \tm\sub{\lvar}{\tmtwo}\sctx$.
  Then:
  \[
    \begin{array}{rlll}
      \tradlsc{(\lam{\lvar}{\tm})\sctx\,\tmtwo}
    & = &
      (\lam{\lvar}{\tradlsc{\tm}})\tradlsc{\sctx}\,\tradlsc{\tmtwo}
    \\
    & \todb &
      \tradlsc{\tm}\esub{\lvar}{\tradlsc{\tmtwo}}\tradlsc{\sctx}
    \\
    & \tols^* &
      \tradlsc{\tm}\sub{\lvar}{\tradlsc{\tmtwo}}\esub{\lvar}{\tradlsc{\tmtwo}}\tradlsc{\sctx}
    \\
    & \tofuse &
      \tradlsc{\tm}\sub{\lvar}{\tradlsc{\tmtwo}}\tradlsc{\sctx}
      & \text{by $\ruleFuseW$}
    \\
    & = &
      \tradlsc{\tm\sub{\lvar}{\tmtwo}\sctx}
      & \text{by \rlem{tradlsc_substitution}}
    \end{array}
  \]
\item
  Root $\toSopen$ step:
  let
    $\open{(\sha{\tm})\sctx}
     \rtoSopen
     \tm\sctx$.
  Then, for $\varthree \notin \fv{\tmthree}$:
  \[
    \begin{array}{rlll}
      \tradlsc{\open{(\sha{\tm})\sctx}}
    & = &
      (\lam{\varthree}{\tradlsc{\tm}})\tradlsc{\sctx}\,\lscunit
    \\
    & \todb &
      \tradlsc{\tm}\esub{\varthree}{\lscunit}\tradlsc{\sctx}
    \\
    & \tofuse &
      \tradlsc{\tm}\tradlsc{\sctx}
      & \text{by $\ruleFuseW$, as $\varthree \notin \fv{\tm} = \fv{\tradlsc{\tm}}$}
    \\
    & = &
      \tradlsc{\tm\sctx}
    \end{array}
  \]
\item
  Root $\toSls$ step:
  let
    $\off{\gctx}{\uvar}\esub{\uvar}{(\ofc{(\sha{\tm})\sctx_1})\sctx_2}
     \rtoSls
     \off{\gctx}{(\sha{\tm})\sctx_1}\esub{\uvar}{\ofc{(\sha{\tm})\sctx_1}}\sctx_2$.
  Then:
  \[
    \begin{array}{rlll}
      \tradlsc{\off{\gctx}{\uvar}\esub{\uvar}{(\ofc{(\sha{\tm})\sctx_1})\sctx_2}}
    & = &
      \off{\tradlsc{\gctx}}{\uvar}\esub{\uvar}{(\lam{\varthree}{\tradlsc{\tm}})\tradlsc{\sctx_1}\tradlsc{\sctx_2}}
    \\
    & \tols &
      \off{\tradlsc{\gctx}}{
        (\lam{\varthree}{\tradlsc{\tm}})\tradlsc{\sctx_1}\tradlsc{\sctx_2}
      }\esub{\uvar}{(\lam{\varthree}{\tradlsc{\tm}})\tradlsc{\sctx_1}\tradlsc{\sctx_2}}
    \\
    & \tofuse &
      \off{\tradlsc{\gctx}}{
        (\lam{\varthree}{\tradlsc{\tm}})\tradlsc{\sctx_1}
      }\esub{\uvar}{(\lam{\varthree}{\tradlsc{\tm}})\tradlsc{\sctx_1}}
       \tradlsc{\sctx_2}
      & \text{by \rlem{properties_of_fusion}}
    \\
    & = &
      \tradlsc{\off{\gctx}{(\sha{\tm})\sctx_1}\esub{\uvar}{\ofc{(\sha{\tm})\sctx_1}}\sctx_2}
    \end{array}
  \]
\item
  Congruence closure below
  contexts of any of the following forms:
  \[
    \lam{\lvar}{\ctxhole}
    \HS
    \ctxhole\,\tmtwo
    \HS
    \tmtwo\,\ctxhole
    \HS
    \sha{\ctxhole}
    \HS
    \open{\ctxhole}
    \HS
    \ofc{\ctxhole}
    \HS
    \ctxhole\,\tmtwo
    \HS
    \tmtwo\,\ctxhole
  \]
  are all immediate by \ih
  because both $\tolsci$ and $\tofuse$
  are closed by compatibility under arbitrary contexts.
  For example,
  suppose that $\sha{\tm} \toSi \sha{\tm'}$
  is derived from $\tm \toSi \tm'$.
  Then by \ih we have that
  $\tradlsc{\tm} \tolsci^+\tofuse \tradlsc{\tm'}$.
  Hence
  $\tradlsc{\sha{\tm}}
   = \lam{\varthree}{\tradlsc{\tm}}
   \tolsci^+\tofuse
     \lam{\varthree}{\tradlsc{\tm'}}
   = \tradlsc{\sha{\tm'}}$.
\end{enumerate}
\end{proof}

\begin{lemma}[Abstract postponement lemma]
\llem{abstract_postponement_lemma}
Let $\ars = (X,\to)$ be an ARS.
Suppose that reduction can be written
as a union ${\to} = {\to_1 \cup \to_2}$,
where $\to_2$ is SN
and it can be postponed in the sense that
${\to_2\to_1} \subseteq {\to_1\to_2^*}$.
Then an arbitrary object $x \in X$ is $\to$-SN if and only if it is $\to_1$-SN.
\end{lemma}
\begin{proof}
$(\Rightarrow)$
  Immediate.
$(\Leftarrow)$
  We prove the statement by first proving two auxiliary claims:
  \begin{enumerate}
  \item
    First claim: ${\to^*_2\to_1} \subseteq {\to_1\to^*_2}$. \\
    Since $\to^*$ is the reflexive--transitive closure of $\to$,
    by definition, this is equivalent to showing that 
    ${\to^n_2\to_1} \subseteq {\to_1\to_2^*}$ holds for all $n \in \Nat_0$.
    We proceed by induction on $n$.
    If $n = 0$, it is immediate to conclude.
    Now assume the property holds for a given $n \in \Nat_0$.
    Then:
    \[
      \begin{array}{rcll}
        {\to^{n+1}_2\to_1}
      & = &
        {\to^{n}_2\to_2\to_1}
      \\
      & \subseteq &
        {\to^{n}_2\to_1\to^*_2}
        & \text{by the hypothesis that ${\to_2\to_1} \subseteq {\to_1\to^*_2}$}
      \\
      & \subseteq &
        \to_1{\to^*_2\to^*_2}
        & \text{by \ih}
      \\
      & \subseteq &
        \to_1\to^*_2
        & \text{by transitivity}
      \end{array}
    \]
  \item
    Second claim:
    if $x \in X$ is not $\to$-SN, there exists $y \in X$
    such that $x \to_1 y$ and $y$ is not $\to$-SN. \\
    Suppose that there is an infinite reduction sequence
    $x = x_0 \to x_1 \to x_2 \to \hdots$.
    By hypothesis, $\to_2$ is SN, so the reduction sequence must contain
    at least one $\to_1$ step.
    Let $i$ be the first index such that $x \to^*_2 x_i \to_1 x_{i+1}$.
    By the first claim we have that ${\to^*_2\to_1} \subseteq {\to_1\to^*_2}$,
    so $x \to_1 y \to^*_2 x_{i+1}$ for some $y \in X$.
    Note there is an infinite $\to$-sequence starting on $y$
    since $y \to^*_2 x_{i+1} \to x_{i+2} \to x_{i+3} \hdots$,
    so $y$ is not $\to$-SN, as required.
  \item
    Finally, we prove the main statement:
    if $x$ is $\to_1$-SN then $x$ is $\to$-SN. \\
    Suppose that $x$ is $\to_1$-SN.
    By contradiction, suppose that $x$ is not $\to$-SN.
    Then by the second claim there exists $x_1 \in X$
    such that $x \to_1 x_1$ and $x_1$ is not $\to$-SN.
    Iterating this argument, we construct an infinite reduction sequence
    $x \to_1 x_1 \to_1 x_2 \hdots$,
    contradicting the fact that $x$ is $\to_1$-SN.
  \end{enumerate}
\end{proof}

\StrongNormalization
\begin{proof}
Suppose that $\tm$ is a typable $\lambdaS$-term,
\ie $\judc{\uenv}{\lenv}{\tm}{\typ}$ holds for some $\uenv,\lenv,\typ$.
By the fact that $\symgc$ steps can be postponed (\rlem{postponement_toSgc})
and~\rlem{abstract_postponement_lemma},
showing that $\tm$ is $\toS$-SN
is equivalent to showing that $\tm$ is $\toSi$-SN,
where we recall that ${\toSi} = {\toS \setminus \toSgc}$.

Suppose that there is an infinite reduction sequence
$\tm \toSi \tm_1 \toSi \tm_2 \hdots$ (without $\toSgc$ steps).
By \rlem{simulation_lambdaS_LSC}
we obtain an infinite reduction sequence
$\tradlsc{\tm} \tolsci^+\tofuse \tradlsc{\tm_1} \tolsci^+\tofuse \tradlsc{\tm_2} \hdots$
where we know that $\tradlsc{\tm}$ is simply typable in $\LSC$,
as the translation preserves typing~(\rlem{tradlsc_preserves_typing}).
Since $\tradlsc{\tm}$ is simply typable in $\LSC$,
then $\tradlsc{\tm}$ in particular $\tolsci$-SN by~\rthm{typed_lsc_sn}.
To conclude, apply~\rlem{fusion_preserves_SN}
to obtain that $\tradlsc{\tm}$ must be $(\tolsci^+\tofuse)$-SN,
in contradiction with the fact that we have an infinite
sequence of $(\tolsci^+\tofuse)$-steps starting from $\tradlsc{\tm}$.
\end{proof}


  \subsection{Normal forms}
  \lsec{appendix:calculus_normal_forms}
  
We present an inductive characterization of the $\toS$-normal forms.
It is used in the proof of preservation of normal forms of our translations.
Let
$\ops\eqdef\{\varSymb,\lamSymb,\appSymb,\openSymb,\shaSymb,\ofcSymb,\oshaSymb\}$.
The set of $\toS$-normal forms is characterized by 
$\nfS{}\eqdef \bigcup_{\nfop\in\ops}\nfS{\nfop}$, where $\nfS{\nfop}$ is defined
as follows:
\[
    \indrule{}
    {\vphantom{\tm\in \nfS{\nfop}}}
    {\lvar\in\nfS{\varSymb}}
  \,\,
    \indrule{}
    {\tm\in \nfS{\nfop}
    \quad
    \tmtwo\in \nfS{\nfoptwo}
    \quad
    \uvar\in\fv{\tm}
    \quad
    \nfoptwo\neq\oshaSymb
    }
    {\tm\esub{\uvar}{\tmtwo}\in\nfS{\nfop}}
\]
\[
    \indrule{}
    {\vphantom{\tm\in \nfS{\nfop}}}
    {\uvar\in\nfS{\varSymb}}
  \,\,
    \indrule{}
    {\tm\in \nfS{\nfop}
    \quad
    \tmtwo\in \nfS{\nfoptwo}
    \quad
    \uvar\notin\fv{\tm}
    \quad
    \nfoptwo\notin\set{\oshaSymb,\ofcSymb}
    }
    {\tm\esub{\uvar}{\tmtwo}\in\nfS{\nfop}}     
\]
\[
    \indrule{}
    {\tm\in \nfS{\nfop}}
    {\lam{\lvar}{\tm}\in\nfS{\lamSymb}}
  \,\,
    \indrule{}
    {\tm\in \nfS{\nfop}}
    {\sha{\tm}\in\nfS{\shaSymb}}
  \,\,
    \indrule{}
    {\tm\in \nfS{\nfop}
    \quad
    \nfop\neq\shaSymb}
    {\ofc{\tm}\in\nfS{\ofcSymb}}
  \,\,
    \indrule{}
    {\tm\in \nfS{\shaSymb}}
    {\ofc{\tm}\in\nfS{\oshaSymb}}
\]
\[
    \indrule{}
    {\tm\in \nfS{\nfop}
    \quad
    \tmtwo\in \nfS{\nfoptwo}
    \quad
    \nfop\neq\lamSymb}
    {\tm\,\tmtwo\in\nfS{\appSymb}}
  \,\,
    \indrule{}
    {\tm\in \nfS{\nfop}
    \quad
    \nfop\neq\shaSymb}
    {\open{\tm}\in\nfS{\openSymb}}
\]

\section{Appendix: Embedding $\CBN$, $\CBV$ and $\CCBNd$}

\subsection{$\CBN$}

\begin{remark}
\lremark{traN_fv}
$\fv{\traN{\tm}} = \fv{\tm}$.
\end{remark}

\begin{remark}
$\traN{\of{\gctx}{\tm}} = \of{\traN{\gctx}}{\traN{\tm}}$
and
$\traN{\off{\gctx}{\tm}} = \off{\traN{\gctx}}{\traN{\tm}}$.
In particular,
$\traN{(\tm\sctx)} = \traN{\tm}\traN{\sctx}$.
\end{remark}

\TypingTheCBNTranslation

\begin{proof}\label{typing_the_CBN_translation:proof}
By induction on the derivation of $\judl{\tctx}{\tm}{\typ}$:
\begin{enumerate}
\item
  $\rulelVar$:
  Let $\judl{\tctx,\var:\typ}{\var}{\typ}$ be derived from the $\rulelVar$ rule.
  Then:
  \[
    \indrule{\rulecOpen}{
      \indrule{\rulecUvar}{
        \emptyPremise
      }{
        \judc{\traN{\tctx},\var:\traN{\typ}}{\noenv}{\var}{\sha{\traN{\typ}}}
      }
    }{
      \judc{\traN{\tctx},\var:\traN{\typ}}{\noenv}{\open{\var}}{\traN{\typ}}
    }
  \]
\item
  $\rulelAbs$:
  Let $\judl{\tctx}{\lam{\var}{\tm}}{\typ\to\typtwo}$
  be derived from $\judl{\tctx,\var:\typ}{\tm}{\typtwo}$.
  Let $\lvar$ be a fresh linear variable,
  such that $\lvar \notin \fv{\traN{\tm}}$.
  Then:
  \[
    \indrule{\rulecAbs}{
      \indrule{\rulecES}{
        \derivih{\judc{\traN{\tctx},\var:\traN{\typ}}{\noenv}{
          \traN{\tm}
        }{\traN{\typtwo}}}
        \indrule{\rulecLvar}{
          \emptyPremise
        }{
          \judc{\traN{\tctx}}{\lvar:\osha{\traN{\typ}}}{
            \lvar
          }{\osha{\traN{\typ}}}
        }
      }{
        \judc{\traN{\tctx}}{\lvar:\osha{\traN{\typ}}}{
          \traN{\tm}\esub{\var}{\lvar}
        }{\traN{\typtwo}}
      }
    }{
      \judc{\traN{\tctx}}{\noenv}{
        \lam{\lvar}{\traN{\tm}\esub{\var}{\lvar}}
      }{\osha{\traN{\typ}}\limp\traN{\typtwo}}
    }
  \]
\item
  $\rulelApp$:
  Let $\judl{\tctx}{\tm\,\tmtwo}{\typtwo}$
  be derived from $\judl{\tctx}{\tm}{\typ\to\typtwo}$
  and $\judl{\tctx}{\tmtwo}{\typ}$.
  Then:
  \[
    \indrule{\rulecApp}{
      \derivih{\judc{\traN{\tctx}}{\noenv}{\traN{\tm}}{\osha{\traN{\typ}}\limp\traN{\typtwo}}}
      \indrule{\rulecProm}{
        \indrule{\rulecSha}{
          \derivih{\judc{\traN{\tctx}}{\noenv}{\traN{\tmtwo}}{\traN{\typ}}}
        }{
          \judc{\traN{\tctx}}{\noenv}{\sha{\traN{\tmtwo}}}{\sha{\traN{\typ}}}
        }
      }{
        \judc{\traN{\tctx}}{\noenv}{\osha{\traN{\tmtwo}}}{\osha{\traN{\typ}}}
      }
    }{
      \judc{\traN{\tctx}}{\noenv}{\traN{\tm}\,\osha{\traN{\tmtwo}}}{\traN{\typtwo}}
    }
  \]
\item
  $\rulelES$:
  Let $\judl{\tctx}{\tm\esub{\var}{\tmtwo}}{\typtwo}$
  be derived from $\judl{\tctx,\var:\typ}{\tm}{\typtwo}$
  and $\judl{\tctx}{\tmtwo}{\typ}$.
  Then:
  \[
    \indrule{}{
      \derivih{\judc{\traN{\tctx},\var:\traN{\typ}}{\noenv}{\traN{\tm}}{\traN{\typtwo}}}
      \indrule{\rulecProm}{
        \indrule{\rulecSha}{
          \derivih{\judc{\traN{\tctx}}{\noenv}{\traN{\tmtwo}}{\traN{\typ}}}
        }{
          \judc{\traN{\tctx}}{\noenv}{\sha{\traN{\tmtwo}}}{\sha{\traN{\typ}}}
        }
      }{
        \judc{\traN{\tctx}}{\noenv}{\osha{\traN{\tmtwo}}}{\osha{\traN{\typ}}}
      }
    }{
      \judc{\traN{\tctx}}{\noenv}{\traN{\tm}\esub{\var}{\osha{\traN{\tmtwo}}}}{\traN{\typtwo}}
    }
  \]
\end{enumerate}
\end{proof}

\begin{definition}
We define a subset $\CtxsSName \subseteq \CtxsS$, called {\em $\CBN$ contexts},
by the following grammar:
\[
  \rgctx ::= \open{\ctxhole}
        \mid \lam{\lvar}{\rgctx\esub{\uvar}{\lvar}}
        \mid \rgctx\,\osha{\rtm}
        \mid \rtm\,\osha{\rgctx}
        \mid \rgctx\esub{\uvar}{\osha{\rtm}}
        \mid \rtm\esub{\uvar}{\osha{\rgctx}}
\]
where, in the production $\rgctx ::= \lam{\lvar}{\rgctx\esub{\uvar}{\lvar}}$
we assume that $\lvar$ is fresh, that is $\lvar \notin \fv{\rgctx}$.
Furthermore, we define a subset $\SCtxsSName \subseteq \SCtxsS$
of the set of substitution contexts,
called {\em $\CBN$ substitution contexts}:
\[
  \rsctx ::= \ctxhole \mid \rsctx\esub{\uvar}{\osha{\rtm}}
\]
The inverse translation can be extended to $\CBN$ contexts
and substitution contexts,
setting $\traNinv{\open{\ctxhole}} \eqdef \ctxhole$ for $\CBN$ contexts,
and $\traNinv{\ctxhole} \eqdef \ctxhole$ for $\CBN$ substitution contexts.
\end{definition}

\begin{remark}
\lremark{traNinv_fv}
$\fv{\traNinv{\rtm}} = \fv{\rtm}$.
\end{remark}

\begin{lemma}[Context decomposition for the inverse $\CBN$ translation]
\llem{traNinv_decompose_contexts}
\quad
\begin{enumerate}
\item
  $\tm\sctx \in \TermsSName$
  if and only if
  $\tm \in \TermsSName$ and $\sctx \in \SCtxsSName$.
\item
  $\off{\gctx}{\var} \in \TermsSName$
  if and only if
  $\gctx \in \CtxsSName$.
\item
  If $\rgctx \in \CtxsSName$
  and $\rtm \in \TermsSName$
  then $\off{\rgctx}{\sha{\rtm}} \in \TermsSName$.
\item
  If $\rtm \in \TermsSName$ and $\rsctx \in \SCtxsSName$,
  then $\traNinv{(\rtm\,\rsctx)} = \traNinv{\rtm}\traNinv{\rsctx}$.
\item
  If $\rgctx \in \CtxsSName$
  then $\traNinv{\off{\rgctx}{\var}} = \off{\traNinv{\rgctx}}{\var}$.
\item
  If $\rgctx \in \CtxsSName$ and $\rtm \in \TermsSName$,
  then $\traNinv{\off{\rgctx}{\sha{\rtm}}} = \off{\traNinv{\rgctx}}{\traNinv{\rtm}}$.
\end{enumerate}
\end{lemma}
\begin{proof}
 By induction on the first judgement in the statement of each item, except the first and fourth items which are by induction on $\sctx$ and $\rsctx \in \SCtxsSName$, resp.
\end{proof}

\traNinvSimulation

\begin{proof}\label{traNinv_simulation:proof}
By induction on the (unique) derivation of $\rtm \in \TermsSName$:
\begin{enumerate}
\item
  $\rtm = \open{\var}$:
  Impossible, as there are no steps $\rtm \toS \tmtwo$.
\item
  $\rtm = \open{\sha{\rtm'}}$:
  We consider two subcases, depending on whether the step is at the root
  of the term or internal to $\sha{\rtm'}$:
  \begin{enumerate}
  \item
    If the step is at the root, we have that
    $\rtm = \open{\sha{\rtm'}} \rtoSopen \rtm' = \tmtwo$,
    so $\tmtwo = \rtm' \in \TermsSName$
    and
    $\traNinv{\rtm}
     = \traNinv{\open{\sha{\rtm'}}}
     = \traNinv{\rtm'}
     = \traNinv{\tmtwo}$.
  \item
    If the step is internal to $\sha{\rtm'}$,
    note that it cannot be at the root of $\sha{\rtm'}$,
    since this term does not match the left-hand side of
    any rewriting rule.
    Then the step must be internal to $\rtm'$,
    that is,
    $\rtm = \open{\sha{\rtm'}} \toS \open{\sha{\tmtwo'}} = \tmtwo$
    with $\rtm' \toS \tmtwo'$.
    By \ih, $\tmtwo' \in \TermsSName$,
    so $\tmtwo = \open{\sha{\tmtwo'}} \in \TermsSName$,
    and
    $\traNinv{\rtm}
     = \traNinv{\open{\sha{\rtm'}}}
     = \traNinv{\rtm'}
     \tocbn^= \traNinv{\tmtwo'}
     = \traNinv{\open{\sha{\tmtwo'}}}
     = \traNinv{\tmtwo}$.
  \end{enumerate}
\item
  $\rtm = \lam{\lvar}{\rtm'\esub{\var}{\lvar}}$:
  Note that the step cannot be at the root,
  and that that there cannot be a $\toSls$ nor a $\toSgc$ step
  involving the substitution $\esub{\var}{\lvar}$,
  since these rules would require that $\lvar$ be of the form
  $(\ofc{\tmthree})\sctx$, but $\lvar$ is a linear variable.
  This means that the step must be internal to $\rtm'$, 
  that is,
  $\rtm = \lam{\lvar}{\rtm'\esub{\var}{\lvar}}
      \toS \lam{\lvar}{\tmtwo'\esub{\var}{\lvar}} = \tmtwo$
  with $\rtm' \toS \tmtwo'$.
  By \ih, $\tmtwo' \in \TermsSName$,
  so $\tmtwo = \lam{\lvar}{\tmtwo'\esub{\var}{\lvar}} \in \TermsSName$,
  and
  $\traNinv{\rtm}
   = \traNinv{(\lam{\lvar}{\rtm'\esub{\var}{\lvar}})}
   = \lam{\var}{\traNinv{\rtm'}}
   \tocbn^= \lam{\var}{\traNinv{\tmtwo'}}
   = \traNinv{(\lam{\lvar}{\tmtwo'\esub{\var}{\lvar}})}
   = \traNinv{\tmtwo}$.
\item
  $\rtm = \rtm_1\,\osha{\rtm_2}$:
  We consider three subcases, depending on whether the step is at the root
  of the term, internal to $\rtm_1$, or internal to $\osha{\rtm_2}$:
  \begin{enumerate}
  \item
    If the step is at the root of the term, it must be a $\symSdb$ step,
    that is, $\rtm_1$ must be of the form $(\lam{\lvar}{\tm})\sctx$.
    By \rlem{traNinv_decompose_contexts},
    this means that $\lam{\lvar}{\tm} \in \TermsSName$
    and $\sctx \in \SCtxsSName$.
    In particular, $\tm$ must be of the form $\tm = \rtm'_1\esub{\var}{\lvar}$.
    Then we have that
    $\rtm = (\lam{\lvar}{\rtm'_1\esub{\var}{\lvar}})\sctx\,\osha{\rtm_2}
     \rtoSdb \rtm'_1\esub{\var}{\osha{\rtm_2}}\sctx = \tmtwo$.
    Note that
    $\tmtwo = \rtm'_1\esub{\var}{\osha{\rtm_2}}\sctx \in \TermsSName$
    by \rlem{traNinv_decompose_contexts}.
    By \ih, we have that
    \[
      \begin{array}{rll}
        \traNinv{\rtm}
      & = &
        \traNinv{((\lam{\lvar}{\rtm'_1\esub{\var}{\lvar}})\sctx\,\osha{\rtm_2})}
      \\
      & = &
        (\lam{\var}{\traNinv{{\rtm'_1}}})\traNinv{\sctx}\,\traNinv{\rtm_2}
      \\
      & \rtodb &
        \traNinv{{\rtm'_1}}\esub{\var}{\traNinv{\rtm_2}}\traNinv{\sctx}
      \\
      & = &
        \traNinv{(\rtm'_1\esub{\var}{\osha{\rtm_2}}\sctx)}
      \\
      & = &
        \traNinv{\tmtwo}
      \end{array}
    \]
  \item
    If the step is internal to $\rtm_1$,
    we have that
    $\rtm = \rtm_1\,\osha{\rtm_2} \toS \tmtwo_1\,\osha{\rtm_2} = \tmtwo$
    with $\rtm_1 \toS \tmtwo_1$.
    By \ih, $\tmtwo_1 \in \TermsSName$,
    so $\tmtwo = \tmtwo_1\,\osha{\rtm_2} \in \TermsSName$
    and
    $\traNinv{\rtm}
     = \traNinv{\rtm_1}\,\traNinv{\rtm_2}
     \tocbn^= \traNinv{\tmtwo_1}\,\traNinv{\rtm_2}
     = \traNinv{\tmtwo}$.
  \item
    If the step is internal to $\osha{\rtm_2}$,
    note that it cannot be at the root of $\osha{\rtm_2}$
    nor at the root of $\sha{\rtm_2}$,
    since these terms do not match the left-hand side of any rewriting rule.
    So the step must be internal to $\rtm_2$,
    that is, we have that
    $\rtm = \rtm_1\,\osha{\rtm_2} \toS \rtm_1\,\osha{\tmtwo_2} = \tmtwo$
    with $\rtm_2 \toS \tmtwo_2$.
    By \ih,
    $\tmtwo_2 \in \TermsSName$,
    so $\tmtwo = \rtm_1\,\osha{\tmtwo_2} \in \TermsSName$
    and
    $\traNinv{\rtm}
     = \traNinv{\rtm_1}\,\traNinv{\rtm_2}
     \tocbn^= \traNinv{\rtm_1}\,\traNinv{\tmtwo_2}
     = \traNinv{\tmtwo}$.
  \end{enumerate}
\item
  $\rtm = \rtm_1\esub{\var}{\osha{\rtm_2}}$:
  We consider three subcases, depending on whether the step is at the root
  of the term, internal to $\rtm_1$, or internal to $\osha{\rtm_2}$:
  \begin{enumerate}
  \item
    If the step is at the root of the term,
    it must be either a $\symSls$ or a $\symSgc$ step.
    We consider two further subcases:
    \begin{enumerate}
    \item
      If the step is a $\symSls$ step, we have that
      $\rtm_1 = \off{\gctx}{\var} \in \TermsSName$
      and
      $\rtm
       = \off{\gctx}{\var}\esub{\var}{\osha{\rtm_2}}
       \rtoSls \off{\gctx}{\sha{\rtm_2}}\esub{\var}{\osha{\rtm_2}}
       = \tmtwo$.
      Note that, by \rlem{traNinv_decompose_contexts},
      $\gctx \in \CtxsSName$,
      so, again by \rlem{traNinv_decompose_contexts}, we have
      $\tmtwo = \off{\gctx}{\sha{\rtm_2}}\esub{\var}{\osha{\rtm_2}} \in \TermsSName$.
      Moreover, also using \rlem{traNinv_decompose_contexts},
      we have that
      $\traNinv{\rtm}
       = \traNinv{(\off{\gctx}{\var}\esub{\var}{\osha{\rtm_2}})}
       = \off{\traNinv{\gctx}}{\var}\esub{\var}{\traNinv{\rtm_2}}
       \rtols \off{\traNinv{\gctx}}{\traNinv{\rtm_2}}\esub{\var}{\traNinv{\rtm_2}}
       = \traNinv{(\off{\gctx}{\sha{\rtm_2}}\esub{\var}{\osha{\rtm_2}})}
       = \traNinv{\tmtwo}$.
    \item
      If the step is a $\symSgc$ step, we have that $\var \notin \fv{\rtm_1}$
      and $\rtm = \rtm_1\esub{\var}{\osha{\rtm_2}} \rtoSgc \rtm_1 = \tmtwo$.
      Note that $\tmtwo = \rtm_1 \in \TermsSName$.
      Moreover,
      $\traNinv{\rtm}
       = \traNinv{\rtm_1\esub{\var}{\osha{\rtm_2}}}
       = \traNinv{\rtm_1}\esub{\var}{\traNinv{\rtm_2}}
       \rtogc \traNinv{\rtm_1}
       = \traNinv{\tmtwo}$.
      Note that $\var \notin \traNinv{\rtm_1}$
      because $\fv{\traNinv{\rtm_1}} = \fv{\rtm_1}$,
      as noted in \rremark{traNinv_fv}.
    \end{enumerate}
  \item
    If the step is internal to $\rtm_1$,
    then
    $\rtm = \rtm_1\esub{\var}{\osha{\rtm_2}}
     \toS \tmtwo_1\esub{\var}{\osha{\rtm_2}} = \tmtwo$
     with $\rtm_1 \toS \tmtwo_1$.
    By \ih, $\tmtwo_1 \in \TermsSName$,
    so $\tmtwo = \tmtwo_1\esub{\var}{\osha{\rtm_2}} \in \TermsSName$
    and
    $\traNinv{\rtm}
     = \traNinv{\rtm_1}\esub{\var}{\traNinv{\rtm_2}}
     \tocbn^= \traNinv{\tmtwo_1}\esub{\var}{\traNinv{\rtm_2}}
     = \traNinv{\tmtwo}$.
  \item
    If the step is internal to $\osha{\rtm_2}$,
    note that it cannot be at the root of $\osha{\rtm_2}$
    nor at the root of $\sha{\rtm_2}$,
    since these terms do not match the left-hand side of any rewriting rule.
    So the step must be internal to $\rtm_2$,
    that is,
    we have that
    $\rtm = \rtm_1\esub{\var}{\osha{\rtm_2}}
     \toS \rtm_1\esub{\var}{\osha{\tmtwo_2}} = \tmtwo$
     with $\rtm_2 \toS \tmtwo_2$.
    By \ih,
    $\tmtwo_2 \in \TermsSName$,
    so $\tmtwo = \rtm_1\esub{\var}{\osha{\tmtwo_2}} \in \TermsSName$ 
    and
    $\traNinv{\rtm}
     = \traNinv{\rtm_1}\esub{\var}{\traNinv{\rtm_2}}
     \tocbn^= \traNinv{\rtm_1}\esub{\var}{\traNinv{\tmtwo_2}}
     = \traNinv{\tmtwo}$.
  \end{enumerate}
\end{enumerate}
\end{proof}

\subsection{$\CBV$}

\TypingTheCBVTranslation

\begin{proof}\label{typing_the_CBV_translation:proof}
By induction on the derivation of $\judl{\tctx}{\tm}{\typ}$:
\begin{enumerate}
\item
  $\rulelVar$:
  Let $\judl{\tctx,\var:\typ}{\var}{\typ}$ be derived from the $\rulelVar$ rule.
  Then:

  \scalebox{\smallDerivation}{\parbox{\textwidth}{
      $
    \indrule{\rulecProm}{
      \indrule{\rulecUvar}{
        \emptyPremise
      }{
        \judc{\traV{\tctx},\var:\traV{\typ}}{\noenv}{\var}{\sha{\traV{\typ}}}
      }
    }{
      \judc{\traV{\tctx},\var:\traV{\typ}}{\noenv}{\ofc{\var}}{\osha{\traV{\typ}}}
    }
    $
  }}

\item
  $\rulelAbs$:
  Let $\judl{\tctx}{\lam{\var}{\tm}}{\typ\to\typtwo}$
  be derived from $\judl{\tctx,\var:\typ}{\tm}{\typtwo}$.
  Let $\lvar$ be a fresh linear variable, such that $\lvar \notin \fv{\traV{\tm}}$.
  Then:
  
\scalebox{\smallDerivation}{\parbox{\textwidth}{
    $
    \indrule{\rulecProm}
    {\indrule{\rulecSha}
     {\indrule{\rulecAbs}{
      \indrule{\rulecES}{
        \derivih{\judc{\traV{\tctx},\var:\traV{\typ}}{\noenv}{\traV{\tm}}{\osha{\traV{\typtwo}}}}
        \HS
        \indrule{\rulecLvar}{
          \emptyPremise
        }{
          \judc{\traV{\tctx}}{\lvar:\osha{\traV{\typ}}}{\lvar}{\osha{\traV{\typ}}}
        }
      }{
        \judc{\traV{\tctx}}{\lvar:\osha{\traV{\typ}}}{\traV{\tm}\esub{\var}{\lvar}}{\osha{\traV{\typtwo}}}
      }
    }{
      \judc{\traV{\tctx}}{\noenv}{\lam{\lvar}{\traV{\tm}\esub{\var}{\lvar}}}{\osha{\traV{\typ}}\limp\osha{\traV{\typtwo}}}
    }
  }{
          \judc{\traV{\tctx}}{\noenv}{\sha{\lam{\lvar}{\traV{\tm}\esub{\var}{\lvar}}}}{\sha{(\osha{\traV{\typ}}\limp\osha{\traV{\typtwo}})}}
        }
      }{
          \judc{\traV{\tctx}}{\noenv}{\osha{\lam{\lvar}{\traV{\tm}\esub{\var}{\lvar}}}}{\osha{(\osha{\traV{\typ}}\limp\osha{\traV{\typtwo}})}}
        }
        $
      }}
\item
  $\rulelApp$:
  Let $\judl{\tctx}{\tm\,\tmtwo}{\typtwo}$
  be derived from $\judl{\tctx}{\tm}{\typ\to\typtwo}$
  and $\judl{\tctx}{\tmtwo}{\typ}$.
  Then:

  \scalebox{\smallDerivation}{\parbox{\textwidth}{
  $
    \indrule{\rulecApp}{
      \indrule{\rulecES}{
        \deriv
      }{
        \judc{\traV{\tctx}}{\noenv}{
          \open{\uvar}\esub{\uvar}{\traV{\tm}}
        }{\osha{\traV{\typ}}\limp\osha{\traV{\typtwo}}}
      }
      \derivih{
        \judc{\traV{\tctx}}{\noenv}{
          \traV{\tmtwo}
        }{\osha{\traV{\typ}}}
      }
    }{
      \judc{\traV{\tctx}}{\noenv}{
        \open{\uvar}\esub{\uvar}{\traV{\tm}}\,\traV{\tmtwo}
      }{\osha{\traV{\typtwo}}}
    }
    $
  }}

  where $\deriv$ is given by:
  
  \scalebox{\smallDerivation}{\parbox{\textwidth}{
      $
    \indrule{\rulecES}{
      \indrule{\rulecOpen}{
        \indrule{\rulecUvar}{
          \emptyPremise
        }{
          \judc{\traV{\tctx},\uvar:\osha{\traV{\typ}}\limp\osha{\traV{\typtwo}}}{\noenv}{
            \uvar
          }{\sha{(\osha{\traV{\typ}}\limp\osha{\traV{\typtwo}})}}
        }
      }{
        \judc{\traV{\tctx},\uvar:\osha{\traV{\typ}}\limp\osha{\traV{\typtwo}}}{\noenv}{
          \open{\uvar}
        }{\osha{\traV{\typ}}\limp\osha{\traV{\typtwo}}}
      }
      \derivih{
        \judc{\traV{\tctx}}{\noenv}{
          \traV{\tm}
        }{\osha{(\osha{\traV{\typ}}\limp\osha{\traV{\typtwo}})}}
      }
    }{
      \judc{\traV{\tctx}}{\noenv}{
        \open{\uvar}\esub{\uvar}{\traV{\tm}}
      }{\osha{\traV{\typ}}\limp\osha{\traV{\typtwo}}}
    }
    $
    }}

\item
  $\rulelES$:
  Let $\judl{\tctx}{\tm\esub{\var}{\tmtwo}}{\typtwo}$
  be derived from $\judl{\tctx,\var:\typ}{\tm}{\typtwo}$
  and $\judl{\tctx}{\tmtwo}{\typ}$.
  Then:
  \[
    \indrule{\rulecES}{
      \derivih{\judc{\traV{\tctx},\var:\traV{\typ}}{\noenv}{\traV{\tm}}{\osha{\traV{\typtwo}}}}
      \derivih{\judc{\traV{\tctx}}{\noenv}{\traV{\tmtwo}}{\osha{\traV{\typ}}}}
    }{
      \judc{\traV{\tctx}}{\noenv}{\traV{\tm}\esub{\var}{\traV{\tmtwo}}}{\osha{\traV{\typtwo}}}
    }
  \]
\end{enumerate}
The translation can be extended to operate on contexts, by declaring
$\traV{\ctxhole} \eqdef \ctxhole$. Note that the translation of a
substitution context is a substitution context:
\[
  \traV{(\ctxhole\esub{\var_1}{\tm_1}\hdots\esub{\var_n}{\tm_n})}
  =
  \ctxhole\esub{\var_1}{\traV{\tm_1}}\hdots\esub{\var_n}{\traV{\tm_n}}
\]
\end{proof}

\begin{remark}
\lremark{traV_fv}
$\fv{\traV{\tm}} = \fv{\tm}$
\end{remark}

\begin{remark}
$\traV{\of{\gctx}{\tm}} = \of{\traV{\gctx}}{\traV{\tm}}$
and
$\traV{\off{\gctx}{\tm}} = \off{\traV{\gctx}}{\traV{\tm}}$.
In particular,
$\traV{(\tm\sctx)} = \traV{\tm}\traV{\sctx}$.
\end{remark}

\traVSimulation

\begin{proof}\label{traV_simulation:proof}
By induction on the derivation of $\tm \tocbv \tmtwo$.
The interesting cases are when
there is a $\symdb$, $\symlsv$, or $\symgclv$ step at the root.
The closure by compatibility under arbitrary contexts is straightforward
by resorting to the \ih:
\begin{enumerate}
\item
  $\rtodb$:
  Let $(\lam{\var}{\tm})\sctx\,\tmtwo
       \rtodb
       \tm\esub{\var}{\tmtwo}\sctx$.
  Then:
  \[
    \begin{array}{rll}
      \traV{((\lam{\var}{\tm})\sctx\,\tmtwo)}
    & = &
      \open{\uvar}\esub{\uvar}{
        (\osha{\lam{\lvar}{\traV{\tm}\esub{\var}{\lvar}}})\traV{\sctx}
      }\,\traV{\tmtwo}
    \\
    & \toSls &
      \open{
        \sha{\lam{\lvar}{\traV{\tm}\esub{\var}{\lvar}}}
      }\esub{\uvar}{
        \osha{\lam{\lvar}{\traV{\tm}\esub{\var}{\lvar}}}
      }\traV{\sctx}\,\traV{\tmtwo}
    \\
    & \toSopen &
      (\lam{\lvar}{\traV{\tm}\esub{\var}{\lvar}})
      \esub{\uvar}{
        \osha{\lam{\lvar}{\traV{\tm}\esub{\var}{\lvar}}}
      }\traV{\sctx}\,\traV{\tmtwo}
    \\
    & \toSdb &
      \traV{\tm}\esub{\var}{\traV{\tmtwo}}
      \esub{\uvar}{
        \osha{\lam{\lvar}{\traV{\tm}\esub{\var}{\lvar}}}
      }\traV{\sctx}
    \\
    & \toSgc &
      \traV{\tm}\esub{\var}{\traV{\tmtwo}}\traV{\sctx}
    \\
    & = &
      \traV{(\tm\esub{\var}{\tmtwo}\sctx)}
    \end{array}
  \]
  To perform the $\symSgc$ step,
  observe that in the second term in the sequence,
  namely the term
  $\open{\uvar}\esub{\uvar}{
    (\osha{\lam{\lvar}{\traV{\tm}\esub{\var}{\lvar}}})\traV{\sctx}
  }\,\traV{\tmtwo}$,
  the subterms $\traV{\tm}$ and $\traV{\tmtwo}$ lie outside the scope
  of the bound variable $\uvar$, so by $\alpha$-conversion we may assume that
  $\uvar \notin \fv{\traV{\tm}\esub{\var}{\traV{\tmtwo}}}$.
\item
  $\rtolsv$:
  Let $\off{\gctx}{\var}\esub{\var}{\val\sctx}
       \rtolsv
       \off{\gctx}{\val}\esub{\var}{\val}\sctx$.
  Recall that values are abstractions,
  so $\val$ is of the form $\val = \lam{\vartwo}{\tm}$,
  which means that in turn $\traV{\val} = \osha{\tmtwo}$,
  where $\tmtwo = \lam{\lvar}{\traV{\tm}\esub{\vartwo}{\lvar}}$
  for a fresh linear variable $\lvar$.
  Then:
  \[
    \begin{array}{rll}
      \traV{(\off{\gctx}{\var}\esub{\var}{\val\sctx})}
    & = &
      \off{\traV{\gctx}}{\ofc{\var}}\esub{\var}{\traV{\val}\traV{\sctx}}
    \\
    & = &
      \off{\traV{\gctx}}{\ofc{\var}}\esub{\var}{(\osha{\tmtwo})\traV{\sctx}}
    \\
    & \toSls &
      \off{\traV{\gctx}}{\osha{\tmtwo}}\esub{\var}{\osha{\tmtwo}}\traV{\sctx}
    \\
    & = &
      \off{\traV{\gctx}}{\traV{\val}}\esub{\var}{\traV{\val}}\traV{\sctx}
    \\
    & = &
      \traV{(\off{\gctx}{\val}\esub{\var}{\val}\sctx)}
    \end{array}
  \]
\item
  $\rtogclv$:
  Let $\tm\esub{\var}{\lval\sctx}
       \rtogclv
       \tm\sctx$,
  where $\var \notin \fv{\tm}$.
  Recall that lax values are either variables abstractions,
  so the lax value $\lval$ is of either of the forms
  $\lval = \vartwo$ or $\lval = \lam{\vartwo}{\tmtwo}$.
  In either case, this means that $\traV{{\lval}}$
  must be of the form $\traV{{\lval}} = \ofc{\tmthree}$
  for some term $\tmthree \in \TermsS$.
  Then:
  \[
    \begin{array}{rll}
      \traV{(\tm\esub{\var}{\lval\sctx})}
    & = &
      \traV{\tm}\esub{\var}{\traV{{\lval}}\traV{\sctx}}
    \\
    & = &
      \traV{\tm}\esub{\var}{(\ofc{\tmthree})\traV{\sctx}}
    \\
    & \toSgc &
      \traV{\tm}\traV{\sctx}
    \\
    & = &
      \traV{(\tm\sctx)}
    \end{array}
  \]
  To perform the $\symSgc$ step, 
  note that $\var\notin\fv{\traV{\tm}}$
  because $\fv{\traV{\tm}} = \fv{\tm}$,
  as noted in \rremark{traV_fv}.
\end{enumerate}
\end{proof}

\begin{definition}
We define a subset $\CtxsSValue \subseteq \CtxsS$, called {\em $\CBV$ contexts}
by the following grammar:
\[
  \begin{array}{rrl}
  \rgctx
  & ::=  &
           \ofc{\ctxhole}
  \\
  & \mid & \osha{\lam{\lvar}{\rgctx\esub{\var}{\lvar}}}
  \\
  & \mid & \open{\uvar}\esub{\uvar}{\rgctx}
  \\
  & \mid & \open{\sha{\lam{\lvar}{\rgctx\esub{\var}{\lvar}}}}
  \\
  & \mid & \lam{\lvar}{\rgctx\esub{\var}{\lvar}}
  \\
  & \mid & \rgctx\,\rtm
    \mid   \rtm\,\rgctx
  \\
  & \mid & \rgctx\esub{\uvar}{\rtm}
    \mid   \rtm\esub{\uvar}{\rgctx}
  \end{array}
\]
\end{definition}

\begin{remark}
\lremark{traVinv_fv}
$\fv{\traVinv{\rtm}} = \fv{\rtm}$.
\end{remark}

\begin{remark}
\lremark{shape_of_values_in_TermsSValue}
\quad
\begin{enumerate}
\item
  If $\rtm \in \TermsSValue$
  is of the form $\rtm = (\lam{\lvar}{\tmtwo})\sctx$,
  then $\tmtwo$ is of the form $\tmtwo = \rtmtwo_1\esub{\uvar}{\lvar}$
  where $\rtmtwo_1 \in \TermsSValue$
  and $\lvar \notin \fv{\rtmtwo_1}$
  and $\sctx \in \SCtxsSValue$.
\item
  If $\rtm \in \TermsSValue$
  is of the form $\rtm = (\ofc{\tmtwo})\sctx$,
  then exactly one of the following two holds:
  \begin{enumerate}
  \item
    $\rtm = (\ofc{\var})\rsctx$, where $\rsctx \in \SCtxsSValue$.
  \item
    $\rtm = (\osha{\lam{\lvar}{\rtm'\esub{\var}{\lvar}}})\rsctx$,
    where $\rtm' \in \TermsSValue$, $\rsctx \in \SCtxsSValue$,
    and $\lvar\notin\fv{\rtm'}$.
  \end{enumerate}
\item
  If $\rtm \in \TermsSValue$
  is of the form $\rtm = (\ofc{(\sha{\tmtwo})\sctx_1})\sctx_2$,
  then it is in fact of the form
  $\rtm = (\osha{\lam{\lvar}{\rtm'\esub{\var}{\lvar}}})\rsctx$,
  where $\rtm' \in \TermsSValue$, $\rsctx \in \SCtxsSValue$,
  and $\lvar\notin\fv{\rtm'}$.
\end{enumerate}
\end{remark}

\begin{lemma}[Context decomposition for the inverse $\CBV$ translation]
\llem{traVinv_decompose_contexts}
\quad
\begin{enumerate}
\item
  $\tm\sctx \in \TermsSValue$
  if and only if
  $\tm \in \TermsSValue$ 
  and
  $\sctx \in \SCtxsSValue$.
\item
  $\off{\gctx}{\var} \in \TermsSValue$
  if and only if
  $\gctx \in \CtxsSValue$.
\item
  If $\rgctx \in \CtxsSValue$
  and $\rtm \in \TermsSValue$
  and $\lvar \notin \fv{\rtm}$,
  then
  $\off{\rgctx}{\sha{\lam{\lvar}{\rtm\esub{\var}{\lvar}}}} \in \TermsSValue$.
\item
  If $\rtm \in \TermsSValue$ and $\rsctx \in \SCtxsSValue$,
  then $\traVinv{(\rtm\rsctx)} = \traVinv{\rtm}\,\traVinv{\rsctx}$.
\item
  If $\rgctx \in \CtxsSValue$
  then $\traVinv{\off{\rgctx}{\var}} = \off{\traVinv{\rgctx}}{\var}$.
\item
  If $\rgctx \in \CtxsSValue$
  and $\rtm \in \TermsSValue$
  and $\lvar \notin \fv{\rtm}$,
  then
  $\traVinv{\off{\rgctx}{\sha{\lam{\lvar}{\rtm\esub{\var}{\lvar}}}}} =
   \off{\traVinv{\rgctx}}{\lam{\var}{\traVinv{\rtm}}}$.
\end{enumerate}
\end{lemma}
\begin{proof}
 By induction on the first judgement in the statement of each item, except the first and fourth items which are by induction on $\sctx$ and $\rsctx \in \SCtxsSName$, resp.
\end{proof}

\traVinvSimulation

\begin{proof}\label{traVinv_simulation:proof}
By induction on the (unique) derivation of $\rtm \in \TermsSValue$:
\begin{enumerate}
\item
  $\rtm = \ofc{\var}$:
  Impossible, as there are no steps $\rtm \toS \tmtwo$.
\item
  $\rtm = \osha{\lam{\lvar}{\rtm'\esub{\var}{\lvar}}}$
      with $\lvar\notin\fv{\rtm'}$:
  Note that the step cannot be at the root of $\rtm$,
  nor at the root of $\sha{\lam{\lvar}{\rtm'\esub{\var}{\lvar}}}$,
  and that there cannot be a $\symSls$ nor a $\symSgc$ step involving
  the substitution $\esub{\var}{\lvar}$,
  since these rules would require that $\lvar$ be of the form
  $(\ofc{\tmthree})\sctx$, but $\lvar$ is a linear variable.
  This means that the step must be internal to $\rtm'$,
  that is,
  $\rtm
   = \osha{\lam{\lvar}{\rtm'\esub{\var}{\lvar}}}
   \toS \osha{\lam{\lvar}{\tmtwo'\esub{\var}{\lvar}}}
   = \tmtwo$
  with $\rtm' \toS \tmtwo'$.
  By \ih, $\tmtwo' \in \TermsSValue$,
  so $\tmtwo = \osha{\lam{\lvar}{\tmtwo'\esub{\var}{\lvar}}} \in \TermsSValue$,
  and
  $\traVinv{\rtm}
   = \lam{\var}{\traVinv{\rtm'}}
   \tocbvUgclvinvs \lam{\var}{\traVinv{\tmtwo'}}
   = \traVinv{\tmtwo}$.
\item
  $\rtm = \open{\uvar}\esub{\uvar}{\rtm_1}$:
  Note that the step cannot be at the root of $\open{\uvar}$,
  since this term does not match the left-hand side of the $\symSopen$ rule.
  We consider two subcases, depending on whether the step is
  at the root of $\open{\uvar}\esub{\uvar}{\rtm_1}$
  or internal to $\rtm_1$:
  \begin{enumerate}
  \item
    If the step is at the root,
    note that it cannot be a $\toSgc$ step,
    because there are free occurrences of $\uvar$ in $\open{\uvar}$.
    Hence the step must be a $\toSls$ step.
    This in turn means that $\rtm_1$
    must be of the form $\rtm_1 = (\ofc{(\sha{\tmthree})\sctx_1})\sctx_2$.
    But by construction of $\rtm$ we know that $\rtm_1 \in \TermsSValue$,
    so by \rremark{shape_of_values_in_TermsSValue}
    we know that it must be of the form
    $\rtm_1 = (\osha{\lam{\lvar}{\rtm'_1\esub{\var}{\lvar}}})\rsctx$.
    Then the step is of the form:
    \[\begin{array}{rl}
      & \rtm\\
      = & \open{\uvar}\esub{\uvar}{
          (\osha{\lam{\lvar}{\rtm'_1\esub{\var}{\lvar}}})\rsctx
        } \\
      \toSls &
        \open{
          \sha{\lam{\lvar}{\rtm'_1\esub{\var}{\lvar}}}
        }\esub{\uvar}{
          \osha{\lam{\lvar}{\rtm'_1\esub{\var}{\lvar}}}
        }\rsctx \\
        = & \tmtwo
        \end{array}
    \]
    Observe that $\tmtwo \in \TermsSValue$ and that:
    \[
      \begin{array}{rlll}
        & &  \traVinv{\rtm} \\
      & = &
        \traVinv{
          (\open{\uvar}\esub{\uvar}{
            (\osha{\lam{\lvar}{\rtm'_1\esub{\var}{\lvar}}})\rsctx
          })}
      \\
      & = &
        (\lam{\var}{\traVinv{{\rtm'_1}}})\traVinv{\rsctx}
        & \text{by \rlem{traVinv_decompose_contexts}}
      \\
      & \togclvinv &
        (\lam{\var}{\traVinv{{\rtm'_1}}})
        \esub{\uvar}{\lam{\var}{\traVinv{{\rtm'_1}}}}
        \traVinv{\rsctx}
      \\
      & = &
        \traVinv{
          (\open{
            \sha{\lam{\lvar}{\rtm'_1\esub{\var}{\lvar}}}
          }\esub{\uvar}{
            \osha{\lam{\lvar}{\rtm'_1\esub{\var}{\lvar}}}
          }\rsctx)
        }
      \\
      & = & \traVinv{\tmtwo}
      \end{array}
    \]
    To be able to perform the $\togclvinv$ step,
    note that in the second term in the sequence,
    namely the term
    $\traVinv{
      (\open{\uvar}\esub{\uvar}{
        (\osha{\lam{\lvar}{\rtm'_1\esub{\var}{\lvar}}})\rsctx
      })}$,
    the subterm $\rtm'_1$ lies outside the scope of the bound variable $\uvar$,
    so by $\alpha$-conversion
    we may assume that $\uvar \notin \fv{\rtm'_1}$.
    This in turn implies that $\uvar \notin \fv{\traVinv{{\rtm'_1}}}$,
    because $\fv{\traVinv{{\rtm'_1}}} = \fv{\rtm'_1}$ by \rremark{traVinv_fv}.
    Moreover, the $\togclvinv$ step can be applied because
    $\lam{\var}{\traVinv{{\rtm'_1}}}$ is indeed a lax value.

    {\em Remark:} this is the only point in the proof in which
    a $\togclvinv$ step is explicitly applied.
  \item
    If the step is internal to $\rtm_1$,
    then
    $\rtm
     = \open{\uvar}\esub{\uvar}{\rtm_1}
     \toS \open{\uvar}\esub{\uvar}{\tmtwo_1} = \tmtwo$
    with $\rtm_1 \toS \tmtwo_1$.
    By \ih, $\tmtwo_1 \in \TermsSValue$,
    so $\tmtwo = \open{\uvar}\esub{\uvar}{\tmtwo_1} \in \TermsSValue$
    and
    $\traVinv{\rtm}
     = \traVinv{(\open{\uvar}\esub{\uvar}{\rtm_1})}
     = \traVinv{\rtm_1}
     \tocbvUgclvinvs \traVinv{\tmtwo_1}
     = \traVinv{(\open{\uvar}\esub{\uvar}{\tmtwo_1})}
     = \traVinv{\tmtwo}$.
  \end{enumerate}
\item
  $\rtm = \open{\sha{\lam{\lvar}{\rtm_1\esub{\var}{\lvar}}}}$,
      with $\lvar\notin\fv{\rtm_1}$:
  Note that there cannot be a $\symSls$ nor a $\symSgc$ step
  involving the substitution $\esub{\var}{\lvar}$ since,
  since these rules would require that $\lvar$ be of the form
  $(\ofc{\tmthree})\sctx$, but $\lvar$ is a linear variable.

  We consider two subcases, depending on whether
  an $\symSopen$ step is performed at the root
  or internal to $\rtm_1$:
  \begin{enumerate}
  \item
    If the step is an $\symSopen$ step at the root,
    then
    $\rtm
     = \open{\sha{\lam{\lvar}{\rtm_1\esub{\var}{\lvar}}}}
     \toSopen \lam{\lvar}{\rtm_1\esub{\var}{\lvar}}
     = \tmtwo$,
    so $\tmtwo \in \TermsSValue$
    and we have that
    \[
      \begin{array}{rll}
        \traVinv{\rtm}
      & = &
        \traVinv{(\open{\sha{\lam{\lvar}{\rtm_1\esub{\var}{\lvar}}}})}
      \\
      & = &
        \lam{\var}{\traVinv{\rtm_1}}
      \\
      & = &
        \traVinv{(\lam{\lvar}{\rtm_1\esub{\var}{\lvar}})}
      \\
      & = &
        \traVinv{\tmtwo}
      \end{array}
    \]
  \item
    If the step is internal to $\rtm_1$,
    then
    $\rtm
     = \open{\sha{\lam{\lvar}{\rtm_1\esub{\var}{\lvar}}}}
     \toS \open{\sha{\lam{\lvar}{\tmtwo_1\esub{\var}{\lvar}}}}
     = \tmtwo$
    with $\rtm_1 \toS \tmtwo_1$.
    By \ih, $\tmtwo_1 \in \TermsSValue$,
    so $\tmtwo
        = \open{\sha{\lam{\lvar}{\tmtwo_1\esub{\var}{\lvar}}}}
        \in \TermsSValue$
    and we have that
      \[
      \begin{array}{rl}
        \traVinv{\rtm} \\
     = & \traVinv{(\open{\sha{\lam{\lvar}{\rtm_1\esub{\var}{\lvar}}}})} \\
     = & \lam{\var}{\traVinv{\rtm_1}} \\
     \tocbvUgclvinvs &
       \lam{\var}{\traVinv{\tmtwo_1}} \\
     = & \traVinv{(\open{\sha{\lam{\lvar}{\tmtwo_1\esub{\var}{\lvar}}}})} \\
            = & \traVinv{\tmtwo}
                  \end{array}
    \]
  \end{enumerate}
\item
  $\rtm = \lam{\lvar}{\rtm_1\esub{\var}{\lvar}}$,
      with $\lvar\notin\fv{\rtm'}$:
  Note that there cannot be a $\symSls$ or a $\symSgc$ step involving
  the substitution $\esub{\var}{\lvar}$,
  since these rules would require that $\lvar$ be of the form
  $(\ofc{\rtmthree})\sctx$, but $\lvar$ is a linear variable.
  Hence the step must be internal to $\rtm_1$.
  Then
  $\rtm
   = (\lam{\lvar}{\rtm_1\esub{\var}{\lvar}})\rsctx\,\rtm_2
   \toS (\lam{\lvar}{\tmtwo_1\esub{\var}{\lvar}})\rsctx\,\rtm_2
   = \rtmtwo$
   with $\rtm_1 \toS \tmtwo_1$
  By \ih, $\tmtwo_1 \in \TermsSValue$,
  so $\rtmtwo
      = (\lam{\lvar}{\tmtwo_1\esub{\var}{\lvar}})\rsctx\,\rtm_2
      \in \TermsSValue$
  and we have that
  \[
      \traVinv{\rtm}
    = \lam{\var}{\traVinv{\rtm_1}}
    \tocbvUgclvinvs \lam{\var}{\traVinv{\tmtwo_1}}
    = \traVinv{\rtmtwo}
  \]
\item
  $\rtm = \rtm_1\,\rtm_2$:
  We consider three subcases, depending on whether the step is a
  $\symSdb$ step at the root, internal to $\rtm_1$ or internal to $\rtm_2$:
  \begin{enumerate}
  \item
    If the step is a $\symSdb$ step at the root of the term,
    then $\rtm_1$ must be of the form $\rtm_1 = \lam{\lvar}{\rtm_{11}\esub{\var}{\lvar}})$
    and
    $\rtm
     = (\lam{\lvar}{\rtm_{11}\esub{\var}{\lvar}})\rsctx\,\rtm_2
     \rtoSdb \rtm_{11}\esub{\var}{\rtm_2}\rsctx
     = \tmtwo$,
    so $\tmtwo \in \TermsSValue$
    and
    \[
      \begin{array}{rl}
      & \traVinv{\rtm} \\
      = & (\lam{\var}{\traVinv{\rtm_{11}}})\traVinv{\rsctx}\,\traVinv{\rtm_2} \\
      \rtodb &
        \traVinv{\rtm_{11}}\esub{\var}{\traVinv{\rtm_2}}\traVinv{\rsctx} \\
      = & \traVinv{(\rtm_{11}\esub{\var}{\rtm_2}\rsctx)} \\
      = & \traVinv{\tmtwo}
      \end{array}
    \]
  \item
    If the step is internal to $\rtm_1$,
    then $\rtm = \rtm_1\,\rtm_2 \toS \tmtwo_1\,\rtm_2 = \tmtwo$
    with $\rtm_1 \toS \tmtwo_1$. 
    By \ih, $\tmtwo_1 \in \TermsSValue$
    and $\traVinv{\rtm_1} \tocbvUgclvinvs \traVinv{\tmtwo_1}$,
    so $\tmtwo = \tmtwo_1\,\rtm_2 \in \TermsSValue$
    and we have that
    \[
      \traVinv{\rtm}
      = \traVinv{\rtm_1}\,\traVinv{\rtm_2}
      \tocbvUgclvinvs
        \traVinv{\tmtwo_1}\,\traVinv{\rtm_2}
      = \traVinv{\tmtwo}
    \]
  \item
    If the step is internal to $\rtm_2$
    then $\rtm = \rtm_1\,\rtm_2 \toS \rtm_1\,\tmtwo_2 = \tmtwo$
    with $\rtm_2 \toS \tmtwo_2$. 
    By \ih, $\tmtwo_2 \in \TermsSValue$
    and $\traVinv{\rtm_2} \tocbvUgclvinvs \traVinv{\tmtwo_2}$,
    so $\tmtwo = \rtm_1\,\tmtwo_2 \in \TermsSValue$
    and we have that
    \[
      \traVinv{\rtm}
      = \traVinv{\rtm_1}\,\traVinv{\rtm_2}
      \tocbvUgclvinvs
        \traVinv{\rtm_1}\,\traVinv{\tmtwo_2}
      = \traVinv{\tmtwo}
    \]
  \end{enumerate}
\item
  $\rtm = \rtm_1\esub{\uvar}{\rtm_2}$:
  We consider four cases, depending on whether
  there is a $\symSls$ step at the root,
  or a $\symSgc$ step at the root,
  or whether the step is internal to $\rtm_1$ or $\rtm_2$:
  \begin{enumerate}
  \item
    If there is a $\symSls$ step at the root,
    then $\rtm_2$ is of the form
    $\rtm_2 = (\ofc{(\sha{\rtm''_2})\sctx_1})\sctx_2$.
    By \rremark{shape_of_values_in_TermsSValue}, this implies that
    $\rtm_2$ must actually be of the form
    $\rtm_2 = (\osha{\lam{\lvar}{\rtm'_2\esub{\var}{\lvar}}})\rsctx$
    where $\rtm'_2 \in \TermsSValue$, $\rsctx \in \SCtxsSValue$,
    and $\lvar \notin \fv{\rtm'_2}$.
    Moreover, $\rtm_1$ must be of the form $\rtm_1 = \off{\gctx}{\uvar}$.
    The step is of the form:
    \[
      \begin{array}{rll}
         & & \rtm \\
        & = &
        \off{\gctx}{\uvar}\esub{\uvar}{
          (\osha{\lam{\lvar}{\rtm'_2\esub{\var}{\lvar}}})\rsctx
        }
      \\
        & \rtoSls &
        \off{\gctx}{
          \sha{\lam{\lvar}{\rtm'_2\esub{\var}{\lvar}}}
        }\esub{\uvar}{
          \osha{\lam{\lvar}{\rtm'_2\esub{\var}{\lvar}}}
        }
        \rsctx
      \\
        & = &
        \tmtwo
      \end{array}
    \]
    Note that $\rtm_1 = \off{\gctx}{\uvar} \in \TermsSValue$
    so, by~\rlem{traVinv_decompose_contexts}, we have that
    $\gctx \in \CtxsSValue$ and, again by~\rlem{traVinv_decompose_contexts},
    $\off{\gctx}{\sha{\lam{\lvar}{\rtm'_2\esub{\var}{\lvar}}}}
     \in \TermsSValue$.
    This in turn implies that $\tmtwo \in \TermsSValue$.
    Moreover, using~\rlem{traVinv_decompose_contexts}, we have that:
    \[
      \begin{array}{rll}
        & & \traVinv{\rtm} \\
        & = &
        \traVinv{
          (\off{\gctx}{\uvar}\esub{\uvar}{
            (\osha{\lam{\lvar}{\rtm'_2\esub{\var}{\lvar}}})\rsctx
          })
        }
      \\
        & = &
        \off{\traVinv{\gctx}}{\uvar}\esub{\uvar}{
          (\lam{\var}{\traVinv{{\rtm'_2}}})
          \traVinv{\rsctx}
        }
      \\
        & \rtolsv &
        \off{\traVinv{\gctx}}{
          \lam{\var}{\traVinv{{\rtm'_2}}}
        }\esub{\uvar}{
          \lam{\var}{\traVinv{{\rtm'_2}}}
        }
        \traVinv{\rsctx}
      \\
        & = &
        \traVinv{
          (\off{\gctx}{
            \sha{\lam{\lvar}{\rtm'_2\esub{\var}{\lvar}}}
          }\esub{\uvar}{
            \osha{\lam{\lvar}{\rtm'_2\esub{\var}{\lvar}}}
          }
          \rsctx)
        }
      \\
        & = &
        \traVinv{\tmtwo}
      \end{array}
    \]
  \item
    If there is a $\symSgc$ step at the root,
    we have that $\uvar \notin \fv{\rtm_1}$,
    and that $\rtm_2$ is of the form $\rtm_2 = (\ofc{\rtm''_2})\sctx$.
    By \rremark{shape_of_values_in_TermsSValue}, this implies that
    $\rtm_2$ must actually be of
    either the form $\rtm_2 = (\ofc{\vartwo})\rsctx$
    with $\rsctx \in \SCtxsSValue$,
    or of the form $\rtm_2 = (\osha{\lam{\lvar}{\rtm'_2\esub{\var}{\lvar}}})\rsctx$
    with $\rtm'_2 \in \TermsSValue$, $\rsctx \in \SCtxsSValue$,
    and $\lvar \notin \fv{\rtm'_2}$.
    In either case, $\rtm_2$ is of the form
    $\rtm_2 = (\ofc{\rtm'_2})\rsctx$ 
    where $\ofc{\rtm'_2} \in \TermsSValue$,
    $\rsctx \in \SCtxsSValue$ ,
    and $\traVinv{(\ofc{\rtm'_2})}$ is a lax value,
    since both $\traVinv{(\ofc{\vartwo})} = \vartwo$
    and
    $\traVinv{(\osha{\lam{\lvar}{\rtm'_2\esub{\var}{\lvar}}})}
     = \lam{\var}{\traVinv{{\rtm'_2}}}$
    are lax values.

    Then the step is of the form:
    \[
      \begin{array}{rll}
        & & \rtm \\
        & = &
        \rtm_1\esub{\uvar}{(\ofc{\rtm'_2})\sctx'}
      \\
        & \rtoSgc &
        \rtm_1\sctx'
      \\
        & = &
        \tmtwo
      \end{array}
    \]
    So $\tmtwo \in \TermsSValue$ and:
    \[
      \begin{array}{rll}
        & & \traVinv{\rtm}\\
        & = &
        \traVinv{(\rtm_1\esub{\uvar}{(\ofc{\rtm'_2})\sctx'})}
      \\
        & = &
        \traVinv{\rtm_1}
        \esub{\uvar}{\traVinv{(\ofc{\rtm'_2})}\traVinv{\sctx'}}
      \\
        & \rtogclv &
        \traVinv{\rtm_1}\traVinv{\sctx'}
      \\
        & = &
        \traVinv{\tmtwo}
      \end{array}
    \]
    To be able to perform the $\symgclv$ step,
    note that $\uvar \notin \fv{\traVinv{\rtm_1}}$
    because we know that $\uvar \notin \fv{\rtm_1} = \fv{\traVinv{\rtm_1}}$
    by \rremark{traVinv_fv}.
    To be able to perform the $\symgclv$ step,
    it must be also noted that, as already remarked,
    $\traVinv{(\ofc{\rtm'_2})}$ is a lax value.
  \item
    If the step is internal to $\rtm_1$,
    then
    $\rtm
     = \rtm_1\esub{\uvar}{\rtm_2}
     \toS \tmtwo_1\esub{\uvar}{\rtm_2}
     = \tmtwo$
    with $\rtm_1 \toS \tmtwo_1$.
    By \ih, $\tmtwo_1 \in \TermsSValue$,
    so $\tmtwo = \tmtwo_1\esub{\uvar}{\rtm_2} \in \TermsSValue$
    and
    $\traVinv{\rtm}
     = \traVinv{\rtm_1}\esub{\uvar}{\traVinv{\rtm_2}}
     \tocbvUgclvinvs \traVinv{\tmtwo_1}\esub{\uvar}{\traVinv{\rtm_2}}
     = \traVinv{\tmtwo}$.
  \item
    If the step is internal to $\rtm_2$,
    then
    $\rtm
     = \rtm_1\esub{\uvar}{\rtm_2}
     \toS \rtm_1\esub{\uvar}{\tmtwo_2}
     = \tmtwo$
    with $\rtm_2 \toS \tmtwo_2$.
    By \ih, $\tmtwo_2 \in \TermsSValue$,
    so $\tmtwo = \rtm_1\esub{\uvar}{\tmtwo_2} \in \TermsSValue$
    and
    $\traVinv{\rtm}
     = \traVinv{\rtm_1}\esub{\uvar}{\traVinv{\rtm_2}}
     \tocbvUgclvinvs \traVinv{\tmtwo_1}\esub{\uvar}{\traVinv{\tmtwo_2}}
     = \traVinv{\tmtwo}$.
  \end{enumerate}
\end{enumerate}
\end{proof}

\subsection{$\CCBNd$}

\TypingTheCCBNdTranslation

\begin{proof}\label{typing_the_CCBNd_translation:proof}
By induction on the derivation of $\judl{\tctx}{\tm}{\typ}$:
\begin{enumerate}
\item
  $\rulelVar$:
  Let $\judl{\tctx,\var:\typ}{\var}{\typ}$ be derived from the $\rulelVar$ rule.
  Then:
  
  \scalebox{\smallDerivation}{\parbox{\textwidth}{
      $
      \indrule{\rulecUvar}{
        \emptyPremise
      }{
        \judc{\traCNd{\tctx},\var:\traCNd{\typ}}{\noenv}{\var}{\sha{\traCNd{\typ}}}
      }
      $
    }}
  
\item
  $\rulelAbs$:
  Let $\judl{\tctx}{\lam{\var}{\tm}}{\typ\to\typtwo}$
  be derived from $\judl{\tctx,\var:\typ}{\tm}{\typtwo}$.
  Let $\lvar$ be a fresh linear variable, such that $\lvar \notin \fv{\traCNd{\tm}}$.
  Then:
  \scalebox{\smallDerivation}{\parbox{\textwidth}{
      $
      \indrule{\rulecSha}{
    \indrule{\rulecAbs}{
      \indrule{\rulecES}{
        \derivih{\judc{\traCNd{\tctx},\var:\traCNd{\typ}}{\noenv}{\traCNd{\tm}}{\sha{\traCNd{\typtwo}}}}
        \HS
        \indrule{\rulecLvar}{
          \emptyPremise
        }{
          \judc{\traCNd{\tctx}}{\lvar:\osha{\traCNd{\typ}}}{\lvar}{\osha{\traCNd{\typ}}}
        }
      }{
        \judc{\traCNd{\tctx}}{\lvar:\osha{\traCNd{\typ}}}{\traCNd{\tm}\esub{\var}{\lvar}}{\sha{\traCNd{\typtwo}}}
      }
    }{
      \judc{\traCNd{\tctx}}{\noenv}{\lam{\lvar}{\traCNd{\tm}\esub{\var}{\lvar}}}{\osha{\traCNd{\typ}}\limp\sha{\traCNd{\typtwo}}}
    }
  }{
           \judc{\traCNd{\tctx}}{\noenv}{\sha{\lam{\lvar}{\traCNd{\tm}\esub{\var}{\lvar}}}}{\sha{(\osha{\traCNd{\typ}}\limp\sha{\traCNd{\typtwo}})}}
  }
  $
  }}

\item
  $\rulelApp$:
  Let $\judl{\tctx}{\tm\,\tmtwo}{\typtwo}$
  be derived from $\judl{\tctx}{\tm}{\typ\to\typtwo}$
  and $\judl{\tctx}{\tmtwo}{\typ}$.
  Then:
  
  \scalebox{\smallDerivation}{\parbox{\textwidth}{
      $
    \indrule{\rulecApp}{
      \indrule{\rulecOpen}{
        \derivih{
          \judc{\traCNd{\tctx}}{\noenv}{
            \traCNd{\tm}
          }{\sha{(\osha{\traCNd{\typ}}\limp\sha{\traCNd{\typtwo}})}}
        }
      }
      {
        \judc{\traCNd{\tctx}}{\noenv}{
            \open{\traCNd{\tm}}
          }{\osha{\traCNd{\typ}}\limp\sha{\traCNd{\typtwo}}}
      }
        \indrule{\rulecProm}{
   \derivih{
        \judc{\traCNd{\tctx}}{\noenv}{
          \traCNd{\tmtwo}
        }{\sha{\traCNd{\typ}}}
      }
   }{
      \judc{\traCNd{\tctx}}{\noenv}{
          \ofc{\traCNd{\tmtwo}}
        }{\osha{\traCNd{\typ}}}
  }
    }{
      \judc{\traCNd{\tctx}}{\noenv}{
        \open{\traCNd{\tm}}\, \ofc{\traCNd{\tmtwo}}
      }{\sha{\traCNd{\typtwo}}}
    }
    $
    }}

\item
  $\rulelES$:
  Let $\judl{\tctx}{\tm\esub{\var}{\tmtwo}}{\typtwo}$
  be derived from $\judl{\tctx,\var:\typ}{\tm}{\typtwo}$
  and $\judl{\tctx}{\tmtwo}{\typ}$.
  Then:

  \scalebox{\smallDerivation}{\parbox{\textwidth}{
      $
    \indrule{\rulecES}{
      \derivih{\judc{\traCNd{\tctx},\var:\traCNd{\typ}}{\noenv}{\traCNd{\tm}}{\sha{\traCNd{\typtwo}}}}
      \indrule{\rulecProm}{
        \derivih{\judc{\traCNd{\tctx}}{\noenv}{\traCNd{\tmtwo}}{\sha{\traCNd{\typ}}}}
      }{
        \judc{\traCNd{\tctx}}{\noenv}{\ofc{\traCNd{\tmtwo}}}{\osha{\traCNd{\typ}}}
      }
    }{
      \judc{\traCNd{\tctx}}{\noenv}{\traCNd{\tm}\esub{\var}{\ofc{\traCNd{\tmtwo}}}}{\sha{\traCNd{\typtwo}}}
    }
    $
  }}

\end{enumerate}
The translation can be extended to operate on contexts, by declaring
$\traCNd{\ctxhole} \eqdef \ctxhole$. Note that the translation of a
substitution context is a substitution context:
\[
  \traCNd{(\ctxhole\esub{\var_1}{\tm_1}\hdots\esub{\var_n}{\tm_n})}
  =
  \ctxhole\esub{\var_1}{\ofc{\traCNd{\tm_1}}}\hdots\esub{\var_n}{\ofc{\traCNd{\tm_n}}}
\]
\end{proof}

\begin{remark}
\lremark{traCNd_fv}
$\fv{\traCNd{\tm}} = \fv{\tm}$
\end{remark}

\begin{remark}
\lremark{traCNd_commutes_with_hold_filling}
$\traCNd{\of{\gctx}{\tm}} = \of{\traCNd{\gctx}}{\traCNd{\tm}}$
and
$\traCNd{\off{\gctx}{\tm}} = \off{\traCNd{\gctx}}{\traCNd{\tm}}$.
In particular,
$\traCNd{(\tm\sctx)} = \traCNd{\tm}\traCNd{\sctx}$.
\end{remark}

\begin{definition}
We define a subset $\CtxsSClumsyNeed \subseteq \CtxsS$,
called {\em $\CCBNd$ contexts}:
\[
  \begin{array}{rrl}
  \rgctx
  & ::=  &
           \ctxhole
  \\
  & \mid & \sha{\lam{\lvar}{\rgctx\esub{\var}{\lvar}}}
  \\
  & \mid & \open{\rgctx}
  \\
  & \mid & \lam{\lvar}{\rgctx\esub{\var}{\lvar}}
  \\
  & \mid & \rgctx\esub{\uvar}{\ofc{\rtm}}
    \mid   \rtm\esub{\uvar}{\ofc{\rgctx}}
  \\
  & \mid & \rgctx\,\ofc{\rtm}
    \mid   \rtm\,\ofc{\rgctx}
  \end{array}
\]
\end{definition}

\begin{remark}
\lremark{traCNdinv_fv}
$\fv{\traCNdinv{\rtm}} = \fv{\rtm}$.
\end{remark}

\begin{remark}
\lremark{shape_of_values_in_TermsNeed}
\quad
\begin{enumerate}
\item
  If $\tm \in \TermsSClumsyNeed$ is such that $\tm = (\lam{\lvar}{\tm'})\sctx$,
  then $\sctx \in \SCtxsSClumsyNeed$
  and $\tm' = \rtmtwo\esub{\uvar}{\lvar}$
  for some $\rtmtwo\in\TermsSClumsyNeed$ and $\lvar \notin \fv{\rtmtwo}$.
\item
  If $\sha{\tm} \in \TermsSClumsyNeed$, then $\tm=\lam{\lvar}{\rtmtwo\esub{\var}{\lvar}}$, for some $\rtmtwo\in\TermsSClumsyNeed$ and $\lvar \notin \fv{\rtmtwo}$.
\end{enumerate}
\end{remark}

\begin{lemma}[Context decomposition for the inverse $\CCBNd$ translation]
\llem{traCNdinv_decompose_contexts}
\quad
\begin{enumerate}
\item
  $\tm\sctx \in \TermsSClumsyNeed$
  if and only if
  $\tm \in \TermsSClumsyNeed$ 
  and
  $\sctx \in \SCtxsSClumsyNeed$.
\item
  $\off{\gctx}{\var} \in \TermsSClumsyNeed$
  if and only if
  $\gctx \in \CtxsSClumsyNeed$.
\item
  If $\rgctx \in \CtxsSClumsyNeed$
  and $\rtm \in \TermsSClumsyNeed$
  then
  $\off{\rgctx}{\rtm} \in \TermsSClumsyNeed$.
\item
  If $\rtm \in \TermsSClumsyNeed$ and $\rsctx \in \SCtxsSClumsyNeed$,
  then $\traCNdinv{(\rtm\rsctx)} = \traCNdinv{\rtm}\,\traCNdinv{\rsctx}$.
\item
  If $\rgctx \in \CtxsSClumsyNeed$
  then $\traCNdinv{\off{\rgctx}{\var}} = \off{\traCNdinv{\rgctx}}{\var}$.
\item
  If $\rgctx \in \CtxsSClumsyNeed$
  and $\rtm \in \TermsSClumsyNeed$
  and $\rsctx \in \SCtxsSClumsyNeed$
  and $\lvar \notin \fv{\rtm}$,
  then
  $\traCNdinv{\off{\rgctx}{(\sha{\lam{\lvar}{\rtm\esub{\var}{\lvar}}})\rsctx}} =
   \off{\traCNdinv{\rgctx}}{(\lam{\var}{\traCNdinv{\rtm}})\traCNdinv{\rsctx}}$.
\end{enumerate}
\end{lemma}
\begin{proof}
 By induction on the first judgement in the statement of each item, except the first and fourth items which are by induction on $\sctx$ and $\rsctx \in \SCtxsSName$, resp.
\end{proof}

\traCNdinvSimulation

\begin{proof}\label{traCNdinv_simulation:proof}
By induction on the (unique) derivation of $\rtm \in \TermsSClumsyNeed$:
\begin{enumerate}
\item
  $\rtm = \var$:
  Impossible, as there are no steps $\rtm \toS \tmtwo$.
\item
  $\rtm = \sha{\lam{\lvar}{\rtm'\esub{\var}{\lvar}}}$
      with $\lvar\notin\fv{\rtm'}$:
  Note that the step cannot be at the root of $\rtm$,
  and that there cannot be a $\symSls$ nor a $\symSgc$ step involving
  the substitution $\esub{\var}{\lvar}$,
  since these rules would require that $\lvar$ be of the form
  $(\ofc{\tmthree})\sctx$, but $\lvar$ is a (linear) variable.
  This means that the step must be internal to $\rtm'$,
  that is,
  $\rtm
   = \sha{\lam{\lvar}{\rtm'\esub{\var}{\lvar}}}
   \toS \sha{\lam{\lvar}{\tmtwo'\esub{\var}{\lvar}}}
   = \tmtwo$
  with $\rtm' \toS \tmtwo'$.
  By \ih, $\tmtwo' \in \TermsSClumsyNeed$,
  so $\tmtwo = \sha{\lam{\lvar}{\tmtwo'\esub{\var}{\lvar}}} \in \TermsSClumsyNeed$,
  and
  $\traCNdinv{\rtm}
   = \lam{\var}{\traCNdinv{\rtm'}}
   \tocbnd^= \lam{\var}{\traCNdinv{\tmtwo'}}
   = \traCNdinv{\tmtwo}$.
\item
  $\rtm = \open{\rtm_1}$:
  We consider two subcases, depending on whether the step is
  at the root or internal to $\rtm_1$:
  \begin{enumerate}
  \item
    If the step is at the root
    of  $\open{\rtm_1}$, then $\rtm_1$ must be of the form $(\sha{\tm_1'})\rsctx$ and, by \rremark{shape_of_values_in_TermsNeed}, $\tm_1'$ must be of the form $\lam{\lvar}{\rtm_1''\esub{\var}{\lvar}}$. Thus the step is
    \[
      \begin{array}{rl}
        & \rtm \\
     = & \open{(\sha{\lam{\lvar}{\rtm_1''\esub{\var}{\lvar}}})\rsctx} \\
     \toSopen & (\lam{\lvar}{\rtm_1''\esub{\var}{\lvar}})\rsctx \\
        = & \tmtwo
      \end{array}
      \]
    so $\tmtwo \in \TermsSClumsyNeed$
    and, using \rlem{traCNdinv_decompose_contexts},
    we have that
    \[
      \begin{array}{rll}
        \traCNdinv{\rtm}
      & = &
        \traCNdinv{(\open{(\sha{\lam{\lvar}{\rtm_1''\esub{\var}{\lvar}}})\rsctx})}
      \\
      & = &
        (\lam{\var}{\traCNdinv{\rtm_1''}})\traCNdinv{\rsctx}
      \\
      & = &
        \traCNdinv{((\lam{\lvar}{\rtm_1''\esub{\var}{\lvar}})\rsctx)}
      \\
      & = &
        \traCNdinv{\tmtwo}
      \end{array}
    \]
  \item
    If the step is internal to $\rtm_1$,
    then
    $\rtm
     = \open{\rtm_1}
     \toS \open{\tmtwo_1} = \tmtwo$
    with $\rtm_1 \toS \tmtwo_1$.
    By \ih, $\tmtwo_1 \in \TermsSClumsyNeed$,
    so $\tmtwo = \open{\tmtwo_1} \in \TermsSClumsyNeed$
    and
    $\traCNdinv{\rtm}
     = \traCNdinv{\open{\rtm_1}}
     = \traCNdinv{\rtm_1}
     \tocbnd^= \traCNdinv{\tmtwo_1}
     = \traCNdinv{\open{\tmtwo_1}}
     = \traCNdinv{\tmtwo}$.
  \end{enumerate}

\item
  $\rtm = \lam{\lvar}{\rtm_1\esub{\var}{\lvar}}$,
      with $\lvar\notin\fv{\rtm_1}$:
  Note that there cannot be a $\symSls$ or a $\symSgc$ step involving
  the substitution $\esub{\var}{\lvar}$,
  since these rules would require that $\lvar$ be of the form
  $\ofc{\rtmthree}\sctx$, but $\lvar$ is a (linear) variable.
  Hence the step must be internal to $\rtm_1$. Then
  $\rtm
   = \lam{\lvar}{\rtm_1\esub{\var}{\lvar}}
   \toS \lam{\lvar}{\tmtwo_1\esub{\var}{\lvar}}
   = \rtmtwo$
   with $\rtm_1 \toS \tmtwo_1$.
  By \ih, $\tmtwo_1 \in \TermsSClumsyNeed$,
  so $\rtmtwo
      = (\lam{\lvar}{\tmtwo_1\esub{\var}{\lvar}})\rsctx\,\ofc{\rtm_2}
      \in \TermsSClumsyNeed$
  and we have that
  $\traCNdinv{\rtm}
   = \lam{\var}{\traCNdinv{\rtm_1}}
   \tocbnd^=
     \lam{\var}{\traCNdinv{\tmtwo_1}}
   = \traCNdinv{\rtmtwo}$.
\item
  $\rtm = \rtm_1\esub{\uvar}{\ofc{\rtm_2}}$: 
 Note that the reduction step cannot be at the root of $\ofc{\rtm_2}$, since there are no rules in $\lambdaS$ that have a $\ofc{}$ in the root of their left-hand sides.
 We consider four cases, depending on whether
  there is a $\symSls$ step at the root,
  or a $\symSgc$ step at the root,
  or whether the step is internal to $\rtm_1$ or $\rtm_2$:
  \begin{enumerate}
  \item
    If there is a $\symSls$ step at the root,
    then $\rtm_2$ is of the form
    $\rtm_2 = (\sha{\tm''_2})\rsctx$.
    By \rremark{shape_of_values_in_TermsNeed}, this implies that 
    $\tm''_2$ must actually be of the form
    $\lam{\lvar}{\rtm'_2\esub{\var}{\lvar}}$
    where $\rtm'_2 \in \TermsSClumsyNeed$
    and $\lvar \notin \fv{\rtm'_2}$.
    Moreover, $\rtm_1$ must be of the form $\rtm_1 = \off{\gctx}{\uvar}$.
    The step is of the form:
    \[
      \begin{array}{rll}
        \rtm
        & = &
        \off{\gctx}{\uvar}\esub{\uvar}{
          \ofc{(\sha{(\lam{\lvar}{\rtm'_2\esub{\var}{\lvar}})}\rsctx)}
        }
      \\
        & \rtoSls &
        \off{\gctx}{
          (\sha{\lam{\lvar}{\rtm'_2\esub{\var}{\lvar}}})
          \rsctx
        }\esub{\uvar}{
           \ofc{
             (\sha{(\lam{\lvar}{\rtm'_2\esub{\var}{\lvar}})})
             \rsctx
           }
        }
      \\
        & = &
        \tmtwo
      \end{array}
    \]
    Note that $\rtm_1 = \off{\gctx}{\uvar} \in \TermsSClumsyNeed$
    so, by~\rlem{traCNdinv_decompose_contexts}, we have that
    $\gctx \in \CtxsSClumsyNeed$ and, again by~\rlem{traCNdinv_decompose_contexts},
    $\off{\gctx}{(\sha{\lam{\lvar}{\rtm'_2\esub{\var}{\lvar}}})\rsctx}
     \in \TermsSClumsyNeed$.
    This in turn implies that $\tmtwo \in \TermsSClumsyNeed$.
    Moreover, using~\rlem{traCNdinv_decompose_contexts}, we have that:
    \[
      \begin{array}{rll}
        \traCNdinv{\rtm}
        & = &
        \traCNdinv{
          (\off{\gctx}{\uvar}\esub{\uvar}{
             \ofc{(\sha{(\lam{\lvar}{\rtm'_2\esub{\var}{\lvar}})}\rsctx)}
          })
        }
      \\
        & = &
        \off{\traCNdinv{\gctx}}{\uvar}\esub{\uvar}{
          (\lam{\var}{\traCNdinv{{\rtm'_2}}})
          \traCNdinv{\rsctx}
        }
      \\
        & \rtolsw &
        \off{\traCNdinv{\gctx}}{
          (\lam{\var}{\traCNdinv{{\rtm'_2}}})
          \traCNdinv{\rsctx}
        }\esub{\uvar}{
          (\lam{\var}{\traCNdinv{{\rtm'_2}}})
          \traCNdinv{\rsctx}
        }
      \\
        & = &
        \traCNdinv{
          (\off{\gctx}{
            (\sha{\lam{\lvar}{\rtm'_2\esub{\var}{\lvar}}})
            \rsctx
          }\esub{\uvar}{
            \ofc{
              (\sha{\lam{\lvar}{\rtm'_2\esub{\var}{\lvar}}})
              \rsctx
            }
          })
        }
      \\
        & = &
        \traCNdinv{\tmtwo}
      \end{array}
    \]
  \item
    If there is a $\symSgc$ step at the root,
    we have that $\uvar \notin \fv{\rtm_1}$,
    and the step is of the form:
    \[
      \begin{array}{rll}
        \rtm
        & = &
        \rtm_1\esub{\uvar}{\ofc{\rtm_2}}
      \\
        & \rtoSgc &
        \rtm_1
      \\
        & = &
        \tmtwo
      \end{array}
    \]
    So $\tmtwo \in \TermsSClumsyNeed$ and:
    \[
      \begin{array}{rll}
        \traCNdinv{\rtm}
        & = &
        \traCNdinv{(\rtm_1\esub{\uvar}{\ofc{\rtm_2}})}
      \\
        & = &
        \traCNdinv{\rtm_1}
        \esub{\uvar}{\traCNdinv{\rtm_2}}
      \\
        & \rtogc &
        \traCNdinv{\rtm_1}
      \\
        & = &
        \traCNdinv{\tmtwo}
      \end{array}
    \]
    To be able to perform the $\symgc$ step,
    note that $\uvar \notin \fv{\traCNdinv{\rtm_1}}$
    because we know that $\uvar \notin \fv{\rtm_1} = \fv{\traCNdinv{\rtm_1}}$
    by \rremark{traCNdinv_fv}.
  \item
    If the step is internal to $\rtm_1$,
    then
    $\rtm
     = \rtm_1\esub{\uvar}{\ofc{\rtm_2}}
     \toS \tmtwo_1\esub{\uvar}{\ofc{\rtm_2}}
     = \tmtwo$
    with $\rtm_1 \toS \tmtwo_1$.
    By \ih, $\tmtwo_1 \in \TermsSClumsyNeed$,
    so $\tmtwo = \tmtwo_1\esub{\uvar}{\ofc{\rtm_2}} \in \TermsSClumsyNeed$
    and
    $\traCNdinv{\rtm}
     = \traCNdinv{\rtm_1}\esub{\uvar}{\traCNdinv{\rtm_2}}
     \tocbnd^= \traCNdinv{\tmtwo_1}\esub{\uvar}{\traCNdinv{\rtm_2}}
     = \traCNdinv{\tmtwo}$.
  \item
    If the step is internal to $\rtm_2$,
    then
    $\rtm
     = \rtm_1\esub{\uvar}{\ofc{\rtm_2}}
     \toS \rtm_1\esub{\uvar}{\tmtwo_2}
     = \tmtwo$
    with $\rtm_2 \toS \tmtwo_2$.
    By \ih, $\tmtwo_2 \in \TermsSClumsyNeed$,
    so $\tmtwo = \rtm_1\esub{\uvar}{\ofc{\tmtwo_2}} \in \TermsSClumsyNeed$
    and
    $\traCNdinv{\rtm}
     = \traCNdinv{\rtm_1}\esub{\uvar}{\traCNdinv{\rtm_2}}
     \tocbnd^= \traCNdinv{\tmtwo_1}\esub{\uvar}{\traCNdinv{\tmtwo_2}}
     = \traCNdinv{\tmtwo}$.
  \end{enumerate}
\item
  $\rtm = \rtm_1\,\ofc{\rtm_2}$:
  We consider three subcases, depending on whether the step is
  at the root, internal to $\rtm_1$, or internal to $\rtm_2$:
  \begin{enumerate}
  \item
    If the step is at the root, it must be a $\symSdb$ step.
    Hence $\rtm_1$ is of the form $(\lam{\lvar}{\tmthree})\sctx$.
    Then by \rremark{shape_of_values_in_TermsNeed}
    $\sctx \in \SCtxsSClumsyNeed$ and $\tmthree = \rtmthree_1\esub{\uvar}{\lvar}$
    for some $\rtmthree_1 \in \TermsSClumsyNeed$ such that $\lvar \notin \fv{\rtmthree_1}$.
    The step is of the form
    $\rtm
    = (\lam{\lvar}{\rtmthree_1\esub{\uvar}{\lvar}})\sctx\,\ofc{\rtm_2}
    \toS \rtmthree_1\esub{\uvar}{\ofc{\rtm_2}}\sctx
    = \tmtwo$.
    Note that $\tmtwo \in \TermsSClumsyNeed$
    and
    $\traVinv{\rtm}
    = (\lam{\uvar}{\traVinv{\rtmthree_1}})\traVinv{\sctx}\,\traVinv{\rtm_2}
    \tocbnd
      \traVinv{\rtmthree_1}\esub{\uvar}{\traVinv{\rtm_2}}\traVinv{\sctx}
    = \traVinv{\tmtwo}$.
  \item
    If the step is internal to $\rtm_1$,
    the step is of the form
    $\rtm = \rtm_1\,\ofc{\rtm_2} \toS \tmtwo_1\,\ofc{\rtm_2} = \tmtwo$,
    with $\rtm_1 \toS \tmtwo_1$.
    By \ih we have that $\tmtwo_1 \in \TermsSClumsyNeed$
    and $\traCNdinv{\rtm_1} \tocbnd^= \traCNdinv{\tmtwo_1}$,
    so $\tmtwo = \tmtwo_1\,\ofc{\rtm_2} \in \TermsSClumsyNeed$
    and
    $\traCNdinv{\rtm}
     = \traCNdinv{\rtm_1}\,\traCNdinv{\rtm_2}
     \tocbnd^= \traCNdinv{\tmtwo_1}\,\traCNdinv{\rtm_2}
     = \traCNdinv{\tmtwo}$.
  \item
    If the step is internal to $\rtm_2$,
    the step is of the form
    $\rtm = \rtm_1\,\ofc{\rtm_2} \toS \rtm_1\,\ofc{\tmtwo_2} = \tmtwo$,
    with $\rtm_2 \toS \tmtwo_2$.
    By \ih we have that $\tmtwo_2 \in \TermsSClumsyNeed$
    and $\traCNdinv{\rtm_2} \tocbnd^= \traCNdinv{\tmtwo_2}$,
    so $\tmtwo = \rtm_1\,\ofc{\tmtwo_2} \in \TermsSClumsyNeed$
    and
    $\traCNdinv{\rtm}
     = \traCNdinv{\rtm_1}\,\traCNdinv{\rtm_2}
     \tocbnd^= \traCNdinv{\rtm_1}\,\traCNdinv{\tmtwo_2}
     = \traCNdinv{\tmtwo}$.
  \end{enumerate}
\end{enumerate}
\end{proof}


\section{Appendix: Simulating Weak Evaluation}
\lsec{appendix:strategies}

\begin{definition}
We write $\tm \tossmany{\sequence{\rulename_1,\hdots,\rulename_n}} \tm'$
if $\tm = \tm_0 \toss{\rulename_1}
          \tm_1 \toss{\rulename_2}
          \hdots
                \toss{\rulename_n}
          \tm_n = \tm'$
holds for terms $\tm_0,\hdots,\tm_n$.
\end{definition}

We write $\RulesS$ for the set of all possible rulenames:
\[
  \RulesS \eqdef \set{\symSdb,\symSls,\symSgc}
            \cup \set{\symSsub{\uvar}{(\sha{\tm})\sctx} \ST \text{varying $\uvar,\tm,\sctx$}}
            \cup \set{\symSid{\uvar} \ST \text{varying $\uvar$}}
\]

\begin{definition}[$\CBN$ Rulenames]
We define a subset $\RulesSName \subseteq \RulesS$
by the following grammar:
\[
  \rrulename ::= \symSdb
            \mid \symSsub{\var}{\sha{\rtm}}
            \mid \symSls
            \mid \symSgc
            \mid \symSopen
\]
where $\rtm$ stands for a term in $\TermsSName$.
The translation of a $\CBN$-rulename is a sequence of $\RulesSName$ rulenames,
given by:
\[
  \begin{array}{rcl}
    \traN{\symdb}
  & \eqdef &
    \sequence{\symSdb}
  \\
    \traN{\symsub{\var}{\tm}}
    & \eqdef &
    \sequence{\symSsub{\var}{\sha{\traN{\tm}}},\symSopen}
  \\
    \traN{\symls}
    & \eqdef &
    \sequence{\symSls,\symSopen}
  \\
    \traN{\symgc}
    & \eqdef &
    \sequence{\symSgc}
  \end{array}
\]
The inverse translation of a $\RulesSName$ rulename 
is a sequence of $\CBN$-rulenames, given by:
\[
  \begin{array}{rcl}
    \traNinv{\symSdb}
  & \eqdef &
    \sequence{\symdb}
  \\
    \traNinv{\symSsub{\var}{\sha{\rtm}}}
    & \eqdef &
    \sequence{\symsub{\var}{\traNinv{\rtm}}}
  \\
    \traNinv{\symSls}
    & \eqdef &
    \sequence{\symls}
  \\
    \traNinv{\symSgc}
    & \eqdef &
    \sequence{\symgc}
  \\
    \traNinv{\symSopen}
    & \eqdef &
    \sequence{}
  \end{array}
\]
\end{definition}

\begin{remark}
\lremark{traNinv_rulename_fv}
$\fv{\traNinv{\rrulename}} = \fv{\rrulename}$
is an immediate consequence of \rremark{traNinv_fv}.
\end{remark}

\begin{lemma}[$\CBN$ Evaluation -- Simulation]
\llem{traN_evaluation_simulation}
If $\tm \tosn{\rulename} \tm'$ then $\traN{\tm} \tossmany{\traN{\rulename}} \traN{\tm'}$.
\end{lemma}
\begin{proof}
By induction on the derivation of the step $\tm \tosn{\rulename} \tm'$:
\begin{enumerate}
\item
  \ruleENdb:
  Let  
    $(\lam{\var}{\tm})\sctx\,\tmtwo \tosn{\symdb} \tm\esub{\var}{\tmtwo}\sctx$.
  Then:
  \[
    \begin{array}{rlll}
      \traN{((\lam{\var}{\tm})\sctx\,\tmtwo)}
    & = &
      (\lam{\lvar}{\traN{\tm}\esub{\var}{\lvar}})\traN{\sctx}\,\osha{\traN{\tmtwo}}
    \\
    & \toss{\symSdb} &
      \traN{\tm}\esub{\var}{\osha{\traN{\tmtwo}}}\traN{\sctx}
      & \text{by $\ruleESdb$}
    \\
    & = &
      \traN{(\tm\esub{\var}{\tmtwo}\sctx)}
    \end{array}
  \]
\item
  \ruleENsub:
  Let $\var \tosn{\symsub{\var}{\tm}} \tm$.
  Then:
  \[
    \begin{array}{rlll}
      \traN{\var}
    & = &
      \open{\var}
    \\
    & \toss{\symSsub{\var}{\sha{\traN{\tm}}}} &
      \open{\sha{\traN{\tm}}}
      & \text{by $\ruleEScopen,\ruleESsub$}
    \\
    & \toss{\symSopen} &
      \traN{\tm}
      & \text{by $\ruleESopen$}
    \end{array}
  \]
\item
  \ruleENls:
  Let
    $\tm\esub{\var}{\tmtwo}
     \tosn{\symls}
     \tm'\esub{\var}{\tmtwo}$
  be derived from
    $\tm \tosn{\symsub{\var}{\tmtwo}} \tm'$.
  By \ih we have that
    $\traN{\tm} \tossmany{\traN{\symsub{\var}{\tmtwo}}} \traN{\tm'}$.
  Since
    $\traN{\symsub{\var}{\tmtwo}} =
     \sequence{\symSsub{\var}{\sha{\traN{\tmtwo}}},\symSopen}$,
  this means there exists a term $\tmthree \in \TermsS$
  such that
    $\traN{\tm}
     \toss{\symSsub{\var}{\sha{\traN{\tmtwo}}}} \tmthree
     \toss{\symSopen} \traN{\tm'}$.
  Hence:
  \[
    \begin{array}{rlll}
      \traN{\tm\esub{\var}{\tmtwo}}
    & = &
      \traN{\tm}\esub{\var}{\osha{\traN{\tmtwo}}}
    \\
    & \toss{\symSls} &
      \tmthree\esub{\var}{\osha{\traN{\tmtwo}}}
      & \text{by $\ruleESls$ since 
                             $\traN{\tm}
                              \toss{\symSsub{\var}{\sha{\traN{\tmtwo}}}} \tmthree$}
    \\
    & \toss{\symSopen} &
      \traN{\tm'}\esub{\var}{\osha{\traN{\tmtwo}}}
      & \text{by $\ruleESsubL$ since $\tmthree \toss{\symSopen} \traN{\tm'}$}
    \\
    & = &
      \traN{(\tm'\esub{\var}{\tmtwo})}
    \end{array}
  \]
\item
  \ruleENgc:
  Let $\tm\esub{\var}{\tmtwo} \tosn{\symgc} \tm$
  where $\var\notin\fv{\tm}$.
  Recall that the translation does not create free variables (\rremark{traN_fv}),
  so $\var\notin\fv{\traN{\tm}}$.
  Hence:
  \[
    \begin{array}{rlll}
      \traN{(\tm\esub{\var}{\tmtwo})}
    & = &
      \traN{\tm}\esub{\var}{\osha{\traN{\tmtwo}}}
    \\
    & \toss{\symSgc} &
      \traN{\tm}
    \end{array}
  \]
\item
  \ruleENapp:
  Let
    $\tm\,\tmtwo \tosn{\rulename} \tm'\,\tmtwo$
  be derived from
    $\tm \tosn{\rulename} \tm'$.
  By \ih
    $\traN{\tm} \tossmany{\traN{\rulename}} \traN{\tm'}$.
  Hence:
  \[
    \begin{array}{rlll}
      \traN{(\tm\,\tmtwo)}
    & = &
      \traN{\tm}\,\osha{\traN{\tmtwo}}
    \\
    & \tossmany{\traN{\rulename}} &
      \traN{\tm'}\,\osha{\traN{\tmtwo}}
      & \text{by $\ruleESapp$ (many times)}
    \\
    & = &
      \traN{(\tm'\,\tmtwo)}
    \end{array}
  \]
\item
  \ruleENsubL:
  Let
    $\tm\esub{\var}{\tmtwo} \tosn{\rulename} \tm'\esub{\var}{\tmtwo}$
  be derived from
    $\tm \tosn{\rulename} \tm'$,
  where
    $\var \notin \fv{\rulename}$.
  By \ih,
    $\traN{\tm} \tossmany{\traN{\rulename}} \traN{\tm'}$.
  Recall that the translation does not create free variables (\rremark{traN_fv}),
  so $\var \notin \fv{\rulename}$
  implies $\var \notin \fv{\traN{\rulename}}$.
  Hence:
  \[
    \begin{array}{rlll}
      \traN{(\tm\esub{\var}{\tmtwo})}
    & = &
      \traN{\tm}\esub{\var}{\osha{\traN{\tmtwo}}}
    \\
    & \tossmany{\traN{\rulename}} &
      \traN{\tm'}\esub{\var}{\osha{\traN{\tmtwo}}}
      & \text{by $\ruleESsubL$ (many times), since $\var \notin \fv{\traN{\rulename}}$}
    \\
    & = &
      \traN{(\tm'\esub{\var}{\tmtwo})}
    \end{array}
  \]
\end{enumerate}
\end{proof}

\begin{lemma}[$\CBN$ Evaluation -- Inverse Simulation]
Let $\rtm \in \TermsSName$, $\rrulename \in \RulesSName$ and $\tmtwo \in \TermsS$
such that $\rtm \toss{\rrulename} \tmtwo$.
Then $\tmtwo \in \TermsSName$
and $\traNinv{\rtm} \tosnmany{\traNinv{\rrulename}} \traNinv{\tmtwo}$.
\end{lemma}
\begin{proof}
By induction on $\rtm$:
\begin{enumerate}
\item
  Translation of a variable before substitution, $\rtm = \open{\uvar}$:
  The step is of the form $\rtm = \open{\uvar} \toss{\rrulename} \tmtwo$
  and must be derived using the $\ruleEScopen$ rule
  from an internal step $\uvar \toss{\rrulename} \tmtwo'$.
  The internal step cannot be derived using the $\ruleESid$ rule
  because $\rrulename \in \RulesSName$ so $\rrulename$
  cannot be of the form $\symid{\uvar}$.
  This means that the internal step can only be derived using and $\ruleESsub$,
  where, again, we know that $\rrulename = \symSsub{\uvar}{\sha{\rtmthree}}$
  because $\rrulename \in \RulesSName$.
  Hence the step must be of the form
  $\open{\uvar} \toss{\symSsub{\uvar}{\sha{\rtmthree}}} \open{\sha{\rtmthree}} = \tmtwo$.
  Note that $\tmtwo = \open{\sha{\rtmthree}} \in \TermsSName$
  and:
  \[
    \traNinv{\rtm}
    = \traNinv{\open{\uvar}}
    = \uvar
    \tosn{\symsub{\uvar}{\traNinv{\rtmthree}}}
    = \traNinv{\rtmthree}
    = \traNinv{\open{\sha{\rtmthree}}}
    = \traNinv{\tmtwo}
  \]
\item
  Translation of a variable after substitution, $\rtm = \open{\sha{\rtmthree}}$:
  Note that the step can not be derived using the $\ruleEScopen$ rule,
  because there are no rules that allow deriving a step of the form
  $\sha{\rtmthree} \toss{\rrulename} \tmtwo'$.
  Hence the step is derived from the $\ruleESopen$ rule and of the form
  $\open{\sha{\rtmthree}} \toss{\symSopen} \rtmthree = \tmtwo$.
  Note that $\tmtwo = \rtmthree \in \TermsSName$
  and:
  \[
    \traNinv{\rtm}
    = \traNinv{\open{\sha{\rtmthree}}}
    = \traNinv{\rtmthree}
    = \traNinv{\tmtwo}
  \]
\item
  Translation of an abstraction, $\rtm = \lam{\var}{\rtm_1\esub{\uvar}{\lvar}}$:
  This case is impossible, as there are no rules
  that allow deriving a step of the form
  $\lam{\var}{\rtm_1\esub{\uvar}{\lvar}} \toss{\rrulename} \tmtwo$.
\item
  Translation of an application, $\rtm = \rtm_1\,\osha{\rtm_2}$:
  We consider two subcases, depending on the inference rule applied
  to conclude $\rtm \toss{\rrulename} \tmtwo$:
  \begin{enumerate}
  \item
    $\ruleESdb$:
    Then $\rtm = (\lam{\lvar}{\rtm_{11}\esub{\uvar}{\lvar}})\rsctx\,\osha{\rtm_2}$
    where $\rtm_{11},\rtm_2 \in \TermsSName$ and $\rsctx \in \SCtxsSName$
    and the step is of the form
    \[
      \rtm
      = (\lam{\lvar}{\rtm_{11}\esub{\uvar}{\lvar}})\rsctx\,\osha{\rtm_2}
      \toss{\symSdb} \rtm_{11}\esub{\uvar}{\osha{\rtm_2}}\rsctx = \tmtwo
    \]
    Note that $\tmtwo \in \TermsSName$ and:
    \[
        \traNinv{\rtm}
      =
        (\lam{\uvar}{\traNinv{\rtm_{11}}})\traNinv{\rsctx}\,\traNinv{\rtm_2}
      \tosn{\symdb}
        \traNinv{\rtm_{11}}\esub{\uvar}{\traNinv{\rtm_2}}\traNinv{\rsctx}
      =
        \traNinv{\tmtwo}
    \]
  \item
    $\ruleESapp$:
    Then the step is of the form
    $\rtm = \rtm_1\,\osha{\rtm_2} \toss{\rrulename} \tmtwo_1\,\osha{\rtm_2} = \tmtwo$
    where $\rtm_1 \toss{\rrulename} \tmtwo_1$.
    By \ih we have that $\tmtwo_1 \in \TermsSName$
    and $\traNinv{\rtm_1} \toss{\traNinv{\rrulename}} \traNinv{\tmtwo_1}$.
    So $\tmtwo = \tmtwo_1\,\osha{\rtm_2} \in \TermsSName$
    and, applying $\ruleENapp$, we conclude:
    \[
      \traNinv{\rtm}
      = \traNinv{\rtm_1}\,\traNinv{\rtm_2}
      \tosn{\traNinv{\rrulename}} \traNinv{\tmtwo_1}\,\traNinv{\rtm_2}
    \]
  \end{enumerate}
\item
  Translation of an explicit substitution, $\rtm = \rtm_1\esub{\uvar}{\osha{\rtm_2}}$:
  We consider five subcases, depending on the inference rule applied to
  conclude that there is a step $\rtm \toss{\rrulename} \tmtwo$.
  \begin{enumerate}
  \item
    $\ruleESls$:
    The step is of the form
    $\rtm
     = \rtm_1\esub{\uvar}{\osha{\rtm_2}}
     \toss{\symSls} \tmtwo_1\esub{\uvar}{\osha{\rtm_2}}
     = \tmtwo$
    where
    $\rtm_1 \toss{\symSsub{\uvar}{\sha{\rtm_2}}} \tmtwo_1$.
    Note that $\symSsub{\uvar}{\sha{\rtm_2}} \in \RulesSName$,
    so by \ih we have that $\tmtwo_1 \in \TermsSName$
    and $\traNinv{\rtm_1} \tosn{\symsub{\uvar}{\traNinv{\rtm_2}}} \traNinv{\tmtwo_1}$.
    So $\tmtwo_1\esub{\uvar}{\rtm_2} \in \TermsSName$ 
    and, applying $\ruleENls$, we conclude:
    \[
      \traNinv{\rtm}
      = \traNinv{\rtm_1}\esub{\uvar}{\traNinv{\rtm_2}}
      \tosn{\symls} \traNinv{\tmtwo_1}\esub{\uvar}{\traNinv{\rtm_2}}
      = \traNinv{\tmtwo}
    \]
  \item
    $\ruleESgc$:
    The step is of the form
    $\rtm
     = \rtm_1\esub{\uvar}{\osha{\rtm_2}}
     \toss{\symSgc} \rtm_1
     = \tmtwo$
    where $\uvar \notin \fv{\rtm_1}$.
    Note that $\tmtwo = \rtm_1 \in \TermsSName$.
    Note also that $\uvar \notin \fv{\traNinv{\rtm_1}}$ by \rremark{traNinv_fv}.
    So:
    \[
      \traNinv{\rtm}
      = \traNinv{\rtm_1}\esub{\uvar}{\traNinv{\rtm_2}}
      \tosn{\symgc} \traNinv{\rtm_1}
      = \traNinv{\tmtwo}
    \]
  \item
    $\ruleESsubL$:
    The step is of the form
    $\rtm
     = \rtm_1\esub{\uvar}{\osha{\rtm_2}}
     \toss{\rrulename} \tmtwo_1\esub{\uvar}{\osha{\rtm_2}}
     = \tmtwo$
    where $\rtm_1 \toss{\rrulename} \tmtwo_1$
    and $\uvar \notin \fv{\rrulename}$.
    By \ih we have that $\tmtwo_1 \in \TermsSName$
    and $\traNinv{\rtm_1} \toss{\traNinv{\rrulename}} \traNinv{\tmtwo_1}$.
    Note that $\tmtwo = \tmtwo_1\esub{\uvar}{\osha{\rtm_2}} \in \TermsSName$.
    Note also that $\uvar \notin \fv{\traNinv{\rrulename}}$ by \rremark{traNinv_rulename_fv}.
    So, applying $\ruleENsubL$, we conclude:
    \[
      \traNinv{\rtm}
      = \traNinv{\rtm_1}\esub{\uvar}{\traNinv{\rtm_2}}
      \toss{\traNinv{\rrulename}} \traNinv{\tmtwo_1}\esub{\uvar}{\traNinv{\rtm_2}}
      = \traNinv{\tmtwo}
    \]
  \item
    $\ruleESsubR$:
    We argue that this case is impossible.
    Indeed, the step must be of the form
    $\rtm
     = \rtm_1\esub{\uvar}{\osha{\rtm_2}}
     \toss{\rrulename} \rtm_1\esub{\uvar}{\tmtwo_2}
     = \tmtwo$
    where $\osha{\rtm_2} \toss{\rrulename} \tmtwo_2$.
    This is impossible because there are no rules that
    allow deriving such a step.
  \item
    $\ruleESsubRofc$:
    We argue that this case is impossible.
    Indeed, the step must be of the form
    $\rtm
     = \rtm_1\esub{\uvar}{\osha{\rtm_2}}
     \toss{\rrulename} \rtm_1\esub{\uvar}{\ofc{\tmtwo_2}}
     = \tmtwo$
    where
    $\rtm_1 \toss{\symSid{\uvar}} \rtm_1$
    and
    $\sha{\rtm_2} \toss{\rrulename} \tmtwo_2$.
    This is impossible because there are no rules that
    allow deriving the step $\sha{\rtm_2} \toss{\rrulename} \tmtwo_2$.
  \end{enumerate}
\end{enumerate}
\end{proof}

\begin{definition}[$\CBV$ Rulenames]
We define a subset $\RulesSValue \subseteq \RulesS$
by the following grammar:
\[
  \rrulename ::= \symSdb
            \mid \symSsub{\var}{\sha{\lam{\lvar}{\rtm\esub{\uvar}{\lvar}}}}
            \mid \symSls
            \mid \symSgc
            \mid \symSopen
\]
where $\rtm$ stands for a term in $\TermsSValue$.
The translation of a $\CBV$-rulename is a sequence of $\RulesSValue$ rulenames,
given by:
\[
  \begin{array}{rcl}
    \traV{\symdb}
  & \eqdef &
    \sequence{\symSls,\symSopen,\symSdb,\symSgc}
  \\
    \traV{\symsub{\var}{\val}}
    & \eqdef &
    \sequence{\symSsub{\var}{\sha{\traVval{\val}}}}
  \\
    \traV{\symlsv}
    & \eqdef &
    \sequence{\symSls}
  \\
    \traV{\symgclv}
    & \eqdef &
    \sequence{\symSgc}
  \end{array}
\]
where we define an auxiliary operation to translate values as follows:
\[
  \traVval{(\lam{\var}{\tm})} \eqdef \lam{\lvar}{\traV{\tm}\esub{\var}{\lvar}}
\]
We define an extended $\CBV$ evaluation relation $\tm \tosvUgclvinv{\rulename} \tmtwo$,
extending the system defining $\tosv{\rulename}$ with a further rulename $\symgclvinv$
and the following rule:
\[
  \indrule{\ruleEVgclv}{
    \var\notin\fv{\tm}
  }{
    \tm
    \tosvUgclvinv{\symgclvinv}
    \tm\esub{\var}{\lval}
  }
\]
The translation of a $\RulesSValue$ rulename is a set of sequences
of $\CBV$-rulenames, given by:
\[
  \begin{array}{rcl}
    \traVinv{\symSdb}
  & \eqdef &
    \set{\sequence{\symdb}}
  \\
    \traVinv{\symSsub{\var}{\sha{\lam{\lvar}{\rtm\esub{\uvar}{\lvar}}}}}
  & \eqdef &
    \set{\sequence{\symsub{\var}{\lam{\uvar}{\traVinv{\rtm}}}}}
  \\
    \traVinv{\symSls}
  & \eqdef &
    \set{\sequence{\symlsv},\sequence{\symgclvinv}}
  \\
    \traVinv{\symSgc}
  & \eqdef &
    \set{\sequence{\symgclv}}
  \\
    \traVinv{\symSopen}
  & \eqdef &
    \set{\sequence{}}
  \end{array}
\]
Note that the result is a set of sequences, which should be understood as
the fact that the translation is ``non-deterministic''.
\end{definition}

\begin{lemma}[$\CBV$ Evaluation -- Simulation]
\llem{traV_evaluation_simulation}
If $\tm \tosv{\rulename} \tm'$
then $\traV{\tm} \tossmany{\traV{\rulename}} \traV{\tm'}$.
\end{lemma}
\begin{proof}
By induction on the derivation of $\tm \tosv{\rulename} \tm'$:
\begin{enumerate}
\item
  \ruleEVdb:
  Let 
    $(\lam{\var}{\tm})\sctx\,\tmtwo \tosv{\symdb} \tm\esub{\var}{\tmtwo}\sctx$.
  Then:
  \[
    \begin{array}{rlll}
      \traV{((\lam{\var}{\tm})\sctx\,\tmtwo)}
    & = &
      \open{
        \uvar\esub{\uvar}{(\osha{\lam{\lvar}{\traV{\tm}\esub{\var}{\lvar}}})\traV{\sctx}}
      }\,\traV{\tmtwo}
    \\
    & \toss{\symSls} &
      \open{
        (\sha{\lam{\lvar}{\traV{\tm}\esub{\var}{\lvar}}})
        \esub{\uvar}{
          \osha{\lam{\lvar}{\traV{\tm}\esub{\var}{\lvar}}}
        }\traV{\sctx}
      }\,\traV{\tmtwo}
      \\&& \HS\text{by $\ruleESapp, \ruleEScopen, \ruleESls, \ruleESsub$}
    \\
    & \toss{\symSopen} &
      (\lam{\lvar}{\traV{\tm}\esub{\var}{\lvar}})
      \esub{\uvar}{
        \osha{\lam{\lvar}{\traV{\tm}\esub{\var}{\lvar}}}
      }\traV{\sctx}
      \,\traV{\tmtwo}
      \\&& \HS\text{by $\ruleESapp, \ruleESopen$}
    \\
    & \toss{\symSdb} &
      \traV{\tm}\esub{\var}{\traV{\tmtwo}}
      \esub{\uvar}{
        \osha{\lam{\lvar}{\traV{\tm}\esub{\var}{\lvar}}}
      }\traV{\sctx}
      \\&& \HS\text{by $\ruleESdb$}
    \\
    & \toss{\symSgc} &
      \traV{\tm}\esub{\var}{\traV{\tmtwo}}\traV{\sctx}
      \\&& \HS\text{by $\ruleESgc$}
    \\
    & = &
      \traV{(\tm\esub{\var}{\tmtwo}\sctx)}
    \end{array}
  \]
  To justify the application of $\ruleESgc$,
  note that by $\alpha$-conversion
  we may assume that $\uvar\notin\fv{\traV{\tm\esub{\var}{\traV{\tmtwo}}}}$.
\item
  \ruleEVsub:
  Let 
    $\var \tosv{\symsub{\var}{\val}} \val$.
  Then:
  \[
    \begin{array}{rlll}
      \traV{\var}
    & = &
      \traV{\ofc{\var}}
    \\
    & \toss{\symSsub{\var}{\sha{\traVval{\val}}}} &
      \ofc{\sha{\traVval{\val}}}
      & \text{by $\ruleESofcsub$}
    \\
    & = &
      \traV{\val}
      & \text{as $\ofc{\sha{\traVval{\val}}} = \traV{\val}$ holds by definition}
    \end{array}
  \]
\item
  \ruleEVlsv:
  Let
    $\tm\esub{\var}{\val\sctx}
     \tosv{\symlsv}
     \tm'\esub{\var}{\val}\sctx$
  be derived from
    $\tm \tosv{\symsub{\var}{\val}} \tm'$.
  By \ih, $\traV{\tm} \tossmany{\traV{\symsub{\var}{\val}}} \traV{\tm'}$,
  that is,
  $\traV{\tm} \toss{\symsub{\var}{\sha{\traVval{\val}}}} \traV{\tm'}$.
  Hence:
  \[
    \begin{array}{rlll}
      \traV{(\tm\esub{\var}{\val\sctx})}
    & = &
      \traV{\tm}\esub{\var}{\traV{\val}\traV{\sctx}}
    \\
    & = &
      \traV{\tm}\esub{\var}{(\osha{\traVval{\val}})\traV{\sctx}}
      & \text{since $\traV{\val} = \osha{\traVval{\val}}$ holds by definition}
    \\
    & \toss{\symls} &
      \traV{\tm'}\esub{\var}{\osha{\traVval{\val}}}\traV{\sctx}
      & \text{by $\ruleESls$, since $\traV{\tm} \toss{\traV{\symsub{\var}{\val}}} \traV{\tm'}$}
    \\
    & = &
      \traV{\tm'}\esub{\var}{\traV{\val}}\traV{\sctx}
    \\
    & = &
      \traV{(\tm'\esub{\var}{\val}\sctx)}
    \end{array}
  \]
\item
  \ruleEVgclv:
  Let
    $\tm\esub{\var}{\lval\sctx}
     \tosv{\symgclv}
     \tm\sctx$,
  where $\var\notin\fv{\tm}$.
  Recall that the translation does not create free variables (\rremark{traV_fv}),
  so $\var\notin\fv{\traV{\tm}}$.
  Note that the translation of a lax value always starts with ``$\ofc{}$'',
  \ie $\traV{(\lval)}$ is of the form $\ofc{\tmtwo}$.
  Hence:
  \[
    \begin{array}{rlll}
      \traV{(\tm\esub{\var}{\lval\sctx})}
    & = &
      \traV{\tm}\esub{\var}{\traV{(\lval)}\traV{\sctx}}
    \\
    & = &
      \traV{\tm}\esub{\var}{(\ofc{\tmtwo})\traV{\sctx}}
    \\
    & \toss{\symSgc} &
      \traV{\tm}\traV{\sctx}
      & \text{by $\ruleESgc$ since $\var\notin\fv{\traV{\tm}}$}
    \\
    & = &
      \traV{(\tm\sctx)}
    \end{array}
  \]
\item
  \ruleEVapp:
  Let
    $\tm\,\tmtwo \tosv{\rulename} \tm'\,\tmtwo$
  be derived from
    $\tm \tosv{\rulename} \tm'$.
  By \ih we have that $\traV{\tm} \tossmany{\traV{\rulename}} \traV{\tm'}$.
  Hence:
  \[
    \begin{array}{rlll}
      \traV{(\tm\,\tmtwo)}
    & = &
      \open{\uvar}\esub{\uvar}{\traV{\tm}}\,\traV{\tmtwo}
    \\
    & \tossmany{\traV{\rulename}} &
      \open{\uvar}\esub{\uvar}{\traV{\tm'}}\,\traV{\tmtwo}
      & \text{by $\ruleESapp,\ruleESsubR$ (many times)}
    \\
    & = &
      \traV{(\tm'\,\tmtwo)}
    \end{array}
  \]
\item
  \ruleEVsubL:
  Let
    $\tm\esub{\var}{\tmtwo} \tosv{\rulename} \tm'\esub{\var}{\tmtwo}$
  be derived from
    $\tm \tosv{\rulename} \tm'$,
  where $\var \notin \fv{\rulename}$.
  By \ih we have that
    $\traV{\tm} \tossmany{\traV{\rulename}} \traV{\tm'}$.
  Recall that the translation does not create free variables (\rremark{traV_fv}),
  so $\var\notin\fv{\traV{\tm}}$,
  which means that $\var\notin\fv{\traV{\rulename}}$.
  Hence:
  \[
    \begin{array}{rlll}
      \traV{(\tm\esub{\var}{\tmtwo})}
    & = &
      \traV{\tm}\esub{\var}{\traV{\tmtwo}}
    \\
    & \tossmany{\traV{\rulename}} &
      \traV{\tm'}\esub{\var}{\traV{\tmtwo}}
      & \text{by $\ruleESsubL$ (many times), since $\var\notin\fv{\traV{\rulename}}$}
    \\
    & = &
      \traV{(\tm'\esub{\var}{\tmtwo})}
    \end{array}
  \]
\item
  \ruleEVsubR:
  Let
    $\tm\esub{\var}{\tmtwo} \tosv{\rulename} \tm\esub{\var}{\tmtwo'}$
  be derived from
    $\tmtwo \tosv{\rulename} \tmtwo'$.
  By \ih we have that
    $\traV{\tmtwo} \tossmany{\traV{\rulename}} \traV{\tmtwo'}$.
  Hence:
  \[
    \begin{array}{rlll}
      \traV{(\tm\esub{\var}{\tmtwo})}
    & = &
      \traV{\tm}\esub{\var}{\traV{\tmtwo}}
    \\
    & \tossmany{\traV{\rulename}} &
      \traV{\tm}\esub{\var}{\traV{\tmtwo'}}
      & \text{by $\ruleESsubR$ (many times)}
    \\
    & = &
      \traV{(\tm\esub{\var}{\tmtwo'})}
    \end{array}
  \]
\end{enumerate}
\end{proof}

\begin{lemma}[$\CBV$ Evaluation -- Inverse Simulation]
Let $\rtm \in \TermsSValue$, $\rrulename \in \RulesSValue$ and $\tmtwo \in \TermsS$
such that $\rtm \tossmany{\rrulename} \tmtwo$.
Then $\tmtwo \in \TermsSValue$ and
there is a sequence of rulenames $\rulenameseq \in \traVinv{\rrulename}$
such that $\traVinv{\tm} \tosvUgclvinvmany{\rulenameseq} \traVinv{\tmtwo}$.
\end{lemma}
\begin{proof}
By induction on $\rtm$:
\begin{enumerate}
\item
  Translation of a variable, $\rtm = \ofc{\uvar}$:
  The step must be derived using the $\ruleESofcsub$ rule
  and thus of the form
  $\rtm
   = \ofc{\uvar}
   \toss{\symSsub{\uvar}{(\sha{\tmthree})\sctx}} \ofc{(\sha{\tmthree})\sctx}
   = \tmtwo$.
  Since $\rrulename = \symSsub{\uvar}{(\sha{\tmthree})\sctx} \in \RulesSValue$,
  we have that $\sctx$ is empty
  and $\tmthree = \lam{\lvar}{\rtmthree_1\esub{\uvartwo}{\lvar}}$.
  Then $\tmtwo = \osha{\lam{\lvar}{\rtmthree_1\esub{\uvartwo}{\lvar}}} \in \TermsSValue$
  and:
  \[
     \traVinv{\rtm}
     = \uvar
     \tosvUgclvinv{\symsub{\uvar}{\lam{\uvartwo}{\traVinv{\rtmthree_1}}}}
     = \lam{\uvartwo}{\traVinv{\rtmthree_1}}
     = \traVinv{\tmtwo}
  \]
\item
  Translation of an abstraction, $\rtm = \osha{\lam{\lvar}{\rtm_1\esub{\uvar}{\lvar}}}$:
  This case is impossible, as there are no rules that allow deriving a step of
  the form
  $\osha{\lam{\lvar}{\rtm_1\esub{\uvar}{\lvar}}} \toss{\rrulename} \tmtwo$.
\item
  Request (1), $\rtm = \open{\uvar}\esub{\uvar}{\rtm_1}$:
  Then the step must be of the form
  $\open{\uvar}\esub{\uvar}{\rtm_1} \toss{\rrulename} \tmtwo$.
  Note that the step cannot be an instance of the $\ruleESgc$ rule because
  $\uvar \in \fv{\open{\uvar}}$.
  We consider four subcases, depending on whether the step
  is derived using
  $\ruleESls$, $\ruleESsubL$, $\ruleESsubR$, or $\ruleESsubRofc$:
  \begin{enumerate}
  \item
    $\ruleESsubL$:
    Then $\rtm_1 = (\ofc{(\sha{\tmthree})\sctx_1})\sctx_2$
    and the step is of the form
    \[
      \rtm
       = \open{\uvar}\esub{\uvar}{(\ofc{(\sha{\tmthree})\sctx_1})\sctx_2}
       \toss{\symSls}
         \open{ (\sha{\tmthree})\sctx_1 }\esub{\uvar}{\ofc{(\sha{\tmthree})\sctx_1}}\sctx_2
       = \tmtwo
    \]
    Since $\rtm_1 = (\ofc{(\sha{\tmthree})\sctx_1})\sctx_2 \in \TermsSValue$
    then by \rremark{shape_of_values_in_TermsSValue}
    we have that
    $\rtm_1 = (\osha{\lam{\lvar}{\rtm_{11}\esub{\uvartwo}{\lvar}}})\sctx_2$
    where $\sctx_2 \in \SCtxsSValue$.
    Hence
    $\tmtwo = 
     \open{ \sha{\lam{\lvar}{\rtm_{11}\esub{\uvartwo}{\lvar}}} }
       \esub{\uvar}{\osha{\lam{\lvar}{\rtm_{11}\esub{\uvartwo}{\lvar}}}}\rsctx_2
     \in \TermsSValue$
    and:
    \[
      \begin{array}{lrl}
        \traVinv{\rtm}
      & = &
        \traVinv{(
          \open{\uvar}
            \esub{\uvar}{(\osha{\lam{\lvar}{\rtm_{11}\esub{\uvartwo}{\lvar}}})\sctx_2}
        )}
      \\
      & = &
        (\lam{\uvartwo}{\traVinv{\rtm_{11}}})\traVinv{\sctx_2}
      \\
      & \tosvUgclvinv{\symgclvinv} &
       (\lam{\uvartwo}{\traVinv{\rtm_{11}}})
         \esub{\uvar}{\lam{\uvartwo}{\traVinv{\rtm_{11}}}}
         \traVinv{\rsctx_2}
      \\
      & = &
       \traVinv{(
         \open{ \sha{\lam{\lvar}{\rtm_{11}\esub{\uvartwo}{\lvar}}} }
           \esub{\uvar}{\osha{\lam{\lvar}{\rtm_{11}\esub{\uvartwo}{\lvar}}}}\rsctx_2
       )}
      \\
      & = &
        \traVinv{\tmtwo}
      \end{array}
    \]
  \item
    $\ruleESsubL$:
    We argue that this case is impossible.
    Indeed, the only way to reduce $\open{\uvar}$
    is by substituting $\uvar$, \ie with a step of the form 
    $\open{\uvar} \toss{\symSsub{\uvar}{(\sha{\tmthree})\sctx}} \open{(\sha{\tmthree})\sctx}$,
    but $\ruleESsubL$ cannot be applied because
    $\uvar \in \fv{\symSsub{\uvar}{(\sha{\tmthree})\sctx}}$.
  \item
    $\ruleESsubR$:
    The step is of the form
    $\open{\uvar}\esub{\uvar}{\rtm_1}
     \toss{\rrulename} \open{\uvar}\esub{\uvar}{\tmtwo_1}
     = \tmtwo$.
    where $\rtm_1 \toss{\rrulename} \tmtwo_1$.
    By \ih we have that $\tmtwo_1 \in \TermsSValue$
    and $\traVinv{\rtm_1} \tosvUgclvinvmany{\rulenameseq} \traVinv{\tmtwo_1}$
    for some $\rulenameseq \in \traVinv{\rrulename}$.
    This means that $\tmtwo = \open{\uvar}\esub{\uvar}{\tmtwo_1} \in \TermsSValue$
    and:
    \[
      \begin{array}{lrl}
        \traVinv{\rtm}
      & = &
        \traVinv{(\open{\uvar}\esub{\uvar}{\rtm_1})}
      \\
      & = &
        \traVinv{\rtm_1}
      \\
      & \tosvUgclvinvmany{\rulenameseq} &
        \traVinv{\tmtwo_1}
      \\
      & = &
        \traVinv{(\open{\uvar}\esub{\uvar}{\tmtwo_1})}
      \\
      & = &
        \traVinv{\tmtwo}
      \end{array}
    \]
  \item
    $\ruleESsubRofc$:
    Then $\rtm_1 = \ofc{\tmthree}$ and the step is of the form
    $\open{\uvar}\esub{\uvar}{\ofc{\tmthree}}
     \toss{\rrulename} \open{\uvar}\esub{\uvar}{\ofc{\tmthree'}}
     = \tmtwo$
    where $\open{\uvar} \toss{\symSid{\uvar}} \open{\uvar}$
    and $\tmthree \toss{\rrulename} \tmthree'$.
    Since $\rtm_1 \in \TermsSValue$
    we have that $\rtm_1$ must be either
    the translation of a variable ($\rtm_1 = \ofc{\uvartwo}$)
    or the translation of an abstraction
    ($\rtm_1 = \osha{\lam{\lvar}{\rtmfour\esub{\var}{\lvar}}}$).
    We consider these two as subcases:
    \begin{enumerate}
    \item
      If $\rtm_1 = \ofc{\uvartwo}$,
      then $\tmthree = \uvartwo$.
      The step $\tmthree = \uvartwo \toss{\rrulename} \tmthree'$
      can only be a substitution step, \ie it must be of the form
      $\tmthree
       = \uvartwo
       \toss{\symSsub{\uvartwo}{(\sha{\tmfour})\sctx}} (\sha{\tmfour})\sctx
       = \tmthree'$.
      Since by hypothesis
      $\rrulename = \symSsub{\uvartwo}{(\sha{\tmfour})\sctx} \in \RulesSValue$,
      we have that $\sctx$ is empty
      and $\tmfour$
      is of the form $\tmfour = \lam{\lvartwo}{\rtmfour_1\esub{\uvarthree}{\lvartwo}}$.
      The step is then of the form
      $\rtm
       = \open{\uvar}\esub{\uvar}{\ofc{\uvartwo}}
       \toss{\symSsub{\uvartwo}{\sha{ \tmfour }}}
         \open{\uvar}\esub{\uvar}{\osha{\tmfour}}
       = \tmtwo
      $.
      To conclude, note that
      $\tmtwo = \open{\uvar}\esub{\uvar}{\osha{\tmfour}}\in \TermsSValue$
      because
      $\osha{\tmfour}
       = \osha{\lam{\lvartwo}{\rtmfour_1\esub{\uvarthree}{\lvartwo}}}
       \in \TermsSValue$
      and note that:
      \[
        \begin{array}{lrl}
          \traVinv{\rtm}
        & = &
          \traVinv{(\open{\uvar}\esub{\uvar}{\ofc{\uvartwo}})}
        \\
        & = &
          \uvartwo
        \\
        & \tosvUgclvinv{\symsub{\uvartwo}{\lam{\uvarthree}{\traVinv{\rtmfour_1}}}} &
          \lam{\uvarthree}{\traVinv{\rtmfour_1}}
        \\
        & = &
          \traVinv{(\osha{\tmfour})}
        \\
        & = &
          \traVinv{(\open{\uvar}\esub{\uvar}{\osha{\tmfour}})}
        \\
        & = &
          \traVinv{\tmtwo}
        \end{array}
      \]
    \item
      If $\rtm_1 = \osha{\lam{\lvar}{\rtmfour\esub{\var}{\lvar}}}$,
      then $\tmthree = \sha{\lam{\lvar}{\rtmfour\esub{\var}{\lvar}}}$.
      This case is impossible, as there are no rules that allow to
      derive a step of the form
      $\tmthree = \sha{\lam{\lvar}{\rtmfour\esub{\var}{\lvar}}} \toss{\rrulename} \tmthree'$.
    \end{enumerate}
  \end{enumerate}
\item
  Request (2), $\rtm = \open{\sha{\lam{\lvar}{\rtm_1\esub{\uvar}{\lvar}}}}$:
  Then the step is of the form
  $\rtm = \open{\sha{\lam{\lvar}{\rtm_1\esub{\uvar}{\lvar}}}}
   \toss{\rrulename} \tmtwo$.
  Note that the step cannot be internal 
  to $\lam{\lvar}{\rtm_1\esub{\uvar}{\lvar}}$
  because there are no rules that allow to reduce inside a ``$\sha{}$''.
  Hence the step must be an instance of the $\ruleESopen$ rule,
  \ie of the form:
  $\rtm
   = \open{\sha{\lam{\lvar}{\rtm_1\esub{\uvar}{\lvar}}}}
   \toss{\symSopen} \lam{\lvar}{\rtm_1\esub{\uvar}{\lvar}}
   = \tmtwo$:
  then note that $\tmtwo \in \TermsSValue$ and:
  \[
    \begin{array}{lrl}
      \traVinv{\rtm}
    & = &
      \traVinv{(\open{\sha{\lam{\lvar}{\rtm_1\esub{\uvar}{\lvar}}}})}
    \\
    & = &
      \lam{\uvar}{\traVinv{\rtm_1}})
    \\
    & = &
      \traVinv{(\lam{\lvar}{\rtm_1\esub{\uvar}{\lvar}})}
    \\
    & = &
      \tmtwo
    \end{array}
  \]
\item
  Abstraction, $\rtm = \lam{\lvar}{\rtm_1\esub{\uvar}{\lvar}}$:
  This case is impossible, as there are no rules that allow deriving
  a step of the form $\lam{\lvar}{\rtm_1\esub{\uvar}{\lvar}} \toss{\rrulename} \tmtwo$.
\item
  Application, $\rtm = \rtm_1\,\rtm_2$:
  We consider two subcases, depending on whether the step is derived
  using the $\ruleESdb$ rule or the $\ruleESapp$ rule:
  \begin{enumerate}
  \item
    $\ruleESdb$:
    Then $\rtm_1$ must be of the form $(\lam{\lvar}{\tmthree})\sctx$.
    and by \rremark{shape_of_values_in_TermsSValue}
    we know in turn that
    $\tmthree = \rtmthree_1\esub{\uvar}{\lvar}$, where $\lvar \notin\fv{\rtmthree_1}$
    and $\rtmthree_1 \in \TermsSValue$ and $\sctx \in \SCtxsSValue$.
    The step is of the form
    $\rtm
     = (\lam{\lvar}{\rtmthree_1\esub{\uvar}{\lvar}})\sctx\,\rtm_2
     \toss{\symSdb} \rtmthree_1\esub{\uvar}{\rtm_2}\sctx
     = \tmtwo$.
    Note that $\tmtwo \in \TermsSValue$ and
    \[
      \begin{array}{lrl}
        \traVinv{\rtm}
      & = &
        \traVinv{(\lam{\lvar}{\rtmthree_1\esub{\uvar}{\lvar}})\sctx\,\rtm_2}
      \\
      & = &
        (\lam{\uvar}{\traVinv{\rtmthree_1}})\traVinv{\sctx}\,\traVinv{\rtm_2}
      \\
      & \tosvUgclvinv{\symdb}
        & \traVinv{\rtmthree_1}\esub{\uvar}{\traVinv{\rtm_2}}\traVinv{\sctx}
      \\
      & = &
        \traVinv{(\rtmthree_1\esub{\uvar}{\rtm_2}\sctx)}
      \\
      & = &
        \traVinv{\tmtwo}
      \end{array}
    \]
  \item
    $\ruleESapp$:
    Then the step is of the form
    $\rtm
     = \rtm_1\,\rtm_2
     \toss{\rrulename} \tmtwo_1\,\rtm_2
     = \tmtwo$
    where $\rtm_1 \toss{\rrulename} \tmtwo_1$.
    By \ih, $\tmtwo_1 \in \TermsSValue$ and
    $\traVinv{\rtm_1} \tosvUgclvinvmany{\rulenameseq} \traVinv{\tmtwo_1}$
    for some $\rulenameseq \in \traVinv{\rrulename}$.
    Then $\tmtwo = \tmtwo_1\,\rtm_2 \in \TermsSValue$
    and, applying $\ruleEVapp$ once per each step, we have:
    \[
        \traVinv{\rtm}
      = \traVinv{(\rtm_1\,\rtm_2)}
      = \traVinv{\rtm_1}\,\traVinv{\rtm_2}
      \tosvUgclvinvmany{\rulenameseq} \traVinv{\tmtwo_1}\,\traVinv{\rtm_2}
      = \traVinv{(\tmtwo_1\,\rtm_2)}
      = \traVinv{\tmtwo}
    \]
  \end{enumerate}
\item
  Explicit substitution, $\rtm = \rtm_1\esub{\uvar}{\rtm_2}$:
  We consider five subcases, depending on the rule used to derive
  the step:
  \begin{enumerate}
  \item
    $\ruleESls$:
    For the $\ruleESls$ rule to be applicable,
    $\rtm_2$ must be of the form $\rtm_2 = (\ofc{(\sha{\tmthree})\sctx_1})\sctx_2$.
    Since $\rtm_2 \in \TermsSValue$, by \rremark{shape_of_values_in_TermsSValue},
    we have that actually $\rtm_2 = (\osha{\lam{\lvar}{\rtmthree_1\esub{\uvar}{\lvar}}})\sctx_2$,
    where $\rtmthree_1 \in \TermsSValue$, $\lvar\notin\fv{\rtmthree_1}$,
    and $\sctx_2 \in \SCtxsSValue$.
    Thus the step is of the form
    $\rtm
     = \rtm_1\esub{\uvar}{(\osha{\lam{\lvar}{\rtmthree_1\esub{\uvar}{\lvar}}})\sctx_2}
     \toss{\symSls} \tmtwo_1\esub{\uvar}{\osha{\lam{\lvar}{\rtmthree_1\esub{\uvar}{\lvar}}}}\sctx_2
     = \tmtwo$,
    where
    $\rtm_1 \toss{\symSsub{\uvar}{\sha{\lam{\lvar}{\rtmthree_1\esub{\uvar}{\lvar}}}}} \tmtwo_1$.
    By \ih, we have that $\tmtwo_1 \in \TermsSValue$
    and
    $\traVinv{\rtm_1}
     \tosvUgclvinv{\symsub{\uvar}{\lam{\uvar}{\traVinv{\rtmthree_1}}}} \traVinv{\tmtwo_1}$.
    Hence $\tmtwo = \tmtwo_1\esub{\uvar}{\osha{\lam{\lvar}{\rtmthree_1\esub{\uvar}{\lvar}}}}\sctx_2
                    \in \TermsSValue$
    and:
    \[
      \begin{array}{lrl}
        \traVinv{\rtm}
      & = &
        \traVinv{(\rtm_1\esub{\uvar}{(\osha{\lam{\lvar}{\rtmthree_1\esub{\uvar}{\lvar}}})\sctx_2})}
      \\
      & = &
        \traVinv{\rtm_1}
          \esub{\uvar}{
            (\lam{\uvar}{\traVinv{\rtmthree_1}})
            \traVinv{\sctx_2}
          }
      \\
      & \tosvUgclvinv{\symlsv} &
        \traVinv{\tmtwo_1}
          \esub{\uvar}{\lam{\uvar}{\traVinv{\rtmthree_1}}}
          \traVinv{\sctx_2}
      \\
      & = &
        \traVinv{(\tmtwo_1\esub{\uvar}{\osha{\lam{\lvar}{\rtmthree_1\esub{\uvar}{\lvar}}}}\sctx_2)}
      \\
      & = &
        \traVinv{\tmtwo}
      \end{array}
    \]
  \item
    $\ruleESgc$:
    For the $\ruleESgc$ rule to be applicable,
    $\rtm_2$ must be of the form $\rtm_2 = (\ofc{\tmthree})\sctx$.
    Since $\rtm_2 \in \TermsSValue$, by \rremark{shape_of_values_in_TermsSValue},
    we have that $\sctx \in \SCtxsSValue$
    and that $\tmthree$ is either a variable ($\tmthree = \uvartwo$)
    or of the form $\tmthree = \sha{\lam{\lvar}{\rtmthree_1\esub{\uvartwo}{\lvar}}}$,
    where $\rtmthree_1 \in \TermsSValue$ and $\lvar\notin\fv{\rtmthree_1}$.
    Note that in both cases $\traVinv{(\ofc{\tmthree})}$ is a lax value,
    because it is either of the form $\uvartwo$
    or of the form $\lam{\uvartwo}{\traVinv{\rtmthree_1}}$.
    Moreover, the step is of the form
    $\rtm
     = \rtm_1\esub{\uvar}{(\sha{\tmthree})\sctx}
     \toss{\symSgc}
     \rtm_1\sctx = \tmtwo$,
    where $\uvar\notin\fv{\rtm_1}$.
    Note that $\tmtwo = \rtm_1\sctx \in \TermsSValue$.
    Moreover, we have that $\uvar\notin\fv{\traVinv{\rtm_1}}$ by \rremark{traVinv_fv},
    so:
    \[
      \begin{array}{lrl}
        \traVinv{\rtm}
      & = &
        \traVinv{(\rtm_1\esub{\uvar}{(\ofc{\tmthree})\sctx})}
      \\
      & = &
        \traVinv{\rtm_1}\esub{\uvar}{\traVinv{\ofc{\tmthree}}\traVinv{\sctx}}
      \\
      & \tosvUgclvinv{\symgclv} &
        \traVinv{\rtm_1}\traVinv{\sctx}
      \\
      & = &
        \traVinv{(\rtm_1\sctx)}
      \\
      & = &
        \traVinv{\tmtwo}
      \end{array}
    \]
  \item
    $\ruleESsubL$:
    The step is of the form
    $\rtm
     = \rtm_1\esub{\uvar}{\rtm_2}
     \toss{\rrulename} \tmtwo_1\esub{\uvar}{\rtm_2}
     = \tmtwo$,
    where $\uvar \notin\fv{\rrulename}$
    and $\rtm_1 \toss{\rrulename} \tmtwo_1$.
    By \ih, we have that $\tmtwo_1 \in \TermsSValue$
    and $\traVinv{\rtm_1} \tosvUgclvinv{\rulenameseq} \traVinv{\tmtwo_1}$
    for some $\rulenameseq \in \traVinv{\rrulename}$.
    Then $\tmtwo = \tmtwo_1\esub{\uvar}{\rtm_2} \in \TermsSValue$
    and:
    \[
      \begin{array}{lrll}
        \traVinv{\rtm}
      & = &
        \traVinv{(\rtm_1\esub{\uvar}{\rtm_2})}
      \\
      & = &
        \traVinv{\rtm_1}\esub{\uvar}{\traVinv{\rtm_2}}
      \\
      & \tosvUgclvinv{\rulenameseq} &
        \traVinv{\tmtwo_1}\esub{\uvar}{\traVinv{\rtm_2}}
        & \text{by $\ruleEVsubL$ (many times)}
      \\
      & = &
        \traVinv{(\tmtwo_1\esub{\uvar}{\rtm_2})}
      \\
      & = &
        \traVinv{\tmtwo}
      \end{array}
    \]
  \item
    $\ruleESsubR$:
    The step is of the form
    $\rtm
     = \rtm_1\esub{\uvar}{\rtm_2}
     \toss{\rrulename} \rtm_1\esub{\uvar}{\tmtwo_2}
     = \tmtwo$,
    where $\rtm_2 \toss{\rrulename} \tmtwo_2$.
    By \ih, we have that $\tmtwo_2 \in \TermsSValue$
    and $\traVinv{\rtm_2} \tosvUgclvinv{\rulenameseq} \traVinv{\tmtwo_2}$
    for some $\rulenameseq \in \traVinv{\rrulename}$.
    Then $\tmtwo = \rtm_1\esub{\uvar}{\tmtwo_2} \in \TermsSValue$
    and:
    \[
      \begin{array}{lrll}
        \traVinv{\rtm}
      & = &
        \traVinv{(\rtm_1\esub{\uvar}{\rtm_2})}
      \\
      & = &
        \traVinv{\rtm_1}\esub{\uvar}{\traVinv{\rtm_2}}
      \\
      & \tosvUgclvinv{\rulenameseq} &
        \traVinv{\rtm_1}\esub{\uvar}{\traVinv{\tmtwo_2}}
        & \text{by $\ruleEVsubR$ (many times)}
      \\
      & = &
        \traVinv{(\rtm_1\esub{\uvar}{\tmtwo_2})}
      \\
      & = &
        \traVinv{\tmtwo}
      \end{array}
    \]
  \item
    $\ruleESsubRofc$:
    For the $\ruleESsubRofc$ to be applicable,
    $\rtm_2$ must be of the form $\rtm_2 = \ofc{\tmthree}$.
    Since $\rtm_2 \in \TermsSValue$, by \rremark{shape_of_values_in_TermsSValue},
    we have $\tmthree$ is either a variable ($\tmthree = \uvartwo$)
    or of the form $\tmthree = \sha{\lam{\lvar}{\rtmthree_1\esub{\uvartwo}{\lvar}}}$,
    where $\rtmthree_1 \in \TermsSValue$ and $\lvar\notin\fv{\rtmthree_1}$.
    Moreover, the step is of the form
    $\rtm
     = \rtm_1\esub{\uvar}{\ofc{\tmthree}}
     \toss{\rrulename} \rtm_1\esub{\uvar}{\tmtwo_2}
     = \tmtwo
    $
    where $\rtm_1 \toss{\symSid{\uvar}} \rtm_1$
    and $\tmthree \toss{\rrulename} \tmtwo_2$.
    Note that $\tmthree$ cannot be of the form
    $\sha{\lam{\lvar}{\rtmthree_1\esub{\uvartwo}{\lvar}}}$,
    because there are no rules that allow
    deriving a step
    $\sha{\lam{\lvar}{\rtmthree_1\esub{\uvartwo}{\lvar}}} \toss{\rrulename} \tmtwo_2$.
    Hence $\tmthree$ must be a variable, \ie $\tmthree = \uvartwo$.
    The step $\tmthree = \uvartwo \toss{\rrulename} \tmtwo_2$
    can only be derived using the $\ruleESsub$ rule,
    so $\rrulename = \symSsub{\uvartwo}{(\sha{\tmfour})\sctx}$
    and $\tmtwo_2 = (\sha{\tmfour})\sctx$.
    Since $\rrulename \in \RulesSValue$,
    we know that $(\sha{\tmfour})\sctx$ must be of the form
    $(\sha{\tmfour})\sctx = \sha{\lam{\lvartwo}{\rtmfour_1\esub{\uvarthree}{\lvartwo}}}$.
    In summary, the step is of the form
    $\rtm
    = \rtm_1\esub{\uvar}{\ofc{\uvartwo}}
    \toss{\symSsub{\uvartwo}{\sha{\lam{\lvartwo}{\rtmfour_1\esub{\uvarthree}{\lvartwo}}}}}
      \rtm_1\esub{\uvar}{\ofc{\sha{\lam{\lvartwo}{\rtmfour_1\esub{\uvarthree}{\lvartwo}}}}}
    = \tmtwo$.
    Note that $\tmtwo \in \TermsSValue$ and:
    \[
      \begin{array}{lrll}
        \traVinv{\rtm}
      & = &
        \traVinv{(\rtm_1\esub{\uvar}{\ofc{\uvartwo}})}
      \\
      & = &
        \traVinv{\rtm_1}\esub{\uvar}{\uvartwo}
      \\
      & \tosvUgclvinv{\symsub{\uvartwo}{\lam{\uvarthree}{\traVinv{\rtmfour_1}}}} &
        \traVinv{\rtm_1}\esub{\uvar}{\lam{\uvarthree}{\traVinv{\rtmfour_1}}}
        & \text{by $\ruleEVsub,\ruleEVsubR$}
      \\
      & = &
        \traVinv{(\rtm_1\esub{\uvar}{\ofc{\sha{\lam{\lvartwo}{\rtmfour_1\esub{\uvarthree}{\lvartwo}}}}})}
      \\
      & = &
        \traVinv{\tmtwo}
      \end{array}
    \]
  \end{enumerate}
\end{enumerate}
\end{proof}

\begin{definition}[$\CCBNd$ Rulenames]
We define a subset $\RulesSClumsyNeed \subseteq \RulesS$
by the following grammar:
\[
  \rrulename ::= \symSdb
            \mid \symSsub{\var}{(\sha{\lam{\lvar}{\rtm\esub{\uvar}{\lvar}}})\rsctx}
            \mid \symSls
            \mid \symSgc
            \mid \symSopen
\]
where $\rtm$ stands for a term in $\TermsSClumsyNeed$.
The translation of a $\CCBNd$-rulename is a sequence
of $\RulesSClumsyNeed$ rulenames, given by:
\[
  \begin{array}{rcl}
    \traD{\symdb}
  & \eqdef &
    \sequence{\symSopen,\symSdb}
  \\
    \traD{\symsub{\var}{\val\sctx}}
    & \eqdef &
    \sequence{\symSsub{\var}{(\sha{\traDval{\val}})\traD{\sctx}}}
  \\
    \traD{\symlsw}
    & \eqdef &
    \sequence{\symSls}
  \\
    \traD{\symgc}
    & \eqdef &
    \sequence{\symSgc}
  \\
    \traD{\symid{\var}}
    & \eqdef &
    \sequence{\symSid{\var}}
  \end{array}
\]
where, moreover, we define an auxiliary operation
to translate values as
follows:
$\traDval{(\lam{\var}{\tm})} \eqdef \lam{\lvar}{\traD{\tm}\esub{\var}{\lvar}}$.

The translation of a $\RulesSClumsyNeed$ rulename is a sequence
of $\CCBNd$-rulenames, given by:
\[
  \begin{array}{rcl}
    \traDinv{\symSdb}
  & \eqdef &
    \sequence{\symdb}
  \\
    \traDinv{\symSsub{\var}{(\sha{\lam{\lvar}{\rtm}\esub{\uvar}{\lvar}})\rsctx}}
  & \eqdef &
    \sequence{\symsub{\var}{(\lam{\uvar}{\traDinv{\rtm}})\traDinv{\rsctx}}}
  \\
    \traDinv{\symSls}
  & \eqdef &
    \sequence{\symlsw}
  \\
    \traDinv{\symSgc}
  & \eqdef &
    \sequence{\symgc}
  \\
    \traDinv{\symSopen}
  & \eqdef &
    \sequence{}
  \\
    \traDinv{\symSid{\var}}
  & \eqdef &
    \sequence{\symid{\var}}
  \end{array}
\]
\end{definition}

\begin{lemma}[$\CCBNd$ Evaluation -- Simulation]
\llem{traSCBNd_evaluation_simulation}
If $\tm \tosd{\rulename} \tm'$ then $\traD{\tm} \tossmany{\traD{\rulename}} \traD{\tm'}$.
\end{lemma}
\begin{proof}
By induction on the derivation of $\tm \tosd{\rulename} \tm'$:
\begin{enumerate}
\item
  \ruleEDdb:
  Let
    $(\lam{\var}{\tm})\sctx\,\tmtwo \tosd{\symdb} \tm\esub{\var}{\tmtwo}\sctx$.
  Then:
  \[
    \begin{array}{rlll}
      \traD{((\lam{\var}{\tm})\sctx\,\tmtwo)}
    & = &
      \open{\sha{(\lam{\lvar}{\traD{\tm}\esub{\var}{\lvar}})}\traD{\sctx}}\,\ofc{\traD{\tmtwo}}
    \\
    & \toss{\symSopen} &
      (\lam{\lvar}{\traD{\tm}\esub{\var}{\lvar}})\traD{\sctx}\,\ofc{\traD{\tmtwo}}
      & \text{by $\ruleESapp$, $\ruleESopen$}
    \\
    & \toss{\symSdb} &
      \traD{\tm}\esub{\var}{\ofc{\traD{\tmtwo}}}\traD{\sctx}
    \\
    & = &
      \traD{(\tm\esub{\var}{\tmtwo}\sctx)}
    \end{array}
  \]
\item
  \ruleEDsub:
  Let
    $\var
     \tosd{\symsub{\var}{\val\sctx}}
     \val\sctx$.
  Then by $\ruleESsub$:
  \[
    \traD{\var} = \var
    \toss{\symSsub{\var}{(\sha{\traDval{\val}})\traD{\sctx}}}
      \sha{\traDval{\val}}\traD{\sctx}
    = \traD{\val}\traD{\sctx}
    = \traD{(\val\sctx)}
  \]
  Note that $\traD{\val} = \sha{\traDval{\val}}$ holds by definition.
\item
  \ruleEDsubES:
  Let
    $\tm\esub{\uvar}{\var}
      \tosd{\symsub{\var}{\val\sctx}}
      \tm\esub{\uvar}{\val\sctx}$.
  Then by $\ruleESsubR$ and $\ruleESofcsub$:
  \[
      \traD{\tm\esub{\uvar}{\var}}
    = \traD{\tm}\esub{\uvar}{\ofc{\var}}
    \tosd{\symSsub{\var}{(\sha{\traDval{\val})\traD{\sctx}}}}
      \traD{\tm}\esub{\uvar}{\ofc{(\sha{\traDval{\val})\traD{\sctx}}}}
    = \traD{(\tm\esub{\uvar}{\val\sctx})}
  \]
\item
  \ruleEDlsw:
  Let
    $\tm\esub{\var}{\val\sctx}
     \tosd{\symlsw}
     \tm'\esub{\var}{\val\sctx}$
  be derived from
    $\tm \tosd{\symsub{\var}{\val\sctx}} \tm'$.
  By \ih we have that
    $\traD{\tm} \tossmany{\traD{\symsub{\var}{\val\sctx}}} \traD{\tm'}$,
  that is,
    $\traD{\tm} \toss{\symsub{\var}{\sha{\traDval{\val}}\traD{\sctx}}} \traD{\tm'}$.
  Hence:
  \[
    \begin{array}{rlll}
      \traD{(\tm\esub{\var}{\val\sctx})}
    & = &
      \traD{\tm}\esub{\var}{\ofc{\traD{(\val\sctx)}}}
    \\
    & = &
      \traD{\tm}\esub{\var}{\ofc{(\sha{\traDval{\val}}\traD{\sctx})}}
      & \text{as $\traD{\val} = \sha{\traDval{\val}}$ by definition}
    \\
    & \toss{\symSls} &
      \traD{\tm'}\esub{\var}{\ofc{(\sha{\traDval{\val}}\traD{\sctx})}}
      & \text{by $\ruleESls$ since $\traD{\tm} \toss{\symsub{\var}{\sha{\traDval{\val}}\traD{\sctx}}} \traD{\tm'}$}
    \\
    & = &
      \traD{\tm'}\esub{\var}{\ofc{(\traD{\val\sctx})}}
    \\
    & = &
      \traD{(\tm'\esub{\var}{\val\sctx})}
    \end{array}
  \]
\item
  \ruleEDgc:
  Let $\tm\esub{\var}{\tmtwo} \tosd{\symgc} \tm$,
  where $\var\notin\fv{\tm}$.
  Recall that the translation does not create free variables (\rremark{traCNd_fv}),
  so $\var\notin\fv{\traD{\tm}}$.
  Then:
  \[
    \begin{array}{rlll}
      \traD{(\tm\esub{\var}{\tmtwo})}
    & = &
      \traD{\tm}\esub{\var}{\ofc{\traD{\tmtwo}}}
    \\
    & \toss{\symSgc} &
      \traD{\tm}
      & \text{by $\ruleESgc$ since $\var\notin\fv{\traD{\tm}}$}
    \end{array}
  \]
\item
  \ruleEDid:
  Let
    $\var \tosd{\symid{\var}} \var$.
  Then $\traD{\var} = \var \toss{\symSid{\var}} \var = \traD{\var}$
  by $\ruleESid$.
\item
  \ruleEDapp:
  Let
    $\tm\,\tmtwo \tosd{\rulename} \tm'\,\tmtwo$
  be derived from
    $\tm \tosd{\rulename} \tm'$.
  By \ih $\traD{\tm} \tossmany{\traD{\rulename}} \traD{\tm'}$.
  Hence:
  \[
    \begin{array}{rlll}
      \traD{(\tm\,\tmtwo)}
    & = &
      \open{\traD{\tm}}\,\ofc{\traD{\tmtwo}}
    \\
    & \tossmany{\traD{\rulename}} &
      \open{\traD{\tm'}}\,\ofc{\traD{\tmtwo}}
      & \text{by $\ruleESapp$, $\ruleEScopen$ (many times)}
    \\
    & = &
      \traD{(\tm'\,\tmtwo)}
    \end{array}
  \]
\item
  \ruleEDsubL:
  Let
    $\tm\esub{\var}{\tmtwo} \tosd{\rulename} \tm'\esub{\var}{\tmtwo}$
  be derived from
    $\tm \tosd{\rulename} \tm'$,
  where
    $\var \notin \fv{\rulename}$.
  By \ih
    $\traD{\tm} \tossmany{\traD{\rulename}} \traD{\tm'}$.
  Recall that the translation does not create free variables (\rremark{traCNd_fv}),
  so $\var\notin\fv{\traD{\rulename}}$.
  Hence:
  \[
    \begin{array}{rlll}
      \traD{(\tm\esub{\var}{\tmtwo})}
    & = &
      \traD{\tm}\esub{\var}{\ofc{\traD{\tmtwo}}}
    \\
    & \toss{\traD{\rulename}} &
      \traD{\tm'}\esub{\var}{\ofc{\traD{\tmtwo}}}
      & \text{by $\ruleESsubL$ (many times), since $\var\notin\fv{\traD{\rulename}}$}
    \\
    & = &
      \traD{(\tm'\esub{\var}{\tmtwo})}
    \end{array}
  \]
\item
  \ruleEDsubR:
  Let
    $\tm\esub{\var}{\tmtwo} \tosd{\rulename} \tm\esub{\var}{\tmtwo'}$
  be derived from
    $\tm \tosd{\symid{\var}} \tm$
  and
    $\tmtwo \tosd{\rulename} \tmtwo'$.
  By \ih we have that
    $\traD{\tm} \toss{\symSid{\var}} \traD{\tm}$
  and
    $\traD{\tmtwo} \tossmany{\traD{\rulename}} \traD{\tmtwo'}$.
  Note that $\traD{\rulename}$ must be a sequence of the form
  $\sequence{\rulename_1,\hdots,\rulename_n}$,
  so there exist terms $\tmtwo_0,\hdots,\tmtwo_n \in \TermsS$
  such that
    $\traD{\tmtwo}
     = \tmtwo_0 \toss{\rulename_1}
       \tmtwo_1 \toss{\rulename_2}
       \hdots
       \tmtwo_n = \traD{\tmtwo'}$.
  Hence:
  \[
    \begin{array}{rlll}
      \traD{\tm\esub{\var}{\tmtwo}}
    & = &
      \traD{\tm}\esub{\var}{\ofc{\traD{\tmtwo}}}
    \\
    & = &
      \traD{\tm}\esub{\var}{\ofc{\tmtwo_0}}
    \\
    & \toss{\rulename_1} &
      \traD{\tm}\esub{\var}{\ofc{\tmtwo_1}}
      & \text{by $\ruleESsubRofc$}
    \\
    &&
      \vdots
    \\
    & \toss{\rulename_n} &
      \traD{\tm}\esub{\var}{\ofc{\tmtwo_n}}
      & \text{by $\ruleESsubRofc$}
    \\
    & = &
      \traD{\tm}\esub{\var}{\ofc{\traD{\tmtwo'}}}
    \\
    & = &
      \traD{\tm\esub{\var}{\tmtwo'}}
    \end{array}
  \]
\end{enumerate}
\end{proof}

\begin{lemma}[$\CCBNd$ Evaluation -- Inverse Simulation]
Let $\rtm \in \TermsSClumsyNeed$, $\rrulename \in \RulesSClumsyNeed$ and $\tmtwo \in \TermsS$
such that $\rtm \toss{\rrulename} \tmtwo$.
Then $\tmtwo \in \TermsSClumsyNeed$
and $\traDinv{\rtm} \tosdmany{\traDinv{\rulename}} \traDinv{\tmtwo}$.
\end{lemma}
\begin{proof}
By induction on $\rtm$:
\begin{enumerate}
\item
  Variable, $\rtm = \uvar$:
  Then the step can only be derived using the $\ruleESsub$ rule,
  \ie of the form
  $\rtm
   = \uvar
   \toss{\symSsub{\uvar}{(\sha{\tmthree})\sctx}} (\sha{\tmthree})\sctx
   = \tmtwo$.
  Note that $\rrulename = \symSsub{\uvar}{(\sha{\tmthree})\sctx} \in \RulesSClumsyNeed$
  so we know that
  $\tmthree$ is of the form $\lam{\lvar}{\rtmthree\esub{\uvartwo}{\lvar}}$
  and $\sctx \in \SCtxsSClumsyNeed$.
  Hence $\tmtwo = (\sha{\lam{\lvar}{\rtmthree\esub{\uvartwo}{\lvar}}})\sctx \in \TermsSClumsyNeed$
  and:
  \[
    \begin{array}{lll}
      \traDinv{\rtm}
    & = &
      \uvar
    \\
    & \tosd{\symsub{\uvar}{(\lam{\uvartwo}{\traDinv{\rtmthree}})\traDinv{\sctx}}} &
      (\lam{\uvartwo}{\traDinv{\rtmthree}})\traDinv{\sctx}
    \\
    & = &
      \traDinv{((\sha{\lam{\lvar}{\rtmthree\esub{\uvartwo}{\lvar}}})\sctx)}
    \\
    & = &
      \traDinv{\tmtwo}
    \end{array}
  \]
\item
  Translation of an abstraction,
  $\rtm = \sha{\lam{\lvar}{\rtm_1\esub{\uvar}{\lvar}}}$
  with $\lvar\notin\fv{\rtm_1}$:
  this case is impossible, as there are no reduction rules that allow
  deriving a step 
  $\rtm = \sha{\lam{\lvar}{\rtm_1\esub{\uvar}{\lvar}}} \toss{\rrulename} \tmtwo$.
\item
  Request, $\rtm = \open{\rtm_1}$:
  we consider two subcases, depending on whether the step is derived using
  the $\ruleESopen$ or the $\ruleEScopen$ rule:
  \begin{enumerate}
  \item
    $\ruleESopen$:
    Note that, for the $\ruleESopen$ rule to be applicable,
    $\rtm_1$ must be of the form $\rtm_1 = (\sha{\tmthree})\sctx$.
    Moreover, since $\rtm_1 \in \TermsSClumsyNeed$
    we have that $\sctx \in \SCtxsSClumsyNeed$
    and that $\tmthree$ is of the form
    $\tmthree = \lam{\lvar}{\rtmthree_1\esub{\uvar}{\lvar}}$ 
    with $\rtmthree_1 \in \TermsSClumsyNeed$ and $\lvar\notin\fv{\rtmthree_1}$.
    Then the step is of the form
    $\rtm
     = \open{(\sha{\lam{\lvar}{\rtmthree_1\esub{\uvar}{\lvar}}})\sctx}
     \toss{\symSopen}
       (\lam{\lvar}{\rtmthree_1\esub{\uvar}{\lvar}})\sctx
     = \tmtwo$.
    Note that
    $\tmtwo = (\lam{\lvar}{\rtmthree_1\esub{\uvar}{\lvar}})\sctx \in \TermsSClumsyNeed$
    and:
    \[
      \begin{array}{lll}
        \traDinv{\rtm}
      & = &
        \traDinv{\open{(\sha{\lam{\lvar}{\rtmthree_1\esub{\uvar}{\lvar}}})\sctx}}
      \\
      & = &
        (\lam{\uvar}{\traDinv{\rtmthree_1}})\traDinv{\sctx}
      \\ 
      & = &
        \traDinv{(\lam{\lvar}{\rtmthree_1\esub{\uvar}{\lvar}})\sctx}
      \\ 
      & = &
        \traDinv{\tmtwo}
      \end{array}
    \]
  \item
    $\ruleEScopen$:
    Then the step is of the form
    $\rtm
     = \open{\rtm_1}
     \toss{\rrulename} \open{\tmtwo_1}
     = \tmtwo$,
    where $\rtm_1 \toss{\rrulename} \tmtwo_1$.
    By \ih, we have that $\tmtwo_1 \in \TermsSClumsyNeed$
    and $\traDinv{\rtm_1} \tosdmany{\traDinv{\rrulename}} \traDinv{\tmtwo_1}$.
    Note that $\tmtwo = \open{\tmtwo_1} \in \TermsSClumsyNeed$
    and:
    \[
      \begin{array}{lll}
        \traDinv{\tm}
      & = &
        \traDinv{\open{\rtm_1}}
      \\
      & = &
        \traDinv{\rtm_1}
      \\
      & \tosdmany{\traDinv{\rrulename}} &
        \traDinv{\tmtwo_1}
      \\
      & = &
        \traDinv{\open{\tmtwo_1}}
      \\
      & = &
        \traDinv{\tmtwo}
      \end{array}
    \]
  \end{enumerate}
\item
  Translation of an explicit substitution,
  $\rtm = \rtm_1\esub{\uvar}{\ofc{\rtm_2}}$:
  We consider five subcases, depending on the rule used to derive the step:
  \begin{enumerate}
  \item
    $\ruleESls$:
    For the $\ruleESls$ rule to be applicable,
    $\rtm_2$ must be of the form $\rtm_2 = (\sha{\tmthree})\sctx$.
    Moreover, since $\rtm_2 \in \TermsSClumsyNeed$, we know that
    $\sctx \in \SCtxsSClumsyNeed$
    and that $\tmthree$ is of the form
    $\tmthree = \lam{\lvar}{\rtmthree_1\esub{\uvartwo}{\lvar}}$
    where $\rtmthree_1 \in \TermsSClumsyNeed$ and $\lvar\notin\fv{\rtmthree_1}$.
    The step is of the form
    $\rtm
     = \rtm_1\esub{\uvar}{\ofc{\rtm_2}}
     \toss{\symSls}
       \tmtwo_1\esub{\uvar}{\ofc{\rtm_2}}
     = \tmtwo$,
    where
    $\rtm_1
     \toss{\symSsub{\uvar}{(\sha{\lam{\lvar}{\rtmthree_1\esub{\uvartwo}{\lvar}}})\sctx}} 
     \tmtwo_1$.
    Note that
    $\symSsub{\uvar}{(\sha{\lam{\lvar}{\rtmthree_1\esub{\uvartwo}{\lvar}}})\sctx}
     \in \RulesSClumsyNeed$
    so we may apply the \ih to obtain that
    $\tmtwo_1 \in \TermsSClumsyNeed$
    and
    $\traDinv{\rtm_1}
     \tosd{\symsub{\uvar}{(\lam{\lvar}{\traDinv{\rtmthree_1}})\traDinv{\sctx}}} 
     \traDinv{\tmtwo_1}$.
    Hence $\tmtwo = \tmtwo_1\esub{\uvar}{\ofc{\rtm_2}} \in \TermsSClumsyNeed$
    and:
    \[
      \begin{array}{lrll}
        \traDinv{\rtm}
      & = &
        \traDinv{(\rtm_1\esub{\uvar}{\ofc{(\sha{\lam{\lvar}{\rtmthree_1\esub{\uvartwo}{\lvar}}})\sctx}})}
      \\
      & = &
        \traDinv{\rtm_1}
          \esub{\uvar}{(\lam{\uvartwo}{\traDinv{\rtmthree_1}})\traDinv{\sctx}}
      \\
      & \tosd{\symlsw} &
        \traDinv{\tmtwo_1}
          \esub{\uvar}{(\lam{\uvartwo}{\traDinv{\rtmthree_1}})\traDinv{\sctx}}
        & \text{by $\ruleEDlsw$}
      \\
      & = &
        \traDinv{(\tmtwo_1\esub{\uvar}{\ofc{\rtm_2}})}
      \\
      & = &
        \traDinv{\tmtwo}
      \end{array}
    \]
  \item
    $\ruleESgc$:
    The step is of the form
    $\rtm
     = \rtm_1\esub{\uvar}{\ofc{\rtm_2}}
     \toSgc \rtm_1
     = \tmtwo$,
    where $\uvar\notin\fv{\rtm_1}$.
    Then $\tmtwo = \rtm_1 \in \TermsSClumsyNeed$.
    Moreover, since $\fv{\rtm_1} = \fv{\traDinv{\rtm_1}}$
    by \rremark{traCNdinv_fv}, we have that $\uvar\notin\fv{\rtm_1}$, so:
    \[
      \begin{array}{lrl}
        \traDinv{\rtm}
      & = &
        \traDinv{(\rtm_1\esub{\uvar}{\ofc{\rtm_2}})}
      \\
      & \tosd{\symgc} &
        \traDinv{\rtm_1}
      \\
      & = &
        \traDinv{\tmtwo}
      \end{array}
    \]
  \item
    $\ruleESsubL$:
    The step is of the form
    $\rtm
     = \rtm_1\esub{\uvar}{\ofc{\rtm_2}}
     \toss{\rrulename} \tmtwo_1\esub{\uvar}{\ofc{\rtm_2}}
     = \tmtwo$,
    derived from an internal step $\rtm_1 \toss{\rrulename} \tmtwo_1$
    where $\uvar\notin\fv{\rrulename}$.
    By \ih, we have that $\tmtwo_1 \in \TermsSClumsyNeed$
    and $\traDinv{\rtm_1} \tosdmany{\traDinv{\rrulename}} \traDinv{\tmtwo_1}$,
    so $\tmtwo = \tmtwo_1\esub{\uvar}{\ofc{\rtm_2}} \in \TermsSClumsyNeed$
    and:
    \[
      \begin{array}{lrll}
        \traDinv{\rtm}
      & = &
        \traDinv{\rtm_1\esub{\uvar}{\ofc{\rtm_2}}}
      \\
      & = &
        \traDinv{\rtm_1}\esub{\uvar}{\traDinv{\rtm_2}}
      \\
      & \tosdmany{\traDinv{\rrulename}} &
        \traDinv{\tmtwo_1}\esub{\uvar}{\traDinv{\rtm_2}}
        & \text{by $\ruleEDsubL$ (many times)}
      \\
      & = &
        \traDinv{(\tmtwo_1\esub{\uvar}{\ofc{\rtm_2}})}
      \\
      & = &
        \traDinv{\tmtwo}
      \end{array}
    \]
  \item
    $\ruleESsubR$:
    The step is of the form
    $\rtm
     = \rtm_1\esub{\uvar}{\ofc{\rtm_2}}
     \toss{\rrulename} \rtm_1\esub{\uvar}{\tmtwo_2}
     = \tmtwo$,
    and derived from an internal step $\ofc{\rtm_2} \toss{\rrulename} \tmtwo_2$.
    The internal step can only be derived using the $\ruleESofcsub$ rule,
    namely $\rtm_2$ must be a variable ($\rtm_2 = \uvartwo$)
    and the internal step is of the form
    $\ofc{\uvartwo}
     \toss{\symSsub{\uvartwo}{(\sha{\lam{\lvar}{\rtmthree\esub{\uvarthree}{\lvar}}})\sctx}}
       \ofc{(\sha{\lam{\lvar}{\rtmthree\esub{\uvarthree}{\lvar}}})\sctx}
     = \tmtwo_2$,
     where
     $\rtmthree \in \TermsSClumsyNeed$,
     $\lvar\notin\fv{\rtmthree}$, and $\sctx \in \SCtxsSClumsyNeed$.
     Then
     $\tmtwo
      = \rtm_1\esub{\uvar}{\ofc{(\sha{\lam{\lvar}{\rtmthree\esub{\uvarthree}{\lvar}}})\sctx}}
      \in \TermsSClumsyNeed$
     and:
     \[
       \begin{array}{lrll}
         \traDinv{\rtm}
       & = &
         \traDinv{(\rtm_1\esub{\uvar}{\ofc{\uvartwo}})}
       \\
       & = &
         \traDinv{\rtm_1}\esub{\uvar}{\uvartwo}
       \\
       & \tosd{\symsub{\uvartwo}{(\lam{\uvarthree}{\traDinv{\rtmthree}})\traDinv{\sctx}}} &
         \traDinv{\rtm_1}\esub{\uvar}{(\lam{\uvarthree}{\traDinv{\rtmthree}})\traDinv{\sctx}}
         & \text{by \ruleEDsubES}
       \\
       & = &
         \traDinv{(\rtm_1\esub{\uvar}{\ofc{(\sha{\lam{\lvar}{\rtmthree\esub{\uvarthree}{\lvar}}})\sctx}})}
       \\
       & = &
         \traDinv{\tmtwo}
       \end{array}
     \]
  \item
    $\ruleESsubRofc$:
    The step is of the form
    $\rtm
     = \rtm_1\esub{\uvar}{\ofc{\rtm_2}}
     \toss{\rrulename} \rtm_1\esub{\uvar}{\ofc{\tmtwo_2}}
     = \tmtwo$,
    where $\rtm_1 \toss{\symSid{\uvar}} \rtm_1$
    and $\rtm_2 \toss{\rrulename} \tmtwo_2$.
    By \ih on the first premise,
    we have that $\traDinv{\rtm_1} \tosd{\symid{\uvar}} \traDinv{\rtm_1}$.
    By \ih on the second premise,
    we have that $\tmtwo_2 \in \TermsSClumsyNeed$
    and $\traDinv{\rtm_2} \tosdmany{\traDinv{\rrulename}} \traDinv{\tmtwo_2}$.
    Then $\tmtwo = \rtm_1\esub{\uvar}{\ofc{\tmtwo_2}} \in \TermsSClumsyNeed$
    and:
    \[
      \begin{array}{lrll}
        \traDinv{\rtm}
      & = &
        \traDinv{(\rtm_1\esub{\uvar}{\ofc{\rtm_2}})}
      \\
      & = &
        \traDinv{\rtm_1}\esub{\uvar}{\traDinv{\rtm_2}}
      \\
      & \tosdmany{\traDinv{\rrulename}} &
        \traDinv{\rtm_1}\esub{\uvar}{\traDinv{\tmtwo_2}}
        & \text{by $\ruleEDsubR$ (many times)}
      \\
      & = &
        \traDinv{(\rtm_1\esub{\uvar}{\ofc{\tmtwo_2}})}
      \\
      & = &
        \traDinv{\tmtwo}
      \end{array}
    \]
  \end{enumerate}
\item
  Abstraction,
  $\rtm = \lam{\lvar}{\rtm_1\esub{\uvar}{\lvar}}$
  with $\lvar\notin\fv{\rtm_1}$:
  this case is impossible, as there are no reduction rules that allow
  deriving a step 
  $\rtm = \lam{\lvar}{\rtm_1\esub{\uvar}{\lvar}} \toss{\rrulename} \tmtwo$.
\item
  Translation of an application,
  $\rtm = \rtm_1\,\ofc{\rtm_2}$:
  We consider two subcases, depending on whether
  the step is derived from $\ruleESdb$ or $\ruleESapp$:
  \begin{enumerate}
  \item
    $\ruleESdb$:
    For the $\ruleESdb$ rule to be applicable,
    $\rtm_1$ must be of the form $(\lam{\lvar}{\tmthree})\sctx$.
    Moreover, since $\rtm_1 \in \TermsSClumsyNeed$,
    we have that $\sctx \in \SCtxsSClumsyNeed$
    and that $\tmthree$ is of the form
    $\tmthree = \rtmthree_1\esub{\uvar}{\lvar}$,
    where $\rtmthree_1 \in \TermsSClumsyNeed$ and $\lvar\notin\fv{\rtmthree_1}$.
    The step is of the form
    $\rtm
     = (\lam{\lvar}{\rtmthree_1\esub{\uvar}{\lvar}})\sctx\,\ofc{\rtm_2}
     \toss{\symSdb}
       \rtmthree_1\esub{\uvar}{\ofc{\rtm_2}}\sctx
     = \tmtwo$.
    So $\tmtwo = \rtmthree_1\esub{\uvar}{\ofc{\rtm_2}}\sctx \in \TermsSClumsyNeed$
    and:
    \[
      \begin{array}{lrl}
        \traDinv{\rtm}
      & = &
        \traDinv{((\lam{\lvar}{\rtmthree_1\esub{\uvar}{\lvar}})\sctx\,\ofc{\rtm_2})}
      \\
      & = &
        (\lam{\uvar}{\traDinv{\rtmthree_1}})\traDinv{\sctx}\,\traDinv{\rtm_2}
      \\
      & \tosd{\symdb} &
        \traDinv{\rtmthree_1}\esub{\uvar}{\traDinv{\rtm_2}}\traDinv{\sctx}
      \\
      & = &
        \traDinv{((\rtmthree_1\esub{\uvar}{\ofc{\rtm_2}}\sctx)}
      \\
      & = &
        \traDinv{\tmtwo}
      \end{array}
    \]
  \item
    $\ruleESapp$:
    The step is of the form
    $\rtm
     = \rtm_1\,\ofc{\rtm_2}
     \toss{\rrulename} \tmtwo_1\,\ofc{\rtm_2}
     = \tmtwo$,
    and derived from an internal step $\rtm_1 \toss{\rrulename} \tmtwo_1$.
    By \ih, we have that $\tmtwo_1 \in \TermsSClumsyNeed$
    and $\traDinv{\tm_1} \tosdmany{\traDinv{\rrulename}} \traDinv{\tmtwo_1}$.
    So $\tmtwo = \tmtwo_1\,\ofc{\rtm_2} \in \TermsSClumsyNeed$
    and:
    \[
      \begin{array}{lrll}
        \traDinv{\rtm}
      & = &
        \traDinv{(\rtm_1\,\ofc{\rtm_2})}
      \\
      & = &
        \traDinv{\rtm_1}\,\traDinv{\rtm_2}
      \\
      & \tosdmany{\traDinv{\rrulename}} &
        \traDinv{\tmtwo_1}\,\traDinv{\rtm_2}
        & \text{by $\ruleEDapp$ (many times)}
      \\
      & = &
        \traDinv{(\tmtwo_1\,\rtm_2)}
      \\
      & = &
        \traDinv{\tmtwo}
      \end{array}
    \]
  \end{enumerate}
\end{enumerate}
\end{proof}

\begin{ifLongAppendix}
  \section{Appendix: The Bang Calculus}

  \subsection{A Simplified Presentation of the Bang Calculus}
  \lsec{appendix:bang}
  
A binary relation  $\derEq\subseteq \TermsBangGM\times \TermsBang$,
called \emph{dereliction unfolding} is defined inductively as:
\[
\begin{array}{c}
    \indrule{}
    {}
    {\var \derEq \var}
  \,\,
    \indrule{}
    {\tm \derEq \tmtwo}
    {\lam{\var}{\tm} \derEq \lam{\var}{\tmtwo}}
  \,\,
    \indrule{}
    { \tm_1 \derEq \tmtwo_1
    \quad
    \tm_2 \derEq \tmtwo_2}
    {\tm_1\, \tm_2 \derEq \tmtwo_1\,\tmtwo_2}
  \,\,
    \indrule{}
    { \tm_1 \derEq \tmtwo_1
    \quad
    \tm_2 \derEq \tmtwo_2
    }
    {\tm_1\esub{\var}{\tm_2} \derEq \tmtwo_1\esub{\var}{\tmtwo_2}}
  \\
  \\
  \,\,
    \indrule{}
    { \tm \derEq \tmtwo}
    {\ofc{\tm} \derEq \ofc{\tmtwo}}
  \,\,
    \indrule{}
    { \tm \derEq \tmtwo}
    {\der{\tm}\derEq \var\esub{\var}{\tmtwo}}
  \,\,
    \indrule{}
    { \tm \derEq \tmtwo
    \quad
    \var\notin\fv{\tmtwo}
    }
    {\tm\derEq \tmtwo\esub{\var}{\ofc{\tmthree}}}
\end{array}
\]
The last rule
is required since a dereliction step
$\der{(\ofc{\tm})\sctx} \toderB \tm\sctx$
is simulated as a $\symlsB$ step
$\var\esub{\var}{(\ofc{\tm})\sctx} \tolsB \tm\esub{\var}{(\ofc{\tm})}\sctx$.

\begin{remark}
\lremark{derEq_is_reflexive_on_simplified_bang_terms}
Note that $\tm\in\TermsBang$ implies $\tm\derEq\tm$.
\end{remark}

\begin{lemma}[Simulation of dereliction
              \proofnote{Proofs on Sec.~\ref{section:appendix:bang}}]
\llem{B_BGM_simulate_each_other}
\quad
\begin{enumerate}
\item
  If $\tm \derEq \tmtwo$ and $\tm \toBangGM \tm'$,
  there exists $\tmtwo'$
  such that $\tmtwo \toBang^* \tmtwo'$ and $\tm' \derEq \tmtwo'$.
  Moreover, if the step $\tm \toBangGM \tm'$ is not a $\symgcB$ step,
  the reduction $\tmtwo \toBang^* \tmtwo'$ consists of exactly one step.
\item
  If $\tm \derEq \tmtwo$ and $\tmtwo \toBangGM \tmtwo'$,
  there exists $\tm'$
  such that $\tm \toBang^= \tm'$ and $\tm' \derEq \tmtwo'$.
\end{enumerate}
\end{lemma}
In the second part of the lemma, the only situation where a step is simulated
by an empty $\symBangGM$-step is when an $\symlsB$-step is made inside the body
of a garbage substitution. For example, $\var\esub{\varthree}{\ofc{\tmthree}}
\derEq \var\esub{\vartwo}{\ofc{\varthree}}\esub{\varthree}{\ofc{\tmthree}}$.
The preceding lemma has the following consequence:

\restate{proposition}{prop:bang_simplified_simulation}{Simplified $\Bang$ simulation}{
$\toBangGM$ and $\toBang$ simulate each other.
}
\begin{proof}
To show that $\toBang$ simulates $\toBangGM$, consider
a reduction sequence $\tm_1 \toBangGM \tm_2 \toBangGM \ldots \tm_{n-1}\toBangGM\tm_n$,
let $\tmtwo_1$ be the smallest term such that $\tm_1 \derEq \tmtwo_1$
and proceed by induction on $n$, resorting to \rlem{B_BGM_simulate_each_other}(1).
To show that $\toBangGM$ simulates $\toBang$,
we resort to \rremark{derEq_is_reflexive_on_simplified_bang_terms}
and \rlem{B_BGM_simulate_each_other}(2).
The $\toBangGM$ step can be taken to never be empty since
${\toBang}\subseteq{\toBangGM}$ and the derivation of
\rremark{derEq_is_reflexive_on_simplified_bang_terms} does not
erase substitutions.
\end{proof}

  \subsection{Embedding the Bang Calculus}
  \lsec{appendix:bang}
  
\begin{remark}
  \lremark{derEq:fv}
  Suppose $\tm\derEq \tmtwo$. Then:
  \begin{enumerate}
  \item\label{rem:derEq:fv:general} $\fv{\tm}\subseteq\fv{\tmtwo}$, and;

  \item\label{rem:derEq:fv:equals} if the rule $\indrulename{\ruleDerEqGc}$ is not used in the derivation, then $\fv{\tm}=\fv{\tmtwo}$.
  \end{enumerate}
  
\end{remark}

\begin{lemma}
\llem{derEq:properties}
  \begin{enumerate}

   \item\label{derEq:properties:lam:prelim} $\lam{\var}{\tm}\derEq \tmtwo$ implies $\tmtwo=(\lam{\var}{\tmtwo_1})\sctx$ with $\tm\derEq\tmtwo_1$ and $\dom{\sctx}\cap\fv{\lam{\var}{\tmtwo_1}}=\emptyset$.
    
  \item\label{derEq:properties:lam} $(\lam{\var}{\tm})\sctx\derEq \tmtwo$ implies $\tmtwo=(\lam{\var}{\tmtwo_1})\sctx'$ with $\tm\derEq\tmtwo_1$ and $\sctx\derEqSctx{\varset} \sctx'$ and $\varset\cap\fv{\lam{\var}{\tmtwo_1}}=\emptyset$.

  \item\label{derEq:properties:gctx}  $\off{\gctx}{\var}\derEq \tmtwo$
    implies $\tmtwo=\off{\gctx'}{\var}$ and
    $\gctx\derEqGctx{\varset}\gctx'$ and $x\notin\varset$.

  \item\label{derEq:properties:ofc}  $(\ofc{\tm})\sctx\derEq \tmtwo$ implies $\tmtwo=(\ofc{\tmtwo_1})\sctx'$ and $\tm\derEq\tmtwo_1$ and $\sctx\derEqSctx{\varset}\sctx'$ and $\varset\cap\fv{\ofc{\tmtwo_1}}=\emptyset$.



      \item\label{derEq:properties:gc} $\tm\derEq \tmtwo$ and
    $\var\notin\fv{\tm}$ and $\var\in\fv{\tmtwo}$ implies
    $\tmtwo=\off{\gctx}{\tmtwo_1\esub{\vartwo}{\ofc{\tmtwo_2}}}$
    and $\vartwo\notin\fv{\tmtwo_1}$ and
    $\var\in\fv{\ofc{\tmtwo_2}}$ and $\tm\derEq\off{\gctx}{\tmtwo_1}$.

       \item\label{derEq:properties:gc:general} $\tm\derEq \tmtwo$ and
    $\var\notin\fv{\tm}$ and $\var\in\fv{\tmtwo}$ implies
    there exists $\tmtwo'$ such that $\tmtwo\togcB^*\tmtwo'$ and
    $\tm\derEq \tmtwo'$ and  $\var\notin\fv{\tmtwo'}$.


  \item\label{derEq:properties:replacement:gctx} $\tm\derEq \tmtwo$ and
    $\gctx \derEqGctx{\varset} \gctxtwo$ and $\varset\cap\fv{\tmtwo}=\emptyset$ implies $\off{\gctx}{\tm}\derEq \off{\gctxtwo}{\tmtwo}$.

      \item\label{derEq:properties:replacement:sctx} $\tm\derEq \tmtwo$ and
    $\sctx\derEqSctx{\varset} \sctxtwo$ and $\varset\cap\fv{\tmtwo}=\emptyset$ implies $\of{\sctx}{\tm}\derEq \of{\sctxtwo}{\tmtwo}$.

  \end{enumerate}
\end{lemma}

\begin{proof}
  Item~\ref{derEq:properties:lam:prelim}. By induction on the derivation of
  $\lam{\var}{\tm}\derEq \tmtwo$. Notice that only two rules apply.
  \begin{enumerate} 
  \item $\tmtwo=\lam{\var}{\tmtwo_1}$ and the derivation ends in:
    \[\indrule{\ruleDerEqLam}
   {\tm \derEq \tmtwo_1}
   {\lam{\var}{\tm} \derEq \lam{\var}{\tmtwo_1}}
   \]
  We set $\sctx'\eqdef\ctxhole$ and conclude.
   
 \item Then $\tmtwo=\tmtwo_1\esub{\vartwo}{\ofc{\tmtwo_2}}$ and the derivation ends in:
   \[   \indrule{\ruleDerEqGc}
   { \lam{\var}{\tm}\derEq \tmtwo_1
   \quad
   \vartwo\notin\fv{\tmtwo_1}
   }
   {\lam{\var}{\tm}\derEq \tmtwo_1\esub{\vartwo}{\ofc{\tmtwo_2}}}
 \]
    From the \ih, $\tmtwo_1=(\lam{\var}{\tmtwo_{11}})\sctx'$ with $\tm\derEq\tmtwo_{11}$ and $\dom{\sctx'}\cap\fv{\lam{\var}{\tmtwo_{11}}}=\emptyset$. We set $\sctx\eqdef \sctx'\esub{\vartwo}{\ofc{\tmtwo_2}}$ and conclude.

  \end{enumerate}
  
  Item~\ref{derEq:properties:lam}. By induction on $\sctx$ using item~\ref{derEq:properties:lam:prelim} for the base case. For the inductive case, only two subcases arise (the derivation ends in $\ruleDerEqESub$ or $\ruleDerEqGc$); in each we use the \ih.

  Item~\ref{derEq:properties:gctx}. By induction on the derivation of $\off{\gctx}{\var}\derEq \tmtwo$.

  Item~\ref{derEq:properties:ofc}. By induction on the derivation of $\tm\derEq \tmtwo$.
    
  Item~\ref{derEq:properties:gc}. By induction on $\tm\derEq \tmtwo$.
 \begin{itemize}

\item $\tm=\var \derEq \var=\tmtwo$. The result holds vacuously since $\var\notin\fv{\tm}$ and $\var\in\fv{\tmtwo}$ is not possible.
  
\item $\tm=\lam{\varthree}{\tm_1} \derEq \lam{\varthree}{\tmtwo_1}=\tmtwo$
  follows from  $\tm_1 \derEq \tmtwo_1$   and  $\var\notin\fv{\lam{\varthree}{\tm_1}}$
  and $\var\in\fv{\lam{\varthree}{\tmtwo_1}}$ and,  w.l.o.g., we assume $\var\neq\varthree$. Note that
    $\var\notin\fv{\tm_1}$ and  $\var\in\fv{\tmtwo_1}$. By the \ih, 
    $\tmtwo_1=\off{\gctx_1}{\tmtwo_{11}\esub{\vartwo}{\ofc{\tmtwo_{12}}}}$
    and $\vartwo\notin\fv{\tmtwo_{11}}$ and
    $\var\in\fv{\ofc{\tmtwo_{12}}}$ and
    $\tm_1\derEq\off{\gctx_1}{\tmtwo_{11}}$. We set $\gctx\eqdef
    \lam{\varthree}{\gctx_1}$. Note that
     $\tmtwo=\off{\gctx}{\tmtwo_{11}\esub{\vartwo}{\ofc{\tmtwo_{12}}}}$
    and $\vartwo\notin\fv{\tmtwo_{11}}$ and
    $\var\in\fv{\ofc{\tmtwo_{12}}}$ and
    $\tm\derEq\off{\gctx}{\tmtwo_{11}}$.  The latter follows from
    $\tm_1\derEq\off{\gctx_1}{\tmtwo_{11}}$  and $\indrulename{\ruleDerEqLam}$.
    Thus we conclude.

\item $\tm=\tm_1\, \tm_2 \derEq \tmtwo_1\,\tmtwo_2=\tmtwo$ follows
  from $\tm_1 \derEq \tmtwo_1$ and $\tm_2 \derEq \tmtwo_2$ and
    $\var\notin\fv{\tm}$ and $\var\in\fv{\tmtwo}$. Note that
    $\var\notin\fv{\tm_1}$ and   $\var\notin\fv{\tm_2}$. Also, either
    $\var\in\fv{\tmtwo_1}$ or $\var\in\fv{\tmtwo_2}$. We consider three
    cases:
    \begin{enumerate}
    \item $\var\in\fv{\tmtwo_1}$ and $\var\notin\fv{\tmtwo_2}$. By the \ih,
          $\tmtwo_1=\off{\gctx_1}{\tmtwo_{11}\esub{\vartwo}{\ofc{\tmtwo_{12}}}}$
    and $\vartwo\notin\fv{\tmtwo_{11}}$ and
    $\var\in\fv{\ofc{\tmtwo_{12}}}$ and
    $\tm_1\derEq\off{\gctx_1}{\tmtwo_{11}}$. We set $\gctx\eqdef
    \gctx_1\,\tmtwo_2$. Note that
     $\tmtwo=\off{\gctx}{\tmtwo_{11}\esub{\vartwo}{\ofc{\tmtwo_{12}}}}$
    and $\vartwo\notin\fv{\tmtwo_{11}}$ and
    $\var\in\fv{\ofc{\tmtwo_{12}}}$ and
    $\tm\derEq\off{\gctx}{\tmtwo_{11}}$. The latter follows from
    $\tm_1\derEq\off{\gctx_1}{\tmtwo_{11}}$ and $\tm_2 \derEq \tmtwo_2$  and $\indrulename{\ruleDerEqApp}$.
    Thus we conclude
    
    \item $\var\notin\fv{\tmtwo_1}$ and $\var\in\fv{\tmtwo_2}$. Similar to the previous case.

    \item $\var\in\fv{\tmtwo_1}$ and $\var\in\fv{\tmtwo_2}$. Similar to the previous case but using the \ih twice.

    \end{enumerate}

\item $\tm=\tm_1\esub{\varthree}{\tm_2} \derEq
  \tmtwo_1\esub{\varthree}{\tmtwo_2}$ follows from $\tm_1 \derEq \tmtwo_1$
  and $\tm_2 \derEq \tmtwo_2$ and
  $\var\notin\fv{\tm}$ and $\var\in\fv{\tmtwo}$ and,  w.l.o.g., we assume $\var\neq\varthree$. Note that
    $\var\notin\fv{\tm_1}$ and   $\var\notin\fv{\tm_2}$. Also, either
    $\var\in\fv{\tmtwo_1}$ or $\var\in\fv{\tmtwo_2}$. We consider two
    cases:
    \begin{enumerate}
    \item $\var\in\fv{\tmtwo_1}$ and $\var\notin\fv{\tmtwo_2}$. By the \ih,
          $\tmtwo_1=\off{\gctx_1}{\tmtwo_{11}\esub{\vartwo}{\ofc{\tmtwo_{12}}}}$
    and $\vartwo\notin\fv{\tmtwo_{11}}$ and
    $\var\in\fv{\ofc{\tmtwo_{12}}}$ and
    $\tm_1\derEq\off{\gctx_1}{\tmtwo_{11}}$. We set $\gctx\eqdef
    \gctx_1\esub{\varthree}{\tmtwo_2}$. Note that
     $\tmtwo=\off{\gctx}{\tmtwo_{11}\esub{\vartwo}{\ofc{\tmtwo_{12}}}}$
    and $\vartwo\notin\fv{\tmtwo_{11}}$ and
    $\var\in\fv{\ofc{\tmtwo_{12}}}$ and
    $\tm\derEq\off{\gctx}{\tmtwo_1}$. The latter follows from
    $\tm_1\derEq\off{\gctx_1}{\tmtwo_{11}}$ and  $\tm_2 \derEq \tmtwo_2$  and $\indrulename{\ruleDerEqApp}$.
    Thus we conclude
    
    \item $\var\notin\fv{\tmtwo_1}$ and $\var\in\fv{\tmtwo_2}$. Similar to the previous case.
    \item $\var\in\fv{\tmtwo_1}$ and $\var\in\fv{\tmtwo_2}$. Similar to the previous case, but using the \ih twice.
        
    \end{enumerate}

\item $\tm=\ofc{\tm_1} \derEq \ofc{\tmtwo_1}=\tmtwo$ follows from
  $\tm_1 \derEq \tmtwo_1$ and
    $\var\notin\fv{\tm}$ and $\var\in\fv{\tmtwo}$. Note that
    $\var\notin\fv{\tm_1}$ and
    $\var\in\fv{\tmtwo_1}$. By the \ih,
          $\tmtwo_1=\off{\gctx_1}{\tmtwo_{11}\esub{\vartwo}{\ofc{\tmtwo_{12}}}}$
    and $\vartwo\notin\fv{\tmtwo_{11}}$ and
    $\var\in\fv{\ofc{\tmtwo_{12}}}$ and
    $\tm_1\derEq\off{\gctx_1}{\tmtwo_{11}}$. We set $\gctx\eqdef
    \ofc{\gctx_1}$. Note that
     $\tmtwo=\off{\gctx}{\tmtwo_{11}\esub{\vartwo}{\ofc{\tmtwo_{12}}}}$
    and $\vartwo\notin\fv{\tmtwo_{11}}$ and
    $\var\in\fv{\ofc{\tmtwo_{12}}}$ and
    $\tm\derEq\off{\gctx}{\tmtwo_1}$. The latter follows from
    $\tm_1\derEq\off{\gctx_1'}{\tmtwo_{11}}$  and $\indrulename{\ruleDerEqOfc}$.
    Thus we conclude.

    \item $\tm=\der{\tm_1}\derEq \varthree\esub{\varthree}{\tmtwo_1}=\tmtwo$
      follows from $\tm_1 \derEq \tmtwo_1$ and
    $\var\notin\fv{\tm}$ and $\var\in\fv{\tmtwo}$. Note that
    $\var\notin\fv{\tm_1}$ and 
    $\var\in\fv{\tmtwo_1}$. By the \ih,
          $\tmtwo_1=\off{\gctx_1'}{\tmtwo_{11}\esub{\vartwo}{\ofc{\tmtwo_{12}}}}$
    and $\vartwo\notin\fv{\tmtwo_{11}}$ and
    $\var\in\fv{\ofc{\tmtwo_{12}}}$ and
    $\tm_1\derEq\off{\gctx_1'}{\tmtwo_{11}}$. We set $\gctx'\eqdef
    \varthree\esub{\varthree}{\gctx_1'}$. Note that
     $\tmtwo=\off{\gctx'}{\tmtwo_{11}\esub{\vartwo}{\ofc{\tmtwo_{12}}}}$
    and $\vartwo\notin\fv{\tmtwo_{11}}$ and
    $\var\in\fv{\ofc{\tmtwo_{12}}}$ and
    $\tm\derEq\off{\gctx'}{\tmtwo_1}$. The latter follows from
    $\tm_1\derEq\off{\gctx_1'}{\tmtwo_{11}}$  and $\indrulename{\ruleDerEqDer}$.
    Thus we conclude.

    \item $\tm\derEq \tmtwo_1\esub{\varthree}{\ofc{\tmtwo_2}}=\tmtwo$ follows from $\tm \derEq \tmtwo_1$ and $
   \varthree\notin\fv{\tmtwo_1}$ and
    $\var\notin\fv{\tm}$ and $\var\in\fv{\tmtwo}$. Note that
   either
    $\var\in\fv{\tmtwo_1}$ or $\var\in\fv{\ofc{\tmtwo_2}}$. We consider two
    cases:
    \begin{enumerate}
    \item $\var\in\fv{\tmtwo_1}$. By the \ih,
          $\tmtwo_1=\off{\gctx_1}{\tmtwo_{11}\esub{\vartwo}{\ofc{\tmtwo_{12}}}}$
    and $\vartwo\notin\fv{\tmtwo_{11}}$ and
    $\var\in\fv{\ofc{\tmtwo_{12}}}$ and
    $\tm_1\derEq\off{\gctx_1}{\tmtwo_{11}}$. We set $\gctx\eqdef
    \gctx_1\esub{\varthree}{\ofc{\tmtwo_2}}$. Note that
     $\tmtwo=\off{\gctx}{\tmtwo_{11}\esub{\vartwo}{\ofc{\tmtwo_{12}}}}$
    and $\vartwo\notin\fv{\tmtwo_{11}}$ and
    $\var\in\fv{\ofc{\tmtwo_{12}}}$ and
    $\tm\derEq\off{\gctx}{\tmtwo_1}$. The latter follows from
    $\tm_1\derEq\off{\gctx_1}{\tmtwo_{11}}$  and $\indrulename{\ruleDerEqGc}$.
    Thus we conclude
    
    \item $\var\in\fv{\tmtwo_2}$. We set $\gctx\eqdef \ctxhole$ and conclude.
        
    \end{enumerate}
   
 \end{itemize}

 Item~\ref{derEq:properties:gc:general}. By induction on the size $n$ of $\tmtwo$. If $n=1$, then the result holds trivially since by the derivation must end in $\indrulename{\ruleDerEqVar}$ and hence  $\var\notin\fv{\tm}$ and $\var\in\fv{\tmtwo}$ is not possible. Suppose $n>0$. Then from item~\ref{derEq:properties:gc}, 
    $\tmtwo=\off{\gctx}{\tmtwo_{11}\esub{\vartwo}{\ofc{\tmtwo_{12}}}}$
    and $\vartwo\notin\fv{\tmtwo_{11}}$ and
    $\var\in\fv{\ofc{\tmtwo_{12}}}$ and $\tm\derEq\off{\gctx}{\tmtwo_{11}}=\tmtwo_1$. Note that  $\tmtwo=\off{\gctx}{\tmtwo_{11}\esub{\vartwo}{\ofc{\tmtwo_{12}}}}\togcB \off{\gctx}{\tmtwo_{11}}$.  If $\var\notin\fv{\tmtwo_1}$, then we conclude. Otherwise, we can apply the \ih on  $\tm\derEq\off{\gctx}{\tmtwo_{11}}$ and conclude from that.

    Item~\ref{derEq:properties:replacement:gctx}. By induction on the derivation of $\gctx \derEqGctx{\varset} \gctxtwo$.
    
Item~\ref{derEq:properties:replacement:sctx}. By induction on the derivation of $\sctx \derEqGctx{\varset} \sctxtwo$.
\end{proof}

\begin{lemma}[Forwards simulation of dereliction]
\llem{B_simulates_BGM}
  \begin{equation}
  \begin{array}{cc}
   \begin{tikzcd}[column sep=0.2em,ampersand replacement=\&]
\tm \arrow[d, "\step"']
  \& \derEq
  \& \tmtwo \arrow[d,dashed, "\symBang"] \\ 
\tm'  
  \& \derEq
  \& \tmtwo'
\end{tikzcd}
    &
\begin{tikzcd}[column sep=0.2em,ampersand replacement=\&]
\tm \arrow[d, "\symgcB"']
  \& \derEq
  \& \tmtwo \arrow[d, twoheadrightarrow, dashed, "\symgcB"] \\ 
\tm'  
  \& \derEq
  \& \tmtwo'
\end{tikzcd}
\end{array}
\end{equation}
where $\step\in\{\symdb, \symlsB,\symderB\}$.
\end{lemma}
\begin{ifShortAppendix}
  \begin{proof}
  Both items are proved by induction on the derivation of  $\tm\derEq \tmtwo$.
  See the extended version~\cite{mells_long} for the detailed proof.
  \end{proof}
\end{ifShortAppendix}
\begin{ifLongAppendix}
  \begin{proof}\label{B_simulates_BGM:proof}  
  We prove both items by induction on the derivation of  $\tm\derEq \tmtwo$.
  \begin{enumerate}

  \item  $\tm\derEq \tmtwo$ is $\var \derEq \var$. Both items are immediate since there are no $\toBangGM$ steps from $\var$.

  \item $\tm\derEq \tmtwo$ is  $\lam{\var}{\tm_1} \derEq \lam{\var}{\tmtwo_1}$ and follows from $\tm_1 \derEq \tmtwo_1$.
    \begin{itemize}
    \item Item 1. The reduction must be internal: $\lam{\var}{\tm_1} \tonogcB\lam{\var}{\tm_1'}=\tm'$ follows from $\tm_1 \tonogcB \tm_1'$. From the \ih there exists $\tmtwo_1'$ such that $\tmtwo_1 \toBang  \tmtwo_1'$ and $\tm_1'\derEq \tmtwo_1'$.  Then $\lam{\var}{\tmtwo_1}\toBang \lam{\var}{\tmtwo_1'}$ and moreover $\lam{\var}{\tm_1'}\derEq \lam{\var}{\tmtwo_1'}$. Thus we conclude by setting $\tmtwo'$ to be $\lam{\var}{\tmtwo_1'}$.

      \item Item 2. Similar to Item 1.
    \end{itemize}

         \item $\tm\derEq \tmtwo$ is $\tm_1\, \tm_2 \derEq \tmtwo_1\,\tmtwo_2$ and follows from $\tm_1 \derEq \tmtwo_1$ and $\tm_2 \derEq \tmtwo_2$
    \begin{itemize}
    \item Item 1. There are three cases.

      \begin{enumerate}
      \item The reduction is internal to $\tm_1$: $\tm=\tm_1 \,\tm_2\tonogcB\tm_1' \tm_2=\tm'$ follows from $\tm_1 \tonogcB \tm_1'$. From the \ih there exists $\tmtwo_1'$ such that $\tmtwo_1 \toBang  \tmtwo_1'$ and $\tm_1'\derEq \tmtwo_1'$.  Then $\tmtwo_1\,\tmtwo_2\toBang \tmtwo_1'\,\tmtwo_2$ and moreover $\tm_1'\,\tm_2\derEq \tmtwo_1'\,\tmtwo_2$. Thus we conclude by setting $\tmtwo'$ to be $\tmtwo_1'\,\tmtwo_2$.
        
      \item The reduction is internal to $\tm_2$: $\tm=\tm_1 \,\tm_2\tonogcB\tm_1 \tm_2'=\tm'$ follows from $\tm_2 \tonogcB \tm_2'$. Similar to the previous case.
        
      \item The reduction is at the root of $\tm$. Then the step is a
        $\symdbB$ step: $\tm=(\lam{\var}{\tm_{11}})\sctx
        \,\tm_2\to_\symdbB \tm_{11}\esub{\var}{\tm_2}\sctx = \tm'$. By
        \rlem{derEq:properties}(\ref{derEq:properties:lam})
        there exists $\tmtwo_{11},\sctx'$ such that
        $\tmtwo_1=(\lam{\var}{\tmtwo_{11}})\sctx'$ with
        $\tm_{11}\derEq\tmtwo_{11}$ and $\sctx\derEqSctx{\varset}
        \sctx'$ and $\varset\cap \fv{\lam{\var}{\tmtwo_{11}}}=\emptyset$.

        We set $\tmtwo'$ to be $\tmtwo_{11}\esub{\var}{\tmtwo_2}\sctx' $.  Note that $\tmtwo
        = (\lam{\var}{\tmtwo_{11}})\sctx'\,\tmtwo_2\to_\symdbB
        \tmtwo_{11}\esub{\var}{\tmtwo_2}\sctx'$. 
        From
   $\tm_{11}\derEq\tmtwo_{11}$  and $\tm_2 \derEq \tmtwo_2$,       $\tm_{11}\esub{\var}{\tm_2}\derEq
        \tmtwo_{11}\esub{\var}{\tmtwo_2}$ holds. Finally, since $\sctx\derEqSctx{\varset}
   \sctx'$ and $\tm_2 \derEq \tmtwo_2$, also 
   $\esub{\var}{\tm_2}\sctx\derEqSctx{\varset}
   \esub{\var}{\tmtwo_2}\sctx'$. We obtain 
              $\tm_{11}\esub{\var}{\tm_2}\sctx \derEq
        \tmtwo_{11}\esub{\var}{\tmtwo_2}\sctx'$ from
        \rlem{derEq:properties}(\ref{derEq:properties:replacement:sctx}).
        
        \end{enumerate}
        
     \item Item 2. There are two cases.

      \begin{enumerate}
      \item The reduction is internal to $\tm_1$: $\tm=\tm_1 \,\tm_2\togcB\tm_1' \tm_2=\tm'$ follows from $\tm_1 \togcB \tm_1'$. From the \ih there exists $\tmtwo_1'$ such that $\tmtwo_1 \togcB^*  \tmtwo_1'$ and $\tm_1'\derEq \tmtwo_1'$.  Then $\tmtwo_1\,\tmtwo_2\togcB^* \tmtwo_1'\,\tmtwo_2$ and moreover $\tm_1'\,\tm_2\derEq \tmtwo_1'\,\tmtwo_2$. Thus we conclude by setting $\tmtwo'$ to be $\tmtwo_1'\,\tmtwo_2$.
        
      \item The reduction is internal to $\tm_2$: $\tm=\tm_1 \,\tm_2\togcB\tm_1 \tm_2'=\tm'$ follows from $\tm_2 \togcB \tm_2'$. Similar to the previous case.

        \end{enumerate}
    \end{itemize}

    \item $\tm\derEq \tmtwo$ is $\tm_1\esub{\var}{\tm_2} \derEq \tmtwo_1\esub{\var}{\tmtwo_2}$ and follows from $\tm_1 \derEq \tmtwo_1$ and $\tm_2 \derEq \tmtwo_2$.
    \begin{itemize}
    \item Item 1. There are three cases.
      \begin{enumerate}
      \item The reduction is internal to $\tm_1$. Then
        $\tm_1\esub{\var}{\tm_2} \tonogcB \tm_1'\esub{\var}{\tm_2}=\tm'$
        follows from $\tm_1 \tonogcB \tm_1'$. From the hypothesis $\tm_1
        \derEq \tmtwo_1$ and the \ih, there exists $\tmtwo_1'$ such that
        $\tmtwo_1 \toBang \tmtwo_1'$ and $\tm_1'\derEq \tmtwo_1'$.  Then
        we set $\tmtwo'$ to be $\tmtwo_1'\esub{\var}{\tmtwo_2}$. Note that
        $\tmtwo_1\esub{\var}{\tmtwo_2} \toBang
        \tmtwo_1'\esub{\var}{\tmtwo_2} $ and $
        \tm_1'\esub{\var}{\tm_2}\derEq  \tmtwo_1'\esub{\var}{\tmtwo_2} $,
        and we conclude.

      \item The reduction is internal to $\tm_2$.  Similar to the
      previous case.
    \item The reduction is at the root of $\tm$. Then the reduction step
      must be a $\symlsB$-step:
      $\tm= \off{\gctx}{\var}\esub{\var}{(\ofc{\tm_{21}})\sctx} \rtolsB
      \off{\gctx}{\tm_{21}}\esub{\var}{\ofc{\tm_{21}}}\sctx=\tm'$ and
      $\var \notin \fv{\tm_{21}}$ and
      $\fv{\gctx} \cap \dom{\sctx} = \emptyset$.  From
      $\off{\gctx}{\var}\derEq \tmtwo_1$ and
      \rlem{derEq:properties}(\ref{derEq:properties:gctx}),
      $\tmtwo_1=\off{\gctx'}{\var}$ and
      $\gctx\derEqGctx{\varset}\gctx'$ and $\var\notin\varset$. Similarly, from
      $(\ofc{\tm_{21}})\sctx\derEq\tmtwo_2$ and
      \rlem{derEq:properties}(\ref{derEq:properties:ofc}),
      $\tmtwo_2=(\ofc{\tmtwo_{21}})\sctx'$ and $\tm_{21}\derEq\tmtwo_{21}$ and
      $\sctx\derEqSctx{\varsettwo} \sctx'$ and $\varsettwo\cap\fv{\ofc{\tmtwo_{21}}}=\emptyset$. Then $\tmtwo
      = \off{\gctx'}{\var}\esub{\var}{(\ofc{\tmtwo_{21}})\sctx'}\toBang
      \off{\gctx'}{\tmtwo_{21}}\esub{\var}{\ofc{\tmtwo_{21}}}\sctx'$. From
      $\tm_{21}\derEq\tmtwo_{21}$ and
      $\gctx\derEqSctx{\varset}\gctx'$, we deduce from 
      \rlem{derEq:properties}(\ref{derEq:properties:replacement:gctx}), that $
      \off{\gctx}{\tm_{21}}\derEq \off{\gctx'}{\tmtwo_{21}}$.  We may assume that $\dom{\sctx'}\cap\fv{\off{\gctx'}{\tmtwo_{21}}}=\emptyset$. 
      Finally, we set $\tmtwo'$ to be
      $\off{\gctx'}{\tmtwo_{21}}\esub{\var}{\ofc{\tmtwo_{21}}}\sctx' $
      and conclude with
      $\off{\gctx}{\tm_{21}}\esub{\var}{\ofc{\tm_{21}}}\sctx\derEq  \off{\gctx'}{\tmtwo_{21}}\esub{\var}{\ofc{\tmtwo_{21}}}\sctx'$ by  \rlem{derEq:properties}(\ref{derEq:properties:replacement:sctx}).
   \end{enumerate}
  \item Item 2. There are three cases.

        \begin{enumerate}
      \item The reduction is internal to $\tm_1$. Same as above.

      \item The reduction is internal to $\tm_2$.  Same as above.

      \item The reduction is at the root of $\tm$. Then the reduction step
      must be a $\symgcB$-step:
      $\tm= \tm_1\esub{\var}{(\ofc{\tm_{21}})\sctx} \rtogcB
       \tm_1\sctx=\tm'$ and
      $\var \notin \fv{\tm_1}$.
      From
      $(\ofc{\tm_{21}})\sctx\derEq\tmtwo_2$ and
      \rlem{derEq:properties}(\ref{derEq:properties:ofc}),
      $\tmtwo_2=(\ofc{\tmtwo_{21}})\sctx'$ and $\tm_{21}\derEq\tmtwo_{21}$ and $\sctx\derEqSctx{\varset} \sctx'$ and $\varset\cap\fv{\ofc{\tmtwo_{21}}}=\emptyset$. Moreover, we may assume that not only $\varset\cap \fv{\ofc{\tmtwo_{21}}}=\emptyset$,  but also  $\varset\cap \fv{\tmtwo_1}=\emptyset$. 
      Consider $\tmtwo_1\esub{\var}{(\ofc{\tmtwo_{21}})\sctx'}$. There are 
      two further cases:
      \begin{enumerate}
        \item $\var\notin\fv{\tmtwo_1}$. Then $\tmtwo=\tmtwo_1\esub{\var}{(\ofc{\tmtwo_{21}})\sctx'} \rtogcB
       \tmtwo_1\sctx'=\tmtwo'$. Note that from  $\sctx\derEqSctx{\varset}
       \sctx'$,
       $\dom{\sctx'}\cap\fv{\tmtwo_1}=\emptyset$ and 
       \rlem{derEq:properties}(\ref{derEq:properties:replacement:sctx}), we have $
       \tm_1\sctx\derEq  \tmtwo_1\sctx'$. We thus set $\tmtwo'$ to be $ \tmtwo_1\sctx'$ and conclude.
       
     \item $\var\in\fv{\tmtwo_1}$. Since  $\var \notin \fv{\tm_1}$ and
       $\tm_1\derEq \tmtwo_1$, then by 
       \rlem{derEq:properties}(\ref{derEq:properties:gc:general}), there exists $\tmtwo_1'$ such that  $\var\notin\fv{\tmtwo_1'}$ and $\tm_1\derEq \tmtwo_1'$ and $\tmtwo_1\togcB^*\tmtwo'$. Then $\tmtwo_1\esub{\var}{(\ofc{\tmtwo_{21}})\sctx'}\togcB^* \tmtwo_1'\esub{\var}{(\ofc{\tmtwo_{21}})\sctx'}\togcB \tmtwo_1' \sctx'$. We thus set $\tmtwo'$ to be $ \tmtwo_1'\sctx'$ and conclude.

        \end{enumerate}
      
    \end{enumerate}
    
      \end{itemize}
      
  \item $\tm\derEq \tmtwo$ is $\ofc{\tm_1} \derEq \ofc{\tmtwo_1}$ and follows from $\tm_1 \derEq \tmtwo_1$.
   
    \begin{itemize}
    \item Item 1.
      \begin{enumerate}
      \item The reduction must be  internal to $\tm_1$. Then
        $\ofc{\tm_1} \tonogcB \ofc{\tm_1'}=\tm'$
        follows from $\tm_1 \tonogcB \tm_1'$. From the hypothesis $\tm_1
        \derEq \tmtwo_1$ and the \ih, there exists $\tmtwo_1'$ such that
        $\tmtwo_1 \toBang \tmtwo_1'$ and $\tm_1'\derEq \tmtwo_1'$.  Then
        we set $\tmtwo'$ to be $\ofc{\tmtwo_1'}$. Note that
        $\ofc{\tmtwo_1} \toBang
        \ofc{\tmtwo_1'} $ and $
        \ofc{\tm_1'}\derEq  \ofc{\tmtwo_1'}$,
        and we conclude.

   \end{enumerate}
  \item Item 2.

        \begin{enumerate}
      \item The reduction must be  internal to $\tm_1$. Then
        $\ofc{\tm_1} \togcB \ofc{\tm_1'}=\tm'$
        follows from $\tm_1 \togcB \tm_1'$. From the hypothesis $\tm_1
        \derEq \tmtwo_1$ and the \ih, there exists $\tmtwo_1'$ such that
        $\tmtwo_1 \togcB^*  \tmtwo_1'$ and $\tm_1'\derEq \tmtwo_1'$.  Then
        we set $\tmtwo'$ to be $\ofc{\tmtwo_1'}$. Note that
        $\ofc{\tmtwo_1} \togcB^* 
        \ofc{\tmtwo_1'} $ and $
        \ofc{\tm_1'}\derEq  \ofc{\tmtwo_1'}$,
        and we conclude.

   \end{enumerate}
  \end{itemize}

  \item $\tm\derEq \tmtwo$ is $\der{\tm_1}\derEq \varthree\esub{\varthree}{\tmtwo_1}$ and follows from $\tm_1 \derEq \tmtwo_1$.

      \begin{itemize}
    \item Item 1. There are two cases.
      \begin{enumerate}
      \item The reduction is internal to $\tm_1$. Then
        $\der{\tm_1} \tonogcB \der{\tm_1'}=\tm'$
        follows from $\tm_1 \tonogcB \tm_1'$. From the hypothesis $\tm_1
        \derEq \tmtwo_1$ and the \ih, there exists $\tmtwo_1'$ such that
        $\tmtwo_1 \toBang \tmtwo_1'$ and $\tm_1'\derEq \tmtwo_1'$.  Then
        we set $\tmtwo'$ to be $\varthree\esub{\varthree}{\tmtwo_1'}$. Note that
        $\varthree\esub{\varthree}{\tmtwo_1} \toBang
        \varthree\esub{\varthree}{\tmtwo_1'} $ and $
        \der{\tm_1'}\derEq  \varthree\esub{\varthree}{\tmtwo_1'}$,
        and we conclude.

    \item The reduction is at the root of $\tm$. Then the reduction step 
      must be a $\symderB$-step:
      $\tm=     \der{(\ofc{\tm_{11}})\sctx}
       \rtoderB 
      \tm_{11}\sctx=\tmtwo$.  From $\tm_1 =(\ofc{\tm_{11}})\sctx \derEq \tmtwo_1$.
      and  
      \rlem{derEq:properties}(\ref{derEq:properties:ofc}),
      $\tmtwo_2=(\ofc{\tmtwo_{21}})\sctx'$ and $\tm_{11}\derEq\tmtwo_{21}$ and $\sctx\derEqSctx{\varset} \sctx'$ and $\varset\cap\fv{\ofc{\tmtwo_{21}}}=\emptyset$.  Then $\varthree\esub{\varthree}{(\ofc{\tmtwo_{21}})\sctx'} \tolsB \tmtwo_{21}\esub{\varthree}{\ofc{\tmtwo_{21}}}\sctx'$. Moreover, $ \tm_{11}\sctx\derEq  \tmtwo_{21}\esub{\varthree}{\ofc{\tmtwo_{21}}}\sctx'$ follows from  $\tm_{11}\derEq\tmtwo_{21}$, then  $\tm_{11}\derEq\tmtwo_{21}\esub{\varthree}{\ofc{\tmtwo_{21}}}$, and finally from the latter and $\sctx\derEqSctx{\varset} \sctx'$ by \rlem{derEq:properties}(\ref{derEq:properties:replacement:sctx}). We set $\tmtwo'$ to be $\tmtwo_{21}\esub{\var}{\ofc{\tmtwo_{21}}}\sctx'$ and conclude.

   \end{enumerate}
  \item Item 2.

        \begin{enumerate}
      \item The reduction is internal to $\tm_1$. Same as above.
      
    \end{enumerate}
    
  \end{itemize}

  \item  $\tm\derEq \tmtwo$ is $\tm\derEq \tmtwo_1\esub{\varthree}{\ofc{\tmtwo_2}}$ follows from $\tm \derEq \tmtwo_1$ and $\varthree\notin\fv{\tmtwo_1}$,
      \begin{itemize}
    \item Item 1.
      \begin{enumerate}
      \item Suppose 
        $\tm \tonogcB \tm'$. 
        From the hypothesis $\tm
        \derEq \tmtwo_1$ and the \ih, there exists $\tmtwo_1'$ such that
        $\tmtwo_1 \toBang \tmtwo_1'$ and $\tm'\derEq \tmtwo_1'$.  Then
        we set $\tmtwo'$ to be $\tmtwo_1'\esub{\var}{\ofc{\tmtwo_2}}$. Note that
        $\tmtwo_1 \esub{\var}{\ofc{\tmtwo_2}}\toBang
        \tmtwo_1'\esub{\var}{\ofc{\tmtwo_2}} $ and $
        \tm'\derEq  \tmtwo_1'\esub{\var}{\ofc{\tmtwo_2}}$,
        and we conclude.

   \end{enumerate}
  \item Item 2.

        \begin{enumerate}
      \item Suppose 
        $\tm \togcB \tm'$. 
        From the hypothesis $\tm
        \derEq \tmtwo_1$ and the \ih, there exists $\tmtwo_1'$ such that
        $\tmtwo_1 \togcB^* \tmtwo_1'$ and $\tm'\derEq \tmtwo_1'$.  Then
        we set $\tmtwo'$ to be $\tmtwo_1'\esub{\var}{\ofc{\tmtwo_2}}$. Note that
        $\tmtwo_1 \esub{\varthree}{\ofc{\tmtwo_2}}\togcB^* 
        \tmtwo_1'\esub{\varthree}{\ofc{\tmtwo_2}} $ and $
        \tm'\derEq  \tmtwo_1'\esub{\varthree}{\ofc{\tmtwo_2}}$,
        and we conclude.

   \end{enumerate}
  \end{itemize}

  \end{enumerate}

  \end{proof}
\end{ifLongAppendix}

\begin{lemma}
\llem{derEq:properties:dual}
  \begin{enumerate}

  \item\label{derEq:properties:lam:dual} $\tm\derEq (\lam{\var}{\tmtwo})\sctx'$ implies $\tm=(\lam{\var}{\tm_1})\sctx$ with $\tm_1\derEq\tmtwo$ and $\sctx\derEqSctx{\varset} \sctx'$ and $\varset\cap\fv{\lam{\var}{\tm_1}}=\emptyset$.

  \item\label{derEq:properties:gctx:dual}  $\tm\derEq \off{\gctx'}{\var}$
    implies $\tm=\off{\gctx}{\var}$ and
    $\gctx\derEqGctx{\varset}\gctx'$ and $x\notin\varset$.

  \item\label{derEq:properties:ofc:dual}  $\tm\derEq (\ofc{\tmtwo})\sctx'$ implies $\tm=(\ofc{\tm_1})\sctx$ and $\tm_1\derEq\tmtwo$ and $\sctx\derEqSctx{\varset}\sctx'$ and $\varset\cap\fv{\ofc{\tmtwo}}=\emptyset$.



  \item\label{derEq:properties:gc:dual} $\tm\derEq \off{\gctxtwo}{\var}$ implies
    \begin{enumerate}
    \item either $\var\in\fv{\tm}$ and $\tm=\off{\gctx}{\var}$ and $\gctx\derEqGctx{\varset}\gctxtwo$;
      \item or  $\var\notin\fv{\tm}$ and $\gctxtwo=\of{\gctxtwo_1}{\tmtwo_1\esub{\vartwo}{\ofc{\off{\gctxtwo_2}{\var}}}}$ and $\vartwo\notin\fv{\tmtwo_1}$ and $\tm=\of{\gctx}{\tm_1}$ and $\of{\gctx}{\tm_1}\derEq \of{\gctxtwo_1}{\tmtwo_1}$. 
      \end{enumerate}
  \end{enumerate}
\end{lemma}

\begin{proof}
  Item~\ref{derEq:properties:lam:dual}. By induction on the derivation of $\tm\derEq (\lam{\var}{\tmtwo})\sctx'$.

  Item~\ref{derEq:properties:gctx:dual}. By induction on the derivation of  $\tm\derEq \off{\gctx'}{\var}$.
  
  Item~\ref{derEq:properties:ofc:dual}. By induction on the derivation of  $\tm\derEq (\ofc{\tmtwo})\sctx'$.
  
  Item~\ref{derEq:properties:gc:dual}. By induction on the derivation of $\tm\derEq \off{\gctxtwo}{\var}$.
\end{proof}

\begin{lemma}[Backwards simulation of dereliction]
\llem{BGM_simulates_B}
\begin{equation}
  \begin{array}{cc}
   \begin{tikzcd}[column sep=0.2em,ampersand replacement=\&]
\tm \arrow[d,dashed, "=" near end, "\symBangGM"']
  \& \derEq
  \& \tmtwo \arrow[d, "\step"] \\ 
\tm'  
  \& \derEq
  \& \tmtwo'
\end{tikzcd}
    &
\begin{tikzcd}[column sep=0.2em,ampersand replacement=\&]
\tm \arrow[d, "\symgcB"', "=" near end, dashed]
  \& \derEq
  \& \tmtwo \arrow[d, "\symgcB"]\\ 
\tm'  
  \& \derEq
  \& \tmtwo'
\end{tikzcd}
\end{array}
\end{equation}   
where $\step\in\{\symdb, \symlsB\}$.
\end{lemma}
\begin{ifShortAppendix}
  \begin{proof}
  Both items are proved by induction on the derivation of  $\tm\derEq \tmtwo$.
  See the extended version~\cite{mells_long} for the detailed proof.
  \end{proof}
\end{ifShortAppendix}
\begin{ifLongAppendix}
  \begin{proof}\label{BGM_simulates_B:proof}  
   We prove both items by induction on the derivation of  $\tm\derEq \tmtwo$.
  \begin{enumerate}

  \item  $\tm\derEq \tmtwo$ is $\var \derEq \var$. Both items are immediate since there are no $\toBang$ steps from $\var$.

  \item $\tm\derEq \tmtwo$ is  $\lam{\var}{\tm_1} \derEq \lam{\var}{\tmtwo_1}$ and follows from $\tm_1 \derEq \tmtwo_1$.
    \begin{itemize} 
    \item Item 1. The reduction must be internal: $\lam{\var}{\tmtwo_1} \tonogcB\lam{\var}{\tmtwo_1'}=\tmtwo'$ follows from $\tmtwo_1 \tonogcB \tmtwo_1'$. From the \ih there exists $\tm_1'$ such that $\tm_1 \toBangGM^=  \tm_1'$ and $\tm_1'\derEq \tmtwo_1'$.  Then $\lam{\var}{\tm_1}\toBangGM^= \lam{\var}{\tm_1'}$ and moreover $\lam{\var}{\tm_1'}\derEq \lam{\var}{\tmtwo_1'}$. Thus we conclude by setting $\tm'$ to be $\lam{\var}{\tm_1'}$.

      \item Item 2. Similar to Item 1.
    \end{itemize}

         \item $\tm\derEq \tmtwo$ is $\tm_1\, \tm_2 \derEq \tmtwo_1\,\tmtwo_2$ and follows from $\tm_1 \derEq \tmtwo_1$ and $\tm_2 \derEq \tmtwo_2$
    \begin{itemize}
    \item Item 1. There are three cases.

      \begin{enumerate}
      \item The reduction is internal to $\tmtwo_1$: $\tmtwo=\tmtwo_1 \,\tmtwo_2\tonogcB\tmtwo_1' \tmtwo_2=\tmtwo'$ follows from $\tmtwo_1 \tonogcB \tmtwo_1'$. From the \ih there exists $\tm_1'$ such that $\tm_1 \toBangGM^=  \tm_1'$ and $\tm_1'\derEq \tmtwo_1'$.  Then $\tm_1\,\tm_2\toBangGM^= \tm_1'\,\tm_2$ and moreover $\tm_1'\,\tm_2\derEq \tmtwo_1'\,\tmtwo_2$. Thus we conclude by setting $\tm'$ to be $\tm_1'\,\tm_2$.
        
      \item The reduction is internal to $\tmtwo_2$: $\tmtwo=\tmtwo_1 \,\tmtwo_2\tonogcB\tmtwo_1 \tmtwo_2'=\tmtwo'$ follows from $\tmtwo_2 \tonogcB \tmtwo_2'$. Similar to the previous case.
        
      \item The reduction is at the root of $\tmtwo$. Then the step is a
        $\symdbB$ step: $\tmtwo=(\lam{\var}{\tmtwo_{11}})\sctx'
        \,\tmtwo_2\to_\symdbB \tmtwo_{11}\esub{\var}{\tmtwo_2}\sctx' = \tmtwo'$. By
        \rlem{derEq:properties:dual}(\ref{derEq:properties:lam:dual})
        there exists $\tm_{11},\sctx$ such that
        $\tm_1=(\lam{\var}{\tm_{11}})\sctx$ with
        $\tm_{11}\derEq\tmtwo_{11}$ and $\sctx\derEqSctx{\varset}
        \sctx'$ and $\varset\cap \fv{\lam{\var}{\tm_{11}}}=\emptyset$.

        We set $\tm'$ to be $\tm_{11}\esub{\var}{\tm_2}\sctx $.  Note that $\tm
        = (\lam{\var}{\tm_{11}})\sctx\,\tm_2\to_\symdbB
        \tm_{11}\esub{\var}{\tm_2}\sctx$. 
        From
   $\tm_{11}\derEq\tmtwo_{11}$  and $\tm_2 \derEq \tmtwo_2$,       $\tm_{11}\esub{\var}{\tm_2}\derEq
        \tmtwo_{11}\esub{\var}{\tmtwo_2}$ holds. Finally, since 
   $\sctx\derEqSctx{\varset}
        \sctx'$ and $\varset\cap \fv{(\lam{\var}{\tmtwo_{11}})\, \tmtwo_2}=\emptyset $, we obtain 
              $\tm_{11}\esub{\var}{\tm_2}\sctx \derEq
        \tmtwo_{11}\esub{\var}{\tmtwo_2}\sctx'$ from
        \rlem{derEq:properties}(\ref{derEq:properties:replacement:sctx}).
        
        \end{enumerate}
        
     \item Item 2. There are two cases.

      \begin{enumerate}
      \item The reduction is internal to $\tmtwo_1$: $\tmtwo=\tmtwo_1
        \,\tmtwo_2\togcB\tmtwo_1' \tmtwo_2=\tm'$ follows from $\tmtwo_1
        \togcB \tmtwo_1'$. From the \ih there exists $\tm_1'$ such that
        $\tm_1 \togcB^=   \tm_1'$ and $\tm_1'\derEq \tmtwo_1'$.  Then
        $\tm_1\,\tm_2\togcB^=\tm_1'\,\tm_2$ and moreover $\tm_1'\,\tm_2\derEq \tmtwo_1'\,\tmtwo_2$. Thus we conclude by setting $\tm'$ to be $\tm_1'\,\tm_2$.
        
      \item The reduction is internal to $\tmtwo_2$: $\tmtwo=\tmtwo_1 \,\tmtwo_2\togcB\tmtwo_1 \tmtwo_2'=\tmtwo'$ follows from $\tmtwo_2 \togcB \tmtwo_2'$. Similar to the previous case.

        \end{enumerate}
    \end{itemize}

    \item $\tm\derEq \tmtwo$ is $\tm_1\esub{\var}{\tm_2} \derEq \tmtwo_1\esub{\var}{\tmtwo_2}$ and follows from $\tm_1 \derEq \tmtwo_1$ and $\tm_2 \derEq \tmtwo_2$.
    \begin{itemize}
    \item Item 1. There are three cases.
      \begin{enumerate}
      \item The reduction is internal to $\tmtwo_1$. Then
        $\tmtwo_1\esub{\var}{\tmtwo_2} \tonogcB \tmtwo_1'\esub{\var}{\tmtwo_2}=\tmtwo'$
        follows from $\tmtwo_1 \tonogcB \tmtwo_1'$. From the hypothesis $\tm_1
        \derEq \tmtwo_1$ and the \ih, there exists $\tm_1'$ such that
        $\tm_1 \toBangGM^= \tm_1'$ and $\tm_1'\derEq \tmtwo_1'$.  Then
        we set $\tm'$ to be $\tm_1'\esub{\var}{\tm_2}$. Note that 
        $\tm_1'\esub{\var}{\tm_2} \toBangGM^=
        \tm_1'\esub{\var}{\tm_2} $ and $
        \tm_1'\esub{\var}{\tm_2}\derEq  \tmtwo_1'\esub{\var}{\tmtwo_2} $,
        and we conclude.

      \item The reduction is internal to $\tmtwo_2$.  Similar to the
      previous case.
    \item The reduction is at the root of $\tmtwo$. Then the reduction step
      must be a $\symlsB$-step:
      $\tmtwo= \off{\gctx'}{\var}\esub{\var}{(\ofc{\tmtwo_{21}})\sctx'} \rtolsB
      \off{\gctx'}{\tmtwo_{21}}\esub{\var}{\ofc{\tmtwo_{21}}}\sctx'=\tmtwo'$ and
      $\var \notin \fv{\tmtwo_{21}}$ and 
      $\fv{\gctx'} \cap \dom{\sctx'} = \emptyset$.  From
      $\tm_1\derEq\off{\gctx'}{\var}$ and
      \rlem{derEq:properties:dual}(\ref{derEq:properties:gc:dual}), there are two possible cases:

      \begin{enumerate}
      \item  $\var\in\fv{\tm_1}$ and $\tm_1=\off{\gctx}{\var}$ and $\gctx\derEqGctx{\varset}\gctx'$. 
      Similarly, from
      $\tm_2\derEq (\ofc{\tmtwo_{21}})\sctx$ and
      \rlem{derEq:properties:dual}(\ref{derEq:properties:ofc:dual}),
      $\tm_2=(\ofc{\tm_{21}})\sctx$ and $\tm_{21}\derEq\tmtwo_{21}$ and
      $\sctx\derEqSctx{\varsettwo} \sctx'$ and $\varsettwo\cap\fv{\ofc{\tmtwo_{21}}}=\emptyset$. Then $\tm
      = \off{\gctx}{\var}\esub{\var}{(\ofc{\tm_{21}})\sctx}\toBangGM^=
      \off{\gctx}{\tm_{21}}\esub{\var}{\ofc{\tm_{21}}}\sctx$. From
      $\tm_{21}\derEq\tmtwo_{21}$ and
      $\gctx\derEqSctx{\varset}\gctx'$, we deduce from 
      \rlem{derEq:properties}(\ref{derEq:properties:replacement:gctx}), that $
      \off{\gctx}{\tm_{21}}\derEq \off{\gctx'}{\tmtwo_{21}}$.  We may assume that $\dom{\sctx'}\cap\fv{\off{\gctx}{\tm_{21}}}=\emptyset$. 
      Finally, we set $\tm'$ to be
      $\off{\gctx}{\tm_{21}}\esub{\var}{\ofc{\tm_{21}}}\sctx$
      and conclude with
      $\off{\gctx}{\tm_{21}}\esub{\var}{\ofc{\tm_{21}}}\sctx\derEq  \off{\gctx'}{\tmtwo_{21}}\esub{\var}{\ofc{\tmtwo_{21}}}\sctx'$ by  \rlem{derEq:properties}(\ref{derEq:properties:replacement:sctx}).
        
    \item  $\var\notin\fv{\tm_1}$ and $\gctx'=\of{\gctx_1'}{\tmtwo_1\esub{\vartwo}{\ofc{\off{\gctx_2'}{\var}}}}$ and $\vartwo\notin\fv{\tmtwo_1}$ and $\tm_1=\of{\gctx}{\tm_{11}}$ and $\of{\gctx}{\tm_{11}}\derEq \of{\gctx_1'}{\tmtwo_1}$.  The step $\tmtwo \rtolsB \tmtwo' $ is thus of the form:
      \[\tmtwo= \of{\gctx_1'}{\tmtwo_1\esub{\vartwo}{\ofc{\off{\gctx_2'}{\var}}}}\esub{\var}{(\ofc{\tmtwo_{21}})\sctx'} \rtolsB
        \of{\gctx_1'}{\tmtwo_1\esub{\vartwo}{\ofc{\off{\gctx_2'}{\tmtwo_{21}}}}}\esub{\var}{\ofc{\tmtwo_{21}}}\sctx'=\tmtwo'
      \]
      Just like in the previous case,   from
      $\tm_2\derEq (\ofc{\tmtwo_{21}})\sctx$ and
      \rlem{derEq:properties:dual}(\ref{derEq:properties:ofc:dual}),
      $\tm_2=(\ofc{\tm_{21}})\sctx$ and $\tm_{21}\derEq\tmtwo_{21}$ and
      $\sctx\derEqSctx{\varsettwo} \sctx'$ and $\varsettwo\cap\fv{\ofc{\tmtwo_{21}}}=\emptyset$.
      Then
         \begin{center}
              \begin{tikzcd}[column sep=3em,row sep=3em]
              \of{\gctx}{\tm_{11}}\esub{\var}{(\ofc{\tm_{21}})\sctx} \arrow[d,equals] \arrow[r, phantom, "\derEq"] &                        \of{\gctx_1'}{\tmtwo_1\esub{\vartwo}{\ofc{\off{\gctx_2'}{\var}}}}\esub{\var}{(\ofc{\tmtwo_{21}})\sctx'}
    \arrow[d, "\symlsB"]  \\
                
             \of{\gctx}{\tm_{11}}\esub{\var}{(\ofc{\tm_{21}})\sctx} \arrow[r, phantom, "\derEq"]   &                         \of{\gctx_1'}{\tmtwo_1\esub{\vartwo}{\ofc{\off{\gctx_2'}{\tmtwo_{21}}}}}\esub{\var}{\ofc{\tmtwo_{21}}}\sctx'
              \end{tikzcd}
            \end{center}
        \end{enumerate}
    \end{enumerate}

  \item Item 2. There are three cases.

        \begin{enumerate} 
      \item The reduction is internal to $\tmtwo_1$: $\tmtwo=\tmtwo_1\esub{\var}{\tmtwo_2}\togcB \tmtwo_1'\esub{\var}{\tmtwo_2}=\tm'$ follows from $\tmtwo_1
        \togcB \tmtwo_1'$. From the \ih there exists $\tm_1'$ such that
        $\tm_1 \togcB^=   \tm_1'$ and $\tm_1'\derEq \tmtwo_1'$.  Then
        $\tm_1\esub{\var}{\tm_2}\togcB^=\tm_1'\esub{\var}{\tm_2}$ and moreover $\tm_1'\esub{\var}{\tm_2}\derEq \tmtwo_1'\esub{\var}{\tmtwo_2}$. Thus we conclude by setting $\tm'$ to be $\tm_1'\esub{\var}{\tm_2}$.

      \item The reduction is internal to $\tmtwo_2$.  Same as above.

      \item The reduction is at the root of $\tmtwo$. Then the reduction step
      must be a $\symgcB$-step:
      $\tmtwo= \tmtwo_1\esub{\var}{(\ofc{\tmtwo_{21}})\sctx} \rtogcB
       \tmtwo_1\sctx=\tm'$ and
      $\var \notin \fv{\tmtwo_1}$.
      From
      $\tm_2\derEq (\ofc{\tmtwo_{21}})\sctx$ and
      \rlem{derEq:properties:dual}(\ref{derEq:properties:ofc:dual}),
      $\tm_2=(\ofc{\tm_{21}})\sctx'$ and $\tm_{21}\derEq\tmtwo_{21}$ and $\sctx\derEqSctx{\varset} \sctx'$ and $\varset\cap\fv{\ofc{\tm_{21}}}=\emptyset$. Moreover, we may assume that not only $\varset\cap \fv{\ofc{\tmtwo_{21}}}=\emptyset$,  but also  $\varset\cap \fv{\tmtwo_1}=\emptyset$. 
      Consider $\tmtwo_1\esub{\var}{(\ofc{\tmtwo_{21}})\sctx'}$. From
  $\var\notin\fv{\tmtwo_1}$ and \rremark{derEq:fv}(\ref{rem:derEq:fv:general}),  also $\var\notin\fv{\tm_1}$.  Then $\tm=\tm_1\esub{\var}{(\ofc{\tm_{21}})\sctx'} \rtogcB
       \tm_1\sctx'=\tm'$. Note that from  $\sctx\derEqSctx{\varset}
       \sctx'$,
       $\dom{\sctx'}\cap\fv{\tmtwo_1}=\emptyset$ and 
       \rlem{derEq:properties}(\ref{derEq:properties:replacement:sctx}), we have $
       \tm_1\sctx\derEq  \tmtwo_1\sctx'$. We thus set $\tm'$ to be $ \tm_1\sctx'$ and conclude.

    \end{enumerate}
    
      \end{itemize}
      
  \item $\tm\derEq \tmtwo$ is $\ofc{\tm_1} \derEq \ofc{\tmtwo_1}$ and follows from $\tm_1 \derEq \tmtwo_1$.
   
    \begin{itemize}
    \item Item 1.
      \begin{enumerate}
      \item The reduction must be  internal to $\tmtwo_1$. Then
        $\ofc{\tmtwo_1} \tonogcB \ofc{\tmtwo_1'}=\tmtwo'$
        follows from $\tmtwo_1 \tonogcB \tmtwo_1'$. From the hypothesis $\tm_1
        \derEq \tmtwo_1$ and the \ih, there exists $\tm_1'$ such that
        $\tm_1 \toBangGM^= \tm_1'$ and $\tm_1'\derEq \tmtwo_1'$.  Then
        we set $\tm'$ to be $\ofc{\tm_1'}$. Note that
        $\ofc{\tm_1} \toBangGM^=
        \ofc{\tm_1'} $ and $ 
        \ofc{\tm_1'}\derEq  \ofc{\tmtwo_1'}$,
        and we conclude.

   \end{enumerate}
  \item Item 2.

        \begin{enumerate}
      \item The reduction must be  internal to $\tmtwo_1$. Then
        $\ofc{\tmtwo_1} \togcB \ofc{\tmtwo_1'}=\tmtwo'$
        follows from $\tmtwo_1 \togcB \tmtwo_1'$. From the hypothesis $\tm_1
        \derEq \tmtwo_1$ and the \ih, there exists $\tm_1'$ such that
        $\tm_1 \togcB^=  \tm_1'$ and $\tm_1'\derEq \tmtwo_1'$.  Then
        we set $\tm'$ to be $\ofc{\tmtwo_1'}$. Note that
        $\ofc{\tm_1} \togcB^= 
        \ofc{\tm_1'} $ and $
        \ofc{\tm_1'}\derEq  \ofc{\tmtwo_1'}$,
        and we conclude.

   \end{enumerate}
  \end{itemize}

  \item $\tm\derEq \tmtwo$ is $\der{\tm_1}\derEq \varthree\esub{\varthree}{\tmtwo_1}$ and follows from $\tm_1 \derEq \tmtwo_1$.
    \begin{itemize}

    \item Item 1. There are two cases.
      
      \begin{enumerate}
      \item The reduction is internal to $\tmtwo_1$. Then
        $\varthree\esub{\varthree}{\tmtwo_1}\tonogcB \varthree\esub{\varthree}{\tmtwo_1'}=\tmtwo'$
        follows from $\tmtwo_1 \tonogcB \tmtwo_1'$. From the hypothesis $\tm_1
        \derEq \tmtwo_1$ and the \ih, there exists $\tm_1'$ such that
        $\tm_1 \toBangGM^= \tm_1'$ and $\tm_1'\derEq \tmtwo_1'$.  Then
        we set $\tm'$ to be $\der{\tm_1'}$. Note that 
        $\der{\tm_1} \toBangGM^=
        \der{\tm_1'} $ and $
        \der{\tm_1'}\derEq  \varthree\esub{\varthree}{\tmtwo_1'}$,
        and we conclude. 

    \item The reduction is at the root of $\tmtwo$. Then the reduction step 
      must be a $\symlsB$-step:
      $\tmtwo = \varthree\esub{\varthree}{(\ofc{\tmtwo_{11}})\sctx'} \tolsB \tmtwo_{11}\esub{\varthree}{\ofc{\tmtwo_{11}}}\sctx'$.  From $\tm_1\derEq (\ofc{\tmtwo_{11}})\sctx' $
      and  
      \rlem{derEq:properties:dual}(\ref{derEq:properties:ofc:dual}),
      $\tm_1=(\ofc{\tm_{12}})\sctx$ and $\tm_{12}\derEq\tmtwo_{11}$ and $\sctx\derEqSctx{\varset} \sctx'$ and $\varset\cap\fv{\ofc{\tm_{12}}}=\emptyset$.  We set $\tm'$ to be $\der{(\ofc{\tm_{12}})\sctx}$. Note that $\der{(\ofc{\tm_{12}})\sctx}
      \rtoderB 
      \tm_{12}\sctx$. Moreover, from  $\tm_{12}\derEq\tmtwo_{11}$  we deduce  $\tm_{12}\derEq\tmtwo_{11}\esub{\varthree}{\ofc{\tmtwo_{11}}}$ using $\indrulename{\ruleDerEqGc}$; and from the latter we obtain  $\tm_{12}\sctx\derEq \tmtwo_{11}\esub{\varthree}{\ofc{\tmtwo_{11}}}\sctx'$ using  \rlem{derEq:properties}(\ref{derEq:properties:replacement:sctx}).

   \end{enumerate}

  \item Item 2. There is one case.
    \begin{enumerate}
          \item The reduction is internal to $\tmtwo_1$. Then
        $\varthree\esub{\varthree}{\tmtwo_1}\togcB \varthree\esub{\varthree}{\tmtwo_1'}=\tmtwo'$
        follows from $\tmtwo_1 \togcB \tmtwo_1'$. From the hypothesis $\tm_1
        \derEq \tmtwo_1$ and the \ih, there exists $\tm_1'$ such that
        $\tm_1 \togcB^= \tm_1'$ and $\tm_1'\derEq \tmtwo_1'$.  Then
        we set $\tm'$ to be $\der{\tm_1'}$. Note that 
        $\der{\tm_1} \togcB^= 
        \der{\tm_1'} $ and $
        \der{\tm_1'}\derEq  \varthree\esub{\varthree}{\tmtwo_1'}$,
        and we conclude. 

      \end{enumerate}
      
  \end{itemize}

  \item  $\tm\derEq \tmtwo$ is $\tm\derEq \tmtwo_1\esub{\varthree}{\ofc{\tmtwo_2}}$ follows from $\tm \derEq \tmtwo_1$ and $\varthree\notin\fv{\tmtwo_1}$,
      \begin{itemize}
    \item Item 1. There are two cases.
      \begin{enumerate}
      \item The reduction is internal to $\tmtwo_1$. Suppose 
        $\tmtwo_1\esub{\varthree}{\ofc{\tmtwo_2}} \tonogcB \tmtwo_1'\esub{\varthree}{\ofc{\tmtwo_2}}$ follows from $\tmtwo_1 \tonogcB \tmtwo_1'$. 
        From the hypothesis $\tm
        \derEq \tmtwo_1$ and the \ih, there exists $\tm'$ such that
        $\tm \toBangGM^= \tm'$ and $\tm'\derEq \tmtwo_1'$.  We conclude since $\tm'\derEq \tmtwo_1'\esub{\varthree}{\ofc{\tmtwo_2 }}$ follows from $\indrulename{\ruleDerEqGc}$. 

            \item The reduction is internal to $\tmtwo_2$. Suppose 
        $\tmtwo_1\esub{\varthree}{\ofc{\tmtwo_2}} \tonogcB \tmtwo_1\esub{\varthree}{\ofc{\tmtwo_2'}}$ follows from $\tmtwo_2 \tonogcB \tmtwo_2'$.  
        We set $\tm'$ to be $\tm$ since $\tm'\derEq \tmtwo_1\esub{\varthree}{\ofc{\tmtwo_2' }}$ follows from $\indrulename{\ruleDerEqGc}$.

   \end{enumerate}
  \item Item 2. There are three cases.

      \begin{enumerate}
      \item The reduction is internal to $\tmtwo_1$. Suppose 
        $\tmtwo_1\esub{\varthree}{\ofc{\tmtwo_2}} \togcB \tmtwo_1'\esub{\varthree}{\ofc{\tmtwo_2}}$ follows from $\tmtwo_1 \togcB \tmtwo_1'$. 
        From the hypothesis $\tm
        \derEq \tmtwo_1$ and the \ih, there exists $\tm'$ such that
        $\tm \togcB^= \tm'$ and $\tm'\derEq \tmtwo_1'$.  We conclude since $\tm'\derEq \tmtwo_1'\esub{\varthree}{\ofc{\tmtwo_2 }}$ follows from $\indrulename{\ruleDerEqGc}$. 

            \item The reduction is internal to $\tmtwo_2$. Suppose 
        $\tmtwo_1\esub{\varthree}{\ofc{\tmtwo_2}} \togcB \tmtwo_1\esub{\varthree}{\ofc{\tmtwo_2'}}$ follows from $\tmtwo_2 \togcB \tmtwo_2'$.  
        We set $\tm'$ to be $\tm$ since $\tm'\derEq \tmtwo_1\esub{\varthree}{\ofc{\tmtwo_2' }}$ follows from $\indrulename{\ruleDerEqGc}$.

            \item The reduction is at the root. Then the reduction step
      must be a $\symgcB$-step:
      $\tmtwo= \tmtwo_1\esub{\varthree}{\ofc{\tmtwo_{21}}} \rtogcB
       \tmtwo_1=\tm'$.
       We set $\tm'$ to be $\tm$ and conclude from $\tm \derEq \tmtwo_1$.

   \end{enumerate}
  \end{itemize}

  \end{enumerate}

  \end{proof}
\end{ifLongAppendix}

\restate{proposition}{prop:typing_the_bang_translation}{Bang typing}{
If $\judl{\tctx}{\tm}{\typ}$
then $\judc{\traB{\tctx}}{\noenv}{\traB{\tm}}{\traB{\typ}}$.
}
\begin{proof}\label{typing_the_bang_translation:proof}
By induction on the derivation of $\judl{\tctx}{\tm}{\typ}$:
\begin{enumerate}
\item
  $\rulebVar$:
  Let $\judl{\tctx,\var:\ofc{\typ}}{\var}{\typ}$ be derived from the $\rulebVar$ rule.
  Then:
  
    \scalebox{\smallDerivation}{\parbox{\textwidth}{
$
    \indrule{\rulecOpen}{
      \indrule{\rulecUvar}{
        \emptyPremise
      }{
        \judc{\traB{\tctx},\var:\traB{\typ}}{\noenv}{\var}{\sha{\traB{\typ}}}
      }
    }{
      \judc{\traB{\tctx},\var:\traB{\typ}}{\noenv}{\open{\var}}{\traB{\typ}}
    }
    $
    }}

\item
  $\rulebAbs$:
  Let $\judl{\tctx}{\lam{\var}{\tm}}{\ofc{\typ}\to\typtwo}$
  be derived from $\judl{\tctx,\var:\ofc{\typ}}{\tm}{\typtwo}$.
  Let $\lvar$ be a fresh linear variable,
  such that $\lvar \notin \fv{\traB{\tm}}$.
  Then:

    \scalebox{\smallDerivation}{\parbox{\textwidth}{
    $
    \indrule{\rulecAbs}{
      \indrule{\rulecES}{
        \derivih{\judc{\traB{\tctx},\var:\traB{\typ}}{\noenv}{
          \traB{\tm}
        }{\traB{\typtwo}}}
        \indrule{\rulecLvar}{
          \emptyPremise
        }{
          \judc{\traB{\tctx}}{\lvar:\osha{\traB{\typ}}}{
            \lvar
          }{\osha{\traB{\typ}}}
        }
      }{
        \judc{\traB{\tctx}}{\lvar:\osha{\traB{\typ}}}{
          \traB{\tm}\esub{\var}{\lvar}
        }{\traB{\typtwo}}
      }
    }{
      \judc{\traB{\tctx}}{\noenv}{
        \lam{\lvar}{\traB{\tm}\esub{\var}{\lvar}}
      }{\osha{\traB{\typ}}\limp\traB{\typtwo}}
    }
    $
  }}

\item
  $\rulebApp$:
  Let $\judl{\tctx}{\tm\,\tmtwo}{\typtwo}$
  be derived from $\judl{\tctx}{\tm}{\ofc{\typ}\to\typtwo}$
  and $\judl{\tctx}{\tmtwo}{\ofc{\typ}}$.
  Then:

    \scalebox{\smallDerivation}{\parbox{\textwidth}{
    $
    \indrule{\rulecApp}{
      \derivih{\judc{\traB{\tctx}}{\noenv}{\traB{\tm}}{\osha{\traB{\typ}}\limp\traB{\typtwo}}}
          \derivih{
        \judc{\traB{\tctx}}{\noenv}{\traB{\tmtwo}}{\osha{\traB{\typ}}}
      }
    }{
      \judc{\traB{\tctx}}{\noenv}{\traB{\tm}\,\traB{\tmtwo}}{\traB{\typtwo}}
    }
    $
    }}

\item $\rulebProm$: Let $\judl{\tctx}{\ofc{\tm}}{\ofc{\typ}}$ be
  derived from $\judl{\tctx}{\tm}{\typ}$. Then:

    \scalebox{\smallDerivation}{\parbox{\textwidth}{
    $
      \indrule{\rulecProm}{
        \indrule{\rulecSha}{
          \derivih{\judc{\traB{\tctx}}{\noenv}{\traB{\tm}}{\traB{\typ}}}
        }{
          \judc{\traB{\tctx}}{\noenv}{\sha{\traB{\tm}}}{\sha{\traB{\typ}}}
        }
      }{
          \judc{\traB{\tctx}}{\noenv}{\osha{\traB{\tm}}}{\osha{\traB{\typ}}}
      }
      $
      }}

  
\item
  $\rulebES$:
  Let $\judl{\tctx}{\tm\esub{\var}{\tmtwo}}{\ofc{\typtwo}}$
  be derived from $\judl{\tctx,\var:\ofc{\typ}}{\tm}{\typtwo}$
  and $\judl{\tctx}{\tmtwo}{\ofc{\typ}}$.
  Then:

    \scalebox{\smallDerivation}{\parbox{\textwidth}{
   $
    \indrule{}{
      \derivih{\judc{\traB{\tctx},\var:\traB{\typ}}{\noenv}{\traB{\tm}}{\osha{\traB{\typtwo}}}}
          \derivih{
        \judc{\traB{\tctx}}{\noenv}{\traB{\tmtwo}}{\osha{\traB{\typ}}}
      }
      }{
      \judc{\traB{\tctx}}{\noenv}{\traB{\tm}\esub{\var}{\traB{\tmtwo}}}{\traB{\typtwo}}
    }
    $
  }}

\end{enumerate}
\end{proof}

\begin{definition}
We define a subset $\CtxsSBang \subseteq \CtxsS$ of the set of contexts,
called {\em Bang contexts}:
\[
  \rgctx ::= \open{\ctxhole}
        \mid \open{\sha{\rgctx}}
        \mid \lam{\lvar}{\rgctx\esub{\uvar}{\lvar}}
        \mid \rgctx\,\rtm
        \mid \rtm\,\rgctx
        \mid \osha{\rgctx}
        \mid \rgctx\esub{\uvar}{\rtm}
        \mid \rtm\esub{\uvar}{\rgctx}
\]
where, in the production $\rgctx ::= \lam{\lvar}{\rgctx\esub{\uvar}{\lvar}}$
we assume that $\lvar$ is fresh, that is $\lvar \notin \fv{\rgctx}$.
Furthermore, we define a subset $\SCtxsSBang \subseteq \SCtxsS$
of the set of substitution contexts,
called {\em Bang substitution contexts}:
\[
  \rsctx ::= \ctxhole \mid \rsctx\esub{\uvar}{\rtm}
\]
The inverse translation can be extended to Bang contexts
and substitution contexts,
setting $\traBinv{\open{\ctxhole}} \eqdef \ctxhole$ for Bang contexts,
and $\traBinv{\ctxhole} \eqdef \ctxhole$ for Bang substitution contexts.
\end{definition}

\begin{lemma}[Context decomposition for the inverse Bang translation]
\llem{traBinv_decompose_contexts}
\quad
\begin{enumerate}
\item
  $\tm\sctx \in \TermsBang$
  if and only if
  $\tm \in \TermsBang$ and $\sctx \in \SCtxsSBang$.
\item
  $\off{\gctx}{\var} \in \TermsBang$
  if and only if
  $\gctx \in \CtxsSBang$.
\item
  If $\rgctx \in \CtxsSBang$
  and $\rtm \in \TermsBang$
  then $\off{\rgctx}{\sha{\rtm}} \in \TermsBang$.
\item
  If $\rtm \in \TermsBang$ and $\rsctx \in \SCtxsSBang$,
  then $\traBinv{(\rtm\,\rsctx)} = \traBinv{\rtm}\traNinv{\rsctx}$.
\item
  If $\rgctx \in \CtxsSBang$
  then $\traBinv{\off{\rgctx}{\var}} = \off{\traBinv{\rgctx}}{\var}$.
\item
  If $\rgctx \in \CtxsSBang$ and $\rtm \in \TermsBang$,
  then $\traBinv{\off{\rgctx}{\sha{\rtm}}} = \off{\traBinv{\rgctx}}{\traBinv{\rtm}}$.
\end{enumerate}
\end{lemma}
\begin{proof}
 By induction on the first judgement in the statement of each item, except the first and fourth items which are by induction on $\sctx$ and $\rsctx \in \SCtxsSBang$, resp.
\end{proof}

\begin{remark}
\lremark{traBinv_fv}
$\fv{\traBinv{\rtm}} = \fv{\rtm}$.
\end{remark}

\restate{lemma}{lemma:traBinv_simulation}{Inverse Bang simulation}{
Let $\rtm \in \TermsBang$ and $\tmtwo \in \TermsS$.
If $\rtm \toS \tmtwo$
then $\tmtwo \in \TermsBang$
and $\traBinv{\rtm} \toBang^= \traBinv{\tmtwo}$.
}
\begin{proof}\label{traBinv_simulation:proof}
By induction on the (unique) derivation of $\rtm \in \TermsBang$:
\begin{enumerate}
\item
  $\rtm = \open{\var}$:
  Impossible, as there are no steps $\rtm \toS \tmtwo$.
\item
  $\rtm = \open{\sha{\rtm_1}}$:
  We consider two subcases, depending on whether the step is at the root
  of the term or internal to $\sha{\rtm_1}$:
  \begin{enumerate}
  \item
    If the step is at the root, we have that
    $\rtm = \open{\sha{\rtm_1}} \rtoSopen \rtm_1 = \tmtwo$,
    so $\tmtwo = \rtm_1 \in \TermsBang$
    and
    $\traBinv{\rtm}
     = \traBinv{\open{\sha{\rtm_1}}}
     = \traBinv{\rtm_1}
     = \traBinv{\tmtwo}$.
  \item
    If the step is internal to $\sha{\rtm_1}$,
    note that it cannot be at the root of $\sha{\rtm_1}$,
    since this term does not match the left-hand side of
    any rewriting rule.
    Then the step must be internal to $\rtm_1$,
    that is,
    $\rtm = \open{\sha{\rtm_1}} \toS \open{\sha{\tmtwo'}} = \tmtwo$
    with $\rtm_1 \toS \tmtwo'$.
    By \ih, $\tmtwo' \in \TermsSName$,
    so $\tmtwo = \open{\sha{\tmtwo'}} \in \TermsSName$,
    and
    $\traBinv{\rtm}
     = \traBinv{\open{\sha{\rtm_1}}}
     = \traBinv{\rtm_1}
     \toBang^= \traBinv{\tmtwo'}
     = \traBinv{\open{\sha{\tmtwo'}}}
     = \traBinv{\tmtwo}$.
  \end{enumerate}
\item
  $\rtm = \lam{\lvar}{\rtm'\esub{\var}{\lvar}}$:
  Note that the step cannot be at the root,
  and that that there cannot be a $\toSls$ nor a $\toSgc$ step
  involving the substitution $\esub{\var}{\lvar}$,
  since these rules would require that $\lvar$ be of the form
  $(\ofc{\tmthree})\sctx$, but $\lvar$ is a linear variable.
  This means that the step must be internal to $\rtm'$, 
  that is,
  $\rtm = \lam{\lvar}{\rtm'\esub{\var}{\lvar}}
      \toS \lam{\lvar}{\tmtwo'\esub{\var}{\lvar}} = \tmtwo$
  with $\rtm' \toS \tmtwo'$.
  By \ih, $\tmtwo' \in \TermsSName$,
  so $\tmtwo = \lam{\lvar}{\tmtwo'\esub{\var}{\lvar}} \in \TermsSName$,
  and
  $\traBinv{\rtm}
   = \traBinv{(\lam{\lvar}{\rtm'\esub{\var}{\lvar}})}
   = \lam{\var}{\traBinv{\rtm'}}
   \tocbn^= \lam{\var}{\traBinv{\tmtwo'}}
   = \traBinv{(\lam{\lvar}{\tmtwo'\esub{\var}{\lvar}})}
   = \traBinv{\tmtwo}$.
\item
  $\rtm = \rtm_1\,\rtm_2$:
  We consider three subcases, depending on whether the step is at the root
  of the term, internal to $\rtm_1$, or internal to $\rtm_2$:
  \begin{enumerate}
  \item
    If the step is at the root of the term, it must be a $\symSdb$ step,
    that is, $\rtm_1$ must be of the form $(\lam{\lvar}{\tm})\sctx$.
    By \rlem{traBinv_decompose_contexts},
    this means that $\lam{\lvar}{\tm} \in \TermsSName$
    and $\sctx \in \SCtxsSName$.
    In particular, $\tm$ must be of the form $\tm = \rtm'_1\esub{\var}{\lvar}$.
    Then we have that
    $\rtm = (\lam{\lvar}{\rtm'_1\esub{\var}{\lvar}})\sctx\,\osha{\rtm_2}
     \rtoSdb \rtm'_1\esub{\var}{\osha{\rtm_2}}\sctx = \tmtwo$.
    Note that
    $\tmtwo = \rtm'_1\esub{\var}{\osha{\rtm_2}}\sctx \in \TermsSName$
    by \rlem{traBinv_decompose_contexts}.
    By \ih, we have that
    \[
      \begin{array}{rll}
        \traBinv{\rtm}
      & = &
        \traBinv{((\lam{\lvar}{\rtm'_1\esub{\var}{\lvar}})\sctx\,\osha{\rtm_2})}
      \\
      & = &
        (\lam{\var}{\traBinv{{\rtm'_1}}})\traBinv{\sctx}\,\traBinv{\rtm_2}
      \\
      & \rtodb &
        \traBinv{{\rtm'_1}}\esub{\var}{\traBinv{\rtm_2}}\traBinv{\sctx}
      \\
      & = &
        \traBinv{(\rtm'_1\esub{\var}{\osha{\rtm_2}}\sctx)}
      \\
      & = &
        \traBinv{\tmtwo}
      \end{array}
    \]
  \item
    If the step is internal to $\rtm_1$,
    we have that
    $\rtm = \rtm_1\,\rtm_2 \toS \tmtwo_1\,\rtm_2 = \tmtwo$
    with $\rtm_1 \toS \tmtwo_1$.
    By \ih, $\tmtwo_1 \in \TermsSName$,
    so $\tmtwo = \tmtwo_1\,\rtm_2 \in \TermsSName$
    and
    $\traBinv{\rtm}
     = \traBinv{\rtm_1}\,\traBinv{\rtm_2}
     \tocbn^= \traBinv{\tmtwo_1}\,\traBinv{\rtm_2}
     = \traBinv{\tmtwo}$.

       \item
    If the step is internal to $\rtm_2$,
    we have that
    $\rtm = \rtm_1\,\rtm_2 \toS \rtm_1\,\tmtwo_2 = \tmtwo$
    with $\rtm_2 \toS \tmtwo_2$. Similar to the previous case.

   \end{enumerate}

\item $\rtm = \osha{\rtm_1}$.  
    If the step is internal to $\osha{\rtm_1}$,
    note that it cannot be at the root of $\osha{\rtm_1}$
    nor at the root of $\sha{\rtm_1}$,
    since these terms do not match the left-hand side of any rewriting rule.
    So the step must be internal to $\rtm_1$,
    that is, we have that
    $\rtm = \osha{\rtm_1} \toS \osha{\tmtwo_1} = \tmtwo$
    with $\rtm_1 \toS \tmtwo_1$.
    By \ih,
    $\tmtwo_1 \in \TermsBang$,
    so $\tmtwo = \osha{\tmtwo_1} \in \TermsBang$
    and
    $\traBinv{\rtm}
     = \traBinv{\rtm_1}
     \toBang^= \traBinv{\tmtwo_1}
     = \traBinv{\tmtwo}$.
\item
  $\rtm = \rtm_1\esub{\var}{\rtm_2}$:
  We consider three subcases, depending on whether the step is at the root
  of the term, internal to $\rtm_1$, or internal to $\rtm_2$:
  \begin{enumerate}
  \item
    If the step is at the root of the term,
    it must be either a $\symSls$ or a $\symSgc$ step.
    We consider two further subcases:
    \begin{enumerate}
    \item
      If the step is a $\symSls$ step, then
      $\rtm_2=(\osha{\rtm_3})\rsctx$ and we have that
      $\rtm_1 = \off{\gctx}{\var} \in \TermsBang$
      and
      $\rtm
       = \off{\gctx}{\var}\esub{\var}{(\osha{\rtm_3})\rsctx}
       \rtoSls \off{\gctx}{\sha{\rtm_2}}\esub{\var}{(\osha{\rtm_3})}\rsctx
       = \tmtwo$.
      Note that, by \rlem{traBinv_decompose_contexts},
      $\gctx \in \CtxsSBang$,
      so, again by \rlem{traBinv_decompose_contexts},
      $\tmtwo = \off{\gctx}{\sha{\rtm_2}}\esub{\var}{\osha{\rtm_2}}\rsctx \in \TermsSName$.
      Moreover, also using \rlem{traBinv_decompose_contexts},
      we have that
      $\traBinv{\rtm}
       = \traBinv{(\off{\gctx}{\var}\esub{\var}{(\osha{\rtm_2})\rsctx})}
       = \off{\traBinv{\gctx}}{\var}\esub{\var}{\traBinv{\rtm_2}}
       \rtols \off{\traBinv{\gctx}}{\traBinv{\rtm_2}}\esub{\var}{\traBinv{\rtm_2}}
       = \traBinv{(\off{\gctx}{\sha{\rtm_2}}\esub{\var}{\osha{\rtm_2}})}
       = \traBinv{\tmtwo}$.
    \item
      If the step is a $\symSgc$ step, we have that $\var \notin \fv{\rtm_1}$
      and $\rtm = \rtm_1\esub{\var}{\osha{\rtm_2}} \rtoSgc \rtm_1 = \tmtwo$.
      Note that $\tmtwo = \rtm_1 \in \TermsSName$.
      Moreover,
      $\traBinv{\rtm}
       = \traBinv{\rtm_1\esub{\var}{\osha{\rtm_2}}}
       = \traBinv{\rtm_1}\esub{\var}{\traBinv{\rtm_2}}
       \rtogc \traBinv{\rtm_1}
       = \traBinv{\tmtwo}$.
      Note that $\var \notin \traBinv{\rtm_1}$
      because $\fv{\traBinv{\rtm_1}} = \fv{\rtm_1}$,
      as noted in \rremark{traBinv_fv}.
    \end{enumerate}
    
  \item
    If the step is internal to $\rtm_1$,
    we have that
    $\rtm = \rtm_1\esub{\var}{\rtm_2}
     \toS \tmtwo_1\esub{\var}{\rtm_2} = \tmtwo$
     with $\rtm_1 \toS \tmtwo_1$.
    By \ih, $\tmtwo_1 \in \TermsBang$,
    so $\tmtwo = \tmtwo_1\esub{\var}{\rtm_2} \in \TermsBang$
    and
    $\traBinv{\rtm}
     = \traBinv{\rtm_1}\esub{\var}{\traBinv{\rtm_2}}
     \tocbn^= \traBinv{\tmtwo_1}\esub{\var}{\traBinv{\rtm_2}}
     = \traBinv{\tmtwo}$.
  \item
    If the step is internal to $\rtm_2$,
    we have that
    $\rtm = \rtm_1\esub{\var}{\rtm_2}
     \toS \rtm_1\esub{\var}{\tmtwo_2} = \tmtwo$
     with $\rtm_2 \toS \tmtwo_2$.
    By \ih,
    $\tmtwo_2 \in \TermsBang$,
    so $\tmtwo = \rtm_1\esub{\var}{\tmtwo_2} \in \TermsBang$ 
    and
    $\traBinv{\rtm}
     = \traBinv{\rtm_1}\esub{\var}{\traBinv{\rtm_2}}
     \toBang^= \traBinv{\rtm_1}\esub{\var}{\traBinv{\tmtwo_2}}
     = \traBinv{\tmtwo}$.
  \end{enumerate}
\end{enumerate}
\end{proof}


\end{ifLongAppendix}

\end{document}
